\documentclass[11pt,a4paper,twoside]{article}

\usepackage[utf8]{inputenc}
\usepackage[T1]{fontenc}
\usepackage[english]{babel}
\usepackage{lmodern}
\usepackage{microtype}
\usepackage{tocloft}
\usepackage[a4paper,margin=2.5cm]{geometry}

\usepackage{amsmath,amssymb,amsthm,mathtools}
\usepackage{mathrsfs}

\usepackage[dvipsnames]{xcolor}
\definecolor{OTBlue}{HTML}{1B4F72}
\definecolor{OTGreen}{HTML}{006D5B}
\definecolor{OTOrange}{HTML}{D9822B}
\definecolor{OTGray}{HTML}{F3F5F7}
\usepackage[colorlinks=true,linkcolor=OTBlue,citecolor=OTBlue,urlcolor=OTBlue]{hyperref}

\usepackage{graphicx}
\usepackage{float}
\usepackage{booktabs}
\usepackage{array}
\usepackage{enumitem}
\usepackage{listings}
\usepackage{tikz}
\usetikzlibrary{arrows.meta,positioning,calc,shapes.geometric}

\usepackage[most]{tcolorbox}
\tcbset{sharp corners,boxrule=.5pt,colback=OTGray,colframe=OTBlue!70!black}
\newtcolorbox{actuarialbox}[1][]{title=Actuarial takeaway,#1}
\newtcolorbox{computebox}[1][]{title=Computational note,colframe=OTGreen!70!black,#1}
\newtcolorbox{readingbox}[1][]{title=Further reading,colframe=GoldenRod!80!black,#1}

\usepackage[backend=bibtex,style=authoryear,maxcitenames=2,maxbibnames=99]{biblatex}
\newcounter{moncompteur}
\usepackage{fancyhdr}
\usepackage{fontawesome5}
\usepackage{lastpage}
\usepackage{titlesec}
\titleformat{\section}{\normalfont\Large\bfseries\color{OTBlue}}{\thesection}{1em}{}
\titleformat{\subsection}{\normalfont\large\bfseries\color{OTBlue!85!black}}{\thesubsection}{1em}{}
\titleformat{\subsubsection}{\normalfont\normalsize\bfseries\color{OTBlue!70!black}}{\thesubsubsection}{1em}{}

\theoremstyle{definition}
\newtheorem{definition}{Definition}[section]
\newtheorem{example}{Example}[section]
\newtheorem{exercise}{Exercise}[section]

\theoremstyle{plain}
\newtheorem{theorem}{Theorem}[section]
\newtheorem{proposition}{Proposition}[section]

\newtheorem{corollary}{Corollary}[section]

\theoremstyle{remark}
\newtheorem{remark}{Remark}[section]

\newcommand{\R}{\mathbb{R}}
\newcommand{\Proba}{\mathbb{P}}
\newcommand{\N}{\mathbb{N}}
\newcommand{\E}{\mathbb{E}}
\newcommand{\Ebb}{\mathbb{E}}
\newcommand{\Pbb}{\mathbb{P}}
\newcommand{\Qbb}{\mathbb{Q}}
\newcommand{\1}{\mathbf{1}}
\newcommand{\dd}{\,\mathrm{d}}
\newcommand{\eps}{\varepsilon}
\newcommand{\Id}{\mathrm{Id}}
\DeclareMathOperator{\Var}{Var}

\DeclareMathOperator{\Corr}{Corr}
\DeclareMathOperator{\Law}{Law}

\DeclareMathOperator*{\argmin}{arg\,min}

\usepackage{xcolor}

\definecolor{couleurA}{HTML}{00A08A}
\definecolor{couleurB}{HTML}{F2AD00}

\definecolor{GoldenRod}{HTML}{DAA520}
\definecolor{Rose}{HTML}{E4A1AC}

\definecolor{OTsource}{HTML}{00A08A}
\definecolor{OTsourceLight}{HTML}{80D0C4}
\definecolor{OTsourceVeryLight}{HTML}{E6F6F3}

\definecolor{OTtarget}{HTML}{F2AD00}
\definecolor{OTtargetLight}{HTML}{F8D680}
\definecolor{OTtargetVeryLight}{HTML}{FEF7E6}

\definecolor{OTplan}{HTML}{4D4D4D}
\definecolor{OTgrid}{HTML}{D9D9D9}
\definecolor{OTwarning}{HTML}{D55E00}
\definecolor{OTbary}{HTML}{7A5DC7}

\newcommand{\Premium}{\Pi}
\newcommand{\Risk}{\rho}
\newcommand{\VaR}{\mathrm{VaR}}
\newcommand{\TVaR}{\mathrm{TVaR}}
\newcommand{\ES}{\mathrm{ES}}
\newcommand{\SCR}{\mathrm{SCR}}
\newcommand{\BEL}{\mathrm{BEL}}

\newcommand{\Prob}{\mathcal{P}}
\newcommand{\Probp}[1]{\mathcal{P}_{#1}}
\newcommand{\Coup}{\Pi}
\newcommand{\OT}{\mathrm{OT}}

\newcommand{\Wp}[1]{\mathsf{W}_{#1}}
\newcommand{\Wtwo}{\mathsf{W}_{2}}
\newcommand{\Wone}{\mathsf{W}_{1}}
\newcommand{\push}{\#}
\newcommand{\KL}{\mathrm{KL}}

\newcommand{\GW}{\mathrm{GW}}

\newcommand{\bK}{\boldsymbol{K}}
\newcommand{\bC}{\boldsymbol{C}}
\newcommand{\bc}{\boldsymbol{c}}
\newcommand{\bA}{\boldsymbol{A}}
\newcommand{\ba}{\boldsymbol{a}}
\newcommand{\bb}{\boldsymbol{b}}
\newcommand{\bm}{\boldsymbol{m}}
\newcommand{\bx}{\boldsymbol{x}}
\newcommand{\bw}{\boldsymbol{w}}
\newcommand{\bu}{\boldsymbol{u}}
\newcommand{\bv}{\boldsymbol{v}}
\newcommand{\bV}{\boldsymbol{V}}
\newcommand{\bH}{\boldsymbol{H}}
\newcommand{\bl}{\boldsymbol{l}}
\newcommand{\bbeta}{\boldsymbol{\beta}}
\newcommand{\bp}{\boldsymbol{p}}
\newcommand{\bP}{\boldsymbol{P}}
\newcommand{\bL}{\boldsymbol{L}}
\newcommand{\by}{\boldsymbol{y}}
\newcommand{\bz}{\boldsymbol{z}}
\newcommand{\bs}{\boldsymbol{s}}
\newcommand{\bt}{\boldsymbol{t}}
\newcommand{\br}{\boldsymbol{r}}
\newcommand{\bB}{\boldsymbol{B}}

\newcommand{\bZ}{\boldsymbol{Z}}
\newcommand{\be}{\boldsymbol{e}}
\newcommand{\bR}{\boldsymbol{R}}
\newcommand{\bT}{\boldsymbol{T}}
\newcommand{\bW}{\boldsymbol{W}}
\newcommand{\bD}{\boldsymbol{D}}

\newcommand{\bX}{\boldsymbol{X}}
\newcommand{\bq}{\boldsymbol{q}}
\newcommand{\bk}{\boldsymbol{k}}
\newcommand{\bY}{\boldsymbol{Y}}
\newcommand{\btheta}{\boldsymbol{\theta}}
\newcommand{\bgamma}{\boldsymbol{\gamma}}
\newcommand{\bzero}{\boldsymbol{0}}

\newcommand{\lecture}[1]{\section{#1}}

\newcommand{\highlight}[2]{\colorbox{#1!30}{$\displaystyle{#2}$}}

\title{\textbf{Optimal Transport for Actuarial Science}\\[.2cm]}
\author{Arthur Charpentier\\\small charpentier.arthur@uqam.ca}
\date{Université du Québec à Montréal (UQAM) \&
Kyoto University\\[.5cm]
Draft version 1, July 2026\\}

\begin{document}

\maketitle
\thispagestyle{empty}

% \begin{abstract}
% These notes introduce optimal transport methods with a focus on actuarial science. The guiding idea is that many actuarial objects are distributional: losses, premiums, reserves, capital requirements, climate scenarios, and group-specific scores. Optimal transport provides a geometry for comparing, interpolating, correcting, and stress-testing such distributions. The notes cover discrete and continuous optimal transport, Wasserstein distances and barycenters, entropic regularization, statistical aspects, and applications to risk measures, pricing, reserving, climate risk, fairness, reinsurance, and capital allocation.
% \end{abstract}

% \vfill
% \noindent\textbf{Suggested prerequisites.} Probability, mathematical statistics, actuarial risk models, and basic convex optimization. Familiarity with risk measures is useful but not required.

\vspace{10cm}

\noindent To cite this document,

\begin{verbatim}
@misc{charpentier2026optimaltransport,
  author        = {Arthur Charpentier},
  title         = {Optimal Transport for Actuarial Science},
  month         = {July},
  year          = {2026},
  publisher     = {Kyoto University Lecture Notes}
}
\end{verbatim}

\newpage
\thispagestyle{empty}
\null
\newpage

\setcounter{tocdepth}{2}
\tableofcontents

\newpage
\lecture{Introduction: Why Optimal Transport in Actuarial Science?}
\label{lec:introduction}

\subsection{Learning objectives}

This introductory lecture explains why optimal transport is a useful language for actuarial science. After reading it, the reader should be able to:
\begin{enumerate}[label=(\roman*)]
\item recognize actuarial quantities as probability measures rather than only as numbers;
\item explain the difference between comparing expectations and comparing full laws;
\item use the probability integral transform and inverse transform sampling as basic rank--quantile tools;
\item interpret a transport cost as an actuarial modelling decision;
\item identify actuarial problems in which distributional movement is the main object;
\item understand, at a high level, the historical path from Monge's earthmoving problem to modern statistical, computational and actuarial uses of optimal transport;
\item read the rest of the notes as a sequence of tools to compare, transform, interpolate and stress actuarial distributions.
\end{enumerate}

\begin{readingbox}
These notes use optimal transport as a geometry for actuarial distributions: losses, premiums, reserves, capital scenarios, climate losses, scores and counterfactual outcomes. General references for optimal transport include \cite{villani2003topics,villani2008oldnew,ambrosio2021lectures,figalli2021invitation}. For computational optimal transport, see \cite{peyre2019computational,solomon2018discrete}. For economic and econometric perspectives, see \cite{galichon2016economics,galichonhenry2026econometrician}. The classical line from Monge, Kantorovich and Brenier is recalled in Section~\ref{sec:intro-history}; see also \cite{monge1781memoire,kantorovich1942translocation,brenier1991polar,mccann1997convexity}. For actuarial background on risk models, dependence and statistical insurance, see \cite{denuit2005actuarial,wuthrich2023statistical,charpentier2024insurance}.
\end{readingbox}

\subsection{Insurance is a distributional business}
\label{sec:intro-distributional-business}

Insurance is often introduced through expectations. A pure premium is written as
\[
  \pi_0 = \E[L],
\]
where \(L\) is a loss. In the rest of these notes, \(\pi(\bX)\) will usually denote a premium or tariff rule depending on rating variables \(\bX\), while \(\pi_0\) denotes a scalar pooled pure premium. A reserve is sometimes described as a best estimate. A solvency capital requirement is sometimes reported as a single number. A pricing model may be summarized by its average loss ratio or by a few validation statistics.

\begin{figure}
    \centering
    \includegraphics[width=0.75\linewidth]{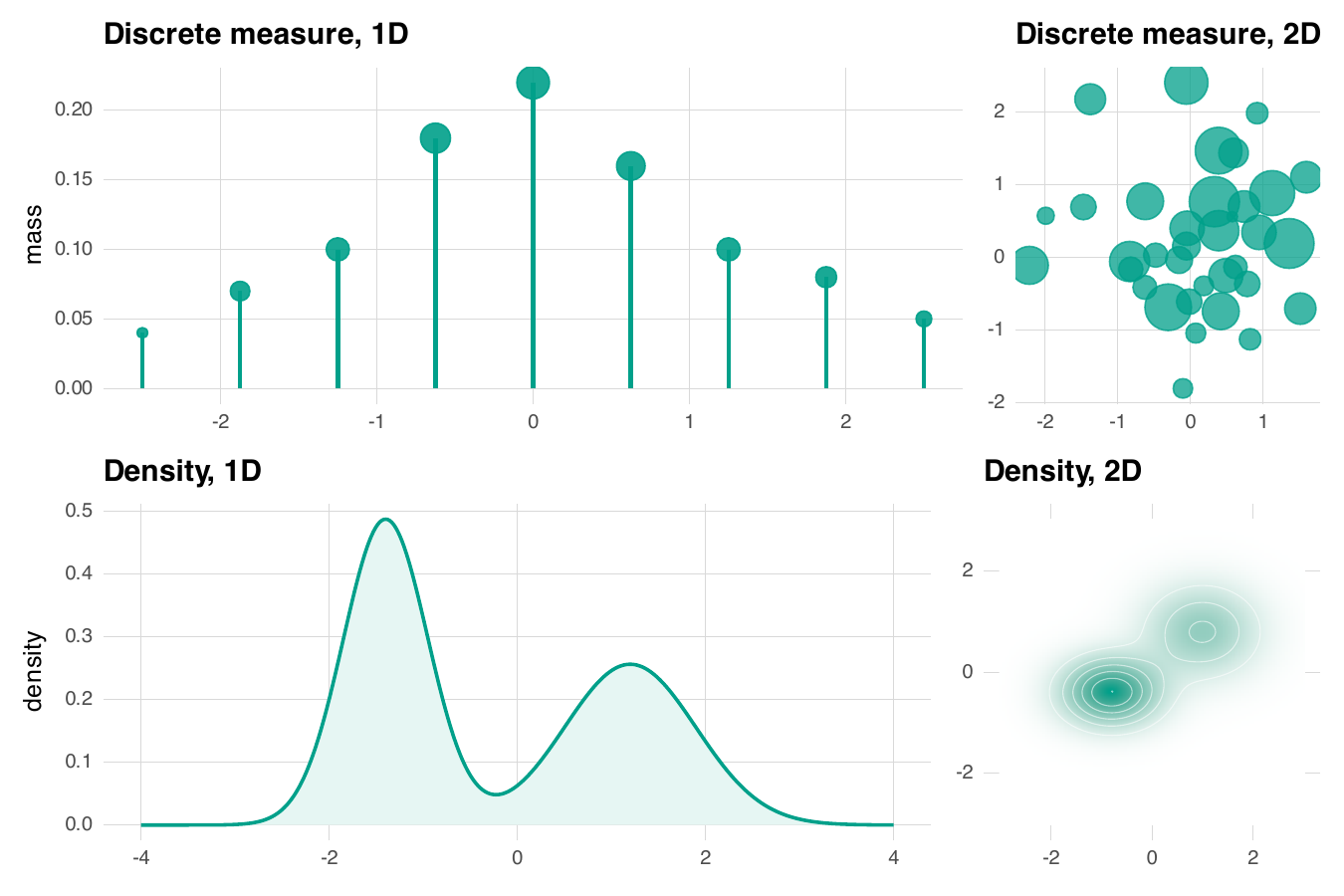}
    \caption{Discrete and continuous actuarial measures. The same actuarial object can be represented either by atoms, point clouds, histograms or densities. In optimal transport, both the weights and the geometry of their support matter.}
    \label{fig:intro-actuarial-measures}
\end{figure}

This scalar view is necessary, but it is incomplete. In actuarial science, the basic object is usually a probability law. A policy does not generate only an expected loss; it generates a full distribution of possible losses. A portfolio is not described only by its average premium; it is described by the distribution of premiums, losses, residuals, reserves and capital contributions across policyholders and scenarios. A climate intervention is not only a change in mean loss; it is a change in the whole loss distribution. A fairness correction is not only a change in group means; it is a transformation of group-specific score distributions.

Throughout the notes, we write laws such as
\[
  \mu_L = \Law(L),
  \qquad
  \mu_{\pi}=\Law(\pi(\bX)),
  \qquad
  \mu_{\bR}=\Law(\bR),\qquad \bR=(R_1,\ldots,R_T).
\]
where \(L\) is a loss, \(\bX\) is a vector of rating variables, \(\pi(\bX)\) is a pricing rule, and \(\bR\) is a vector of future cash-flows. Many actuarial questions can then be written as questions about how one law is transformed into another.

\begin{definition}[Actuarial law]
\label{def:intro-actuarial-law}
An \emph{actuarial law} is a probability measure \(\mu\in\Prob(\mathcal X)\) representing the distribution of an actuarial object on a chosen state space \(\mathcal X\). The state space may be monetary, such as \(\mathcal X=\R_+\) for claim severities or premiums; multivariate, such as \(\mathcal X=\R_+^T\) for cash-flow vectors; or mixed, such as a product of numerical, categorical and geographical covariates. The modelling choice is therefore twofold: the law \(\mu\) and the geometry put on \(\mathcal X\).
\end{definition}

Figure~\ref{fig:intro-actuarial-measures} illustrates this point: a law may be represented by atoms, a point cloud, a histogram or a density, but optimal transport keeps track of both probability mass and location.

\begin{actuarialbox}
The basic unit of analysis in these notes is often a probability measure, not a scalar. A change in underwriting, a tariff revision, a prevention policy, a climate scenario or a fairness repair can be represented as a movement from one distribution to another.
\end{actuarialbox}

\subsection{A first transport problem}
\label{sec:intro-first-transport}

Let \(\mu\) and \(\nu\) be two probability measures on a space \(\mathcal X\). For example, \(\mu\) may be the current distribution of claim severities and \(\nu\) the distribution after inflation, prevention or climate stress. A transport plan describes how mass from \(\mu\) is matched to mass from \(\nu\). If \(c(\bx,\by)\) is the cost of moving an actuarial state \(\bx\) into an actuarial state \(\by\), the optimal transport problem is
\[
\highlight{GoldenRod}{
  \inf_{\gamma\in\Coup(\mu,\nu)}
  \int_{\mathcal X\times\mathcal X} c(\bx,\by)\,\dd\gamma(\bx,\by)}.
\]
Here \(\Coup(\mu,\nu)\) denotes the set of all couplings of \(\mu\) and \(\nu\), that is, all joint laws with first marginal \(\mu\) and second marginal \(\nu\). The coupling \(\gamma\) says which parts of the source distribution are matched to which parts of the target distribution.

This formula is deliberately abstract. Its actuarial meaning depends on the choice of the state space and of the cost. The same notation will be made more formal in Lecture~2, where couplings are introduced in detail, and in Lecture~4, where the continuous Kantorovich problem is stated.

\begin{definition}[Ground cost]
\label{def:intro-ground-cost}
A \emph{ground cost} is a measurable function \(c:\mathcal X\times\mathcal X\to[0,+\infty]\) such that \(c(\bx,\by)\) measures the cost, effort or inadmissibility of changing state \(\bx\) into state \(\by\). When \(c(\bx,\by)=d(\bx,\by)^p\) for a metric \(d\) and \(p\ge1\), the resulting optimal value defines the Wasserstein distance \(W_p\). In actuarial applications, however, \(c\) may encode business, regulatory or physical constraints and need not always be a pure metric cost.
\end{definition}

\begin{itemize}
\item If \(x\) and \(y\) are claim amounts, \(c(x,y)=|x-y|\) measures average monetary displacement.
\item If \(x\) and \(y\) are premium levels, \(c(x,y)=|x-y|^2\) penalizes large tariff movements.
\item If \(\bx\) and \(\by\) are policyholder profiles, \(c\) encodes which covariate changes are actuarially meaningful.
\item If \(\bx\) and \(\by\) are cash-flow vectors, \(c\) may combine timing and amount discrepancies.
\item If \(\bx\) and \(\by\) are geographical locations, \(c\) may be built from physical distance, hazard distance or administrative distance.
\end{itemize}

The cost is therefore not a technical detail. It is the modelling language of the problem.

\begin{actuarialbox}
Optimal transport is not only a way to compute a distance. It is a way to state which distributional changes are considered small, large, reasonable, costly or inadmissible.
\end{actuarialbox}

\begin{actuarialbox}%[Distributional, not causal]
A transport map or a coupling aligns two distributions. It does not by itself identify what would have happened to a given policyholder, contract, claim or scenario under an intervention. Such an interpretation requires a structural model, an intervention design, or additional identifying assumptions. This distinction will matter later for prevention, fairness and counterfactual analysis.
\end{actuarialbox}

\subsection{Two rank--quantile facts used throughout}
\label{sec:intro-quantile-facts}

A large part of these notes relies on the link between ranks, quantiles and one-dimensional transport. The following two elementary facts explain why a quantile map sends one distribution to another.

Let \(F\) be a distribution function on \(\R\). We use the lower generalized inverse
\[
  F^{-1}(u)=\inf\{x\in\R:F(x)\ge u\},
  \qquad u\in(0,1).
\]
When the distribution is denoted by \(\mu\), we write interchangeably \(Q_\mu(u)=F_\mu^{-1}(u)\) for its quantile function. This convention is used throughout the notes.

\begin{actuarialbox}%[Quantile convention]
Throughout the notes, \(Q_\mu\) denotes the lower generalized inverse. In particular, \(\VaR_\alpha\) is read in this lower-quantile sense unless stated otherwise. With atoms, ties or software quantile conventions, an implementation may not coincide exactly with this mathematical convention. Whenever empirical quantiles are used in a lab, the convention should be stated.
\end{actuarialbox}

\begin{proposition}[Probability integral transform]
\label{prop:intro-probability-integral-transform}
Let \(X\) be a real-valued random variable with continuous distribution function \(F\). Then
\[
  U=F(X)
\]
has the uniform distribution on \([0,1]\).
\end{proposition}

\begin{proof}
For \(u\in[0,1]\), continuity of \(F\) gives \(F(F^{-1}(u))=u\), with the usual endpoint convention. Hence, up to null boundary sets,
\[
  \Proba(F(X)\le u)=\Proba(X\le F^{-1}(u))=F(F^{-1}(u))=u.
\]
Thus \(F(X)\sim \mathrm{Unif}(0,1)\).
\end{proof}

\begin{proposition}[Inverse transform sampling]
\label{prop:intro-inverse-transform-sampling}
Let \(U\sim\mathrm{Unif}(0,1)\), and let \(F\) be a distribution function on \(\R\). Then
\[
  X=F^{-1}(U)
\]
has distribution function \(F\).
\end{proposition}

\begin{proof}
For any \(x\in\R\), by definition of the generalized inverse,
\[
  F^{-1}(U)\le x
  \quad\Longleftrightarrow\quad
  U\le F(x)
\]
up to endpoint sets of Lebesgue measure zero. Therefore
\[
  \Proba(F^{-1}(U)\le x)=\Proba(U\le F(x))=F(x).
\]
Thus \(F^{-1}(U)\) has distribution function \(F\).
\end{proof}

Combining the two facts gives the basic one-dimensional transport mechanism.

\begin{definition}[Rank-preserving quantile map]
\label{def:intro-rank-preserving-quantile-map}
Let \(\mu\) and \(\nu\) be probability measures on \(\R\), with distribution functions \(F_\mu\) and \(F_\nu\). Their \emph{quantile coupling} is
\[
  \bigl(Q_\mu(U),Q_\nu(U)\bigr),\qquad U\sim\mathrm{Unif}(0,1),
\]
where \(Q_\mu=F_\mu^{-1}\) and \(Q_\nu=F_\nu^{-1}\). If \(F_\mu\) is continuous, this coupling can be represented by the rank-preserving map
\[
  T_{\mu\to\nu}(x)=Q_\nu(F_\mu(x)).
\]
\end{definition}

The map in Definition~\ref{def:intro-rank-preserving-quantile-map} sends a source observation to the target observation with the same percentile. With atoms, the quantile coupling remains well defined, while the deterministic map \(Q_\nu\circ F_\mu\) should be read with care; a pointwise deterministic representation may require a convention for ties, and randomized ranks give the cleanest formulation. This distinction will reappear in one-dimensional transport, fairness repair and longevity stresses.

\subsection{A short history of optimal transport}
\label{sec:intro-history}

Optimal transport began as a problem about moving physical material. The classical origin is Monge's eighteenth-century problem \cite{monge1781memoire}: given a pile of soil and a hole, how should one move the soil so as to minimize total work? In modern notation, one looks for a map \(T\) that sends a source distribution \(\mu\) to a target distribution \(\nu\), and solves
\[
 \highlight{GoldenRod}{\min_{T} \int c(\bx,T(\bx))\,\dd\mu(\bx)}.
\]
This is now called the Monge problem. Its key feature is determinism: each source point \(\bx\) is sent to a single destination \(T(\bx)\).

The Monge formulation is intuitive, but mathematically difficult. The constraint \(T_\#\mu=\nu\) is nonlinear, admissible maps may fail to exist, and mass cannot be split. These difficulties also arise in actuarial work. A rating cell may need to be split across several target cells. An atom at zero claim may need to be partially moved and partially kept. A scenario may not have a unique natural successor. Such situations are awkward in Monge's deterministic formulation.

A decisive shift came with Kantorovich's relaxation \cite{kantorovich1942translocation}. Instead of requiring a map, one allows a transport plan \(\gamma\), a joint distribution with prescribed marginals. The problem becomes
\[
  \inf_{\gamma\in\Coup(\mu,\nu)} \int c(\bx,\by)\,\dd\gamma(\bx,\by).
\]
This is linear in \(\gamma\). In finite spaces, it is a linear program. Its dual introduces potentials, shadow prices and sensitivity quantities. At this point, optimal transport becomes an optimization problem as much as a geometric problem. For this reason, OT sits naturally between probability, convex analysis, operations research and economics.

The next major step was to understand when Kantorovich's relaxed problem returns a deterministic map. For the quadratic cost on \(\R^d\), Brenier's theorem \cite{brenier1991polar} shows that, under absolute continuity of the source measure, the optimal plan is induced by a map of the form
\[
  T(\bx)=\nabla\varphi(\bx),
\]
where \(\varphi\) is convex. This result links optimal transport with convex analysis, partial differential equations and the geometry of probability measures. In one dimension, the same phenomenon is simpler: the optimal map is the monotone quantile map
\[
  T(x)=F_\nu^{-1}(F_\mu(x)).
\]
This formula is one of the practical facts used most often in these notes.

During the late twentieth century, Wasserstein distances became central in probability theory, analysis and geometry. The optimal value for the cost \(d(\bx,\by)^p\) defines
\[
  W_p(\mu,\nu)
  =
  \left(
  \inf_{\gamma\in\Coup(\mu,\nu)}
  \int d(\bx,\by)^p\,\dd\gamma(\bx,\by)
  \right)^{1/p}.
\]
This turns spaces of probability measures into metric spaces. Probability laws can then be compared, interpolated, averaged and studied geometrically. This is why one can speak of Wasserstein barycenters, Wasserstein balls, Wasserstein geodesics and Wasserstein gradients.

A second renaissance came from computation. Entropic regularization, revived in large-scale data analysis by \cite{cuturi2013sinkhorn}, replaces the linear transport problem by a smoother problem, typically of the form
\[
  \inf_{\gamma\in\Coup(\mu,\nu)}
  \left\{
    \int c\,\dd\gamma
    + \varepsilon \sum_{i,j}\gamma_{ij}(\log\gamma_{ij}-1)
  \right\}
\]
in the discrete case. This leads to Sinkhorn algorithms, fast iterative scaling methods that can handle much larger problems than generic linear programming. The computational viewpoint made OT widely usable in machine learning, statistics, imaging, economics and data science.

For actuarial science, the timing is natural. Modern insurance increasingly works with distributions produced by models: simulated capital scenarios, reserving predictive distributions, climate loss distributions, telematics profiles, premium distributions, fairness audit distributions and stress-test outputs. These objects are not only scalar summaries. They are probability masses on structured spaces. Optimal transport provides tools to compare and transform them.

\begin{actuarialbox}
The historical movement from Monge to Kantorovich to Wasserstein to Sinkhorn is also a movement from deterministic transformation to probabilistic matching, from geometry to optimization, and from abstract theory to scalable computation. This combination is useful for modern actuarial applications.
\end{actuarialbox}

\subsection{Why not only classical distances?}
\label{sec:intro-why-wasserstein}

Before using optimal transport, one might ask why classical distances between probability distributions are not enough. Total variation, Hellinger distance, Kullback--Leibler divergence and Kolmogorov distance are important. They answer useful questions. But they often ignore the geometry of the state space.

\begin{definition}[Geometry-aware distributional distance]
\label{def:intro-geometry-aware-distance}
A distance or divergence between laws is \emph{geometry-aware} if its value depends not only on how much mass differs, but also on where the mass is located in the underlying state space. Wasserstein distances are geometry-aware because they are built from the ground cost of Definition~\ref{def:intro-ground-cost}.
\end{definition}

For example, a small shift in all losses by one monetary unit may have a large impact under some non-geometric distances if the distributions are singular or empirical. Wasserstein distance recognizes that moving a claim from \(1000\) to \(1001\) is small, while moving it from \(1000\) to \(100000\) is large. This matters because the magnitude of the move is part of the actuarial question.

The following toy example is useful. Let
\[
  \mu=\delta_0,
  \qquad
  \nu=\delta_\varepsilon.
\]
These two distributions are mutually singular whenever \(\varepsilon\neq 0\), so total variation sees them as far apart. But Wasserstein gives
\[
  W_p(\mu,\nu)=|\varepsilon|.
\]
Thus Wasserstein distance is continuous with respect to physical or economic movement in the underlying space.

\begin{actuarialbox}
For losses, premiums, scores, covariates and cash-flows, the state space has meaning. A distance between distributions should often know whether two states are close or far. This is the central advantage of optimal transport.
\end{actuarialbox}

\subsection{Running actuarial examples}
\label{sec:intro-running-examples}

The same mathematical objects will appear repeatedly in different actuarial contexts. The following examples will guide the notes.

\paragraph{Severity and inflation.}
Let \(L\) be a claim severity and let \(\mu=\Law(L)\). Inflation, legal changes or claims handling changes may move \(\mu\) to a new distribution \(\nu\). In one dimension, the rank-preserving stress map is
\[
  T(x)=F_\nu^{-1}(F_\mu(x)).
\]
This sends each baseline claim to the target claim at the same percentile. This example is developed later in the lectures on continuous OT, Wasserstein quantiles and risk measures.

\paragraph{Risk measures and robustness.}
Risk measures such as \(\VaR\), \(\TVaR\) and spectral risk measures are functionals of loss distributions. Wasserstein balls
\[
  \{\nu: W_p(\nu,\widehat\mu)\le r\}
\]
allow one to ask how large a risk measure can become under a controlled perturbation of the baseline model. This is the subject of the lecture on risk measures and Wasserstein robustness.

\iffalse
% To be restored in v2, after submission of the portfolio demutualization paper.
\paragraph{Pricing and mutualization.}
A pricing rule \(m(\bX)\) induces a premium distribution \(\Law(m(\bX))\). A pooled tariff corresponds to a concentrated distribution, while a highly segmented tariff is more dispersed. Wasserstein distances can quantify the movement from pooling toward profiling. This is developed in the lecture on pricing, segmentation and mutualization, and in the premium-drift lab.
\fi

\paragraph{Portfolio drift and domain adaptation.}
Let \(\mu_t=\Law(\bX_t)\) and \(\mu_{t+1}=\Law(\bX_{t+1})\) be portfolio covariate distributions at two dates. Wasserstein distance measures how much the portfolio has moved in covariate space. Transport plans can also be used to reweight or match portfolios. This is developed in the lecture on portfolio drift, calibration and domain adaptation.

\paragraph{Reserving and cash-flow distributions.}
A stochastic reserving model produces a distribution of future cash-flow vectors \((R_1,\ldots,R_T)\). OT compares complete predictive distributions, not only best estimates. Costs can account for amount differences and timing differences. This is developed in the lectures and labs on reserving and cash-flow distributions.

\paragraph{Climate risk and prevention.}
A prevention measure may move a baseline loss distribution \(\mu_0\) to a post-prevention distribution \(\mu_1\). Wasserstein distance measures the size of the whole distributional change. Dynamic OT and Schr\"odinger bridges then add a path from \(\mu_0\) to \(\mu_1\). This connects to the lectures on reserving, climate prevention and dynamic transport.

\paragraph{Fairness and discrimination.}
Let \(S\) be a sensitive group variable and let \(Z\) be a score or predicted premium. The group-specific laws \(\Law(Z\mid S=s)\) can be compared using Wasserstein distances. Transport maps can align these distributions with a reference law, producing a distributional post-processing rule. This is developed in the fairness lecture and in the Wasserstein fairness repair lab.

\paragraph{Reinsurance and dependence uncertainty.}
Aggregate losses depend on the coupling between lines of business. With fixed marginal laws, different dependence structures lead to different capital requirements. Multi-marginal transport provides a language for such problems; this appears in the lecture on reinsurance, capital allocation and multi-marginal transport.

\subsection{What optimal transport adds}
\label{sec:intro-what-ot-adds}

Optimal transport adds six ingredients to the actuarial toolkit.

\paragraph{A geometry of distributions.}
The Wasserstein distance \(W_p(\mu,\nu)\) compares full distributions while respecting the geometry of the state space. This is useful when the state is monetary, geographical, temporal or multivariate.

\paragraph{A transformation map.}
When an optimal map exists, OT does not only give a distance. It gives a transformation \(T\) such that \(T_\#\mu=\nu\). This is useful for stress testing, calibration and counterfactual analysis, subject to the caution above about causal interpretation.

\paragraph{A coupling or audit trail.}
When no deterministic map is appropriate, OT gives a plan \(\gamma\). This plan records how source mass is matched to target mass. In actuarial applications, this can be read as an audit trail for portfolio migration, tariff change, reserve scenario transformation or climate stress.

\paragraph{A robustness device.}
Wasserstein balls define neighbourhoods around empirical or fitted laws. They allow actuaries to compute worst-case expectations, prudent premiums and robust capital over plausible distributional perturbations; see, for example, \cite{esfahani2018data,blanchet2019quantifying} for distributionally robust optimization with Wasserstein ambiguity sets.

\paragraph{A barycenter.}
Wasserstein barycenters provide a way to average probability laws without washing out their geometry \cite{agueh2011barycenters}. They will be used for consensus scenarios, common reference distributions and mortality-scenario summaries.

\paragraph{A path.}
Displacement interpolation, dynamic optimal transport and Schr\"odinger bridges provide trajectories between laws \cite{mccann1997convexity,benamou2000computational,leonard2014survey}. They are useful when the actuarial question is not only how far two distributions are, but how one distribution could evolve into another over time.

\begin{actuarialbox}
Optimal transport is useful because it gives a distance, a map, a coupling, a path, a barycenter and a robustness neighbourhood. These become actuarial objects once losses, premiums, reserves and scores are treated as distributions.
\end{actuarialbox}

\subsection{How the notes are organized}
\label{sec:intro-structure}

The notes are organized in three layers.

\paragraph{Mathematical foundations.}
The first lectures introduce couplings, push-forwards, discrete optimal transport, continuous Kantorovich duality, Wasserstein distances, barycenters, entropic regularization and statistical aspects. These lectures provide the common language used throughout the applications.

\paragraph{Actuarial applications.}
The middle part studies risk measures, robust capital, portfolio drift, reserving, climate prevention, fairness, reinsurance, capital allocation and longevity. Each application uses the same mathematical objects in a different state space with a different cost.

\paragraph{Extensions and modelling choices.}
The later lectures and appendices cover counterfactual transport, dynamic OT, unbalanced OT, longevity transport, sliced Wasserstein distances, cost engineering and computational labs. These sections emphasize modelling choices: which cost, which radius, which reference distribution, which path and which diagnostic.

\subsection{A small dictionary}
\label{sec:intro-dictionary}

\begin{table}[ht]
\centering
\begin{tabular}{ll}
\toprule
Mathematical object & Actuarial interpretation \\
\midrule
\(\mu\in\Prob(\mathcal X)\) & Distribution of losses, premiums, scores, reserves or scenarios \\
\(T_\#\mu\) & Law after applying a tariff, stress, deductible or post-processing map \\
\(Q_\mu=F_\mu^{-1}\) & Quantile function of a one-dimensional law \\
\(\gamma\in\Coup(\mu,\nu)\) & Matching, dependence model or scenario audit trail \\
\(c(\bx,\by)\) & Cost of changing actuarial state \(\bx\) into state \(\by\) \\
\(W_p(\mu,\nu)\) & Size of a distributional displacement \\
Kantorovich potentials & Shadow prices or sensitivity functions \\
Quantile map & Rank-preserving severity, score or premium transformation \\
Barycenter & Consensus distribution or common reference law \\
Wasserstein ball & Model-uncertainty or stress neighbourhood \\
Dynamic OT path & Trajectory of distributions over time \\
Unbalanced OT & Distributional comparison with changing exposure mass \\
\bottomrule
\end{tabular}
\caption{A first dictionary between optimal transport objects and actuarial interpretations.}
\label{tab:intro-dictionary}
\end{table}

\subsection{What to remember}
\label{sec:intro-remember}

The guiding message of these notes is that actuarial science is increasingly distributional. We do not only estimate means; we compare laws. We do not only report capital numbers; we analyze loss distributions. We do not only revise premiums; we study premium distributions. We do not only state fairness metrics; we compare group-specific score laws. We do not only compare climate scenarios; we construct paths between them.

Optimal transport provides a coherent mathematical language for these tasks. It starts from a simple question: how should one move mass from one distribution to another? The rest of the notes explore how far this question can go in actuarial science.

\subsection{Exercises}
\label{sec:intro-exercises}

\begin{exercise}[Distributional objects]
\label{ex:intro-distributional-objects}
Give three actuarial quantities that are often reported as scalars but are more naturally viewed as probability distributions. For each quantity, specify the underlying state space.
\end{exercise}

\begin{exercise}[Cost as modelling choice]
\label{ex:intro-cost-choice}
Using the language of ground costs from Definition~\ref{def:intro-ground-cost}, suppose \(x\) and \(y\) are two premium levels. Compare the actuarial meaning of \(c(x,y)=|x-y|\), \(c(x,y)=|x-y|^2\), and \(c(x,y)=|\log x-\log y|\). Which one is most appropriate for measuring premium disruption?
\end{exercise}

\begin{exercise}[Wasserstein versus total variation]
\label{ex:intro-wasserstein-tv}
Let \(\mu=\delta_0\) and \(\nu=\delta_\varepsilon\) on \(\R\). Compute \(W_1(\mu,\nu)\). Compare with total variation distance. Explain which answer is more natural for a monetary displacement.
\end{exercise}

\begin{exercise}[A first quantile stress]
\label{ex:intro-quantile-stress}
Let \(\mu\) be an exponential distribution with rate \(\lambda\) and \(\nu\) an exponential distribution with rate \(\theta\). Derive the rank-preserving map \(T(x)=F_\nu^{-1}(F_\mu(x))\). Interpret the case \(\theta<\lambda\) as a severity stress.
\end{exercise}

\begin{exercise}[Scalar versus distributional comparison]
\label{ex:intro-scalar-vs-distributional}
Construct two distributions with the same mean but different dispersion. Explain why comparing only expected losses misses an actuarial difference that Wasserstein distance can detect.
\end{exercise}

\begin{exercise}[Running example]
\label{ex:intro-running-example}
Choose one of the running examples in Section~\ref{sec:intro-running-examples}. Identify the source distribution, the target distribution, the state space and a reasonable first choice of ground cost.
\end{exercise}

\newpage
\lecture{Couplings, Push-Forwards and Insurance Dependence}
\label{lec:couplings}

\subsection*{Learning objectives}

This lecture sets up the probabilistic language used throughout the notes. After this lecture, the reader should be able to:
\begin{enumerate}[label=(\roman*)]
  \item describe actuarial transformations as push-forwards of probability measures;
  \item distinguish marginal modelling from dependence modelling;
  \item manipulate the set of couplings $\Coup(\mu,\nu)$ and interpret it as a feasible set;
  \item use deterministic maps, stochastic kernels, quantile couplings and copulas as different representations of dependence;
  % To be restored in v2 after the fairness material is submitted.
  % \item explain why aggregation, capital, reinsurance and fairness problems are often problems over couplings.
  \item explain why aggregation, capital and reinsurance problems are often problems over couplings.
\end{enumerate}

\begin{readingbox}
The probabilistic notation used here is standard in optimal transport; see \cite{villani2003topics,villani2008oldnew,ambrosio2021lectures,figalli2021invitation}. For background on couplings, weak convergence and regular conditional probabilities, see also \cite{billingsley1999convergence,kallenberg2021foundations}. For computational and statistical perspectives, see \cite{peyre2019computational,chewi2025statistical}. The actuarial motivation from dependence, risk orders and aggregation is closely related to \cite{denuit2005actuarial,dhaene2002concept,embrechts2002correlation}. For copulas and Fr\'echet classes, see \cite{sklar1959fonctions,nelsen2006introduction,joe1997multivariate}. These notes use couplings rather than copulas as the primary language because couplings are the feasible objects in Kantorovich's formulation of optimal transport, and because couplings keep the original actuarial state space on which a ground cost is defined.
\end{readingbox}

\subsection{From marginal distributions to joint models}
\label{sec:couplings-marginal-to-joint}

Insurance modelling often begins with marginal distributions. A motor book may have a severity distribution for bodily injury, another distribution for property damage, a frequency model for claims counts and a distribution of policyholder covariates. Reserving models provide predictive distributions for future cash-flows. Climate models provide distributions of event intensities and losses. Pricing models provide distributions of premiums across a portfolio; if \(\Premium(\bX)\) denotes the tariff assigned to covariates \(\bX\), the premium law is typically \(\Law(\Premium(\bX))\).

Optimal transport treats a distribution not only as an object to be estimated, but also as an object that can be transformed, compared, interpolated and constrained. The first operation we need is the push-forward: a probability law is transported through a measurable map. The second is the coupling: two probability laws are embedded into a joint law. Push-forwards describe how distributions change under transformations; couplings describe how marginal laws can be made jointly compatible.

The distinction is actuarially important. Knowing the marginal distributions of two lines of business does not determine the distribution of their aggregate loss. Knowing the marginal distributions of current and future losses does not determine a scenario path. All these missing pieces are couplings.
% To be restored in v2 after the fairness material is submitted: knowing the distribution of scores in two protected groups does not determine how individuals are matched under a post-processing rule.

\subsection{Measurable spaces and push-forwards}
\label{sec:couplings-pushforwards}

Let $(\mathcal X,\mathcal B(\mathcal X))$ and $(\mathcal Y,\mathcal B(\mathcal Y))$ be measurable spaces. In most of these notes, $\mathcal X$ and $\mathcal Y$ will be Polish spaces, that is, complete separable metric spaces, equipped with their Borel $\sigma$-fields. This assumption covers most actuarial applications on $\R^d$ and gives the regular conditional probabilities and compactness properties used below.

\begin{definition}[Push-forward]
\label{def:push-forward}
Let $\mu\in\Prob(\mathcal X)$, and let $T:\mathcal X\to\mathcal Y$ be measurable. The push-forward of $\mu$ by $T$, denoted $T_{\push}\mu$, is the probability measure on $\mathcal Y$ defined by
\[
  T_{\push}\mu(B)=\mu(T^{-1}(B)),
  \qquad B\in\mathcal B(\mathcal Y).
\]
Equivalently, if $X\sim\mu$, then $T(X)\sim T_{\push}\mu$.
\end{definition}

Push-forwards formalize the basic statistical operation ``apply a model to a random input''. If $\bX$ is a vector of rating variables and $\Premium(\bX)$ is a premium, then $\Law(\Premium(\bX))=\Premium_{\push}\Law(\bX)$. If $L$ is a gross loss and $T(L)=(L-d)_+\wedge u$ is an insured payment with deductible $d$ and limit $u$, then $\Law(T(L))=T_{\push}\Law(L)$.

\begin{proposition}[Change of variables for push-forwards]
\label{prop:pushforward-cov}
Let $T:\mathcal X\to\mathcal Y$ be measurable and let $\mu\in\Prob(\mathcal X)$. For every nonnegative measurable function $\varphi:\mathcal Y\to[0,\infty]$,
\[
  \int_{\mathcal Y}\varphi(y)\,(T_{\push}\mu)(\dd y)
  =
  \int_{\mathcal X}\varphi(T(x))\,\mu(\dd x).
\]
The same identity holds for any integrable real-valued $\varphi$.
\end{proposition}

\begin{proof}
This is the standard change-of-variables formula for image measures; see, for example, the measure-theoretic preliminaries in \cite{villani2008oldnew} and the introductory material in \cite{ambrosio2021lectures}. Sketch of the proof: first prove the identity for indicators $\varphi=\1_B$, where it is exactly the definition of $T_{\push}\mu$. Extend by linearity to nonnegative simple functions. Then use monotone convergence for nonnegative measurable functions, and finally decompose an integrable real-valued function into positive and negative parts.
\end{proof}

\begin{example}[Deductible and policy limit]
Let $L\ge0$ be a claim amount with law $\mu$, deductible $d\ge0$, and policy limit $u\in(0,\infty]$. The insurer's payment is
\[
  T(L)=(L-d)_+\wedge u.
\]
The distribution of the insured payment is $T_{\push}\mu$. This distribution may have atoms at $0$ and at $u$, even when $\mu$ is continuous. A policy transformation changes the full probability measure, not only its mean or tail summaries.
\end{example}

\begin{example}[Distribution of premiums]
Let $\bX$ be policyholder covariates and let $\Premium:\mathcal X\to\R_+$ be a pricing rule. The premium distribution in the portfolio is
\[
  \mu_{\mathrm{prem}}=\Law(\Premium(\bX))=\Premium_{\push}\Law(\bX).
\]
Replacing \(\Premium_0\) by \(\Premium_1\) moves the portfolio from \((\Premium_0)_{\push}\Law(\bX)\) to \((\Premium_1)_{\push}\Law(\bX)\). Later, Wasserstein distances will measure how far this tariff change moves the whole premium distribution, not just its mean.
\end{example}

\subsection{Couplings}
\label{sec:couplings-definition}

Let $\mu\in\Prob(\mathcal X)$ and $\nu\in\Prob(\mathcal Y)$. Let $p_1(x,y)=x$ and $p_2(x,y)=y$ be the canonical projections from $\mathcal X\times\mathcal Y$.

\begin{definition}[Coupling]
\label{def:coupling}
A coupling of $\mu$ and $\nu$ is a probability measure $\gamma\in\Prob(\mathcal X\times\mathcal Y)$ such that
\[
  (p_1)_{\push}\gamma=\mu,
  \qquad
  (p_2)_{\push}\gamma=\nu.
\]
The set of all such couplings is denoted by
\[
  \Coup(\mu,\nu)
  =
  \bigl\{\gamma\in\Prob(\mathcal X\times\mathcal Y):
  (p_1)_{\push}\gamma=\mu,
  (p_2)_{\push}\gamma=\nu\bigr\}.
\]
\end{definition}

\begin{definition}[Product coupling]
\label{def:product-coupling}
The \emph{product coupling} of \(\mu\) and \(\nu\) is the product measure \(\mu\otimes\nu\in\Coup(\mu,\nu)\). If \((X,Y)\sim\mu\otimes\nu\), then \(X\sim\mu\), \(Y\sim\nu\), and \(X\) and \(Y\) are independent.
\end{definition}

Equivalently, $\gamma\in\Coup(\mu,\nu)$ if
\[
  \gamma(A\times\mathcal Y)=\mu(A),
  \qquad
  \gamma(\mathcal X\times B)=\nu(B)
\]
for all measurable $A\subseteq\mathcal X$, $B\subseteq\mathcal Y$. If $(X,Y)\sim\gamma$, then $X\sim\mu$ and $Y\sim\nu$. Conversely, the joint law of any pair $(X,Y)$ with these marginals is a coupling.

The terminology varies across the literature. Villani uses \emph{transference plan}; probabilists often say \emph{joint distribution with fixed marginals}; actuaries may say \emph{dependence model} or \emph{Fr{\'e}chet class element}. These are the same objects.

\begin{proposition}[Non-emptiness and convexity]
\label{prop:coupling-nonempty-convex}
For any $\mu\in\Prob(\mathcal X)$ and $\nu\in\Prob(\mathcal Y)$, the set $\Coup(\mu,\nu)$ is non-empty and convex.
\end{proposition}

\begin{proof}
Non-emptiness follows because the product measure $\mu\otimes\nu$ has marginals $\mu$ and $\nu$. For convexity, let $\gamma_0,\gamma_1\in\Coup(\mu,\nu)$ and $\theta\in[0,1]$. Then $\gamma_\theta=(1-\theta)\gamma_0+\theta\gamma_1$ is a probability measure, and for every measurable $A\subseteq\mathcal X$,
\[
  \gamma_\theta(A\times\mathcal Y)
  =(1-\theta)\mu(A)+\theta\mu(A)=\mu(A).
\]
The same argument gives the second marginal $\nu$.
\end{proof}

\begin{remark}[Independence is one coupling, not the coupling]
The product coupling $\mu\otimes\nu$ corresponds to independence. It is often chosen for convenience, but it is only one feasible element of $\Coup(\mu,\nu)$. In capital modelling, replacing independence by another coupling can leave all marginal calibrations unchanged while substantially changing the tail of the aggregate loss.
\end{remark}

\begin{proposition}[Compactness under weak convergence]
\label{prop:coupling-compact}
Assume that $\mathcal X$ and $\mathcal Y$ are Polish spaces. Then $\Coup(\mu,\nu)$ is tight and closed in $\Prob(\mathcal X\times\mathcal Y)$ for the topology of weak convergence. Together with Prokhorov's theorem, this implies that $\Coup(\mu,\nu)$ is compact for weak convergence.
\end{proposition}

\begin{proof}
This is Lemma~4.4 and the compactness argument in the proof of Theorem~4.1 of \cite{villani2008oldnew}. Sketch of the proof: the singleton sets $\{\mu\}$ and $\{\nu\}$ are tight. Hence, for every $\eps>0$, there exist compact sets $K_\eps\subseteq\mathcal X$ and $L_\eps\subseteq\mathcal Y$ such that $\mu(K_\eps^c)\le\eps$ and $\nu(L_\eps^c)\le\eps$. For any $\gamma\in\Coup(\mu,\nu)$,
\[
  \gamma\bigl((K_\eps\times L_\eps)^c\bigr)
  \le \gamma(K_\eps^c\times\mathcal Y)+\gamma(\mathcal X\times L_\eps^c)
  \le 2\eps.
\]
This proves tightness. Closedness follows because weak convergence preserves marginals: if $\gamma_n\Rightarrow\gamma$, then for every bounded continuous $f$ on $\mathcal X$,
\[
  \int f(x)\,\gamma(\dd x,\dd y)
  =\lim_{n\to\infty}\int f(x)\,\gamma_n(\dd x,\dd y)
  =\int f\,\dd\mu,
\]
and similarly for $\nu$.
\end{proof}

\begin{actuarialbox}
A coupling is a dependence model with fixed marginals. In actuarial language, $\Coup(\mu,\nu)$ is the feasible set of all joint models compatible with the two marginal calibrations. Optimal transport adds a cost function and selects one coupling by optimization.
\end{actuarialbox}

\subsection{Deterministic couplings and transport maps}
\label{sec:couplings-deterministic}

Recall the push-forward notation from Definition~\ref{def:push-forward}. A special class of couplings is induced by measurable maps. If $T:\mathcal X\to\mathcal Y$ satisfies $T_{\push}\mu=\nu$, then the joint law of $(X,T(X))$, with $X\sim\mu$, is a coupling. Such couplings are concentrated on the graph of $T$. They represent deterministic transport: once the source point is known, the target point is known.

\begin{proposition}[Graph coupling]
\label{prop:graph-coupling}
Let $T:\mathcal X\to\mathcal Y$ be measurable. Then
\[
  T_{\push}\mu=\nu
  \quad\Longleftrightarrow\quad
  (\Id,T)_{\push}\mu\in\Coup(\mu,\nu).
\]
Moreover, $(\Id,T)_{\push}\mu$ is concentrated on the graph $\{(x,T(x)):x\in\mathcal X\}$.
\end{proposition}

\begin{proof}
This is the standard graph-plan construction used in Monge's problem; see Chapter~1 of \cite{villani2003topics}, Chapter~1 of \cite{figalli2021invitation}, and \cite{ambrosio2021lectures}. Sketch of the proof: set $\gamma=(\Id,T)_{\push}\mu$. For measurable $A\subseteq\mathcal X$,
\[
  \gamma(A\times\mathcal Y)=\mu\{x:(x,T(x))\in A\times\mathcal Y\}=\mu(A).
\]
For measurable $B\subseteq\mathcal Y$,
\[
  \gamma(\mathcal X\times B)=\mu\{x:T(x)\in B\}=T_{\push}\mu(B).
\]
Thus the second marginal is $\nu$ exactly when $T_{\push}\mu=\nu$. Concentration on the graph follows from $(x,T(x))\in\operatorname{graph}(T)$ for $\mu$-almost every $x$.
\end{proof}

Not every coupling is induced by a map. A map cannot split one source point into several possible target points. Kantorovich's relaxation replaces deterministic maps by arbitrary couplings, allowing mass splitting.

\begin{example}[Splitting a risk class]
Suppose a group of policyholders with current premium $x$ is split into two possible adjusted premiums $y_1$ and $y_2$, with probabilities depending on $x$. Unless one probability is zero, this is not a deterministic map $T(x)$. It is naturally a kernel $K(x,\dd y)$, hence a coupling $\gamma(\dd x,\dd y)=\mu(\dd x)K(x,\dd y)$.
\end{example}

\subsection{Couplings as stochastic kernels}
\label{sec:couplings-kernels}

A coupling can be read globally as a joint distribution or locally as a transition rule. The local description uses Markov kernels and disintegration.

\begin{definition}[Markov kernel]
\label{def:markov-kernel}
Let \((\mathcal X,\mathcal B(\mathcal X))\) and \((\mathcal Y,\mathcal B(\mathcal Y))\) be measurable spaces. A \emph{Markov kernel} from \(\mathcal X\) to \(\mathcal Y\) is a map
\[
  K:\mathcal X\times\mathcal B(\mathcal Y)\to[0,1]
\]
such that, for each \(x\in\mathcal X\), \(K(x,\cdot)\) is a probability measure on \(\mathcal Y\), and for each \(B\in\mathcal B(\mathcal Y)\), the map \(x\mapsto K(x,B)\) is measurable.
\end{definition}

\begin{theorem}[Disintegration of a coupling]
\label{thm:disintegration}
Let $\mathcal X$ and $\mathcal Y$ be Polish spaces, and let $\gamma\in\Coup(\mu,\nu)$. Then there exists a Markov kernel $K:\mathcal X\times\mathcal B(\mathcal Y)\to[0,1]$ such that
\[
  \gamma(\dd x,\dd y)=\mu(\dd x)K(x,\dd y),
\]
meaning that for every nonnegative measurable $h$,
\[
  \int_{\mathcal X\times\mathcal Y} h(x,y)\,\gamma(\dd x,\dd y)
  =
  \int_{\mathcal X}\left[\int_{\mathcal Y}h(x,y)K(x,\dd y)\right]\mu(\dd x).
\]
The second marginal condition becomes
\[
  \nu(B)=\int_{\mathcal X}K(x,B)\,\mu(\dd x),
  \qquad B\in\mathcal B(\mathcal Y).
\]
\end{theorem}

\begin{proof}
This is the standard disintegration theorem for probability measures on Polish spaces; see the measure-theoretic appendix of \cite{villani2008oldnew} and the preliminaries in \cite{ambrosio2021lectures}. Sketch of the proof: since $\mathcal X$ and $\mathcal Y$ are standard Borel spaces, regular conditional probabilities exist. Apply this to the coordinate projection $(X,Y)\mapsto X$ under $\gamma$. The conditional law of $Y$ given $X=x$ is the kernel $K(x,\cdot)$. Integrating conditional expectations gives the displayed identity.
\end{proof}

The deterministic case is recovered by $K(x,\dd y)=\delta_{T(x)}(\dd y)$. Thus kernels interpolate between deterministic transport maps and arbitrary stochastic reallocations.

\begin{actuarialbox}
A coupling may be read as an audit trail. The kernel $K(x,\dd y)$ says how mass starting at actuarial state $x$ is redistributed toward possible target states $y$. This is useful for portfolio migration, climate scenario transformation and stress testing. % To be restored in v2 after the fairness material is submitted: fairness post-processing.
\end{actuarialbox}

\subsection{Basic couplings on the real line}
\label{sec:couplings-real-line}

The real line is special because quantile functions provide explicit couplings. As in Definition~\ref{def:intro-rank-preserving-quantile-map}, let \(F_\mu(x)=\mu(({-\infty},x])\) be the distribution function of \(\mu\in\Prob(\R)\). We use the generalized inverse
\[
  Q_\mu(u)=F_\mu^{-1}(u)=\inf\{x\in\R:F_\mu(x)\ge u\},
  \qquad u\in(0,1),
\]
as its quantile function. This is the same lower-quantile convention as in Lecture~\ref{lec:introduction}; it is important when atoms are present.

\begin{proposition}[Quantile representation]
\label{prop:quantile-representation}
Let $U\sim\mathrm{Unif}(0,1)$. Then $Q_\mu(U)\sim\mu$. Consequently, for $\mu,\nu\in\Prob(\R)$,
\[
  \bigl(Q_\mu(U),Q_\nu(U)\bigr)
\]
induces a coupling of $\mu$ and $\nu$.
\end{proposition}

\begin{proof}
This is the classical quantile transform, used repeatedly in one-dimensional optimal transport; see Chapter~2 of \cite{villani2003topics}, Chapter~2 of \cite{figalli2021invitation}, and \cite{ambrosio2021lectures}. Sketch of the proof: by the definition of the generalized inverse,
\[
  Q_\mu(u)\le x
  \quad\Longleftrightarrow\quad
  u\le F_\mu(x)
\]
with the usual endpoint convention for the lower generalized inverse. Hence
\[
  \Pbb(Q_\mu(U)\le x)=\Pbb(U\le F_\mu(x))=F_\mu(x).
\]
The two coordinates of the displayed pair therefore have marginal laws $\mu$ and $\nu$.
\end{proof}

The coupling
\[
  X=Q_\mu(U),
  \qquad
  Y=Q_\nu(U)
\]
is called the comonotonic or increasing coupling. High quantiles are matched with high quantiles. The coupling
\[
  X=Q_\mu(U),
  \qquad
  Y=Q_\nu(1-U)
\]
is the countermonotonic or decreasing coupling in dimension two. It matches high quantiles of one marginal with low quantiles of the other.

\begin{definition}[Comonotonicity]
\label{def:comonotonicity}
A pair $(X,Y)$ is comonotonic if there exists a random variable $U$ and nondecreasing functions $f,g$ such that $X=f(U)$ and $Y=g(U)$. For one-dimensional marginals, the quantile construction with a common uniform random variable is the canonical comonotonic coupling.
\end{definition}

\begin{example}[Severity dependence]
Let $X$ be a bodily injury severity and $Y$ a property damage severity. The marginal fits $\Law(X)=\mu$, $\Law(Y)=\nu$ do not determine the law of $X+Y$. The independent, comonotonic and countermonotonic couplings have the same marginal models but different aggregate tails. This is a basic reason dependence modelling matters for solvency capital.
\end{example}

\subsection{Aggregation as a push-forward of a coupling}
\label{sec:couplings-aggregation}

Let $A:\R^2\to\R$ be the aggregation map $A(x,y)=x+y$. If $\gamma\in\Coup(\mu,\nu)$, then the aggregate loss distribution is
\[
  \Law(X+Y)=A_{\push}\gamma,
  \qquad (X,Y)\sim\gamma.
\]
Thus the set of aggregate-loss distributions compatible with two fixed marginal models is
\[
  \mathcal A(\mu,\nu)
  =
  \{A_{\push}\gamma:\gamma\in\Coup(\mu,\nu)\}.
\]

\begin{proposition}[Marginals do not identify aggregate risk]
\label{prop:aggregate-not-identified}
Let $\mu$ and $\nu$ be probability measures on $\R_+$. The law of $X+Y$ is generally not determined by $\mu$ and $\nu$. More precisely, it is determined by the choice of $\gamma\in\Coup(\mu,\nu)$ through $A_{\push}\gamma$.
\end{proposition}

\begin{proof}
This is a direct consequence of the push-forward formalism, and it is the dependence viewpoint developed in actuarial risk theory; see \cite{denuit2005actuarial,dhaene2002concept,embrechts2002correlation}. Sharp aggregation bounds under fixed marginals are also related to \cite{makarov1981estimates,rueschendorf1982random}. Sketch of the proof: if $(X,Y)\sim\gamma$, then Proposition~\ref{prop:pushforward-cov} gives $\Law(X+Y)=A_{\push}\gamma$. Different couplings can give different push-forward laws. For instance, if both marginals are Bernoulli$(1/2)$, then under the comonotonic coupling $X=Y$, $X+Y\in\{0,2\}$, whereas under the countermonotonic coupling $Y=1-X$, $X+Y=1$ almost surely.
\end{proof}

If $\Risk$ is a law-invariant risk measure, dependence uncertainty can be written as an optimization problem over couplings:
\[
  \sup_{\gamma\in\Coup(\mu,\nu)}\Risk(A_{\push}\gamma),
  \qquad
  \inf_{\gamma\in\Coup(\mu,\nu)}\Risk(A_{\push}\gamma).
\]
These problems have the same feasible set as optimal transport. They become optimal transport problems when the objective can be written as $\int c(x,y)\,\gamma(\dd x,\dd y)$ for a cost function $c$.

\begin{example}[Capital with fixed line marginals]
Suppose $L_1\sim\mu_1$ and $L_2\sim\mu_2$ are calibrated line-of-business losses. The solvency capital for $L_1+L_2$, such as $\VaR_{99.5\%}(L_1+L_2)$, is not determined by $\mu_1$ and $\mu_2$ alone. It depends on $\gamma\in\Coup(\mu_1,\mu_2)$. Independence is a modelling assumption. Comonotonicity is a conservative benchmark. Neither is forced by the marginal calibrations.
\end{example}

\subsection{Fr{\'e}chet classes and copulas}
\label{sec:couplings-frechet-copulas}

For fixed marginals \(\mu_1,\ldots,\mu_d\), the coupling notation extends directly.

\begin{definition}[Fr{\'e}chet class]
\label{def:frechet-class}
Let \(\mu_k\in\Prob(\mathcal X_k)\), \(k=1,\ldots,d\). The \emph{Fr{\'e}chet class} with marginals \(\mu_1,\ldots,\mu_d\) is
\[
  \Coup(\mu_1,\ldots,\mu_d)
  =
  \left\{\gamma\in\Prob(\mathcal X_1\times\cdots\times\mathcal X_d):
  (p_k)_{\push}\gamma=\mu_k,\; k=1,\ldots,d\right\}.
\]
It is the set of all joint laws compatible with the prescribed marginal models.
\end{definition}

\begin{definition}[Copula]
\label{def:copula}
A \(d\)-dimensional \emph{copula} is a distribution function on \([0,1]^d\) whose one-dimensional marginals are uniform on \([0,1]\). Equivalently, it is the joint law of a random vector \((U_1,\ldots,U_d)\) such that each \(U_k\sim\mathrm{Unif}(0,1)\).
\end{definition}

Copulas provide another representation of elements of the Fr{\'e}chet class when \(\mathcal X_k=\R\). The representation is unique when the marginal distribution functions are continuous; without continuity, copulas still exist but need not be unique.

\begin{theorem}[Sklar representation, informal form]
\label{thm:sklar-informal}
Let $H$ be a joint distribution function on $\R^d$ with continuous marginals $F_1,\ldots,F_d$. Then there exists a unique copula $C$ such that
\[
  H(x_1,\ldots,x_d)
  =
  C(F_1(x_1),\ldots,F_d(x_d)).
\]
Conversely, any copula $C$ combined with marginals $F_1,\ldots,F_d$ defines a joint distribution with those marginals.
\end{theorem}

\begin{proof}
This is Sklar's theorem \cite{sklar1959fonctions}; in an actuarial dependence context, see \cite{denuit2005actuarial,nelsen2006introduction,joe1997multivariate}. Sketch of the proof: when the marginals are continuous, define $U_k=F_k(X_k)$. Each $U_k$ is uniform on $[0,1]$, and the joint law of $(U_1,\ldots,U_d)$ is the copula $C$. Conversely, if $(U_1,\ldots,U_d)\sim C$ and $Q_k=F_k^{-1}$ denotes the generalized inverse, then $(Q_1(U_1),\ldots,Q_d(U_d))$ has marginals $F_1,\ldots,F_d$. Uniqueness follows from continuity of the marginals.
\end{proof}

\begin{remark}[Couplings versus copulas]
Copulas separate marginal distributions from dependence by transforming variables to ranks. Couplings keep the original state spaces and therefore allow a cost $c(x,y)$ to depend on actual levels. This matters in insurance: moving a loss from $100$ to $120$ is not the same actuarial operation as moving a loss from $10{,}000$ to $10{,}020$, even if the rank displacement is similar.
\end{remark}

\subsection{Fr{\'e}chet--Hoeffding bounds}
\label{sec:couplings-frechet-hoeffding}

For two real-valued random variables with distribution functions $F$ and $G$, every joint distribution function $H$ with these marginals satisfies the Fr{\'e}chet--Hoeffding bounds
\[
  \max\{F(x)+G(y)-1,0\}
  \le H(x,y)
  \le \min\{F(x),G(y)\}.
\]
The upper bound is attained by the comonotonic coupling. For two marginals, the lower bound is attained by the countermonotonic coupling. In dimensions larger than two, the lower Fr{\'e}chet--Hoeffding expression is not always itself a copula, so pairwise intuition should not be transferred mechanically to all multivariate settings.

\begin{theorem}[Fr{\'e}chet--Hoeffding bounds]
\label{thm:frechet-hoeffding}
Let $(X,Y)$ have marginal distribution functions $F$ and $G$, and joint distribution function $H$. Then, for all $(x,y)\in\R^2$,
\[
  \max\{F(x)+G(y)-1,0\}
  \le H(x,y)
  \le \min\{F(x),G(y)\}.
\]
Both bounds are sharp.
\end{theorem}

\begin{proof}
This is the classical Fr{\'e}chet--Hoeffding bound for bivariate distributions; see \cite{frechet1951sur,hoeffding1940maszstabinvariante} and the dependence treatment in \cite{denuit2005actuarial,nelsen2006introduction}. Sketch of the proof: since $\{X\le x,Y\le y\}\subseteq\{X\le x\}$ and similarly for $Y$, we have $H(x,y)\le F(x)\wedge G(y)$. For the lower bound, use
\[
  \Pbb(A\cap B)=1-\Pbb(A^c\cup B^c)
  \ge 1-\Pbb(A^c)-\Pbb(B^c)
  =\Pbb(A)+\Pbb(B)-1
\]
with $A=\{X\le x\}$ and $B=\{Y\le y\}$, and also $H\ge0$. Sharpness follows from the quantile couplings $(F^{-1}(U),G^{-1}(U))$ and $(F^{-1}(U),G^{-1}(1-U))$, with the usual generalized-inverse conventions.
\end{proof}

\subsection{Expectation objectives over couplings}
\label{sec:couplings-expectation-objectives}

Many dependence problems have the form
\[
  \sup_{\gamma\in\Coup(\mu,\nu)}\int_{\mathcal X\times\mathcal Y} h(x,y)\,\gamma(\dd x,\dd y)
\]
or the corresponding infimum. This is the Kantorovich problem with cost $-h$ or $h$. The next statement records the elementary variational structure.

\begin{proposition}[Linear objectives over couplings]
\label{prop:linear-coupling-objective}
Let $h:\mathcal X\times\mathcal Y\to\R$ be bounded measurable. The map
\[
  \gamma\longmapsto \int h\,\dd\gamma
\]
is affine on $\Coup(\mu,\nu)$. If $h$ is upper semicontinuous and bounded above, then the supremum over $\Coup(\mu,\nu)$ is attained under the Polish-space assumptions of Proposition~\ref{prop:coupling-compact}; if $h$ is lower semicontinuous and bounded below, then the infimum is attained.
\end{proposition}

\begin{proof}
This is the same compactness and semicontinuity mechanism as Theorem~4.1 of \cite{villani2008oldnew}; see also the Kantorovich formulation in \cite{villani2003topics}. Sketch of the proof: affinity follows from linearity of integration. For existence, apply compactness of $\Coup(\mu,\nu)$ from Proposition~\ref{prop:coupling-compact}. If $h$ is lower semicontinuous and bounded below, $\gamma\mapsto\int h\,\dd\gamma$ is lower semicontinuous by the portmanteau theorem. Thus a minimizing sequence has a weakly convergent subsequence, and the limit attains the infimum. The upper semicontinuous case follows by applying the same argument to $-h$.
\end{proof}

\begin{example}[Maximum covariance with fixed marginals]
For square-integrable real marginals $\mu$ and $\nu$, maximizing $\E[XY]$ over $\Coup(\mu,\nu)$ is an optimal transport problem with reward $h(x,y)=xy$. In dimension one, the increasing quantile coupling maximizes this expectation, while the decreasing quantile coupling minimizes it under suitable integrability. This is the rearrangement principle behind comonotonic extremality.
\end{example}

\begin{proposition}[Rearrangement principle on the line]
\label{prop:rearrangement}
Let $\mu,\nu\in\Prob(\R)$ have finite second moments. Then
\[
  \sup_{\gamma\in\Coup(\mu,\nu)}\int xy\,\gamma(\dd x,\dd y)
  =
  \int_0^1 Q_\mu(u)Q_\nu(u)\,\dd u,
\]
and
\[
  \inf_{\gamma\in\Coup(\mu,\nu)}\int xy\,\gamma(\dd x,\dd y)
  =
  \int_0^1 Q_\mu(u)Q_\nu(1-u)\,\dd u.
\]
\end{proposition}

\begin{proof}
This is the Hardy--Littlewood rearrangement inequality \cite{hardy1952inequalities}, which is also the one-dimensional optimal transport structure discussed in \cite{villani2003topics,figalli2021invitation}. Sketch of the proof: for the supremum, approximate the marginals by empirical distributions with equal weights. The finite-dimensional result is the classical rearrangement inequality: two sorted sequences maximize the scalar product, while opposite sorting minimizes it. Passing to quantile limits gives the formulas. An alternative proof uses cyclical monotonicity for the cost $c(x,y)=-xy$: optimality requires that mass is supported on a monotone set, which is exactly the increasing rearrangement in one dimension.
\end{proof}

\subsection{A complete two-point example}
\label{sec:couplings-two-point}

Let $X\sim\mathrm{Bernoulli}(p)$ and $Y\sim\mathrm{Bernoulli}(q)$, with state space $\{0,1\}$. A coupling is a $2\times2$ table
\[
\begin{array}{c|cc|c}
 &Y=0&Y=1& \\\hline
X=0&\gamma_{00}&\gamma_{01}&1-p\\
X=1&\gamma_{10}&\gamma_{11}&p\\\hline
 &1-q&q&1
\end{array}
\]
The table is determined by $a=\gamma_{11}$. The constraints imply
\[
  \gamma_{10}=p-a,
  \qquad
  \gamma_{01}=q-a,
  \qquad
  \gamma_{00}=1-p-q+a.
\]
Nonnegativity gives
\[
  \max\{0,p+q-1\}\le a\le \min\{p,q\}.
\]
Thus $\Coup(\mu,\nu)$ is an interval. The lower endpoint is the strongest negative dependence compatible with the marginals; the upper endpoint is the comonotonic endpoint.

If $S=X+Y$, then
\[
  \Pbb(S=2)=a,
  \qquad
  \Pbb(S=1)=p+q-2a,
  \qquad
  \Pbb(S=0)=1-p-q+a.
\]
The aggregate distribution varies linearly with the coupling parameter $a$. The marginals are fixed throughout.

\begin{actuarialbox}
This two-point example is the smallest possible model of dependence uncertainty. It is also the finite-dimensional picture behind discrete optimal transport: the feasible set is a transportation polytope, the objective is linear in the entries of $\gamma$, and optimal solutions often live on faces or extreme points of that polytope.
\end{actuarialbox}

\subsection{Empirical couplings}
\label{sec:couplings-empirical}

In applications we often work with empirical measures
\[
  \hat\mu_n=\sum_{i=1}^n a_i\delta_{x_i},
  \qquad
  \hat\nu_m=\sum_{j=1}^m b_j\delta_{y_j},
\]
where $a_i,b_j\ge0$, $\sum_i a_i=\sum_j b_j=1$. A coupling is then a nonnegative matrix $\Gamma=(\gamma_{ij})$ satisfying
\[
  \sum_{j=1}^m \gamma_{ij}=a_i,
  \qquad
  \sum_{i=1}^n \gamma_{ij}=b_j.
\]
This is the transportation polytope. It is the object studied in detail in the next lecture.

\begin{proposition}[Empirical couplings are matrices]
\label{prop:empirical-couplings}
Let $\hat\mu_n=\sum_i a_i\delta_{x_i}$ and $\hat\nu_m=\sum_j b_j\delta_{y_j}$. Every coupling $\gamma\in\Coup(\hat\mu_n,\hat\nu_m)$ can be written as
\[
  \gamma=\sum_{i=1}^n\sum_{j=1}^m \gamma_{ij}\delta_{(x_i,y_j)},
\]
where $\Gamma=(\gamma_{ij})$ is a nonnegative matrix with row sums $a_i$ and column sums $b_j$. Conversely, every such matrix defines a coupling.
\end{proposition}

\begin{proof}
This is the finite-support specialization of Kantorovich's transference plans; see the discrete formulation in \cite{peyre2019computational} and \cite{solomon2018discrete}. Sketch of the proof: since the first marginal is supported on $\{x_i\}$ and the second on $\{y_j\}$, the coupling is supported on $\{x_i\}\times\{y_j\}$. Set $\gamma_{ij}=\gamma(\{(x_i,y_j)\})$. The marginal constraints are exactly the row and column sum equations. The converse follows by direct summation.
\end{proof}

\subsection{Mini-lab: three couplings, three aggregate losses}
\label{sec:couplings-mini-lab-three-couplings}

The following short R code illustrates the same marginals under three different couplings.

\begin{lstlisting}[language=R]
set.seed(123)
n <- 20000

# two severity marginals
qX <- function(u) qlnorm(u, meanlog = 8.0, sdlog = 0.7)
qY <- function(u) qlnorm(u, meanlog = 7.5, sdlog = 0.9)

u <- runif(n)

# comonotonic
X_com <- qX(u)
Y_com <- qY(u)
S_com <- X_com + Y_com

# countermonotonic
X_ctr <- qX(u)
Y_ctr <- qY(1-u)
S_ctr <- X_ctr + Y_ctr

# independent
X_ind <- qX(runif(n))
Y_ind <- qY(runif(n))
S_ind <- X_ind + Y_ind

rbind(
  independent     = quantile(S_ind, c(.95, .99, .995)),
  comonotonic     = quantile(S_com, c(.95, .99, .995)),
  countermonotone = quantile(S_ctr, c(.95, .99, .995))
)
\end{lstlisting}

The output shows that the high quantiles of the aggregate are typically much larger under comonotonicity than under countermonotonicity. The marginal quantiles of $X$ and $Y$ are unchanged. This small experiment is a preview of the later sections on risk aggregation, reinsurance and capital allocation, and of the computational labs where empirical couplings are represented by matrices.

\subsection{Couplings as scenario design}
\label{sec:couplings-scenario-design}

Couplings are not only tools for bivariate dependence. They also provide a language for scenario design.

\begin{itemize}
  \item A coupling between today's portfolio distribution and next year's portfolio distribution describes a migration scenario.
  \item A coupling between baseline climate losses and post-prevention losses describes how losses are displaced by an intervention.
  % To be restored in v2 after the fairness material is submitted.
  % \item A coupling between raw scores and fair scores describes an audit trail for a post-processing algorithm.
  \item A coupling between simulated reserving cash-flow vectors and stress cash-flow vectors describes how a stress test moves the distribution of liabilities.
\end{itemize}

In each case, the source and target marginal distributions are not enough. The coupling records which source states are matched to which target states. Optimal transport will select this matching by minimizing a cost. These examples are developed later in the lectures on portfolio drift and domain adaptation, reserving and cash-flow distributions, counterfactual worlds, and reinsurance/capital allocation. % To be restored in v2 after the fairness and demutualization material is submitted: the premium drift and fairness labs reuse exactly this before--after coupling logic.

\subsection{A warning: coupling is not causality}
\label{sec:couplings-not-causality}

A coupling is a joint probability measure with fixed marginals. It does not, by itself, define a structural causal mechanism. For instance, a coupling between a baseline loss $L_0$ and a post-intervention loss $L_1$ may be useful for measuring distributional displacement, but it does not prove that a given policyholder's loss would have changed from $L_0$ to $L_1$. Causal interpretation requires additional assumptions, such as a structural model, ignorability, rank preservation or an explicit intervention design; see, for example, the structural-causal perspective in \cite{pearl2009causality,peters2017elements}.

\begin{actuarialbox}
A coupling aligns distributions. It does not identify unit-level effects unless additional causal assumptions are supplied.
\end{actuarialbox}

This warning will matter later for prevention and counterfactual modelling. Wasserstein maps can provide useful distributional attributions, but individual counterfactual claims require more structure than a coupling alone. % To be restored in v2 after the fairness material is submitted: the same warning applies to fairness repairs.

\subsection{Small dictionary for actuarial use}
\label{sec:couplings-dictionary}

\begin{table}
\centering
\begin{tabular}{p{.32\textwidth}p{.55\textwidth}}
\toprule
Mathematical object & Actuarial interpretation \\
\midrule
$\mu\in\Prob(\mathcal X)$ & Distribution of losses, premiums, reserves, scores or scenarios \\
$T_{\push}\mu$ & Distribution after applying a deductible, tariff or stress map; see Definition~\ref{def:push-forward} \\ % To be restored in v2 after the fairness material is submitted: or post-processing rule.
$\gamma\in\Coup(\mu,\nu)$ & Dependence model or scenario matching compatible with two marginals; see Definition~\ref{def:coupling} \\
$\mu\otimes\nu$ & Independence benchmark \\
Graph coupling $(\Id,T)_{\push}\mu$ & Deterministic transformation or Monge map; see Proposition~\ref{prop:graph-coupling} \\
Kernel $K(x,\dd y)$ & Randomized transformation or stochastic transition rule; see Definition~\ref{def:markov-kernel} \\
Quantile coupling & Rank-preserving matching in one dimension; see Proposition~\ref{prop:quantile-representation} \\
Transportation matrix & Empirical or discrete coupling; see Proposition~\ref{prop:empirical-couplings} \\
\bottomrule
\end{tabular}
\caption{Coupling terminology used throughout the actuarial applications.}
\label{tab:couplings-dictionary}
\end{table}

\subsection{From couplings to optimal transport}
\label{sec:couplings-to-ot}

The next lecture adds a ground cost, in the sense of Definition~\ref{def:intro-ground-cost}, $c:\mathcal X\times\mathcal Y\to\R\cup\{+\infty\}$ and studies
\[
  \inf_{\gamma\in\Coup(\mu,\nu)}
  \int_{\mathcal X\times\mathcal Y} c(x,y)\,\gamma(\dd x,\dd y).
\]
The coupling set is the feasible set; the cost selects the preferred dependence or matching. In actuarial language, optimal transport asks: among all ways to match two actuarial distributions, which one has the smallest total transformation cost?

\subsection{Exercises}

\begin{exercise}[Push-forward of a deductible]\label{ex:couplings-deductible-pushforward}\label{ex:2:1}
Let $L$ have distribution function $F$, and define $T(L)=(L-d)_+$ with $d\ge0$. Derive the distribution function of $T(L)$. Identify any atom created at zero.
\end{exercise}

\begin{exercise}[Graph couplings]\label{ex:couplings-graph}\label{ex:2:2}
Let $X\sim\mu$, and let $T:\mathcal X\to\mathcal Y$ be measurable. Show directly that $\Law(X,T(X))$ is concentrated on the graph of $T$. Give an example of a coupling that cannot be concentrated on the graph of any measurable map.
\end{exercise}

\begin{exercise}[Two-point Fr{\'e}chet class]\label{ex:couplings-two-point-frechet}\label{ex:2:3}
Work out the full two-point example when $p=.3$ and $q=.6$. Determine the interval of possible values of $a=\Pbb(X=1,Y=1)$. Compute the possible distributions of $X+Y$.
\end{exercise}

\begin{exercise}[Aggregation under three couplings]\label{ex:couplings-three-aggregate}\label{ex:2:4}
Let $X$ and $Y$ be exponential random variables with means $1$ and $2$. Simulate the independent, comonotonic and countermonotonic couplings. Compare the empirical $.95$, $.99$ and $.995$ quantiles of $X+Y$.
\end{exercise}

\begin{exercise}[Convexity of couplings]\label{ex:couplings-convexity}\label{ex:2:5}
Let $\gamma_0$ be the independent coupling and $\gamma_1$ the comonotonic coupling of two continuous marginals on $\R$. Interpret $(1-\theta)\gamma_0+\theta\gamma_1$ probabilistically. Does it correspond to a copula? Does it correspond to a deterministic map?
\end{exercise}

\begin{exercise}[Worst-case expectation]\label{ex:couplings-worst-case-expectation}\label{ex:2:6}
Let $\mu=\nu=\frac12\delta_0+\frac12\delta_1$. Compute
\[
  \sup_{\gamma\in\Coup(\mu,\nu)}\E_\gamma[XY]
  \qquad\text{and}\qquad
  \inf_{\gamma\in\Coup(\mu,\nu)}\E_\gamma[XY].
\]
Relate the answer to the two endpoints of the Fr{\'e}chet class.
\end{exercise}

\iffalse
% To be restored in v2 after the fairness material is submitted.
\begin{exercise}[Coupling and fairness post-processing]\label{ex:couplings-fairness-postprocessing}\label{ex:2:7}
Let $S\in\{0,1\}$ denote a protected group, and let $Z\mid S=s\sim\mu_s$ be a score distribution. Suppose a post-processed score $\widetilde Z\mid S=s\sim\bar\mu$ is obtained from a coupling $\gamma_s\in\Coup(\mu_s,\bar\mu)$. Explain why $\gamma_s$ contains more information than the two marginal laws $\mu_s$ and $\bar\mu$. What would be lost if only the marginal before/after distributions were reported?
\end{exercise}
\fi

\newpage
\lecture{Discrete Optimal Transport}
\label{lec:discrete-ot}

\subsection*{Learning objectives}

This lecture develops optimal transport in its most concrete form: a finite-dimensional linear program. This is the right entry point for actuarial applications because many data sets are empirical, grouped, tabulated or discretized. Claims are observed as samples; rating variables are grouped into cells; reserving triangles are arrays; scenario models are simulated particles.

After this lecture, the reader should be able to:
\begin{enumerate}[label=(\roman*)]
  \item write the transportation problem between two discrete probability measures;
  \item identify transport plans with matrices satisfying row and column constraints;
  \item interpret the transportation polytope as a set of dependence models;
  \item derive the linear-programming dual and interpret the dual variables;
  \item use complementary slackness to understand the support of an optimal plan;
  \item solve by hand small examples, especially the \(2\times2\) case;
  \item recognize special cases such as assignment, matching, one-dimensional monotone transport and empirical Wasserstein distances;
  \item connect discrete OT to actuarial questions: portfolio drift, reserving scenarios and risk aggregation.
\end{enumerate}

\begin{readingbox}
This chapter is the finite-dimensional entry point into optimal transport. For computational and geometric introductions to discrete OT, see \cite{solomon2018discrete,peyre2019computational,santambrogio2015optimal}. The classical transportation and assignment problems are rooted in linear programming and network-flow theory; see \cite{hitchcock1941distribution,kantorovich1942translocation,ahuja1993networkflows,schrijver2003combinatorial}. The assignment special case relies on the Birkhoff--von Neumann theorem and is closely related to the Hungarian and auction algorithms \cite{birkhoff1946tres,kuhn1955hungarian,munkres1957algorithms,bertsekas1992auction}. Actuarial links with dependence uncertainty and aggregation are developed in \cite{denuit2005actuarial,dhaene2002concept,embrechts2002correlation,puccetti2012computation}.
\end{readingbox}

\begin{actuarialbox}
Discrete optimal transport is not merely a numerical approximation to a continuous theory. It is also a natural actuarial object. A portfolio is a finite population, a rating grid is finite, a reserving triangle is a finite array, and a simulation output is a finite cloud of particles. In all these cases, OT begins as a matrix problem.
\end{actuarialbox}

\subsection{Why start with the discrete problem?}
\label{sec:discrete-why}

Let \(\mu\) and \(\nu\) be two probability measures. In the previous lecture, we introduced the set of couplings \(\Coup(\mu,\nu)\); see Definition~\ref{def:coupling}. A coupling specifies how mass from \(\mu\) is paired with mass from \(\nu\). Optimal transport attaches a cost to each possible pairing and selects a coupling with minimal total cost.

When \(\mu\) and \(\nu\) have finite support, this becomes a linear program. This is useful for three reasons.

First, the notation is transparent. A transport plan is a matrix. Its row sums and column sums are prescribed. The cost is the scalar product between a cost matrix and the transport matrix.

Second, the economic interpretation is immediate. The rows are origins, the columns are destinations, and \(\gamma_{ij}\) is the amount of mass moved from origin \(i\) to destination \(j\). In insurance, origins and destinations can be loss levels, rating classes, reserve scenarios, premium bins, climate-risk zones or risk classes.

Third, the finite problem contains the main ideas of the general theory: primal and dual formulations, shadow prices, complementary slackness, geometry of feasible couplings, and computational algorithms. More advanced topics, such as entropic regularization, Sinkhorn scaling and Wasserstein barycenters, are easier to understand once the discrete problem is clear. Standard presentations of computational OT follow this route; see for example \cite{solomon2018discrete,peyre2019computational}.

\subsection{Discrete probability measures}
\label{sec:discrete-measures}

\begin{definition}[Discrete probability measure]
\label{def:discrete-probability-measure}
A probability measure \(\mu\) is discrete with finite support if there are support points \(\bx_1,\ldots,\bx_n\) and weights \(a_1,\ldots,a_n\ge0\), with \(\sum_i a_i=1\), such that
\[
  \mu=\sum_{i=1}^n a_i\delta_{\bx_i}.
\]
The vector \(\ba=(a_1,\ldots,a_n)^\top\) records probability masses, whereas the support points \(\bx_i\) carry the geometry of the actuarial state space.
\end{definition}

Let \(\mathcal X=\{\bx_1,\ldots,\bx_n\}\) and \(\mathcal Y=\{\by_1,\ldots,\by_m\}\) be two finite subsets of a space, often \(\R\) or \(\R^d\). We consider two discrete probability measures, in the sense of Definition~\ref{def:discrete-probability-measure},
\[
  \mu=\sum_{i=1}^n a_i\delta_{\bx_i},
  \qquad
  \nu=\sum_{j=1}^m b_j\delta_{\by_j},
\]
where
\[
  a_i\ge 0,\quad b_j\ge 0,\qquad
  \sum_{i=1}^n a_i=1,
  \qquad
  \sum_{j=1}^m b_j=1.
\]
We write \(\ba=(a_1,\ldots,a_n)^\top\in\R^n\) and \(\bb=(b_1,\ldots,b_m)^\top\in\R^m\). Thus \(\ba\) and \(\bb\) are probability vectors.

The support points \(\bx_i\) and \(\by_j\) contain the geometry or actuarial meaning of the problem. The weights \(a_i\) and \(b_j\) contain the probability masses. For instance:
\begin{itemize}
  \item \(x_i\) and \(y_j\) may be observed claim amounts in two years;
  \item they may be representative rating-cell centroids in a finite rating grid;
  \item they may be simulated aggregate losses under two climate scenarios;
% \item score values for two demographic groups; % To be restored in v2.
  \item they may be reserve cash-flow vectors produced by two stochastic reserving methods.
\end{itemize}

\begin{example}[Empirical distributions]
\label{ex:discrete-empirical-distributions}
If \(\bX_1,\ldots,\bX_n\) and \(\bY_1,\ldots,\bY_m\) are two samples, their empirical laws are
\[
  \hat\mu_n=\frac1n\sum_{i=1}^n \delta_{\bX_i},
  \qquad
  \hat\nu_m=\frac1m\sum_{j=1}^m \delta_{\bY_j}.
\]
This corresponds to \(a_i=1/n\) and \(b_j=1/m\). In actuarial work, the observations may be individual claims, policies, simulated losses, reserve outcomes or other simulated actuarial quantities.
\end{example}

\begin{example}[Grouped loss distributions]
\label{ex:discrete-grouped-losses}
Suppose claim amounts are grouped into bins \(I_1,\ldots,I_n\), with bin midpoints \(x_i\). If the proportion of claims in bin \(i\) is \(a_i\), we approximate the severity distribution by \(\sum_i a_i\delta_{x_i}\). Comparing two such grouped distributions is a natural discrete OT problem.
\end{example}

\subsection{Transport plans as matrices}
\label{sec:discrete-plans}

A coupling \(\gamma\in\Coup(\mu,\nu)\) is a probability measure on \(\mathcal X\times\mathcal Y\). Since \(\mathcal X\times\mathcal Y\) is finite, \(\gamma\) is fully described by the numbers
\[
  \gamma_{ij}=\gamma(\{\bx_i\}\times\{\by_j\}),
  \qquad i=1,
  \ldots,n,\quad j=1,\ldots,m.
\]
Thus \(\gamma\) is an \(n\times m\) matrix \(\Gamma=(\gamma_{ij})\).

The marginal constraints are simply
\[
  \sum_{j=1}^m \gamma_{ij}=a_i,
  \qquad i=1,
  \ldots,n,
\]
and
\[
  \sum_{i=1}^n \gamma_{ij}=b_j,
  \qquad j=1,
  \ldots,m.
\]
Together with \(\gamma_{ij}\ge 0\), these equations define the transportation polytope of Definition~\ref{def:transportation-polytope}.

\begin{definition}[Transportation polytope]
\label{def:transportation-polytope}
For probability vectors \(\ba\in\R^n\) and \(\bb\in\R^m\), the transportation polytope is
\[
  \mathcal U(\ba,\bb)
  =\left\{\Gamma\in\R_+^{n\times m}:
  \Gamma\mathbf 1_m=\ba,
  \Gamma^\top\mathbf 1_n=\bb\right\}.
\]
Equivalently, \(\mathcal U(\ba,\bb)\) is the set of all couplings between
\(\mu=\sum_i a_i\delta_{\bx_i}\) and \(\nu=\sum_j b_j\delta_{\by_j}\).
\end{definition}

The notation \(\mathcal U(\ba,\bb)\) is common in computational OT. It emphasizes that the feasible set depends only on the weights, not on the locations \(\bx_i\), \(\by_j\), nor on the cost function.

\begin{proposition}[Basic properties]
\label{prop:transportation-polytope-properties}
The set \(\mathcal U(\ba,\bb)\) is non-empty, convex and compact. If all weights are strictly positive, its affine dimension is
\[
  (n-1)(m-1).
\]
\end{proposition}

\begin{proof}
Non-emptiness follows from the independent coupling \(\Gamma=\ba\bb^\top\), whose entries are \(\gamma_{ij}=a_i b_j\). Convexity follows because the constraints defining \(\mathcal U(\ba,\bb)\) are linear and the inequalities are convex. Compactness follows because \(0\le \gamma_{ij}\le 1\) and the set is closed.

There are \(nm\) variables. The row and column constraints give \(n+m\) linear equations, but one is redundant because both total masses equal one. Hence the affine hull has codimension \(n+m-1\), and the dimension is \(nm-(n+m-1)=(n-1)(m-1)\), provided no marginal constraint is degenerate.
\end{proof}

\begin{remark}[Dependence, not only transportation]
In actuarial risk aggregation, \(\mathcal U(\ba,\bb)\) is the set of all joint distributions with fixed marginals. Thus a transport plan is also a dependence model. The same matrix \(\Gamma\) may be read as a shipping plan, a matching, a contingency table, a copula discretization, or a joint risk model.
\end{remark}

\subsection{The discrete optimal transport problem}
\label{sec:discrete-problem}

Let \(c:\mathcal X\times\mathcal Y\to\R\) be a cost function. Recall from Definition~\ref{def:intro-ground-cost} that the ground cost encodes the modelling meaning of moving one actuarial state to another.

\begin{definition}[Cost matrix]
\label{def:cost-matrix}
Given support points \(\bx_1,\ldots,\bx_n\) and \(\by_1,\ldots,\by_m\), the cost matrix associated with a ground cost \(c\) is the matrix \(C\in\R^{n\times m}\) defined by
\[
  C_{ij}=c(\bx_i,\by_j),
  \qquad i=1,\ldots,n,\quad j=1,\ldots,m.
\]
\end{definition}

The expected transportation cost under plan \(\Gamma\) is
\[
  \sum_{i=1}^n\sum_{j=1}^m C_{ij}\gamma_{ij}
  =\langle C,\Gamma\rangle.
\]

\begin{definition}[Discrete optimal transport]
\label{def:discrete-ot}
The discrete optimal transport problem between \(\mu\) and \(\nu\), with cost matrix \(C\), is
\[
  \highlight{GoldenRod}{\OT_C(\ba,\bb)
  =\min_{\Gamma\in\mathcal U(\ba,\bb)}\langle C,\Gamma\rangle}
\]
Any minimizer \(\Gamma^\star\in\mathcal U(\ba,\bb)\) is called an optimal transport plan. This is the finite-support version of the Kantorovich problem introduced at the end of Lecture~\ref{lec:couplings}.
\end{definition}

When \(c(\bx,\by)=d(\bx,\by)^p\) for a metric \(d\) and \(p\ge1\), the value of the problem gives the \(p\)-Wasserstein distance:
\[
  \Wp{p}(\mu,\nu)^p
  =\min_{\Gamma\in\mathcal U(\ba,\bb)}
  \sum_{i,j} d(\bx_i,\by_j)^p\gamma_{ij}.
\]
For \(p=1\), the cost is linear in distance. For \(p=2\), the quadratic cost penalizes large movements more strongly and is central in the geometry of Wasserstein space.

\begin{example}[Portfolio drift]
\label{ex:discrete-portfolio-drift}
Let \(\bx_i\in\R^d\) be representative profiles in a portfolio at time \(t\), and let \(\by_j\in\R^d\) be representative profiles at time \(t+1\). We may take
\[
  c(\bx_i,\by_j)=\|\bx_i-\by_j\|^2.
\]
Then \(\OT_C(\ba,\bb)\) measures how much the portfolio distribution has moved. If coordinates include age, exposure, insured value, past claims and territory risk, the value is a distributional drift metric.
\end{example}

\iffalse
% To be restored in v2, together with the material on portfolio mutualization and demutualization.
\begin{example}[Tariff disruption]
\label{ex:discrete-tariff-disruption}
Let \(x_i\) be premium levels under an old tariff and \(y_j\) premium levels under a new tariff. With \(c(x,y)=|x-y|\) or \(|x-y|^2\), OT measures the minimum aggregate premium displacement required to transform the old premium distribution into the new one. This is different from comparing average premiums: it compares the full distribution.
\end{example}
\fi

\subsection{A network-flow interpretation}
\label{sec:discrete-network}

The discrete transport problem can be represented as a minimum-cost flow on a bipartite network. We have source nodes \(i=1,\ldots,n\), target nodes \(j=1,\ldots,m\), and an arc from every source \(i\) to every target \(j\). Source \(i\) supplies mass \(a_i\), target \(j\) demands mass \(b_j\), and sending one unit of mass from \(i\) to \(j\) costs \(C_{ij}\).

\begin{center}
\begin{tikzpicture}[scale=1, every node/.style={font=\small}]
  \node[circle,draw,minimum size=6mm] (x1) at (0,2) {\(x_1\)};
  \node[circle,draw,minimum size=6mm] (x2) at (0,1) {\(x_2\)};
  \node[circle,draw,minimum size=6mm] (x3) at (0,0) {\(x_3\)};
  \node[circle,draw,minimum size=6mm] (y1) at (5,1.7) {\(y_1\)};
  \node[circle,draw,minimum size=6mm] (y2) at (5,.7) {\(y_2\)};
  \foreach \i in {1,2,3}{
    \foreach \j in {1,2}{
      \draw[-{Latex[length=2mm]},gray] (x\i) -- (y\j);
    }
  }
  \node[left=3mm of x1] {\(a_1\)};
  \node[left=3mm of x2] {\(a_2\)};
  \node[left=3mm of x3] {\(a_3\)};
  \node[right=3mm of y1] {\(b_1\)};
  \node[right=3mm of y2] {\(b_2\)};
  \node at (2.5,2.4) {arc cost \(C_{ij}\)};
\end{tikzpicture}
\end{center}

This interpretation is especially useful for algorithms. The transportation problem is a special case of minimum-cost flow, which explains why network simplex and related combinatorial algorithms are often efficient. It also clarifies the connection with classical operations research and with the assignment problem. For background on finite-dimensional optimization and network-flow algorithms, see \cite{ahuja1993networkflows,schrijver2003combinatorial,galichon2016economics,solomon2018discrete}.

\subsection{Existence of an optimal plan}
\label{sec:discrete-existence}

In the discrete setting, existence is immediate.

\begin{proposition}[Existence]
\label{prop:discrete-ot-existence}
For any finite cost matrix \(C\in\R^{n\times m}\), the discrete OT problem admits at least one optimal plan.
\end{proposition}

\begin{proof}
The map \(\Gamma\mapsto \langle C,\Gamma\rangle\) is continuous and linear. The feasible set \(\mathcal U(\ba,\bb)\) is compact and non-empty. Therefore the minimum is attained.
\end{proof}

Uniqueness is more delicate. The optimal value is unique, but the optimal plan need not be. Because the problem is linear, if two different plans are optimal, any convex combination of them is also optimal. Non-uniqueness occurs naturally when the cost contains symmetries or when mass can be split at the same cost.

\begin{example}[Non-uniqueness]
\label{ex:discrete-nonuniqueness}
Let \(\ba=\bb=(1/2,1/2)\) and
\[
  C=\begin{pmatrix}
  0 & 0\\
  0 & 0
  \end{pmatrix}.
\]
Every feasible transport plan is optimal. The cost does not distinguish among dependence structures.
\end{example}

\subsection{The \texorpdfstring{\(2\times2\)}{2x2} transportation problem}
\label{sec:discrete-two-by-two}

The \(2\times2\) case is worth solving explicitly. Let
\[
  \ba=(\alpha,1-\alpha),\qquad \bb=(\beta,1-\beta),
\]
with \(0\le \alpha,\beta\le1\), and let
\[
  C=\begin{pmatrix}
  c_{11} & c_{12}\\
  c_{21} & c_{22}
  \end{pmatrix}.
\]
A feasible transport plan has the form
\[
  \Gamma(t)=
  \begin{pmatrix}
    t & \alpha-t\\
    \beta-t & 1-\alpha-\beta+t
  \end{pmatrix},
\]
where non-negativity imposes
\[
  \max\{0,\alpha+\beta-1\}\le t\le \min\{\alpha,\beta\}.
\]
The cost is
\[
\begin{aligned}
  L(t)
  &=c_{11}t+c_{12}(\alpha-t)+c_{21}(\beta-t)+c_{22}(1-\alpha-\beta+t)\\
  &=\text{constant}+t\,(c_{11}-c_{12}-c_{21}+c_{22}).
\end{aligned}
\]
Thus the optimum is always at one endpoint, unless
\[
  c_{11}+c_{22}=c_{12}+c_{21},
\]
in which case every feasible \(t\) is optimal.

\begin{actuarialbox}
In a two-risk aggregation problem with fixed marginals, all dependence structures are indexed by a single number \(t\). The sign of \(c_{11}-c_{12}-c_{21}+c_{22}\) decides whether the optimal coupling pushes toward positive or negative dependence. This is the finite version of the rearrangement principle that later appears as monotone transport on the real line.
\end{actuarialbox}

\subsection{The dual problem}
\label{sec:discrete-dual}

The primal problem is
\[
\begin{array}{ll}
\text{minimize} & \displaystyle \sum_{i=1}^n\sum_{j=1}^m C_{ij}\gamma_{ij}\\[.2cm]
\text{subject to} & \displaystyle \sum_{j=1}^m \gamma_{ij}=a_i,\quad i=1,\ldots,n,\\[.2cm]
& \displaystyle \sum_{i=1}^n \gamma_{ij}=b_j,\quad j=1,\ldots,m,\\[.2cm]
& \gamma_{ij}\ge0.
\end{array}
\]
We now derive its dual. Introduce Lagrange multipliers \(u_i\) for the row constraints and \(v_j\) for the column constraints. Formally, the dual variables are constrained by
\[
  u_i+v_j\le C_{ij}.
\]
The dual objective is
\[
  \sum_{i=1}^n a_i u_i+
  \sum_{j=1}^m b_j v_j.
\]

\begin{theorem}[Kantorovich duality: finite case]
\label{thm:finite-kantorovich-duality}
The value of the discrete OT problem is equal to
\[
  \max_{\bu\in\R^n,\,\bv\in\R^m}
  \left\{
  \sum_{i=1}^n a_i u_i+\sum_{j=1}^m b_j v_j:
  u_i+v_j\le C_{ij}\quad\forall i,j
  \right\}.
\]
Moreover, the maximum is attained.
\end{theorem}

\begin{proof}
This is the linear-programming dual of the primal transportation problem. Weak duality follows directly: for any feasible \(\Gamma\) and any feasible \((\bu,\bv)\),
\[
  \sum_i a_i u_i+\sum_j b_jv_j
  =\sum_{i,j}\gamma_{ij}(u_i+v_j)
  \le \sum_{i,j}\gamma_{ij}C_{ij}.
\]
Taking the supremum over feasible \((\bu,\bv)\) and the infimum over feasible \(\Gamma\) gives dual value \(\le\) primal value. Since the primal is a feasible finite-dimensional linear program with finite optimum, strong linear-programming duality yields equality and existence of a dual optimizer.
\end{proof}

The components \(u_i\) and \(v_j\) of the dual vectors \(\bu\) and \(\bv\) are called Kantorovich potentials. They are not unique: adding a constant \(\alpha\) to all \(u_i\) and subtracting it from all \(v_j\) leaves both constraints and objective unchanged.

\begin{remark}[Insurance interpretation of potentials]
Suppose \(\bx_i\) are source portfolio states and \(\by_j\) are target portfolio states. The potential \(u_i\) can be interpreted as the shadow value of one additional unit of mass at origin \(\bx_i\); \(v_j\) is the shadow value of one additional unit of demand at destination \(\by_j\). In reserving or portfolio comparison, these potentials measure marginal distributional pressure: where mass is expensive to remove and where mass is expensive to create.
\end{remark}

\subsection{Complementary slackness}
\label{sec:discrete-slackness}

The dual constraints say
\[
  u_i+v_j\le C_{ij}.
\]
At an optimal pair \((\Gamma^\star,\bu^\star,\bv^\star)\), complementary slackness states that mass can move only along tight edges.

\begin{proposition}[Complementary slackness]
\label{prop:finite-complementary-slackness}
Let \(\Gamma^\star\) be primal optimal and \((\bu^\star,\bv^\star)\) dual optimal. Then
\[
  \gamma^\star_{ij}>0
  \quad\Longrightarrow\quad
  u_i^\star+v_j^\star=C_{ij}.
\]
Equivalently,
\[
  \gamma^\star_{ij}\bigl(C_{ij}-u_i^\star-v_j^\star\bigr)=0
  \qquad\forall i,j.
\]
\end{proposition}

\begin{proof}
By strong duality,
\[
  \sum_{i,j}\gamma^\star_{ij}C_{ij}
  =\sum_i a_i u_i^\star+\sum_j b_jv_j^\star
  =\sum_{i,j}\gamma^\star_{ij}(u_i^\star+v_j^\star).
\]
Therefore
\[
  \sum_{i,j}\gamma^\star_{ij}
  \bigl(C_{ij}-u_i^\star-v_j^\star\bigr)=0.
\]
Each term is non-negative because \(\Gamma^\star\ge0\) and \((\bu^\star,\bv^\star)\) is dual feasible. Hence every term must vanish.
\end{proof}

Complementary slackness is often the simplest way to understand an optimal plan. The potentials define a set of admissible active edges, and the optimal plan routes mass only through those edges.

\subsection{The discrete \texorpdfstring{\(c\)}{c}-transform}
\label{sec:discrete-ctransform}

Given a vector \(\bu\in\R^n\), define its discrete \(c\)-transform by
\[
  (\bu^c)_j=\min_{1\le i\le n}\{C_{ij}-u_i\},
  \qquad j=1,\ldots,m.
\]
Then \(u_i+(\bu^c)_j\le C_{ij}\) for all \(i,j\), so \((\bu,\bu^c)\) is dual feasible. Conversely, if \((\bu,\bv)\) is feasible, then \(v_j\le (\bu^c)_j\) for all \(j\). Since the dual objective is increasing in the relevant components of \(\bv\), at optimum one may take \(\bv=\bu^c\).

Similarly, for \(\bv\in\R^m\), define
\[
  (\bv^c)_i=\min_{1\le j\le m}\{C_{ij}-v_j\}.
\]
At optimum, potentials can be chosen to satisfy the mutual relations
\[
  \bv=\bu^c,
  \qquad
  \bu=\bv^c.
\]
These are the finite analogues of \(c\)-concavity in the general Kantorovich theory.

\begin{computebox}
In computations, the potentials are often as useful as the transport plan. For large problems, storing \(\Gamma\) may be expensive, while the potentials \(\bu\) and \(\bv\) contain information about active constraints, sensitivity and marginal values.
\end{computebox}

\subsection{Special case: the assignment problem}
\label{sec:discrete-assignment}

Suppose \(n=m\), and both measures are uniform:
\[
  a_i=b_j=\frac1n.
\]
A transport plan \(\Gamma\in\mathcal U(\ba,\bb)\) is a doubly stochastic matrix scaled by \(1/n\). The next classical result explains why this case reduces to matching.

\begin{theorem}[Birkhoff--von Neumann theorem, scaled form]
\label{thm:birkhoff-von-neumann}
The extreme points of the polytope of doubly stochastic \(n\times n\) matrices are the permutation matrices. Equivalently, the extreme points of \(\mathcal U(\ba,\bb)\), when \(a_i=b_j=1/n\), are the permutation matrices scaled by \(1/n\).
\end{theorem}

Since the objective is linear, at least one optimum is attained at an extreme point. Hence there exists an optimal transport plan corresponding to a matching:
\[
  \gamma_{i,\sigma(i)}=\frac1n,
  \qquad
  \gamma_{ij}=0\quad\text{otherwise}.
\]
The OT problem becomes
\[
  \min_{\sigma\in\mathfrak S_n}
  \frac1n\sum_{i=1}^n C_{i,\sigma(i)}.
\]
This is the classical linear assignment problem; see \cite{kuhn1955hungarian,munkres1957algorithms,bertsekas1992auction,burkard2009assignment}.

\begin{example}[Matching simulated scenarios]
\label{ex:discrete-scenario-matching}
Suppose two stochastic capital models produce \(n\) equally weighted scenarios \(\bx_1,\ldots,\bx_n\) and \(\by_1,\ldots,\by_n\), where each scenario is a vector of line-of-business losses. With \(C_{ij}=\|\bx_i-\by_j\|^2\), the assignment problem matches scenarios from the first model to scenarios from the second model. The resulting cost measures the discrepancy between the two simulated distributions.
\end{example}

\begin{remark}[Splitting versus matching]
When weights are not uniform or when \(n\ne m\), mass may split. In insurance this is natural: one rating cell may be redistributed across several new cells, or one loss bin may contribute mass to several target bins.
\end{remark}

\subsection{One-dimensional discrete transport}
\label{sec:discrete-oned}

When \(x_i,y_j\in\R\) and \(c(x,y)=|x-y|^p\), \(p\ge1\), the optimal plan is monotone: mass should not cross. This fact is the discrete form of monotone rearrangement and will be used again for quantile formulas and risk measures.

Assume \(x_1\le\cdots\le x_n\) and \(y_1\le\cdots\le y_m\). A plan \(\Gamma\) has a crossing if there exist \(i<k\) and \(j>\ell\) such that \(\gamma_{ij}>0\) and \(\gamma_{k\ell}>0\). In words, some lower source point is matched to a higher target point while a higher source point is matched to a lower target point.

\begin{proposition}[No-crossing property]
\label{prop:discrete-no-crossing}
Let \(c(x,y)=|x-y|^p\), \(p\ge1\), on \(\R\). There exists an optimal transport plan with no crossings.
\end{proposition}

\begin{proof}[Sketch]
Suppose \(i<k\) and \(j>\ell\). Move a small mass \(\eps>0\) from the crossed pairs \((x_i,y_j)\), \((x_k,y_\ell)\) to the uncrossed pairs \((x_i,y_\ell)\), \((x_k,y_j)\). The row and column sums are unchanged. Convexity of \(z\mapsto |z|^p\), or the Monge property of the cost matrix, gives
\[
  |x_i-y_j|^p+|x_k-y_\ell|^p
  \ge
  |x_i-y_\ell|^p+|x_k-y_j|^p.
\]
Thus uncrossing does not increase cost. Repeating this operation eliminates crossings.
\end{proof}

This gives a simple greedy algorithm: match the smallest remaining \(x_i\) with the smallest remaining \(y_j\), move as much mass as possible, then continue.

\begin{computebox}
For one-dimensional empirical distributions, exact \(W_p\) can be computed by sorting and matching quantiles. This is much faster and more stable than solving a generic linear program. It is also the bridge between OT and actuarial quantile methods.
\end{computebox}

\subsection{Empirical Wasserstein distances}
\label{sec:discrete-empirical}

\begin{definition}[Empirical Wasserstein distance]
\label{def:empirical-wasserstein-discrete}
Let \(\bX_1,\ldots,\bX_n\) and \(\bY_1,\ldots,\bY_m\) be two samples in \(\R^d\), and let
\[
  \hat\mu_n=\frac1n\sum_{i=1}^n\delta_{\bX_i},
  \qquad
  \hat\nu_m=\frac1m\sum_{j=1}^m\delta_{\bY_j}.
\]
The empirical Wasserstein distance is the Wasserstein distance between these empirical probability measures.
\end{definition}

With the notation of Definition~\ref{def:empirical-wasserstein-discrete}, it is obtained by solving
\[
  \Wp{p}(\hat\mu_n,\hat\nu_m)^p
  =\min_{\Gamma\in\mathcal U(\ba,\bb)}
  \sum_{i=1}^n\sum_{j=1}^m \|\bX_i-\bY_j\|^p\gamma_{ij},
\]
where \(a_i=1/n\) and \(b_j=1/m\). If \(n=m\), this is closely related to matching; if \(n\ne m\), mass splitting is generally needed.

For actuarial data, empirical OT distances can be used to compare:
\begin{itemize}
  \item claim severity samples between accident years;
  \item loss-ratio distributions between regions;
% \item predicted premium distributions before and after a tariff revision; % To be restored in v2.
  \item simulated aggregate-loss distributions under baseline and stress scenarios;
  \item reserve cash-flow scenarios from two reserving models;
% \item score distributions across protected groups. % To be restored in v2.
\end{itemize}

\begin{example}[Two years of claim severities]
\label{ex:discrete-severity-years}
Let \(X_1,\ldots,X_n\) be claims from year \(t\) and \(Y_1,\ldots,Y_m\) claims from year \(t+1\). If claims are one-dimensional and \(p=1\), \(\Wone\) measures an average absolute displacement between quantiles. It is sensitive to distributional shifts, not only to changes in mean severity.
\end{example}

\iffalse
% To be restored in v2.
\begin{example}[Score distributions in fairness auditing]
\label{ex:discrete-fairness-scores}
Let \(X_i\) be scores for group \(A=0\) and \(Y_j\) scores for group \(A=1\). A large Wasserstein distance indicates that the score distributions differ. Later, we will use optimal maps and barycenters to post-process scores so that group-specific score distributions are aligned.
\end{example}
\fi

\subsection{Discrete OT and risk aggregation}
\label{sec:discrete-risk-aggregation}

A transport plan can be pushed forward through an aggregation map. Let \((X,Y)\sim\Gamma\), where \(X\sim\mu\) and \(Y\sim\nu\). The aggregate loss is
\[
  S=X+Y.
\]
The distribution of \(S\) depends on \(\Gamma\), even when \(\mu\) and \(\nu\) are fixed. In finite form,
\[
  \Law(S)=\sum_{i=1}^n\sum_{j=1}^m
  \gamma_{ij}\delta_{x_i+y_j}.
\]
Thus the set \(\mathcal U(\ba,\bb)\) represents all possible aggregate-loss distributions consistent with the marginal severity models.

This is a classical actuarial theme: dependence uncertainty can dominate marginal uncertainty. Optimal transport offers a unified language for optimization over dependence structures.

\begin{example}[Worst-case expected stop-loss]
\label{ex:discrete-worstcase-stoploss}
Let \(S=X+Y\) and consider the stop-loss function
\[
  \ell_d(S)=(S-d)_+.
\]
Among all couplings \(\Gamma\in\mathcal U(\ba,\bb)\), the largest expected stop-loss is
\[
  \max_{\Gamma\in\mathcal U(\ba,\bb)}
  \sum_{i,j}(x_i+y_j-d)_+\gamma_{ij}.
\]
This is a linear program over the transportation polytope. It is not written as a minimum transport problem, but it is the same object with cost
\[
  C_{ij}=-(x_i+y_j-d)_+.
\]
The corresponding minimization gives the worst case.
\end{example}

\begin{actuarialbox}
Discrete OT is a flexible way to optimize actuarial quantities under fixed marginals. Depending on the cost, it can compute best-case or worst-case dependence, minimum distributional displacement, or robust risk bounds.
\end{actuarialbox}

\subsection{Discrete OT as a linear program in standard form}
\label{sec:discrete-lp}

It is sometimes useful to vectorize the transport matrix. Let
\[
  \bgamma=\operatorname{vec}(\Gamma)\in\R^{nm},
  \qquad
  \bc=\operatorname{vec}(C)\in\R^{nm}.
\]
The constraints can be written as
\[
  A\bgamma=\br,
  \qquad
  \br=\begin{pmatrix}
  \ba\\ \bb
  \end{pmatrix},
  \qquad
  \bgamma\ge0,
\]
where \(A\) is the node-arc incidence matrix of the complete bipartite graph, except that one redundant row may be removed. The primal problem becomes
\[
  \min_{\bgamma\ge0}\ \bc^\top\bgamma
  \quad\text{subject to}\quad A\bgamma=\br.
\]

This form is helpful for implementing the problem using generic linear-programming solvers. It also connects the problem to convex optimization and duality, as in standard treatments of linear programming and convex optimization \cite{galichon2016economics,peyre2019computational}.

\subsection{Computational approaches}
\label{sec:discrete-computation}

There are several ways to solve the discrete OT problem.

\subsubsection*{Generic linear programming}

For small and medium problems, one may use a generic linear-programming solver. This is convenient and reliable, but the number of variables is \(nm\). With two samples of size \(10{,}000\), the transport matrix has \(10^8\) entries, which is already too large for naive dense methods.

\subsubsection*{Network simplex and minimum-cost flow}

Because the transportation problem is a network-flow problem, specialized algorithms can exploit the bipartite structure. Network simplex methods are often effective for exact solutions on sparse or moderate-size problems.

\subsubsection*{Assignment algorithms}

When \(n=m\) and all weights are equal, the problem becomes an assignment problem. Specialized algorithms such as the Hungarian algorithm or auction algorithms can be used.

\subsubsection*{One-dimensional sorting}

On the real line with convex distance cost, sorting gives the solution. This is the fastest and most interpretable case, and it is especially relevant for actuarial quantities based on quantiles.

\subsubsection*{Entropic regularization}

For large-scale problems, one often solves the regularized problem
\[
\highlight{GoldenRod}{  \min_{\Gamma\in\mathcal U(\ba,\bb)}
  \langle C,\Gamma\rangle
  +\eps\sum_{i,j}\gamma_{ij}(\log\gamma_{ij}-1)}
\]
with \(\eps>0\), using the convention \(0\log 0=0\). This leads to Sinkhorn's algorithm, which will be studied in the lecture on entropic regularization. Regularization makes the problem smooth and computationally scalable, at the cost of introducing bias; see \cite{cuturi2013sinkhorn,peyre2019computational}.

\begin{computebox}
Exact OT is often best for interpretation and small examples. Entropic OT is often best for large actuarial portfolios, simulations and repeated computations. The two should be viewed as complementary tools.
\end{computebox}

\subsection{Mini-lab: computing discrete OT in R}
\label{sec:discrete-mini-lab}

The following code computes a small discrete OT problem using a generic linear-programming solver. It is intentionally written in transparent matrix form rather than optimized form. This template will reappear in the later labs on empirical Wasserstein noise and reserving cash-flow distributions.

\begin{lstlisting}[language=R]
# install.packages("lpSolve")
library(lpSolve)

x <- c(0, 1, 4)
y <- c(0, 2)
a <- c(0.30, 0.40, 0.30)
b <- c(0.55, 0.45)

C <- outer(x, y, function(u, v) (u - v)^2)
n <- length(x)
m <- length(y)

# variables are gamma_11, gamma_21, ..., gamma_n1,
# then gamma_12, ..., gamma_nm under R's column-major order
obj <- as.vector(C)

# row constraints: sum_j gamma_ij = a_i
A_row <- matrix(0, nrow = n, ncol = n*m)
for(i in 1:n){
  for(j in 1:m){
    A_row[i, i + (j-1)*n] <- 1
  }
}

# column constraints: sum_i gamma_ij = b_j
A_col <- matrix(0, nrow = m, ncol = n*m)
for(j in 1:m){
  for(i in 1:n){
    A_col[j, i + (j-1)*n] <- 1
  }
}

A <- rbind(A_row, A_col)
rhs <- c(a, b)
dir <- rep("=", n + m)

ans <- lp("min", obj, A, dir, rhs)
Gamma <- matrix(ans$solution, nrow = n, ncol = m)

ans$objval
Gamma
rowSums(Gamma)
colSums(Gamma)
\end{lstlisting}

For one-dimensional empirical distributions with equal sample sizes, we can compute \(W_p\) by sorting.

\begin{lstlisting}[language=R]
set.seed(123)
X <- rgamma(1000, shape = 2, rate = 1)
Y <- rgamma(1000, shape = 3, rate = 1)

p <- 2
Wp_p <- mean(abs(sort(X) - sort(Y))^p)
Wp <- Wp_p^(1/p)
Wp
\end{lstlisting}

\subsection{What the discrete problem teaches us}
\label{sec:discrete-summary}

The discrete problem gives several lessons that will remain valid throughout the notes.

\begin{enumerate}[label=(\roman*)]
  \item OT is optimization over couplings.
  \item A coupling is a dependence model with fixed marginals.
  \item The primal object is a transport plan; the dual object is a pair of potentials.
  \item Geometry enters through the cost matrix \(C\), not through the feasible set \(\mathcal U(\ba,\bb)\).
  \item In one dimension, optimal transport is monotone and connected to quantiles.
  \item In high dimension, computation becomes the central difficulty.
  \item Actuarial applications often begin with empirical or grouped distributions, so the discrete formulation is not optional; it is the working model.
\end{enumerate}

\begin{readingbox}
For a computational introduction to discrete OT, see \cite{solomon2018discrete,peyre2019computational,santambrogio2015optimal}. For the economic interpretation of duality and assignment, see \cite{galichon2016economics}. For network-flow and assignment algorithms, see \cite{ahuja1993networkflows,schrijver2003combinatorial,burkard2009assignment}. For actuarial dependence and risk aggregation, see \cite{denuit2005actuarial,dhaene2002concept,embrechts2002correlation,puccetti2012computation}.
\end{readingbox}

\subsection{Exercises}
\label{sec:discrete-exercises}

\begin{exercise}[The \(2\times2\) polytope]\label{ex:3:1}
Let \(\ba=(\alpha,1-\alpha)\), \(\bb=(\beta,1-\beta)\). Show that every feasible transport plan has the form
\[
  \Gamma(t)=
  \begin{pmatrix}
  t & \alpha-t\\
  \beta-t & 1-\alpha-\beta+t
  \end{pmatrix}
\]
for
\[
  \max\{0,\alpha+\beta-1\}\le t\le \min\{\alpha,\beta\}.
\]
For a generic cost matrix \(C\), determine which endpoint is optimal.
\end{exercise}

\begin{exercise}[Duality by hand]\label{ex:3:2}
Consider
\[
  \ba=\bb=(1/2,1/2),
  \qquad
  C=\begin{pmatrix}
  0 & 1\\
  1 & 0
  \end{pmatrix}.
\]
Find an optimal primal plan and optimal dual potentials \((\bu,\bv)\). Verify complementary slackness.
\end{exercise}

\begin{exercise}[Independence is usually not optimal]\label{ex:3:3}
Let \(\Gamma^{\mathrm{ind}}=\ba\bb^\top\). Give a numerical example with \(n=m=2\) where \(\Gamma^{\mathrm{ind}}\) is not optimal for the cost \(C_{ij}=|x_i-y_j|^2\). Compute the cost gap between the independent plan and the optimal plan.
\end{exercise}

\begin{exercise}[No-crossing argument]\label{ex:3:4}
Let \(x_1<x_2\) and \(y_1<y_2\). Show that for \(c(x,y)=|x-y|^p\), \(p\ge1\),
\[
  c(x_1,y_1)+c(x_2,y_2)
  \le
  c(x_1,y_2)+c(x_2,y_1).
\]
Interpret this inequality as saying that crossing transport is suboptimal.
\end{exercise}

\begin{exercise}[Worst-case stop-loss]\label{ex:3:5}
Let
\[
  \mu=0.5\delta_0+0.5\delta_{10},
  \qquad
  \nu=0.5\delta_0+0.5\delta_{10}.
\]
Among all couplings of \(\mu\) and \(\nu\), maximize
\[
  \E[(X+Y-12)_+].
\]
Solve the problem by enumerating the \(2\times2\) transportation plans. Which dependence structure is worst?
\end{exercise}

\iffalse
% To be restored in v2, together with the material on portfolio mutualization and demutualization.
\begin{exercise}[Tariff distribution versus policy disruption]\label{ex:3:6}
Construct a simple portfolio with four policies for which the old and new premium distributions are identical, so that their Wasserstein distance is zero, but every policy experiences a non-zero premium change. Explain the actuarial meaning of the example.
\end{exercise}
\fi

\begin{exercise}[Empirical computation]\label{ex:3:7}
Simulate two samples of size \(n=500\) from lognormal distributions with different parameters. Compute the one-dimensional \(W_1\) and \(W_2\) distances by sorting. Then multiply the second sample by an inflation factor \(1+r\). Plot the Wasserstein distance as a function of \(r\).
\end{exercise}

\begin{exercise}[Dual potentials as sensitivity]\label{ex:3:8}
Take the numerical example from the mini-lab. Solve the primal problem and then solve the dual linear program. Perturb one source weight \(a_i\) slightly, adjusting another source weight so that the total remains one. Compare the observed change in optimal value with the difference in the corresponding dual potentials.
\end{exercise}

\newpage
\lecture{Continuous Optimal Transport and Kantorovich Duality}
\label{lec:continuous-ot}

\subsection*{Learning objectives}

The previous lecture presented optimal transport as a finite-dimensional linear program. This lecture moves to the continuous setting. The main point is not to abandon the discrete viewpoint, but to understand the population-level object that the discrete problem approximates. In actuarial science, the continuous formulation is natural when losses, covariates, capital scenarios or cash-flow vectors are modelled by probability laws rather than by finite histograms.

After this lecture, the reader should be able to:
\begin{enumerate}[label=(\roman*)]
\item distinguish the Monge formulation from the Kantorovich relaxation;
\item interpret push-forward constraints as distributional constraints;
\item understand why mass splitting is necessary in general;
\item state the Kantorovich primal and dual problems;
\item prove weak duality and interpret dual potentials;
\item use the \(c\)-transform and complementary slackness in continuous form;
\item identify the one-dimensional optimal map through quantiles;
\item state the quadratic-cost Brenier theorem and understand its actuarial meaning;
\item recognize continuous OT models in severity modelling, portfolio drift, climate stress testing, reserving and dependence analysis.
\end{enumerate}

\begin{readingbox}
This chapter is the population-level counterpart of the discrete formulation in Lecture~\ref{lec:discrete-ot}. The Monge and Kantorovich formulations go back to \cite{monge1781memoire,kantorovich1942translocation}; modern accounts of existence, duality and optimal maps include \cite{villani2003topics,villani2008oldnew,ambrosio2021lectures,santambrogio2015optimal,figalli2021invitation}. The quadratic-cost map is tied to Brenier's theorem \cite{brenier1991polar}, McCann's displacement interpolation \cite{mccann1997convexity}, and convex analysis in the sense of \cite{rockafellar1970convex}. Computational and statistical viewpoints are developed in \cite{peyre2019computational,panaretos2020invitation}. Actuarial uses below should be read together with the later chapters on risk measures, drift, reserving, reinsurance and dependence.
\end{readingbox}

\begin{actuarialbox}
Discrete OT is the matrix version of optimal transport. Continuous OT is the modelling version: it tells us what it means to transform one risk law into another, independently of a particular discretization. For an actuary, the key question is often not ``which individual claim moves where?'' but ``what is the least costly distributional transformation from one risk environment to another?''
\end{actuarialbox}

\subsection{From finite matrices to probability laws}
\label{sec:continuous-from-discrete}

In Lecture~\ref{lec:discrete-ot}, two probability measures were written as
\[
  \mu=\sum_{i=1}^n a_i\delta_{x_i},
  \qquad
  \nu=\sum_{j=1}^m b_j\delta_{y_j},
\]
and a coupling was an \(n\times m\) matrix \(\Gamma=(\gamma_{ij})\) with prescribed row and column sums. The continuous version replaces the matrix by a probability measure on a product space.

Recall from Definition~\ref{def:coupling} that a coupling is a joint law with prescribed marginals. Let \(\mathcal X\) and \(\mathcal Y\) be measurable spaces. Usually, they are subsets of \(\R^d\), or more generally Polish spaces. Let \(\mu\in\Prob(\mathcal X)\) and \(\nu\in\Prob(\mathcal Y)\). A coupling is a probability measure \(\gamma\in\Prob(\mathcal X\times\mathcal Y)\) whose first marginal is \(\mu\) and whose second marginal is \(\nu\). We write
\[
  \gamma\in\Coup(\mu,\nu).
\]
Equivalently, if \((X,Y)\sim\gamma\), then
\[
  X\sim\mu,
  \qquad
  Y\sim\nu.
\]
Thus \(\Coup(\mu,\nu)\) is the set of all joint laws with fixed marginals.

This already has an actuarial interpretation. If \(\mu\) is the marginal law of motor losses and \(\nu\) is the marginal law of home losses, then \(\Coup(\mu,\nu)\) is the set of all possible dependence structures between the two lines, keeping the marginal models unchanged.

% If \(\mu\) is the current distribution of predicted premiums and \(\nu\) the distribution after a tariff revision, then \(\Coup(\mu,\nu)\) is the set of all possible distributional matchings between the two premium populations. % To be restored in v2.

\begin{example}[Continuous severity models]
\label{ex:continuous-severity-models}
Let \(\mu\) be a lognormal severity distribution and \(\nu\) be a gamma severity distribution calibrated to another portfolio. A coupling \(\gamma\in\Coup(\mu,\nu)\) is a joint model \((X,Y)\) where \(X\) has the first severity law and \(Y\) has the second. OT selects the joint model minimizing an average transformation cost, for example \(\E|X-Y|\) or \(\E|X-Y|^2\).
\end{example}

\begin{example}[Scenario vectors]
\label{ex:continuous-scenario-vectors}
Let \(\mu\) and \(\nu\) be laws on \(\R^d\), where each vector contains losses by line of business:
\[
  \bx=(x^{(1)},\ldots,x^{(d)}).
\]
A cost such as \(c(\bx,\by)=\|\bx-\by\|^2\) compares complete multivariate loss scenarios. The resulting transport problem measures distributional discrepancy between two capital models, two climate scenarios, or two reserving models.
\end{example}

\subsection{Push-forwards and deterministic transformations}
\label{sec:continuous-push-forwards}

Recall the push-forward notation of Definition~\ref{def:push-forward}. A measurable map \(T:\mathcal X\to\mathcal Y\) pushes \(\mu\) forward to \(\nu\), written
\[
  T_{\push}\mu=\nu,
\]
if
\[
  \nu(B)=\mu(T^{-1}(B))
\]
for every measurable set \(B\subseteq\mathcal Y\). Equivalently, if \(X\sim\mu\), then
\[
  T(X)\sim\nu.
\]

A deterministic map \(T\) induces a coupling
\[
  \gamma_T=(\Id,T)_{\push}\mu,
\]
meaning that under \(\gamma_T\), the pair is \((X,T(X))\). Such a coupling is concentrated on the graph of \(T\).

\begin{definition}[Monge admissible map]
\label{def:monge-admissible-map}
A measurable map \(T:\mathcal X\to\mathcal Y\) is admissible for the Monge problem from \(\mu\) to \(\nu\) if
\[
  T_{\push}\mu=\nu.
\]
The set of admissible maps is denoted informally by
\[
  \mathcal T(\mu,\nu)=\{T:T_{\push}\mu=\nu\}.
\]
\end{definition}

In actuarial modelling, a push-forward map is a deterministic distributional transformation. Examples include:
\begin{itemize}
\item applying inflation to claims: \(T(x)=(1+i)x\);
\item applying a deductible or cap: \(T(x)=\min\{(x-d)_+,M\}\);
\item transforming baseline climate losses into stressed losses;
% \item mapping raw scores into calibrated scores; % To be restored in v2.
% \item mapping group-specific score distributions onto a common fair distribution. % To be restored in v2.
\end{itemize}

\begin{example}[Inflation as push-forward]
\label{ex:inflation-push-forward}
Let \(X\sim\mu\) be a severity distribution and let \(T(x)=(1+i)x\). The inflated severity law is
\[
  \nu=T_{\push}\mu.
\]
If \(F_\mu\) is continuous and strictly increasing, then
\[
  F_\nu(y)=F_\mu\left(\frac{y}{1+i}\right).
\]
This is a deterministic transformation of the full distribution, not only of the mean.
\end{example}

\begin{example}[Deductibles]
\label{ex:deductible-push-forward-continuous}
Let \(T(x)=(x-d)_+\). Even if \(\mu\) is continuous, the push-forward \(T_{\push}\mu\) has an atom at zero, with mass \(\mu((-\infty,d])\). This example is useful because it shows how deterministic maps can create atoms. Conversely, deterministic maps cannot always split atoms, which is why Monge's formulation can be too restrictive.
\end{example}

\subsection{The Monge problem}
\label{sec:continuous-monge}

Let \(c:\mathcal X\times\mathcal Y\to\R\cup\{+\infty\}\) be a ground cost, in the sense of Definition~\ref{def:intro-ground-cost}. The Monge problem is
\[
  \inf_{T:T_{\push}\mu=\nu}
  \int_{\mathcal X} c(x,T(x))\dd\mu(x).
\]
When an admissible map \(T\) exists and attains the infimum, it is called an optimal transport map.

This is the original formulation: each point \(x\) is sent to exactly one point \(T(x)\). In finite insurance language, each policyholder, claim amount or scenario is matched deterministically to a new one. No splitting is allowed.

\begin{definition}[Monge problem]
\label{def:monge-problem}
The Monge value is
\[
  M_c(\mu,\nu)
  =\inf\left\{\int c(x,T(x))\dd\mu(x):T_{\push}\mu=\nu\right\},
\]
with the convention that the infimum is \(+\infty\) if there is no admissible map.
\end{definition}

\begin{example}[No admissible Monge map]
\label{ex:no-monge-map-dirac}
Let \(\mu=\delta_0\) and \(\nu=\frac12\delta_{-1}+\frac12\delta_1\). Any map \(T\) sends the single point \(0\) to one point \(T(0)\). Hence \(T_{\push}\mu=\delta_{T(0)}\), which is a Dirac mass. It cannot equal \(\nu\). Thus the Monge problem has no feasible map.
\end{example}

The previous example is not a pathology. In insurance, atoms are common: zero claims, deductible effects, policy limits, discretized rating classes, and grouped reserving states. A formulation that forbids splitting mass is therefore too rigid for many actuarial purposes.

\begin{remark}[Monge is nonlinear]
\label{rem:monge-nonlinear}
The constraint \(T_{\push}\mu=\nu\) is nonlinear in \(T\), and the set of admissible maps is not convex. If \(T_1\) and \(T_2\) both push \(\mu\) to \(\nu\), the map \((T_1+T_2)/2\) usually does not. This is the first reason why the Kantorovich relaxation is mathematically and computationally easier.
\end{remark}

\subsection{The Kantorovich relaxation}
\label{sec:continuous-kantorovich}

Kantorovich's idea is to optimize over all couplings, not only deterministic ones. Instead of choosing a map \(T\), one chooses a joint law \(\gamma\) with marginals \(\mu\) and \(\nu\). The problem becomes
\[
  \inf_{\gamma\in\Coup(\mu,\nu)}
  \int_{\mathcal X\times\mathcal Y} c(x,y)\dd\gamma(x,y).
\]

\begin{definition}[Kantorovich problem]
\label{def:kantorovich-problem}
The Kantorovich optimal transport value is
\[
  \OT_c(\mu,\nu)
  =\inf_{\gamma\in\Coup(\mu,\nu)}
  \int c(x,y)\dd\gamma(x,y).
\]
Any minimizer \(\gamma^\star\in\Coup(\mu,\nu)\) is called an optimal transport plan.
\end{definition}

The Monge problem embeds into the Kantorovich problem. The following elementary result is often the safest way to remember the relation between deterministic maps and transport plans.

\begin{proposition}[Monge plans are Kantorovich plans]
\label{prop:monge-plans-kantorovich-plans}
If \(T_{\push}\mu=\nu\), then \(\gamma_T=(\Id,T)_{\push}\mu\) belongs to \(\Coup(\mu,\nu)\) and
\[
  \int c(x,y)\dd\gamma_T(x,y)
  =\int c(x,T(x))\dd\mu(x).
\]
Consequently, whenever the Monge problem is feasible,
\[
  \OT_c(\mu,\nu)\le M_c(\mu,\nu).
\]
\end{proposition}

\begin{proof}
The first marginal of \(\gamma_T\) is \(\mu\), and its second marginal is \(T_{\push}\mu=\nu\), so \(\gamma_T\in\Coup(\mu,\nu)\). The equality of costs is the change-of-variables formula for push-forwards, applied to the function \((x,y)\mapsto c(x,y)\) and the map \(x\mapsto (x,T(x))\). The inequality follows by minimizing over the larger Kantorovich feasible set.
\end{proof}

\begin{proposition}[Convexity of the Kantorovich formulation]
\label{prop:kantorovich-convexity}
The set \(\Coup(\mu,\nu)\) is convex. The objective
\[
  \gamma\mapsto \int c\dd\gamma
\]
is linear in \(\gamma\). Hence the Kantorovich problem is an infinite-dimensional linear optimization problem.
\end{proposition}

\begin{proof}
If \(\gamma_1,\gamma_2\in\Coup(\mu,\nu)\) and \(\lambda\in[0,1]\), then \(\lambda\gamma_1+(1-\lambda)\gamma_2\) is a probability measure. Its first marginal is
\[
  \lambda\mu+(1-\lambda)\mu=\mu,
\]
and similarly its second marginal is \(\nu\). Linearity of the objective follows from linearity of integration with respect to the measure.
\end{proof}

\begin{actuarialbox}
The Kantorovich relaxation is well suited to aggregate distributions, cells and scenarios. It allows one unit of source mass to be split across several target states. This is natural for grouped claims, rating cells, reserve outcomes and dependence models.
\end{actuarialbox}

\subsection{A picture: deterministic map versus transport plan}
\label{sec:continuous-map-vs-plan}

In a Monge solution, all mass starting from \(x\) goes to one destination \(T(x)\). In a Kantorovich plan, mass starting from \(x\) can be spread according to a conditional distribution \(\gamma(\dd y\mid x)\).

\begin{figure}
    \centering
    \includegraphics[width=0.75\linewidth]{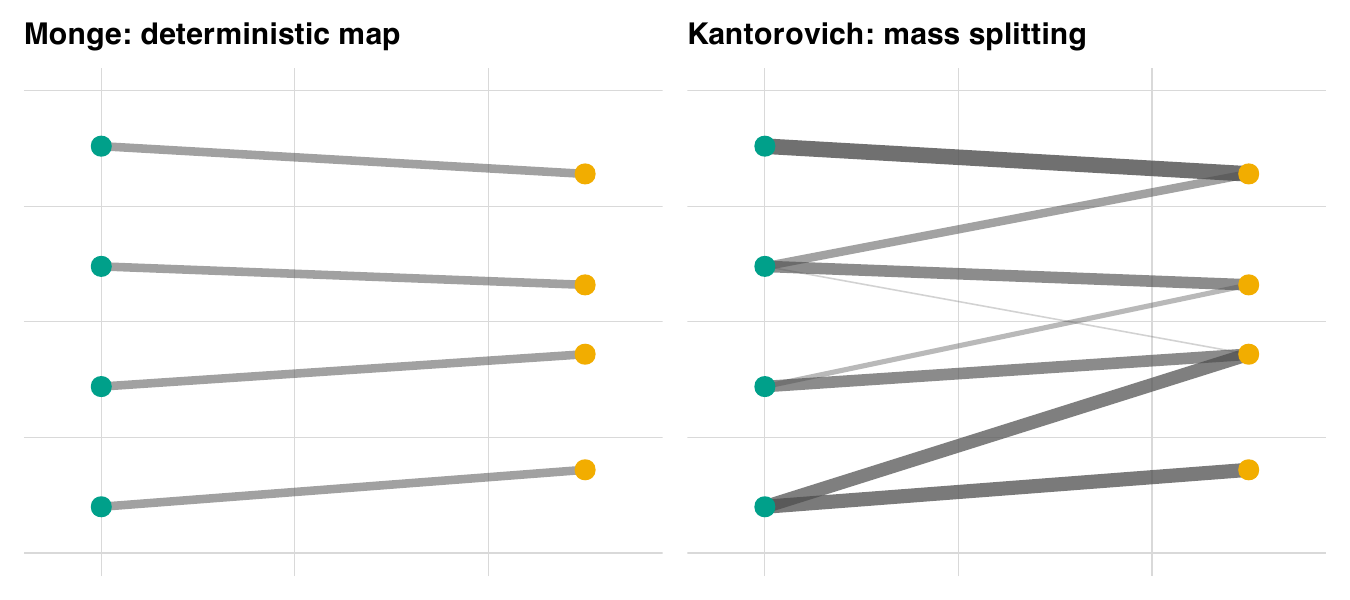}
    \caption{Monge map versus Kantorovich plan. In Monge's formulation, each source point is sent to a single target point. In Kantorovich's relaxation, mass can be split across several targets, producing a coupling or transport plan.}
    \label{fig:monge-kantorovich-plan}
\end{figure}

% \begin{center}
% \begin{tikzpicture}[scale=1, every node/.style={font=\small}]
%   \node at (1.5,2.6) {Monge map};
%   \draw[->,thick] (.2,2) -- (2.8,2);
%   \draw[->,thick] (.2,1.3) -- (2.8,1.7);
%   \draw[->,thick] (.2,.6) -- (2.8,1.1);
%   \foreach \y in {2,1.3,.6}{\fill (.2,\y) circle (2pt);}
%   \foreach \y in {2,1.7,1.1}{\fill (2.8,\y) circle (2pt);}
%   \node at (.2,.2) {\(\mu\)};
%   \node at (2.8,.2) {\(\nu\)};

%   \node at (6.5,2.6) {Kantorovich plan};
%   \foreach \y in {2,1.3,.6}{\fill (5.2,\y) circle (2pt);}
%   \foreach \y in {2.1,1.6,1.0,.5}{\fill (7.8,\y) circle (2pt);}
%   \draw[->,gray] (5.2,2) -- (7.8,2.1);
%   \draw[->,gray] (5.2,2) -- (7.8,1.6);
%   \draw[->,gray] (5.2,1.3) -- (7.8,1.6);
%   \draw[->,gray] (5.2,1.3) -- (7.8,1.0);
%   \draw[->,gray] (5.2,.6) -- (7.8,1.0);
%   \draw[->,gray] (5.2,.6) -- (7.8,.5);
%   \node at (5.2,.2) {\(\mu\)};
%   \node at (7.8,.2) {\(\nu\)};
% \end{tikzpicture}
% \end{center}

This distinction matters in applications. A deterministic map may be desirable when the transformation must act directly on an observed state, while a non-deterministic plan may be more natural for scenario matching or dependence uncertainty.

\subsection{Existence of optimal transport plans}
\label{sec:continuous-existence}

The finite problem always has a minimizer because the transportation polytope is compact. In the continuous setting, compactness is replaced by tightness and weak convergence of probability measures.

The following statement is a standard existence result; it is the continuous analogue of Proposition~\ref{prop:discrete-ot-existence}. The precise hypotheses can be relaxed or strengthened depending on the space and cost; the formulation below is enough for most actuarial uses.

\begin{theorem}[Existence, standard form]
\label{thm:continuous-ot-existence}
Let \(\mathcal X\) and \(\mathcal Y\) be Polish spaces, let \(\mu\in\Prob(\mathcal X)\) and \(\nu\in\Prob(\mathcal Y)\), and let \(c:\mathcal X\times\mathcal Y\to[0,+\infty]\) be lower semicontinuous. Then the Kantorovich problem admits an optimal plan, possibly with value \(+\infty\). If at least one coupling has finite cost, then the optimal value is finite and attained.
\end{theorem}

\begin{proof}[Proof sketch]
The set \(\Coup(\mu,\nu)\) is tight because its marginals are fixed. By Prokhorov's theorem, every minimizing sequence admits a weakly convergent subsequence. The limit has marginals \(\mu\) and \(\nu\), so it remains feasible. Lower semicontinuity of \(c\) implies lower semicontinuity of \(\gamma\mapsto\int c\dd\gamma\) under weak convergence. Thus the limit is optimal.
\end{proof}

\begin{remark}[Why lower semicontinuity matters]
\label{rem:lsc-cost}
If \(c\) is lower semicontinuous, jumps in the limiting plan cannot artificially reduce cost. This is the continuous analogue of having a well-defined finite cost matrix in the discrete problem.
\end{remark}

\begin{readingbox}
For rigorous treatments of existence, duality and optimal maps, see \cite{villani2003topics,villani2008oldnew,ambrosio2021lectures,figalli2021invitation}. For an economics-oriented presentation close to the finite-dimensional linear-programming intuition, see \cite{galichon2016economics}. For a computational perspective, see \cite{peyre2019computational}.
\end{readingbox}

\subsection{Kantorovich duality}
\label{sec:continuous-duality}

The continuous dual is the analogue of the finite dual from the previous lecture. In the finite case, the dual constraints were
\[
  u_i+v_j\le C_{ij}.
\]
In the continuous case, the vectors \(u\) and \(v\) become functions \(\varphi\) and \(\psi\), and the constraint becomes
\[
  \varphi(x)+\psi(y)\le c(x,y)
  \qquad\text{for all }x,y.
\]

\begin{theorem}[Kantorovich duality, standard version]
\label{thm:kantorovich-duality-continuous}
Under standard regularity assumptions on \(\mathcal X\), \(\mathcal Y\), \(\mu\), \(\nu\) and \(c\),
\[
 \highlight{GoldenRod}{ \OT_c(\mu,\nu)
  =
  \sup_{\varphi,\psi}
  \left\{
    \int \varphi\dd\mu+\int \psi\dd\nu:
    \varphi(x)+\psi(y)\le c(x,y)
  \right\}}
\]
The functions \(\varphi\) and \(\psi\) are called Kantorovich potentials.
\end{theorem}

The exact functional class can vary. For example, if \(c\) is lower semicontinuous and bounded below on Polish spaces, one can formulate the dual with suitable integrable potentials; in compact settings, continuous potentials are often enough. For the purpose of these notes, the important point is the structure: primal variables are couplings, dual variables are prices or potentials. This is the continuous counterpart of Theorem~\ref{thm:finite-kantorovich-duality}.

\begin{proposition}[Weak duality]
\label{prop:continuous-weak-duality}
For any \(\gamma\in\Coup(\mu,\nu)\) and any measurable functions \(\varphi,\psi\) such that \(\varphi(x)+\psi(y)\le c(x,y)\),
\[
  \int \varphi\dd\mu+\int \psi\dd\nu
  \le
  \int c\dd\gamma.
\]
\end{proposition}

\begin{proof}
Since \(\gamma\) has marginals \(\mu\) and \(\nu\),
\[
  \int \varphi(x)\dd\gamma(x,y)=\int \varphi\dd\mu,
  \qquad
  \int \psi(y)\dd\gamma(x,y)=\int \psi\dd\nu.
\]
Therefore
\[
  \int \varphi\dd\mu+\int \psi\dd\nu
  =\int \{\varphi(x)+\psi(y)\}\dd\gamma(x,y)
  \le \int c(x,y)\dd\gamma(x,y).
\]
\end{proof}

Weak duality has a useful actuarial interpretation. If \(\varphi\) and \(\psi\) are any functions satisfying \(\varphi(x)+\psi(y)\le c(x,y)\), then
\[
  \int \varphi\dd\mu+\int \psi\dd\nu
\]
is a certified lower bound on the cost of any transformation from \(\mu\) to \(\nu\). In risk aggregation problems where one maximizes a payoff over all couplings, the same argument provides certified upper or lower bounds on worst-case risk.

\begin{remark}[Shadow prices]
\label{rem:shadow-prices-potentials}
The dual potential \(\varphi(x)\) is the shadow value of mass at source state \(x\), while \(\psi(y)\) is the shadow value of mass at target state \(y\). In a portfolio drift problem, they identify states where distributional movement is costly.
\end{remark}

\subsection{The \texorpdfstring{\(c\)}{c}-transform}
\label{sec:continuous-c-transform}

The dual problem can be simplified by the \(c\)-transform. The following definition is the continuous counterpart of the discrete \(c\)-transform in Section~\ref{sec:discrete-ctransform}.

\begin{definition}[\(c\)-transform and \(c\)-concavity]
\label{def:c-transform-c-concavity}
Given a function \(\varphi:\mathcal X\to\R\cup\{-\infty\}\), its \(c\)-transform is the function on \(\mathcal Y\) defined by
\[
  \varphi^c(y)=\inf_{x\in\mathcal X}\{c(x,y)-\varphi(x)\}.
\]
Similarly, for \(\psi:\mathcal Y\to\R\cup\{-\infty\}\), define
\[
  \psi^c(x)=\inf_{y\in\mathcal Y}\{c(x,y)-\psi(y)\}.
\]
A function \(\varphi\) is called \(c\)-concave if there exists a function \(\psi\) such that
\[
  \varphi=\psi^c.
\]
Equivalently, \(\varphi\) can be written as an infimum of shifted cost functions.
\end{definition}

By construction,
\[
  \varphi(x)+\varphi^c(y)\le c(x,y)
\]
for all \(x,y\). Thus \((\varphi,\varphi^c)\) is dual feasible whenever the integrals are meaningful.

In the finite problem, replacing \(v\) by \(u^c\) can only improve the dual objective. The same intuition holds in the continuous case. At optimality, one may often choose potentials satisfying
\[
  \psi=\varphi^c,
  \qquad
  \varphi=\psi^c.
\]

\begin{example}[Squared Euclidean cost]
\label{ex:squared-cost-convex-analysis}
Let \(c(\bx,\by)=\frac12\|\bx-\by\|^2\) on \(\R^d\). Then
\[
  \varphi(\bx)+\psi(\by)\le \frac12\|\bx-\by\|^2.
\]
By rewriting
\[
  \frac12\|\bx-\by\|^2=\frac12\|\bx\|^2+\frac12\|\by\|^2-\bx\cdot \by,
\]
one can relate \(c\)-concavity to ordinary convex analysis. This is the route leading to Brenier's theorem.
\end{example}

\begin{proposition}[Quadratic cost and convex conjugacy]
\label{prop:quadratic-cost-convex-conjugacy}
Let \(c(\bx,\by)=\frac12\|\bx-\by\|^2\) on \(\R^d\). If \((\varphi,\psi)\) is dual feasible and
\[
  u(\bx)=\frac12\|\bx\|^2-\varphi(\bx),
  \qquad
  v(\by)=\frac12\|\by\|^2-\psi(\by),
\]
then
\[
  u(\bx)+v(\by)\ge \bx\cdot \by.
\]
In particular, at optimality one may interpret \(v\) as a convex-conjugate-type partner of \(u\), and the contact condition becomes
\[
  \by\in \partial u(\bx).
\]
If \(u\) is differentiable, this gives the Brenier map \(T(\bx)=\nabla u(\bx)\).
\end{proposition}

\begin{proof}
The dual inequality \(\varphi(\bx)+\psi(\by)\le\frac12\|\bx-\by\|^2\) is equivalent to
\[
  \frac12\|\bx\|^2-\varphi(\bx)+\frac12\|\by\|^2-\psi(\by)\ge \bx\cdot \by.
\]
This is precisely \(u(\bx)+v(\by)\ge \bx\cdot \by\). Equality on the contact set is the equality condition in the Fenchel inequality, hence \(\by\in\partial u(\bx)\). If \(u\) is differentiable, \(\partial u(\bx)=\{\nabla u(\bx)\}\).
\end{proof}

This is the first place where convex analysis enters explicitly. A short appendix insert on convex conjugacy, subgradients and Fenchel equality is proposed in the revision notes for the mathematical toolbox.

\subsection{Optimality and complementary slackness}
\label{sec:continuous-complementary-slackness}

In the discrete problem, mass moves only on edges where
\[
  u_i+v_j=C_{ij}.
\]
In the continuous problem, the same principle holds. If \(\gamma^\star\) is optimal and \((\varphi,\psi)\) are optimal potentials, then
\[
  \varphi(x)+\psi(y)=c(x,y)
\]
for \(\gamma^\star\)-almost every \((x,y)\).

\begin{proposition}[Continuous complementary slackness]
\label{prop:continuous-complementary-slackness}
Assume strong duality holds and let \(\gamma^\star\) be primal optimal and \((\varphi,\psi)\) be dual optimal. Then
\[
  c(x,y)-\varphi(x)-\psi(y)=0
  \qquad \gamma^\star\text{-a.s.}
\]
\end{proposition}

\begin{proof}
By feasibility,
\[
  c(x,y)-\varphi(x)-\psi(y)\ge0.
\]
By equality of primal and dual values,
\[
  \int \{c(x,y)-\varphi(x)-\psi(y)\}\dd\gamma^\star(x,y)=0.
\]
A non-negative random variable with expectation zero is zero almost surely.
\end{proof}

This result says that the support of an optimal plan is contained in the contact set
\[
  \{(x,y):\varphi(x)+\psi(y)=c(x,y)\}.
\]
When this contact set is the graph of a function, the optimal plan is induced by a Monge map. When it is not, mass splitting may occur.

\subsection{Cyclical monotonicity}
\label{sec:continuous-cyclical-monotonicity}

There is another way to describe the support of optimal plans.

\begin{definition}[\(c\)-cyclical monotonicity]
\label{def:c-cyclical-monotonicity}
Let \(\Gamma\subseteq\mathcal X\times\mathcal Y\). The set \(\Gamma\) is called \(c\)-cyclically monotone if for every finite family \((x_1,y_1),\ldots,(x_N,y_N)\in\Gamma\) and every permutation \(\sigma\) of \(\{1,\ldots,N\}\),
\[
  \sum_{k=1}^N c(x_k,y_k)
  \le
  \sum_{k=1}^N c(x_k,y_{\sigma(k)}).
\]
\end{definition}

This means that once the pairs in \(\Gamma\) have been chosen, no finite reshuffling of destinations can reduce the total cost. In the discrete assignment problem, this is exactly the condition that no cycle exchange improves the matching. For the quadratic cost, this condition is closely related to cyclic monotonicity of the graph of a subgradient of a convex function.

\begin{theorem}[Optimality and \(c\)-cyclical monotonicity, standard version]
\label{thm:c-cyclical-monotonicity-optimality}
Under standard assumptions, a transport plan \(\gamma\) is optimal only if its support is \(c\)-cyclically monotone. Conversely, a plan concentrated on a suitable \(c\)-cyclically monotone set is optimal.
\end{theorem}

\begin{example}[Two-pair exchange]
\label{ex:two-pair-cyclical-monotonicity}
For two pairs \((x_1,y_1)\), \((x_2,y_2)\), \(c\)-cyclical monotonicity gives
\[
  c(x_1,y_1)+c(x_2,y_2)
  \le
  c(x_1,y_2)+c(x_2,y_1).
\]
In dimension one with convex distance cost, this inequality rules out crossings. Thus the monotone rearrangement is optimal.
\end{example}

\subsection{The one-dimensional case: quantile transport}
\label{sec:continuous-quantile-transport}

A central actuarial special case is dimension one. Let \(\mu\) and \(\nu\) be probability distributions on \(\R\), with distribution functions \(F_\mu\) and \(F_\nu\). Following the convention introduced in Definition~\ref{def:rank-preserving-quantile-map}, write
\[
  Q_\mu(u)=F_\mu^{-1}(u)=\inf\{x:F_\mu(x)\ge u\},
  \qquad
  Q_\nu(u)=F_\nu^{-1}(u)=\inf\{y:F_\nu(y)\ge u\}.
\]
Let \(U\sim\mathrm{Uniform}(0,1)\). Then
\[
  X=Q_\mu(U)\sim\mu,
  \qquad
  Y=Q_\nu(U)\sim\nu.
\]
The coupling \((X,Y)\) is comonotonic. It matches equal probability levels.

\begin{theorem}[Monotone rearrangement]
\label{thm:monotone-rearrangement}
Let \(c(x,y)=h(x-y)\), where \(h\) is convex. Then the coupling
\[
  (X,Y)=\bigl(Q_\mu(U),Q_\nu(U)\bigr)
\]
is optimal. In particular, for \(p\ge1\),
\[
  \Wp{p}(\mu,\nu)^p
  =\int_0^1
  \left|Q_\mu(u)-Q_\nu(u)\right|^p\dd u.
\]
\end{theorem}

If \(F_\mu\) is continuous and strictly increasing, the optimal map from \(\mu\) to \(\nu\) is
\[
  T(x)=Q_\nu(F_\mu(x)).
\]
This is the quantile-to-quantile transformation.

\begin{actuarialbox}
The formula
\[
  T(x)=Q_\nu(F_\mu(x))
\]
is a useful practical formula in OT for actuaries. It transforms a severity observation into the corresponding severity level under another distribution, preserving its rank. This is the basis for distributional calibration, stress transformations and quantile-based risk comparisons.
\end{actuarialbox}

\begin{example}[Severity stress by quantile mapping]
\label{ex:severity-stress-quantile-map}
Let \(\mu\) be the baseline severity law and \(\nu\) a stressed severity law. A claim at percentile \(u=F_\mu(x)\) is mapped to the stressed claim
\[
  T(x)=Q_\nu(u).
\]
This produces a stress transformation that preserves ranks. High baseline claims remain high stressed claims, but the scale and tail thickness are changed according to \(\nu\).
\end{example}

\iffalse
% To be restored in v2.
\begin{example}[Premium distribution correction]
\label{ex:premium-fairness-quantile-correction}
Suppose \(\mu_s\) is the score or premium distribution for group \(s\), and \(\bar\mu\) is a target distribution. In one dimension, the Wasserstein correction is
\[
  T_s(x)=Q_{\bar\mu}(F_{\mu_s}(x)).
\]
This is exactly the monotone map sending each group-specific distribution to the target distribution. This construction is used again in the fairness chapter and in Lab~3 on Wasserstein fairness repair.
\end{example}
\fi

\begin{remark}[Atoms and randomized ranks]
\label{rem:atoms-randomized-ranks}
The coupling \((Q_\mu(U),Q_\nu(U))\) is always well defined. The deterministic map \(x\mapsto Q_\nu(F_\mu(x))\), however, should be used with care when \(\mu\) has atoms: ordinary ranks are then not uniform without a randomization convention inside flat parts of the distribution function. In actuarial applications, atoms arise naturally at zero claims, deductibles and policy limits.
\end{remark}

\subsection{Wasserstein distances}
\label{sec:continuous-wasserstein-distances}

The following definition fixes the convention used throughout the rest of the notes.

\begin{definition}[\(p\)-Wasserstein distance]
\label{def:p-wasserstein-distance}
Let \((\mathcal X,d)\) be a metric space and let \(p\ge1\). Define
\[
  \Probp{p}(\mathcal X)
  =\left\{\mu\in\Prob(\mathcal X):
  \int d(x_0,x)^p\dd\mu(x)<\infty
  \text{ for some }x_0\in\mathcal X
  \right\}.
\]
For \(\mu,\nu\in\Probp{p}(\mathcal X)\), the \(p\)-Wasserstein distance is
\[
  \Wp{p}(\mu,\nu)
  =\left(
  \inf_{\gamma\in\Coup(\mu,\nu)}
  \int d(x,y)^p\dd\gamma(x,y)
  \right)^{1/p}.
\]
The quantity inside the parentheses is often denoted \(\Wp{p}(\mu,\nu)^p\).
\end{definition}

The condition defining \(\Probp{p}(\mathcal X)\) does not depend on the particular choice of \(x_0\), as soon as \(d\) is a metric.

\begin{proposition}[Metric property]
\label{prop:wasserstein-metric}
For \(p\ge1\), \(\Wp{p}\) from Definition~\ref{def:p-wasserstein-distance} defines a metric on \(\Probp{p}(\mathcal X)\), under standard assumptions on \((\mathcal X,d)\).
\end{proposition}

\begin{proof}[Proof idea]
Non-negativity and symmetry are immediate from the definition. If \(\Wp{p}(\mu,\nu)=0\), then any optimal coupling must be concentrated on \(x=y\), so \(\mu=\nu\). The triangle inequality is obtained by gluing an optimal coupling between \(\mu\) and \(\nu\) with one between \(\nu\) and \(\eta\), producing a joint law \((X,Y,Z)\), and then using Minkowski's inequality:
\[
  \left(\E d(X,Z)^p\right)^{1/p}
  \le
  \left(\E d(X,Y)^p\right)^{1/p}
  +
  \left(\E d(Y,Z)^p\right)^{1/p}.
\]
\end{proof}

\subsection{The Kantorovich--Rubinstein dual for \texorpdfstring{\(W_1\)}{W1}}
\label{sec:continuous-kr-dual}

For \(p=1\), the dual has a particularly important form.

\begin{theorem}[Kantorovich--Rubinstein duality]
\label{thm:kantorovich-rubinstein-duality}
If \(c(x,y)=d(x,y)\) and \(\mu,\nu\in\Probp{1}(\mathcal X)\), then
\[
  \Wone(\mu,\nu)
  =\sup_{f:\operatorname{Lip}(f)\le1}
  \left\{\int f\dd\mu-\int f\dd\nu\right\}.
\]
Here \(\operatorname{Lip}(f)\le1\) means
\[
  |f(x)-f(y)|\le d(x,y)
  \qquad\forall x,y.
\]
\end{theorem}

\begin{actuarialbox}
The \(W_1\) distance compares distributions through their worst possible difference in expectation over all 1-Lipschitz test functions. Thus it asks: among all stable actuarial indicators, what is the largest change in expected value between two distributions?
\end{actuarialbox}

\begin{example}[Stable portfolio indicators]
\label{ex:stable-portfolio-indicators}
A function \(f\) of policyholder characteristics is 1-Lipschitz if similar policyholders have similar indicator values. The \(W_1\) distance between two portfolio distributions is the largest possible difference in expected value over all such stable indicators.
\end{example}

\subsection{Quadratic cost and Brenier maps}
\label{sec:continuous-brenier}

A central multivariate cost is the quadratic cost on \(\R^d\):
\[
  c(\bx,\by)=\frac12\|\bx-\by\|^2.
\]
It leads to \(\Wtwo\), Hilbert-like geometry, barycenters, gradient flows, and many applications in statistics and machine learning.

The central structural result is Brenier's theorem.

\begin{theorem}[Brenier theorem, standard version]
\label{thm:brenier}
Let \(\mu,\nu\in\Probp{2}(\R^d)\). Suppose that \(\mu\) is absolutely continuous with respect to Lebesgue measure. Then for the quadratic cost \(c(\bx,\by)=\frac12\|\bx-\by\|^2\), there exists a unique optimal transport plan. It is induced by a \(\mu\)-a.e. unique map
\[
  T(\bx)=\nabla \phi(\bx),
\]
where \(\phi:\R^d\to\R\) is a convex function. Thus
\[
  \gamma^\star=(\Id,T)_{\push}\mu,
  \qquad
  T_{\push}\mu=\nu.
\]
\end{theorem}

The theorem says that, under regularity of the source distribution, the Kantorovich relaxation recovers a deterministic Monge map. Moreover, this map has a special form: it is the gradient of a convex potential.

\begin{remark}[Why convex gradients?]
For quadratic cost,
\[
  \frac12\|\bx-\by\|^2=\frac12\|\bx\|^2+\frac12\|\by\|^2-\bx\cdot \by.
\]
Optimality becomes linked to convex conjugacy. The map \(T=\nabla\phi\) arises from the subdifferential of a convex function, and cyclical monotonicity becomes ordinary monotonicity of a convex gradient.
\end{remark}

\begin{example}[Multivariate risk transformation]
\label{ex:multivariate-risk-transformation}
Let \(\bX\in\R^d\) represent a vector of losses by line of business under a baseline model, and let \(\nu\) be a stressed multivariate loss distribution. Brenier's map, when it exists, gives the least-squares distributional transformation
\[
  T_{\push}\Law(\bX)=\nu.
\]
Unlike componentwise quantile mapping, Brenier's map respects multivariate dependence.
\end{example}

\subsection{The Gaussian case}
\label{sec:continuous-gaussian}

A useful explicit example is the transport between multivariate Gaussian distributions. Closed-form Gaussian formulas for the quadratic Wasserstein distance go back at least to \cite{dowson1982frechet,gelbrich1990formula} and are now standard in OT and statistics.

\begin{theorem}[Gaussian \(W_2\) formula]
\label{thm:gaussian-w2-formula}
Let
\[
  \mu=\mathcal N(\bm_0,\Sigma_0),
  \qquad
  \nu=\mathcal N(\bm_1,\Sigma_1),
\]
where \(\Sigma_0\) and \(\Sigma_1\) are positive definite. For quadratic cost, the optimal map is affine:
\[
  T(\bx)=\bm_1+A(\bx-\bm_0),
\]
where
\[
  A=\Sigma_0^{-1/2}
  \left(\Sigma_0^{1/2}\Sigma_1\Sigma_0^{1/2}\right)^{1/2}
  \Sigma_0^{-1/2}.
\]
The squared Wasserstein distance is
\[
  \Wtwo^2(\mu,\nu)
  =\|\bm_0-\bm_1\|^2
  +\operatorname{tr}\left(
    \Sigma_0+\Sigma_1
    -2(\Sigma_0^{1/2}\Sigma_1\Sigma_0^{1/2})^{1/2}
  \right).
\]
\end{theorem}

\begin{figure}[t]
\centering
\includegraphics[width=.82\textwidth]{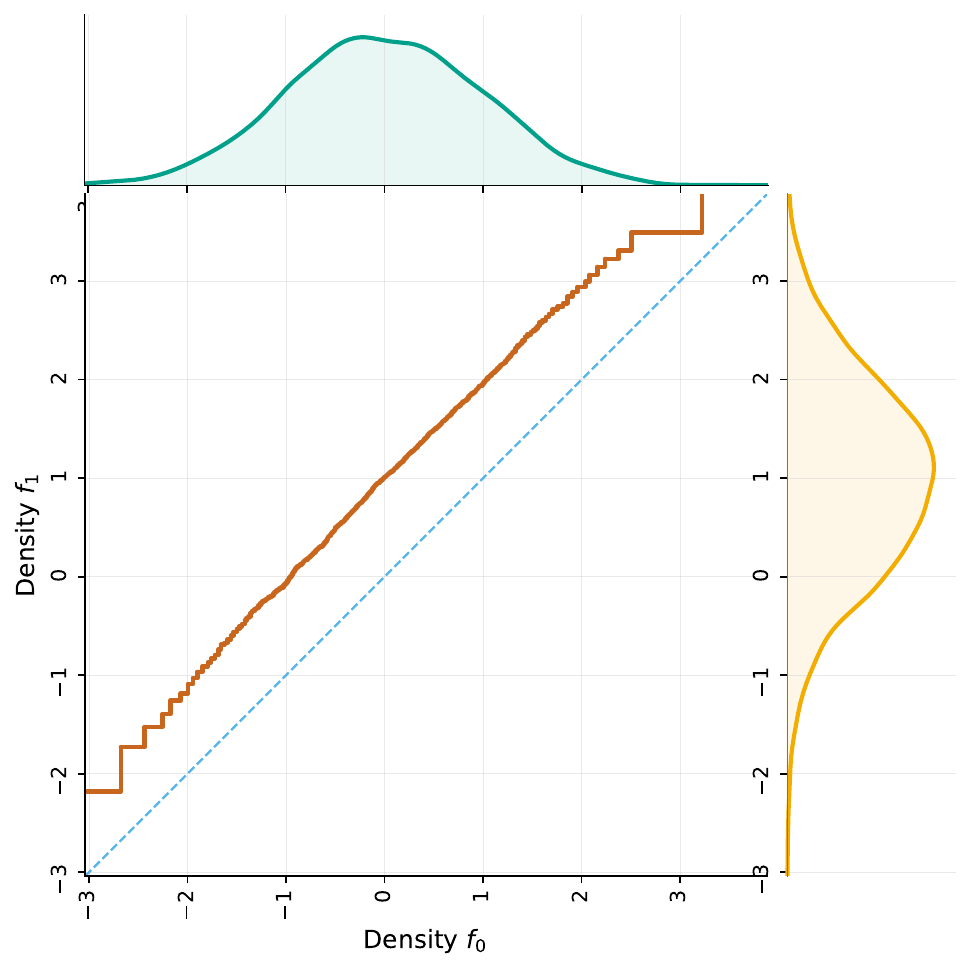}
\caption{One-dimensional transport with a Gaussian mean shift. The source and target distributions have the same variance, so the monotone transport map is approximately a translation: ranks are preserved and each value is shifted by the change in mean.}
\label{fig:transport-gaussian-mean-shift}
\end{figure}

\begin{figure}[t]
\centering
\includegraphics[width=.82\textwidth]{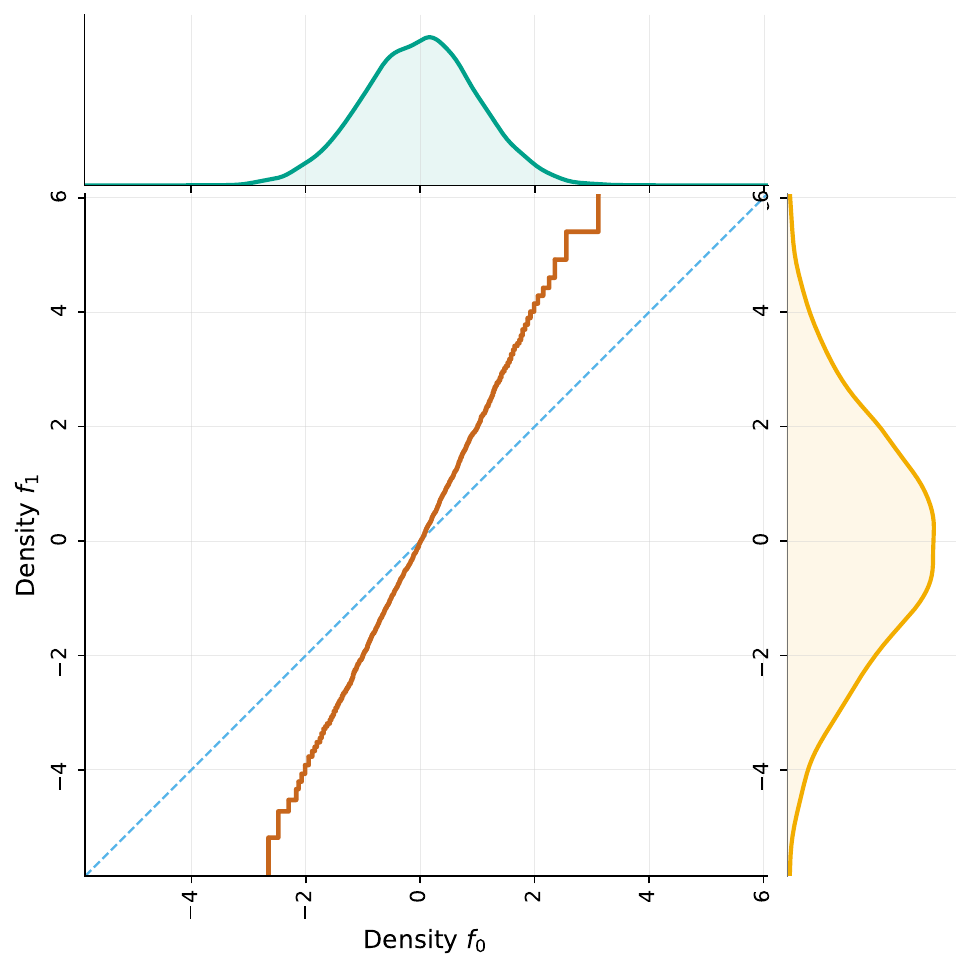}
\caption{One-dimensional transport with a Gaussian variance shift. The two distributions have the same mean but different variances. The transport map expands the tails and contracts the centre according to the ratio of standard deviations.}
\label{fig:transport-gaussian-variance-shift}
\end{figure}

\begin{figure}[t]
\centering
\includegraphics[width=.82\textwidth]{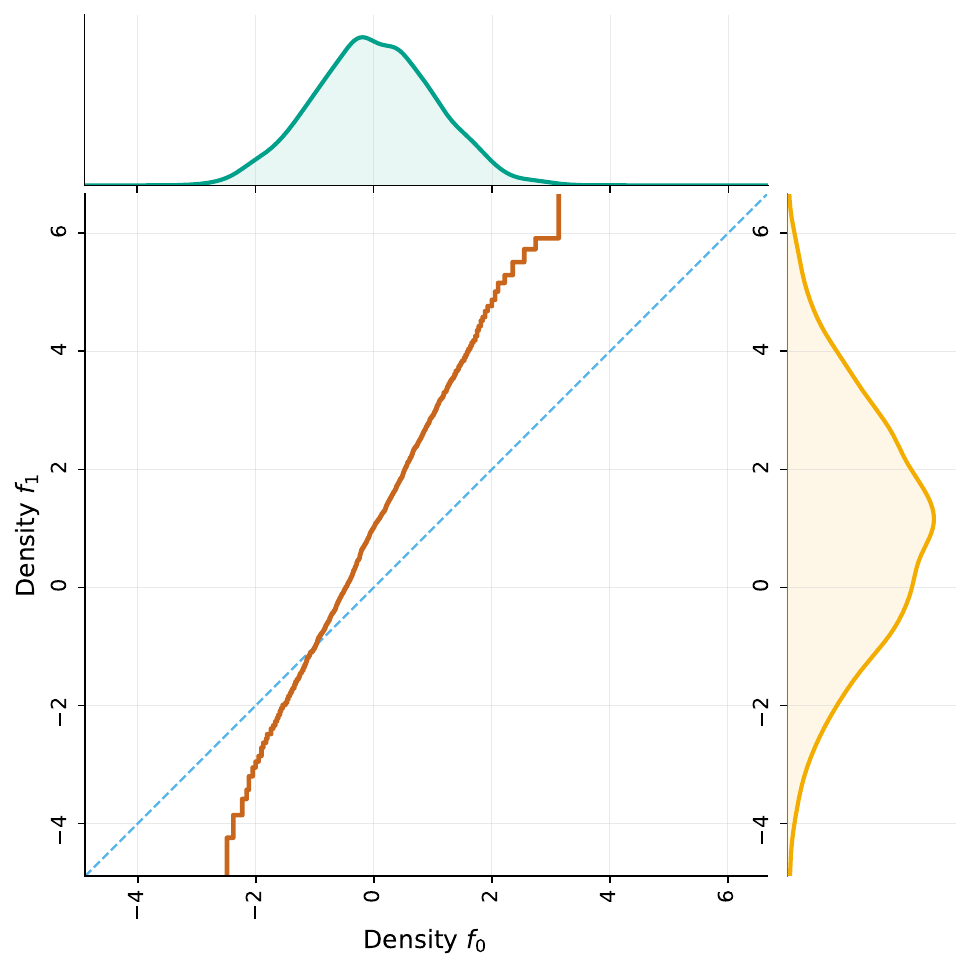}
\caption{Gaussian transport with simultaneous mean and variance change, displayed with common axes. For Gaussian laws, the optimal one-dimensional map is affine; the intercept reflects the mean shift and the slope reflects the standard-deviation ratio.}
\label{fig:transport-gaussian-mean-variance-equal-axes}
\end{figure}

\begin{example}[Capital-model comparison]
\label{ex:gaussian-capital-model-comparison}
Suppose two internal capital models approximate the vector of line-of-business losses by Gaussian distributions with different mean vectors and covariance matrices. The formula separates a mean shift from a covariance-structure shift. This is useful when distinguishing a change in expected losses from a change in diversification.
\end{example}

\subsection{Beyond Brenier: triangular and sliced Gaussian maps}
\label{sec:continuous-gaussian-triangular-sliced-maps}

The Gaussian formula in the previous section gives the quadratic optimal transport map between two multivariate normal laws. It is the Brenier map: it is the gradient of a convex function and minimizes the mean squared displacement. However, it is not the only useful map sending one Gaussian law to another.

In actuarial applications, the way the transformation is constructed may matter. For instance, when a bivariate vector represents two guarantees, two lines of business, two cash-flow dates, or two risk components, one may want either:
\begin{itemize}
\item a symmetric least-squares transformation of the joint law;
\item a sequential transformation, where one coordinate is transformed first and the second is adjusted conditionally;
\item a projection-wise transformation, where selected one-dimensional directions are matched.
\end{itemize}
These lead to different maps, even when the source and target distributions are the same.

\begin{readingbox}
Triangular transports are closely related to the Knothe--Rosenblatt rearrangement \cite{knothe1957contributions,rosenblatt1952remarks}. Projection-wise and sliced constructions are related to sliced Wasserstein distances and their computational variants \cite{bonneel2015sliced,kolouri2019sliced,nadjahi2020statistical}. These maps are useful modelling devices, but only the Brenier map is quadratic-cost optimal under the assumptions of Theorem~\ref{thm:brenier}.
\end{readingbox}

Let
\[
  \mu_0=\mathcal N(\bm_0,\Sigma_0),
  \qquad
  \mu_1=\mathcal N(\bm_1,\Sigma_1)
\]
be two Gaussian measures on \(\R^2\), with
\[
  \bm_i=
  \begin{pmatrix}
  \bar x_i\\
  \bar y_i
  \end{pmatrix},
  \qquad
  \Sigma_i=
  \begin{pmatrix}
  a_i & c_i\\
  c_i & b_i
  \end{pmatrix},
  \qquad i\in\{0,1\}.
\]
We assume throughout this section that \(\Sigma_0\) and \(\Sigma_1\) are positive definite.

\subsubsection{The Brenier Gaussian map}
\label{sec:continuous-gaussian-brenier-map}

For the quadratic cost, the optimal transport map from \(\mu_0\) to \(\mu_1\) is affine:
\[
  T_{\mathrm{OT}}(\bz)
  =
  \bm_1 + A(\bz-\bm_0),
\]
where
\[
  A
  =
  \Sigma_0^{-1/2}
  \left(
    \Sigma_0^{1/2}\Sigma_1\Sigma_0^{1/2}
  \right)^{1/2}
  \Sigma_0^{-1/2}.
\]
Equivalently, \(A\) is the unique symmetric positive definite matrix satisfying
\[
  A=A^\top,
  \qquad
  A\succ 0,
  \qquad
  A\Sigma_0 A=\Sigma_1.
\]
This symmetry condition is important. Among all affine maps sending \(\mathcal N(\bm_0,\Sigma_0)\) to \(\mathcal N(\bm_1,\Sigma_1)\), the Brenier map is the one compatible with a convex potential. It is therefore the quadratic optimal transport map.

\begin{actuarialbox}
If \(\bZ=(Z_1,Z_2)\) represents two dependent actuarial quantities, the Brenier map gives the least-squares joint stress from the baseline Gaussian model to the target Gaussian model. It treats the two coordinates symmetrically through the covariance matrices.
\end{actuarialbox}

\subsubsection{A lower-triangular Knothe--Rosenblatt map}
\label{sec:continuous-kr-lower-map}

Another natural construction is sequential. First transport the first marginal, then transport the conditional law of the second coordinate given the first.

For \(i\in\{0,1\}\), define
\[
  \beta_i=\frac{c_i}{a_i},
  \qquad
  v_i=b_i-\frac{c_i^2}{a_i}.
\]
Then
\[
  X_i\sim \mathcal N(\bar x_i,a_i),
  \qquad
  Y_i\mid X_i=x
  \sim
  \mathcal N\left(
    \bar y_i+\beta_i(x-\bar x_i),
    v_i
  \right).
\]
The monotone transport of the \(x\)-marginal is
\[
  T_x(x)
  =
  \bar x_1
  +
  \sqrt{\frac{a_1}{a_0}}\,(x-\bar x_0).
\]
Given this transformed value \(T_x(x)\), the conditional Gaussian transport in the second coordinate is
\[
  T_{y\mid x}(x,y)
  =
  \bar y_1
  +
  \beta_1\bigl(T_x(x)-\bar x_1\bigr)
  +
  \sqrt{\frac{v_1}{v_0}}\,
  \left[
    y-\bar y_0-\beta_0(x-\bar x_0)
  \right].
\]
The lower-triangular Knothe--Rosenblatt map is therefore
\[
  T_{\mathrm{KR},\downarrow}(x,y)
  =
  \begin{pmatrix}
  T_x(x)\\
  T_{y\mid x}(x,y)
  \end{pmatrix}.
\]
It can be written in affine form as
\[
  T_{\mathrm{KR},\downarrow}(\bz)
  =
  \bm_1+L(\bz-\bm_0),
\]
with
\[
  L=
  \begin{pmatrix}
  \sqrt{a_1/a_0} & 0\\
  \beta_1\sqrt{a_1/a_0}
  -
  \sqrt{v_1/v_0}\,\beta_0
  &
  \sqrt{v_1/v_0}
  \end{pmatrix}.
\]

\begin{proposition}[Lower-triangular Gaussian transport]
\label{prop:lower-triangular-gaussian-transport}
The map \(T_{\mathrm{KR},\downarrow}\) pushes \(\mu_0\) forward to \(\mu_1\).
\end{proposition}

\begin{proof}
The first coordinate \(T_x(X_0)\) has law \(\mathcal N(\bar x_1,a_1)\). Conditional on \(X_0=x\), the residual
\[
  Y_0-\bar y_0-\beta_0(x-\bar x_0)
\]
has law \(\mathcal N(0,v_0)\). Multiplying it by \(\sqrt{v_1/v_0}\) gives a residual with law \(\mathcal N(0,v_1)\). Adding the conditional mean
\[
  \bar y_1+\beta_1(T_x(x)-\bar x_1)
\]
therefore gives the conditional law of \(Y_1\) given \(X_1=T_x(x)\). Hence the joint law is \(\mathcal N(\bm_1,\Sigma_1)\).
\end{proof}

This construction is the Gaussian version of the conditional quantile transform. It is exact, but it is not symmetric in the two coordinates: the first coordinate is transformed first, and the second one follows conditionally.

\subsubsection{An upper-triangular Knothe--Rosenblatt map}
\label{sec:continuous-kr-upper-map}

Changing the order gives another exact transport map. Now one first transports the \(y\)-marginal and then the conditional law of \(x\) given \(y\).

Define
\[
  \gamma_i=\frac{c_i}{b_i},
  \qquad
  w_i=a_i-\frac{c_i^2}{b_i},
  \qquad i\in\{0,1\}.
\]
Then
\[
  Y_i\sim \mathcal N(\bar y_i,b_i),
  \qquad
  X_i\mid Y_i=y
  \sim
  \mathcal N\left(
    \bar x_i+\gamma_i(y-\bar y_i),
    w_i
  \right).
\]
The monotone transport of the \(y\)-marginal is
\[
  T_y(y)
  =
  \bar y_1
  +
  \sqrt{\frac{b_1}{b_0}}\,(y-\bar y_0).
\]
The conditional transport in the first coordinate is
\[
  T_{x\mid y}(x,y)
  =
  \bar x_1
  +
  \gamma_1\bigl(T_y(y)-\bar y_1\bigr)
  +
  \sqrt{\frac{w_1}{w_0}}\,
  \left[
    x-\bar x_0-\gamma_0(y-\bar y_0)
  \right].
\]
Thus the upper-triangular Knothe--Rosenblatt map is
\[
  T_{\mathrm{KR},\uparrow}(x,y)
  =
  \begin{pmatrix}
  T_{x\mid y}(x,y)\\
  T_y(y)
  \end{pmatrix},
\]
or equivalently
\[
  T_{\mathrm{KR},\uparrow}(\bz)
  =
  \bm_1+U(\bz-\bm_0),
\]
with
\[
  U=
  \begin{pmatrix}
  \sqrt{w_1/w_0}
  &
  \gamma_1\sqrt{b_1/b_0}
  -
  \sqrt{w_1/w_0}\,\gamma_0\\
  0
  &
  \sqrt{b_1/b_0}
  \end{pmatrix}.
\]

\begin{proposition}[Upper-triangular Gaussian transport]
\label{prop:upper-triangular-gaussian-transport}
The map \(T_{\mathrm{KR},\uparrow}\) pushes \(\mu_0\) forward to \(\mu_1\).
\end{proposition}

\begin{proof}
The proof is the same conditional-Gaussian argument as in Proposition~\ref{prop:lower-triangular-gaussian-transport}, with the roles of the two coordinates reversed. The transformed marginal \(T_y(Y_0)\) has law \(\mathcal N(\bar y_1,b_1)\), and the conditional residual in the first coordinate is rescaled from variance \(w_0\) to variance \(w_1\). The resulting conditional law is the conditional law of \(X_1\) given \(Y_1=T_y(y)\), hence the joint law is \(\mu_1\).
\end{proof}

\begin{remark}[Order matters]
\label{rem:kr-order-matters}
Both \(T_{\mathrm{KR},\downarrow}\) and \(T_{\mathrm{KR},\uparrow}\) transport \(\mu_0\) to \(\mu_1\), but they are generally different. The lower map gives priority to the first coordinate; the upper map gives priority to the second coordinate. Neither is usually equal to the Brenier map.
\end{remark}

\begin{actuarialbox}
Triangular maps are useful when the actuarial vector has a natural order. For example, one may first transport the main guarantee and then the rider conditional on it; or first transport the early cash-flow and then later cash-flows conditionally. The price paid for this sequential interpretation is that the map depends on the chosen ordering.
\end{actuarialbox}

\subsubsection{A projection-wise sliced Gaussian map}
\label{sec:continuous-sliced-gaussian-map}

A third construction uses one-dimensional Gaussian projections. This is related to sliced Wasserstein diagnostics, which are developed later in Lab~2 on sliced and projected Wasserstein distances. Let \(\theta\in S^{d-1}\) be a unit direction. If \(\bZ_i\sim \mathcal N(\bm_i,\Sigma_i)\), then
\[
  \theta^\top \bZ_i
  \sim
  \mathcal N\left(
    \theta^\top \bm_i,
    \theta^\top\Sigma_i\theta
  \right).
\]
The one-dimensional monotone Gaussian transport along direction \(\theta\) is therefore
\[
  t_\theta(s)
  =
  \theta^\top \bm_1
  +
  \sqrt{
    \frac{\theta^\top\Sigma_1\theta}
         {\theta^\top\Sigma_0\theta}
  }
  \left(
    s-\theta^\top \bm_0
  \right).
\]

Now fix an orthonormal basis
\[
  B=(\theta_1,\ldots,\theta_d).
\]
For each \(k\), set
\[
  \lambda_{i,k}=\theta_k^\top\Sigma_i\theta_k,
  \qquad
  r_k=\sqrt{\frac{\lambda_{1,k}}{\lambda_{0,k}}},
  \qquad
  i\in\{0,1\}.
\]
A projection-wise Gaussian scaling map is
\[
  T_B(\bz)
  =
  \bm_1
  +
  B\operatorname{diag}(r_1,\ldots,r_d)B^\top(\bz-\bm_0).
\]
In dimension two, with \(B=(\theta,\theta^\perp)\), this becomes
\[
  T_B(\bz)
  =
  \bm_1
  +
  B
  \begin{pmatrix}
  \sqrt{
    \dfrac{\theta^\top\Sigma_1\theta}
          {\theta^\top\Sigma_0\theta}
  }
  & 0\\[1.2em]
  0 &
  \sqrt{
    \dfrac{(\theta^\perp)^\top\Sigma_1\theta^\perp}
          {(\theta^\perp)^\top\Sigma_0\theta^\perp}
  }
  \end{pmatrix}
  B^\top(\bz-\bm_0).
\]

\begin{remark}[Projection-wise matching is not full transport]
\label{rem:projection-wise-not-full-transport}
The map \(T_B\) matches means and variances along the selected orthonormal directions. However, it does not generally push \(\mu_0\) exactly to \(\mu_1\). Indeed, the covariance of \(T_B(\bZ_0)\) is
\[
  B\operatorname{diag}(r_1,\ldots,r_d)B^\top
  \Sigma_0
  B\operatorname{diag}(r_1,\ldots,r_d)B^\top,
\]
which need not equal \(\Sigma_1\). Equality holds in particular when \(\Sigma_0\) and \(\Sigma_1\) are diagonal in the same basis \(B\). Thus \(T_B\) should be interpreted as a projection-wise stress map, not as the full Brenier map.
\end{remark}

\subsubsection{Comparison}
\label{sec:continuous-gaussian-map-comparison}

The three constructions answer different questions:
\[
\begin{array}{lll}
T_{\mathrm{OT}} & \text{is symmetric and quadratic-cost optimal;} \\
T_{\mathrm{KR},\downarrow} & \text{is exact and sequential in the order } x\to y;\\
T_{\mathrm{KR},\uparrow} & \text{is exact and sequential in the order } y\to x;\\
T_B & \text{matches selected projected variances.}
\end{array}
\]
Only the first one is the Brenier map. The two Knothe--Rosenblatt maps are exact transport maps but depend on the ordering of the coordinates. The sliced map is a diagnostic or approximate stress map unless the covariance matrices are compatible with the chosen basis.

\begin{example}[Two guarantees]
\label{ex:two-guarantees-gaussian-maps}
Let \(\bZ=(Z_1,Z_2)\) represent the present values of two guarantees in a life or P\&C contract. A Gaussian approximation may be used for the baseline law and for a stressed law. The Brenier map gives the least-squares joint stress. A lower Knothe--Rosenblatt map can be used when \(Z_1\) is the primary guarantee and \(Z_2\) is adjusted conditionally. An upper map reverses this priority. A sliced map may be used to stress the total \(Z_1+Z_2\) and the difference \(Z_1-Z_2\) by taking directions proportional to \((1,1)\) and \((1,-1)\).
\end{example}

\begin{computebox}
In Gaussian stress testing, one should report which map is used. Saying that a baseline Gaussian model is transported to a stressed Gaussian model is incomplete: Brenier, triangular and projection-wise maps may produce different transformed scenarios, even when they share the same endpoint distributions.
\end{computebox}

\subsection{Monge--Ampere equation, briefly}
\label{sec:continuous-monge-ampere}

If \(\mu\) and \(\nu\) have densities \(f\) and \(g\) on \(\R^d\), and if the Brenier map \(T=\nabla\phi\) is smooth and invertible, the constraint \(T_{\push}\mu=\nu\) can be written as a change-of-variables formula:
\[
\highlight{GoldenRod}{  f(\bx)=g(\nabla\phi(\bx))\det(D^2\phi(\bx))}
\]
This is the Monge--Ampere equation; see, for example, \cite{caffarelli1992regularity,figalli2017mongeampere} for regularity-oriented treatments.

For actuarial purposes, the main message is not that one should solve this PDE in routine applications. Rather, it clarifies that multivariate OT is a nonlinear distributional calibration problem. The Jacobian of the map adjusts density: where target density is high, the map compresses mass; where target density is low, it expands mass.

\begin{actuarialbox}%[Distributional, not causal]
A transport map or coupling aligns two probability laws. It does not by itself identify what would have happened to a given policyholder, claim or scenario under an intervention. Such an interpretation requires a structural model, an intervention design or additional identifying assumptions.
\end{actuarialbox}

\subsection{Actuarial interpretations of continuous OT}
\label{sec:continuous-actuarial-interpretations}

\subsubsection*{Severity models}
\label{sec:continuous-severity-models}

If \(\mu\) and \(\nu\) are two severity laws, \(\Wp{p}(\mu,\nu)\) measures the full distributional discrepancy. It is sensitive to differences in spread and tails, not only to mean severity. In dimension one, the quantile formula makes this especially transparent:
\[
  \Wp{p}(\mu,\nu)^p
  =\int_0^1 |Q_\mu(u)-Q_\nu(u)|^p\dd u.
\]

\iffalse
% To be restored in v2.
\subsubsection*{Tariff change}
\label{sec:continuous-tariff-change}

Let \(\mu\) be the distribution of old premiums \(\Law(\Premium_{\mathrm{old}}(\bX))\) and \(\nu\) the distribution of new premiums \(\Law(\Premium_{\mathrm{new}}(\bX))\). The Wasserstein distance measures distributional tariff movement. A paired policy-level change measures disruption for individuals; OT measures the change between premium distributions.

\fi

\subsubsection*{Portfolio drift}
\label{sec:continuous-portfolio-drift}

Let \(\mu_t\) be the law of policyholder covariates at time \(t\), and \(\mu_{t+1}\) the law one year later. A Wasserstein distance between \(\mu_t\) and \(\mu_{t+1}\) gives a geometrically meaningful drift measure. If the coordinates include age, region, insured value and claim history, the distance captures joint movement in the portfolio mix.

\subsubsection*{Climate stress}
\label{sec:continuous-climate-stress}

Let \(\mu\) be the distribution of annual losses under a baseline climate and \(\nu\) the distribution under a stressed climate. The transport map describes how baseline scenarios are transformed into stressed scenarios. In multivariate settings, this may include changes in both marginal hazard intensities and dependence between regions.

\subsubsection*{Reserving distributions}
\label{sec:continuous-reserving-distributions}

Let \(\mu\) and \(\nu\) be predictive distributions of future claim payments produced by two reserving methods. OT compares the full cash-flow distributions. If the state space is \(\R^d\), each coordinate may represent a future development year.

\iffalse
% To be restored in v2.
\subsubsection*{Fairness}
\label{sec:continuous-fairness}

Let \(\mu_s\) be the distribution of scores or premiums in group \(s\). A distributional fairness constraint such as demographic parity asks that these distributions be equal across groups. OT provides the minimal transformation needed to align each \(\mu_s\) with a target distribution, often a Wasserstein barycenter.

\fi

\begin{actuarialbox}
Continuous OT gives a common language for several actuarial transformations: inflation, stress testing, calibration, model comparison and scenario alignment. In each case, the mathematical object is the same: transform one law into another at minimum cost.
\end{actuarialbox}

\subsection{A mini-lab: quantile transport for severity distributions}
\label{sec:continuous-mini-lab}

The following R code illustrates one-dimensional quantile transport. It also prepares the later labs on empirical Wasserstein noise and longevity transport. It maps a baseline lognormal severity distribution to a heavier-tailed Pareto-type distribution through ranks.

\begin{lstlisting}[language=R]
set.seed(1)

n <- 5000
x <- rlnorm(n, meanlog = 8, sdlog = 0.8)   # baseline losses

# Target distribution: shifted Pareto-type quantile function
q_target <- function(u, xm = 1000, alpha = 2.2) {
  xm * (1 - u)^(-1 / alpha)
}

# Empirical rank map
u <- rank(x, ties.method = "average") / (n + 1)
y <- q_target(u)

# Empirical W2 between baseline and transported target in 1D
w2_sq <- mean((sort(x) - sort(y))^2)
sqrt(w2_sq)

par(mfrow = c(1, 2))
plot(sort(x), sort(y), pch = 19, cex = .25,
     xlab = "baseline quantiles", ylab = "target quantiles",
     main = "Quantile transport map")
abline(0, 1, lty = 2)

plot(quantile(x, seq(.01, .99, by = .01)),
     quantile(y, seq(.01, .99, by = .01)), type = "l",
     xlab = "Q_baseline(u)", ylab = "Q_target(u)",
     main = "Severity stress by ranks")
abline(0, 1, lty = 2)
\end{lstlisting}

The map is monotone: larger baseline losses are sent to larger target losses. The rank of each claim is preserved, but the tail is reshaped.

\begin{computebox}
In one dimension, empirical \(W_p\) is best computed by sorting rather than by solving a generic transport linear program. In higher dimensions, exact OT becomes substantially harder, which motivates discrete algorithms, entropic regularization and sliced Wasserstein approximations.
\end{computebox}

\subsection{What changes from discrete to continuous?}
\label{sec:continuous-discrete-comparison}

The following table summarizes the parallel between the discrete and continuous theories.

\begin{table}[t]
\centering
\caption{Discrete and continuous optimal transport in parallel.}
\label{tab:discrete-continuous-ot-parallel}
\begin{tabular}{lll}
\toprule
Object & Discrete case & Continuous case \\
\midrule
Source law & \(a_i\) on \(x_i\) & \(\mu\in\Prob(\mathcal X)\) \\
Target law & \(b_j\) on \(y_j\) & \(\nu\in\Prob(\mathcal Y)\) \\
Coupling & matrix \(\Gamma=(\gamma_{ij})\) & measure \(\gamma\in\Prob(\mathcal X\times\mathcal Y)\) \\
Marginals & row/column sums & projections of \(\gamma\) \\
Cost & \(\sum C_{ij}\gamma_{ij}\) & \(\int c\dd\gamma\) \\
Dual potentials & vectors \(u_i,v_j\) & functions \(\varphi,\psi\) \\
Dual constraint & \(u_i+v_j\le C_{ij}\) & \(\varphi(x)+\psi(y)\le c(x,y)\) \\
Optimal support & tight edges & contact set \\
\bottomrule
\end{tabular}
\end{table}

The conceptual structure is the same. The main technical differences are compactness, measurability, lower semicontinuity and the possibility of optimal maps.

\subsection{Summary}
\label{sec:continuous-summary}

This lecture introduced continuous optimal transport. The Monge problem searches for a deterministic map pushing \(\mu\) to \(\nu\), but may be infeasible or nonconvex. The Kantorovich formulation relaxes the problem to all couplings and becomes a linear optimization problem over probability measures. Duality introduces Kantorovich potentials, whose contact set determines where optimal mass can move. In one dimension, optimal transport is monotone and given by quantile matching. For quadratic cost in \(\R^d\), Brenier's theorem says that, under absolute continuity, the optimal plan is induced by the gradient of a convex function.

For actuarial science, continuous OT is a theory of distributional transformation. It gives a coherent way to compare and align loss laws, reserving scenarios, dependence models and climate stress distributions.

\subsection{Exercises}
\label{sec:continuous-exercises}

\begin{exercise}[Monge plans are Kantorovich plans]\label{ex:4:1}
Let \(T:\mathcal X\to\mathcal Y\) be measurable and suppose \(T_{\push}\mu=\nu\). Define \(\gamma_T=(\Id,T)_{\push}\mu\). Show that \(\gamma_T\in\Coup(\mu,\nu)\) and that
\[
  \int c(x,y)\dd\gamma_T(x,y)
  =\int c(x,T(x))\dd\mu(x).
\]
Deduce that \(\OT_c(\mu,\nu)\le M_c(\mu,\nu)\).
\end{exercise}

\begin{exercise}[No deterministic splitting]\label{ex:4:2}
Let \(\mu=\delta_0\) and \(\nu=p\delta_0+(1-p)\delta_1\), where \(0<p<1\). Show that no deterministic map \(T\) can push \(\mu\) to \(\nu\). Then construct an admissible Kantorovich coupling.
\end{exercise}

\begin{exercise}[Weak duality]\label{ex:4:3}
Let \(\gamma\in\Coup(\mu,\nu)\), and let \(\varphi,\psi\) satisfy \(\varphi(x)+\psi(y)\le c(x,y)\). Prove
\[
  \int \varphi\dd\mu+\int \psi\dd\nu
  \le
  \int c\dd\gamma.
\]
How would the inequality change if one maximized a payoff \(h(x,y)\) over all couplings rather than minimized a cost?
\end{exercise}

\begin{exercise}[The \(c\)-transform]\label{ex:4:4}
Let \(\varphi^c(y)=\inf_x\{c(x,y)-\varphi(x)\}\). Show that \((\varphi,\varphi^c)\) is dual feasible. If \((\varphi,\psi)\) is dual feasible, show that \(\psi(y)\le\varphi^c(y)\) for all \(y\). Interpret this result as a continuous version of improving the vector \(v\) in the finite dual.
\end{exercise}

\begin{exercise}[Quantile transport]\label{ex:4:5}
Let \(\mu\) and \(\nu\) be continuous distributions on \(\R\) with strictly increasing distribution functions. Show that
\[
  T(x)=Q_\nu(F_\mu(x))
\]
pushes \(\mu\) to \(\nu\). Then compute this map explicitly when \(\mu\) is exponential with rate \(\lambda\) and \(\nu\) is exponential with rate \(\theta\).
\end{exercise}

\begin{exercise}[One-dimensional \(W_1\)]\label{ex:4:6}
Let \(\mu\) and \(\nu\) be probability measures on \(\R\) with finite first moments. Use the quantile formula to show that
\[
  \Wone(\mu,\nu)=\int_0^1 |F_\mu^{-1}(u)-F_\nu^{-1}(u)|\dd u.
\]
Compute \(\Wone\) between two uniform distributions \(\mathcal U[a,b]\) and \(\mathcal U[c,d]\).
\end{exercise}

\begin{exercise}[Gaussian transport]\label{ex:4:7}
Let \(\mu=\mathcal N(m_0,\sigma_0^2)\) and \(\nu=\mathcal N(m_1,\sigma_1^2)\) on \(\R\). Derive the optimal map using the quantile formula. Show that it is affine and compute \(\Wtwo^2(\mu,\nu)\). Compare your answer with the multivariate Gaussian formula in dimension one.
\end{exercise}

\begin{exercise}[Actuarial severity stress]\label{ex:4:8}
Let \(X\) have distribution \(\mu\) and suppose a stress scenario is specified by a target distribution \(\nu\). Define \(T(x)=Q_\nu(F_\mu(x))\). Explain why this is rank-preserving. Discuss one advantage and one limitation of using this map to stress individual claim severities.
\end{exercise}

\newpage
\lecture{Wasserstein Distances, Quantiles and Barycenters}
\label{lec:wasserstein-barycenters}

% Local compatibility macros for this lecture. They do not override the manuscript preamble.
\providecommand{\Prob}{\mathcal P}
\providecommand{\Probp}[1]{\mathcal P_{#1}}
\providecommand{\Coup}{\Pi}
\providecommand{\Wp}[1]{W_{#1}}
\providecommand{\Wone}{W_1}
\providecommand{\Wtwo}{W_2}
\providecommand{\Law}{\operatorname{Law}}
\providecommand{\Pbb}{\mathbb P}
\providecommand{\Ebb}{\mathbb E}
\providecommand{\Rbb}{\mathbb R}
\providecommand{\E}{\Ebb}
\providecommand{\R}{\Rbb}
\providecommand{\dd}{\,\mathrm d}
\providecommand{\push}{\#}
\providecommand{\bX}{\mathbf X}
\providecommand{\bY}{\mathbf Y}
\providecommand{\bU}{\mathbf U}
\providecommand{\bx}{\mathbf x}
\providecommand{\by}{\mathbf y}
\providecommand{\bu}{\mathbf u}
\providecommand{\bm}{\mathbf m}
\providecommand{\VaR}{\operatorname{VaR}}
\providecommand{\TVaR}{\operatorname{TVaR}}
\providecommand{\argmin}{\operatorname*{arg\,min}}

\subsection*{Learning objectives}

The previous lectures introduced optimal transport first as a finite-dimensional linear program, then as a continuous optimization problem over couplings. This lecture turns the optimal value itself into a distance between probability laws. The resulting Wasserstein distances are among the most useful objects in actuarial applications, because they compare entire distributions while respecting the geometry of the underlying risk variable.

After this lecture, the reader should be able to:
\begin{enumerate}[label=(\roman*)]
\item define the Wasserstein distance \(\Wp{p}\) on a metric space;
\item understand the finite-moment condition and why it is needed;
\item prove the main metric properties at a working level;
\item compute \(\Wp{p}\) in dimension one using quantile functions;
\item relate Wasserstein distances to VaR, TVaR and distributional changes in actuarial risk;
\item compute simple examples, including Dirac masses, uniform laws and Gaussian laws;
\item define displacement interpolation and Wasserstein barycenters;
\item use one-dimensional barycenters through averages of quantile functions;
\item understand actuarial uses of barycenters, such as expert aggregation, model blending and climate-scenario summaries.
\end{enumerate}

\begin{readingbox}
This chapter builds on the definitions of couplings and push-forwards in Lecture~\ref{lec:couplings}, the discrete formulation in Lecture~\ref{lec:discrete-ot}, and the continuous Kantorovich problem in Lecture~\ref{lec:continuous-ot}. Wasserstein distances and their geometric properties are standard in \cite{villani2003topics,villani2008oldnew,ambrosio2021lectures,santambrogio2015optimal,figalli2021invitation}. One-dimensional quantile representations connect the theory to actuarial quantiles, VaR and TVaR; see also \cite{denuit2005actuarial,mcneil2015quantitative}. Barycenters are developed in \cite{agueh2011barycenters,peyre2019computational,panaretos2020invitation}. Vector ranks and vector quantiles are related to Brenier maps and to the econometric formulation of multivariate quantiles in \cite{brenier1991polar,chernozhukov2017monge,galichon2016economics}.
\end{readingbox}

\begin{actuarialbox}
A Wasserstein distance does not only ask whether two loss distributions are different. It asks how much work is needed to turn one into the other. This gives a useful geometry for actuarial objects whose changes have economic meaning: inflation, prevention, climate stress, portfolio drift and reserve-scenario changes.
\end{actuarialbox}

\subsection{From optimal values to distances}
\label{sec:wasserstein-definition}

Let \((\mathcal X,d)\) be a metric space. In most actuarial examples, \(\mathcal X\) is \(\Rbb\), \(\Rbb^d\), a finite set of risk cells, or a space of cash-flow vectors. Given two probability measures \(\mu\) and \(\nu\) on \(\mathcal X\), the optimal transport problem with cost
\[
  c(x,y)=d(x,y)^p,
  \qquad p\ge 1,
\]
is
\[
  \inf_{\gamma\in\Coup(\mu,\nu)}
  \int_{\mathcal X\times\mathcal X} d(x,y)^p\dd\gamma(x,y).
\]
The Wasserstein distance is the \(p\)-th root of this optimal value. It is the metric-cost specialization of the Kantorovich problem in Definition~\ref{def:kantorovich-problem}.

\begin{definition}[Finite \(p\)-th moment]\label{def:finite-p-moment}
Let \((\mathcal X,d)\) be a metric space and let \(x_0\in\mathcal X\). A probability measure \(\mu\) has finite \(p\)-th moment if
\[
  \int_{\mathcal X} d(x,x_0)^p\dd\mu(x)<\infty.
\]
The set of such probability measures is denoted by \(\Probp{p}(\mathcal X)\).
\end{definition}

The definition does not depend on the choice of \(x_0\). Indeed, by the triangle inequality,
\[
  d(x,x_1)^p\le 2^{p-1}\{d(x,x_0)^p+d(x_0,x_1)^p\},
\]
so if the moment is finite around one point, it is finite around any other point.

\begin{definition}[Wasserstein distance]\label{def:wasserstein-distance}
For \(p\ge1\) and \(\mu,\nu\in\Probp{p}(\mathcal X)\), the \(p\)-Wasserstein distance is
\[
  \Wp{p}(\mu,\nu)
  =\left(
  \inf_{\gamma\in\Coup(\mu,\nu)}
  \int d(x,y)^p\dd\gamma(x,y)
  \right)^{1/p}.
\]
When \(p=1\), we write \(\Wone\). When \(p=2\), we write \(\Wtwo\).
\end{definition}

The interpretation is straightforward. The notation \(\Coup(\mu,\nu)\) is the coupling set introduced in Definition~\ref{def:coupling}. If \(\mu\) is the distribution before a change and \(\nu\) the distribution after a change, then \(\Wp{p}(\mu,\nu)^p\) is the minimum average \(p\)-cost required to transform \(\mu\) into \(\nu\). The coupling describes which parts of the source distribution are matched to which parts of the target distribution.

\begin{example}[Dirac masses]\label{ex:wasserstein-dirac-masses}
Let \(\mu=\delta_x\) and \(\nu=\delta_y\). There is only one coupling, namely \(\gamma=\delta_{(x,y)}\). Hence
\[
  \Wp{p}(\delta_x,\delta_y)=d(x,y).
\]
Thus Wasserstein distance extends the ground distance from points to probability laws.
\end{example}

\begin{example}[A deterministic shift]\label{ex:wasserstein-deterministic-shift}
Let \(\mathcal X=\Rbb\), \(d(x,y)=|x-y|\), and let \(\nu\) be the law of \(X+a\) when \(X\sim\mu\). The deterministic coupling \((X,X+a)\) gives
\[
  \Wp{p}(\mu,\nu)^p\le |a|^p.
\]
In fact equality holds on the real line by Theorem~\ref{thm:wasserstein-quantile-formula}, because quantiles are shifted by the same amount. Thus a pure inflation shift by \(a\) has Wasserstein size \(|a|\), independently of the shape of the distribution.
\end{example}

\begin{example}[Scaling a severity distribution]\label{ex:wasserstein-scaling-severity}
Let \(X\sim\mu\) on \(\Rbb_+\) and let \(Y=(1+i)X\), where \(i>0\) represents claims inflation. The coupling \((X,(1+i)X)\) yields
\[
  \Wp{p}(\mu,\Law(Y))^p
  \le i^p\Ebb[X^p].
\]
In dimension one, equality holds by the monotone rearrangement theorem, provided \(\Ebb[X^p]<\infty\). This gives a direct distributional measure of inflation stress.
\end{example}

\subsection{Why Wasserstein rather than only moments?}
\label{sec:wasserstein-vs-moments}

Actuarial practice often compares distributions through a few summaries: mean, standard deviation, VaR, TVaR, skewness, probability of zero claim, or selected quantiles. These summaries are important, but they can miss the geometry of the full distribution.

Two distributions can have the same mean and variance but different tail shapes. Two stochastic reserving models can have the same best estimate but very different predictive tails. Two prevention strategies can reduce expected loss by the same amount but have different effects on extreme losses. A Wasserstein distance keeps track of the full movement of mass.

\begin{example}[Same mean, different shape]\label{ex:same-mean-different-shape}
Let
\[
  \mu=\frac12\delta_0+\frac12\delta_2,
  \qquad
  \nu=\delta_1.
\]
Both distributions have mean \(1\). But \(\nu\) is concentrated at the mean, while \(\mu\) is dispersed. For \(p=2\),
\[
  \Wtwo^2(\mu,\nu)=\frac12(0-1)^2+\frac12(2-1)^2=1.
\]
This quantity is the amount of squared displacement needed to collapse the distribution to its mean.
\end{example}

\iffalse
% To be restored in v2, after the portfolio demutualization paper is submitted.
\begin{actuarialbox}
In pricing, the movement from a pooled premium to an individualized premium can be represented as a movement from a concentrated distribution to a dispersed one. Wasserstein distance is therefore a natural measure of segmentation or demutualization.
\end{actuarialbox}
\fi

\subsection{Basic metric properties}
\label{sec:wasserstein-metric-properties}

The fact that \(\Wp{p}\) is a distance is not only a formal mathematical property. It is what allows us to compare portfolio shifts, climate scenarios, model outputs and reserving distributions in a coherent geometry.

\begin{theorem}[Metric property]\label{thm:wasserstein-metric}
Let \((\mathcal X,d)\) be a Polish metric space and \(p\ge1\). Then \(\Wp{p}\) is a metric on \(\Probp{p}(\mathcal X)\). That is, for \(\mu,\nu,\eta\in\Probp{p}(\mathcal X)\),
\begin{enumerate}[label=(\roman*)]
\item \(\Wp{p}(\mu,\nu)\ge0\);
\item \(\Wp{p}(\mu,\nu)=0\) if and only if \(\mu=\nu\);
\item \(\Wp{p}(\mu,\nu)=\Wp{p}(\nu,\mu)\);
\item \(\Wp{p}(\mu,\eta)\le \Wp{p}(\mu,\nu)+\Wp{p}(\nu,\eta)\).
\end{enumerate}
\end{theorem}

\begin{proof}[Proof sketch]
Non-negativity follows from non-negativity of \(d^p\). Symmetry follows because if \(\gamma\) couples \(\mu\) and \(\nu\), the swapped coupling couples \(\nu\) and \(\mu\), and \(d(x,y)=d(y,x)\).

If \(\mu=\nu\), the diagonal coupling \((X,X)\) has zero cost. Conversely, if \(\Wp{p}(\mu,\nu)=0\), there are couplings with arbitrarily small expected \(d(X,Y)^p\). In the limit, the mass is concentrated on \(x=y\), implying equal marginals.

For the triangle inequality, take nearly optimal couplings \(\gamma_{12}\in\Coup(\mu,\nu)\) and \(\gamma_{23}\in\Coup(\nu,\eta)\). By the gluing lemma, there exists a joint law of \((X,Y,Z)\) such that \((X,Y)\sim\gamma_{12}\) and \((Y,Z)\sim\gamma_{23}\). Then
\[
  d(X,Z)\le d(X,Y)+d(Y,Z).
\]
Taking the \(L^p\)-norm and using Minkowski's inequality gives
\[
  \left(\Ebb d(X,Z)^p\right)^{1/p}
  \le
  \left(\Ebb d(X,Y)^p\right)^{1/p}
  +
  \left(\Ebb d(Y,Z)^p\right)^{1/p}.
\]
Minimizing over the two couplings yields the result.
\end{proof}

\begin{remark}[Gluing]\label{rem:gluing-wasserstein-triangle}
The gluing step is a formal version of the following actuarial idea: if portfolio A can be matched to portfolio B, and portfolio B can be matched to portfolio C, then one can construct a joint scenario linking A, B and C. This is what allows Wasserstein distances to compare distributional paths consistently.
\end{remark}

\subsection{The order in \(p\)}
\label{sec:wasserstein-order-p}

For a fixed pair of distributions, \(\Wp{p}\) tends to increase with \(p\). The following simple inequality is often useful when deciding how tail-sensitive the metric should be.

\begin{proposition}[Monotonicity in \(p\)]\label{prop:wasserstein-monotonicity-p}
If \(1\le p\le q\) and \(\mu,\nu\in\Probp{q}(\mathcal X)\), then
\[
  \Wp{p}(\mu,\nu)
  \le
  \Wp{q}(\mu,\nu).
\]
\end{proposition}

\begin{proof}
For any coupling \((X,Y)\), Jensen's inequality gives
\[
  \left(\Ebb d(X,Y)^p\right)^{1/p}
  \le
  \left(\Ebb d(X,Y)^q\right)^{1/q}.
\]
Taking the infimum over couplings proves the inequality.
\end{proof}

The choice of \(p\) is therefore not neutral.
\begin{itemize}
\item \(\Wone\) gives a linear cost of displacement. It is robust and has a useful dual representation through Lipschitz functions.
\item \(\Wtwo\) penalizes long movements more strongly and gives a rich geometric structure, including barycenters and geodesics.
\item Higher \(p\) values place increasing emphasis on tail mismatches and large distributional moves.
\end{itemize}

\begin{actuarialbox}
For severity distributions, \(\Wone\) often has a natural monetary interpretation: average amount of claim displacement. \(\Wtwo\) is more geometric and often better suited for barycenters and displacement interpolation. For tail-sensitive applications, the choice of \(p\) should be aligned with the risk question.
\end{actuarialbox}

\subsection{The one-dimensional quantile formula}
\label{sec:wasserstein-quantile-formula}

The most useful explicit formula in actuarial applications is the one-dimensional quantile formula. Let \(F_\mu\) denote the distribution function of \(\mu\). Throughout these notes we use the lower generalized inverse
\[
  Q_\mu(u)=F_\mu^{-1}(u)=\inf\{x\in\Rbb:F_\mu(x)\ge u\},
  \qquad u\in(0,1).
\]
In actuarial language, \(Q_\mu(u)\) is \(\VaR_u\) for the loss distribution \(\mu\). The notation \(Q_\mu\) will be used below whenever quantile functions are manipulated as functions in \(L^p(0,1)\). With atoms or ties, this convention is the mathematical object; empirical implementations must state how ranks are discretized.

\begin{theorem}[Quantile formula]\label{thm:wasserstein-quantile-formula}
Let \(\mu,\nu\in\Probp{p}(\Rbb)\). Then
\[
  \Wp{p}(\mu,\nu)^p
  =
  \int_0^1
  \left|Q_\mu(u)-Q_\nu(u)\right|^p
  \dd u.
\]
An optimal coupling is obtained by taking \(U\sim\mathcal U(0,1)\) and setting
\[
  X=Q_\mu(U),
  \qquad
  Y=Q_\nu(U).
\]
\end{theorem}

\begin{proof}[Proof idea]
The coupling above is comonotonic: the same rank \(U\) is used in both distributions. It has the required marginals by the probability integral transform. For convex costs of the form \(|x-y|^p\), matching equal ranks is optimal. Intuitively, crossings are inefficient: if \(x_1<x_2\) but \(y_1>y_2\), swapping the destinations reduces the total cost because the cost is Monge in one dimension. Passing from finite samples to general distributions yields the integral formula.
\end{proof}

\begin{remark}[Quantiles as coordinates]\label{rem:quantiles-as-coordinates}
In dimension one, the map
\[
  \mu\mapsto Q_\mu
\]
embeds \(\Probp{p}(\Rbb)\) into \(L^p(0,1)\). In these coordinates,
\[
  \Wp{p}(\mu,\nu)=\|Q_\mu-Q_\nu\|_{L^p(0,1)}.
\]
For \(p=2\), this means that one-dimensional \(\Wtwo\)-geometry is Hilbertian after the quantile transform. This is one reason why univariate actuarial examples are so transparent.
\end{remark}

\subsection{Connection with VaR and TVaR}
\label{sec:wasserstein-var-tvar}

For a loss variable \(L\sim\mu\), the Value-at-Risk at level \(\alpha\) is
\[
 \highlight{GoldenRod}{ \VaR_\alpha(L)=F_\mu^{-1}(\alpha)=Q_\mu(\alpha)}
\]
with the usual convention for generalized quantiles. Thus the quantile function used in Wasserstein distance is the complete VaR curve.

For an integrable loss, the Tail Value-at-Risk can be written as an average of tail quantiles. We use the notation \(\TVaR_\alpha(\mu)\) when the argument is the law rather than a random variable:
\[
  \highlight{GoldenRod}{\TVaR_\alpha(L)=\TVaR_\alpha(\mu)
  =\frac{1}{1-\alpha}\int_\alpha^1 Q_\mu(u)\dd u,
  \qquad 0<\alpha<1}
\]
under the integrated-quantile convention. With atoms, this convention should be stated explicitly. Hence \(\Wone\) and \(\Wtwo\) compare distributions through differences of full quantile curves, while VaR and TVaR extract local or tail averages from those curves.

\begin{proposition}[Average quantile stability]\label{prop:average-quantile-stability}
Let \(\mu,\nu\in\Probp{p}(\Rbb)\). For any interval \([a,b]\subset(0,1)\),
\[
  \int_a^b
  \left|Q_\mu(u)-Q_\nu(u)\right|
  \dd u
  \le
  (b-a)^{1-1/p}\Wp{p}(\mu,\nu).
\]
In particular, average quantile differences over a range of confidence levels are controlled by Wasserstein distance.
\end{proposition}

\begin{proof}
By H{\"o}lder's inequality,
\[
  \int_a^b |Q_\mu(u)-Q_\nu(u)|\dd u
  \le
  (b-a)^{1-1/p}
  \left(\int_a^b |Q_\mu(u)-Q_\nu(u)|^p\dd u\right)^{1/p},
\]
where \(Q_\mu=F_\mu^{-1}\) and \(Q_\nu=F_\nu^{-1}\). The last term is bounded by the full \(L^p(0,1)\) norm, which is \(\Wp{p}(\mu,\nu)\).
\end{proof}

\begin{proposition}[TVaR stability]\label{cor:tvar-stability-wasserstein}
For \(p\ge1\),
\[
  |\TVaR_\alpha(\mu)-\TVaR_\alpha(\nu)|
  \le
  (1-\alpha)^{-1/p}\Wp{p}(\mu,\nu),
\]
provided both TVaR quantities are defined by integrated quantiles.
\end{proposition}

\begin{proof}
Use the previous proposition with \(a=\alpha\), \(b=1\), and divide by \(1-\alpha\).
\end{proof}

A small \(\Wp{p}\) distance does not automatically control \(|\VaR_\alpha(\mu)-\VaR_\alpha(\nu)|\) at a single level \(\alpha\). A small amount of mass can shift around a point and change a discontinuous quantile. Pointwise VaR stability requires additional assumptions, such as positive density near the quantile.

\begin{actuarialbox}
A small Wasserstein distance implies that average tail quantiles cannot move too much. This is useful for stress testing: it gives a distributional notion of closeness that has consequences for capital measures such as \(\TVaR\).
\end{actuarialbox}

\subsection{Explicit examples in dimension one}
\label{sec:wasserstein-explicit-examples}

\subsubsection{Uniform distributions}
\label{sec:uniform-wasserstein-examples}

Let \(\mu=\mathcal U[a,b]\) and \(\nu=\mathcal U[c,d]\). Their quantile functions are
\[
  Q_\mu(u)=a+(b-a)u,
  \qquad
  Q_\nu(u)=c+(d-c)u.
\]
Thus
\[
  \Wtwo^2(\mu,\nu)
  =\int_0^1
  \left\{(a-c)+[(b-a)-(d-c)]u\right\}^2\dd u.
\]
Writing \(A=a-c\) and \(B=(b-a)-(d-c)\),
\[
  \Wtwo^2(\mu,\nu)=A^2+AB+\frac{B^2}{3}.
\]
If the two intervals have the same length, then \(B=0\), and \(\Wtwo=|a-c|\): only the location shift matters. If they have the same midpoint but different lengths, then the distance measures the scale mismatch.

\subsubsection{Exponential distributions}
\label{sec:exponential-wasserstein-examples}

Let \(\mu=\mathrm{Exp}(\lambda)\) and \(\nu=\mathrm{Exp}(\theta)\). The quantile functions are
\[
  Q_\mu(u)=-\frac{1}{\lambda}\log(1-u),
  \qquad
  Q_\nu(u)=-\frac{1}{\theta}\log(1-u).
\]
Therefore
\[
  \Wtwo^2(\mu,\nu)
  =\left(\frac1\lambda-\frac1\theta\right)^2
  \int_0^1[-\log(1-u)]^2\dd u.
\]
Since \( -\log(1-U)\sim\mathrm{Exp}(1)\), the integral is \(2\), and
\[
  \Wtwo^2(\mu,\nu)=2\left(\frac1\lambda-\frac1\theta\right)^2.
\]
Similarly,
\[
  \Wone(\mu,\nu)=\left|\frac1\lambda-\frac1\theta\right|.
\]

\subsubsection{Gaussian distributions on the real line}
\label{sec:gaussian-wasserstein-oned}

Let \(\mu=\mathcal N(m_0,\sigma_0^2)\) and \(\nu=\mathcal N(m_1,\sigma_1^2)\). If \(\Phi\) is the standard normal distribution function, then
\[
  Q_\mu(u)=m_0+\sigma_0\Phi^{-1}(u),
  \qquad
  Q_\nu(u)=m_1+\sigma_1\Phi^{-1}(u).
\]
Hence
\[
  \Wtwo^2(\mu,\nu)
  =\int_0^1
  \left\{(m_0-m_1)+(\sigma_0-\sigma_1)\Phi^{-1}(u)\right\}^2\dd u.
\]
Since \(\int_0^1\Phi^{-1}(u)\dd u=0\) and \(\int_0^1[\Phi^{-1}(u)]^2\dd u=1\),
\[
  \Wtwo^2(\mu,\nu)
  =(m_0-m_1)^2+(\sigma_0-\sigma_1)^2.
\]
The optimal map is affine:
\[
  T(x)=m_1+\frac{\sigma_1}{\sigma_0}(x-m_0).
\]

\begin{example}[Severity model comparison]\label{ex:severity-model-gaussian-w2}
Suppose two actuarial teams fit normal approximations to transformed losses. If one model differs mainly in location and the other mainly in dispersion, the formula
\[
  \Wtwo^2=(\Delta m)^2+(\Delta\sigma)^2
\]
separates these two contributions. This is often more interpretable than comparing likelihoods or means alone.
\end{example}

\subsection{The multivariate Gaussian formula}
\label{sec:wasserstein-gaussian-formula}

In \(\Rbb^d\) with quadratic cost, the Wasserstein distance between multivariate Gaussian distributions has a closed form. If
\[
  \mu=\mathcal N(\bm_0,\Sigma_0),
  \qquad
  \nu=\mathcal N(\bm_1,\Sigma_1),
\]
with positive definite covariance matrices, then
\[
  \Wtwo^2(\mu,\nu)
  =
  \|\bm_0-\bm_1\|^2
  +
  \operatorname{tr}\left(
    \Sigma_0+\Sigma_1
    -2(\Sigma_0^{1/2}\Sigma_1\Sigma_0^{1/2})^{1/2}
  \right).
\]
Equivalent symmetric forms are also common. When \(d=1\), this reduces to the formula above.

\begin{remark}[Why actuaries may care]
Multivariate Gaussian approximations appear in capital aggregation, reserving uncertainty, large-portfolio approximations and scenario generation. The Wasserstein formula separates mean displacement from covariance displacement. Thus two capital models can be compared not only by total SCR, but by how their entire scenario cloud has moved.
\end{remark}

\subsection{Wasserstein convergence and distributional stability}
\label{sec:wasserstein-convergence}

Wasserstein distance metrizes more than weak convergence. It also controls moments. The standard characterization is the following; see, for instance, \cite[Chapter~6]{villani2008oldnew} or \cite[Chapter~5]{santambrogio2015optimal}.

\begin{theorem}[Wasserstein convergence]\label{thm:wasserstein-convergence}
Let \((\mathcal X,d)\) be a Polish metric space and let \(p\ge1\). For \(\mu_n,\mu\in\Probp{p}(\mathcal X)\),
\[
  \Wp{p}(\mu_n,\mu)\to0
\]
if and only if \(\mu_n\Rightarrow\mu\) weakly and
\[
  \int d(x,x_0)^p\dd\mu_n(x)\to
  \int d(x,x_0)^p\dd\mu(x)
\]
for one, hence every, \(x_0\in\mathcal X\).
\end{theorem}

In actuarial terms, this is important because weak convergence alone may ignore tail mass escaping to infinity. A sequence of models may converge weakly but still have unstable capital if small probabilities of very large losses move outward. Wasserstein convergence with moment control rules out that behaviour at the chosen moment order.

\begin{example}[Weak convergence can miss tail escape]\label{ex:tail-escape-wasserstein}
Let
\[
  \mu_n=\left(1-\frac1n\right)\delta_0+\frac1n\delta_n.
\]
Then \(\mu_n\) converges weakly to \(\delta_0\), because the mass at \(n\) has probability \(1/n\). But
\[
  \Wone(\mu_n,\delta_0)=1,
\]
and
\[
  \Wtwo^2(\mu_n,\delta_0)=n.
\]
The weak limit misses a small probability of a large loss; Wasserstein distance detects it.
\end{example}

\begin{actuarialbox}
This example is a warning for capital modelling. Rare large claims may be invisible under weak convergence but highly visible under Wasserstein distances, especially for larger \(p\).
\end{actuarialbox}

\subsection{Wasserstein balls and robust actuarial thinking}
\label{sec:wasserstein-balls}

\begin{definition}[Wasserstein ball]\label{def:wasserstein-ball}
A Wasserstein ball around a baseline distribution \(\hat\mu\) is
\[
\highlight{GoldenRod}{  \mathcal B_p(\hat\mu,r)
  =\{\nu\in\Probp{p}(\mathcal X):\Wp{p}(\nu,\hat\mu)\le r\}}
\]
The radius \(r\) has the same units as the ground distance \(d\).
\end{definition}

This object is central in distributionally robust optimization. Instead of trusting a single empirical or fitted distribution, one considers all distributions within transport distance \(r\).

In actuarial science, this has several interpretations:
\begin{itemize}
\item model uncertainty around an empirical severity distribution;
\item robustness of capital to tail perturbations;
\item uncertainty around climate scenario distributions;
\item ambiguity in portfolio drift after a change in underwriting;
\end{itemize}

\begin{example}[Robust expected loss]\label{ex:robust-expected-loss-w1}
Let \(\ell:\mathcal X\to\Rbb\) be a loss functional. A robust expectation takes the form
\[
  \sup_{\nu:\Wp{p}(\nu,\hat\mu)\le r}
  \int \ell(x)\dd\nu(x).
\]
If \(\ell\) is Lipschitz and \(p=1\), the Kantorovich--Rubinstein duality gives a direct bound:
\[
  \left|\int \ell\dd\nu-\int \ell\dd\hat\mu\right|
  \le
  \mathrm{Lip}(\ell)\Wone(\nu,\hat\mu).
\]
Thus the worst-case expectation over a \(\Wone\)-ball is controlled by the Lipschitz sensitivity of the quantity of interest.
\end{example}

\subsection{Displacement interpolation}
\label{sec:displacement-interpolation}

A Wasserstein distance does not only give a number. When an optimal map exists, it also gives a path between distributions.

Suppose \(\mathcal X=\Rbb^d\), \(p=2\), and an optimal map \(T\) transports \(\mu_0\) to \(\mu_1\). For \(t\in[0,1]\), define
\[
  T_t(x)=(1-t)x+tT(x),
\]
and
\[
  \mu_t=(T_t)_{\push}\mu_0.
\]

\begin{definition}[Displacement interpolation]\label{def:displacement-interpolation}
The curve \((\mu_t)_{0\le t\le1}\) defined above is called the displacement interpolation, or constant-speed \(\Wtwo\)-geodesic, between \(\mu_0\) and \(\mu_1\), when the optimal map formulation applies.
\end{definition}

\begin{example}[Interpolating severity distributions]\label{ex:interpolating-severity-distributions}
Let \(\mu_0\) be a baseline severity distribution and \(\mu_1\) a stressed severity distribution. In dimension one, the interpolation has quantile function
\[
  Q_t(u)=(1-t)Q_0(u)+tQ_1(u).
\]
Thus each rank \(u\) moves linearly from its baseline quantile to its stressed quantile. This gives a full path of intermediate stress scenarios.
\end{example}

\begin{remark}[Displacement interpolation is not mixture interpolation]\label{rem:displacement-not-mixture}
Displacement interpolation is different from mixture interpolation. The mixture path is
\[
  (1-t)\mu_0+t\mu_1.
\]
It puts some mass from the baseline distribution and some mass from the stressed distribution. The displacement path moves the distribution itself. For location shifts, displacement interpolation shifts the whole law gradually, while mixture interpolation creates two overlapping populations.
\end{remark}

\begin{actuarialbox}
For climate stress testing, displacement interpolation is often more plausible than mixture interpolation. It should still be read as a scenario path, not as an identified physical or causal transition.
\end{actuarialbox}

\subsection{Wasserstein barycenters}
\label{sec:wasserstein-barycenters}

A barycenter is an average. In Euclidean space, the weighted average of points \(x_1, \ldots,x_K\) minimizes the weighted sum of squared distances:
\[
  \bar x=\argmin_x \sum_{k=1}^K\lambda_k\|x-x_k\|^2.
\]
The Wasserstein barycenter extends this idea from points to probability distributions.

\begin{definition}[Wasserstein barycenter]\label{def:wasserstein-barycenter}
Let \(\mu_1, \ldots,\mu_K\in\Probp{2}(\mathcal X)\) and let \(\lambda_k\ge0\) with \(\sum_k\lambda_k=1\). A \(\Wtwo\)-barycenter is any solution of
\[
  \highlight{GoldenRod}{\bar\mu\in
  \argmin_{\mu
  \in\Probp{2}(\mathcal X)}
  \sum_{k=1}^K\lambda_k\Wtwo^2(\mu,\mu_k)}
\]
\end{definition}

Barycenters are useful because they average distributions in a geometry-aware way. If the \(\mu_k\) are severity laws, a barycenter is not obtained by averaging means or parameters; it is the distribution that minimizes total squared transport discrepancy.

\begin{example}[Expert aggregation]\label{ex:wasserstein-barycenter-expert-aggregation}
Suppose three experts provide predictive loss distributions \(\mu_1,\mu_2,\mu_3\), and the credibility weights are \(\lambda_1,\lambda_2,\lambda_3\). The Wasserstein barycenter is a consensus distribution. It keeps distributional information, unlike averaging only expected losses or capital figures.
\end{example}

\iffalse
% To be restored in v2, after the fairness paper is submitted.
\begin{example}[Fairness target]\label{ex:wasserstein-barycenter-fairness-target}
Let \(\mu_s=\Law(Z\mid S=s)\) be the distribution of a score or predicted premium in group \(s\), with group weight \(\lambda_s=\Pbb(S=s)\). A barycenter
\[
  \bar\mu\in\argmin_\mu\sum_s\lambda_s\Wtwo^2(\mu,\mu_s)
\]
can be used as a common target distribution for Wasserstein fairness repair. Each group distribution is transported toward \(\bar\mu\).
\end{example}
\fi

\subsection{One-dimensional barycenters}
\label{sec:wasserstein-oned-barycenters}

In dimension one, barycenters are especially simple because of quantiles.

\begin{theorem}[Quantile formula for barycenters]\label{thm:oned-wasserstein-barycenter}
Let \(\mu_1,\ldots,\mu_K\in\Probp{2}(\Rbb)\) and let \(\lambda_k\ge0\), \(\sum_k\lambda_k=1\). A \(\Wtwo\)-barycenter \(\bar\mu\) has quantile function
\[
  Q_{\bar\mu}(u)
  =
  \sum_{k=1}^K\lambda_k Q_{\mu_k}(u),
  \qquad u\in(0,1).
\]
\end{theorem}

\begin{proof}
Using the quantile embedding, the barycenter problem becomes
\[
  \min_Q
  \sum_{k=1}^K\lambda_k
  \int_0^1 |Q(u)-Q_k(u)|^2\dd u,
\]
where \(Q_k=Q_{\mu_k}\), and \(Q\) must be a nondecreasing left-continuous function. Pointwise in \(u\), the minimizer of \(\sum_k\lambda_k|q-Q_k(u)|^2\) is \(q=\sum_k\lambda_k Q_k(u)\). Since a weighted average of nondecreasing functions is nondecreasing, this pointwise minimizer is a valid quantile function. Hence it defines the barycenter.
\end{proof}

\begin{example}[Barycenter of Gaussian laws on \(\Rbb\)]\label{ex:barycenter-gaussian-oned}
Let \(\mu_k=\mathcal N(m_k,\sigma_k^2)\). The quantile function is
\[
  Q_k(u)=m_k+\sigma_k\Phi^{-1}(u).
\]
Thus the barycenter has quantile function
\[
  \bar Q(u)=\bar m+\bar\sigma\Phi^{-1}(u),
\]
where
\[
  \bar m=\sum_k\lambda_k m_k,
  \qquad
  \bar\sigma=\sum_k\lambda_k\sigma_k.
\]
Therefore the barycenter is
\[
  \bar\mu=\mathcal N(\bar m,\bar\sigma^2).
\]
Notice that standard deviations, not variances, are averaged in one-dimensional \(\Wtwo\)-geometry.
\end{example}

\iffalse
% To be restored in v2, after the portfolio demutualization paper is submitted.
\begin{example}[Barycenter of tariff distributions]\label{ex:barycenter-tariff-distributions}
Suppose \(\mu_1\) is the distribution of premiums under a fully pooled tariff, \(\mu_2\) under a coarse tariff, and \(\mu_3\) under a highly individualized tariff. The quantile barycenter provides an intermediate tariff distribution. It can be interpreted as an average on the pooling--profiling scale.
\end{example}
\fi

\subsection{Barycenters versus mixtures}
\label{sec:barycenters-vs-mixtures}

A Wasserstein barycenter should not be confused with a mixture distribution. Given \(\mu_1, \ldots,\mu_K\), their mixture is
\[
  \mu_{\mathrm{mix}}=\sum_{k=1}^K\lambda_k\mu_k.
\]
This corresponds to first selecting a model \(k\) with probability \(\lambda_k\), then sampling from \(\mu_k\). It preserves model heterogeneity.

A Wasserstein barycenter instead seeks a central distribution. In dimension one, it averages quantiles. It is therefore closer to aligning ranks across models and averaging corresponding risk levels.

\begin{example}[Two shifted distributions]\label{ex:mixture-versus-barycenter-shift}
Let \(\mu_1=\mathcal N(-m,\sigma^2)\) and \(\mu_2=\mathcal N(m,\sigma^2)\), with equal weights. The mixture is bimodal when \(m\) is large. The Wasserstein barycenter is \(\mathcal N(0,\sigma^2)\). The mixture represents uncertainty about which regime applies; the barycenter represents a central regime.
\end{example}

\begin{actuarialbox}
For model risk, mixtures and barycenters answer different questions. A mixture says: one of these models may be true. A barycenter says: construct a consensus model that is central among them. Both are useful, but they should not be confused.
\end{actuarialbox}

\subsection{Barycentric projection of a transport plan}
\label{sec:barycentric-projection}

In applications, especially in higher dimension or in discrete problems, an optimal coupling may not be induced by a deterministic map. A common way to extract a map-like object from a plan is the barycentric projection.

\begin{definition}[Barycentric projection]\label{def:barycentric-projection}
Let \(\gamma\in\Coup(\mu,\nu)\) on \(\Rbb^d\times\Rbb^d\), and assume the conditional first moment in \(\by\) is finite. If \(\gamma(\dd \by\mid \bx)\) is a conditional distribution, the barycentric projection of \(\gamma\) is
\[
  T_\gamma(\bx)=\int \by\,\gamma(\dd \by\mid \bx).
\]
This is the conditional mean destination of mass starting at \(\bx\).
\end{definition}

\iffalse
% To be restored in v2, after the fairness paper is submitted.
\begin{example}[Fair score repair]\label{ex:barycentric-projection-fair-score-repair}
Suppose \(Z\) is a score and \(S\) a group. A discrete transport plan from the empirical distribution of scores in group \(s\) to a target barycenter may split mass from one observed score across several target scores. The barycentric projection assigns to each original score the average target score under that conditional transport. This gives a deterministic post-processing rule.
\end{example}
\fi

\begin{remark}[Projection is not exact transport]\label{rem:barycentric-projection-not-exact}
The barycentric projection usually does not push \(\mu\) exactly to \(\nu\). It summarizes a possibly randomized plan by a deterministic map. This may be useful for interpretability, but one should not confuse it with an exact Monge map.
\end{remark}

\subsection{Computing one-dimensional Wasserstein distances}
\label{sec:computing-wasserstein-oned}

For empirical one-dimensional data, computation is simple. The statistical behaviour of empirical Wasserstein distances is treated later, but one-dimensional sorting is the computational backbone already here. Let \(x_1, \ldots,x_n\) and \(y_1, \ldots,y_n\) have equal weights. Sort both samples:
\[
  x_{(1)}\le\cdots\le x_{(n)},
  \qquad
  y_{(1)}\le\cdots\le y_{(n)}.
\]
Then
\[
  \Wp{p}(\hat\mu_n,\hat\nu_n)^p
  =\frac1n\sum_{i=1}^n|x_{(i)}-y_{(i)}|^p.
\]
If sample sizes differ, one can compute quantiles on a common grid.

\begin{computebox}
For one-dimensional actuarial data, no linear programming solver is needed. Sorting or empirical quantiles are enough. This makes Wasserstein distances easy to add to severity, reserving and drift diagnostics; this code is reused conceptually in Lab~1.
\end{computebox}

\begin{lstlisting}[language=R]
# Empirical Wp between two one-dimensional samples
emp_quantile_left <- function(x, u) {
  xs <- sort(x)
  n <- length(xs)
  xs[pmax(1, pmin(n, ceiling(n * u)))]
}

wasserstein_1d <- function(x, y, p = 2, grid_size = 1000) {
  u <- seq(1 / (grid_size + 1), grid_size / (grid_size + 1), length.out = grid_size)
  qx <- emp_quantile_left(x, u)
  qy <- emp_quantile_left(y, u)
  mean(abs(qx - qy)^p)^(1 / p)
}

# Example: compare two loss samples
# W2 <- wasserstein_1d(losses_2025, losses_2026, p = 2)
# W1 <- wasserstein_1d(losses_2025, losses_2026, p = 1)
\end{lstlisting}

\subsection{Computing a one-dimensional barycenter}
\label{sec:computing-barycenter-oned}

The same quantile idea gives a very simple empirical barycenter.

\begin{lstlisting}[language=R]
# One-dimensional Wasserstein barycenter by quantile averaging
barycenter_1d <- function(samples, weights = NULL, grid_size = 1000) {
  K <- length(samples)
  if (is.null(weights)) weights <- rep(1 / K, K)
  weights <- weights / sum(weights)
  u <- seq(1 / (grid_size + 1), grid_size / (grid_size + 1), length.out = grid_size)
  Q <- sapply(samples, function(x) emp_quantile_left(x, u))
  qbar <- as.numeric(Q %*% weights)
  data.frame(u = u, qbar = qbar)
}

# Example:
# b <- barycenter_1d(list(loss_model_A, loss_model_B, loss_model_C),
#                    weights = c(.25, .50, .25))
\end{lstlisting}

The output is a quantile function. One can sample approximately from the barycenter by drawing \(U\sim\mathcal U(0,1)\) and interpolating \(\bar Q(U)\).

\subsection{Actuarial applications}
\label{sec:wasserstein-actuarial-applications}

\subsubsection{Comparing loss distributions}
\label{sec:app-comparing-loss-distributions}

Given two severity distributions \(\mu\) and \(\nu\), \(\Wp{p}(\mu,\nu)\) gives a full-distribution discrepancy. In dimension one, this discrepancy is simply an \(L^p\)-distance between VaR curves.

Possible uses include:
\begin{itemize}
\item comparing claim severity before and after inflation adjustment;
\item validating a fitted distribution against the empirical distribution;
\item comparing experience between two territories;
\item monitoring distribution drift in claim sizes over time.
\end{itemize}

\iffalse
% To be restored in v2, after the portfolio demutualization paper is submitted.
\subsubsection{Pricing and mutualization}
\label{sec:app-pricing-mutualization-wasserstein}

Let \(\Premium(\bX)\) be a predicted premium over a portfolio. The law \(\Law(\Premium(\bX))\) describes the distribution of premiums. A pooled premium corresponds to a Dirac law at the mean premium. Therefore
\[
  \Wtwo^2\left(\Law(\Premium(\bX)),\delta_{\Ebb[\Premium(\bX)]}\right)
\]
measures how dispersed the tariff is around its pooled value. More generally, comparing the distribution of an individualized tariff to the distribution of a coarse tariff gives a Wasserstein measure of segmentation.

\fi

\subsubsection{Portfolio drift}
\label{sec:app-portfolio-drift-wasserstein}

Let \(\mu_t=\Law(\bX_t)\) and \(\mu_{t+1}=\Law(\bX_{t+1})\) be covariate distributions at two dates. A Wasserstein distance between them measures how much the portfolio has shifted. If the ground metric weights risk-relevant variables, this provides a model-monitoring diagnostic.

\subsubsection{Climate and prevention}
\label{sec:app-climate-prevention-wasserstein}

Let \(\mu_0\) be the distribution of losses without a prevention measure and \(\mu_1\) the distribution after prevention. A reduction in expected loss is
\[
  \int x\dd\mu_0(x)-\int x\dd\mu_1(x).
\]
A Wasserstein distance adds another question: how much did the entire loss distribution move? Two prevention measures may have the same expected benefit but different distributional effects. One may reduce frequent moderate losses; another may reduce tail losses. Their Wasserstein paths and quantile differences reveal this distinction.

\iffalse
% To be restored in v2, after the fairness paper is submitted.
\subsubsection{Fairness}
\label{sec:app-fairness-wasserstein}

Let \(Z=\widehat m(\bX)\) be a score or predicted premium, and let \(S\) denote a sensitive group. The group distributions are
\[
  \mu_s=\Law(Z\mid S=s).
\]
A Wasserstein barycenter of the \(\mu_s\)'s is a natural common target. Transporting each \(\mu_s\) to the barycenter gives a distributional fairness repair. This approach is especially transparent in dimension one, where the repair is a quantile map. The empirical implementation is developed later in the Wasserstein fairness repair lab, while the conceptual actuarial discussion appears in the fairness chapter.

\fi

\begin{readingbox}
The applications in this subsection are expanded later: portfolio drift in the drift and domain-adaptation chapter, and climate and prevention in the reserving/climate chapter. The one-dimensional computational routines in Sections~\ref{sec:computing-wasserstein-oned}--\ref{sec:computing-barycenter-oned} are intended to be reused in the empirical Wasserstein labs.
\end{readingbox}

\subsection{From univariate ranks to vector ranks}
\label{sec:vector-ranks}

In one dimension, optimal transport is almost synonymous with quantile matching. If \(X\sim\mu\), \(Y\sim\nu\), and \(U\sim \mathcal U(0,1)\), then
\[
  X = Q_\mu(U), \qquad Y = Q_\nu(U)
\]
defines the increasing coupling. The scalar \(U=F_\mu(X)\) is the rank of \(X\), and the map
\[
  T(x)=Q_\nu(F_\mu(x))
\]
sends \(\mu\) to \(\nu\). Thus, in dimension one, transport can be described as follows: transform observations into ranks, then transform ranks into the target distribution.

\begin{figure}[t]
\centering
\includegraphics[width=.82\textwidth]{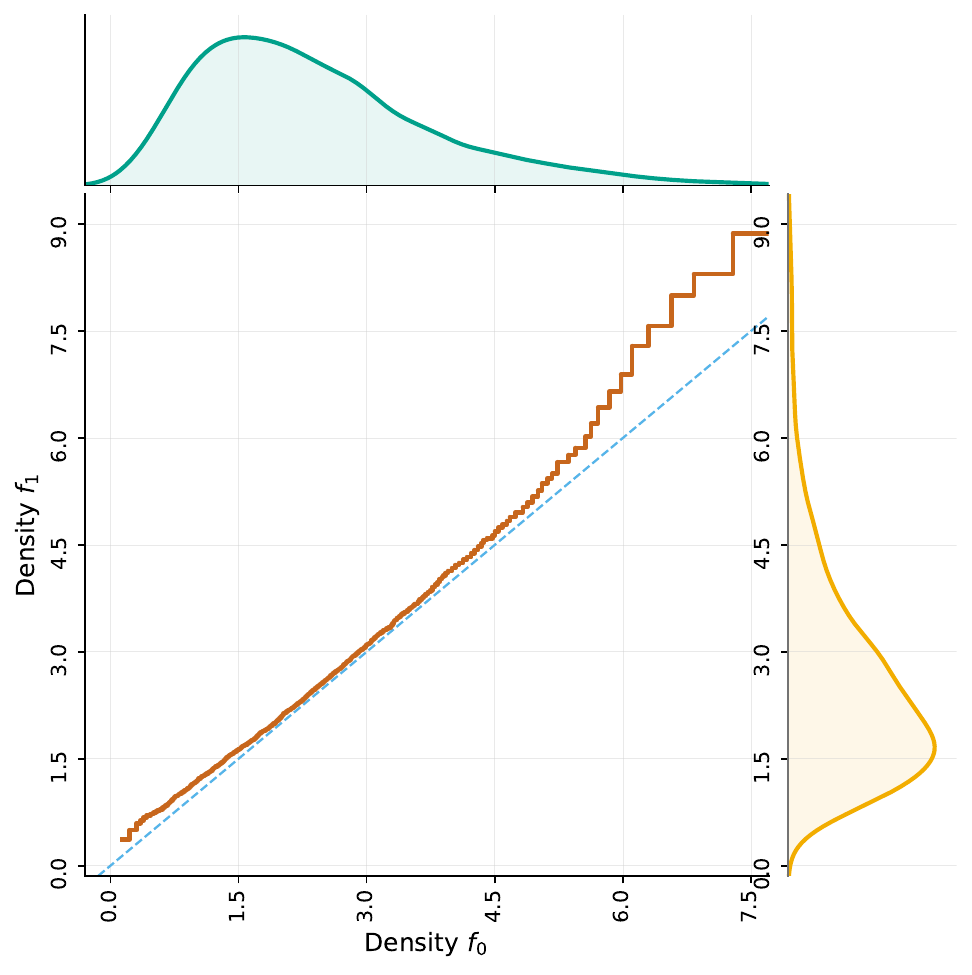}
\caption{Transport from a Gamma severity distribution to a Lognormal severity distribution. The monotone map is nonlinear: moderate losses and tail losses are displaced differently because the two distributions have different skewness and tail behaviour.}
\label{fig:transport-gamma-lognormal}
\end{figure}

\begin{figure}[t]
\centering
\includegraphics[width=.82\textwidth]{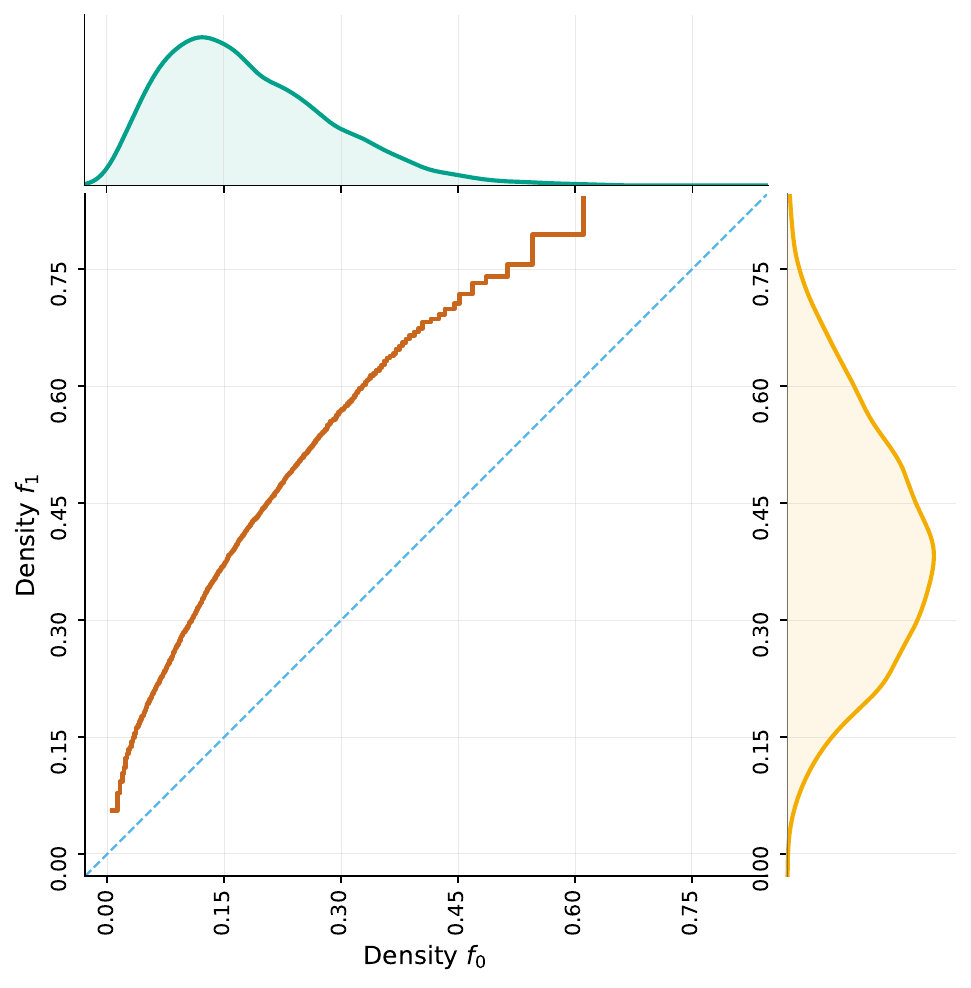}
\caption{Transport between two bounded distributions. The red curve shows the empirical quantile map from the source law to the target law; deviations from the diagonal reveal where ranks are compressed or expanded.}
\label{fig:transport-beta-beta}
\end{figure}

In higher dimension, there is no canonical total order on \(\Rbb^d\). Hence there is no unique distribution function \(F(x)\) whose inverse could play the role of a multivariate quantile. Nevertheless, quadratic optimal transport provides a useful substitute. The idea is to replace the scalar uniform rank by a reference probability measure on \(\Rbb^d\), and to define ranks and quantiles through optimal transport maps.

\begin{readingbox}
The construction below is closely related to vector quantiles and vector ranks in \cite{carlier2016vector,chernozhukov2017monge}. It should also be read together with Brenier's theorem in Lecture~\ref{lec:continuous-ot} and with the convex-analysis insert in the mathematical appendix.
\end{readingbox}

\begin{definition}[Reference measure and vector rank]\label{def:vector-rank}
Let \(\mu_0\in\Probp{2}(\Rbb^d)\) be a reference measure, for instance the uniform distribution on \([0,1]^d\) or the standard Gaussian distribution \(\mathcal N(0,I_d)\). Let \(\bX\sim\mu\in\Probp{2}(\Rbb^d)\). A vector rank map for \(\bX\), relative to \(\mu_0\), is a map
\[
  R_\mu:\Rbb^d\to\Rbb^d
\]
such that
\[
  (R_\mu)_\#\mu=\mu_0.
\]
When \(R_\mu\) is chosen as the quadratic optimal transport map from \(\mu\) to \(\mu_0\), it is called the optimal transport rank map. Brenier uniqueness for this map requires the source law \(\mu\) to be absolutely continuous; otherwise a rank map may be non-unique or plan-based.
\end{definition}

The analogy with the univariate case is direct. In dimension one, if \(\mu_0\) is the uniform distribution on \((0,1)\), then
\[
  R_\mu(x)=F_\mu(x)
\]
is exactly the usual rank transform. In dimension \(d\), \(R_\mu(\bx)\) is a vector-valued rank: it places the point \(\bx\) inside the reference geometry \(\mu_0\).

\begin{definition}[Vector quantile]\label{def:vector-quantile}
Let \(\mu_0\in\Probp{2}(\Rbb^d)\) be absolutely continuous and let \(\mu\in\Probp{2}(\Rbb^d)\). A vector quantile map for \(\mu\), relative to \(\mu_0\), is a map
\[
  Q_\mu:\Rbb^d\to\Rbb^d
\]
such that
\[
  (Q_\mu)_\#\mu_0=\mu.
\]
When \(Q_\mu\) is chosen as the quadratic optimal transport map from \(\mu_0\) to \(\mu\), it is called the optimal transport vector quantile map.
\end{definition}

Thus the vector quantile map plays the role of \(F_\mu^{-1}\), while the vector rank map plays the role of \(F_\mu\). Under the hypotheses of Brenier's theorem, the vector quantile has the form
\[
  Q_\mu(\bu)=\nabla\varphi(\bu),
\]
where \(\varphi\) is a convex function. Similarly, when \(\mu\) is absolutely continuous, the optimal rank map is the gradient of a convex function transporting \(\mu\) back to \(\mu_0\).

\begin{proposition}[Univariate ranks as a special case]\label{prop:univariate-ranks-special-case}
Let \(d=1\), let \(\mu_0\) be the uniform distribution on \((0,1)\), and let \(\mu\) be a probability distribution on \(\Rbb\) with continuous strictly increasing distribution function \(F_\mu\). Then the optimal transport rank and quantile maps are
\[
  R_\mu(x)=F_\mu(x),
  \qquad
  Q_\mu(u)=F_\mu^{-1}(u).
\]
\end{proposition}

\begin{proof}
This is the one-dimensional monotone rearrangement theorem; see \cite{villani2003topics,villani2008oldnew,figalli2021invitation}. The increasing map from \(\mu\) to the uniform distribution is \(F_\mu\), because \(F_\mu(X)\sim \mathcal U(0,1)\) when \(F_\mu\) is continuous. Conversely, the increasing map from the uniform distribution to \(\mu\) is \(Q_\mu=F_\mu^{-1}\). Since the increasing coupling is optimal for convex costs on the line, these are the optimal transport rank and quantile maps.
\end{proof}

\begin{proposition}[Vector quantiles from Brenier maps]\label{prop:vector-quantiles-brenier}
Let \(\mu_0,\mu\in\Probp{2}(\Rbb^d)\), and assume that \(\mu_0\) is absolutely continuous. For the quadratic cost \(c(u,x)=\|u-x\|^2/2\), there exists a convex function \(\varphi\) such that
\[
  Q_\mu(u)=\nabla\varphi(u)
\]
pushes \(\mu_0\) forward to \(\mu\). Moreover, \(Q_\mu\) is the unique quadratic optimal transport map from \(\mu_0\) to \(\mu\), up to \(\mu_0\)-null sets.
\end{proposition}

\begin{proof}
This is Brenier's theorem; see \cite{brenier1991polar,villani2003topics,villani2008oldnew}. Because the source measure \(\mu_0\) is absolutely continuous and the cost is quadratic, the optimal Kantorovich plan is induced by a deterministic map. This map is the gradient of a convex function. Since it pushes \(\mu_0\) to \(\mu\), it plays the role of a multivariate quantile map.
\end{proof}

\begin{actuarialbox}
A scalar quantile ranks a claim, cash-flow value or univariate risk amount along one risk dimension. A vector rank embeds a multivariate actuarial profile into a common reference distribution. This matters when risk is genuinely multivariate: losses by line of business, reserving cash-flow vectors, telematics summaries, or climate hazard vectors.
\end{actuarialbox}

\paragraph{Rank-preserving transport.}
Once vector ranks are available, one can define a multivariate analogue of quantile mapping. Suppose \(\bX\sim\mu\) is a baseline actuarial vector and \(\nu\) is a target distribution. Let \(R_\mu\) be the rank map from \(\mu\) to \(\mu_0\), and let \(Q_\nu\) be the vector quantile map from \(\mu_0\) to \(\nu\). Then
\[
  T_{\mu\to\nu}(\bx)=Q_\nu(R_\mu(\bx))
\]
pushes \(\mu\) to \(\nu\). This is the multivariate counterpart of
\[
  Q_\nu\circ F_\mu.
\]
It transports a point \(\bx\) by first finding its vector rank in the reference space, then reading off the corresponding target quantile.

\begin{example}[Multivariate severity stress]\label{ex:multivariate-severity-stress}
Let \(\bX=(X_1,\ldots,X_d)\) be a vector of losses by line of business under a baseline model, and let \(\nu\) be a stressed multivariate distribution. A componentwise quantile stress would transform each coordinate separately:
\[
  X_k \mapsto Q_{\nu_k}(F_{\mu_k}(X_k)).
\]
This preserves marginal ranks but may distort dependence. A vector-quantile stress instead uses
\[
  T_{\mu\to\nu}=Q_\nu\circ R_\mu,
\]
and therefore transports the joint distribution. It can modify marginal tails and dependence simultaneously.
\end{example}

\begin{example}[Portfolio profiles]\label{ex:portfolio-profile-vector-ranks}
Let \(\bX\in\Rbb^d\) contain policyholder-level features such as age, insured value, exposure, past claims, territory indicator and telematics summaries. The rank map \(R_\mu(\bX)\) embeds each policyholder into a common reference space. Two portfolios can then be compared not only through marginal distributions of each feature, but through the geometry of their joint profiles.
\end{example}

\paragraph{Vector ranks and multivariate comonotonicity.}
In dimension one, two random variables \(X\) and \(Y\) are comonotonic if they can be written as increasing functions of the same scalar rank:
\[
  X=F_X^{-1}(U),
  \qquad
  Y=F_Y^{-1}(U).
\]
Vector ranks suggest the following extension.

\begin{definition}[Reference comonotonicity]\label{def:reference-comonotonicity}
Fix a reference measure \(\mu_0\) and let \(\bU\sim\mu_0\). Two random vectors \(\bX\sim\mu\) and \(\bY\sim\nu\) are \(\mu_0\)-comonotonic if there exist convex functions \(\varphi\) and \(\psi\) such that
\[
  \bX=\nabla\varphi(\bU),
  \qquad
  \bY=\nabla\psi(\bU).
\]
Thus both vectors are generated from the same vector rank \(\bU\) through optimal transport maps.
\end{definition}

\begin{proposition}[Maximal correlation interpretation]\label{prop:maximal-correlation-vector-quantiles}
Let \(\bU\sim\mu_0\) and \(\bX\sim\mu\). Under the quadratic transport setting, the vector quantile map \(Q_\mu=\nabla\varphi\) solves
\[
  \sup_{\widetilde{\bX}\sim\mu} \Ebb[\widetilde{\bX}\cdot \bU],
\]
and the optimizer is \(\widetilde{\bX}=Q_\mu(\bU)\).
\end{proposition}

\begin{proof}
The problem is equivalent to minimizing
\[
  \Ebb\|\widetilde{\bX}-\bU\|^2
\]
over all \(\widetilde{\bX}\sim\mu\), because
\[
  \Ebb\|\widetilde{\bX}-\bU\|^2
  =
  \Ebb\|\widetilde{\bX}\|^2+\Ebb\|\bU\|^2-2\Ebb[\widetilde{\bX}\cdot \bU],
\]
and the first two terms are fixed by the marginal laws. Hence maximizing correlation with \(\bU\) is the same as solving the quadratic optimal transport problem between \(\mu_0\) and \(\mu\). Brenier's theorem gives the optimizer as \(Q_\mu(\bU)=\nabla\varphi(\bU)\).
\end{proof}

This interpretation is useful in risk theory. In one dimension, comonotonicity corresponds to maximal positive dependence and is closely related to worst-case aggregation for many law-invariant risk measures. In higher dimension, \(\mu_0\)-comonotonicity provides an analogous notion, but it depends on the chosen reference distribution \(\mu_0\). There is no reference-free total order on \(\Rbb^d\).

\begin{example}[Gaussian vector ranks]\label{ex:gaussian-vector-ranks}
Let \(\mu_0=\mathcal N(0,I_d)\) and let \(\mu=\mathcal N(\bm,\Sigma)\), where \(\Sigma\) is positive definite. The quadratic optimal vector quantile map is affine:
\[
  Q_\mu(\bu)=\bm+\Sigma^{1/2}\bu.
\]
More generally, if \(\mu_0=\mathcal N(0,\Sigma_0)\), then
\[
  Q_\mu(\bu)
  =
  \bm+
  \Sigma_0^{-1/2}
  \left(\Sigma_0^{1/2}\Sigma\Sigma_0^{1/2}\right)^{1/2}
  \Sigma_0^{-1/2}\bu.
\]
Thus, for Gaussian laws, vector ranks and vector quantiles reduce to linear transformations. This is useful for multivariate capital models, where changes in covariance encode changes in diversification.
\end{example}

\paragraph{Why this matters for actuarial science.}
Vector ranks are useful whenever componentwise quantiles are too poor to describe the risk.

\begin{itemize}
\item In capital modelling, the vector \(\bX\) may contain losses by line of business. Componentwise quantile matching ignores diversification; vector transport preserves the joint law.

\item In reserving, the vector \(\bX\) may contain future cash-flows by development year. Vector ranks compare complete cash-flow scenarios rather than marginal payments year by year.

\item In climate risk, the vector \(\bX\) may contain losses in several regions or hazards. Vector transport can represent changes in spatial dependence.

\iffalse
% To be restored in v2, after the fairness paper is submitted.
\item In fairness auditing, the vector \(\bX\) may contain several scores or premium components. A vector repair can align multivariate score distributions, while componentwise repairs may fail to preserve dependence.
\fi

\item In portfolio monitoring, vector ranks allow one to compare the joint distribution of policyholder profiles across time.
\end{itemize}

\begin{computebox}
In practice, exact vector rank maps are difficult to estimate in high dimension. Common computational approximations include discrete optimal transport, entropic regularization, semi-discrete transport, normal-score Gaussian approximations, and sliced Wasserstein methods. Thus vector ranks are conceptually clean, but their empirical use requires regularization and careful validation.
\end{computebox}

\paragraph{Limitations.}
The vector-rank construction depends on the reference measure \(\mu_0\). In dimension one, the uniform distribution on \((0,1)\) is canonical. In higher dimension, the choice is part of the modelling decision. A uniform reference on a cube emphasizes bounded ranks; a Gaussian reference emphasizes elliptical geometry; an empirical reference may better reflect a portfolio benchmark. Different choices of \(\mu_0\) may lead to different notions of multivariate rank and comonotonicity.

There is also no simple ordering interpretation. A large scalar rank means ``large loss'' in a clear sense. A vector rank does not say that one point is larger than another; it says where the point lies in a reference geometry. Thus vector ranks are best interpreted geometrically, not ordinally.

\begin{actuarialbox}
The univariate formula \(Q_\nu\circ F_\mu\) is rank-preserving because there is a natural order on the real line. In several dimensions, the analogue is \(Q_\nu\circ R_\mu\), where ranks and quantiles are defined by optimal transport to and from a reference distribution. This construction is useful, but it makes the choice of reference geometry an actuarial modelling decision.
\end{actuarialbox}

\subsection{Summary}
\label{sec:wasserstein-summary}

This lecture introduced Wasserstein distances as optimal transport costs with metric ground cost. Unlike scalar summaries, Wasserstein distances compare complete distributions. In one dimension, they reduce to \(L^p\)-distances between quantile functions, making them directly connected to VaR curves and TVaR. The lecture also introduced displacement interpolation and Wasserstein barycenters. In dimension one, barycenters are computed by averaging quantile functions. For actuarial science, these tools provide a geometry for losses, reserves, scenario distributions, climate stresses and multivariate actuarial profiles.

\subsection{Further reading}
\label{sec:wasserstein-further-reading}

\begin{readingbox}
The metric and geometric foundations of Wasserstein distances are treated in depth in \cite{villani2003topics,villani2008oldnew,santambrogio2015optimal}. A concise introduction emphasizing Wasserstein distances and gradient flows is \cite{figalli2021invitation}; the statistical viewpoint is developed in \cite{panaretos2020invitation}, with empirical convergence questions treated for example in \cite{delbarrio1999tests,fournier2015rate}. Computational aspects, including barycenters, are covered in \cite{peyre2019computational}. Wasserstein barycenters originate in \cite{agueh2011barycenters}; vector quantiles and vector ranks are developed in \cite{chernozhukov2017monge,carlier2016vector,galichon2016economics}.
\end{readingbox}

\subsection{Exercises}
\label{sec:wasserstein-exercises}

\begin{exercise}[Dirac and two-point laws]\label{ex:5:1}
Let \(\mu=\delta_0\) and \(\nu=p\delta_0+(1-p)\delta_a\), where \(a>0\). Compute \(\Wp{p}(\mu,\nu)\) for a general exponent \(p\ge1\). Interpret the result as a rare-loss perturbation.
\end{exercise}

\begin{exercise}[Uniform laws]\label{ex:5:2}
Let \(\mu=\mathcal U[0,1]\) and \(\nu=\mathcal U[0,a]\), with \(a>0\). Compute \(\Wone(\mu,\nu)\) and \(\Wtwo(\mu,\nu)\). What happens when \(a\to\infty\)?
\end{exercise}

\begin{exercise}[TVaR stability]\label{ex:5:3}
Let \(\mu,\nu\in\Probp{1}(\Rbb)\). Prove that
\[
  |\TVaR_\alpha(\mu)-\TVaR_\alpha(\nu)|
  \le
  \frac{1}{1-\alpha}\Wone(\mu,\nu).
\]
Then improve the bound using \(\Wtwo\) and H{\"o}lder's inequality.
\end{exercise}

\begin{exercise}[Gaussian laws]\label{ex:5:4}
Let \(\mu=\mathcal N(m_0,\sigma_0^2)\) and \(\nu=\mathcal N(m_1,\sigma_1^2)\). Use quantiles to derive
\[
  \Wtwo^2(\mu,\nu)=(m_0-m_1)^2+(\sigma_0-\sigma_1)^2.
\]
What is \(\Wone(\mu,\nu)\) when \(\sigma_0=\sigma_1\)? What if \(m_0=m_1\)?
\end{exercise}

\begin{exercise}[Barycenter of two distributions]\label{ex:5:5}
Let \(\mu_0,\mu_1\in\Probp{2}(\Rbb)\) and let \(\lambda\in[0,1]\). Show that the barycenter of \(\mu_0\) and \(\mu_1\) with weights \(1-\lambda\) and \(\lambda\) has quantile function
\[
  Q_\lambda(u)=(1-\lambda)Q_0(u)+\lambda Q_1(u).
\]
Explain why this is also the displacement interpolation between \(\mu_0\) and \(\mu_1\) in dimension one.
\end{exercise}

\begin{exercise}[Mixture versus barycenter]\label{ex:5:6}
Let \(\mu_1=\delta_0\) and \(\mu_2=\delta_2\), with equal weights. Compute the mixture and the \(\Wtwo\)-barycenter. Repeat with \(\mu_1=\mathcal N(0,1)\) and \(\mu_2=\mathcal N(2,1)\). Discuss the actuarial difference between model averaging by mixture and model averaging by barycenter.
\end{exercise}

\begin{exercise}[Empirical implementation]\label{ex:5:7}
Take two samples of claim severities \(x_1,\ldots,x_n\) and \(y_1,\ldots,y_m\). Write pseudo-code or R code to compute an empirical approximation of \(\Wtwo\) using quantiles on a common grid. How would you modify the code to focus only on the upper tail, say \(u\in[0.9,0.995]\)?
\end{exercise}

\begin{exercise}[Vector-rank stress]\label{ex:5:8}
Let \(\bX\sim\mu\in\Probp{2}(\Rbb^d)\) be a baseline vector of losses and let \(\nu\in\Probp{2}(\Rbb^d)\) be a stressed law. Fix an absolutely continuous reference measure \(\mu_0\). Suppose that the optimal rank map \(R_\mu\) from \(\mu\) to \(\mu_0\), and the optimal vector quantile map \(Q_\nu\) from \(\mu_0\) to \(\nu\), exist. Show that \(T_{\mu\to\nu}=Q_\nu\circ R_\mu\) pushes \(\mu\) to \(\nu\). Explain why the choice of \(\mu_0\) is a modelling decision in dimension larger than one.
\end{exercise}

\newpage
\lecture{Entropic Regularization and Sinkhorn Algorithms}
\label{lec:entropic-sinkhorn}

\subsection*{Learning objectives}

The previous lecture introduced Wasserstein distances as a geometry on probability distributions. This lecture explains how these distances can be computed at scale. Exact discrete optimal transport is a linear program. It is conceptually clean, but it becomes expensive when the number of support points is large. Entropic regularization replaces the original linear program by a strictly convex problem whose solution can be computed by simple matrix scaling.

After this lecture, the reader should be able to:
\begin{enumerate}[label=(\roman*)]
  \item formulate the entropically regularized transport problem for two discrete distributions;
  \item derive the scaling form of the regularized optimal plan;
  \item state and implement Sinkhorn's algorithm;
  \item understand the role of the regularization parameter \(\eps\);
  \item explain the numerical bias introduced by entropic regularization;
  \item define the Sinkhorn divergence and explain why it corrects part of that bias;
  \item use log-domain stabilization in simple computations;
  \item identify actuarial situations where fast approximate transport is more useful than exact transport;
  \item connect entropic transport to portfolio drift, reserving scenarios and climate model aggregation.
\end{enumerate}

\begin{readingbox}
This lecture builds on the discrete formulation of Lecture~\ref{lec:discrete-ot}, the continuous Kantorovich problem of Lecture~\ref{lec:continuous-ot}, and the Wasserstein geometry of Lecture~\ref{lec:wasserstein-barycenters}. Entropic regularization and Sinkhorn scaling are standard tools in computational optimal transport; see \cite{cuturi2013sinkhorn,peyre2019computational}. The matrix-scaling viewpoint goes back to \cite{sinkhorn1967concerning,franklin1989scaling,knight2008sinkhorn}, and the entropy-projection interpretation is closely related to \cite{csiszar1975divergence,bregman1967relaxation,benamou2015iterative}. Sinkhorn divergences and their interpolation between optimal-transport and kernel-type behavior are developed in \cite{feydy2019interpolating}. Stabilized and large-scale implementations are discussed in \cite{schmitzer2019stabilized,altschuler2017near}.
\end{readingbox}

\begin{actuarialbox}
Entropic regularization turns optimal transport from a linear programming problem into a smooth scaling problem. In actuarial language, it replaces a sparse cheapest matching by a cheap matching penalized for excessive concentration. The parameter \(\eps\) controls how much diversification is imposed on the coupling.
\end{actuarialbox}

\subsection{Why regularize optimal transport?}
\label{sec:entropic-why-regularize}

Let
\[
  \mu=\sum_{i=1}^n a_i\delta_{\bx_i},
  \qquad
  \nu=\sum_{j=1}^m b_j\delta_{\by_j},
\]
where \(\ba=(a_1,\ldots,a_n)^\top\in\R_+^n\), \(\bb=(b_1,\ldots,b_m)^\top\in\R_+^m\), and \(\sum_i a_i=\sum_j b_j=1\). For a cost matrix \(C\in\R^{n\times m}\), with \(C_{ij}=c(\bx_i,\by_j)\), the discrete optimal transport problem, introduced in Definition~\ref{def:discrete-ot}, is
\[
  \OT_C(\ba,\bb)
  =
  \min_{\Gamma\in U(\ba,\bb)} \langle C,\Gamma\rangle,
\]
where the transportation polytope, defined in Definition~\ref{def:transportation-polytope}, is
\[
  U(\ba,\bb)=\left\{\Gamma\in\R_+^{n\times m}:\Gamma\mathbf 1_m=\ba,
  \Gamma^\top\mathbf 1_n=\bb\right\},
\]
and
\[
  \langle C,\Gamma\rangle=\sum_{i=1}^n\sum_{j=1}^m C_{ij}\Gamma_{ij}.
\]
This problem is a linear program. It has a clear economic interpretation: \(\Gamma_{ij}\) is the mass transported from \(\bx_i\) to \(\by_j\). But several computational and statistical difficulties appear immediately.

First, the number of variables is \(nm\). If \(n=m=10^4\), the transport plan has \(10^8\) entries. Even storing it can be difficult. Second, exact optimal plans are often sparse and non-smooth. This is useful if the goal is an exact combinatorial matching, but it can be unstable if the empirical distributions are noisy. Third, many actuarial uses of optimal transport require repeated computation: monthly drift monitoring, bootstrap inference, scenario aggregation, or optimization over transport losses.

Entropic regularization addresses these issues by adding a convex penalty to the transport plan. The result is no longer an exact optimal transport cost, but it is smooth, stable, and computable by Sinkhorn scaling. The main reference for this computational viewpoint is \cite{peyre2019computational}; the same ideas are also closely connected with matrix scaling, entropy minimization and the static Schr\"odinger bridge problem.

\subsection{Entropy of a transport plan}
\label{sec:entropic-entropy-plan}

For a nonnegative matrix \(\Gamma\), define the entropy-type functional
\[
  H(\Gamma)=\sum_{i,j}\Gamma_{ij}(\log \Gamma_{ij}-1),
\]
with the convention \(0\log 0=0\). Some authors use \(-\sum_{ij}\Gamma_{ij}\log\Gamma_{ij}\) and call this entropy. Here the sign is chosen so that \(H\) is convex. Indeed, the function \(t\mapsto t\log t-t\) is strictly convex on \(\R_+\).

The entropically regularized transport problem is
\begin{equation}
\label{eq:eot-primal}
  \OT_\eps(\ba,\bb;C)
  =
  \min_{\Gamma\in U(\ba,\bb)}
  \left\{
  \langle C,\Gamma\rangle
  +\eps\sum_{i,j}\Gamma_{ij}(\log\Gamma_{ij}-1)
  \right\},
\end{equation}
where \(\eps>0\).

The additional term makes the objective strictly convex on the positive orthant. Since \(U(\ba,\bb)\) is compact when \(\ba\) and \(\bb\) are probability vectors, the minimum exists. If \(a_i>0\) and \(b_j>0\) for all \(i,j\), the minimizer is unique and strictly positive.

\begin{definition}[Entropic optimal transport]\label{def:entropic-optimal-transport}
Let \(\ba\in\R_+^n\) and \(\bb\in\R_+^m\) be probability vectors and let \(C\in\R^{n\times m}\). For \(\eps>0\), the entropic transport cost is the value of \eqref{eq:eot-primal}. The unique minimizer is denoted by \(\Gamma_\eps\). When the cost matrix needs to be shown explicitly, we write \(\OT_{\eps,C}(\ba,\bb)\).
\end{definition}

\begin{definition}[Discrete Kullback--Leibler divergence]\label{def:discrete-kl-divergence}
For nonnegative matrices \(P=(p_{ij})\) and \(Q=(q_{ij})\) with \(q_{ij}>0\), define
\[
  \KL(P\mid Q)
  =\sum_{i,j}\left\{p_{ij}\log\frac{p_{ij}}{q_{ij}}-p_{ij}+q_{ij}\right\},
\]
with the convention \(0\log(0/q)=0\). This is sometimes called the generalized relative entropy or \(I\)-divergence.
\end{definition}

\begin{proposition}[Entropic OT as a KL projection]\label{prop:eot-kl-projection}
Let \(K_{ij}=\exp(-C_{ij}/\eps)\). Up to the additive constant \(\eps\sum_{i,j}K_{ij}\), the objective in \eqref{eq:eot-primal} is
\[
  \eps\KL(\Gamma\mid K).
\]
Consequently, the entropic optimal plan is the \(\KL\)-projection of the Gibbs kernel \(K\) onto the transportation polytope \(U(\ba,\bb)\).
\end{proposition}

\begin{proof}
Using \(\log K_{ij}=-C_{ij}/\eps\),
\[
  \eps\KL(\Gamma\mid K)
  =\sum_{i,j}C_{ij}\Gamma_{ij}
   +\eps\sum_{i,j}\Gamma_{ij}(\log\Gamma_{ij}-1)
   +\eps\sum_{i,j}K_{ij}.
\]
The last term does not depend on \(\Gamma\), so it does not change the minimizer.
\end{proof}

\begin{remark}[A sign convention]\label{rem:entropic-sign-convention}
The term \(\sum_{ij}\Gamma_{ij}(\log\Gamma_{ij}-1)\) is negative for many entries smaller than one. This is not a problem: adding the constant \(\eps\) or using \(\sum\Gamma_{ij}\log\Gamma_{ij}\) instead would change the value by a fixed amount on \(U(\ba,\bb)\), because \(\sum_{ij}\Gamma_{ij}=1\). The optimizer is unchanged.
\end{remark}

\begin{actuarialbox}
The entropy penalty discourages deterministic or overly concentrated matchings. For noisy empirical portfolios, this is often desirable: the resulting transport plan is less sensitive to a few individual policies or claims. In this sense, \(\eps\) acts like a smoothing parameter.
\end{actuarialbox}

\subsection{First-order conditions and the scaling form}
\label{sec:entropic-scaling-form}

The central reason why entropic transport is computationally attractive is that the optimal plan has a multiplicative scaling form. We derive it in the finite setting.

Assume \(a_i>0\) and \(b_j>0\). Consider the Lagrangian
\[
  \mathcal L(\Gamma,\alpha,\beta)
  =
  \sum_{i,j}C_{ij}\Gamma_{ij}
  +\eps\sum_{i,j}\Gamma_{ij}(\log\Gamma_{ij}-1)
  +\sum_i\alpha_i\left(a_i-\sum_j\Gamma_{ij}\right)
  +\sum_j\beta_j\left(b_j-\sum_i\Gamma_{ij}\right).
\]
The derivative with respect to \(\Gamma_{ij}\) is
\[
  C_{ij}+\eps\log\Gamma_{ij}-\alpha_i-\beta_j.
\]
At the optimum,
\[
  C_{ij}+\eps\log\Gamma_{ij}-\alpha_i-\beta_j=0.
\]
Hence
\[
  \Gamma_{ij}
  =
  \exp\left(\frac{\alpha_i}{\eps}\right)
  \exp\left(-\frac{C_{ij}}{\eps}\right)
  \exp\left(\frac{\beta_j}{\eps}\right).
\]
Define
\[
  u_i=\exp(\alpha_i/\eps),
  \qquad
  v_j=\exp(\beta_j/\eps),
  \qquad
  K_{ij}=\exp(-C_{ij}/\eps).
\]
Then
\begin{equation}
\label{eq:sinkhorn-scaling-form}
  \Gamma_\eps=\operatorname{diag}(\bu)K\operatorname{diag}(\bv),
\end{equation}
where \(\bu\in\R_{++}^n\) and \(\bv\in\R_{++}^m\) are chosen so that the row and column sums are \(\ba\) and \(\bb\):
\[
  \bu\odot (K\bv)=\ba,
  \qquad
  \bv\odot (K^\top \bu)=\bb.
\]
Here \(\odot\) denotes componentwise multiplication.

\begin{theorem}[Scaling form of the entropic plan]\label{thm:entropic-scaling-form}
Assume \(a_i>0\), \(b_j>0\), and \(K_{ij}=\exp(-C_{ij}/\eps)>0\). Then the unique minimizer of \eqref{eq:eot-primal} has the form
\[
  \Gamma_\eps=\operatorname{diag}(\bu)K\operatorname{diag}(\bv),
\]
for two positive vectors \(\bu\) and \(\bv\). These vectors are unique up to multiplication of \(\bu\) by a positive constant and division of \(\bv\) by the same constant.
\end{theorem}

\begin{proof}[Proof sketch]
The first-order conditions above show that any interior optimum must have the stated form. The entropy term forces the optimizer to be strictly positive when all marginals are positive and all entries of \(K\) are positive. The scaling vectors are then determined by the marginal constraints. If \((\bu,\bv)\) and \((\tilde u,\tilde v)\) yield the same positive matrix with the same kernel \(K\), then \(\widetilde u_i/u_i\) must be constant across rows and \(\widetilde v_j/v_j\) the reciprocal constant across columns.
\end{proof}

\begin{remark}[Computational meaning]
The original linear program has \(nm\) unknowns. The regularized problem still has a transport plan with \(nm\) entries, but the optimizer is determined by only \(n+m\) scaling parameters. Sinkhorn's algorithm exploits exactly this reduction.
\end{remark}

\subsection{Sinkhorn's algorithm}
\label{sec:sinkhorn-algorithm}

From the marginal constraints, the scaling vectors must satisfy
\[
  \bu=\frac{\ba}{K\bv},
  \qquad
  \bv=\frac{\bb}{K^\top \bu},
\]
where divisions are componentwise. This suggests the iterative scheme
\begin{equation}
\label{eq:sinkhorn-updates}
  \bu^{(\ell+1)}=\frac{\ba}{K\bv^{(\ell)}},
  \qquad
  \bv^{(\ell+1)}=\frac{\bb}{K^\top \bu^{(\ell+1)}}.
\end{equation}
This is Sinkhorn's algorithm, also known as iterative proportional fitting or matrix balancing.

\begin{definition}[Sinkhorn iteration]\label{def:sinkhorn-iteration}
Let \(K\in\R_{++}^{n\times m}\), \(\ba\in\R_{++}^n\), \(\bb\in\R_{++}^m\), with equal total mass. Starting from \(\bv^{(0)}=\mathbf 1_m\), Sinkhorn's algorithm alternates the two updates in \eqref{eq:sinkhorn-updates}. The approximate transport plan after \(\ell\) iterations is
\[
  \Gamma^{(\ell)}=\operatorname{diag}(\bu^{(\ell)})K\operatorname{diag}(\bv^{(\ell)}).
\]
\end{definition}

\begin{computebox}
Each Sinkhorn step requires one multiplication by \(K\) and one multiplication by \(K^\top\). If \(K\) is dense, this costs \(O(nm)\) per iteration. If the cost has special structure, for example a grid or a convolution kernel, the multiplication can sometimes be accelerated.
\end{computebox}

\begin{theorem}[Sinkhorn convergence, informal statement]\label{thm:sinkhorn-convergence-informal}
If \(\ba\) and \(\bb\) are positive probability vectors and \(K\) has strictly positive entries, then Sinkhorn's algorithm converges to scaling vectors \((\bu,\bv)\) such that \(\operatorname{diag}(\bu)K\operatorname{diag}(\bv)\in U(\ba,\bb)\). Consequently, \(\Gamma^{(\ell)}\to\Gamma_\eps\).
\end{theorem}

\begin{proof}[Proof idea]
The algorithm alternately enforces the row sums and the column sums. In entropy geometry, these two operations are projections onto affine marginal constraint sets. The convergence follows from classical results on alternating projections in Kullback--Leibler divergence or, equivalently, from matrix scaling theory.
\end{proof}

\subsection{A small actuarial example}
\label{sec:sinkhorn-small-actuarial-example}

Suppose an insurer compares two empirical severity distributions on three levels:
\[
  \bx=(0,1,4),
  \qquad
  \by=(0,2,5),
\]
with weights
\[
  \ba=(0.50,0.30,0.20),
  \qquad
  \bb=(0.40,0.40,0.20).
\]
Use squared cost \(C_{ij}=(x_i-y_j)^2\), where \(x_i\) and \(y_j\) are the components of \(\bx\) and \(\by\). Then
\[
C=
\begin{pmatrix}
0&4&25\\
1&1&16\\
16&4&1
\end{pmatrix}.
\]
For a small value of \(\eps\), \(K_{ij}=\exp(-C_{ij}/\eps)\) puts almost all weight on low-cost movements. For a larger value of \(\eps\), even relatively expensive movements remain possible. The optimizer is therefore more diffuse.

With \(\eps=0.5\), entries with cost \(25\) or \(16\) are almost zero. The plan resembles exact optimal transport. With \(\eps=5\), the plan keeps more cross-matching between severity levels. This may be computationally helpful, but the value differs more from the unregularized OT value.

\begin{actuarialbox}
In a severity comparison, a small \(\eps\) emphasizes sharp tail matching. A larger \(\eps\) produces smoother comparisons and is less sensitive to individual observations. The choice depends on whether the actuarial question is tail attribution, routine monitoring, or exploratory visualization.
\end{actuarialbox}

\subsection{The dual problem}
\label{sec:entropic-dual-problem}

Entropic regularization also gives a smooth dual problem. Starting from the Lagrangian above and minimizing over \(\Gamma_{ij}\ge0\), one obtains
\begin{equation}
\label{eq:eot-dual}
  \highlight{GoldenRod}{\OT_\eps(\ba,\bb;C)
  =
  \max_{\boldsymbol{\alpha}\in\R^n,\boldsymbol{\beta}\in\R^m}
  \left\{
  \boldsymbol{\alpha}^\top \ba+\boldsymbol{\beta}^\top \bb
  -\eps\sum_{i,j}\exp\left(\frac{\alpha_i+\beta_j-C_{ij}}{\eps}\right)
  \right\}}
\end{equation}
up to the harmless additive convention associated with the entropy definition. At optimality,
\[
  \Gamma_{ij}=\exp\left(\frac{\alpha_i+\beta_j-C_{ij}}{\eps}\right).
\]
The dual variables \(\boldsymbol{\alpha}\) and \(\boldsymbol{\beta}\) are entropic analogues of Kantorovich potentials. As \(\eps\to0\), they approach optimal transport potentials under suitable conditions.

A useful way to express the dual updates is through a soft minimum. Given \(\boldsymbol{\beta}\), maximizing over \(\alpha_i\) yields the exact update
\[
  \alpha_i
  =
  \eps\log a_i
  -\eps\log\left(\sum_j \exp\left(\frac{\beta_j-C_{ij}}{\eps}\right)\right).
\]
Equivalently,
\[
  \alpha_i
  =\eps\log a_i
  +\operatorname{softmin}_\eps\{C_{ij}-\beta_j:j=1,\ldots,m\}.
\]
The corresponding update for \(\beta_j\) is obtained by exchanging rows and columns. Thus the hard \(c\)-transform of classical optimal transport, introduced in Definition~\ref{def:c-transform-c-concavity}, is replaced by a log-sum-exp transform.

\begin{definition}[Soft minimum]\label{def:soft-minimum}
For numbers \(z_1,\ldots,z_m\) and \(\eps>0\), define
\[
  \operatorname{softmin}_\eps(z_1,
  \ldots,z_m)
  =
  -\eps\log\left(\sum_{j=1}^m e^{-z_j/\eps}\right).
\]
As \(\eps\to0\), \(\operatorname{softmin}_\eps(z_1,\ldots,z_m)\to \min_j z_j\).
\end{definition}

\begin{remark}[Smooth duality]
The log-sum-exp smoothing is a recurrent theme in convex optimization. Entropic OT replaces a non-smooth linear programming dual by a smooth concave maximization problem. This is one reason why entropic costs can be differentiated and used inside larger statistical or machine-learning procedures.
\end{remark}

\subsection{What does \(\eps\) do?}
\label{sec:sinkhorn-epsilon-effect}

The parameter \(\eps\) controls the compromise between transport cost and entropy. This can be understood in three complementary ways.

\begin{figure}
    \centering
    \includegraphics[width=0.95\linewidth]{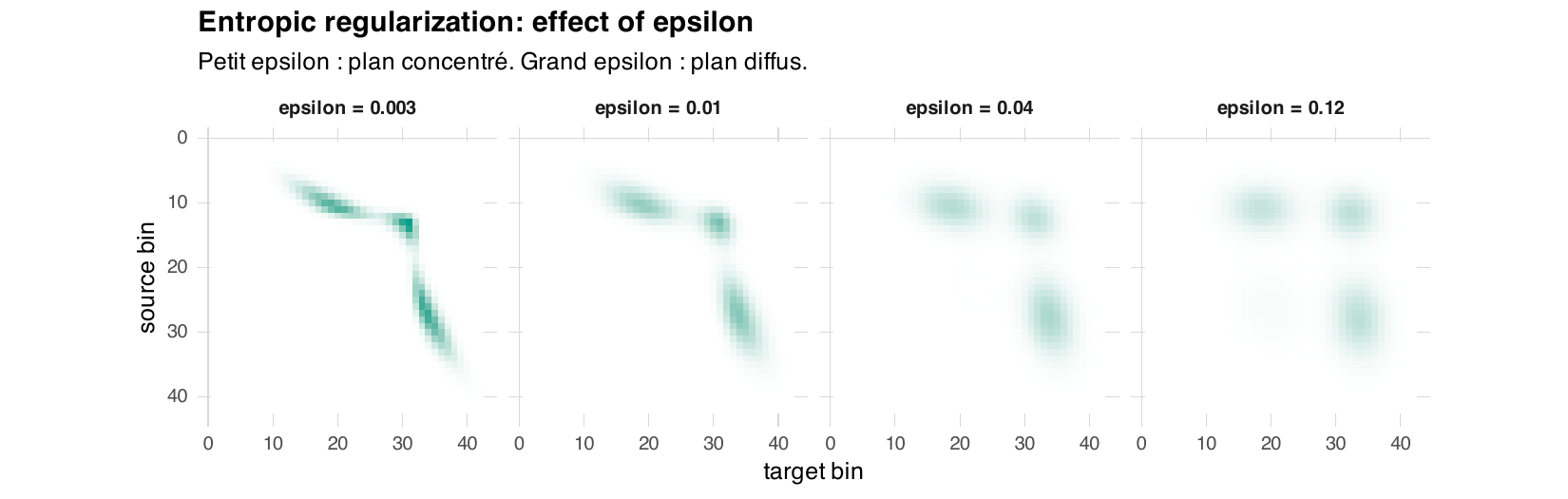}
    \caption{{\bf Effect of entropic regularization}. As $\varepsilon$ increases, the transport plan becomes more diffuse. Small values approximate sparse optimal transport plans, while larger values produce smoother couplings that are easier to compute but more biased.}
    \label{fig:sinkhorn-epsilon-effect}
\end{figure}

\subsubsection*{1. The small-\(\eps\) regime}

When \(\eps\) is small, the penalty is weak. The optimizer is close to an exact OT plan. The kernel
\[
  K_{ij}=e^{-C_{ij}/\eps}
\]
becomes sharply concentrated on low-cost pairs. Numerically, this is the most challenging regime: many entries of \(K\) may underflow to zero.

In actuarial terms, small \(\eps\) is useful when the exact geometry matters: tail matching, stress testing, precise decomposition of distributional differences.

\subsubsection*{2. The large-\(\eps\) regime}

When \(\eps\) is large, the entropy term dominates. The optimizer approaches the independent coupling \(\ba\bb^\top\), because the independent coupling has maximal entropy among all couplings with marginals \(\ba\) and \(\bb\). The transport cost then resembles an average cost under random matching rather than optimal matching.

In actuarial terms, large \(\eps\) gives a diffuse comparison. It is useful for exploratory analysis and robust smoothing, but it can overstate the amount of economically meaningful movement.

\subsubsection*{3. The intermediate regime}

In practice, one often chooses \(\eps\) so that the plan is neither too sparse nor too diffuse. A common rule of thumb is to scale the cost matrix so that meaningful movements have costs of order one, then choose \(\eps\) in a range where Sinkhorn converges stably and the plan remains interpretable.

\begin{computebox}
The value of \(\eps\) is not dimensionless. If costs are multiplied by \(\lambda\), then \(\eps\) should also be multiplied by \(\lambda\) to obtain the same kernel. It is therefore usually preferable to standardize costs before comparing results across portfolios.
\end{computebox}

\subsection{Bias of the regularized cost}
\label{sec:sinkhorn-bias}

The entropic cost \(\OT_{\eps,C}(\ba,\bb)\) is not the same as \(\OT_C(\ba,\bb)\). In the convention used in \eqref{eq:eot-primal}, the value can be shifted by entropy constants; more importantly, the optimizer is different. The regularization introduces bias.

There are two related effects.

First, because the entropy term favors diffuse plans, the minimizer may use transport routes that are not used by exact OT. The expected transport cost \(\langle C,\Gamma_\eps\rangle\) is therefore typically larger than the exact optimal value \(\OT_C(\ba,\bb)\). Second, if one uses the full regularized objective, the entropy contribution changes the scale and can make self-comparisons nonzero:
\[
  \OT_\eps(\ba,\ba;C)
\]
need not be zero. This is undesirable if one wants a divergence between distributions.

\begin{example}[Self-cost]
Let \(\ba=\bb\) and suppose \(C_{ii}=0\) while \(C_{ij}>0\) for \(i\ne j\). Exact OT can use the diagonal plan and has zero cost. Entropic OT spreads some mass away from the diagonal unless \(\eps\) is extremely small. Therefore the regularized plan has positive transport cost, and the full regularized value includes a nontrivial entropy contribution.
\end{example}

\subsection{Sinkhorn divergence}
\label{sec:sinkhorn-divergence}

A common correction is the Sinkhorn divergence. It is best written at the level of measures, because the self-comparisons generally use different cost matrices from the cross-comparison. Let
\[
  \hat\mu=\sum_{i=1}^n a_i\delta_{\bx_i},
  \qquad
  \hat\nu=\sum_{j=1}^m b_j\delta_{\by_j}.
\]
Write \(\OT_\eps(\hat\mu,\hat\nu)\) for the entropic value computed with the cross-cost \(c(\bx_i,\by_j)\). Then define
\begin{equation}
\label{eq:sinkhorn-divergence}
  S_\eps(\hat\mu,\hat\nu)
  =
  \OT_\eps(\hat\mu,\hat\nu)
  -\frac12\OT_\eps(\hat\mu,\hat\mu)
  -\frac12\OT_\eps(\hat\nu,\hat\nu),
\end{equation}
where the two self-terms are computed with the costs \(c(\bx_i,\bx_k)\) and \(c(\by_j,\by_\ell)\), respectively.

\begin{definition}[Sinkhorn divergence]\label{def:sinkhorn-divergence}
Given a regularized transport cost \(\OT_\eps\), the Sinkhorn divergence between two empirical measures \(\hat\mu\) and \(\hat\nu\) is the debiased quantity in \eqref{eq:sinkhorn-divergence}.
\end{definition}

The correction ensures that
\[
  S_\eps(\hat\mu,\hat\mu)=0.
\]
It also reduces the entropic bias while keeping the favorable computational and differentiability properties of entropic transport. In many applications, \(S_\eps\) interpolates between Wasserstein-like behavior for small \(\eps\) and kernel/MMD-like behavior for large \(\eps\).

\begin{remark}[Notation for self-costs]\label{rem:sinkhorn-self-costs}
In matrix notation, \(\OT_\eps(\ba,\bb;C^{xy})\), \(\OT_\eps(\ba,\ba;C^{xx})\), and \(\OT_\eps(\bb,\bb;C^{yy})\) generally use three different cost matrices. Writing the Sinkhorn divergence only as a function of \(\ba\), \(\bb\), and a single matrix \(C\) can be misleading unless all measures live on the same support.
\end{remark}

\begin{actuarialbox}
For monitoring a portfolio over time, a self-comparison should be zero. This makes Sinkhorn divergences more natural than raw entropic costs when the goal is drift detection or dashboard reporting. The same warning applies to reserving-model comparisons: the debiased divergence is usually easier to interpret than the raw regularized cost.
\end{actuarialbox}

\subsection{Numerical stability and log-domain Sinkhorn}
\label{sec:sinkhorn-log-domain}

When \(\eps\) is small or the costs are large, entries of
\[
  K_{ij}=e^{-C_{ij}/\eps}
\]
can underflow to zero. Then the simple updates
\[
  \bu\leftarrow \ba/(K\bv),
  \qquad
  \bv\leftarrow \bb/(K^\top \bu)
\]
can become unstable. The standard remedy is to work in the log domain.

Let
\[
  f_i=\eps\log u_i,
  \qquad
  g_j=\eps\log v_j.
\]
Then
\[
  \Gamma_{ij}=\exp\left(\frac{f_i+g_j-C_{ij}}{\eps}\right).
\]
The row constraint \(\sum_j\Gamma_{ij}=a_i\) becomes
\[
  \sum_j\exp\left(\frac{f_i+g_j-C_{ij}}{\eps}\right)=a_i.
\]
Solving for \(f_i\),
\[
  f_i
  =
  \eps\log a_i
  -\eps\log\left(
  \sum_j\exp\left(\frac{g_j-C_{ij}}{\eps}\right)
  \right).
\]
Similarly,
\[
  g_j
  =
  \eps\log b_j
  -\eps\log\left(
  \sum_i\exp\left(\frac{f_i-C_{ij}}{\eps}\right)
  \right).
\]
These updates can be evaluated stably using the log-sum-exp trick:
\[
  \log\sum_k e^{z_k}
  =
  z_{\max}+\log\sum_k e^{z_k-z_{\max}},
  \qquad z_{\max}=\max_k z_k.
\]

\begin{computebox}
For teaching examples, ordinary Sinkhorn updates are often enough. For real actuarial data, especially with costs based on monetary losses, log-domain stabilization is essential. Otherwise the algorithm may fail for exactly the small values of \(\eps\) that are most Wasserstein-like.
\end{computebox}

\subsection{Stopping criteria}
\label{sec:sinkhorn-stopping-criteria}

A simple stopping criterion is based on marginal violations:
\[
  r_{\mathrm{row}}=\|\Gamma\mathbf 1_m-\ba\|_1,
  \qquad
  r_{\mathrm{col}}=\|\Gamma^\top\mathbf 1_n-\bb\|_1.
\]
The algorithm stops when
\[
  r_{\mathrm{row}}+r_{\mathrm{col}}\le \tau,
\]
for a prescribed tolerance \(\tau\).

Another criterion uses the relative change in the objective:
\[
  \frac{|J^{(\ell+1)}-J^{(\ell)}|}{1+|J^{(\ell)}|}\le \tau.
\]
For most actuarial reporting tasks, marginal feasibility is the most interpretable criterion: the transport plan should move the correct amount of mass from each source cell and into each target cell.

\begin{remark}[Approximate feasibility]
When the plan is used only to compute a scalar divergence, small marginal errors may be acceptable. When the plan is interpreted economically, for example as a redistribution of losses, marginal errors should be monitored carefully.
\end{remark}

\subsection{Differentiability and sensitivity}
\label{sec:sinkhorn-differentiability}

One major advantage of entropic OT is differentiability. The unregularized OT cost is typically non-smooth as a function of the weights, support points, or model parameters. Entropic regularization smooths it.

Suppose the source locations \(\bx_i(\theta)\) depend on a parameter \(\theta\), and the cost matrix is
\[
  C_{ij}(\theta)=c(\bx_i(\theta),\by_j).
\]
Under regularity conditions, the derivative of the entropic value with respect to \(C_{ij}\) is
\[
  \frac{\partial \OT_\eps}{\partial C_{ij}}=\Gamma_{\eps,ij}.
\]
By the chain rule,
\[
  \frac{\dd}{\dd\theta}\OT_\eps(\ba,\bb;C(\theta))
  =
  \sum_{i,j}\Gamma_{\eps,ij}\frac{\partial C_{ij}(\theta)}{\partial\theta}.
\]
This is an envelope theorem: the derivative of the optimized value is computed by differentiating the cost at the optimal plan.

\begin{actuarialbox}
This differentiability is useful when a reserving model, severity model or climate adjustment is calibrated by minimizing a transport loss. Exact OT can be non-smooth; entropic OT is easier to optimize inside larger estimation problems.
\end{actuarialbox}

\subsection{Implementation in R}
\label{sec:sinkhorn-r-implementation}

The following minimal R code implements Sinkhorn's algorithm for two empirical one-dimensional distributions. It is deliberately simple. It is suitable for lecture notes, but not for production use.

\begin{lstlisting}[language=R]
sinkhorn <- function(x, y, a = NULL, b = NULL, eps = 0.1,
                     p = 2, maxit = 5000, tol = 1e-9) {
  n <- length(x)
  m <- length(y)
  if (is.null(a)) a <- rep(1 / n, n)
  if (is.null(b)) b <- rep(1 / m, m)
  stopifnot(abs(sum(a) - 1) < 1e-10, abs(sum(b) - 1) < 1e-10)
  stopifnot(all(a > 0), all(b > 0), eps > 0)

  C <- outer(x, y, function(u, v) abs(u - v)^p)
  K <- exp(-C / eps)

  u <- rep(1, n)
  v <- rep(1, m)

  for (it in seq_len(maxit)) {
    u_old <- u
    u <- a / as.vector(K %*% v)
    v <- b / as.vector(t(K) %*% u)

    if (max(abs(u - u_old)) < tol) break
  }

  Gamma <- K * outer(u, v)
  value_transport <- sum(C * Gamma)
  value_regularized <- value_transport +
    eps * sum(Gamma * (log(pmax(Gamma, 1e-300)) - 1))

  list(plan = Gamma,
       cost = C,
       transport_value = value_transport,
       regularized_value = value_regularized,
       iterations = it,
       row_error = sum(abs(rowSums(Gamma) - a)),
       col_error = sum(abs(colSums(Gamma) - b)))
}

set.seed(123)
x <- rgamma(80, shape = 2, rate = 1)
y <- rgamma(100, shape = 2.2, rate = 0.9)
fit <- sinkhorn(x, y, eps = 0.2, p = 2)
fit$transport_value
fit$row_error + fit$col_error
\end{lstlisting}

For large datasets, one should avoid forming dense diagonal matrices. The expression
\begin{lstlisting}[language=R]
Gamma <- K * outer(u, v)
\end{lstlisting}
is clearer and less error-prone than explicitly multiplying by two diagonal matrices.

\subsection{A log-domain R implementation}
\label{sec:sinkhorn-r-log-implementation}

A more stable version updates the dual potentials. The helper function below computes \(\log\sum_j\exp(z_j)\) stably.

\begin{lstlisting}[language=R]
logsumexp <- function(z) {
  zmax <- max(z)
  zmax + log(sum(exp(z - zmax)))
}

sinkhorn_log <- function(x, y, a = NULL, b = NULL, eps = 0.1,
                         p = 2, maxit = 2000, tol = 1e-8) {
  n <- length(x)
  m <- length(y)
  if (is.null(a)) a <- rep(1 / n, n)
  if (is.null(b)) b <- rep(1 / m, m)

  C <- outer(x, y, function(u, v) abs(u - v)^p)
  f <- rep(0, n)
  g <- rep(0, m)
  loga <- log(a)
  logb <- log(b)

  for (it in seq_len(maxit)) {
    f_old <- f
    for (i in seq_len(n)) {
      z <- (g - C[i, ]) / eps
      f[i] <- eps * (loga[i] - logsumexp(z))
    }
    for (j in seq_len(m)) {
      z <- (f - C[, j]) / eps
      g[j] <- eps * (logb[j] - logsumexp(z))
    }
    if (max(abs(f - f_old)) < tol) break
  }

  eps_logGamma <- outer(f, g, "+") - C
  logGamma <- eps_logGamma / eps
  Gamma <- exp(logGamma)
  list(plan = Gamma,
       transport_value = sum(C * Gamma),
       iterations = it,
       marginal_error = sum(abs(rowSums(Gamma) - a)) +
                        sum(abs(colSums(Gamma) - b)))
}
\end{lstlisting}

This version is slower because it uses explicit loops, but it is much more stable. It is useful for teaching the dual-potential interpretation.

\subsection{Actuarial application 1: portfolio drift monitoring}
\label{sec:sinkhorn-portfolio-drift}

Suppose a company observes a portfolio at two dates. Each policyholder is represented by a vector
\[
  \bz=(\text{age},\text{vehicle value},\text{region score},\text{predicted frequency},\text{predicted severity}).
\]
Let \(\mu_t\) and \(\mu_{t+1}\) be the empirical distributions of these feature vectors at two dates. A Wasserstein or Sinkhorn divergence between \(\mu_t\) and \(\mu_{t+1}\) measures portfolio drift.

This is more informative than comparing coordinate-wise means. A portfolio can keep the same average age and average vehicle value while changing the association between age and vehicle value. A transport divergence captures the joint distributional movement.

This application is developed further in the chapter on portfolio drift and domain adaptation, and it is close to the exercises in Lab~1 on empirical Wasserstein distances. A practical workflow is:
\begin{enumerate}[label=(\roman*)]
  \item choose a standardized feature vector \(\bz\);
  \item define a ground cost \(c(\bz,\bz')\), often squared Euclidean distance after scaling;
  \item compute \(S_\eps(\mu_t,\mu_{t+1})\) monthly;
  \item decompose the movement by feature groups or by projected one-dimensional transports;
  \item trigger actuarial review when the divergence exceeds a threshold.
\end{enumerate}

\begin{actuarialbox}
A portfolio drift indicator based on Sinkhorn divergence detects changes in distribution, not only changes in averages. This is useful for portfolio monitoring, reserving assumptions and model governance.
\end{actuarialbox}

\subsection{Actuarial application 3: reserving distributions}
\label{sec:sinkhorn-reserving-distributions}

Reserving applications are developed later in the chapter on cash-flow distributions and in Lab~5. Reserving models produce predictive distributions for future payments. Suppose several models are fitted to the same triangle: Mack chain ladder, overdispersed Poisson GLM, bootstrap chain ladder, Bayesian model and individual claims model. Each model yields a predictive distribution \(\mu_k\) for the outstanding loss.

Pairwise Sinkhorn divergences can be used to compare these predictive distributions. A barycenter can summarize them. In one dimension, a Wasserstein barycenter is often preferable and easier to compute through quantiles. In higher-dimensional cash-flow reserving, where each observation is a payment vector over future development years, entropic transport becomes useful.

The cost may be
\[
  c(\bx,\by)=\sum_{t=1}^T w_t(x_t-y_t)^2,
\]
where \(x_t\) and \(y_t\) are future payments at development time \(t\), and \(w_t\) are discounting or materiality weights.

\begin{actuarialbox}
For reserving, the object of interest is often not only the ultimate claim, but the cash-flow pattern. Entropic OT can compare full predictive cash-flow distributions, not just best estimates or one-year risk measures.
\end{actuarialbox}

\subsection{Actuarial application 4: climate and catastrophe scenarios}
\label{sec:sinkhorn-climate-catastrophe}

Catastrophe and climate applications are developed later in the reserving-and-prevention chapter and in the chapter on unbalanced transport. Catastrophe models often generate spatial distributions of losses. A scenario can be represented by a discrete distribution over geographic cells:
\[
  \mu=\sum_i a_i\delta_{\bx_i},
\]
where \(\bx_i\) is a location and \(a_i\) is normalized expected loss, probability of claim, or exposure-weighted loss. Comparing two climate scenarios then becomes a transport problem on geography.

Entropic regularization is particularly useful here because spatial grids can be large. The regularized plan can be interpreted as a smooth displacement of loss intensity from one region to another. A small \(\eps\) detects localized shifts; a larger \(\eps\) detects broad regional changes.

Possible uses include:
\begin{enumerate}[label=(\roman*)]
  \item comparing current and future flood risk maps;
  \item measuring the effect of adaptation on the spatial distribution of losses;
  \item aggregating catastrophe model outputs from several vendors;
  \item identifying whether risk is shifted geographically or simply reduced in total.
\end{enumerate}

If total mass changes, balanced OT may not be appropriate. A prevention measure may reduce total expected loss. A climate scenario may increase it. This motivates unbalanced transport.

\subsection{Unbalanced entropic transport}
\label{sec:sinkhorn-unbalanced-entropic}

The classical problem requires equal total masses and exact marginal constraints. In some actuarial situations, this is too restrictive. A prevention action may remove expected loss mass. A new hazard may create additional mass. A portfolio may grow or shrink.

Unbalanced OT relaxes the marginal constraints and penalizes deviations from the prescribed marginals. A common formulation is
\[
  \min_{\Gamma\ge0}
  \left\{
    \langle C,\Gamma\rangle
    +\eps\sum_{ij}\Gamma_{ij}(\log\Gamma_{ij}-1)
    +\lambda_a\KL(\Gamma\mathbf 1_m\mid \ba)
    +\lambda_b\KL(\Gamma^\top\mathbf 1_n\mid \bb)
  \right\}.
\]
Here \(\KL(\br\mid \ba)\) is the vector version of Definition~\ref{def:discrete-kl-divergence},
\[
  \KL(\br\mid \ba)=\sum_i\left\{r_i\log\frac{r_i}{a_i}-r_i+a_i\right\},
\]
and it penalizes the discrepancy between the transported source marginal \(\br\) and the nominal source mass \(\ba\). The parameters \(\lambda_a\) and \(\lambda_b\) control how strongly the marginals are enforced. This is a preview of the dedicated chapter on unbalanced transport.

When \(\lambda_a,\lambda_b\to\infty\), one recovers balanced entropic OT. When they are small, mass can be created or destroyed at lower penalty.

\begin{actuarialbox}
Balanced OT is natural for redistributing a fixed amount of probability mass. Unbalanced OT is natural for comparing expected loss maps before and after prevention, because the total expected loss may change.
\end{actuarialbox}

\subsection{Choosing the ground cost}
\label{sec:sinkhorn-ground-cost}

Entropic regularization makes computation easier, but it does not remove the need to choose a meaningful ground cost. The cost matrix \(C\) is the actuarial modeling choice. It defines what it means for two policies, losses, cells or scenarios to be close.

Some typical choices are:
\begin{enumerate}[label=(\roman*)]
  \item \textbf{Severity cost:} \(C_{ij}=|x_i-y_j|^p\), for scalar losses.
  \item \textbf{Feature cost:} \(C_{ij}=\|\bz_i-\bz'_j\|^2\), after standardizing features.
  % \item \textbf{Premium cost:} \(C_{ij}=|\Premium(\bX_i)-\Premium'(\bX'_j)|^2\), for tariff comparisons. To be restored in v2.
  \item \textbf{Geographic cost:} \(C_{ij}=d_{\mathrm{geo}}(x_i,y_j)^2\), for spatial risk.
  \item \textbf{Cash-flow cost:} \(C_{ij}=\sum_t w_t(x_{it}-y_{jt})^2\), for reserving distributions.
\end{enumerate}

A poor ground cost can make a transport analysis misleading. If age, income, claim frequency and region are mixed in a Euclidean cost without scaling, the variable with the largest numerical scale may dominate. In actuarial work, cost design should be documented as part of model governance.

\begin{remark}[Cost calibration]
A useful practice is to choose the scaling of each coordinate so that a one-unit standardized movement has comparable actuarial meaning. For instance, moving one standard deviation in predicted frequency should be comparable to moving one standard deviation in predicted severity only if the analyst believes these movements have similar relevance.
\end{remark}

\subsection{Exact OT versus entropic OT}
\label{sec:sinkhorn-exact-vs-entropic}

Exact and entropic transport answer slightly different questions. Exact OT asks for the cheapest feasible coupling. Entropic OT asks for a coupling that is cheap but also entropically spread.

\begin{table}[h]
\centering
\begin{tabular}{lll}
\toprule
Feature & Exact OT & Entropic OT \\
\midrule
Optimization & Linear program & Strictly convex smooth problem \\
Plan & Often sparse & Positive and diffuse \\
Computation & LP/network methods & Matrix scaling \\
Stability & Can be non-smooth & Smooth in inputs \\
Self-cost & Zero & Usually biased \\
Best for & exact geometry & large-scale repeated computation \\
\bottomrule
\end{tabular}
\caption{Exact optimal transport versus entropic optimal transport.}
\label{tab:exact-vs-entropic-ot}
\end{table}

Neither is universally better. Exact OT is conceptually closer to the mathematical Wasserstein distance. Entropic OT is often more practical and statistically smoother.

\begin{actuarialbox}
For a one-off regulatory calculation, exact OT or one-dimensional quantile OT may be preferable. For repeated dashboards, high-dimensional portfolios, bootstrap loops or optimization-based calibration, entropic OT is often the practical choice.
\end{actuarialbox}

\subsection{Mini-lab: severity distribution drift}
\label{sec:sinkhorn-lab-severity-drift}

Consider two samples of claim severities. The following code simulates a mild distributional drift and computes a debiased Sinkhorn quantity.

\begin{lstlisting}[language=R]
set.seed(1)
n <- 200
severity_sample_1 <- exp(rnorm(n, mean = log(800), sd = 0.35))
severity_sample_2 <- exp(rnorm(n, mean = log(850), sd = 0.38))

fit12 <- sinkhorn(severity_sample_1, severity_sample_2, eps = 5000, p = 2)
fit11 <- sinkhorn(severity_sample_1, severity_sample_1, eps = 5000, p = 2)
fit22 <- sinkhorn(severity_sample_2, severity_sample_2, eps = 5000, p = 2)

sinkhorn_div <- fit12$regularized_value -
  0.5 * fit11$regularized_value -
  0.5 * fit22$regularized_value

sqrt(max(sinkhorn_div, 0))
\end{lstlisting}

The square root is reported only for interpretability, as an analogue of a distance. The value depends heavily on the scaling of the cost and on \(\eps\). In a real monitoring system, one would compute the same indicator consistently over time and compare it to historical variability.

\subsection{Mini-lab: visualizing the plan}
\label{sec:sinkhorn-lab-visualizing-plan}

The transport plan can be visualized as an image. Rows correspond to source severities and columns to target severities. Sorting the supports makes the structure easier to read.

\begin{lstlisting}[language=R]
o1 <- order(severity_sample_1)
o2 <- order(severity_sample_2)
G <- fit12$plan[o1, o2]
image(log(G + 1e-12), axes = FALSE,
      main = "Log Sinkhorn transport plan")
\end{lstlisting}

For small \(\eps\), the image concentrates near a monotone band in one dimension. For larger \(\eps\), the band becomes wider. This is a useful diagnostic: if the plan is nearly uniform, \(\eps\) is probably too large for a Wasserstein-like interpretation.

\subsection{Reading guide}
\label{sec:sinkhorn-reading-guide}

\begin{readingbox}
For computational optimal transport, including entropic regularization, Sinkhorn algorithms, Sinkhorn divergences and barycenters, see \cite{cuturi2013sinkhorn,peyre2019computational}. For discrete OT and algorithmic intuition, see \cite{solomon2018discrete}. The matrix-scaling and iterative-proportional-fitting literature includes \cite{sinkhorn1967concerning,franklin1989scaling,rueschendorf1995convergence,knight2008sinkhorn}. For stabilized algorithms and computational complexity, see \cite{altschuler2017near,schmitzer2019stabilized}. For debiased Sinkhorn divergences, see \cite{feydy2019interpolating}. The convex optimization background behind smooth duality, log-sum-exp regularization, Bregman projections and envelope arguments is closely related to \cite{rockafellar1970convex,boyd2004convex,bregman1967relaxation,csiszar1975divergence}.
\end{readingbox}

\subsection{Summary}
\label{sec:sinkhorn-summary}

Entropic regularization modifies optimal transport by adding a convex entropy penalty to the transport plan. The regularized optimizer is unique, positive and has the scaling form
\[
  \Gamma_\eps=\operatorname{diag}(\bu)K\operatorname{diag}(\bv),
  \qquad
  K_{ij}=e^{-C_{ij}/\eps}.
\]
The scaling vectors are computed by Sinkhorn iterations. The algorithm is simple, fast and differentiable, but the resulting cost is biased relative to exact OT. The Sinkhorn divergence corrects self-similarity bias and is often better suited for comparing distributions.

For actuarial science, entropic OT is especially useful when distributional comparisons must be repeated many times or performed in high dimension: portfolio drift, climate scenario comparison, reserving cash-flow distributions and model aggregation.

\subsection{Exercises}
\label{sec:sinkhorn-exercises}

\begin{exercise}[Scaling form]\label{ex:6:1}
Let \(\ba\in\R_{++}^n\), \(\bb\in\R_{++}^m\), and \(C\in\R^{n\times m}\). Derive the first-order conditions of the entropic transport problem
\[
  \min_{\Gamma\in U(\ba,\bb)}
  \sum_{ij}C_{ij}\Gamma_{ij}
  +\eps\sum_{ij}\Gamma_{ij}(\log\Gamma_{ij}-1).
\]
Show that the optimizer has the form
\[
  \Gamma_{ij}=u_i e^{-C_{ij}/\eps}v_j.
\]
Explain why \(\bu\) and \(\bv\) are not separately unique.
\end{exercise}

\begin{exercise}[Two-by-two Sinkhorn]\label{ex:6:2}
Let
\[
  \ba=\bb=(1/2,1/2),
  \qquad
  C=\begin{pmatrix}0&1\\1&0\end{pmatrix}.
\]
Compute the entropic optimizer explicitly. What happens as \(\eps\to0\)? What happens as \(\eps\to\infty\)?
\end{exercise}

\begin{exercise}[Independent coupling limit]\label{ex:6:3}
Show heuristically that, as \(\eps\to\infty\), the entropic optimizer tends to the independent coupling \(\ba\bb^\top\). Hint: the entropy term dominates the cost term, so the problem becomes entropy maximization over \(U(\ba,\bb)\).
\end{exercise}

\begin{exercise}[Bias and self-cost]\label{ex:6:4}
Take three ordered points \(x_1<x_2<x_3\), equal weights \(\ba=(1/3,1/3,1/3)\), and squared cost \(C_{ij}=(x_i-x_j)^2\). Explain why exact OT from \(\ba\) to itself has zero cost. Then explain why the entropic optimizer has positive off-diagonal entries for every \(\eps>0\), and hence positive transport cost \(\langle C,\Gamma_\eps\rangle\).
\end{exercise}

\begin{exercise}[Implementation]\label{ex:6:5}
Implement Sinkhorn's algorithm for two empirical samples of losses. Compare the approximate transport cost for \(\eps\in\{0.01,0.1,1,10\}\), after scaling losses to have unit standard deviation. Plot the transport plan for each value of \(\eps\). Comment on the sparsity or diffuseness of the plan.
\end{exercise}

\iffalse
% To be restored in v2, after submission of the paper on portfolio mutualization and demutualization.
\begin{exercise}[Premium drift]\label{ex:6:6}
Let \(\Premium^{(0)}(\bX_i)\) and \(\Premium^{(1)}(\bX_i)\) be predicted premiums for two versions of a tariff on the same portfolio. Propose a Sinkhorn-based indicator of tariff drift. Should the cost be based only on premium differences, or should it also include policyholder features? Discuss the actuarial interpretation of both choices.
\end{exercise}

\fi

\begin{exercise}[Unbalanced prevention]\label{ex:6:7}
A prevention program reduces expected losses in some geographic cells and leaves others unchanged. Explain why balanced OT may be inappropriate for comparing the before and after loss maps. Write an unbalanced entropic formulation with a KL penalty on marginal deviations. Interpret the parameters \(\lambda_a\) and \(\lambda_b\).
\end{exercise}

\iffalse
% To be restored in v2, after submission of the fairness paper with several sensitive variables.
\begin{exercise}[Fairness correction]\label{ex:6:8}
Suppose two groups have empirical score distributions \(\mu_0\) and \(\mu_1\) on \(\R^2\), where the two coordinates are predicted frequency and predicted severity. Describe how entropic OT could be used to map both groups toward a common barycenter. What are the advantages and drawbacks of using a barycentric projection of the entropic plan?
\end{exercise}
\fi

\newpage
\lecture{Statistical Optimal Transport}
\label{lec:statistical-ot}

\subsection*{Learning objectives}

The previous lectures introduced optimal transport as a modelling and computational tool: couplings, optimal plans, Wasserstein distances, barycenters, and entropic regularization. This lecture asks a statistical question. In actuarial work, the laws \(\mu\) and \(\nu\) are rarely known. They are estimated from claims, exposure data, simulated scenarios, posterior predictive samples, or model outputs. Thus any Wasserstein distance, map or barycenter computed from data is itself a statistical object.

After this lecture, the reader should be able to:
\begin{enumerate}[label=(\roman*)]
  \item define empirical measures and plug-in Wasserstein estimators;
  \item distinguish population quantities from sample quantities;
  \item explain why empirical Wasserstein distances converge slowly in high dimension;
  \item understand the special role of the one-dimensional case;
  \item use bootstrap and permutation methods for uncertainty quantification;
  \item design a Wasserstein two-sample test for portfolio drift;
  \item understand why entropic regularization can improve statistical stability;
  \item discuss uncertainty for barycenters and transport maps;
  \item translate these ideas into actuarial workflows: claims monitoring, reserving and climate scenarios.
\end{enumerate}

\begin{actuarialbox}
A Wasserstein number computed from data is not automatically evidence of a structural difference. It contains both signal and sampling noise. Statistical optimal transport is the set of tools that helps us decide whether an observed distributional movement is material, stable and interpretable.
\end{actuarialbox}

This lecture builds on the Wasserstein distance of Definition~\ref{def:wasserstein-distance}, the empirical transport matrix of Definition~\ref{def:transportation-polytope}, and the entropic regularization discussed in Lecture~\ref{lec:entropic-sinkhorn}. It also prepares the empirical labs in Appendix~D and Appendix~E.
% The fairness repair lab in Appendix~F and the premium drift lab in Appendix~G are kept for v2.

\subsection{Population distances and empirical distances}\label{subsec:stat-ot-population-empirical}

Let \(\mu\) and \(\nu\) be two probability measures on a metric space \((\mathcal X,d)\). The population Wasserstein distance is
\[
  \Wp{p}(\mu,\nu)
  =\left(
  \inf_{\gamma\in\Coup(\mu,\nu)}
  \int d(\bx,\by)^p\dd\gamma(\bx,\by)
  \right)^{1/p}.
\]
In most actuarial applications, \(\mu\) and \(\nu\) are unknown. We observe samples
\[
  \bX_1,\ldots,\bX_n\sim \mu,
  \qquad
  \bY_1,\ldots,\bY_m\sim \nu,
\]
possibly independent, and estimate the laws by empirical measures
\[
  \widehat\mu_n=\frac1n\sum_{i=1}^n\delta_{\bX_i},
  \qquad
  \widehat\nu_m=\frac1m\sum_{j=1}^m\delta_{\bY_j}.
\]

\begin{definition}[Empirical measure]\label{def:empirical-measure}
Given observations \(\bx_1,\ldots,\bx_n\) in a measurable space \((\mathcal X,\mathcal B(\mathcal X))\), the empirical measure is
\[
  \widehat\mu_n=\frac1n\sum_{i=1}^n\delta_{\bx_i}.
\]
If the observations carry exposure weights \(w_i\ge0\) with \(\sum_i w_i=1\), the weighted empirical measure is \(\widehat\mu_w=\sum_i w_i\delta_{\bx_i}\). In actuarial applications, weights may represent exposure volume, claim count or scenario probability.
\end{definition}

The most direct estimator is the plug-in estimator.

\begin{definition}[Plug-in Wasserstein estimator]\label{def:plugin-wasserstein-estimator}
For samples from \(\mu\) and \(\nu\), the plug-in estimator of \(\Wp{p}(\mu,\nu)\) is
\[
  \widehat W_{p,n,m}=\Wp{p}(\widehat\mu_n,\widehat\nu_m).
\]
It is obtained by replacing the unknown laws by empirical laws and solving the same transport problem.
\end{definition}

This object is conceptually simple and is exactly what many software packages compute. Its interpretation, however, is statistical: it estimates a population distance and inherits sampling variability from both samples.

\begin{example}[Claim severity drift]
Let \(X_i\) be claim severities from year \(t\), and \(Y_j\) claim severities from year \(t+1\), after exposure normalization. The empirical distance \(\Wone(\widehat\mu_n,\widehat\nu_m)\) measures the minimal average absolute shift needed to transform the observed severity distribution in year \(t\) into the observed severity distribution in year \(t+1\). A large value can indicate inflation, changes in claim handling, different mix of business, or genuine portfolio drift.
\end{example}

\iffalse
% To be restored in v2.
\begin{example}[Predicted premium distributions]
Let \(X_i=\Premium_{\mathrm{old}}(\bX_i)\) be predicted premiums under a current pricing model and \(Y_i=\Premium_{\mathrm{new}}(\bX_i)\) predicted premiums under a proposed model, evaluated on the same portfolio. If we ignore the pairing of policies, \(\Wp{p}(\widehat\mu_n,\widehat\nu_n)\) measures the distributional change in the premium book. If we keep the pairing, the average \(n^{-1}\sum_i |X_i-Y_i|^p\) measures policy-level displacement under the identity coupling. The optimal coupling gives the smallest possible distributional displacement and is therefore a lower bound on any policy-level implementation cost.
\end{example}
\fi

\subsection{The empirical transport problem}\label{subsec:stat-ot-empirical-problem}

Suppose for simplicity that \(n=m\) and empirical weights are uniform. Then
\[
  \widehat\mu_n=\frac1n\sum_{i=1}^n\delta_{\bX_i},
  \qquad
  \widehat\nu_n=\frac1n\sum_{j=1}^n\delta_{\bY_j}.
\]
The empirical \(p\)-Wasserstein problem is
\[
  \Wp{p}^p(\widehat\mu_n,\widehat\nu_n)
  =
  \min_{\Gamma\in\mathcal U_n}
  \sum_{i=1}^n\sum_{j=1}^n d(\bX_i,\bY_j)^p\Gamma_{ij},
\]
where
\[
  \mathcal U_n=\left\{
  \Gamma\in\R_+^{n\times n}:
  \Gamma\mathbf 1_n=\frac1n\mathbf 1_n,
  \quad
  \Gamma^\top\mathbf 1_n=\frac1n\mathbf 1_n
  \right\}.
\]
When the weights are equal on both sides, extreme points of \(\mathcal U_n\) correspond to permutation matrices divided by \(n\). Therefore, in many non-degenerate cases, the problem reduces to an assignment problem:
\[
  \Wp{p}^p(\widehat\mu_n,\widehat\nu_n)
  =
  \frac1n\min_{\sigma\in\mathfrak S_n}
  \sum_{i=1}^n d(\bX_i,\bY_{\sigma(i)})^p,
\]
where \(\mathfrak S_n\) is the set of permutations of \(\{1,\ldots,n\}\). This formula is useful for intuition, but it should not obscure the fact that unequal weights, repeated observations, bins and entropic regularization lead back to general transport matrices.

\begin{remark}[Matched and unmatched samples]
Two empirical distributions can arise from matched or unmatched observations. If every policy has two measurements, before and after an update, then the data are paired. If two periods have different portfolios, the data are not paired. OT treats both as distributional comparisons. The distinction matters for interpretation: matched comparisons estimate the implementation change; unmatched OT estimates a minimal distributional change.
\end{remark}

\subsection{Consistency}\label{subsec:stat-ot-consistency}

The first statistical question is whether the plug-in estimator converges to the population target. The key deterministic step is the following stability inequality.

\begin{proposition}[Stability of Wasserstein distances]\label{prop:wasserstein-distance-stability}
Let \(\mu,\nu,\mu',\nu'\in\Probp{p}(\mathcal X)\). Then
\[
  \left|\Wp{p}(\mu,\nu)-\Wp{p}(\mu',\nu')\right|
  \le
  \Wp{p}(\mu,\mu')+\Wp{p}(\nu,\nu').
\]
In particular, estimating the two marginal laws accurately in \(W_p\) is enough to estimate their distance accurately.
\end{proposition}

\begin{proof}
By the triangle inequality,
\[
  \Wp{p}(\mu,\nu)
  \le
  \Wp{p}(\mu,\mu')+\Wp{p}(\mu',\nu')+\Wp{p}(\nu',\nu).
\]
Rearrange and repeat the argument with \((\mu,\nu)\) and \((\mu',\nu')\) interchanged.
\end{proof}

Under standard moment conditions, the empirical version is therefore consistent.

\begin{proposition}[Consistency of empirical Wasserstein distances]\label{prop:empirical-wasserstein-consistency}
Let \((\mathcal X,d)\) be a Polish metric space and let \(p\ge1\). Suppose \(\mu,\nu\in\Probp{p}(\mathcal X)\). If \(\bX_1,\bX_2,\ldots\) are i.i.d. with law \(\mu\) and \(\bY_1,\bY_2,\ldots\) are i.i.d. with law \(\nu\), then, almost surely,
\[
  \Wp{p}(\widehat\mu_n,\mu)
  \to 0,
  \qquad
  \Wp{p}(\widehat\nu_m,\nu)
  \to 0,
\]
Consequently,
\[
  \Wp{p}(\widehat\mu_n,\widehat\nu_m)
  \to
  \Wp{p}(\mu,\nu)
\]
whenever \(n,m\to\infty\).
\end{proposition}

\begin{proof}[Proof sketch]
A standard characterization of Wasserstein convergence states that \(\Wp{p}(\mu_n,\mu)\to0\) if and only if \(\mu_n\) converges weakly to \(\mu\) and the \(p\)-th moments converge. The empirical measure converges weakly to the population law almost surely, and the empirical \(p\)-th moment converges by the law of large numbers. The second statement follows from the triangle inequality:
\[
  \left|\Wp{p}(\widehat\mu_n,\widehat\nu_m)-\Wp{p}(\mu,\nu)\right|
  \le
  \Wp{p}(\widehat\mu_n,\mu)+\Wp{p}(\widehat\nu_m,\nu).
\]
\end{proof}

Consistency is reassuring but insufficient. It says that, with enough data, empirical transport distances are correct. It does not say how much data are needed, nor whether a distance observed in a finite portfolio is statistically meaningful.

\subsection{Rates and the curse of dimensionality}\label{subsec:stat-ot-rates}

The main difficulty in statistical OT is that empirical Wasserstein distances can converge slowly, especially when the underlying dimension is larger than one. A useful informal message is:
\[
  \E\left[\Wp{p}(\widehat\mu_n,\mu)\right]
  \quad\text{often behaves like}\quad
  n^{-1/d}
\]
when \(\mu\) is a reasonably regular distribution on \(\R^d\), at least in dimensions where the geometry dominates the central-limit scaling. The precise rate depends on \(p\), \(d\), tails and regularity. A representative compact-support statement is
\[
  \E\left[\Wp{p}^p(\widehat\mu_n,\mu)\right]
  =
  \begin{cases}
    O(n^{-1/2}), & d<2p,\\
    O(n^{-1/2}\log n), & d=2p,\\
    O(n^{-p/d}), & d>2p,
  \end{cases}
\]
under regularity conditions. Written for \(\Wp{p}\) rather than \(\Wp{p}^p\), the high-dimensional regime gives the heuristic \(n^{-1/d}\). See \cite{fournier2015rate,weed2019sharp,bobkov2019one} for precise conditions, constants and tail assumptions. The key point for actuaries is that \(d\) matters.

This is not just a technical nuisance. Many actuarial objects are high-dimensional:
\begin{itemize}
  \item a risk profile may contain many rating variables;
  \item a reserve scenario may be a vector of future cash flows;
  \item a capital model scenario may contain losses by line, region and peril;
  \item a climate exposure may contain spatial coordinates and hazard summaries;
%  \item a fairness audit may compare distributions of multivariate scores or outcomes. % To be restored in v2.
\end{itemize}

In such cases, a naive empirical Wasserstein distance may be dominated by sampling variability. This is why one often uses dimension reduction, sliced Wasserstein distances, projections, regularization, parametric models, or problem-specific summaries.

\begin{actuarialbox}
In one dimension, empirical OT is often statistically friendly. In high dimension, empirical OT is statistically expensive. Before using a Wasserstein distance as a KPI, ask: what is the effective dimension of the object being transported?
\end{actuarialbox}

\subsection{The one-dimensional advantage}\label{subsec:stat-ot-one-dimensional}

In dimension one, Wasserstein distances have the quantile representation
\[
  \Wp{p}^p(\mu,\nu)
  =
  \int_0^1
  \left|F_\mu^{-1}(u)-F_\nu^{-1}(u)\right|^p\dd u.
\]
The empirical version is obtained by replacing quantile functions by empirical quantile functions. If \(n=m\), uniform weights, and sorted samples
\[
  X_{(1)}\le\cdots\le X_{(n)},
  \qquad
  Y_{(1)}\le\cdots\le Y_{(n)},
\]
then
\[
  \Wp{p}^p(\widehat\mu_n,\widehat\nu_n)
  =
  \frac1n\sum_{i=1}^n |X_{(i)}-Y_{(i)}|^p.
\]
This simple formula is one of the reasons Wasserstein distances are particularly attractive for severity distributions, reserve summaries and affordability ratios. The formula uses the generalized inverse convention fixed earlier in the notes.

\begin{proposition}[Empirical formula on the real line]\label{prop:empirical-wasserstein-1d}
Let \(\widehat\mu_n=n^{-1}\sum_i\delta_{X_i}\) and \(\widehat\nu_n=n^{-1}\sum_i\delta_{Y_i}\) be empirical measures on \(\R\) with equal weights. Then, for \(p\ge1\),
\[
  \Wp{p}^p(\widehat\mu_n,\widehat\nu_n)
  =
  \frac1n\sum_{i=1}^n |X_{(i)}-Y_{(i)}|^p.
\]
For unequal sample sizes or exposure weights, the same statement holds by matching the two weighted empirical quantile functions.
\end{proposition}

\begin{example}[Severity inflation through quantiles]
Suppose \(Y\) is a severity after inflation and \(X\) is before inflation. If the quantile functions satisfy approximately
\[
  F_Y^{-1}(u)\approx a+bF_X^{-1}(u),
  \qquad b>0,
\]
then \(\Wtwo^2(\mu,\nu)\) decomposes approximately as
\[
  \int_0^1\left(a+(b-1)F_X^{-1}(u)\right)^2\dd u.
\]
This separates a level effect \(a\) from a scale effect \(b\). In practice, plotting \(F_Y^{-1}(u)-F_X^{-1}(u)\) against \(u\) is often more informative than reporting a single number.
\end{example}

\iffalse
% To be restored in v2.
\begin{example}[Fairness audit for scores]
Let \(S\in\{0,1\}\) be a sensitive group indicator and let \(R\) be a risk score. If \(\mu_s=\Law(R\mid S=s)\), then \(\Wtwo(\mu_0,\mu_1)\) measures distributional disparity between groups. In one dimension, this is the \(L^2\)-distance between group-specific quantile functions. This makes the audit interpretable: disparities can be localized at low, median or high risk levels.
\end{example}
\fi

\subsection{Bias under equality of distributions}\label{subsec:stat-ot-noise-floor}

A common mistake is to interpret \(\Wp{p}(\widehat\mu_n,\widehat\nu_m)>0\) as evidence that \(\mu\ne\nu\). But even when \(\mu=\nu\), two independent empirical samples are different. Thus
\[
  \Wp{p}(\widehat\mu_n,\widehat\nu_m)>0
\]
with probability one in continuous models.

\begin{definition}[Wasserstein noise floor]\label{def:wasserstein-noise-floor}
For a sampling design under a null model \(\mu=\nu\), the Wasserstein noise floor is the distribution, or a high quantile, of
\[
  \Wp{p}(\widehat\mu_n,\widehat\nu_m)
\]
when the two empirical measures are generated from the same population law. It is a baseline for deciding whether an observed empirical distance is material.
\end{definition}

The relevant question is not whether the empirical distance is positive. It is whether it is large relative to the noise floor.

\begin{example}[Two samples from the same portfolio]
Split one homogeneous portfolio randomly into two halves. Compute \(\Wone\) between the severity distributions of the two halves. The result is positive, but it should not be interpreted as a business drift. It estimates the noise floor of the Wasserstein monitoring procedure.
\end{example}

This observation motivates permutation tests and bootstrap calibration.

\subsection{Permutation tests for two-sample problems}\label{subsec:stat-ot-permutation}

Consider the two-sample testing problem
\[
  H_0:\mu=\nu
  \qquad\text{versus}\qquad
  H_1:\mu\ne\nu.
\]
A natural statistic is
\[
  T_{n,m}=\Wp{p}(\widehat\mu_n,\widehat\nu_m).
\]
Under \(H_0\), the labels ``sample \(X\)'' and ``sample \(Y\)'' are exchangeable. This leads to a permutation test; see also \cite{ramdas2017wasserstein} for Wasserstein two-sample testing.

\begin{definition}[Permutation-calibrated Wasserstein test]\label{def:permutation-wasserstein-test}
A permutation-calibrated Wasserstein test uses \(T_{n,m}=\Wp{p}(\widehat\mu_n,\widehat\nu_m)\) as the statistic and estimates its null distribution by repeatedly permuting the labels of the pooled sample. The procedure is exact under exchangeability of the labels, up to Monte Carlo error from using finitely many permutations.
\end{definition}

\begin{enumerate}[label=Step \arabic*.]
  \item Pool the data into \(Z_1,\ldots,Z_{n+m}\).
  \item Compute the observed statistic \(T_{\mathrm{obs}}=\Wp{p}(\widehat\mu_n,\widehat\nu_m)\).
  \item Randomly permute the labels, split the pooled data into pseudo-samples of sizes \(n\) and \(m\), and compute \(T^{(b)}\).
  \item Repeat for \(b=1,\ldots,B\).
  \item Estimate the \(p\)-value by
  \[
    \widehat p
    =
    \frac{1+\sum_{b=1}^B\mathbf 1\{T^{(b)}\ge T_{\mathrm{obs}}\}}{B+1}.
  \]
\end{enumerate}

The procedure has a direct interpretation: it asks how often a distance as large as the observed one would appear if the two portfolios were randomly relabelled.

\begin{actuarialbox}
Permutation tests are useful for governance because they turn a Wasserstein dashboard into a statistical monitoring procedure: the point is not only to report a distance, but to compare it with the distance produced by random portfolio splitting.
\end{actuarialbox}

\begin{remark}[Choice of cost and test sensitivity]
The choice of \(p\) and the ground distance matters. With \(p=1\), the test is often robust and interpretable in monetary units. With \(p=2\), large discrepancies receive more weight. For multivariate vectors, the ground distance should reflect actuarial materiality: dollars, log-dollars, normalized scores, standardized risk factors, or business-weighted cash-flow vectors.
\end{remark}

\subsection{Bootstrap confidence intervals}\label{subsec:stat-ot-bootstrap}

A bootstrap procedure estimates the sampling variability of a statistic by resampling from the observed data. For a Wasserstein distance between two independent samples, a simple nonparametric bootstrap is:

\begin{enumerate}[label=Step \arabic*.]
  \item Resample \(X_1^*,\ldots,X_n^*\) with replacement from \(X_1,\ldots,X_n\).
  \item Resample \(Y_1^*,\ldots,Y_m^*\) with replacement from \(Y_1,\ldots,Y_m\).
  \item Compute \(T^*=\Wp{p}(\widehat\mu_n^*,\widehat\nu_m^*)\).
  \item Repeat many times and use the empirical quantiles of \(T^*\) to summarize uncertainty.
\end{enumerate}

For reporting, one may use percentile intervals, bias-corrected intervals, or simply a bootstrap standard error. In actuarial communication, a bootstrap band around a Wasserstein trend is often more useful than a single point estimate.

\begin{remark}[Bootstrap caution]
Bootstrap methods for Wasserstein quantities are not uniformly regular in all dimensions and all null settings. In particular, under equality of distributions, the limiting law of empirical OT can be nonstandard and the naive bootstrap may be misleading. For two-sample testing under \(H_0\), the permutation calibration in Definition~\ref{def:permutation-wasserstein-test} is often the safer default. Bootstrap bands remain useful as descriptive uncertainty diagnostics, especially in one-dimensional or smoothed settings.
\end{remark}

\subsection{Regularization as statistical smoothing}\label{subsec:stat-ot-regularization}

Entropic optimal transport was introduced in the previous lecture for computational reasons. It also has statistical consequences. The entropic problem
\[
\highlight{GoldenRod}{  \OT_\eps(a,b;C)
  =
  \min_{\Gamma\in U(a,b)}
  \left\{\langle C,\Gamma\rangle
  +\eps\sum_{i,j}\Gamma_{ij}(\log\Gamma_{ij}-1)\right\}}
\]
produces a smoother plan than exact OT. This can reduce sensitivity to small perturbations in the data.

There is a trade-off:
\begin{itemize}
  \item small \(\eps\): closer to exact OT, less bias, more variance and less numerical stability;
  \item large \(\eps\): smoother, more stable, more biased toward independent or diffuse couplings.
\end{itemize}

The Sinkhorn divergence corrects part of the regularization bias by subtracting self-costs:
\[
  \operatorname{Sink}_{\eps,c}(\mu,\nu)
  =
  \OT_{\eps,c}(\mu,\nu)
  -\frac12\OT_{\eps,c}(\mu,\mu)
  -\frac12\OT_{\eps,c}(\nu,\nu).
\]
In the empirical discrete case, the self-costs must be computed with the within-sample cost matrices, not by reusing the cross-cost matrix. The Sinkhorn divergence is often preferable to \(\OT_{\eps,c}\) when comparing distributions statistically, because \(\OT_{\eps,c}(\mu,\mu)\) need not be zero.

\begin{remark}[Bias--variance language]
Exact OT is not automatically the best statistical estimator. Entropic OT can be understood as a regularized estimator: it introduces bias, but may reduce variance and improve stability. This is familiar to actuaries from credibility, penalized GLMs, smoothing splines and reserving models.
\end{remark}

\subsection{Sliced and projected Wasserstein distances}\label{subsec:stat-ot-sliced}

High-dimensional OT can be statistically and computationally difficult. A common workaround is to project the data to one dimension and average one-dimensional Wasserstein distances.

\begin{definition}[Sliced Wasserstein distance]\label{def:sliced-wasserstein-distance}
For \(\btheta\in\R^d\) with \(\|\btheta\|=1\), define the projection
\[
  \pi_{\btheta}(\bx)=\btheta^\top \bx.
\]
The sliced Wasserstein distance is
\[
  \operatorname{SW}_p^p(\mu,\nu)
  =
  \int_{\mathbb S^{d-1}}
  \Wp{p}^p((\pi_{\btheta})_\push\mu, (\pi_{\btheta})_\push\nu)
  \dd\btheta,
\]
where \(\dd\btheta\) denotes the uniform probability measure on the unit sphere. In practice, one replaces the integral by an average over finitely many random or actuarially chosen directions.
\end{definition}

For actuarial science, projections are not only numerical tricks. They can encode business questions:
\begin{itemize}
  \item total loss projection: \(\btheta^\top\bx=\sum_k x_k\);
  \item line-of-business stress: \(\btheta\) emphasizes selected lines;
  \item tail-factor projection: \(\btheta\) points toward adverse scenarios;
  \item affordability projection: \(\btheta\) maps features to affordability ratios.
\end{itemize}

\begin{example}[Cash-flow vectors]
Let \(\bX\in\R^d\) be a discounted cash-flow vector over \(d\) future years. Exact \(\Wtwo\) between two empirical laws on \(\R^d\) may be unstable. Sliced distances compare many one-dimensional summaries of cash-flow trajectories. Some directions can be chosen randomly; others can be chosen actuarially, such as total discounted cost, early development years, late tail years, or regulatory one-year view.
\end{example}

\subsection{Distributional confidence bands in one dimension}\label{subsec:stat-ot-confidence-bands}

In one-dimensional actuarial problems, it is often useful to accompany a Wasserstein number with a plot of quantile differences.

\begin{definition}[Quantile-difference function]\label{def:quantile-difference-function}
For two laws \(\mu\) and \(\nu\) on \(\R\), the quantile-difference function is
\[
  \Delta_{\mu,\nu}(u)=Q_\nu(u)-Q_\mu(u),
  \qquad 0<u<1,
\]
where \(Q_\mu=F_\mu^{-1}\) and \(Q_\nu=F_\nu^{-1}\). It localizes the Wasserstein discrepancy across ranks.
\end{definition}

The empirical version is
\[
  \widehat\Delta(u)=Q_{\widehat\nu}(u)-Q_{\widehat\mu}(u).
\]
Bootstrap resampling can produce pointwise or simultaneous bands for \(\Delta(u)\). This is more informative than a scalar distance because it shows where the distributions differ.

\begin{example}[Tail inflation]
A severity distribution can have almost no change below the median but strong inflation above the 95th percentile. A single \(\Wone\) value aggregates this effect. A quantile-difference plot reveals whether the drift is central, tail-driven or uniform across the distribution.
\end{example}

\subsection{Uncertainty for barycenters}\label{subsec:stat-ot-barycenters}

Suppose \(K\) probability measures \(\mu_1,\ldots,\mu_K\) are combined through a Wasserstein barycenter
\[
  \bar\mu
  \in
  \argmin_{\mu}
  \sum_{k=1}^K \lambda_k\Wtwo^2(\mu,\mu_k).
\]
In practice, the \(\mu_k\)'s are estimated. We may have empirical laws \(\widehat\mu_k\), simulated samples, posterior predictive distributions, or expert scenarios. The empirical barycenter is
\[
  \widehat{\bar\mu}
  \in
  \argmin_{\mu}
  \sum_{k=1}^K \lambda_k\Wtwo^2(\mu,\widehat\mu_k).
\]
Uncertainty comes from several sources:
\begin{itemize}
  \item sampling uncertainty within each source distribution;
  \item model uncertainty across the \(K\) sources;
  \item uncertainty in the weights \(\lambda_k\);
  \item numerical uncertainty from discretization or regularization.
\end{itemize}

In one dimension, if \(Q_k(u)=F_{\mu_k}^{-1}(u)\), the barycenter quantile is
\[
  \bar Q(u)=\sum_{k=1}^K \lambda_k Q_k(u).
\]
Thus bootstrap uncertainty is straightforward: resample each source, recompute each empirical quantile, average the quantiles, and build bands for \(\bar Q\).

\begin{example}[Expert aggregation for climate loss]
Suppose \(K\) catastrophe models produce distributions of annual aggregate flood losses for the same portfolio. A one-dimensional Wasserstein barycenter gives a consensus distribution whose quantile function is the weighted average of model quantile functions. Bootstrap or scenario resampling provides uncertainty bands around the consensus quantile curve.
\end{example}

\subsection{Uncertainty for transport maps}\label{subsec:stat-ot-maps}

Transport maps are attractive because they provide transformations, not only distances. In one dimension, the optimal increasing map from \(\mu\) to \(\nu\) is
\[
  T(x)=F_\nu^{-1}(F_\mu(x)).
\]
The empirical map is obtained by replacing \(F_\mu\) and \(F_\nu^{-1}\) by empirical versions. With atoms or ties, the rank convention should be stated explicitly, or randomized ranks should be used. The map can be used for quantile mapping, inflation adjustment or distributional calibration.

However, empirical maps can be unstable in the tails. A small number of extreme claims can create large changes in \(F_{\widehat\nu}^{-1}(u)\) when \(u\) is close to one. This is especially relevant in actuarial severity modelling.

\begin{actuarialbox}
Transport maps are often more sensitive than transport distances. A distance averages the cost over the distribution. A map estimates a pointwise transformation. Tail maps should be treated with the same caution as tail quantiles.
\end{actuarialbox}

\subsection{Application: portfolio drift monitoring}\label{subsec:stat-ot-portfolio-drift}

Let \(Z_{i,t}\) denote an actuarial quantity for policy \(i\) at time \(t\): claim frequency score, predicted severity, exposure value or affordability ratio. Define the empirical law
\[
  \widehat\mu_t=\frac1{n_t}\sum_{i=1}^{n_t}\delta_{Z_{i,t}}.
\]
A Wasserstein drift monitor tracks
\[
  D_t=\Wp{p}(\widehat\mu_t,\widehat\mu_{t-1})
\]
or a distance to a reference period,
\[
  D_t^{\mathrm{ref}}=\Wp{p}(\widehat\mu_t,\widehat\mu_{0}).
\]
A practical dashboard should include:
\begin{enumerate}[label=(\alph*)]
  \item the point estimate \(D_t\);
  \item a bootstrap or permutation noise benchmark;
  \item quantile-difference plots when \(Z\) is one-dimensional;
  \item decomposition by line of business, geography or channel;
  \item documentation of the ground metric and scale transformation.
\end{enumerate}
This workflow is implemented in the empirical uncertainty lab of Appendix~D.
% The premium drift version is kept for v2.

\iffalse
% To be restored in v2.
\subsection{Application: fairness auditing}\label{subsec:stat-ot-fairness}

Let \(R\) be a score, predicted premium or model output, and let \(S\) be a sensitive attribute. For groups \(s=1,\ldots,K\), write
\[
  \mu_s=\Law(R\mid S=s).
\]
A distributional fairness audit may report pairwise distances
\[
  \Wp{p}(\widehat\mu_s,\widehat\mu_{s'}),
\]
or distances to a barycenter
\[
  \Wp{p}(\widehat\mu_s,\widehat{\bar\mu}).
\]
Statistical uncertainty matters because small groups can have noisy empirical distributions. A group with few observations may appear far from the barycenter simply because its empirical distribution is variable.

A sound audit should therefore report group sample sizes, uncertainty intervals, and sensitivity to regularization. In one dimension, quantile-difference plots by group are often the clearest diagnostic. The distributional repair built from the same empirical ingredients is developed later in Section~12.4 and in Appendix~F.

\begin{example}[Small protected group]
Suppose a protected group represents \(3\%\) of the portfolio. Its score distribution is compared with the portfolio barycenter. The observed \(\Wtwo\) distance may be large, but a bootstrap interval may also be wide. The audit conclusion should distinguish statistical uncertainty from substantive disparity.
\end{example}

\fi

\subsection{Application: reserving and predictive distributions}\label{subsec:stat-ot-reserving}

Modern reserving increasingly produces predictive distributions rather than single best estimates. Let \(\mu_A\) and \(\mu_B\) be predictive laws of ultimate losses or discounted cash flows under two reserving methods. Empirical samples may come from bootstrap, Bayesian MCMC, or simulation.

A Wasserstein comparison can answer questions such as:
\begin{itemize}
  \item Are two reserving methods distributionally close, even if their means differ slightly?
  \item Is the difference concentrated in the tail?
  \item How far is a new model from the current approved model?
  \item Does the one-year reserve risk distribution drift materially after data updates?
\end{itemize}

Because reserving samples are often generated by simulation, Monte Carlo error should be included. If the model output is a sample from a predictive distribution, increasing the simulation size reduces numerical Monte Carlo error but not model uncertainty. The cash-flow version of this comparison is revisited in Appendix~H.

\subsection{Application: climate scenarios and prevention}\label{subsec:stat-ot-climate}

Let \(\mu_{\mathrm{before}}\) and \(\mu_{\mathrm{after}}\) denote loss distributions before and after a prevention measure. The Wasserstein distance
\[
  \Wp{p}(\mu_{\mathrm{before}},\mu_{\mathrm{after}})
\]
measures the distributional displacement between the two loss laws. It does not, by itself, identify the causal effect of prevention on individual risks. If both distributions are estimated from simulations, the empirical distance also has sampling or Monte Carlo uncertainty.

A prevention study should ideally report several quantities:
\[
  \Delta\E[L],
  \qquad
  \Delta\VaR_\alpha,
  \qquad
  \Delta\TVaR_\alpha,
  \qquad
  \Wp{p}(\widehat\mu_{\mathrm{before}},\widehat\mu_{\mathrm{after}}),
\]
with uncertainty intervals. Wasserstein distances complement classical risk measures because they reflect the whole distribution, not only a mean or one tail level. This connects to the prevention discussion in Section~11.10 and the climate-scenario workflow in Section~15.13.

\subsection{Mini-lab: Wasserstein uncertainty in R}\label{subsec:stat-ot-mini-lab}

For one-dimensional samples, empirical \(\Wp{p}\) is easy to compute by sorting. The following code gives a minimal implementation for one-dimensional equal-weight samples, including unequal sample sizes.

\begin{lstlisting}[language=R]
wass1d <- function(x, y, p = 2) {
  x <- sort(x)
  y <- sort(y)
  n <- length(x)
  m <- length(y)
  wx <- rep(1 / n, n)
  wy <- rep(1 / m, m)
  i <- 1
  j <- 1
  val <- 0
  while (i <= n && j <= m) {
    mass <- min(wx[i], wy[j])
    val <- val + mass * abs(x[i] - y[j])^p
    wx[i] <- wx[i] - mass
    wy[j] <- wy[j] - mass
    if (wx[i] <= .Machine$double.eps) i <- i + 1
    if (wy[j] <= .Machine$double.eps) j <- j + 1
  }
  val^(1 / p)
}

set.seed(123)
n <- 500
x <- rlnorm(n, meanlog = 0, sdlog = 1)
y <- rlnorm(n, meanlog = 0.1, sdlog = 1.05)

wass1d(x, y, p = 1)
wass1d(x, y, p = 2)
\end{lstlisting}

A simple two-sample permutation test is:

\begin{lstlisting}[language=R]
perm_test_wass <- function(x, y, p = 2, B = 999) {
  n <- length(x)
  z <- c(x, y)
  t_obs <- wass1d(x, y, p = p)
  t_perm <- numeric(B)
  for (b in seq_len(B)) {
    ind <- sample(seq_along(z), n, replace = FALSE)
    xb <- z[ind]
    yb <- z[-ind]
    t_perm[b] <- wass1d(xb, yb, p = p)
  }
  (1 + sum(t_perm >= t_obs)) / (B + 1)
}

perm_test_wass(x, y, p = 2, B = 499)
\end{lstlisting}

\iffalse
% To be restored in v2.
A paired bootstrap for old and new premiums is:

\begin{lstlisting}[language=R]
paired_boot_wass <- function(old, new, p = 1, B = 999) {
  stopifnot(length(old) == length(new))
  n <- length(old)
  stat <- numeric(B)
  for (b in seq_len(B)) {
    ind <- sample.int(n, n, replace = TRUE)
    stat[b] <- wass1d(old[ind], new[ind], p = p)
  }
  quantile(stat, c(0.025, 0.5, 0.975))
}
\end{lstlisting}
\fi

\begin{computebox}
The function \texttt{wass1d} computes exact one-dimensional empirical \(W_p\) for equal-weight samples, including unequal sample sizes. For weighted samples, use weighted empirical quantile functions or a package implementation. For multivariate data, exact OT requires solving a transport problem; for large samples, use entropic, sliced or approximate methods.
\end{computebox}

\subsection{Practical guidelines}\label{subsec:stat-ot-guidelines}

The following checklist is useful when reporting empirical OT results in actuarial work.

\begin{enumerate}[label=(\arabic*)]
  \item \textbf{Define the object.} Are we comparing losses, ratios, cash-flow vectors, covariates or simulated scenarios?
  \item \textbf{Choose the scale.} Dollars, log-dollars, standardized variables and ranks lead to different distances.
  \item \textbf{Choose the ground metric.} It should encode actuarial materiality.
  \item \textbf{Report sample sizes.} Small samples produce noisy empirical distributions.
  \item \textbf{Calibrate noise.} Use bootstrap, permutation or repeated random splits.
  \item \textbf{Inspect where the difference lies.} In one dimension, use quantile-difference plots.
  \item \textbf{Check robustness.} Vary \(p\), trimming, regularization, groups and time windows.
  \item \textbf{Avoid overinterpreting maps.} Maps are pointwise objects and can be unstable in tails.
\end{enumerate}

\begin{readingbox}
For mathematical background on statistical optimal transport, see \cite{panaretos2020invitation,chewi2025statistical}. For empirical rates and concentration, see \cite{fournier2015rate,weed2019sharp,bobkov2019one}. For limit distributions and inference in Wasserstein spaces, see \cite{delbarrio1999central,sommerfeld2018inference}. For computational aspects and entropic regularization, see \cite{peyre2019computational,genevay2019sample}. For the general theory of Wasserstein convergence, see \cite{villani2008oldnew}. The actuarial uses in these notes emphasize model validation, reserving and prevention rather than asymptotic theory for its own sake.
\end{readingbox}

\subsection{Exercises}\label{subsec:stat-ot-exercises}

\begin{exercise}[Empirical consistency]\label{ex:7:1}
Let \(\mu_n\) and \(\nu_n\) be probability measures in \(\Probp{p}(\mathcal X)\). Prove that
\[
  \left|\Wp{p}(\mu_n,\nu_n)-\Wp{p}(\mu,\nu)\right|
  \le
  \Wp{p}(\mu_n,\mu)+\Wp{p}(\nu_n,\nu).
\]
Use this inequality to explain why consistency of empirical measures implies consistency of the plug-in Wasserstein distance.
\end{exercise}

\begin{exercise}[One-dimensional computation]\label{ex:7:2}
Let
\[
  x=(1,2,7,10),
  \qquad
  y=(0,4,6,12).
\]
Compute \(\Wone\) and \(\Wtwo\) between the two empirical distributions with equal weights. Interpret the answer as a comparison of two small claim samples.
\end{exercise}

\begin{exercise}[Noise floor under the null]\label{ex:7:3}
Simulate \(1000\) observations from a lognormal distribution. Randomly split the sample into two groups of size \(500\), and compute \(\Wone\) between the two empirical distributions. Repeat this experiment many times. What is the distribution of the distance under no portfolio drift?
\end{exercise}

\iffalse
% To be restored in v2.
\begin{exercise}[Permutation test for premium drift]\label{ex:7:4}
Design a permutation test for comparing the distribution of predicted premiums between two years. Specify the test statistic, the resampling scheme, and the interpretation of the resulting \(p\)-value. How would your procedure change if the two years contain exactly the same policies?
\end{exercise}
\fi

\iffalse
% To be restored in v2.
\begin{exercise}[Bootstrap for a tariff revision]\label{ex:7:5}
Suppose each policy has an old premium \(P_i^{\mathrm{old}}\) and a new premium \(P_i^{\mathrm{new}}\). Write a paired bootstrap algorithm for a confidence interval of
\[
  \Wone(\widehat\mu_{\mathrm{old}},\widehat\mu_{\mathrm{new}}).
\]
Explain why resampling the two marginal samples independently would be inappropriate.
\end{exercise}
\fi

\iffalse
% To be restored in v2.
\begin{exercise}[Small group fairness]\label{ex:7:6}
A score distribution is compared across two groups, with sample sizes \(n_0=5000\) and \(n_1=120\). The empirical \(\Wtwo\) distance is larger for the small group than for other groups. Propose a statistical workflow to decide whether this reflects a genuine distributional disparity or sampling variability.
\end{exercise}
\fi

\begin{exercise}[Sliced Wasserstein for reserve cash flows]\label{ex:7:7}
Let \(\bX\in\R^{10}\) be a ten-year cash-flow vector generated by reserving model A, and \(\bY\in\R^{10}\) by model B. Propose three actuarially meaningful projection directions \(\btheta\). For each direction, explain what the one-dimensional Wasserstein comparison measures.
\end{exercise}

\begin{exercise}[Entropic smoothing]\label{ex:7:8}
Consider an empirical OT comparison with a small sample size and several repeated values. Explain qualitatively how increasing the entropic regularization parameter \(\eps\) changes the optimal plan. What kind of bias does this introduce? Why might it still be useful in a monitoring dashboard?
\end{exercise}

\newpage
\lecture{Risk Measures, Quantiles and Wasserstein Robustness}
\label{lec:risk-measures-robustness}

\subsection*{Learning objectives}

The preceding lectures developed optimal transport as a geometry for probability measures and introduced the statistical issues that arise when the relevant measures are estimated from data. We now turn to one of the most natural actuarial uses of that geometry: measuring, comparing and stress-testing risk.

After this lecture, the reader should be able to:
\begin{enumerate}[label=(\roman*)]
  \item express classical actuarial risk measures in terms of quantile functions;
  \item explain why Wasserstein distances are particularly well adapted to quantile-based risk measures;
  \item prove Lipschitz stability bounds for expectations, TVaR and spectral risk measures;
  \item understand the instability of VaR under small tail perturbations;
  \item formulate Wasserstein ambiguity sets and robust risk measures;
  \item derive one-dimensional robust formulas using quantile functions;
  \item connect dependence uncertainty to couplings and multi-marginal optimal transport;
  \item interpret Wasserstein robustness in insurance capital, reinsurance and stress testing.
\end{enumerate}

\begin{actuarialbox}
Risk measures are functionals of loss distributions. Optimal transport gives a way to ask: how much can a risk measure change if the loss distribution is moved by a small amount? This question is central for model risk, capital uncertainty, stress testing and regulatory robustness.
\end{actuarialbox}

This lecture uses the lower generalized inverse convention introduced in Definition~\ref{def:intro-rank-preserving-quantile-map}, the Wasserstein distance of Definition~\ref{def:wasserstein-distance}, and the statistical plug-in viewpoint of Definition~\ref{def:plugin-wasserstein-estimator}. It prepares the robust-capital discussion in Section~\ref{subsec:risk-robustness-actuarial-dro}, the tail-aware modelling tools in Section~\ref{subsec:risk-robustness-tail-aware}, the dependence and reinsurance material in Lecture~\ref{lec:reinsurance-capital-mmot}, and the empirical lab on robust VaR and TVaR in Section~\ref{subsec:risk-robustness-mini-lab}.

\subsection{Loss distributions and quantile functions}\label{subsec:risk-robustness-loss-quantiles}

Let $L$ denote a non-negative insurance loss. Its distribution function is
\[
  F_L(x)=\Pbb(L\le x).
\]

\begin{definition}[Loss quantile function]\label{def:loss-quantile-function}
For a loss law $\mu\in\Prob(\R)$ with distribution function $F_\mu$, the generalized quantile function is
\[
  Q_\mu(u)=F_\mu^{-1}(u)=\inf\{x\in\R:F_\mu(x)\ge u\},
  \qquad u\in(0,1).
\]
If $L\sim\mu$, we also write $Q_L=Q_\mu$. This is the same convention as in Definition~\ref{def:intro-rank-preserving-quantile-map}; the notation $Q_\mu$ will be preferred in the risk-measure chapters because quantiles are actuarial objects such as VaR and TVaR.
\end{definition}

The quantile function is the basic bridge between optimal transport and actuarial risk. In one dimension,
\[
  \Wp{p}^p(\mu,\nu)
  =\int_0^1 |Q_\mu(u)-Q_\nu(u)|^p\dd u.
\]
Thus, in dimension one, $\Wp{p}$ is simply the $L^p$-distance between quantile functions.

\begin{example}[Severity models]
Suppose $\mu$ is a fitted severity distribution and $\widehat\mu_n$ is the empirical severity distribution. Then $\Wp{p}(\widehat\mu_n,\mu)$ measures the average vertical discrepancy between empirical and fitted quantiles. This is more distributional than comparing only the mean, variance or a few selected percentiles.
\end{example}

\begin{example}[Inflation and claim development]
If $L_0$ is a loss before inflation and $L_1$ after inflation, an approximate multiplicative inflation model corresponds to
\[
  Q_{L_1}(u)\approx (1+i)Q_{L_0}(u).
\]
Then
\[
  \Wp{p}(\Law(L_0),\Law(L_1))
  \approx i\left(\int_0^1 Q_{L_0}(u)^p\dd u\right)^{1/p}.
\]
The same inflation rate produces a larger Wasserstein movement for heavy-tailed portfolios.
\end{example}

\subsection{Classical actuarial risk measures}\label{subsec:risk-robustness-classical-risk-measures}

\begin{definition}[Law-invariant actuarial risk measure]\label{def:law-invariant-risk-measure}
A law-invariant actuarial risk measure is a functional $\Risk$ that assigns a number to a loss distribution and depends on a random loss $L$ only through its law $\mu=\Law(L)$. We write either $\Risk(L)$ or $\Risk(\mu)$ depending on whether the random variable or its distribution is the object of interest.
\end{definition}

In actuarial science, the most common examples are the mean, Value-at-Risk and Tail Value-at-Risk.

\subsubsection{Mean loss}

The net premium principle uses
\[
  \Risk(L)=\E[L].
\]
When $L\ge0$, this can also be written in quantile form as
\[
  \E[L]=\int_0^1 Q_L(u)\dd u,
\]
provided $L$ is integrable.

\subsubsection{Value-at-Risk}

For $\alpha\in(0,1)$, the Value-at-Risk is
\[
  \VaR_\alpha(L)=Q_L(\alpha).
\]
It is the loss threshold exceeded with probability at most $1-\alpha$. In regulatory capital, $\alpha$ is often very high, such as $99\%$, $99.5\%$, or $99.9\%$, depending on the context.

\subsubsection{Tail Value-at-Risk and Expected Shortfall}

The Tail Value-at-Risk, also called Expected Shortfall, is
\[
  \TVaR_\alpha(L)=\ES_\alpha(L)
  =\frac{1}{1-\alpha}\int_\alpha^1 Q_L(u)\dd u.
\]
If the distribution is continuous at the relevant quantile, this equals the usual conditional tail expectation
\[
  \E[L\mid L\ge \VaR_\alpha(L)].
\]
The quantile definition is the safer convention in these notes because it remains well defined with atoms.

\begin{remark}[TVaR as a tail average]
VaR reads one point of the quantile curve. TVaR averages the whole tail of the quantile curve. This distinction is crucial for Wasserstein stability: averages of quantiles are smoother than point evaluations of quantiles.
\end{remark}

\subsubsection{Spectral and distortion risk measures}

A spectral risk measure has the form
\[
  \Risk_\phi(L)=\int_0^1 Q_L(u)\phi(u)\dd u,
\]
where $\phi\ge0$ and $\int_0^1\phi(u)\dd u=1$. The weight $\phi$ determines which parts of the loss distribution matter most. The mean corresponds to $\phi(u)=1$, while TVaR corresponds to
\[
  \phi_\alpha(u)=\frac{1}{1-\alpha}\1_{\{u\ge\alpha\}}.
\]
More general distortion risk measures can be written as weighted integrals of quantiles, or as integrals with respect to a distortion function. For these notes, the spectral form is the most convenient because it connects directly to $L^p$-geometry.

\begin{readingbox}
For actuarial background on dependence, stochastic orders and risk measures, see \cite{denuit2005actuarial}. For coherent, convex and spectral risk measures, see \cite{artzner1999coherent,acerbi2002spectral,follmer2016stochastic,rockafellar2000optimization}. For extreme-value methods in insurance and finance, see \cite{embrechts1997modelling,coles2001introduction,beirlant2004statistics,dehaan2006extreme}. For Wasserstein distributionally robust optimization, see \cite{esfahani2018data,blanchet2019quantifying,gao2016distributionally}. For the optimal transport background behind Wasserstein distances and quantile couplings, see \cite{villani2003topics,villani2008oldnew,peyre2019computational}.
\end{readingbox}

\subsection{Wasserstein stability of expectations and spectral risks}\label{subsec:risk-robustness-stability}

A central actuarial question is: if two loss distributions are close in Wasserstein distance, are their risk measures close?

\subsubsection{Expectation and $\Wone$}

Let $\mu,\nu\in\Probp{1}(\R)$. Since $x\mapsto x$ is 1-Lipschitz on $\R$, Kantorovich--Rubinstein duality implies
\[
  \left|\int x\dd\mu(x)-\int x\dd\nu(x)\right|
  \le \Wone(\mu,\nu).
\]
Equivalently, using quantiles,
\[
  |\E_\mu[L]-\E_\nu[L]|
  =\left|\int_0^1 \{Q_\mu(u)-Q_\nu(u)\}\dd u\right|
  \le \int_0^1 |Q_\mu(u)-Q_\nu(u)|\dd u.
\]
The right-hand side is $\Wone(\mu,\nu)$.

\begin{proposition}[Expectation is Wasserstein-Lipschitz]\label{prop:expectation-wasserstein-lipschitz}
For any $p\ge1$ and any $\mu,\nu\in\Probp{p}(\R)$,
\[
  |\E_\mu[L]-\E_\nu[L]|
  \le \Wp{p}(\mu,\nu).
\]
\end{proposition}

\begin{proof}
By the quantile representation,
\[
  |\E_\mu[L]-\E_\nu[L]|
  \le \int_0^1 |Q_\mu(u)-Q_\nu(u)|\dd u.
\]
Since the interval $(0,1)$ has unit mass, Jensen's inequality or H\"older's inequality gives
\[
  \int_0^1 |Q_\mu(u)-Q_\nu(u)|\dd u
  \le
  \left(\int_0^1 |Q_\mu(u)-Q_\nu(u)|^p\dd u\right)^{1/p}.
\]
The last term is $\Wp{p}(\mu,\nu)$.
\end{proof}

\subsubsection{Spectral risks}

Let $\phi\in L^q(0,1)$, with $q$ conjugate to $p$:
\[
  \frac1p+\frac1q=1,
\]
with the convention $q=\infty$ when $p=1$. Then
\[
  \Risk_\phi(\mu)-\Risk_\phi(\nu)
  =\int_0^1 \{Q_\mu(u)-Q_\nu(u)\}\phi(u)\dd u.
\]
H\"older's inequality yields the basic stability bound.

\begin{proposition}[Wasserstein stability of spectral risks]\label{prop:spectral-risk-wasserstein-stability}
Let $p\ge1$, and let $q$ be the conjugate exponent. If $\phi\in L^q(0,1)$, then
\[
  |\Risk_\phi(\mu)-\Risk_\phi(\nu)|
  \le
  \|\phi\|_{L^q}\Wp{p}(\mu,\nu).
\]
\end{proposition}

\begin{proof}
By the quantile representation and H\"older's inequality,
\[
\begin{aligned}
  |\Risk_\phi(\mu)-\Risk_\phi(\nu)|
  &\le
  \int_0^1 |Q_\mu(u)-Q_\nu(u)|\phi(u)\dd u\\
  &\le
  \left(\int_0^1 |Q_\mu(u)-Q_\nu(u)|^p\dd u\right)^{1/p}
  \left(\int_0^1 \phi(u)^q\dd u\right)^{1/q}.
\end{aligned}
\]
This is the desired inequality.
\end{proof}

\subsubsection{TVaR stability}

For TVaR, the spectral weight is
\[
  \phi_\alpha(u)=\frac{1}{1-\alpha}\1_{\{u\ge\alpha\}}.
\]
Its $L^q$-norm is
\[
  \|\phi_\alpha\|_{L^q}
  =\left(\int_\alpha^1 \frac{1}{(1-\alpha)^q}\dd u\right)^{1/q}
  =(1-\alpha)^{-1/p}.
\]
Therefore,
\begin{proposition}[Wasserstein stability of TVaR]\label{prop:tvar-wasserstein-stability}
For $p\ge1$ and one-dimensional loss laws $\mu,\nu\in\Probp{p}(\R)$,
\[
  |\TVaR_\alpha(\mu)-\TVaR_\alpha(\nu)|
  \le
  (1-\alpha)^{-1/p}\Wp{p}(\mu,\nu).
\]
\end{proposition}

\begin{actuarialbox}
The factor $(1-\alpha)^{-1/p}$ is the price of tail focus. At $\alpha=99.5\%$, the factor is large, especially for $p=1$. Tail risk is necessarily more sensitive to distributional perturbations than mean risk.
\end{actuarialbox}

\subsection{VaR instability}\label{subsec:risk-robustness-var-instability}

VaR behaves differently. Since $\VaR_\alpha$ is a point evaluation of the quantile function, it is not Lipschitz with respect to $L^p$-distance between quantile functions. Small perturbations in probability mass can produce large changes in a high quantile.

\begin{example}[A small tail contamination]
Let
\[
  L=0 \quad\text{almost surely},
\]
and let
\[
  L_M=\begin{cases}
  0, & \text{with probability }1-\eps,\\
  M, & \text{with probability }\eps.
  \end{cases}
\]
Then
\[
  \Wp{p}(\Law(L),\Law(L_M))=M\eps^{1/p}.
\]
If $\eps$ is small, this can be made moderate even for large $M$, provided $p>1$ or $\eps$ is tiny. However,
\[
  \VaR_\alpha(L_M)=M
\]
whenever $\alpha>1-\eps$. Thus a tiny amount of tail mass can dominate very high VaR.
\end{example}

The example does not mean that VaR is useless. It means that VaR is a local functional of the quantile curve, and local functionals require local regularity. If $F_L$ has a positive density near $\VaR_\alpha(L)$, then small perturbations of the distribution can lead to controlled perturbations of the quantile. But without such regularity, VaR is fragile.

\begin{remark}[Density near the quantile]
Suppose $F$ is continuous and strictly increasing near $q_\alpha=F^{-1}(\alpha)$, with density $f(q_\alpha)>0$. Then a small vertical error in the distribution function translates into a horizontal error of approximate size $1/f(q_\alpha)$. Quantile stability is therefore poor when the density is small near the target quantile, which is common in far tails.
\end{remark}

\subsection{Wasserstein ambiguity sets}\label{subsec:risk-robustness-ambiguity-sets}

\begin{definition}[Wasserstein ambiguity set]\label{def:wasserstein-ambiguity-set}
Let $\widehat\mu\in\Probp{p}(\mathcal X)$ be a reference law on a metric space $(\mathcal X,d)$ and let $\rho\ge0$. The $p$-Wasserstein ambiguity set with radius $\rho$ is
\[
  \mathcal B_p(\widehat\mu,\rho)
  =\{\nu\in\Probp{p}(\mathcal X):\Wp{p}(\nu,\widehat\mu)\le\rho\}.
\]
The reference law $\widehat\mu$ may be empirical, parametric, simulated or expert-based.
\end{definition}

\begin{definition}[Wasserstein robust risk measure]\label{def:wasserstein-robust-risk-measure}
For a law-invariant risk measure $\Risk$, its robust version over $\mathcal B_p(\widehat\mu,\rho)$ is
\[
  \Risk^{\mathrm{rob}}_{p,\rho}(\widehat\mu)
  =\sup_{\nu\in\mathcal B_p(\widehat\mu,\rho)}\Risk(\nu).
\]
It asks for the worst risk value among all loss distributions within Wasserstein distance $\rho$ of the reference model. The value may be $+\infty$ if the risk functional grows too fast relative to the cost.
\end{definition}

\begin{example}[Capital model uncertainty]
Suppose an internal model produces simulated annual losses $L_1,\ldots,L_n$ and the empirical law $\widehat\mu_n$. A regulator or risk committee may not trust $\widehat\mu_n$ exactly. Instead, it considers all distributions within $\Wp{p}$-distance $\rho$. The robust capital is then
\[
  \sup_{\nu:\Wp{p}(\nu,\widehat\mu_n)\le\rho}\TVaR_{99\%}(\nu).
\]
The parameter $\rho$ encodes model ambiguity, estimation error, climate non-stationarity, or expert conservatism.
\end{example}

\subsection{Robust expectations and TVaR in one dimension}\label{subsec:risk-robustness-one-dimensional}

The quantile representation makes Wasserstein robustness transparent in dimension one.

\subsubsection{Robust expectation}

Let $Q$ be the quantile function of $\widehat\mu$. Any alternative distribution $\nu$ has quantile $\widetilde Q$, and the Wasserstein constraint becomes
\[
  \|\widetilde Q-Q\|_{L^p}\le \rho.
\]
The expected loss is
\[
  \E_\nu[L]=\int_0^1 \widetilde Q(u)\dd u.
\]
By H\"older's inequality,
\[
  \int_0^1 \widetilde Q(u)\dd u
  \le
  \int_0^1 Q(u)\dd u+\rho.
\]
The upper bound is attained by shifting the whole quantile function upward by $\rho$, namely $\widetilde Q(u)=Q(u)+\rho$. Hence
\begin{proposition}[Robust mean over a one-dimensional Wasserstein ball]\label{prop:robust-mean-quantile-ball}
For any $p\ge1$ and any one-dimensional reference loss law $\widehat\mu\in\Probp{p}(\R)$,
\[
  \sup_{\nu:\Wp{p}(\nu,\widehat\mu)\le\rho}
  \E_\nu[L]
  =
  \E_{\widehat\mu}[L]+\rho.
\]
\end{proposition}
This formula is independent of $p$ because the mean averages all quantile levels equally.

\subsubsection{Robust TVaR}

For TVaR,
\[
  \TVaR_\alpha(\nu)
  =\frac1{1-\alpha}\int_\alpha^1 \widetilde Q(u)\dd u.
\]
By the stability bound,
\[
  \TVaR_\alpha(\nu)
  \le
  \TVaR_\alpha(\widehat\mu)
  +(1-\alpha)^{-1/p}\rho.
\]
The upper bound is attained by shifting only the upper tail:
\[
  \widetilde Q(u)
  =Q(u)+a\1_{\{u\ge\alpha\}},
  \qquad
  a=\rho(1-\alpha)^{-1/p}.
\]
Indeed,
\[
  \|\widetilde Q-Q\|_{L^p}
  =a(1-\alpha)^{1/p}=\rho,
\]
and the TVaR increases by $a$. Therefore,
\begin{proposition}[Robust TVaR over a one-dimensional Wasserstein ball]\label{prop:robust-tvar-quantile-ball}
For any $p\ge1$ and any one-dimensional reference loss law $\widehat\mu\in\Probp{p}(\R)$,
\[
  \sup_{\nu:\Wp{p}(\nu,\widehat\mu)\le\rho}
  \TVaR_\alpha(\nu)
  =
  \TVaR_\alpha(\widehat\mu)
  +\rho(1-\alpha)^{-1/p}.
\]
\end{proposition}

\begin{actuarialbox}
A Wasserstein robust TVaR has a simple interpretation: start from the estimated TVaR, then add a tail uncertainty loading. The loading increases as the confidence level approaches one and decreases as $p$ increases.
\end{actuarialbox}

\begin{example}[Numerical scale of the loading]
Take $\alpha=99\%$ and $\rho=1000$. Then
\[
  \rho(1-\alpha)^{-1}=100000
  \quad\text{for }p=1,
\]
while
\[
  \rho(1-\alpha)^{-1/2}=10000
  \quad\text{for }p=2.
\]
The choice of ground metric and exponent $p$ has a material impact on robust capital. A $\Wone$-ball permits very concentrated tail perturbations at relatively low average cost, while a $\Wtwo$-ball penalizes large tail movements more strongly.
\end{example}

\subsection{Tail-aware transport and extreme-value modelling}
\label{subsec:risk-robustness-tail-aware}

Wasserstein distances compare whole distributions. This is an advantage when one wants a global geometry of losses, reserves or capital. It can also be a limitation in tail problems. Two loss distributions may be close in $W_p$ because most of the mass lies in the body of the distribution, while being materially different for high quantiles, excess losses or reinsurance layers.

This is particularly important for quantities such as
\[
  \VaR_\alpha(L),
  \qquad
  \TVaR_\alpha(L),
  \qquad
  \E[(L-u)_+],
  \qquad
  \E[(L-d)_+\wedge \ell],
\]
where the upper tail, or a specific layer of the loss distribution, drives the actuarial conclusion.

\begin{actuarialbox}
Standard Wasserstein distances are global distributional distances. Extreme-value modelling is local in the tail. Tail-aware optimal transport combines the two viewpoints: it keeps the geometric interpretation of transport, but focuses the cost, the radius or the comparison on the part of the distribution that matters for tail risk.
\end{actuarialbox}

\subsubsection{Tail conditional laws}

\begin{definition}[Tail mass and excess law]\label{def:tail-mass-excess-law}
Let $L$ be a non-negative loss with distribution $\mu$. For a high threshold $u$, the exceedance probability is
\[
  p_\mu(u)=\mu((u,\infty))=\Pbb(L>u),
\]
and the conditional excess law is
\[
  \mu_u^+
  =
  \Law(L-u\mid L>u).
\]
Thus $\mu_u^+$ is the law of the excess loss above $u$.
\end{definition}

A tail-aware comparison between two loss laws $\mu$ and $\nu$ should compare both:
\[
  p_\mu(u) \quad \text{and} \quad p_\nu(u),
\]
and
\[
  \mu_u^+ \quad \text{and} \quad \nu_u^+.
\]
\begin{definition}[Tail-aware exceedance discrepancy]\label{def:tail-aware-exceedance-discrepancy}
A simple tail-aware discrepancy is
\[
  D_{p,u}^p(\mu,\nu)
  =
  \lambda |p_\mu(u)-p_\nu(u)|^p
  +
  \bar p_u
  W_p^p(\mu_u^+,\nu_u^+),
\]
where $\lambda>0$ is a scaling parameter and $\bar p_u$ is a reference tail mass, for example
\[
  \bar p_u=\frac{p_\mu(u)+p_\nu(u)}{2}.
\]
The first term compares tail frequencies; the second term compares tail severities conditional on exceedance. This definition assumes positive exceedance probabilities when conditional excess laws are used; if one tail probability is zero, the tail mass discrepancy should be reported separately.
\end{definition}

\begin{remark}
The quantity $D_{p,u}$ is best viewed as an actuarial diagnostic, not necessarily as a canonical mathematical distance. It depends on the threshold $u$, the scaling $\lambda$, and the choice of the tail mass $\bar p_u$. These choices should be documented as modelling decisions.
\end{remark}

\subsubsection{Tail quantile Wasserstein distance}

In one dimension, a particularly transparent construction uses quantile functions. Let
\[
  Q_\mu(t)=F_\mu^{-1}(t),
  \qquad
  Q_\nu(t)=F_\nu^{-1}(t).
\]
\begin{definition}[Tail quantile Wasserstein discrepancy]\label{def:tail-quantile-wasserstein}
For $\alpha\in(0,1)$, define the upper-tail Wasserstein discrepancy
\[
  W_{p,\alpha}^{\mathrm{tail}}(\mu,\nu)^p
  =
  \frac{1}{1-\alpha}
  \int_\alpha^1
  |Q_\mu(t)-Q_\nu(t)|^p\,\dd t.
\]
This is simply a Wasserstein comparison restricted to the upper tail quantiles.
\end{definition}

For $p=1$,
\[
  W_{1,\alpha}^{\mathrm{tail}}(\mu,\nu)
  =
  \frac{1}{1-\alpha}
  \int_\alpha^1
  |Q_\mu(t)-Q_\nu(t)|\,\dd t.
\]
This is closely aligned with $\TVaR_\alpha$, since
\[
  \TVaR_\alpha(\mu)
  =
  \frac{1}{1-\alpha}
  \int_\alpha^1 Q_\mu(t)\,\dd t.
\]

\begin{proposition}[Tail TVaR stability]\label{prop:tail-tvar-stability}
Let $\mu$ and $\nu$ be one-dimensional loss distributions with finite first moments. Then
\[
  |\TVaR_\alpha(\mu)-\TVaR_\alpha(\nu)|
  \le
  W_{1,\alpha}^{\mathrm{tail}}(\mu,\nu).
\]
\end{proposition}

\begin{proof}
Using the quantile representation of TVaR,
\[
  \TVaR_\alpha(\mu)-\TVaR_\alpha(\nu)
  =
  \frac{1}{1-\alpha}
  \int_\alpha^1
  \bigl(Q_\mu(t)-Q_\nu(t)\bigr)\,\dd t.
\]
Taking absolute values and applying the triangle inequality gives
\[
  |\TVaR_\alpha(\mu)-\TVaR_\alpha(\nu)|
  \le
  \frac{1}{1-\alpha}
  \int_\alpha^1
  |Q_\mu(t)-Q_\nu(t)|\,\dd t
  =
  W_{1,\alpha}^{\mathrm{tail}}(\mu,\nu).
\]
\end{proof}

\begin{actuarialbox}
The usual $W_1$ distance controls differences in expectations of Lipschitz functions. The tail-restricted version controls differences in tail averages. This is why $W_{1,\alpha}^{\mathrm{tail}}$ is often more informative than global $W_1$ when the actuarial question concerns $\TVaR_\alpha$, excess-of-loss layers or capital.
\end{actuarialbox}

\subsubsection{Peaks-over-threshold modelling}

Extreme-value theory provides a parametric approximation for the conditional excess law $\mu_u^+$. For a high threshold $u$, the peaks-over-threshold principle suggests approximating the law of
\[
  L-u \mid L>u
\]
by a generalized Pareto distribution.

\begin{definition}[Generalized Pareto excess model]\label{def:gpd-excess-model}
For shape parameter $\xi\in\R$ and scale parameter $\beta>0$, the generalized Pareto distribution has distribution function
\[
  G_{\xi,\beta}(y)
  =
  1-\left(1+\frac{\xi y}{\beta}\right)^{-1/\xi},
  \qquad y\ge 0,
\]
when $\xi\ne 0$, with the usual support condition
\[
  1+\frac{\xi y}{\beta}>0.
\]
For $\xi=0$, the limit is the exponential distribution
\[
  G_{0,\beta}(y)
  =
  1-\exp(-y/\beta).
\]
Here $\xi$ is the tail index and $\beta>0$ is a scale parameter.
\end{definition}

If
\[
  p_u=\Pbb(L>u),
\]
then for $\alpha>F(u)$, the high quantile can be approximated by
\[
  \VaR_\alpha(L)
  \approx
  u+
  \frac{\beta}{\xi}
  \left[
    \left(\frac{p_u}{1-\alpha}\right)^\xi
    -1
  \right],
  \qquad \xi\ne 0.
\]
For $\xi=0$,
\[
  \VaR_\alpha(L)
  \approx
  u+
  \beta
  \log\left(\frac{p_u}{1-\alpha}\right).
\]
If $\xi<1$, the corresponding tail mean approximation is
\[
  \TVaR_\alpha(L)
  \approx
  \frac{\VaR_\alpha(L)+\beta-\xi u}{1-\xi}.
\]

\begin{remark}[Tail index and actuarial sensitivity]
The parameter $\xi$ is crucial. When $\xi>0$, the tail is heavy and high quantiles are very sensitive to parameter uncertainty. When $\xi\ge 1$, the mean of the fitted generalized Pareto tail is infinite, which makes expectation-based risk quantities unstable or undefined under the fitted tail model.
\end{remark}

\subsubsection{Transport between fitted tails}

Suppose two tail models are fitted above the same threshold $u$:
\[
  L_0-u\mid L_0>u \sim G_{\xi_0,\beta_0},
  \qquad
  L_1-u\mid L_1>u \sim G_{\xi_1,\beta_1}.
\]
The one-dimensional monotone transport between the two excess laws is
\[
  T_u^{+}(e)
  =
  G_{\xi_1,\beta_1}^{-1}
  \left(
    G_{\xi_0,\beta_0}(e)
  \right),
  \qquad e\ge 0.
\]
Thus a tail-only stress map can be written as
\[
  T^{\mathrm{tail}}(l)
  =
  \begin{cases}
  l, & l\le u,\\[0.4em]
  u+T_u^+(l-u), & l>u.
  \end{cases}
\]
This map leaves the body unchanged and transforms the conditional tail.

\begin{example}[Tail-thickening stress]
Let a baseline excess distribution above $u$ be fitted by $G_{\xi_0,\beta_0}$, and suppose the stress scenario increases the tail index from $\xi_0$ to $\xi_1>\xi_0$. The map
\[
  e\mapsto G_{\xi_1,\beta_1}^{-1}(G_{\xi_0,\beta_0}(e))
\]
moves high excesses further into the tail. This gives an explicit scenario-level stress, not only a change in a fitted parameter.
\end{example}

\begin{remark}[Changing the tail mass]
The map $T^{\mathrm{tail}}$ above preserves the exceedance probability $\Pbb(L>u)$. If the target scenario also changes the tail mass, for example by increasing the probability of crossing $u$, one needs either a mixture adjustment or an unbalanced transport formulation. This connects tail-aware transport with unbalanced OT.
\end{remark}

\subsubsection{Layer-aware costs}

Many insurance contracts do not respond to the full loss $L$, but to a layer. For a deductible $d$ and limit $\ell$, the layer payment is
\[
  h_{d,\ell}(L)
  =
  (L-d)_+-(L-d-\ell)_+.
\]
\begin{definition}[Layer-aware transport cost]\label{def:layer-aware-transport-cost}
A transport cost can be designed to emphasize the layer:
\[
  c_{d,\ell}(x,y)
  =
  |h_{d,\ell}(x)-h_{d,\ell}(y)|^p.
\]
Then the corresponding transport discrepancy compares distributions through the lens of the layer payment rather than through raw loss amounts:
\[
  \inf_{\gamma\in\Coup(\mu,\nu)}
  \int c_{d,\ell}(x,y)\,\dd\gamma(x,y).
\]
This is useful for excess-of-loss reinsurance, catastrophe layers, stop-loss covers and capital layers.
\end{definition}

More generally, one can define a weighted tail cost
\[
  c_\tau(x,y)
  =
  w_\tau(x,y)|x-y|^p,
\]
where $w_\tau(x,y)$ is larger when $x$ or $y$ lies in the tail. For instance,
\[
  w_\tau(x,y)
  =
  1+\tau
  \left(
    \frac{\max(x,y)}{u}
  \right)^\kappa
  \1_{\{\max(x,y)>u\}},
  \qquad
  \tau\ge 0,\;\kappa\ge 0.
\]
Increasing $\tau$ makes the distance more sensitive to tail movements.

\begin{computebox}
Tail-aware costs are modelling devices. They make the transport problem answer a specific actuarial question, such as ``how different are these two models for an excess-of-loss layer?'' or ``how much of the model change is concentrated above the capital threshold?''
\end{computebox}

\subsubsection{Tail-aware Wasserstein DRO}

Tail-aware transport also gives a natural ambiguity set for robust capital. Instead of perturbing the whole loss distribution with a global Wasserstein ball, one can define
\[
  \mathcal B_{\mathrm{tail}}(\widehat\mu,r)
  =
  \{\nu:D_{p,u}(\nu,\widehat\mu)\le r\}.
\]
A tail-robust capital diagnostic is then
\[
  \mathsf{Cap}^{\mathrm{tail}}_r
  =
  \sup_{\nu\in\mathcal B_{\mathrm{tail}}(\widehat\mu,r)}
  \Risk(\nu),
\]
where $\Risk$ may be $\VaR_\alpha$, $\TVaR_\alpha$, or a layer-based risk measure.

Alternatively, one can use the tail quantile distance and define
\[
  \mathcal B_{\alpha}^{\mathrm{tail}}(\widehat\mu,r)
  =
  \left\{
    \nu:
    W_{p,\alpha}^{\mathrm{tail}}(\nu,\widehat\mu)\le r
  \right\}.
\]
This ambiguity set is explicitly focused on upper quantiles.

\begin{actuarialbox}
A global Wasserstein ball asks: how bad can the model become after a transport perturbation of size $r$? A tail-aware Wasserstein ball asks: how bad can the capital layer become after a tail perturbation of size $r$?
\end{actuarialbox}

\subsubsection{Threshold choice and validation}

Tail-aware transport inherits the main practical difficulty of extreme-value modelling: the threshold $u$. If $u$ is too low, the generalized Pareto approximation may be biased. If $u$ is too high, there are too few exceedances and the estimate becomes unstable.

A practical workflow is:
\begin{enumerate}
\item choose a range of candidate thresholds $u$;
\item fit a tail model above each threshold;
\item compute tail-aware distances and capital quantities across thresholds;
\item report sensitivity to $u$, not only one selected value;
\item compare parametric EVT transport with empirical tail-quantile transport.
\end{enumerate}

\begin{remark}[No automatic tail truth]
Extreme-value theory provides asymptotic guidance, not automatic truth. A tail-aware OT analysis should therefore report threshold sensitivity, parameter uncertainty and the effect of the chosen transport cost.
\end{remark}

\subsubsection{A small R illustration}

The following code compares two samples through a global Wasserstein distance and a tail-restricted Wasserstein distance.

\begin{lstlisting}[language=R]
emp_q_left <- function(x, probs) {
  x <- sort(x)
  n <- length(x)
  idx <- pmax(1, ceiling(probs * n))
  x[idx]
}

W1_quantile <- function(x, y, probs) {
  qx <- emp_q_left(x, probs)
  qy <- emp_q_left(y, probs)
  mean(abs(qx - qy))
}

emp_tvar <- function(x, alpha = 0.99) {
  q <- sort(x)
  n <- length(q)
  k <- pmax(1, ceiling(alpha * n))
  mean(q[k:n])
}

set.seed(123)

n <- 5000
x <- rlnorm(n, meanlog = 8, sdlog = 0.8)

# Same body, heavier tail through a mixture perturbation
y <- x
idx <- runif(n) < 0.04
y[idx] <- y[idx] * rlnorm(sum(idx), meanlog = 0.5, sdlog = 0.4)

grid_global <- seq(0.001, 0.999, length.out = 1000)
grid_tail   <- seq(0.95, 0.999, length.out = 300)

W1_global <- W1_quantile(x, y, grid_global)
W1_tail95 <- W1_quantile(x, y, grid_tail)

c(
  W1_global = W1_global,
  W1_tail_95 = W1_tail95,
  VaR99_x = emp_q_left(x, 0.99),
  VaR99_y = emp_q_left(y, 0.99),
  TVaR99_x = emp_tvar(x, 0.99),
  TVaR99_y = emp_tvar(y, 0.99)
)
\end{lstlisting}

The global distance may look moderate because most observations lie in the body. The tail-restricted distance is more sensitive to the perturbation, and is more aligned with high-quantile and tail-mean changes.

\subsubsection{What to remember}

Tail-aware transport is useful when the actuarial object of interest is driven by rare but severe losses. It can be implemented in several complementary ways:
\begin{itemize}
\item compare conditional excess laws above a high threshold;
\item compare only upper quantile functions;
\item use EVT models for the tail and transport fitted excess distributions;
\item design costs that emphasize reinsurance or capital layers;
\item define ambiguity sets directly on tail discrepancies.
\end{itemize}
The purpose is not to replace extreme-value theory, but to give EVT a transport geometry.

\begin{readingbox}
The peaks-over-threshold approximation is classical in extreme-value theory; see \cite{balkema1974residual,pickands1975statistical,coles2001introduction,dehaan2006extreme}. Tail dependence and insurance aggregation are discussed in \cite{embrechts1997modelling,mcneil2015quantitative}. Layer-based interpretations connect naturally with reinsurance pricing; see the reinsurance chapter, especially Section~\ref{subsec:reinsurance-treaties-as-maps}.
\end{readingbox}

\subsection{Robust spectral risk measures}\label{subsec:risk-robustness-spectral}

For a spectral risk measure
\[
  \Risk_\phi(\mu)=\int_0^1 Q_\mu(u)\phi(u)\dd u,
\]
the same argument gives
\[
  \Risk_\phi(\nu)
  \le
  \Risk_\phi(\widehat\mu)+\rho\|\phi\|_{L^q}.
\]
When the extremal perturbation is compatible with a nondecreasing quantile function, this upper bound is attained. This is the case for many standard spectral weights, including the mean and TVaR.

Formally, for $1<p<\infty$, if $\phi\ge0$, $\int_0^1\phi=1$, and the function
\[
  h(u)=\phi(u)^{q-1}
\]
is nondecreasing, then setting
\[
  \widetilde Q(u)=Q(u)+\rho\frac{h(u)}{\|\phi\|_{L^q}^{q-1}}
\]
produces the worst-case direction in quantile space. This shows why the robust loading depends on the $L^q$-norm of the spectral weight. The monotonicity condition is not cosmetic: without it, the formal worst-case direction in $L^p$ need not be a nondecreasing quantile function.

\begin{proposition}[Robust spectral loading, monotone case]\label{prop:robust-spectral-loading}
Under the monotonicity condition above,
\[
  \sup_{\nu:\Wp{p}(\nu,\widehat\mu)\le\rho}
  \Risk_\phi(\nu)
  =\Risk_\phi(\widehat\mu)+\rho\|\phi\|_{L^q}.
\]
\end{proposition}

\begin{remark}[Robustness as a premium principle]
The formula
\[
  \Risk^{\mathrm{rob}}_\phi(\widehat\mu)
  =\Risk_\phi(\widehat\mu)+\rho\|\phi\|_{L^q}
\]
looks like a premium principle: best estimate plus risk loading. The loading is not imposed externally; it is induced by the geometry of the ambiguity set and the tail sensitivity of the risk measure.
\end{remark}

\subsection{Robust VaR through quantile constraints}\label{subsec:risk-robustness-var-constraints}

VaR is not Lipschitz in $\Wp{p}$, but one can still describe its robust version in one dimension.

Let $Q$ be the reference quantile function. To make $\VaR_\alpha$ at least $z$, it is enough to move the upper $(1-\alpha)$-tail of the quantile curve above $z$. The minimal $L^p$-cost of achieving this is
\[
  C_p(z)^p=
  \int_\alpha^1 \{z-Q(u)\}_+^p\dd u.
\]
Therefore, the worst-case upper VaR is characterized as follows.
\begin{proposition}[Robust VaR by tail lifting]\label{prop:robust-var-tail-lifting}
With the quantile convention above,
\[
  \sup_{\nu:\Wp{p}(\nu,\widehat\mu)\le\rho}
  \VaR_\alpha(\nu)
  =
  \sup\left\{z\in\R:
  \int_\alpha^1 \{z-Q(u)\}_+^p\dd u\le\rho^p
  \right\}.
\]
\end{proposition}
This formula is useful computationally: one can solve for $z$ by one-dimensional root finding.

\begin{example}[Empirical robust VaR]
Let the empirical losses be sorted as
\[
  L_{(1)}\le\cdots\le L_{(n)}.
\]
The empirical quantile function is constant on intervals $((i-1)/n,i/n]$. The robust VaR formula becomes a finite sum over the upper tail:
\[
  \frac1n\sum_{i>\alpha n}
  \{z-L_{(i)}\}_+^p\le \rho^p,
\]
with the obvious convention when $\alpha n$ is not an integer. This gives a direct stress-testing algorithm for empirical VaR.
\end{example}

\begin{computebox}
Robust VaR under a one-dimensional Wasserstein ball can be computed by bisection over $z$. Robust TVaR does not require bisection: it is the empirical TVaR plus the explicit loading $\rho(1-\alpha)^{-1/p}$.
\end{computebox}

\subsection{Dependence uncertainty and couplings}\label{subsec:risk-robustness-dependence}

So far the loss distribution itself was uncertain. A different source of model risk arises when marginal loss distributions are known, but dependence is uncertain.

Let $L_1,\ldots,L_d$ be losses for $d$ lines of business, regions, perils or policy types. Suppose the marginal laws $\mu_1,\ldots,\mu_d$ are known or estimated, but the joint distribution is not.

\begin{definition}[Multi-marginal coupling set]\label{def:risk-multimarginal-coupling-set}
The admissible joint laws with fixed marginals $\mu_1,\ldots,\mu_d$ are
\[
  \Coup(\mu_1,\ldots,\mu_d)
  =\left\{\gamma\in\Prob(\R^d):
  (\mathrm{proj}_k)_\#\gamma=\mu_k,
  \ k=1,\ldots,d
  \right\}.
\]
\end{definition}
The aggregate loss is
\[
  S=L_1+\cdots+L_d.
\]
A worst-case aggregate risk problem is
\[
  \sup_{\gamma\in\Coup(\mu_1,\ldots,\mu_d)}
  \Risk_\gamma(S).
\]
This is a multi-marginal optimal transport problem whenever $\Risk$ can be represented as an expectation of a function of $S$. For example, the stop-loss premium
\[
  \E[(S-r)_+]
\]
leads to
\[
  \sup_{\gamma
  \in\Coup(\mu_1,\ldots,\mu_d)}
  \int (x_1+\cdots+x_d-r)_+\dd\gamma(x_1,
  \ldots,x_d).
\]
The cost function is
\[
  c(x_1,\ldots,x_d)=(x_1+\cdots+x_d-r)_+.
\]

\begin{example}[Two lines of business]
An insurer models motor liability and property losses separately. Marginal severity models may be credible, but dependence during inflationary or climate-stressed years may be uncertain. The set of all couplings describes all admissible dependence structures with those marginals. Worst-case stop-loss or TVaR over this set gives a dependence-robust capital measure.
\end{example}

\subsection{Comonotonicity and worst-case aggregation}\label{subsec:risk-robustness-comonotonicity}

In dimension one, the comonotonic coupling is obtained by taking a single uniform variable $U\sim\mathrm{Unif}(0,1)$ and setting
\[
  L_k=Q_k(U),
  \qquad k=1,\ldots,d.
\]
The aggregate comonotonic loss is
\[
  S^{\mathrm{comon}}=
  Q_1(U)+\cdots+Q_d(U).
\]
Its quantile function is
\[
  Q_{S^{\mathrm{comon}}}(u)=Q_1(u)+\cdots+Q_d(u),
\]
at continuity points. This gives the additivity of VaR and TVaR under comonotonicity:
\[
  \VaR_\alpha(S^{\mathrm{comon}})
  =\sum_{k=1}^d\VaR_\alpha(L_k),
\]
and
\[
  \TVaR_\alpha(S^{\mathrm{comon}})
  =\sum_{k=1}^d\TVaR_\alpha(L_k).
\]

For convex increasing functions $\ell$, the expectation
\[
  \E[\ell(L_1+\cdots+L_d)]
\]
is maximized by comonotonic dependence among all couplings with fixed marginals. Thus, for many stop-loss and tail-sensitive objectives, comonotonicity is the adverse dependence benchmark.

\begin{actuarialbox}
Comonotonicity is not a realistic dependence model in most portfolios, but it is a useful upper benchmark. It answers the question: what if all marginal risks materialize in the same quantile state?
\end{actuarialbox}

\subsection{Wasserstein balls versus coupling ambiguity}\label{subsec:risk-robustness-balls-versus-couplings}

There are two distinct but related robustness problems:
\begin{enumerate}[label=(\roman*)]
  \item \emph{Distributional ambiguity}: the entire loss law may move within a Wasserstein ball,
  \[
    \nu\in\mathcal B_p(\widehat\mu,\rho).
  \]
  \item \emph{Dependence ambiguity}: the marginal laws are fixed, but the joint law may vary over couplings,
  \[
    \gamma\in\Coup(\mu_1,\ldots,\mu_d).
  \]
\end{enumerate}

Both are optimal transport problems. In the first, the constraint is a distance from a reference measure. In the second, the constraint is a set of fixed marginals. In actuarial applications, both may appear simultaneously:
\[
  \sup_{\substack{\nu_k:\Wp{p}(\nu_k,\widehat\mu_k)\le\rho_k\\
  \gamma\in\Coup(\nu_1,\ldots,\nu_d)}}
  \Risk_\gamma(L_1+\cdots+L_d).
\]
This formulation allows marginal model uncertainty and dependence uncertainty at the same time.

\begin{example}[Climate and catastrophe risk]
A catastrophe model may produce marginal loss distributions by region and peril. Climate uncertainty may perturb each marginal distribution, while physical dependence between regions or perils may also change. A combined Wasserstein--coupling ambiguity set captures both effects: the marginal distributions can move, and the dependence structure can vary.
\end{example}

\subsection{Choosing the radius $\rho$}\label{subsec:risk-robustness-choosing-rho}

The robustness radius $\rho$ is not a purely mathematical parameter. It should be interpreted and calibrated. Common strategies include:
\begin{enumerate}[label=(\roman*)]
  \item \emph{Statistical calibration}: choose $\rho$ so that $\mathcal B_p(\widehat\mu_n,\rho)$ contains the unknown law with a prescribed confidence level.
  \item \emph{Backtesting calibration}: choose $\rho$ from historical model errors, out-of-sample Wasserstein discrepancies, or forecast residuals.
  \item \emph{Expert calibration}: choose $\rho$ to reflect plausible shifts in severity, inflation, climate hazard or claims handling.
  \item \emph{Regulatory calibration}: choose $\rho$ to obtain a conservative but explainable capital buffer.
  \item \emph{Scenario calibration}: set $\rho$ equal to the Wasserstein distance between a baseline scenario and a specified adverse scenario.
\end{enumerate}

\begin{remark}[Units]
If the ground metric is measured in dollars, then $\rho$ is measured in dollars. This is an advantage of Wasserstein ambiguity sets: the radius has a physical interpretation as an average transportation budget. But it also means that scaling and normalization choices matter.
\end{remark}

\subsection{Ground cost and actuarial meaning}\label{subsec:risk-robustness-ground-cost}

The Wasserstein distance depends on the ground cost introduced in Definition~\ref{def:intro-ground-cost}. For one-dimensional losses, the default choice is
\[
  c(x,y)=|x-y|^p.
\]
For multivariate actuarial vectors, the choice is less obvious. If
\[
  \bx=(x_1,\ldots,x_d)
\]
contains losses by line of business, one may use
\[
  c(\bx,\by)=\sum_{k=1}^d w_k|x_k-y_k|^p,
\]
where $w_k$ reflects line-specific materiality. If $\bx$ contains future cash flows, one may discount:
\[
  c(\bx,\by)=\sum_{t=1}^T v_t|x_t-y_t|^p.
\]
If $\bx$ contains spatial losses, one may combine financial and geographical displacement.

\begin{actuarialbox}
The ground cost is a modelling choice. It determines which distributional changes are considered small. A Wasserstein robust capital model is only as meaningful as the cost used to define the Wasserstein ball.
\end{actuarialbox}

\subsection{Mini-lab: empirical robust TVaR and VaR}\label{subsec:risk-robustness-mini-lab}

The following R code computes empirical VaR, TVaR, robust TVaR and robust VaR for a vector of simulated losses.

\begin{lstlisting}[language=R]
emp_var <- function(x, alpha = 0.99) {
  as.numeric(quantile(x, probs = alpha, type = 1))
}

emp_tvar <- function(x, alpha = 0.99) {
  q <- sort(x)
  n <- length(q)
  k <- ceiling(alpha * n)
  mean(q[k:n])
}

robust_tvar <- function(x, alpha = 0.99, rho = 1, p = 2) {
  emp_tvar(x, alpha) + rho * (1 - alpha)^(-1 / p)
}

robust_var <- function(x, alpha = 0.99, rho = 1, p = 2) {
  q <- sort(x)
  n <- length(q)
  idx <- seq_len(n) / n >= alpha
  tail_q <- q[idx]
  cost <- function(z) mean(pmax(z - tail_q, 0)^p) * length(tail_q) / n
  lower <- emp_var(x, alpha)
  upper <- lower + 10 * rho * (1 - alpha)^(-1 / p) + sd(x)
  while (cost(upper) <= rho^p) upper <- 2 * upper
  uniroot(function(z) cost(z) - rho^p, c(lower, upper))$root
}

set.seed(1)
x <- rlnorm(5000, meanlog = 9, sdlog = 1)
alpha <- 0.995
rho <- 1000
p <- 2

emp_var(x, alpha)
emp_tvar(x, alpha)
robust_tvar(x, alpha, rho, p)
robust_var(x, alpha, rho, p)
\end{lstlisting}

\begin{computebox}
The code is intentionally simple. For production work, pay attention to quantile conventions, interpolation, tail sample size, numerical bracketing and the treatment of policy limits or deductibles.
\end{computebox}

\subsection{Insurance interpretations}\label{subsec:risk-robustness-insurance-interpretations}

\subsubsection{Capital and solvency}

A solvency capital requirement is often a high-quantile or tail-average functional of annual loss. Wasserstein robustness turns
\[
  \Risk(\widehat\mu)
\]
into
\[
  \sup_{\nu:\Wp{p}(\nu,\widehat\mu)\le\rho}\Risk(\nu).
\]
This can be interpreted as a capital requirement under distributional model uncertainty. The additional term for TVaR gives a transparent uncertainty loading.

\subsubsection{Reinsurance pricing}

Stop-loss reinsurance depends on
\[
  \E[(L-r)_+].
\]
The function $x\mapsto (x-r)_+$ is 1-Lipschitz. Therefore
\[
  \left|
  \E_\mu[(L-r)_+]-\E_\nu[(L-r)_+]
  \right|
  \le \Wone(\mu,\nu).
\]
This gives a direct Wasserstein stability bound for stop-loss pure premiums.

\begin{example}[Layer pricing]
A layer payoff
\[
  \ell(L)=\min\{(L-a)_+,b-a\}
\]
is also 1-Lipschitz. Hence its expected value changes by at most $\Wone(\mu,\nu)$ when the severity distribution moves from $\mu$ to $\nu$. This is a useful diagnostic when comparing two fitted severity models for reinsurance pricing.
\end{example}

\iffalse
% To be restored in v2.
\subsubsection{Tariff revisions}

If a new pricing model shifts the distribution of premiums from $\mu$ to $\nu$, Wasserstein distances measure the distributional displacement. Risk measures applied to affordability ratios, such as
\[
  R=\frac{\Premium}{\text{income}},
\]
can be stress-tested under Wasserstein perturbations. For instance, TVaR of affordability ratios focuses on the most burdened policyholders.

\subsubsection{Fairness and group risk}

If loss or premium distributions differ across groups, a Wasserstein correction may map each group distribution toward a barycenter. Risk measures of group-specific premium changes can then be bounded by the Wasserstein cost of the correction. This connects fairness post-processing with stability of risk and affordability measures.
\fi

\subsection{Actuarial DRO: prudent pricing, prudent capital and radius selection}\label{subsec:risk-robustness-actuarial-dro}

The previous sections introduced Wasserstein ambiguity sets as local neighbourhoods around a baseline distribution. We now make the actuarial interpretation more explicit. In actuarial work, a fitted distribution is rarely the final object. It is an estimate built from finite data, modelling choices, portfolio assumptions and expert judgement. A Wasserstein ball represents a family of plausible alternatives around that baseline model.

Let $\widehat\mu$ be a fitted or empirical distribution of a loss, a covariate vector, or a future cash-flow vector. For a transportation cost $c(\bx,\by)$, define the transport discrepancy
\[
  \mathsf W_c(\mu,\nu)
  =
  \inf_{\gamma\in\Coup(\mu,\nu)}
  \int c(\bx,\by)\,\dd\gamma(\bx,\by).
\]
For instance, $c(\bx,\by)=d(\bx,\by)^p$ gives $\mathsf W_c=W_p^p$. A Wasserstein ambiguity set is
\[
  \mathcal B_c(\widehat\mu,r)
  =
  \{\nu\in\Prob(\mathcal X):\mathsf W_c(\nu,\widehat\mu)\le r\},
\]
where $r\ge 0$ is the radius of ambiguity.

\begin{actuarialbox}
A Wasserstein ball should not be read only as a statistical confidence set. It is also a stress-testing device. It says: among all distributions that can be obtained from the baseline law by spending at most $r$ units of transportation cost, what is the worst actuarial quantity?
\end{actuarialbox}

\subsubsection{Robust expectations and the basic dual formula}

Let $h:\mathcal X\to\R$ be an actuarial functional of the state. Examples include a claim severity $h(x)=x$, a pure premium $h(\bx)=\E[L\mid \bX=\bx]$, a loss-ratio contribution, or a discounted cash-flow value. The distributionally robust expectation is
\[
  \mathcal E_r(h;\widehat\mu)
  =
  \sup_{\nu\in\mathcal B_c(\widehat\mu,r)}
  \int h(\by)\,\dd\nu(\by).
\]
This quantity is the largest expected value of $h$ over all distributions within transport radius $r$ of the baseline.

\begin{proposition}[Dual formula for Wasserstein robust expectations]\label{prop:dro-expectation-dual-formula}
Under standard lower-semicontinuity, compactness or growth assumptions,
\[
  \mathcal E_r(h;\widehat\mu)
  =
  \inf_{\lambda\ge 0}
  \left\{
    \lambda r
    +
    \int
    \sup_{\by\in\mathcal X}
    \bigl[h(\by)-\lambda c(\bx,\by)\bigr]
    \,\dd\widehat\mu(\bx)
  \right\}.
\]
\end{proposition}
This formula is the DRO analogue of Kantorovich duality.

\begin{proof}
This is the standard Lagrangian argument for Wasserstein distributionally robust optimization; see the optimal-transport duality background in \cite{villani2008oldnew,peyre2019computational,chewi2025statistical}. Sketch of the proof: introduce a Lagrange multiplier $\lambda\ge 0$ for the constraint $\mathsf W_c(\nu,\widehat\mu)\le r$. The robust expectation is bounded above by
\[
  \sup_{\nu}
  \left\{
    \int h\,\dd\nu
    +
    \lambda\bigl(r-\mathsf W_c(\nu,\widehat\mu)\bigr)
  \right\}.
\]
Using the definition of $\mathsf W_c$, this can be rewritten as an optimization over couplings between $\widehat\mu$ and the adversarial law $\nu$. After maximizing over the second coordinate, one obtains
\[
  \lambda r
  +
  \int
  \sup_{\by} \{h(\by)-\lambda c(\bx,\by)\}
  \,\dd\widehat\mu(\bx).
\]
Taking the infimum over $\lambda\ge 0$ gives the dual formula. Under the usual lower semicontinuity, compactness or growth assumptions, there is no duality gap.
\end{proof}

The inner term
\[
  h^\lambda_c(\bx)
  =
  \sup_{\by}\{h(\by)-\lambda c(\bx,\by)\}
\]
is a penalized worst-case perturbation of state $\bx$. The adversary can move $\bx$ to another state $\by$, improving the payoff $h$, but pays the transport penalty $\lambda c(\bx,\by)$. The multiplier $\lambda$ is therefore the shadow price of robustness.

\subsubsection{Prudent pure premiums}

Let $\bX$ denote policyholder covariates and let $m(\bX)$ be the predicted pure premium, for example $m(\bx)=\E[L\mid \bX=\bx]$. The empirical portfolio law is $\widehat\mu_{\bX}$. The usual average technical premium is
\[
  \widehat\Pi
  =
  \int m(\bx)\,\dd\widehat\mu_{\bX}(\bx).
\]
A Wasserstein-prudent premium is
\[
  \widehat\Pi_r^{\mathrm{DRO}}
  =
  \sup_{\nu\in\mathcal B_c(\widehat\mu_{\bX},r)}
  \int m(\bx)\,\dd\nu(\bx).
\]
This is not a loading applied uniformly to all policies. It is a distributional loading: it asks how large the average premium would become if the portfolio distribution were perturbed within radius $r$.

If $c(\bx,\by)=d(\bx,\by)$ and $m$ is $L_m$-Lipschitz, then by the Kantorovich--Rubinstein duality,
\[
  \widehat\Pi_r^{\mathrm{DRO}}
  \le
  \widehat\Pi + L_m r.
\]
In many regular cases this bound is sharp:
\[
  \sup_{\nu:W_1(\nu,\widehat\mu_{\bX})\le r}
  \int m\,\dd\nu
  =
  \int m\,\dd\widehat\mu_{\bX}
  +
  r\,\mathrm{Lip}(m).
\]

\begin{actuarialbox}
The robust premium decomposes into a baseline premium and a sensitivity loading. The loading is large when the pricing function $m$ changes sharply with respect to the portfolio metric. Thus DRO penalizes not only uncertainty in the portfolio distribution, but also instability of the pricing rule.
\end{actuarialbox}

\begin{example}[Territorial drift]
Suppose $\bX$ contains a territory-risk score and $m(\bX)$ is highly sensitive to that score. A small Wasserstein movement of the portfolio toward higher-risk territories can increase the robust pure premium. The term $\mathrm{Lip}(m)r$ makes this dependence explicit: a more unstable tariff requires a larger prudence margin for the same radius.
\end{example}

\subsubsection{Prudent capital}

Let $L$ denote an annual aggregate loss with baseline law $\widehat\mu_L$. For a law-invariant capital functional $\mathcal R$, such as $\VaR_\alpha$, $\TVaR_\alpha$ or a spectral risk measure, define the robust capital
\[
  \mathsf{Cap}_r(\mathcal R)
  =
  \sup_{\nu\in\mathcal B_c(\widehat\mu_L,r)}
  \mathcal R(\nu).
\]
For expected loss, this reduces to the robust expectation above. For tail risk, the interpretation is more conservative: the adversary is allowed to move probability mass toward worse loss levels, subject to the transport budget $r$.

When $\mathcal R=\TVaR_\alpha$, and losses are one-dimensional with finite first moment, the quantile representation gives
\[
  \TVaR_\alpha(\nu)
  =
  \frac{1}{1-\alpha}
  \int_\alpha^1 F_\nu^{-1}(u)\,\dd u.
\]
Hence robust TVaR over a Wasserstein ball can be interpreted as a worst-case tail-quantile average over all quantile functions within transport distance $r$ of the baseline.

For $W_1$, the following useful bound follows from the TVaR stability inequality:
\[
  \TVaR_\alpha(\nu)
  \le
  \TVaR_\alpha(\widehat\mu_L)
  +
  \frac{1}{1-\alpha} W_1(\nu,\widehat\mu_L).
\]
Therefore
\[
  \sup_{\nu:W_1(\nu,\widehat\mu_L)\le r}
  \TVaR_\alpha(\nu)
  \le
  \TVaR_\alpha(\widehat\mu_L)
  +
  \frac{r}{1-\alpha}.
\]
The multiplier $1/(1-\alpha)$ shows why tail capital is more sensitive to distributional ambiguity than the mean.

\begin{remark}[VaR requires care]
Robust $\VaR_\alpha$ is more delicate than robust $\TVaR_\alpha$. A small amount of probability mass moved across the $\alpha$-quantile can change $\VaR_\alpha$ discontinuously. Wasserstein balls control the amount and distance of such movements, but pointwise quantile stability usually requires additional regularity, such as a positive density near the quantile.
\end{remark}

\begin{example}[Capital add-on from transport ambiguity]
Suppose a baseline loss model gives $\TVaR_{99\%}=100$ million. If the actuary accepts a $W_1$-radius $r=0.5$ million in the loss scale, the bound above gives
\[
  \TVaR_{99\%}^{\mathrm{robust}}
  \le
  100 + \frac{0.5}{0.01}
  =
  150
\]
million. The example is intentionally simple, but it highlights the scale effect: a small average transportation budget can imply a substantial tail capital margin at very high confidence levels.
\end{example}

\subsubsection{Robust pricing versus robust capital}

Pricing and capital use the same ambiguity set, but they answer different questions.

For pricing, the object is often a conditional expectation:
\[
  m(\bx)=\E[L\mid \bX=\bx].
\]
The robust premium asks how the average technical premium changes when the portfolio mix changes within radius $r$.

For capital, the object is a tail functional of the aggregate loss distribution:
\[
  \mathcal R(\Law(L)).
\]
The robust capital asks how the tail of the aggregate law can deteriorate within the same transport radius.

Thus the same Wasserstein ball can be used twice:
\[
  \widehat\Pi_r^{\mathrm{DRO}}
  =
  \sup_{\nu\in\mathcal B(\widehat\mu_{\bX},r)}
  \int m(\bx)\,\dd\nu(\bx),
\]
for portfolio-mix uncertainty, and
\[
  \mathsf{Cap}_r
  =
  \sup_{\nu\in\mathcal B(\widehat\mu_L,r)}
  \mathcal R(\nu),
\]
for loss-distribution uncertainty. The first lives on the covariate space; the second lives on the loss space.

\begin{actuarialbox}
A robust premium is not automatically a robust capital charge. The premium problem perturbs the distribution of risks or policyholder profiles. The capital problem perturbs the distribution of aggregate losses. They may use different ground costs and different radii.
\end{actuarialbox}

\subsubsection{Choosing the radius}

The radius $r$ is the most important modelling parameter in Wasserstein DRO. If $r=0$, the model reduces to the baseline distribution. If $r$ is too large, the robust quantity may become excessively conservative or even infinite when the payoff grows faster than the transportation cost. A useful radius should therefore combine statistical, actuarial and governance considerations.

\paragraph{1. Statistical radius.}
When $\widehat\mu_n$ is an empirical law, one can choose $r$ so that
\[
  \Pbb\bigl(W_p(\widehat\mu_n,\mu_0)\le r\bigr)\approx 1-\delta,
\]
where $\mu_0$ is the unknown data-generating law. In practice, $\mu_0$ is unknown, so one may use bootstrap or subsampling:
\[
  r_{\mathrm{stat}}
  =
  \mathrm{Quantile}_{1-\delta}
  \left(
    W_p(\widehat\mu_n^\ast,\widehat\mu_n)
  \right),
\]
where $\widehat\mu_n^\ast$ is a bootstrap empirical distribution.

\paragraph{2. Backtesting radius.}
For time-indexed portfolios, one can compute historical distributional changes
\[
  W_p(\widehat\mu_t,\widehat\mu_{t+1}),
\]
and choose $r$ as a high quantile of these observed shifts. This makes the radius reflect historically observed portfolio drift.

\paragraph{3. Stress radius.}
The radius can be calibrated to a named stress. For instance, if a regulatory or internal stress distribution $\mu^{\mathrm{stress}}$ is available, define
\[
  r_{\mathrm{stress}}
  =
  W_p(\widehat\mu,\mu^{\mathrm{stress}}).
\]
Then fractions of $r_{\mathrm{stress}}$ correspond to partial stresses, while larger multiples correspond to more severe scenarios.

\paragraph{4. Materiality radius.}
One may choose $r$ by specifying an acceptable change in an actuarial quantity. For example, choose $r$ so that
\[
  \widehat\Pi_r^{\mathrm{DRO}}-\widehat\Pi
\]
equals a target prudence margin, or so that robust capital increases by a pre-specified percentage. This reverses the DRO calculation: the radius is chosen to match a governance or materiality threshold.

\paragraph{5. Combined radius.}
In practice, the radius often combines several sources:
\[
  r
  =
  r_{\mathrm{stat}}
  +
  r_{\mathrm{model}}
  +
  r_{\mathrm{stress}},
\]
or, when the components are viewed as roughly independent,
\[
  r
  =
  \bigl(
    r_{\mathrm{stat}}^2
    +
    r_{\mathrm{model}}^2
    +
    r_{\mathrm{stress}}^2
  \bigr)^{1/2}.
\]
The first form is conservative; the second is closer to a root-sum-square aggregation of uncertainty budgets.

\begin{computebox}
A practical radius-selection workflow is:
\begin{enumerate}
\item compute a statistical radius by bootstrap;
\item compute historical Wasserstein drift between successive portfolio periods;
\item compute the distance to at least one named stress scenario;
\item report robust premium and robust capital as functions of $r$;
\item choose the operational radius through governance, not only through a
statistical formula.
\end{enumerate}
\end{computebox}

\subsubsection{A mini-lab: robust mean and radius selection in R}

The following code illustrates a simple one-dimensional $W_1$-DRO calculation. For a Lipschitz payoff $h(x)=x$, the robust expectation over a $W_1$-ball is the empirical mean plus the radius.

\begin{lstlisting}[language=R]
set.seed(123)

# Baseline empirical losses
n <- 1000
loss <- rlnorm(n, meanlog = 8, sdlog = 0.8)

# Empirical W1 by sorting on a common grid and using the left empirical quantile
emp_q_left <- function(x, probs) {
  x <- sort(x)
  n <- length(x)
  idx <- pmax(1, ceiling(probs * n))
  x[idx]
}

W1_1d <- function(x, y, grid_size = 1000) {
  u <- seq(1/(grid_size + 1), grid_size/(grid_size + 1),
           length.out = grid_size)
  qx <- emp_q_left(x, u)
  qy <- emp_q_left(y, u)
  mean(abs(qx - qy))
}

# Bootstrap distribution of W1 distance to the empirical law
B <- 300
w_boot <- numeric(B)

for (b in 1:B) {
  loss_b <- sample(loss, size = n, replace = TRUE)
  w_boot[b] <- W1_1d(loss_b, loss)
}

r_stat <- as.numeric(quantile(w_boot, 0.95))

baseline_mean <- mean(loss)
robust_mean <- baseline_mean + r_stat

c(
  baseline_mean = baseline_mean,
  radius_95 = r_stat,
  robust_mean_W1 = robust_mean
)
\end{lstlisting}

For a tail functional such as $\TVaR_\alpha$, one can combine the empirical TVaR with the stability bound:
\[
  \TVaR_\alpha^{\mathrm{robust}}
  \le
  \widehat{\TVaR}_\alpha
  +
  \frac{r}{1-\alpha}.
\]

\begin{lstlisting}[language=R]
TVaR_emp <- function(x, alpha = 0.99) {
  q <- sort(x)
  n <- length(q)
  k <- pmax(1, ceiling(alpha * n))
  mean(q[k:n])
}

alpha <- 0.99
baseline_tvar <- TVaR_emp(loss, alpha)
robust_tvar_bound <- baseline_tvar + r_stat / (1 - alpha)

c(
  baseline_TVaR = baseline_tvar,
  robust_TVaR_bound = robust_tvar_bound
)
\end{lstlisting}

\subsubsection{Interpretation and limitations}

Wasserstein DRO is attractive because it is geometrically meaningful. The adversarial distribution cannot be arbitrary: it must be reachable from the baseline distribution at a controlled transportation cost. This makes the method more interpretable than worst-case analysis over very broad model classes.

However, the approach is not automatic. The result depends on:
\begin{itemize}
\item the ground cost $c(\bx,\by)$;
\item the radius $r$;
\item the state space on which the ambiguity is defined;
\item the growth of the actuarial payoff $h$ or risk functional $\mathcal R$;
\item the distinction between perturbing covariates, individual losses, aggregate losses, or cash-flow vectors.
\end{itemize}

A good DRO analysis should therefore report a sensitivity curve:
\[
  r\longmapsto
  \widehat\Pi_r^{\mathrm{DRO}},
  \qquad
  r\longmapsto
  \mathsf{Cap}_r.
\]
The curve is often more informative than a single robust value. It shows how quickly the actuarial conclusion deteriorates as one moves away from the baseline model.

\begin{actuarialbox}
The radius $r$ is not a technical nuisance. It is the actuarial statement of how much model uncertainty, portfolio drift or stress displacement the insurer is willing to consider. In that sense, Wasserstein DRO is as much a governance tool as a mathematical one.
\end{actuarialbox}

\begin{actuarialbox}
\textbf{Professional lens: Solvency II and Wasserstein capital.}

Under Solvency II, capital is framed as a one-year risk measure on the economic balance sheet. The Solvency Capital Requirement is calibrated as a high-quantile loss of own funds over a one-year horizon, and the regulatory architecture separates risk modules such as life underwriting, non-life underwriting, health underwriting, market, credit and operational risk.

This makes Wasserstein robustness a natural actuarial diagnostic. A baseline internal model provides a distribution of one-year own-fund changes,
\[
  \Delta OF = OF_1-OF_0.
\]
A Wasserstein ambiguity set around its fitted law,
\[
  \mathcal B_p(\widehat\mu,r)
  =
  \{\nu:W_p(\nu,\widehat\mu)\le r\},
\]
does not replace the regulatory calculation. Rather, it asks a model-risk question: how much would the capital requirement change if the one-year loss distribution were perturbed within a controlled transportation budget?

For example, a robust capital diagnostic may report
\[
  SCR_r
  =
  \sup_{\nu\in\mathcal B_p(\widehat\mu,r)}
  \VaR_{99.5\%}^{\nu}(-\Delta OF),
\]
or a tail-average analogue based on $\TVaR$. The radius $r$ can be interpreted as a model uncertainty, portfolio drift, parameter uncertainty or stress-displacement budget.
\end{actuarialbox}

\begin{readingbox}
The dual formula in Proposition~\ref{prop:dro-expectation-dual-formula} is the standard Wasserstein-DRO duality; see \cite{esfahani2018data,blanchet2019quantifying,gao2016distributionally}. Dependence uncertainty and comonotonic bounds are connected to the Fr\'echet-class and copula literature; see \cite{dhaene2002concept,embrechts2002correlation,puccetti2013computation}.
\end{readingbox}

\subsection{Summary}\label{subsec:risk-robustness-summary}

This lecture introduced Wasserstein robustness for actuarial risk measures. The main messages are:
\begin{enumerate}[label=(\roman*)]
  \item Quantile-based risk measures are naturally expressed in Wasserstein geometry.
  \item Expectations, TVaR and spectral risk measures are Lipschitz with respect to $\Wp{p}$, with constants determined by tail weights.
  \item VaR is less stable because it reads a single quantile level.
  \item Wasserstein ambiguity sets produce robust risk measures with interpretable loadings.
  \item Dependence uncertainty is a coupling problem and leads to multi-marginal optimal transport.
  \item Actuarial interpretation depends critically on the ground cost and the radius $\rho$.
\end{enumerate}

\subsection{Exercises}\label{subsec:risk-robustness-exercises}

\begin{exercise}[Robust mean under $W_1$]\label{ex:8:1}
Let $\widehat\mu$ be a probability distribution on $\R$ with finite first moment. Show that
\[
  \sup_{\nu:W_1(\nu,\widehat\mu)\le r}
  \int x\,\dd\nu(x)
  =
  \int x\,\dd\widehat\mu(x)+r.
\]
What changes if the payoff is $h(x)=ax+b$?
\end{exercise}

\begin{exercise}[Sensitivity of robust TVaR]\label{ex:8:2}
Using the inequality
\[
  |\TVaR_\alpha(\mu)-\TVaR_\alpha(\nu)|
  \le
  \frac{1}{1-\alpha}W_1(\mu,\nu),
\]
derive a robust upper bound for $\TVaR_\alpha$ over a $W_1$-ball. Comment on the behaviour as $\alpha\to 1$.
\end{exercise}

\begin{exercise}[Radius by bootstrap]\label{ex:8:3}
Simulate $n=500$ losses from a lognormal distribution. Use the bootstrap procedure in the mini-lab to estimate a 95\% radius for $W_1$. Repeat for $n=1000$ and $n=5000$. How does the radius change with sample size?
\end{exercise}

\begin{exercise}[Pricing versus capital ambiguity]\label{ex:8:4}
Let $\bX$ be a covariate vector and $L$ a loss. Explain why a Wasserstein ball for $\Law(\bX)$ and a Wasserstein ball for $\Law(L)$ answer different actuarial questions. Give an example where portfolio drift changes the robust premium but not the conditional severity model.
\end{exercise}

\begin{exercise}[Ground cost]\label{ex:8:5}
Suppose $\bX=(A,V)$, where $A$ is driver age and $V$ is vehicle value. Propose two different ground costs $c(\bx,\by)$: one that treats both coordinates symmetrically, and one that gives more weight to changes in $V$. Explain how the robust premium may differ under the two choices.
\end{exercise}

\begin{exercise}[Expectation and $\Wone$]\label{ex:8:6}
Let $\mu,\nu\in\Probp{1}(\R)$. Prove directly from the primal coupling formulation that
\[
  \left|\int x\dd\mu(x)-\int x\dd\nu(x)\right|
  \le \Wone(\mu,\nu).
\]
Compare your proof with the proof based on quantile functions.
\end{exercise}

\begin{exercise}[TVaR Lipschitz constant]\label{ex:8:7}
Let $p\ge1$. Show that
\[
  |\TVaR_\alpha(\mu)-\TVaR_\alpha(\nu)|
  \le (1-\alpha)^{-1/p}\Wp{p}(\mu,\nu).
\]
Then compute the constant for $\alpha=0.95,0.99,0.995$ and $p=1,2$. Interpret the result actuarially.
\end{exercise}

\begin{exercise}[Robust TVaR]\label{ex:8:8}
Let $Q$ be a nondecreasing quantile function and define
\[
  \widetilde Q(u)=Q(u)+a\1_{\{u\ge\alpha\}}.
\]
Find $a$ such that $\|\widetilde Q-Q\|_{L^p}=\rho$. Deduce the formula
\[
  \sup_{\nu:\Wp{p}(\nu,\mu)\le\rho}\TVaR_\alpha(\nu)
  =\TVaR_\alpha(\mu)+\rho(1-\alpha)^{-1/p}.
\]
What changes if losses are capped at a policy limit $M$?
\end{exercise}

\begin{exercise}[VaR sensitivity]\label{ex:8:9}
Let $L=0$ almost surely and let $L_M=M$ with probability $\eps$, and $0$ otherwise. Compute $\Wp{p}(\Law(L),\Law(L_M))$, $\VaR_\alpha(L_M)$, and $\TVaR_\alpha(L_M)$. Discuss the regimes $\eps<1-\alpha$ and $\eps>1-\alpha$.
\end{exercise}

\begin{exercise}[Stop-loss stability]\label{ex:8:10}
For fixed retention $r\ge0$, prove that
\[
  \left|
  \E_\mu[(L-r)_+]-\E_\nu[(L-r)_+]
  \right|
  \le \Wone(\mu,\nu).
\]
Extend the argument to an insurance layer
\[
  \ell(L)=\min\{(L-a)_+,b-a\}.
\]
\end{exercise}

\begin{exercise}[Dependence-robust stop-loss]\label{ex:8:11}
Let $L_1,L_2$ have marginal quantile functions $Q_1,Q_2$. Show that the comonotonic coupling gives aggregate quantile
\[
  Q_{L_1+L_2}(u)=Q_1(u)+Q_2(u)
\]
when both losses are increasing functions of the same uniform random variable. Use this to compute the comonotonic TVaR of $L_1+L_2$. Why does this provide a natural adverse dependence benchmark?
\end{exercise}

\begin{exercise}[Computational exercise]\label{ex:8:12}
Simulate $n=10{,}000$ losses from a lognormal distribution. Compute empirical VaR and TVaR at levels $0.95$, $0.99$, and $0.995$. For $p=1$ and $p=2$, add the Wasserstein robust TVaR loading for several values of $\rho$. Plot the robust TVaR as a function of $\rho$. Comment on the difference between $p=1$ and $p=2$.
\end{exercise}

\newpage
\lecture{Pricing and Premium Distributions}
\label{lec:pricing-mutualization}

\subsection*{Learning objectives}

The preceding lecture studied risk measures as functionals of loss distributions. This lecture applies the same distributional viewpoint to insurance pricing. A tariff is a measurable function on a portfolio, and therefore induces a probability distribution of premiums. The objective is to compare premium books, separate policy-level changes from distributional changes, and relate classical pricing dispersion measures to Wasserstein geometry.

After this lecture, the reader should be able to:
\begin{enumerate}[label=(\roman*)]
  \item distinguish individual risk prediction from the distribution of premiums induced by a tariff;
  \item represent tariffs as measurable functions and premium books as push-forward measures;
  \item formalize pooled, class-based and individualized pure premiums through sigma-algebras;
  \item interpret variance and Wasserstein distances as complementary measures of premium dispersion;
  \item prove that $\Wtwo^2(\Law(\Premium),\delta_{\E[\Premium]})=\Var(\Premium)$;
  \item compare tariff revisions using both policy-level displacement and distributional transport distance;
  \item understand shrinkage paths between a pooled benchmark and a technical premium;
  \item use one-dimensional quantile formulas to compute distances between empirical premium books;
  \item study affordability through the distribution of premium burdens.
\end{enumerate}

\begin{actuarialbox}
A tariff is a random variable on a portfolio. Once this is accepted, pricing has a distributional component: not only ``what premium should each policyholder pay?'', but also ``what premium distribution does the tariff create?''
\end{actuarialbox}

This lecture uses the push-forward notation of Definition~\ref{def:push-forward}, the coupling notation of Definition~\ref{def:coupling}, the discrete OT formulation of Definition~\ref{def:discrete-ot}, and the Wasserstein distance of Definition~\ref{def:wasserstein-distance}. It prepares the portfolio-drift material in Lecture~\ref{lec:portfolio-drift-domain-adaptation} and the empirical tariff-comparison computations in Section~\ref{subsec:pricing-mini-lab-r}.

\subsection{Why pricing is a transport problem}\label{subsec:pricing-why-transport}

Let $L\ge 0$ denote the future claim cost of a policy over a fixed coverage period, and let $\bX$ denote the information used for pricing. A standard pure-premium model estimates
\[
  m(x)=\E[L\mid \bX=x].
\]
A premium may then be written as
\[
  \Premium(x)=\ell(m(x),x),
\]
where $\ell$ includes expenses, profit loading, reinsurance cost, taxes, caps, floors, credibility adjustments, regulatory constraints, or commercial smoothing.

If $\bX$ is random under the portfolio distribution, then the charged premium
\[
  \Pi=\Premium(\bX)
\]
is a random variable. Its law is
\[
  \mu_\Premium=\Law(\Pi)=\Law(\Premium(\bX)).
\]
This is the premium distribution induced by the tariff. More generally, if the tariff uses multiple outputs, such as predicted frequency, predicted severity and a risk loading, one may consider the joint law
\[
  \Law\bigl(\widehat\lambda(\bX),\widehat s(\bX),\Premium(\bX)\bigr),
\]
or even the law of vectors of premiums across coverages.

This is the key viewpoint of the lecture: pricing is not only an estimation problem, but also a transformation problem. The portfolio distribution $\Law(\bX)$ is pushed forward by the premium function $\Premium$ into a distribution on $\R_+$:
\[
  \mu_\Premium = \Premium_{\push}\Law(\bX).
\]
When a tariff changes from $\Premium_0$ to $\Premium_1$, the corresponding premium distribution changes from
\[
  \mu_0=\Law(\Premium_0(\bX))
  \qquad\text{to}\qquad
  \mu_1=\Law(\Premium_1(\bX)).
\]
Optimal transport compares these two distributions.

\begin{example}[A tariff revision]\label{ex:pricing-tariff-revision}
An insurer replaces a generalized linear model by a gradient-boosted model. On the same portfolio, the old premium is $\Premium_0(\bX_i)$, and the new premium is $\Premium_1(\bX_i)$. The average change
\[
  \frac1n\sum_{i=1}^n |\Premium_1(\bX_i)-\Premium_0(\bX_i)|
\]
measures policy-level disruption under the identity matching of policies. The Wasserstein distance
\[
  \Wone\bigl(\widehat\mu_{\Premium_0},\widehat\mu_{\Premium_1}\bigr)
\]
measures the smallest possible average distributional displacement between the old and new premium books. The first is an implementation quantity; the second is a distributional lower bound.
\end{example}

\iffalse
% To be restored in v2, after the portfolio-demutualization paper has been submitted.

\begin{example}[Segmentation as a change of distribution]\label{ex:pricing-segmentation-change-distribution}
Suppose all policyholders used to pay the same pooled premium $\bar\pi$, while a new model charges $\Premium(\bX)$. The pooled premium distribution is $\delta_{\bar\pi}$, whereas the individualized tariff has distribution $\Law(\Premium(\bX))$. The movement from $\delta_{\bar\pi}$ to $\Law(\Premium(\bX))$ is a distributional measure of distance-to-pooling.
\end{example}
\fi

\subsection{Pure premium, loaded premium and tariff random variables}\label{subsec:pricing-pure-loaded-tariff}

A basic actuarial pricing model may decompose total claim cost across coverages or perils as
\[
  L = L^{(1)}+\cdots+L^{(K)},
\]
and, for a single coverage, more commonly as a frequency--severity process
\[
  L=\sum_{j=1}^{N} Z_j.
\]
Given covariates $\bX=x$, the conditional pure premium is
\[
  m(x)=\E[L\mid \bX=x]
  =\E[N\mid \bX=x]\E[Z\mid \bX=x]
\]
when, conditionally on $\bX=x$, claim counts and individual severities are independent and severities are identically distributed with finite conditional mean.

\begin{definition}[Pure premium and charged premium]\label{def:pure-and-charged-premium}
Let \(L\in L^1\) be the future loss over the coverage period and let \(\mathcal G\subseteq\mathcal F\) be the information used for pricing. The \(\mathcal G\)-pure premium is
\[
  m_{\mathcal G}=\E[L\mid\mathcal G].
\]
A charged premium is a non-negative random variable of the form
\[
  \Pi=\Premium(\bX),
\]
where the tariff map \(\Premium\) may add loadings, expenses, capital charges, caps, floors or smoothing to the pure-premium estimate.
\end{definition}

A loaded premium may take the form
\[
  \Premium(x)=(1+\eta)m(x)+e(x),
\]
where $\eta$ is a proportional loading and $e(x)$ is an expense or fixed charge. More generally, a tariff can include a risk margin:
\[
  \Premium(x)=\E[L\mid \bX=x]+\theta\,r(x),
\]
where $r(x)$ may be a conditional standard deviation, a conditional TVaR, a capital allocation, or a reinsurance-cost allocation.

In all cases, once $\bX$ is random under the portfolio, $\Premium(\bX)$ is random. Its distribution may have policy, regulatory and social meaning. For example:
\begin{itemize}
  \item a concentrated distribution indicates limited premium differentiation;
  \item a dispersed distribution indicates stronger segmentation of the premium book;
  \item a heavy upper tail indicates that a small set of policyholders receives very high premiums;
  \item a tariff revision can be viewed as a transport from the old premium distribution to the new one;
  \item the distribution of $\Premium(\bX)/Y$, where $Y$ is income or wealth, is an affordability distribution.
\end{itemize}

\subsection{Push-forward formulation of a tariff}\label{subsec:pricing-pushforward-tariff}

Let $(\Omega,\mathcal F,\Pbb)$ be the probability space describing a randomly selected policy from the portfolio. A pricing model is a measurable map
\[
  \Premium:\Omega\to\R_+.
\]
Usually, $\Premium$ is written as $\Premium(\bX)$, where $\bX:\Omega\to\mathcal X$ is the vector of rating variables.

The induced premium distribution is
\[
  \mu_\Premium = \Premium_{\push}\Pbb.
\]
Equivalently, for every Borel set $A\subseteq\R_+$,
\[
  \mu_\Premium(A)=\Pbb(\Premium\in A).
\]
Thus all distributional questions about the premium book can be written as questions about a push-forward measure.

\begin{definition}[Premium book]\label{def:premium-book}
Given a portfolio probability measure $P_{\bX}=\Law(\bX)$ and a tariff $\Premium:\mathcal X\to\R_+$, the premium book induced by $\Premium$ is the probability measure
\[
  \mu_\Premium=\Premium_{\push}P_{\bX}.
\]
If $\bX_1,\ldots,\bX_n$ is a finite portfolio, the empirical premium book is
\[
  \widehat\mu_\Premium
  =\frac1n\sum_{i=1}^n \delta_{\Premium(\bX_i)}.
\]
\end{definition}

\begin{remark}[Portfolio weights]
In practice, policies may have exposure weights $w_i\ge0$, with $\sum_i w_i=1$. Then the empirical premium book is
\[
  \widehat\mu_\Premium
  =\sum_{i=1}^n w_i\delta_{\Premium(\bX_i)}.
\]
All Wasserstein definitions apply without change. The weights may represent exposure units, earned premium, policy duration, insured value, or sampling weights.
\end{remark}

\begin{definition}[Policy-level and distributional tariff comparison]\label{def:policy-level-distributional-tariff-comparison}
Let \(\Pi_0=\Premium_0(\bX)\) and \(\Pi_1=\Premium_1(\bX)\) be two premium random variables on the same portfolio probability space. For \(p\ge1\), the policy-level displacement is
\[
  \Delta_p^{\mathrm{id}}(\Pi_0,\Pi_1)
  =\left(\E|\Pi_1-\Pi_0|^p\right)^{1/p},
\]
where the same policyholder is compared before and after the tariff change. The distributional displacement is
\[
  \Delta_p^{\mathrm{dist}}(\Pi_0,\Pi_1)
  =\Wp{p}\bigl(\Law(\Pi_0),\Law(\Pi_1)\bigr),
\]
where policy identities are ignored and only the two marginal premium books are compared. Proposition~\ref{prop:distributional-lower-bound-pricing} shows that \(\Delta_p^{\mathrm{dist}}\le \Delta_p^{\mathrm{id}}\).
\end{definition}

\subsection{Pooled, class-based and individualized pure premiums}\label{subsec:pricing-pooling-class-profiling}

Pricing uses information at different levels of granularity. At one extreme, all policyholders pay the same pure premium. At the other, the price is individualized using all admissible rating variables. Between the two lie class tariffs, credibility tariffs, regulated tariffs and smoothed commercial tariffs.

A useful formalization uses sigma-algebras. Let $L$ be the loss and let $\mathcal G\subseteq\mathcal F$ denote the information used for pricing. The $\mathcal G$-based pure premium is
\[
  m_{\mathcal G}=\E[L\mid \mathcal G].
\]
If $\mathcal G$ is trivial, then
\[
  m_{\mathcal G}=\E[L],
\]
and the tariff is constant. If $\mathcal G=\sigma(\bX)$, then
\[
  m_{\mathcal G}=\E[L\mid \bX],
\]
and the tariff uses all rating variables in $\bX$. If $\mathcal G$ is generated by a coarse classification, the tariff is a class tariff.

\begin{definition}[Pricing information]\label{def:pricing-information}
Let $\mathcal G\subseteq\mathcal F$ be a sigma-algebra. The corresponding pure-premium tariff is
\[
  \Premium_{\mathcal G}=\E[L\mid\mathcal G].
\]
The induced premium distribution is
\[
  \mu_{\mathcal G}=\Law(\Premium_{\mathcal G}).
\]
\end{definition}

If $\mathcal G_1\subseteq\mathcal G_2$, then $\mathcal G_2$ is more informative. The tariff $\Premium_{\mathcal G_2}$ may be more segmented than $\Premium_{\mathcal G_1}$. The classical variance decomposition gives
\[
  \Var(\E[L\mid\mathcal G_2])
  =\Var(\E[L\mid\mathcal G_1])
   +\E\left[\Var\bigl(\E[L\mid\mathcal G_2]\mid\mathcal G_1\bigr)\right]
\]
when $\mathcal G_1\subseteq\mathcal G_2$. Thus increasing the pricing information cannot decrease the variance of the pure premium.

\begin{proposition}[Segmentation increases premium variance]\label{prop:segmentation-increases-variance}
Let $\mathcal G_1\subseteq\mathcal G_2\subseteq\mathcal F$, and assume $L\in L^2$. Then
\[
  \Var(\E[L\mid\mathcal G_1])
  \le
  \Var(\E[L\mid\mathcal G_2]).
\]
\end{proposition}

\begin{proof}
By the tower property,
\[
  \E[L\mid\mathcal G_1]
  =\E[\E[L\mid\mathcal G_2]\mid\mathcal G_1].
\]
Applying the law of total variance to $Z=\E[L\mid\mathcal G_2]$ conditional on $\mathcal G_1$,
\[
  \Var(Z)
  =\Var(\E[Z\mid\mathcal G_1])+
    \E[\Var(Z\mid\mathcal G_1)].
\]
Since the second term is nonnegative, the result follows.
\end{proof}

\begin{corollary}[Nested information and distance to pooling]\label{cor:nested-information-distance-pooling}
Under the assumptions of Proposition~\ref{prop:segmentation-increases-variance},
\[
  \Wtwo^2\bigl(\Law(\Premium_{\mathcal G_1}),\delta_{\E[L]}\bigr)
  \le
  \Wtwo^2\bigl(\Law(\Premium_{\mathcal G_2}),\delta_{\E[L]}\bigr).
\]
Thus nested refinement of the pricing information cannot decrease the squared \(\Wtwo\)-distance from the constant pure-premium benchmark.
\end{corollary}

\begin{actuarialbox}
More information does not necessarily mean better pricing in every social or regulatory sense, but it does mechanically increase the dispersion of the conditional pure premium, unless the additional information is irrelevant for the conditional pure premium.
\end{actuarialbox}

\subsection{Wasserstein distance to a pooled benchmark}\label{subsec:pricing-wasserstein-to-pooling}

Let $\Premium\in L^2$ be a premium random variable. A natural pooled benchmark is the constant premium
\[
  \bar\pi=\E[\Premium].
\]
The corresponding premium distribution is $\delta_{\bar\pi}$. The squared $2$-Wasserstein distance to this pooled benchmark is exactly the variance.

\begin{proposition}[Variance as Wasserstein distance to pooling]\label{prop:variance-wasserstein-pooling}
Let $\Premium\in L^2$, with law $\mu=\Law(\Premium)$ and mean $\bar\pi=\E[\Premium]$. Then
\[
  \Wtwo^2(\mu,\delta_{\bar\pi})=\Var(\Premium).
\]
More generally, for $p\ge1$,
\[
  \Wp{p}^p(\mu,\delta_a)=\E[|\Premium-a|^p].
\]
\end{proposition}

\begin{proof}
There is only one coupling between $\mu$ and $\delta_a$: it is the law of $(\Premium,a)$. Therefore
\[
  \Wp{p}^p(\mu,\delta_a)
  =\E[|\Premium-a|^p].
\]
Taking $p=2$ and $a=\E[\Premium]$ gives the variance.
\end{proof}

Proposition~\ref{prop:variance-wasserstein-pooling} is useful because it connects a classical actuarial measure of segmentation to Wasserstein geometry. It also clarifies what the variance misses. Variance measures distance to a single pooled premium. It does not compare two non-degenerate tariff distributions. Wasserstein distances do.

\begin{definition}[Distance-to-pooling index]\label{def:wasserstein-demutualization-index}
Let $\Premium\in L^2$ be a tariff and let $\bar\pi=\E[\Premium]$. Define
\[
  D_2(\Premium)
  =\Wtwo^2\bigl(\Law(\Premium),\delta_{\bar\pi}\bigr)
  =\Var(\Premium).
\]
More generally, for $p\ge1$, define
\[
  D_p(\Premium)
  =\Wp{p}^p\bigl(\Law(\Premium),\delta_{\bar\pi}\bigr)
  =\E[|\Premium-\bar\pi|^p].
\]
\end{definition}

In Definition~\ref{def:wasserstein-demutualization-index}, the case $p=1$ gives a mean absolute deviation from the constant benchmark. The case $p=2$ gives variance. Larger $p$ emphasizes extreme premium deviations.

\subsection{Shrinkage paths between a benchmark and a technical tariff}\label{subsec:pricing-shrinkage-paths}

Let $\Premium$ be an individualized tariff with mean $\bar\pi$. A simple path from full pooling to full profiling is
\[
  \Premium_\lambda
  =\bar\pi+\lambda(\Premium-\bar\pi),
  \qquad \lambda\in[0,1].
\]
At $\lambda=0$, all policyholders pay $\bar\pi$. At $\lambda=1$, the technical tariff is used. Intermediate values shrink premiums toward the constant benchmark.

\begin{proposition}[Scaling along a shrinkage path]\label{prop:shrinkage-demutualization-scaling}
Let $\Premium\in L^2$, $\bar\pi=\E[\Premium]$, and
\[
  \Premium_\lambda=\bar\pi+\lambda(\Premium-\bar\pi).
\]
Then
\[
  D_2(\Premium_\lambda)
  =\lambda^2D_2(\Premium).
\]
More generally,
\[
  D_p(\Premium_\lambda)
  =\lambda^pD_p(\Premium).
\]
\end{proposition}

\begin{proof}
Since $\E[\Premium_\lambda]=\bar\pi$,
\[
  D_p(\Premium_\lambda)
  =\E[|\Premium_\lambda-\bar\pi|^p]
  =\E[|\lambda(\Premium-\bar\pi)|^p]
  =\lambda^pD_p(\Premium).
\]
\end{proof}

This path is also the one-dimensional displacement interpolation from $\delta_{\bar\pi}$ to $\Law(\Premium)$, as in Definition~\ref{def:displacement-interpolation}, when the map $x\mapsto \bar\pi+\lambda(x-\bar\pi)$ is applied to the premium distribution. In one dimension, the quantile function of $\Premium_\lambda$ is
\[
  Q_\lambda(u)=\bar\pi+\lambda(Q(u)-\bar\pi),
\]
where $Q$ is the quantile function of $\Premium$. Hence
\[
  \Wp{p}^p\bigl(\Law(\Premium_\lambda),\Law(\Premium_{\lambda'})\bigr)
  =|\lambda-\lambda'|^p\E[|\Premium-\bar\pi|^p].
\]

\begin{example}[Commercial smoothing]\label{ex:pricing-commercial-smoothing}
A company may decide not to deploy the full technical premium $\Premium$, but instead apply
\[
  \Premium_\lambda=(1-\lambda)\bar\pi+\lambda\Premium.
\]
For $p=2$, the distance-to-pooling index is multiplied by $\lambda^2$. Thus $\lambda=0.5$ uses half the technical displacement but only one quarter of the variance from the constant benchmark.
\end{example}

\subsection{Class tariffs and Wasserstein comparison}\label{subsec:pricing-class-tariffs}

Suppose the portfolio is partitioned into $K$ rating classes $C_1,\ldots,C_K$. A class tariff charges a constant premium $\pi_k$ to all policies in class $k$. The premium random variable is
\[
  \Premium_C=\sum_{k=1}^K \pi_k\1_{\{\bX\in C_k\}}.
\]
Its distribution is discrete:
\[
  \Law(\Premium_C)=\sum_{k=1}^K p_k\delta_{\pi_k},
  \qquad p_k=\Pbb(\bX\in C_k).
\]
If a more refined tariff uses subclasses $C_{k\ell}$, the new distribution has more support points. Wasserstein distances can compare the coarse and refined tariffs.

Using the terminology of Definition~\ref{def:policy-level-distributional-tariff-comparison}, there are two distinct comparisons.

First, the policy-level distance uses the actual same policy under the two tariffs:
\[
  \Delta_p^{\mathrm{id}}(\Premium_C,\Premium_R)
  =\left(\E|\Premium_C(\bX)-\Premium_R(\bX)|^p\right)^{1/p}.
\]
Second, the distributional distance ignores the identity of policies and compares only the two premium distributions:
\[
  \Delta_p^{\mathrm{dist}}(\Premium_C,\Premium_R)
  =\Wp{p}\left(\Law(\Premium_C),\Law(\Premium_R)\right).
\]
The second is always no larger than the first.

\begin{proposition}[Distributional lower bound]\label{prop:distributional-lower-bound-pricing}
Let $\Premium_0,\Premium_1\in L^p$. Then
\[
  \Wp{p}\bigl(\Law(\Premium_0),\Law(\Premium_1)\bigr)
  \le
  \left(\E|\Premium_1-\Premium_0|^p\right)^{1/p}.
\]
\end{proposition}

\begin{proof}
The joint law of $(\Premium_0,\Premium_1)$ is a coupling of $\Law(\Premium_0)$ and $\Law(\Premium_1)$. The Wasserstein distance is the infimum over all couplings, hence no larger than the cost under this particular coupling.
\end{proof}

\begin{actuarialbox}
Policy-level premium change is what customers experience. Wasserstein distance is the smallest distributional change compatible with the two premium books. The gap between them is informative: it measures how much of the disruption comes from policy identities rather than from the marginal premium distributions alone.
\end{actuarialbox}

\subsection{A two-class example}\label{subsec:pricing-two-class-example}

Consider a portfolio with two equally weighted groups. The pooled premium is $\bar\pi$. A class tariff charges
\[
  \bar\pi-d\quad\text{to group 1},
  \qquad
  \bar\pi+d\quad\text{to group 2}.
\]
The premium distribution is
\[
  \mu_d=\frac12\delta_{\bar\pi-d}+\frac12\delta_{\bar\pi+d}.
\]
Then
\[
  \Wtwo^2(\mu_d,\delta_{\bar\pi})=d^2.
\]
If a refined model further splits each group symmetrically by an amount $e$, the premium distribution becomes
\[
  \nu_{d,e}=\frac14\delta_{\bar\pi-d-e}
  +\frac14\delta_{\bar\pi-d+e}
  +\frac14\delta_{\bar\pi+d-e}
  +\frac14\delta_{\bar\pi+d+e}.
\]
Assuming $0\le e\le d$, the monotone coupling gives
\[
  \Wtwo^2(\mu_d,\nu_{d,e})=e^2.
\]
Thus the first split creates segmentation $d^2$ from pooling, while the refinement creates an additional distributional movement $e^2$.

This example is deliberately simple, but it illustrates how OT separates levels of pricing granularity. Each new set of rating variables changes the premium distribution and can increase its distance from a constant benchmark.

\subsection{Segmentation, prediction and the variance decomposition}\label{subsec:pricing-segmentation-variance}

Let $m(\bX)=\E[L\mid \bX]$ be the full technical pure premium. Let $\mathcal G\subseteq\sigma(\bX)$ be the information used by a coarser tariff. The corresponding pure premium is
\[
  m_{\mathcal G}=\E[L\mid\mathcal G]
  =\E[m(\bX)\mid\mathcal G].
\]
The total predictive heterogeneity of the full model is
\[
  \Var(m(\bX)).
\]
The heterogeneity captured by the coarser tariff is
\[
  \Var(m_{\mathcal G}).
\]
The remaining within-class heterogeneity is
\[
  \E\left[\Var(m(\bX)\mid\mathcal G)\right].
\]
The variance decomposition gives
\[
  \Var(m(\bX))
  =\Var(m_{\mathcal G})+
   \E\left[\Var(m(\bX)\mid\mathcal G)\right].
\]
This identity has a direct pricing interpretation:
\begin{itemize}
  \item $\Var(m(\bX))$ is the dispersion of fully individualized pure premiums;
  \item $\Var(m_{\mathcal G})$ is the dispersion actually introduced by the class tariff;
  \item the residual term is predictive heterogeneity left pooled within classes.
\end{itemize}

\begin{definition}[Captured profiling ratio]\label{def:captured-profiling-ratio}
Assume $m(\bX)\in L^2$ and $\Var(m(\bX))>0$. The captured profiling ratio of $\mathcal G$ is
\[
  R_{\mathrm{prof}}(\mathcal G)
  =\frac{\Var(\E[L\mid\mathcal G])}{\Var(\E[L\mid \bX])}.
\]
It lies in $[0,1]$ when $\mathcal G\subseteq\sigma(\bX)$.
\end{definition}

This ratio is not a Wasserstein distance by itself, but by the previous proposition it is the ratio of squared Wasserstein distances to the pooled mean:
\[
  R_{\mathrm{prof}}(\mathcal G)
  =
  \frac{
  \Wtwo^2(\Law(\E[L\mid\mathcal G]),\delta_{\E[L]})
  }{
  \Wtwo^2(\Law(\E[L\mid \bX]),\delta_{\E[L]})
  }.
\]

\subsection{From technical premium to market premium}\label{subsec:pricing-technical-market-premium}

The technical premium $m(\bX)$ is not always the charged premium. Insurers may apply transformations:
\[
  \Premium(\bX)=T(m(\bX),\bX),
\]
where $T$ may include:
\begin{itemize}
  \item minimum premiums;
  \item maximum premiums or caps;
  \item bonus--malus smoothing;
  \item credibility shrinkage;
  \item territorial smoothing;
  \item regulatory constraints;
  \item commercial discounts;
  \item expense and commission loadings.
\end{itemize}

Each transformation changes the premium distribution. If $T$ is a deterministic map on premiums, for example $\Premium=T(m)$, then
\[
  \Law(\Premium)=T_{\push}\Law(m).
\]
A cap $T(z)=\min(z,M)$ moves all mass above $M$ to $M$. A floor $T(z)=\max(z,m_0)$ moves all mass below $m_0$ to $m_0$. A shrinkage $T(z)=\bar\pi+\lambda(z-\bar\pi)$ contracts the distribution around the mean.

\begin{example}[Premium cap]
Let $\Premium$ be a technical premium and define the capped premium
\[
  \Premium^M=\min(\Premium,M).
\]
Then
\[
  \E[|\Premium-\Premium^M|^p]
  =\E[(\Premium-M)_+^p].
\]
Therefore
\[
  \Wp{p}\bigl(\Law(\Premium),\Law(\Premium^M)\bigr)
  \le
  \left(\E[(\Premium-M)_+^p]\right)^{1/p}.
\]
In one dimension, if the cap is applied monotonically to the same premium variable, the right-hand side is the cost of the induced monotone coupling and is often the natural implementation cost.
\end{example}

\subsection{Tariff revisions as transport plans}\label{subsec:pricing-tariff-revisions-plans}

Suppose an insurer moves from $\Premium_0$ to $\Premium_1$. The joint law
\[
  \gamma_{\mathrm{id}}=
  \Law(\Premium_0(\bX),\Premium_1(\bX))
\]
is the identity-policy coupling between the old and new premium distributions. Its cost is
\[
  C_p^{\mathrm{id}}
  =\E|\Premium_1(\bX)-\Premium_0(\bX)|^p.
\]
The optimal transport cost is
\[
  C_p^{\mathrm{OT}}
  =\Wp{p}^p(\Law(\Premium_0),\Law(\Premium_1)).
\]
Since $C_p^{\mathrm{OT}}\le C_p^{\mathrm{id}}$, the difference
\[
  C_p^{\mathrm{id}}-C_p^{\mathrm{OT}}
\]
measures how much extra movement is caused by the identity of policyholders rather than by the marginal distributional change.

\begin{definition}[Reallocation gap]\label{def:reallocation-gap}
For $\Premium_0,\Premium_1\in L^p$, define
\[
  G_p(\Premium_0,\Premium_1)
  =\E|\Premium_1-\Premium_0|^p
  -\Wp{p}^p(\Law(\Premium_0),\Law(\Premium_1)).
\]
The quantity $G_p\ge0$ is the reallocation gap between the policy-level implementation and the distributionally optimal matching.
\end{definition}

A small reallocation gap means that the actual policy-level changes are close to the optimal distributional rearrangement. A large gap means that two tariffs may have similar marginal premium distributions while charging very different policies.

\begin{example}[Same distribution, different people]
Two tariffs can have the same premium distribution but assign high premiums to different policyholders. Then
\[
  \Wp{p}(\Law(\Premium_0),\Law(\Premium_1))=0,
\]
but
\[
  \E|\Premium_1-\Premium_0|^p
\]
may be large. This is important in tariff governance: distributional stability alone is not enough to guarantee customer-level stability.
\end{example}

\subsection{One-dimensional formulas for premium books}\label{subsec:pricing-one-dimensional-formulas}

Because premiums are one-dimensional, the quantile formula is especially useful. Let $Q_0,Q_1$ be the quantile functions of $\Premium_0$ and $\Premium_1$. Then
\[
  \Wp{p}^p(\Law(\Premium_0),\Law(\Premium_1))
  =\int_0^1 |Q_1(u)-Q_0(u)|^p\dd u.
\]
This allows premium distributions to be compared without solving a large linear program. In a finite portfolio with equal weights, sort the premiums:
\[
  \Premium_{0,(1)}\le\cdots\le\Premium_{0,(n)},
  \qquad
  \Premium_{1,(1)}\le\cdots\le\Premium_{1,(n)}.
\]
Then
\[
  \Wp{p}^p(\widehat\mu_0,\widehat\mu_1)
  =\frac1n\sum_{i=1}^n
  |\Premium_{1,(i)}-\Premium_{0,(i)}|^p.
\]

\begin{computebox}
For one-dimensional premium distributions, Wasserstein distances are computationally cheap. Sorting is enough. This is one reason why OT is practical for tariff governance: a pricing team can compute premium-book movements for millions of policies using quantiles rather than a full transport matrix.
\end{computebox}

\subsection{Ground costs and actuarial meaning}\label{subsec:pricing-ground-costs}

The default cost for premiums is often $|x-y|^p$, but other costs may be more meaningful.

A relative premium change can be measured by
\[
  c(x,y)=\left|\log y-\log x\right|^p,
\]
when premiums are positive. This treats a movement from $100$ to $120$ similarly to a movement from $500$ to $600$, since both are roughly $20\%$ increases.

An affordability cost may compare premium-to-income ratios:
\[
  c\bigl((\pi,y),(\pi',y')\bigr)
  =\left|\frac{\pi'}{y'}-\frac{\pi}{y}\right|^p.
\]
A customer-disruption cost may be asymmetric:
\[
  c(x,y)=a(y-x)_+^p+b(x-y)_+^p,
\]
with $a>b$ if increases are viewed as more disruptive than decreases.

\begin{remark}[Metric versus loss function]
The function $c$ need not always be a metric for the Kantorovich problem to make sense. However, if one wants $\Wp{p}$ to be a genuine distance, one uses a metric cost $d(x,y)^p$. In pricing governance, non-metric costs can still be useful as decision losses.
\end{remark}

\iffalse
% To be restored in v2, after the fairness paper on multiple sensitive variables has been submitted.

\subsection{Segmentation and protected attributes}\label{subsec:pricing-protected-attributes}

Let $S$ denote a sensitive or protected attribute. A pricing model may or may not use $S$ explicitly. Even when $S$ is excluded, other variables may act as proxies. The premium distribution conditional on $S=s$ is
\[
  \mu_s=\Law(\Premium(\bX)\mid S=s).
\]
The collection $(\mu_s)_s$ describes group-level premium distributions. A natural group disparity measure is
\[
  \Wp{p}(\mu_s,\mu_{s'}),
\]
or the dispersion of group distributions around a Wasserstein barycenter. This links pricing segmentation to the fairness tools developed later in the notes.

However, the interpretation is delicate. Differences in $\mu_s$ may reflect differences in expected losses, differences in exposure, proxy effects, historical inequities, or modelling artifacts. OT provides a geometry for the audit; it does not by itself decide which differences are justified.

\begin{example}[Proxy segmentation]\label{ex:pricing-proxy-segmentation}
Suppose a protected attribute $S$ is not used, but territory and vehicle type are strongly associated with $S$. A model using territory and vehicle type can still produce different conditional premium distributions $\Law(\Premium\mid S=s)$. The Wasserstein distance between these group distributions is a distributional audit statistic.
\end{example}

\fi

\subsection{Affordability and welfare distributions}\label{subsec:pricing-affordability-welfare}

Premiums matter not only in absolute terms. A premium of $1000$ may be affordable for one household and unaffordable for another. Let $Y>0$ denote income or an affordability denominator. Define the burden ratio
\[
  A=\frac{\Premium}{Y}.
\]
The induced affordability distribution is
\[
  \Law(A)=\Law\left(\frac{\Premium}{Y}\right).
\]
A tariff revision changes $A_0=\Premium_0/Y$ into $A_1=\Premium_1/Y$. Wasserstein distances can compare the distributions $\Law(A_0)$ and $\Law(A_1)$, while risk measures such as $\TVaR_\alpha(A)$ focus on the most burdened policyholders.

\begin{definition}[Affordability transport]\label{def:affordability-transport}
Let $Y>0$ and let $\Premium_0,\Premium_1$ be two tariffs. The affordability displacement is
\[
  \Wp{p}\left(
    \Law\left(\frac{\Premium_0}{Y}\right),
    \Law\left(\frac{\Premium_1}{Y}\right)
  \right).
\]
The policy-level affordability displacement is
\[
  \left(\E\left|\frac{\Premium_1-\Premium_0}{Y}\right|^p\right)^{1/p}.
\]
\end{definition}

This distinction mirrors the earlier distinction between distributional and policy-level premium changes. A tariff may have limited distributional impact on affordability but still strongly affect particular households.

\subsection{Micro-insurance and affordability distributions}\label{subsec:microinsurance-affordability}

Micro-insurance raises a different actuarial question from standard technical pricing. The premium must be sufficient to finance expected claims and expenses, but it must also remain affordable for low-income households or small businesses. Thus the actuarial object is not only the loss distribution. It is also the joint distribution of premiums, income, wealth, liquidity and welfare.

Let \(P\) denote the premium charged to an insured and let \(Y\) denote a measure of ability to pay, such as income, disposable income or consumption capacity. A simple affordability ratio is
\[
  R=\frac{P}{Y}.
\]
An affordability constraint may require
\[
  \Pbb(R>\tau)\le \varepsilon,
\]
for a tolerance threshold \(\tau\) and a small exceedance probability \(\varepsilon\). Equivalently, one may study the distribution of the premium burden \(R\).

Optimal transport enters because pricing reform moves the distribution of premium burdens. Let \(R_0\) be the burden under a baseline tariff and \(R_1\) the burden under a new tariff. Then
\[
  \Wp{p}\bigl(\Law(R_0),\Law(R_1)\bigr)
\]
measures how far the burden distribution is displaced. This is different from comparing average premiums. Two tariffs can have the same average premium but very different affordability distributions.

\begin{actuarialbox}
For micro-insurance, the relevant distribution is often not the premium distribution alone, but the distribution of premium burden:
\[
  R=\frac{\text{premium}}{\text{ability to pay}}.
\]
Optimal transport provides a way to compare tariff designs by asking how they move the whole affordability distribution, including its upper tail.
\end{actuarialbox}

A useful benchmark is an affordability target distribution. Let \(\eta\) be a target law for acceptable premium burdens. For example, \(\eta\) may be concentrated below a threshold \(\tau\), or may represent a socially acceptable burden distribution. The distance
\[
  \Wp{p}\bigl(\Law(R),\eta\bigr)
\]
then measures how far a tariff is from the affordability benchmark.

This leads to a constrained pricing problem. Let \(P_{\btheta}\) be a tariff indexed by parameters \(\btheta\), and let \(L\) be the loss. One may seek
\[
  \min_{\btheta}
  \E\left[\ell(P_{\btheta},L)\right]
  \quad
  \text{subject to}
  \quad
  \Wp{p}\bigl(\Law(P_{\btheta}/Y),\eta\bigr)\le r.
\]
The first term is actuarial or financial. The constraint is distributional and social: it restricts how far the premium burden can be from an affordability target.

Another formulation uses a barycentric compromise between technical adequacy and affordability. Let \(\mu_{\mathrm{tech}}\) be the distribution of technical premiums and \(\mu_{\mathrm{aff}}\) a distribution of affordable premiums. A Wasserstein barycenter
\[
  \mu_\lambda
  \in
  \operatorname*{argmin}_{\mu}
  \left\{
    (1-\lambda)\Wtwo^2(\mu,\mu_{\mathrm{tech}})
    +
    \lambda \Wtwo^2(\mu,\mu_{\mathrm{aff}})
  \right\}
\]
gives a continuum between actuarial adequacy and affordability. In one dimension, its quantile function is
\[
  Q_{\mu_\lambda}(t)
  =
  (1-\lambda)Q_{\mu_{\mathrm{tech}}}(t)
  +
  \lambda Q_{\mu_{\mathrm{aff}}}(t).
\]

\begin{remark}[Affordability is not average premium alone]
An affordability audit compares premiums to resources. Average premium income can remain unchanged while the upper tail of the burden distribution increases. For this reason, the distribution of \(\Premium/Y\) is often more informative than a scalar premium summary.
\end{remark}

Micro-insurance also involves demand. Let \(D(p,y)\) be the probability that an individual with resources \(y\) purchases coverage at premium \(p\). The insured portfolio law is then a selected distribution:
\[
  \Law(\bX,L,Y\mid \text{purchase}).
\]
A tariff reform can improve affordability while changing the composition of the insured pool. This creates an adverse-selection feedback:
\[
  P_{\btheta}
  \longrightarrow
  D(P_{\btheta},Y)
  \longrightarrow
  \Law(L\mid \text{purchase})
  \longrightarrow
  \text{technical premium}.
\]
Transport diagnostics can be used to monitor each movement: premium burden, portfolio composition and claim distribution.

\begin{computebox}
A micro-insurance dashboard may report:
\[
  \E[P],\qquad
  \E[P/Y],\qquad
  \Pbb(P/Y>\tau),\qquad
  \TVaR_\alpha(P/Y),
  \qquad
  \Wp{p}(\Law(P/Y),\eta).
\]
The Wasserstein term summarizes the distributional distance to an affordability target, while the exceedance and tail-average terms remain directly interpretable.
\end{computebox}

\subsection{Barycentric tariff benchmarks}\label{subsec:pricing-barycentric-benchmarks}

Suppose several tariffs are available:
\[
  \Premium_1,\ldots,\Premium_K.
\]
They may come from different models, different parameter choices, different regulatory regimes, or different companies. Let
\[
  \mu_k=\Law(\Premium_k(\bX)).
\]
A Wasserstein barycenter of $\mu_1,\ldots,\mu_K$ solves
\[
  \bar\mu\in\operatorname*{argmin}_{\mu}
  \sum_{k=1}^K w_k\Wtwo^2(\mu,\mu_k).
\]
The barycenter is a distributional compromise between tariffs.

\begin{proposition}[One-dimensional barycentric tariff benchmark]\label{prop:one-dimensional-barycentric-tariff-benchmark}
Let \(\mu_1,\ldots,\mu_K\in\Probp{2}(\R)\) be premium distributions with quantile functions \(Q_1,\ldots,Q_K\), and let \(w_k\ge0\), \(\sum_k w_k=1\). The one-dimensional \(\Wtwo\)-barycenter has quantile function
\[
  Q_{\bar\mu}(u)=\sum_{k=1}^K w_k Q_k(u),\qquad u\in(0,1).
\]
Consequently, a barycentric benchmark averages tariff distributions by ranks, not by policy identities.
\end{proposition}

Thus a barycentric premium benchmark can be constructed by averaging sorted premium books. This is different from averaging premiums policy by policy. It averages positions in the distribution rather than identities.

\begin{example}[Benchmarking several models]\label{ex:pricing-benchmarking-models}
Suppose three pricing models produce premium distributions $\mu_1,\mu_2,\mu_3$. Their one-dimensional Wasserstein barycenter has quantile function
\[
  Q_{\bar\mu}(u)=\frac13(Q_1(u)+Q_2(u)+Q_3(u)).
\]
This benchmark may be used to identify which model produces the most extreme premium distribution at each quantile level.
\end{example}

\subsection{Discrete implementation}\label{subsec:pricing-discrete-implementation}

For a finite portfolio $i=1,\ldots,n$, let $w_i$ be exposure weights and $\pi_i$ the premium. The empirical premium distribution is
\[
  \widehat\mu_\pi=\sum_{i=1}^n w_i\delta_{\pi_i}.
\]
If the portfolio is grouped into rating cells $k=1,\ldots,K$, with cell weights $a_k$ and cell premiums $\pi_k$, then
\[
  \widehat\mu_\pi=\sum_{k=1}^K a_k\delta_{\pi_k}.
\]
Comparing two tariffs with old cell weights $\ba=(a_1,\ldots,a_K)$, new cell weights $\bb=(b_1,\ldots,b_M)$, old premiums $\pi_k^{(0)}$ and new premiums $\pi_\ell^{(1)}$ leads to the discrete OT problem
\[
  \min_{\Gamma\in U(\ba,\bb)}
  \sum_{k=1}^K\sum_{\ell=1}^M
  |\pi_k^{(0)}-\pi_\ell^{(1)}|^p\Gamma_{k\ell}.
\]
The optimal plan $\Gamma$ matches old premium cells to new premium cells.

\begin{example}[Tariff-cell transition table]
An insurer can compute a transition matrix between old and new premium deciles. The empirical transition matrix based on policy identities shows where policyholders move. The optimal transport matrix between decile distributions shows the minimal movement needed to transform the old decile distribution into the new one. Comparing the two matrices reveals whether the tariff revision is mostly a smooth distributional shift or a large reshuffling of individual policyholders.
\end{example}

\subsection{A mini-lab in R}\label{subsec:pricing-mini-lab-r}

The following code simulates a portfolio, constructs a pooled premium, a coarse class tariff, a technical individualized tariff and a shrinkage tariff. It then compares their premium distributions using one-dimensional Wasserstein distances.

\begin{lstlisting}[language=R]
set.seed(123)

n <- 5000
age <- runif(n, 18, 80)
veh <- rbinom(n, 1, 0.35)
terr <- sample(1:5, n, replace = TRUE)

# Synthetic conditional pure premium
eta <- 6.5 + 0.015 * (age - 45) + 0.35 * veh + 0.08 * (terr - 3)
m <- exp(eta)

# Pooled tariff
pi_pool <- rep(mean(m), n)

# Coarse class tariff: territory only
pi_class <- ave(m, terr, FUN = mean)

# Individualized tariff
pi_ind <- m

# Shrinkage tariff
lambda <- 0.5
pi_shrink <- mean(m) + lambda * (pi_ind - mean(m))

Wp1d <- function(x, y, p = 2, grid_size = 1000) {
  # Equal-size shortcut; otherwise compare empirical quantiles on a common grid.
  if (length(x) == length(y)) {
    return(mean(abs(sort(x) - sort(y))^p)^(1/p))
  }
  u <- (seq_len(grid_size) - 0.5) / grid_size
  qx <- quantile(x, probs = u, type = 1, names = FALSE)
  qy <- quantile(y, probs = u, type = 1, names = FALSE)
  mean(abs(qx - qy)^p)^(1/p)
}

D_pool_class <- Wp1d(pi_pool, pi_class, p = 2)
D_pool_ind   <- Wp1d(pi_pool, pi_ind, p = 2)
D_class_ind  <- Wp1d(pi_class, pi_ind, p = 2)
D_pool_shr   <- Wp1d(pi_pool, pi_shrink, p = 2)

c(pool_to_class = D_pool_class,
  pool_to_individual = D_pool_ind,
  class_to_individual = D_class_ind,
  pool_to_shrinkage = D_pool_shr)

# Check the variance identity
sqrt(var(pi_ind) * (n - 1) / n)
Wp1d(pi_pool, pi_ind, p = 2)
\end{lstlisting}

The last two quantities differ only by the convention for empirical variance. The Wasserstein distance to the pooled empirical distribution is the empirical standard deviation around the mean.

\subsection{Pricing governance dashboard}\label{subsec:pricing-governance-dashboard}

A practical pricing governance dashboard could report several complementary quantities:
\begin{enumerate}[label=(\roman*)]
  \item average premium and total premium income;
  \item $D_1$ and $D_2$ from the pooled benchmark;
  \item $\Wone$ and $\Wtwo$ between old and new premium distributions;
  \item policy-level average absolute and squared changes;
  \item the reallocation gap $G_p$;
  \item premium changes by quantile level;
  \item affordability distances and tail affordability risk measures.
\end{enumerate}

None of these numbers is sufficient alone. Together, they separate different notions: premium income, dispersion, distributional movement, individual disruption and affordability.

\begin{readingbox}
The convex-analytic and optimization viewpoint behind these comparisons is close in spirit to the general convex optimization framework in \cite{galichon2016economics,peyre2019computational}. For actuarial pricing, credibility and GLM-based tariff construction, see \cite{buhlmann1967experience,buhlmann2005course,ohlsson2010nonlife,wuthrich2023statistical}. For adverse selection and insurance-market segmentation, see \cite{rothschild1976equilibrium,chiappori2006asymmetric}. For micro-insurance and affordability, see \cite{churchill2006protecting,churchill2012protecting,biener2012insurability}. Dependence and actuarial ordering ideas are treated in \cite{denuit2005actuarial}.
\end{readingbox}

\subsection{Summary}\label{subsec:pricing-summary}

This lecture developed a Wasserstein view of pricing and premium distributions. The main messages are:
\begin{enumerate}[label=(\roman*)]
  \item A tariff induces a premium distribution by push-forward of the portfolio distribution.
  \item A constant benchmark corresponds to a Dirac premium distribution; a segmented tariff corresponds to a more dispersed premium distribution.
  \item The squared $\Wtwo$ distance to the pooled mean is the variance of premiums.
  \item Increasing pricing information increases the variance of conditional pure premiums.
  \item Shrinkage paths provide interpretable routes between a constant benchmark and a technical tariff.
  \item A tariff revision has both a policy-level displacement and a distributional OT displacement.
  \item The gap between these quantities reveals reshuffling beyond marginal distributional change.
  \item Premium affordability can also be studied as a distributional transport problem.
\end{enumerate}

\subsection{Exercises}\label{subsec:pricing-exercises}

\begin{exercise}[Variance and Wasserstein distance]\label{ex:9:1}
Let $Z\in L^2$, $\mu=\Law(Z)$, and $\bar z=\E[Z]$. Prove that
\[
  \Wtwo^2(\mu,\delta_{\bar z})=\Var(Z).
\]
Then compute the corresponding formula for $\Wp{p}^p(\mu,\delta_a)$ for arbitrary $a\in\R$ and $p\ge1$.
\end{exercise}

\begin{exercise}[Nested rating information]\label{ex:9:2}
Let $\mathcal G_1\subseteq\mathcal G_2$ and $L\in L^2$. Prove that
\[
  \Var(\E[L\mid\mathcal G_1])
  \le
  \Var(\E[L\mid\mathcal G_2]).
\]
Interpret the result for a tariff moving from territory-only rating to territory plus vehicle plus claims-history rating.
\end{exercise}

\begin{exercise}[Shrinkage and distance to pooling]\label{ex:9:3}
Let $\Premium_\lambda=\bar\pi+\lambda(\Premium-\bar\pi)$, $\lambda\in[0,1]$. Show that
\[
  \Wp{p}\bigl(\Law(\Premium_\lambda),\delta_{\bar\pi}\bigr)
  =\lambda\Wp{p}\bigl(\Law(\Premium),\delta_{\bar\pi}\bigr).
\]
Then compute $\Wp{p}(\Law(\Premium_\lambda),\Law(\Premium_{\lambda'}))$.
\end{exercise}

\begin{exercise}[Policy-level versus distributional movement]\label{ex:9:4}
Let $\Premium_0,\Premium_1\in L^p$. Prove that
\[
  \Wp{p}^p(\Law(\Premium_0),\Law(\Premium_1))
  \le
  \E|\Premium_1-\Premium_0|^p.
\]
Construct an example where the left-hand side is zero but the right-hand side is positive.
\end{exercise}

\begin{exercise}[Two-class tariff]\label{ex:9:5}
Consider the distribution
\[
  \mu_d=\frac12\delta_{\bar\pi-d}+\frac12\delta_{\bar\pi+d}.
\]
Compute $\Wone(\mu_d,\delta_{\bar\pi})$ and $\Wtwo(\mu_d,\delta_{\bar\pi})$. Then compare $\mu_d$ with
\[
  \nu_e=\frac12\delta_{\bar\pi-e}+\frac12\delta_{\bar\pi+e}
\]
for $0\le d,e$. Interpret the result as a change in segmentation strength.
\end{exercise}

\begin{exercise}[Affordability distribution]\label{ex:9:6}
Let $Y>0$ be income and let $\Premium_0,\Premium_1$ be two tariffs. Show that
\[
  \Wp{p}\left(\Law\left(\frac{\Premium_0}{Y}\right),
  \Law\left(\frac{\Premium_1}{Y}\right)\right)
  \le
  \left(\E\left|\frac{\Premium_1-\Premium_0}{Y}\right|^p\right)^{1/p}.
\]
When might the right-hand side be much larger than the left-hand side?
\end{exercise}

\begin{exercise}[Premium caps]\label{ex:9:7}
Let $\Premium\ge0$ and $\Premium^M=\min(\Premium,M)$. Prove that
\[
  \Wp{p}\bigl(\Law(\Premium),\Law(\Premium^M)\bigr)
  \le
  \left(\E[(\Premium-M)_+^p]\right)^{1/p}.
\]
Assuming \(\Premium\in L^p\), show that equality holds because the cap map is increasing and hence induces the monotone optimal coupling on the real line.
\end{exercise}

\iffalse
% To be restored in v2, after the fairness paper on multiple sensitive variables has been submitted.

\begin{exercise}[Computational exercise]\label{ex:9:8}
Using the mini-lab above, add a protected binary variable $S$ correlated with territory. Compute the conditional premium distributions $\Law(\Premium\mid S=0)$ and $\Law(\Premium\mid S=1)$ for the pooled, class, individualized and shrinkage tariffs. Compare the groups using $\Wone$ and $\Wtwo$. Discuss whether shrinkage reduces group distributional disparities.
\end{exercise}
\fi

\newpage
\lecture{Portfolio Drift, Calibration and Domain Adaptation}
\label{lec:portfolio-drift-domain-adaptation}

\subsection*{Learning objectives}

The previous lectures described actuarial models as push-forward maps: a function transforms a portfolio distribution into a distribution of outputs. This lecture studies what happens when the portfolio itself changes. An actuarial model is rarely used forever on the same population on which it was trained. The distribution of policyholders changes, the mix of regions and coverages changes, inflation affects claim severities, climate modifies exposure patterns, and underwriting or regulatory decisions may alter the book. Optimal transport provides a natural way to compare the old and new portfolios, to diagnose model drift, and to design calibration or adaptation procedures.

After this lecture, the reader should be able to:
\begin{enumerate}[label=(\roman*)]
  \item distinguish covariate shift, label shift, concept drift and portfolio drift in an actuarial setting;
  \item represent training and target portfolios as probability measures on covariates, losses, scores or residuals;
  \item use Wasserstein distances to monitor drift in covariates, predicted losses, scores and residuals;
  \item derive simple stability bounds for expected losses and risk scores under distributional perturbations;
  \item interpret transport plans as soft matchings between portfolios;
  \item use a transport plan to construct target-calibrated portfolio weights;
  \item understand barycentric transport maps and their use in portfolio transfer;
  \item separate model adaptation from distributional validation;
  \item design actuarial diagnostics based on Wasserstein distances, transport maps and residual distributions.
\end{enumerate}

\begin{actuarialbox}
A pricing or reserving model is not only a function. It is a function used on a portfolio. When the portfolio changes, the law of the model output changes even if the formula does not. Optimal transport gives a geometry for monitoring this change and for deciding whether recalibration is enough or whether the model must be rebuilt.
\end{actuarialbox}

This lecture uses the push-forward notation of Definition~\ref{def:push-forward}, the coupling notation of Definition~\ref{def:coupling}, the Wasserstein distance of Definition~\ref{def:wasserstein-distance}, and the empirical viewpoint of Lecture~\ref{lec:statistical-ot}. It also prepares the reserving cash-flow lab in Appendix~\ref{lab:reserving-cashflow} and the climate-prevention discussion in Lecture~\ref{lec:reserving-climate-prevention}. The domain-adaptation perspective is close to the statistical learning literature on covariate shift, sample selection bias and dataset shift; see, for instance, \cite{shimodaira2000improving,quinonero2009dataset,sugiyama2012machine,pan2010survey,ben2010theory}. For optimal transport approaches to domain adaptation, see \cite{courty2014domain,courty2017optimal,redko2017theoretical,flamary2021pot}. For concept drift and monitoring, see \cite{gama2014survey,webb2016characterizing}.

\subsection{The covariate-shift problem}\label{subsec:drift-covariate-shift-problem}

Let $(\bX,L)$ denote covariates and future loss. A model is usually estimated on a source or training portfolio,
\[
  P_s = \Law_s(\bX,L),
\]
and deployed on a target portfolio,
\[
  P_t = \Law_t(\bX,L).
\]
The notation may refer to two time periods, two countries, two distribution channels, two regions, or two underwriting books. The most common actuarial examples are:
\begin{itemize}
  \item a motor pricing model trained on last year's policies and used on this year's renewals;
  \item a homeowner model transferred from one province to another;
  \item a cyber model built on large firms and used on small firms;
  \item a climate-risk module estimated on historical exposure and used under a future scenario;
  \item a reserving model trained on mature accident years and used on more recent accident years.
\end{itemize}

A useful first decomposition, written formally in terms of conditional laws, is
\[
  P_s(\bX,L)=P_s(L\mid \bX)P_s(\bX),
  \qquad
  P_t(\bX,L)=P_t(L\mid \bX)P_t(\bX).
\]
The simplest form of domain adaptation assumes that
\[
  P_s(L\mid \bX)=P_t(L\mid \bX),
  \qquad
  P_s(\bX)\ne P_t(\bX).
\]
This is usually called covariate shift. It says that the actuarial relationship between risk characteristics and losses is stable, but the mix of policyholders has changed. For instance, a model may remain valid at the individual level but be applied to a portfolio containing more young drivers, more urban risks, or more high insured values.

A more difficult situation is concept drift:
\[
  P_s(L\mid \bX)\ne P_t(L\mid \bX).
\]
This may occur when repair costs inflate, when claim behavior changes, when climate changes the distribution of hazards conditional on exposure, or when legal inflation modifies liability losses. Transport methods can diagnose such changes, but cannot by themselves guarantee that a model trained under $P_s$ remains valid under $P_t$.

\begin{definition}[Source and target portfolios]\label{def:source-target-portfolios}
Let $\mathcal X$ be the covariate space. A source portfolio is a probability measure $\mu\in\Prob(\mathcal X)$, and a target portfolio is a probability measure $\nu\in\Prob(\mathcal X)$. If losses are included, the source and target joint laws are probability measures on $\mathcal X\times\R_+$.
\end{definition}

\begin{definition}[Covariate shift, label shift and concept drift]\label{def:covariate-label-concept-drift}
Let $P_s$ and $P_t$ denote source and target joint laws of $(\bX,L)$. There is covariate shift from $P_s$ to $P_t$ if
\[
  P_s(\bX)\ne P_t(\bX)
  \quad\text{but}\quad
  P_s(L\mid \bX)=P_t(L\mid \bX).
\]
There is label shift if the marginal law of the response changes while the conditional law of covariates given the response is stable, for example
\[
  P_s(L)\ne P_t(L)
  \quad\text{but}\quad
  P_s(\bX\mid L)=P_t(\bX\mid L),
\]
when this conditional law is meaningful; see \cite{lipton2018detecting} for statistical treatments of label shift. There is concept drift if the actuarial mechanism changes, that is, if the conditional law $P(L\mid \bX)$ changes.
\end{definition}

\begin{definition}[Portfolio drift]\label{def:portfolio-drift}
A portfolio drift is a change in the law of a chosen actuarial representation $Z$: covariates, scores, losses, residuals, cash-flow vectors or hazard variables. If $Z_s\sim\mu_Z$ and $Z_t\sim\nu_Z$, a drift diagnostic is any functional of the pair $(\mu_Z,\nu_Z)$, such as a Wasserstein distance, a sliced Wasserstein distance, a tail quantile difference, or a residual-distribution comparison.
\end{definition}

\begin{remark}[Drift is scale-dependent]
Whether a shift is large or small depends on the representation of $\bX$. A change in a standardized log-income variable is not comparable to a change in raw income. Before computing a Wasserstein distance, one must choose a meaningful metric on the covariate space. In actuarial applications, this choice is part of the modelling problem.
\end{remark}

\subsection{From portfolio drift to output drift}\label{subsec:drift-output-drift}

Let $f:\mathcal X\to\R$ be a fitted actuarial model: a predicted frequency, a predicted severity, a risk score, a cancellation probability, or a reserve factor. Even if $f$ is unchanged, the distribution of $f(\bX)$ changes when $\Law(\bX)$ changes.

Under the source portfolio,
\[
  \mu_f = f_{\push}\mu,
\]
where $\mu=\Law_s(\bX)$. Under the target portfolio,
\[
  \nu_f = f_{\push}\nu,
\]
where $\nu=\Law_t(\bX)$. Thus we can monitor drift at several levels:
\begin{align*}
  \text{covariate drift:}      && \Wp{p}(\mu,\nu), \\
  \text{score drift:}          && \Wp{p}(f_{\push}\mu,f_{\push}\nu), \\
% To be restored in v2 with the premium-distribution material.
%   \text{premium drift:}        && \Wp{p}(\Premium_{\push}\mu,\Premium_{\push}\nu), \\
  \text{loss drift:}           && \Wp{p}(\Law_s(L),\Law_t(L)), \\
  \text{residual drift:}       && \Wp{p}(\Law_s(L-f(\bX)),\Law_t(L-f(\bX))).
\end{align*}

The first quantity asks whether the population has changed. The second asks whether the model sees the new population as different. The remaining quantities ask whether claim behavior and errors have changed.

\begin{example}[A stable model with unstable outputs]
Suppose $f(x)=x$, and the source covariate has distribution $\mu=N(0,1)$, while the target covariate has distribution $\nu=N(1,1)$. The model formula is unchanged, but
\[
  f_{\push}\mu=N(0,1),
  \qquad
  f_{\push}\nu=N(1,1).
\]
Thus the score distribution moves by one unit in $\Wtwo$. This means that the book has drifted toward higher predicted risk, even if the model has not changed.
\end{example}

\begin{example}[A drift invisible to a model]
Let $\bX=(X_1,X_2)$, and suppose the fitted score depends only on $X_1$:
\[
  f(\bX)=\exp(\beta_0+\beta_1X_1).
\]
If the portfolio drifts only in $X_2$, then $\Wp{p}(\Law_s(\bX),\Law_t(\bX))$ may be large while
\[
  \Wp{p}(\Law_s(f(\bX)),\Law_t(f(\bX)))=0.
\]
This does not mean the drift is harmless. If $X_2$ is predictive of losses but absent from the model, residual drift may reveal a problem.
\end{example}

\subsection{Wasserstein diagnostics for drift}\label{subsec:drift-wasserstein-diagnostics}

Let $(\mathcal X,d)$ be a metric space, and let $\mu,\nu\in\Probp{p}(\mathcal X)$. The Wasserstein distance
\[
  \Wp{p}^p(\mu,\nu)
  =\inf_{\gamma\in\Coup(\mu,\nu)}
     \int_{\mathcal X\times\mathcal X} d(\bx,\by)^p\,\gamma(\dd \bx,\dd \by)
\]
measures the minimal average cost of moving the source portfolio into the target portfolio. If $\mathcal X\subseteq\R^d$, one often starts with
\[
  d(\bx,\by)=\|S(\bx-\by)\|_2,
\]
where $S$ standardizes or weights covariates.

\begin{definition}[Actuarial drift index]\label{def:actuarial-drift-index}
Let $Z_s\sim\mu_Z$ and $Z_t\sim\nu_Z$ denote a chosen representation of the source and target books: covariates, scores, losses or residuals. For $p\ge1$, the order-$p$ Wasserstein drift index is
\[
  D_p(Z;s,t)=\Wp{p}(\mu_Z,
  \nu_Z).
\]
When $Z$ is one-dimensional, this becomes
\[
  D_p^p(Z;s,t)
  =\int_0^1 |Q_s(u)-Q_t(u)|^p\dd u.
\]
\end{definition}

The one-dimensional formula is particularly useful in actuarial diagnostics, because many outputs are scalar: predicted claim frequency, expected severity, loss ratio, reserve residual, or risk score. The quantile representation also gives interpretable plots: the function
\[
  u\mapsto Q_t(u)-Q_s(u)
\]
shows where the shift occurs in the distribution.

\begin{proposition}[Lipschitz propagation of drift]\label{prop:lipschitz-propagation-drift}
Let $f:(\mathcal X,d_\mathcal X)\to(\mathcal Y,d_\mathcal Y)$ be $K$-Lipschitz:
\[
  d_\mathcal Y(f(\bx),f(\by))\le K d_\mathcal X(\bx,\by).
\]
Then, for every $p\ge1$,
\[
  \Wp{p}(f_{\push}\mu,f_{\push}\nu)
  \le K\Wp{p}(\mu,\nu).
\]
\end{proposition}

\begin{proof}
Let $\gamma\in\Coup(\mu,\nu)$. The push-forward $(f,f)_{\push}\gamma$ is a coupling of $f_{\push}\mu$ and $f_{\push}\nu$. Therefore
\[
  \Wp{p}^p(f_{\push}\mu,f_{\push}\nu)
  \le \int d_\mathcal Y(f(\bx),f(\by))^p\gamma(\dd \bx,\dd \by)
  \le K^p\int d_\mathcal X(\bx,\by)^p\gamma(\dd \bx,\dd \by).
\]
Taking the infimum over $\gamma$ gives the result.
\end{proof}

\iffalse
% To be restored in v2 with the premium-distribution and portfolio demutualization material.
\begin{proposition}[Portfolio-mix and model-change decomposition]\label{prop:portfolio-mix-model-change-decomposition}
Let $\Premium_0,\Premium_1:\mathcal X\to\R$ be two tariff maps, let $\mu$ and $\nu$ be two portfolio laws on $\mathcal X$, and assume $\Premium_0$ is $K$-Lipschitz. Then
\[
\begin{aligned}
  \Wone\bigl((\Premium_0)_{\push}\mu,(\Premium_1)_{\push}\nu\bigr)
  &\le
  \Wone\bigl((\Premium_0)_{\push}\mu,(\Premium_0)_{\push}\nu\bigr)
  +\Wone\bigl((\Premium_0)_{\push}\nu,(\Premium_1)_{\push}\nu\bigr) \\
  &\le K\Wone(\mu,\nu)
  +\E_\nu\bigl[|\Premium_1(\bX)-\Premium_0(\bX)|\bigr].
\end{aligned}
\]
\end{proposition}

\begin{proof}
The first inequality is the triangle inequality for $\Wone$. The first term is bounded by Proposition~\ref{prop:lipschitz-propagation-drift}. For the second term, use the coupling induced by the same target policy $\bX\sim\nu$ under the two tariffs, namely $(\Premium_0(\bX),\Premium_1(\bX))$.
\end{proof}
\fi

\begin{actuarialbox}
A large covariate drift may have little output impact if the model is insensitive to the drifting variables. Conversely, a moderate covariate drift in a highly leveraged variable may create a large score or residual drift. The relevant diagnostic is often a combination of covariate drift and output drift.
\end{actuarialbox}

\subsection{Stability of expected losses under covariate shift}\label{subsec:drift-expected-loss-stability}

Assume covariate shift:
\[
  P_s(L\mid \bX)=P_t(L\mid \bX).
\]
Let
\[
  m(\bx)=\E[L\mid \bX=\bx]
\]
be the stable conditional mean. Then the difference between source and target expected losses is
\[
  \E_t[L]-\E_s[L]
  =\int m(\bx)\nu(\dd \bx)-\int m(\bx)\mu(\dd \bx).
\]
This difference can be controlled by Wasserstein distance if $m$ is regular.

\begin{proposition}[Expected-loss stability under covariate shift]\label{prop:expected-loss-stability-covariate-shift}
Let $(\mathcal X,d)$ be a metric space. Suppose that covariate shift holds and that $m(\bx)=\E[L\mid \bX=\bx]$ is $K$-Lipschitz. Then
\[
  |\E_t[L]-\E_s[L]|\le K\Wone(\mu,\nu).
\]
More generally, since $\Wone\le\Wp{p}$ on probability spaces with finite $p$-moments,
\[
  |\E_t[L]-\E_s[L]|\le K\Wp{p}(\mu,\nu).
\]
\end{proposition}

\begin{proof}
By Kantorovich--Rubinstein duality for $\Wone$,
\[
  \left|\int m\dd\nu-\int m\dd\mu\right|
  \le K\Wone(\mu,\nu).
\]
Under covariate shift, the two integrals are exactly the target and source expected losses.
\end{proof}

\begin{remark}[What the bound does not say]
The bound does not validate the conditional model. It says that if the conditional mean is stable and Lipschitz, then portfolio-level expected loss cannot change faster than the covariate distribution moves in $\Wone$. If inflation, legal environment or claim behavior changes $m$, the bound no longer captures the full problem.
\end{remark}

A similar bound holds for any Lipschitz functional of the covariates. If $h:\mathcal X\to\R$ is $K_h$-Lipschitz, then
\[
  |\E_t[h(\bX)]-\E_s[h(\bX)]|\le K_h\Wone(\mu,\nu).
\]
This is useful for quick impact assessments: a portfolio can drift in many variables, but if the functional is smooth and the transport distance is small, its mean impact is controlled.

\subsection{Transport plans as portfolio matchings}\label{subsec:drift-transport-matchings}

In a finite portfolio, the source book is
\[
  \widehat\mu=\sum_{i=1}^n a_i\delta_{\bx_i},
  \qquad a_i\ge0,
  \qquad \sum_i a_i=1,
\]
and the target book is
\[
  \widehat\nu=\sum_{j=1}^m b_j\delta_{\by_j},
  \qquad b_j\ge0,
  \qquad \sum_j b_j=1.
\]
A transport plan is a nonnegative matrix $\Gamma=(\Gamma_{ij})$ satisfying
\[
  \sum_{j=1}^m\Gamma_{ij}=a_i,
  \qquad
  \sum_{i=1}^n\Gamma_{ij}=b_j.
\]
The transport cost is
\[
  \sum_{i=1}^n\sum_{j=1}^m c(\bx_i,\by_j)\Gamma_{ij}.
\]
Actuarially, $\Gamma_{ij}$ is a soft matching between source policy $i$ and target policy $j$. If $\Gamma_{ij}$ is large, the two policies are similar under the chosen metric.

\begin{figure}
    \centering
    \includegraphics[width=0.75\linewidth]{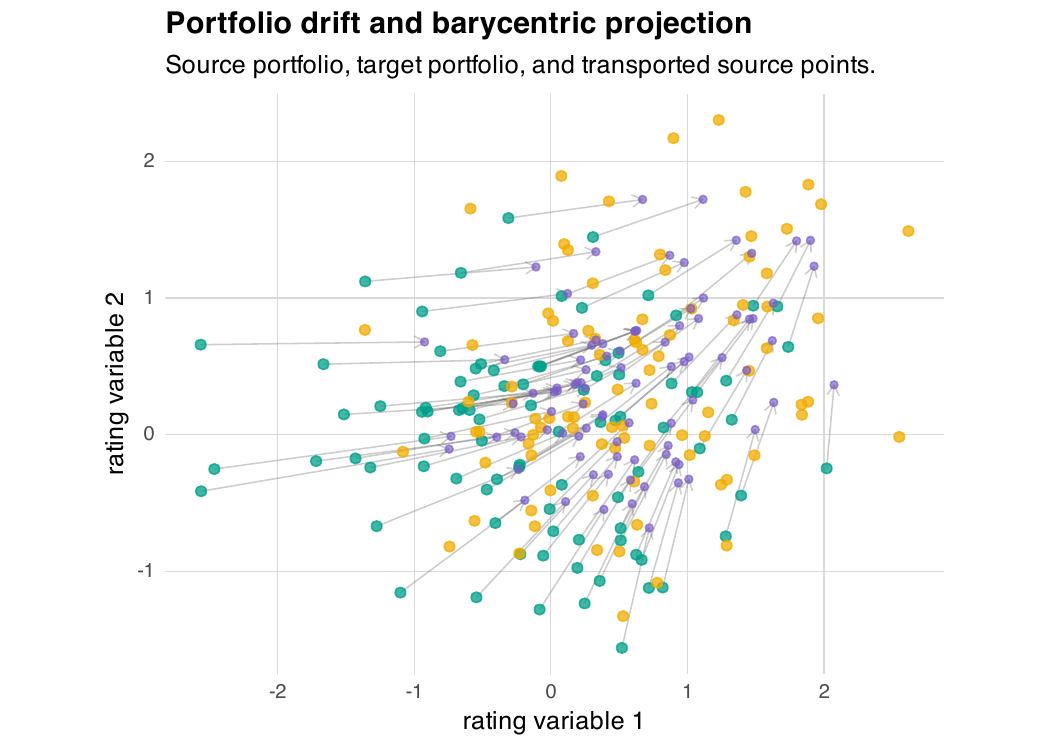}
    \caption{{\bf Portfolio drift and barycentric projection}. The source portfolio is matched to the target portfolio through a transport plan. The barycentric projection moves each source profile toward the average location of its matched target profiles, providing an interpretable drift map.}
    \label{fig:portfolio-drift-barycentric-projection}
\end{figure}

\begin{definition}[Transport matching]\label{def:transport-matching}
A transport matching between two empirical portfolios is any feasible transport matrix $\Gamma\in\R_+^{n\times m}$ satisfying the marginal constraints above. An optimal transport matching minimizes the total cost
\[
  \sum_{i,j} c(\bx_i,\by_j)\Gamma_{ij}.
\]
\end{definition}

This matching can be used in three ways:
\begin{enumerate}[label=(\roman*)]
  \item to explain which source profiles are closest to which target profiles;
  \item to transfer predictions or residuals from source to target;
  \item to build calibrated source weights that mimic the target book.
\end{enumerate}

\begin{example}[Matching an old motor book to a new one]
Let $\bx_i$ be standardized rating variables for last year's motor policies and $\by_j$ the same variables for this year's renewals. An optimal transport plan identifies which old policies represent the new portfolio. If young urban drivers increased in the target book, mass is transported toward source profiles with similar characteristics. The unmatched or expensive movements identify regions of the covariate space where the old model has little support.
\end{example}

\subsection{Transport weights for calibration}\label{subsec:drift-transport-weights}

Classical covariate-shift correction uses importance weights
\[
  w(\bx)=\frac{\dd\nu}{\dd\mu}(\bx),
\]
when $\nu\ll\mu$. These weights transform source expectations into target expectations and are the standard starting point in covariate-shift correction \cite{shimodaira2000improving,sugiyama2012machine}:
\[
  \int h(\bx)\nu(\dd \bx)=\int h(\bx)w(\bx)\mu(\dd \bx).
\]
In high-dimensional or mixed actuarial data, density ratios may be hard to estimate. Transport provides an alternative: instead of estimating a pointwise ratio, one estimates a matching between source and target masses, as in OT-based domain adaptation \cite{courty2014domain,courty2017optimal,redko2017theoretical}.

Given an empirical transport plan $\Gamma$, define the target-calibrated version of a source function $h(\bx_i)$ by
\[
  \widehat E_t^{\Gamma}[h]
  =\sum_{j=1}^m\sum_{i=1}^n \Gamma_{ij}h(\bx_i).
\]
Since $\sum_j\Gamma_{ij}=a_i$, this is also
\[
  \widehat E_t^{\Gamma}[h]
  =\sum_{i=1}^n a_i h(\bx_i),
\]
which does not change the total source average if $h$ is kept only at the source points. To obtain target-specific predictions, one instead uses conditional transport weights for each target point:
\[
  \widehat h_\Gamma(\by_j)
  =\frac{1}{b_j}\sum_{i=1}^n\Gamma_{ij}h(\bx_i),
  \qquad b_j>0.
\]
This is a local transport average of source values around the target point $\by_j$. The target average is then
\[
  \sum_{j=1}^m b_j\widehat h_\Gamma(\by_j)
  =\sum_{i=1}^n a_i h(\bx_i).
\]
Thus the global mean is preserved, but the values are redistributed over the target book.

\begin{definition}[Transport-transfer predictor]\label{def:transport-transfer-predictor}
Let $h:\{\bx_1,\ldots,\bx_n\}\to\R$ be a source quantity, and let $\Gamma$ be a coupling between source weights $a_i$ and target weights $b_j$. The transport-transfer predictor at target point $\by_j$ is
\[
  h_\Gamma(\by_j)
  =\frac{1}{b_j}\sum_{i=1}^n\Gamma_{ij}h(\bx_i).
\]
\end{definition}

\begin{proposition}[Mean preservation by transport transfer]\label{prop:transport-transfer-mean-preservation}
If $b_j>0$ for all $j$, then
\[
  \sum_{j=1}^m b_j h_\Gamma(\by_j)
  =\sum_{i=1}^n a_i h(\bx_i).
\]
\end{proposition}

\begin{proof}
By definition,
\[
  \sum_{j=1}^m b_j h_\Gamma(\by_j)
  =\sum_{j=1}^m\sum_{i=1}^n\Gamma_{ij}h(\bx_i)
  =\sum_{i=1}^n\left(\sum_{j=1}^m\Gamma_{ij}\right)h(\bx_i)
  =\sum_{i=1}^n a_i h(\bx_i).
\]
\end{proof}

\begin{proposition}[Local transfer error for Lipschitz quantities]\label{prop:transport-transfer-lipschitz-error}
Assume that $h:\mathcal X\to\R$ is $K$-Lipschitz and that the same function can be evaluated on source and target covariates. Then
\[
  \left|\sum_{j=1}^m b_j h(\by_j)-\sum_{j=1}^m b_j h_\Gamma(\by_j)\right|
  \le
  K\sum_{i=1}^n\sum_{j=1}^m \Gamma_{ij}d(\bx_i,\by_j).
\]
For a $p$-cost, H\"older's inequality gives the weaker but scale-compatible bound
\[
  \left|\sum_{j=1}^m b_j h(\by_j)-\sum_{j=1}^m b_j h_\Gamma(\by_j)\right|
  \le K\left(\sum_{i,j}\Gamma_{ij}d(\bx_i,\by_j)^p\right)^{1/p}.
\]
\end{proposition}

\begin{proof}
Using the marginal constraints and the definition of $h_\Gamma$,
\[
\begin{aligned}
  \sum_j b_j h(\by_j)-\sum_j b_j h_\Gamma(\by_j)
  &=\sum_{i,j}\Gamma_{ij}\{h(\by_j)-h(\bx_i)\}.
\end{aligned}
\]
Taking absolute values and using the Lipschitz property gives the first inequality. The second follows from H\"older's inequality because $\sum_{i,j}\Gamma_{ij}=1$.
\end{proof}

\begin{remark}[When mean preservation is undesirable]
If the target book is genuinely riskier, mean preservation may understate target losses. The transport-transfer predictor is a redistribution of source information, not a substitute for modelling the conditional mean. It is most useful for comparing like-for-like portfolios or for diagnostic transfer. For final target prediction, it should be combined with a conditional model.
\end{remark}

\subsection{Barycentric maps and transferred portfolios}\label{subsec:drift-barycentric-maps}

A transport plan can be summarized by a barycentric map. For each source point $\bx_i$, define
\[
  T_\Gamma(\bx_i)
  =\frac{1}{a_i}\sum_{j=1}^m\Gamma_{ij}\by_j,
  \qquad a_i>0.
\]
This maps each source point to the weighted average of target points receiving mass from it. Conversely, the reverse barycentric map is
\[
  S_\Gamma(\by_j)
  =\frac{1}{b_j}\sum_{i=1}^n\Gamma_{ij}\bx_i.
\]

\begin{definition}[Empirical barycentric projection of a transport plan]\label{def:empirical-barycentric-projection}
Let $\Gamma$ be a transport plan between empirical measures $\widehat\mu$ and $\widehat\nu$. The source-to-target barycentric projection is the map
\[
  T_\Gamma(\bx_i)=\frac{1}{a_i}\sum_j\Gamma_{ij}\by_j.
\]
The target-to-source barycentric projection is
\[
  S_\Gamma(\by_j)=\frac{1}{b_j}\sum_i\Gamma_{ij}\bx_i.
\]
\end{definition}

Barycentric projections are useful when the transport plan splits mass. They create a deterministic approximation to a probabilistic matching. In actuarial work, they can be used to:
\begin{itemize}
  \item visualize how the old book moves toward the new one;
  \item identify which variables account for the largest shift;
  \item transfer source residuals to target profiles;
  \item construct synthetic target books under a changed portfolio mix;
  \item compare real target exposure to synthetic climate or economic scenarios.
\end{itemize}

\begin{example}[Barycentric shift in a household portfolio]
Suppose $\bX=(\log\text{insured value},\text{hazard score})$. A target book has higher insured values and higher hazard scores. The barycentric displacement
\[
  T_\Gamma(\bx_i)-\bx_i
\]
can be averaged componentwise to quantify whether the shift is mainly due to property values, hazard exposure, or their interaction.
\end{example}

\subsection{Distributional validation}\label{subsec:drift-distributional-validation}

A model can have a good average error and still fail distributionally. In actuarial science, the tails matter. A risk model may be well calibrated on average but underpredict high-risk policyholders. A reserving model may be unbiased at the aggregate level but produce residuals with a changing tail. A climate model may match expected annual loss while missing the upper tail of catastrophe loss.

Let $R=L-f(\bX)$ be a residual or scaled residual. Distributional validation compares
\[
  \Law_s(R)
  \quad\text{and}\quad
  \Law_t(R).
\]
In one dimension, the diagnostic
\[
  \Wtwo^2(\Law_s(R),\Law_t(R))
  =\int_0^1 |Q_s(u)-Q_t(u)|^2\dd u
\]
can be decomposed visually through the quantile-difference curve.

For count data, one may work with randomized or deviance residuals; randomized quantile residuals are a useful reference construction \cite{dunn1996randomized}. For severity data, one may work with log-residuals:
\[
  R=\log(L+1)-\log(f(\bX)+1).
\]
For reserving, one may use cell-level Pearson residuals or accident-year residual summaries.

\begin{definition}[Distributional calibration]\label{def:distributional-calibration}
A fitted model $f$ is distributionally stable between source and target if the residual laws are close:
\[
  \Wp{p}\bigl(\Law_s(R),\Law_t(R)\bigr)
\]
is small relative to a chosen benchmark, where $R$ is an appropriate residual or transformed residual.
\end{definition}

\begin{remark}[Choice of residual]
There is no universal residual. For Poisson frequency models, Pearson or deviance residuals are natural. For Gamma severity models, log-ratios may be more interpretable. For models producing positive fitted values, one may compare ratios $L/f(\bX)$, capped ratios, or randomized quantile residuals. The residual must be chosen before the Wasserstein diagnostic is interpreted.
\end{remark}

\subsection{A transport view of model recalibration}\label{subsec:drift-model-recalibration}

Suppose a fitted model produces scores $S=f(\bX)$. The target score distribution may differ from the source score distribution. Sometimes this is expected because the target portfolio is different. Sometimes it indicates miscalibration. A recalibration map is a function
\[
  r:\R\to\R
\]
so that $r(S)$ has a desired target distribution.

In one dimension, the monotone transport map from $\mu_S$ to $\nu_S$ is
\[
  T(s)=Q_\nu(F_\mu(s)),
\]
where $F_\mu$ is the source score distribution function and $Q_\nu$ is the target quantile function. Applying $T$ to source scores transforms the source score distribution into the target score distribution.

This is useful for post-processing, and is closely related to calibration and post-processing ideas in predictive modelling \cite{zadrozny2002transforming,guo2017calibration}, but it must be interpreted carefully. If the target distribution is a benchmark, monotone recalibration can enforce distributional agreement. If the target distribution reflects new risk, forcing equality to an old distribution can hide genuine drift.

\iffalse
% To be restored in v2 with the premium-distribution material.
\begin{example}[Premium scale recalibration]
An insurer updates expenses and inflation factors. The new premium distribution is expected to be higher but should preserve ranking. A monotone map $T$ applied to old premiums can implement a distributional recalibration:
\[
  \Premium_{\mathrm{new}}=T(\Premium_{\mathrm{old}}).
\]
If $T$ is increasing, policyholder ranking is preserved. If $T$ is smooth and close to the identity, the recalibration is commercially stable.
\end{example}
\fi

\begin{proposition}[Monotone one-dimensional recalibration]\label{prop:monotone-one-dimensional-recalibration}
Let $\mu$ and $\nu$ be probability measures on $\R$, and assume $\mu$ is continuous. Define
\[
  T(x)=Q_\nu(F_\mu(x)).
\]
If $X\sim\mu$, then $T(X)\sim\nu$. Moreover, $T$ is nondecreasing.
\end{proposition}

\begin{proof}
If $\mu$ is continuous, then $F_\mu(X)$ is uniform on $(0,1)$. Therefore $Q_\nu(F_\mu(X))$ has distribution $\nu$. Since both $F_\mu$ and $Q_\nu$ are nondecreasing, their composition is nondecreasing.
\end{proof}

\begin{remark}[Atoms and randomized ranks]\label{rem:recalibration-atoms-randomized-ranks}
If $\mu$ has atoms, $F_\mu(X)$ is not uniform. One can either use the generalized inverse construction directly at the level of quantiles, or introduce randomized ranks $F_\mu(X-)+U\bigl(F_\mu(X)-F_\mu(X-)\bigr)$, where $U\sim\mathrm{Unif}(0,1)$ is independent. This is the same convention as in the rank--quantile discussion in Lecture~\ref{lec:introduction-ot-actuarial}.
\end{remark}

\subsection{Support, extrapolation and lack of overlap}\label{subsec:drift-support-extrapolation}

Transport distances do not only give a number. An optimal plan identifies where support is missing. If target points $\by_j$ are far from all source points, then any transport plan must move mass over a large distance. This is a warning sign: the model is extrapolating.

A simple diagnostic is the nearest-source distance
\[
  d_s(\by_j)=\min_i d(\bx_i,\by_j).
\]
A transport-based version uses the reverse barycentric source representative
\[
  S_\Gamma(\by_j)=\frac{1}{b_j}\sum_i\Gamma_{ij}\bx_i,
\]
and the displacement
\[
  \Delta_j=d(\by_j,S_\Gamma(\by_j)).
\]
Large values of $\Delta_j$ indicate target profiles that are poorly represented in the source book.

\begin{definition}[Transport extrapolation score]\label{def:transport-extrapolation-score}
Given a transport plan $\Gamma$, the transport extrapolation score of target policy $j$ is
\[
  e_j=d\bigl(\by_j,S_\Gamma(\by_j)\bigr).
\]
The portfolio-level extrapolation score may be summarized by
\[
  \left(\sum_{j=1}^m b_j e_j^p\right)^{1/p}
\]
or by upper quantiles of the distribution of $e_j$.
\end{definition}

\begin{actuarialbox}
A high Wasserstein distance is not only evidence that the portfolio changed. The optimal plan tells where it changed, especially through Definition~\ref{def:transport-extrapolation-score}. In model governance, the local transport displacement is often more useful than the scalar distance.
\end{actuarialbox}

\subsection{Scenario transport and climate adaptation}\label{subsec:drift-climate-scenario-transport}

Climate applications often involve a baseline distribution and one or several scenario distributions. Let $\bH$ denote a hazard or exposure vector: flood depth, wildfire score, heat stress, wind speed, or a combination of physical and insured-value variables. A baseline climate-exposure distribution is $\mu_{\bH}$, and a scenario distribution is $\nu_{\bH}$.

A transport plan from $\mu_{\bH}$ to $\nu_{\bH}$ can be interpreted as a scenario displacement. If losses are generated by a vulnerability function
\[
  L=g(\bH,\bV),
\]
where $\bV$ denotes vulnerability or insured value, then the change in loss distribution can be studied by pushing forward the transported exposures through $g$.

For example, if a prevention measure modifies the hazard distribution from $\nu_{\bH}$ to $\nu_{\bH}^{\mathrm{prev}}$, one can measure the avoided distributional shift by
\[
  \Wp{p}(\mu_{\bH},\nu_{\bH})-\Wp{p}(\mu_{\bH},\nu_{\bH}^{\mathrm{prev}}),
\]
or, more directly, by comparing the induced loss laws
\[
  \Wp{p}\bigl(g_{\push}\nu_{\bH},
  g_{\push}\nu_{\bH}^{\mathrm{prev}}\bigr).
\]

\begin{example}[Avoided drift]
A coastal portfolio faces a future flood scenario $\nu_{\bH}$. A prevention investment changes the scenario distribution to $\nu_{\bH}^{\mathrm{prev}}$. If the induced annual-loss laws are $\mu_L^{\mathrm{future}}$ and $\mu_L^{\mathrm{prev}}$, then
\[
  \Wone(\mu_L^{\mathrm{future}},\mu_L^{\mathrm{prev}})
\]
measures the average distributional loss displacement avoided by prevention. Tail-specific diagnostics may use quantile differences or TVaR differences in addition to Wasserstein distance.
\end{example}

\begin{actuarialbox}
The quantities above compare baseline, scenario and prevention distributions. They do not identify the loss avoided by a particular insured unit unless a causal or physical intervention model justifies that interpretation.
\end{actuarialbox}

\subsection{Domain adaptation in reserving}\label{subsec:drift-domain-adaptation-reserving}

In reserving, drift appears across accident years, development years and calendar years. Let $C_{ij}$ denote incremental paid or incurred claims for accident year $i$ and development year $j$. A reserving model may be trained on older cells and used on recent incomplete triangles. Covariates include
\[
  \bX_{ij}=(\text{accident year},\text{development year},\text{calendar year},\text{line},\text{exposure},\ldots).
\]
The source distribution is the empirical distribution of observed mature cells. The target distribution is the empirical distribution of cells to be predicted.

Transport diagnostics can ask:
\begin{itemize}
  \item Are target cells close to observed cells in accident-development-calendar space?
  \item Does the distribution of fitted residuals change by calendar year?
  \item Are recent accident years a transport extrapolation from older years?
  \item Does inflation create a systematic displacement in severity residuals?
\end{itemize}

A simple target-cell extrapolation score is
\[
  e_{ij}=d\bigl(\bX_{ij},S_\Gamma(\bX_{ij})\bigr),
\]
where $S_\Gamma(\bX_{ij})$ is the source representative transported to the target cell. Large $e_{ij}$ suggests that the reserve prediction relies on extrapolation rather than interpolation.

\subsection{Algorithmic workflow}\label{subsec:drift-algorithmic-workflow}

A practical Wasserstein drift audit can be organized as follows.

\begin{enumerate}[label=\textbf{Step \arabic*.}]
  \item \textbf{Choose the representation.} Decide whether to compare covariates, model scores, losses, residuals, or several of them.
  \item \textbf{Choose the metric.} Standardize continuous variables, encode categorical variables, and choose weights reflecting actuarial relevance.
  \item \textbf{Compute scalar distances.} Compute $\Wone$, $\Wtwo$, sliced Wasserstein or entropic OT distances.
  \item \textbf{Inspect local displacements.} Use an optimal or regularized plan to identify where mass moves.
  \item \textbf{Compare output drift.} Check whether covariate drift propagates to score and residual drift.
  \item \textbf{Decide the action.} No action, monitoring, recalibration, reweighting, local model update, or full refit.
\end{enumerate}

\begin{computebox}
In high dimension, exact empirical $\Wp{p}$ can be unstable and computationally expensive. For monitoring, sliced Wasserstein distances, random projections, entropic regularization, or low-dimensional score summaries are often more robust. Exact transport plans are most useful for small or medium diagnostic samples.
\end{computebox}

\subsection{Mini-lab: score drift and residual drift in R}\label{subsec:drift-mini-lab-r}

The following pseudo-code illustrates a minimal workflow. It assumes a fitted model \texttt{fit}, a source data frame \texttt{source}, a target data frame \texttt{target}, and a scalar response \texttt{loss}. The function \texttt{quantile\_w2} computes a one-dimensional empirical $\Wtwo$ distance using quantiles.

\begin{lstlisting}[language=R]
emp_q_left <- function(x, probs) {
  x <- sort(x[is.finite(x)])
  n <- length(x)
  if (n == 0) stop("empty sample")
  x[pmax(1, ceiling(probs * n))]
}

quantile_w2 <- function(x, y, grid = seq(0.001, 0.999, length.out = 999)) {
  qx <- emp_q_left(x, grid)
  qy <- emp_q_left(y, grid)
  sqrt(mean((qx - qy)^2))
}

source$score <- predict(fit, newdata = source, type = "response")
target$score <- predict(fit, newdata = target, type = "response")

D_score <- quantile_w2(source$score, target$score)

source$resid <- log(source$loss + 1) - log(source$score + 1)
target$resid <- log(target$loss + 1) - log(target$score + 1)

D_resid <- quantile_w2(source$resid, target$resid)

list(score_drift = D_score, residual_drift = D_resid)
\end{lstlisting}

A useful plot is the quantile-difference curve.

\begin{lstlisting}[language=R]
grid <- seq(0.01, 0.99, length.out = 99)
qs <- emp_q_left(source$score, grid)
qt <- emp_q_left(target$score, grid)
plot(grid, qt - qs, type = "l",
     xlab = "Quantile level",
     ylab = "Target score quantile - source score quantile")
abline(h = 0, lty = 2)
\end{lstlisting}

If the curve is positive only at high quantiles, the target book has a heavier high-score tail. If it is approximately flat and positive, the target distribution is mostly shifted upward. If it crosses zero, the drift changes the shape rather than only the level.

\subsection{Governance interpretation}\label{subsec:drift-governance}

Transport diagnostics can be attached to model governance thresholds. For example, an insurer may define:
\begin{itemize}
  \item a green zone: score and residual drifts are below historical monitoring thresholds;
  \item an orange zone: score drift is high but residual drift is stable, suggesting portfolio mix change but no immediate evidence of conditional failure;
  \item a red zone: residual drift or tail residual drift is high, suggesting model degradation;
  \item a black zone: target extrapolation scores are high, suggesting the model is used outside its support.
\end{itemize}

The thresholds should be calibrated on historical backtests. The absolute value of $\Wp{p}$ is not meaningful without a scale. A distance of $100$ monetary units may be small for commercial property losses and large for small personal-lines payments.

\begin{readingbox}
For mathematical foundations of Wasserstein distances and empirical transport, see Lectures~\ref{lec:wasserstein-barycenters} and~\ref{lec:statistical-ot}, together with \cite{villani2008oldnew,peyre2019computational,panaretos2020invitation}. For domain adaptation and dataset shift, see \cite{quinonero2009dataset,sugiyama2012machine,pan2010survey,ben2010theory}. For OT-based domain adaptation, see \cite{courty2014domain,courty2017optimal,redko2017theoretical}. For concept drift, see \cite{gama2014survey,webb2016characterizing}. In actuarial applications, the relevant connection is not the transport algorithm alone but the choice of representation: covariates, scores, losses, residuals or cash-flow vectors.
\end{readingbox}

\subsection{Exercises}\label{subsec:drift-exercises}

\begin{exercise}[Covariate shift and expected loss]\label{ex:10:1}
Let $X\in\R$, and suppose covariate shift holds with stable conditional mean $m(x)=\alpha+\beta x$. Show that
\[
  \E_t[L]-\E_s[L]=\beta\{\E_t[X]-\E_s[X]\}.
\]
Then compute $\Wtwo(\Law_s(X),\Law_t(X))$ when $X_s\sim N(\mu_s,\sigma^2)$ and $X_t\sim N(\mu_t,\sigma^2)$. Relate the result to the expected-loss difference.
\end{exercise}

\begin{exercise}[Push-forward drift]\label{ex:10:2}
Let $f:\mathcal X\to\R$ be $K$-Lipschitz. Prove directly, using couplings, that
\[
  \Wp{p}(f_{\push}\mu,f_{\push}\nu)
  \le K\Wp{p}(\mu,\nu).
\]
Give an actuarial example where the inequality is strict.
\end{exercise}

\iffalse
% To be restored in v2 with the premium-distribution and portfolio demutualization material.
\begin{exercise}[Premium drift decomposition]\label{ex:10:3}
Let $\Premium_0$ and $\Premium_1$ be two tariffs applied to the same target book $X\sim\nu$. Define
\[
  \mu_0=\Law(\Premium_0(X)),
  \qquad
  \mu_1=\Law(\Premium_1(X)).
\]
Show that
\[
  \Wone(\mu_0,\mu_1)
  \le \E_\nu[|\Premium_1(X)-\Premium_0(X)|].
\]
Interpret the inequality as the difference between distributional change and policy-level disruption.
\end{exercise}
\fi

\begin{exercise}[Transport-transfer predictor]\label{ex:10:4}
Consider source points $\bx_1,\bx_2$, target points $\by_1,\by_2,\by_3$, source weights $\ba=(1/2,1/2)$, target weights $\bb=(1/3,1/3,1/3)$, and a feasible coupling $\Gamma$. Write the constraints on $\Gamma$. Given source predictions $h(\bx_1)=10$ and $h(\bx_2)=20$, compute $h_\Gamma(\by_j)$ for a coupling of your choice.
\end{exercise}

\begin{exercise}[Residual drift]\label{ex:10:5}
Suppose a Poisson frequency model produces fitted means $\widehat\lambda_i$. Define Pearson residuals
\[
  R_i=\frac{N_i-\widehat\lambda_i}{\sqrt{\widehat\lambda_i}}.
\]
Explain how you would compare the residual distribution between two calendar years using $\Wtwo$. What patterns in the quantile-difference curve would suggest inflation in claim frequency? What patterns would suggest a tail problem?
\end{exercise}

\begin{exercise}[Choice of metric]\label{ex:10:6}
A motor portfolio contains age, vehicle value, region and prior-claims count. Propose a distance $d(\bx,\by)$ for comparing two policies. Discuss how changing the weight of region affects the resulting Wasserstein drift and the interpretation of the optimal transport plan.
\end{exercise}

\begin{exercise}[Scenario transport]\label{ex:10:7}
Let $H$ be a scalar hazard index and let $L=g(H)=\exp(H)$. Suppose a baseline scenario has $H_0\sim N(0,1)$, a future scenario has $H_1\sim N(\delta,1)$, and a prevention scenario has $H_p\sim N(\delta/2,1)$. Compute $\Wtwo(\Law(H_0),\Law(H_1))$ and $\Wtwo(\Law(H_0),\Law(H_p))$. Discuss why the corresponding Wasserstein distances between loss distributions are not simply divided by two.
\end{exercise}

\newpage
\lecture{Reserving, Cash-Flow Distributions and Climate Prevention}
\label{lec:reserving-climate-prevention}

\subsection*{Learning objectives}

The previous lectures used optimal transport to compare portfolios, to monitor distributional drift, and to adapt actuarial models across domains. This lecture turns to two actuarial objects where the same distributional point of view is especially natural: outstanding claims reserves and climate-related loss distributions. In both cases, the actuary is not only interested in one number. A best estimate is useful, but it hides timing, tail risk, dependence between future payments, model uncertainty, and the effect of prevention or adaptation measures. Optimal transport provides a language for comparing entire predictive distributions, interpolating between reserving models, and describing how a specified prevention model moves mass in a loss distribution.

After this lecture, the reader should be able to:
\begin{enumerate}[label=(\roman*)]
  \item represent a reserve as a predictive distribution of future cash-flows rather than as a single scalar;
  \item compare reserving models by Wasserstein distances between predictive laws;
  \item use quantile formulas to compare one-dimensional reserve distributions;
  \item define barycenters of predictive reserving distributions and interpret them as consensus forecasts;
  \item distinguish calendar-time inflation, accident-year heterogeneity and development uncertainty in a transport framework;
  \item use transport maps to describe the distributional effect of a prevention measure;
  \item summarize prevention through changes in mean, quantiles, tail risk and Wasserstein displacement;
  \item understand why unbalanced transport is relevant when exposure mass changes;
  \item design diagnostic plots and computational experiments for reserving and climate applications.
\end{enumerate}

\begin{actuarialbox}
A reserve is often reported as a number, but it is produced by a distribution. Climate prevention is often reported as an avoided expected loss, but a prevention model changes an entire loss distribution. Optimal transport is useful because it keeps the distribution visible.
\end{actuarialbox}

This lecture uses the push-forward notation of Definition~\ref{def:push-forward}, the coupling notation of Definition~\ref{def:coupling}, the Wasserstein distance of Definition~\ref{def:wasserstein-distance}, the barycenter construction of Definition~\ref{def:wasserstein-barycenter}, and the empirical viewpoint of Lecture~\ref{lec:statistical-ot}. It also prepares the reserving cash-flow lab in Appendix~\ref{lab:reserving-cashflow} and the later unbalanced-transport chapter, where changing exposure mass is treated directly.
% The reference to the premium-drift lab is postponed to v2, together with the material on portfolio demutualization.

Classical and stochastic reserving references include \cite{mack1993distribution,renshaw1998stochastic,england2002stochastic,wuthrich2008stochastic}. The accounting language around future cash flows and uncertainty connects naturally with IFRS~17; see \cite{ifrs2017insurance}. For climate risk, prevention, catastrophe risk financing and index insurance, useful complements are \cite{kunreuther2009weather,kousky2019role,cummins2009catastrophe,barnett2008poverty,miranda1991area,clarke2016theory}.

\subsection{From best estimate reserves to predictive laws}\label{subsec:reserve-predictive-laws}

Let $R$ denote the total outstanding liability for a line of business. Classical reserving reports a best estimate,
\[
  \BEL = \E[R \mid \mathcal D],
\]
where $\mathcal D$ denotes the observed claims development data. The best estimate is central for accounting and solvency reporting, but it is only one functional of the predictive distribution
\[
  \mu_R = \Law(R\mid \mathcal D).
\]
The law $\mu_R$ contains information about uncertainty, skewness, tail risk and model sensitivity. Two reserving models can have the same best estimate but very different tail distributions.

In many applications the future liability is not a single payment but a vector of future cash-flows. Let
\[
  \bR=(R_1,\ldots,R_T)\in\R_+^T,
\]
where $R_t$ is the future payment during development or calendar period $t$. A predictive reserving model produces a probability measure
\[
  \mu_{\bR}=\Law(\bR\mid\mathcal D)
  \quad\text{on }\R_+^T.
\]
The total outstanding reserve is the push-forward of this distribution by the aggregation map
\[
  A(\br)=A(r_1,\ldots,r_T)=\sum_{t=1}^T r_t.
\]
Thus
\[
  \mu_R = A\push\mu_{\bR}.
\]
This simple identity is useful. It says that the total reserve distribution is obtained by pushing the cash-flow distribution through a deterministic map. It also makes clear that two models can agree on $\mu_R$ but disagree on timing, liquidity, calendar-year capital needs or reinsurance recoveries.

\begin{definition}[Predictive reserving law]\label{def:predictive-reserving-law}
A predictive reserving law is a probability measure $\mu\in\Prob(\R_+^T)$ describing the conditional distribution of future payments $(R_1,\ldots,R_T)$ given the observed claims data. The induced total-reserve law is $A\push\mu$, where $A(\br)=\sum_t r_t$.
\end{definition}

\begin{example}[Same best estimate, different risk]
Consider two predictive laws for a total reserve $R$. Under model A,
\[
  R_A = 100 + 10Z,
\]
where $Z$ is centered and moderately dispersed. Under model B,
\[
  R_B = 100 + Y,
\]
where $Y$ is centered but has a heavy right tail. Both models may have $\E[R_A]=\E[R_B]=100$, but $\TVaR_{0.995}(R_B)$ can be much larger than $\TVaR_{0.995}(R_A)$. A best-estimate comparison would miss this difference; a distributional comparison would not.
\end{example}

\begin{remark}[Cash-flow risk is not only reserve risk]
The scalar total $R=\sum_t R_t$ loses information about timing. A model with early payments and a model with late payments may have the same total reserve law but different liquidity consequences. Conversely, models may have similar payment timing but different total tail risk. For this reason, it is often useful to compare both $\mu_{\bR}$ and $A\push\mu_{\bR}$.
\end{remark}

\begin{actuarialbox}
\textbf{Professional lens: IFRS 17 and distributional cash flows.}

IFRS 17 shifts the accounting view of insurance contracts toward current estimates of future cash flows, discounting, uncertainty and the emergence of insurance service results over time. In these notes, this is naturally expressed through random cash-flow vectors rather than through a single deterministic reserve.

Let
\[
  \bC=(C_1,\ldots,C_T)
\]
denote the future cash-flow vector of a group of contracts, including claims, expenses, premiums, recoveries or benefits. A best estimate is a functional of $\Law(\bC)$, for example
\[
  BEL=\E\left[\sum_{t=1}^T v_t C_t\right].
\]
A risk adjustment, by contrast, depends on the uncertainty around this estimate. Optimal transport provides a geometry for comparing two projected cash-flow laws:
\[
  W_p\bigl(\Law(\bC^{\mathrm{old}}),\Law(\bC^{\mathrm{new}})\bigr),
\]
or for comparing a baseline projection to a stressed projection.

This is useful when model changes, assumption updates or portfolio changes affect not only the mean cash flow but also the timing, dependence and tail shape of the cash-flow distribution.
\end{actuarialbox}

\subsection{Triangles as empirical measures}\label{subsec:triangles-empirical-measures}

A run-off triangle can be viewed as an array of incremental payments
\[
  C_{i,j},
  \qquad i=1,\ldots,I,
  \quad j=1,\ldots,J,
\]
where $i$ is accident year and $j$ is development year. The observed region is typically
\[
  i+j\le I+1.
\]
Classical methods impose structure on the conditional expectation of unobserved cells. For example, a multiplicative development model may write
\[
  \E[C_{i,j}] = \alpha_i\beta_j,
\]
possibly with an exposure offset or a calendar-year effect. GLM versions replace this by a link function and a variance function.

The transport viewpoint starts by turning triangle information into a measure. There are several choices, depending on the question.

\paragraph{A measure over cells.}
Let $\bx_{i,j}\in\R^d$ encode the location and features of cell $(i,j)$: accident year, development year, calendar year, line of business, inflation index, exposure and possibly covariates. Define
\[
  \widehat\mu_C
  = \frac{1}{M}\sum_{(i,j)\in\mathcal O} C_{i,j}\,\delta_{\bx_{i,j}},
  \qquad
  M=\sum_{(i,j)\in\mathcal O} C_{i,j},
\]
where $\mathcal O$ is the observed region. This measure describes where paid mass has appeared in the triangle.

\begin{definition}[Triangle payment measure]\label{def:triangle-payment-measure}
Let $\mathcal O$ be the observed part of a run-off triangle and let $\bx_{i,j}$ encode the actuarial location of cell $(i,j)$. The normalized triangle payment measure is
\[
  \widehat\mu_C
  =\frac{1}{\sum_{(i,j)\in\mathcal O} C_{i,j}}
  \sum_{(i,j)\in\mathcal O} C_{i,j}\,\delta_{\bx_{i,j}}.
\]
It is a finite measure normalized to total mass one. If the total paid amount itself matters, the unnormalized version is a finite measure and should be compared with unbalanced OT.
\end{definition}

\paragraph{A measure over development patterns.}
For each accident year, define a normalized development profile
\[
  \bp_i = \left(\frac{C_{i,1}}{\sum_j C_{i,j}},\ldots,
  \frac{C_{i,J}}{\sum_j C_{i,j}}\right),
\]
whenever the full row is observed or completed. The empirical law
\[
  \widehat\mu_p = \frac{1}{n}\sum_i \delta_{\bp_i}
\]
is a distribution over development patterns. A distance between rows can then measure changes in reporting or settlement speed.

\paragraph{A measure over predictive samples.}
If a stochastic reserving model produces simulations
\[
  \bR^{(1)},\ldots,\bR^{(N)}\in\R_+^T,
\]
then the predictive law is represented empirically as
\[
  \widehat\mu_R=\frac1N\sum_{k=1}^N \delta_{\bR^{(k)}}.
\]
This is the representation most directly connected to Wasserstein distances.

\begin{actuarialbox}
There is no unique ``transport measure'' associated with a triangle. The measure must be chosen according to the actuarial question: movement of paid mass across cells, changes in development patterns, or uncertainty of future cash-flows.
\end{actuarialbox}

\subsection{Comparing predictive reserve distributions}\label{subsec:reserve-distribution-comparison}

Suppose two reserving models produce predictive laws $\mu$ and $\nu$ on $\R_+^T$. Examples include:
\begin{itemize}
  \item chain-ladder bootstrap versus Bayesian over-dispersed Poisson;
  \item lognormal reserving versus Tweedie GLM reserving;
  \item model with calendar-year inflation versus model without calendar-year inflation;
  \item model trained on all historical years versus model trained on recent years only;
  \item baseline climate scenario versus stressed climate scenario.
\end{itemize}

A Wasserstein distance compares these laws by solving
\[
  \Wp{p}^p(\mu,\nu)
  =\inf_{\pi\in\Coup(\mu,\nu)}
  \int \|\br-\bs\|^p\,\pi(\dd \br,\dd \bs).
\]
Here $\br,\bs\in\R_+^T$ are future cash-flow vectors. The norm should reflect the actuarial meaning of differences across payment years. The simplest choice is Euclidean:
\[
  \|\br-\bs\|^2 = \sum_{t=1}^T (r_t-s_t)^2.
\]
A discounted choice is often more meaningful:
\[
  \|\br-\bs\|_{\bv}^2 = \sum_{t=1}^T v_t^2(r_t-s_t)^2,
\]
where $v_t$ is a discount factor or an economic weight. A tail-oriented choice may put more weight on late liability years or high-volatility payment periods.

\begin{definition}[Cash-flow ground cost]\label{def:cash-flow-ground-cost}
A cash-flow ground cost is a function $c:\R_+^T\times\R_+^T\to[0,\infty]$ that measures the actuarial discrepancy between two future payment vectors. Common choices are
\[
  c_p(\br,\bs)=\sum_{t=1}^T |r_t-s_t|^p,
  \qquad
  c_{p,\bv}(\br,\bs)=\sum_{t=1}^T v_t^p |r_t-s_t|^p,
\]
with economic weights $v_t$. The choice of $c$ determines whether the comparison emphasizes nominal amount, discounted value, timing, liquidity or capital relevance.
\end{definition}

\begin{definition}[Discounted Wasserstein distance for cash-flows]\label{def:discounted-wasserstein-cashflows}
Let $\bv=(v_1,\ldots,v_T)\in\R_+^T$. The discounted Wasserstein distance between predictive cash-flow laws $\mu,\nu\in\Probp{p}(\R_+^T)$ is
\[
  \Wp{p,\bv}^p(\mu,\nu)
  =\inf_{\pi\in\Coup(\mu,\nu)}
  \int \left(\sum_{t=1}^T v_t^p |r_t-s_t|^p\right)\pi(\dd \br,\dd \bs).
\]
\end{definition}

\begin{proposition}[Aggregation is Lipschitz]\label{prop:reserve-aggregation-lipschitz}
Let $A(\br)=\sum_{t=1}^T r_t$. If $\mu,\nu\in\Probp{p}(\R_+^T)$, then
\[
  \Wp{p}(A\push\mu,A\push\nu)
  \le T^{1-1/p}\Wp{p}(\mu,\nu),
\]
with the convention $T^{1-1/p}=1$ for $p=1$.
\end{proposition}

\begin{proof}
For any $\br,\bs\in\R_+^T$, H\"older's inequality gives
\[
  |A(\br)-A(\bs)|
  =\left|\sum_{t=1}^T(r_t-s_t)\right|
  \le T^{1-1/p}\|\br-\bs\|_p.
\]
Thus $A$ is $T^{1-1/p}$-Lipschitz. The contraction property of Wasserstein distances under Lipschitz maps gives the result.
\end{proof}

\begin{proposition}[Present-value stability]\label{prop:present-value-lipschitz}
Let $\bv=(v_1,\ldots,v_T)\in\R_+^T$ and define the present-value map
\[
  \operatorname{PV}_{\bv}(\br)=\sum_{t=1}^T v_t r_t.
\]
If $\mu,\nu\in\Probp{p}(\R_+^T)$ and the ground metric on $\R_+^T$ is the $\ell_p$ norm, then
\[
  \Wp{p}\bigl((\operatorname{PV}_{\bv})\push\mu,(\operatorname{PV}_{\bv})\push\nu\bigr)
  \le \|\bv\|_q\,\Wp{p}(\mu,\nu),
\]
where $q$ is the conjugate exponent of $p$. In particular, discounted best-estimate distributions are stable under small Wasserstein perturbations of cash-flow laws.
\end{proposition}

\begin{proof}
For all $\br,\bs\in\R_+^T$, H\"older's inequality gives
\[
  |\operatorname{PV}_{\bv}(\br)-\operatorname{PV}_{\bv}(\bs)|
  \le \|\bv\|_q\|\br-\bs\|_p.
\]
Thus $\operatorname{PV}_{\bv}$ is $\|\bv\|_q$-Lipschitz. The conclusion follows from the contraction property of Wasserstein distances under Lipschitz maps.
\end{proof}

This proposition says that if two predictive cash-flow laws are close, then their induced total reserve laws are close. Proposition~\ref{prop:present-value-lipschitz} gives the analogous statement for discounted present values. The converse is false: total reserves or present values can be close while the time pattern of payments differs substantially.

\subsection{One-dimensional reserve comparison}\label{subsec:one-dimensional-reserve-comparison}

For scalar reserve distributions, the Wasserstein distance has a particularly transparent quantile representation. If $R\sim\mu$ and $S\sim\nu$, then
\[
  \Wp{p}^p(\mu,\nu)
  =\int_0^1 |F_R^{-1}(u)-F_S^{-1}(u)|^p\,\dd u.
\]
Here, as throughout the notes, $F^{-1}$ denotes the lower generalized inverse. With atoms or tied empirical values, the formula should be read with this convention or with an explicit randomized-rank convention. This identity is useful for communicating model differences. It converts an abstract optimal coupling into a quantile-by-quantile comparison.

Let $q_R(u)=Q_{\mu}(u)=F_R^{-1}(u)$ and $q_S(u)=Q_{\nu}(u)=F_S^{-1}(u)$. The function
\[
  d(u)=q_S(u)-q_R(u)
\]
shows where the model difference lies. If $d(u)$ is nearly constant, then one model is essentially a shifted version of the other. If $d(u)$ is small near the center and large near $u=1$, then the models mostly differ in the tail. If $d(u)$ changes sign, the models differ in shape, not merely in level.

\begin{example}[Tail disagreement]
Suppose two reserving models have the same median and the same mean, but one has a heavier tail. Then $q_R(u)\approx q_S(u)$ for central $u$, but $|q_R(u)-q_S(u)|$ becomes large near $u=1$. The $\Wtwo$ distance captures this disagreement, but a tail-weighted diagnostic such as
\[
  \int_\alpha^1 |q_R(u)-q_S(u)|^2\,\dd u
\]
may be more directly connected to solvency concerns.
\end{example}

\begin{definition}[Tail Wasserstein discrepancy]\label{def:tail-wasserstein-discrepancy}
For $\alpha\in(0,1)$, define the tail discrepancy
\[
  \Wp{p,\alpha}^p(\mu,\nu)
  =\frac{1}{1-\alpha}\int_\alpha^1
  |Q_\mu(u)-Q_\nu(u)|^p\,\dd u.
\]
This is not a Wasserstein distance on the full distribution, but it is a useful diagnostic for tail model uncertainty.
\end{definition}

\begin{remark}[Quantiles are actuarially legible]
A quantile plot is often easier to explain than a transport plan. For scalar reserves, plotting $q_R(u)$, $q_S(u)$, and $q_S(u)-q_R(u)$ is one of the simplest ways to diagnose whether two predictive models disagree in level, dispersion or tail.
\end{remark}

\subsection{Barycenters of reserving models}\label{subsec:reserve-barycenters}

Reserving often involves several plausible models rather than one objectively correct model. Let
\[
  \mu_1,\ldots,\mu_K
\]
be predictive laws produced by $K$ reserving methods, expert assumptions, inflation scenarios or climate scenarios. A Wasserstein barycenter is a distribution $\bar\mu$ solving
\[
  \bar\mu\in\operatorname*{arg\,min}_{\eta\in\Probp{2}(\R_+^T)}
  \sum_{k=1}^K \omega_k \Wtwo^2(\eta,\mu_k),
\]
where $\omega_k\ge0$ and $\sum_k\omega_k=1$.

\begin{definition}[Predictive reserve barycenter]\label{def:predictive-reserve-barycenter}
A predictive reserve barycenter is a Wasserstein barycenter of several predictive reserving laws. It represents a consensus distribution, not an average of best estimates.
\end{definition}

In one dimension the barycenter is especially simple. If $q_k(u)=F_k^{-1}(u)$, then the $\Wtwo$-barycenter has quantile function
\[
  \bar q(u)=\sum_{k=1}^K\omega_k q_k(u).
\]
Therefore the barycenter preserves the quantile structure of the models. It is not the distribution of a mixture that first selects a model and then simulates from it. A mixture has quantile function that is generally not the weighted average of quantile functions.

\begin{example}[Mixture versus barycenter]
Let $\mu_1$ and $\mu_2$ be two reserve distributions. The mixture
\[
  \mu_{\mathrm{mix}}=\frac12\mu_1+
  \frac12\mu_2
\]
represents model uncertainty: one of the two models is true, but we do not know which. The barycenter represents a geometric compromise between the two predictive laws. In scalar problems, its quantile is the average of the two quantiles. The mixture can be more dispersed than either model, while the barycenter interpolates between corresponding quantile levels.
\end{example}

\begin{actuarialbox}
Averaging best estimates gives one number. Mixing predictive distributions gives model uncertainty. Taking a Wasserstein barycenter gives a consensus predictive distribution. These are three different operations and should not be confused.
\end{actuarialbox}

\subsection{Model uncertainty and reserve capital}\label{subsec:reserve-model-uncertainty}

A predictive law can be summarized by risk measures. For a scalar reserve $R$, one may compute
\[
  \E[R],\qquad \VaR_\alpha(R),\qquad \TVaR_\alpha(R),
\]
or a cost-of-capital margin based on a future solvency capital requirement. If several predictive models are available, a distributional approach can separate three effects:
\begin{enumerate}[label=(\alph*)]
  \item variation in best estimates $\E_{\mu_k}[R]$;
  \item variation in tail risk $\VaR_\alpha(\mu_k)$, $\TVaR_\alpha(\mu_k)$;
  \item geometric disagreement between predictive laws $\Wtwo(\mu_k,\mu_\ell)$.
\end{enumerate}

A model-risk diagnostic can be built from pairwise distances
\[
  D_{k\ell}=\Wtwo(\mu_k,\mu_\ell).
\]
Large values indicate that models produce substantially different predictive laws. A model may have a best estimate close to the consensus while still being far in Wasserstein distance because it has a different tail or timing structure.

\begin{definition}[Wasserstein model dispersion]\label{def:wasserstein-model-dispersion}
Given predictive laws $\mu_1,\ldots,\mu_K$ and weights $\omega_k$, define the Wasserstein model dispersion around a barycenter $\bar\mu$ by
\[
  \mathcal D_W
  =\sum_{k=1}^K \omega_k \Wtwo^2(\mu_k,\bar\mu).
\]
\end{definition}

\begin{remark}[A distributional analogue of variance]
The quantity $\mathcal D_W$ plays the role of a variance of predictive distributions around their barycenter. It is not a variance of losses under one model. It is a measure of disagreement across models.
\end{remark}

\subsection{Transporting cash-flow scenarios}\label{subsec:cash-flow-scenario-transport}

Suppose $\mu$ is a baseline predictive cash-flow law and $\nu$ is a stressed predictive law. An optimal coupling $\pi\in\Coup(\mu,\nu)$ pairs baseline scenarios $\br$ with stressed scenarios $\bs$. If $\pi$ is induced by a map $T$, so that $\nu=T\push\mu$, then
\[
  \bs=T(\br)
\]
may be interpreted as the stressed version of the baseline cash-flow scenario $\br$.

In practice, empirical predictive laws are finite samples:
\[
  \widehat\mu=\frac1N\sum_{i=1}^N \delta_{\br_i},
  \qquad
  \widehat\nu=\frac1M\sum_{j=1}^M \delta_{\bs_j}.
\]
A discrete transport plan $\pi_{ij}$ is a soft matching between baseline scenarios and stressed scenarios. The barycentric projection
\[
  \widehat T(\br_i)
  =\frac{\sum_{j=1}^M \pi_{ij}\bs_j}{\sum_{j=1}^M \pi_{ij}}
\]
gives a transported stressed scenario associated with baseline scenario $\br_i$.

This can be used diagnostically. For each baseline scenario $\br_i$, the displacement
\[
  \Delta_i=\widehat T(\br_i)-\br_i
\]
indicates how stress affects payment timing. If $\Delta_i$ is concentrated in early development years, the stress accelerates payments. If it is concentrated in late years, the stress increases long-tail uncertainty.

\begin{computebox}
In finite samples, exact optimal transport can be unstable when $N$ and $M$ are large or when many scenarios are close. Entropic regularization gives smoother transport plans and more stable barycentric projections. For scenario diagnostics, this smoothing is often a feature rather than a defect.
\end{computebox}

\subsection{Calendar-year inflation as a transport effect}\label{subsec:calendar-inflation-transport}

Inflation is one of the main reasons why reserve distributions move over time. Let $R_t$ be a future payment in calendar year $t$, and suppose an inflation scenario multiplies payments by factors $a_t>0$. The inflated cash-flow vector is
\[
  T_{\ba}(\br)=T_{\ba}(r_1,\ldots,r_T)=(a_1r_1,\ldots,a_Tr_T).
\]
The stressed predictive law is
\[
  \mu_{\ba}=T_{\ba}\push\mu.
\]
If all factors are equal to $a$, then scalar total reserve satisfies
\[
  R_a=aR,
\]
and the Wasserstein distance between $\Law(R)$ and $\Law(aR)$ can be expressed through quantiles:
\[
  \Wp{p}^p(\Law(R),\Law(aR))
  =\int_0^1 |q_R(u)-a q_R(u)|^p\,\dd u
  =|a-1|^p\E[|R|^p].
\]
For nonconstant inflation by payment year, the total reserve is
\[
  A(T_{\ba}(\br))=\sum_{t=1}^T a_t r_t,
\]
and timing matters. A long-tailed line of business is more exposed to late inflation factors than a short-tailed line, even if both have the same undiscounted best estimate.

\begin{example}[Inflation and payment timing]
Consider two portfolios with the same total reserve $R=100$. Portfolio A pays everything next year. Portfolio B pays everything in ten years. If the annual inflation shock affects only late years, then $A$ and $B$ have the same current best estimate but different stressed distributions. The cash-flow law detects this; the scalar current reserve does not.
\end{example}

\subsection{Climate loss distributions}\label{subsec:climate-loss-distributions}

Climate risk is naturally distributional. Let $L$ denote annual aggregate loss for a portfolio exposed to a climate hazard. A baseline climate or exposure scenario produces a loss law
\[
  \mu_0=\Law(L_0),
\]
whereas a future scenario produces
\[
  \mu_1=\Law(L_1).
\]
Classical summaries include
\[
  \Delta \E[L]=\E[L_1]-\E[L_0],
  \qquad
  \Delta\VaR_\alpha=\VaR_\alpha(L_1)-\VaR_\alpha(L_0),
\]
\[
  \Delta\TVaR_\alpha=\TVaR_\alpha(L_1)-\TVaR_\alpha(L_0).
\]
These summaries are important, but they do not describe how the whole distribution changes. Wasserstein distance provides a complementary measure:
\[
  \Wp{p}(\mu_0,\mu_1).
\]
In one dimension, the quantile displacement
\[
  d(u)=Q_{\mu_1}(u)-Q_{\mu_0}(u)
\]
is especially informative. It reveals whether climate change shifts the whole distribution, increases moderate losses, or mainly thickens the tail.

\begin{definition}[Climate distributional shift]\label{def:climate-distributional-shift}
Let $\mu_0$ be a baseline loss law and $\mu_1$ a scenario loss law. The climate distributional shift at order $p$ is $\Wp{p}(\mu_0,\mu_1)$, possibly supplemented by the quantile displacement function $Q_{\mu_1}-Q_{\mu_0}$.
\end{definition}

\begin{actuarialbox}
The same change in expected annual loss can correspond to very different changes in insurability. A small increase in mean with a large increase in tail risk may matter more for capital and reinsurance than a larger increase in mean concentrated in attritional losses.
\end{actuarialbox}

\subsection{Prevention as a map between loss distributions}\label{subsec:prevention-map-loss-distributions}

A prevention or adaptation measure changes the loss distribution. Examples include:
\begin{itemize}
  \item flood barriers reducing the frequency and severity of inundation losses;
  \item roof strengthening reducing wind losses;
  \item wildfire defensible space reducing total-loss probability;
  \item cyber hygiene reducing ransomware frequency and recovery cost;
  \item road safety interventions reducing bodily injury severity.
\end{itemize}

Let $\mu_{\mathrm{before}}$ and $\mu_{\mathrm{after}}$ denote loss distributions before and after the measure. A deterministic prevention model may be written as
\[
  L_{\mathrm{after}}=T(L_{\mathrm{before}}),
\]
so that
\[
  \mu_{\mathrm{after}}=T\push\mu_{\mathrm{before}}.
\]
The map $T$ may reduce large losses more than small losses. For example,
\[
  T(x)=x-\delta(x),
\]
where $\delta(x)\ge0$ is increasing in $x$, represents a prevention measure whose avoided loss is larger in severe events.

In many cases the prevention effect is not deterministic. Then the measure is transformed by a Markov kernel $K$:
\[
  \mu_{\mathrm{after}}(B)
  =\int K(x,B)\,\mu_{\mathrm{before}}(\dd x).
\]
The corresponding coupling
\[
  \pi(\dd x,\dd y)=K(x,\dd y)\mu_{\mathrm{before}}(\dd x)
\]
pairs pre-intervention losses with post-intervention losses. It need not be the optimal coupling. It supports a causal or physical interpretation only when the kernel is derived from a credible intervention model, experimental design or structural assumption.

\begin{actuarialbox}
A transport map compares two distributions. It does not by itself identify avoided losses for the same event or policyholder. Causal avoided-loss claims require a baseline, an intervention model and evidence linking the intervention to the observed or projected shift.
\end{actuarialbox}

\begin{definition}[Avoided-loss map]\label{def:avoided-loss-map}
A prevention map is a measurable map $T:\R_+\to\R_+$ such that $T(x)\le x$. The avoided-loss function is
\[
  a(x)=x-T(x).
\]
The post-prevention loss law is $T\push\mu$, where $\mu$ is the pre-prevention law.
\end{definition}

\begin{proposition}[Prevention and expected avoided loss]\label{prop:prevention-expected-avoided-loss}
If $T(x)\le x$ and $L\sim\mu$, then
\[
  \E[L]-\E[T(L)]=\E[a(L)].
\]
Moreover, if $T$ is nondecreasing, then
\[
  \VaR_\alpha(T(L))=T(\VaR_\alpha(L))
\]
for all levels $\alpha$ at which the lower quantile convention and the monotone transformation are compatible, in particular at continuity levels of the distribution of $L$.
\end{proposition}

\begin{proof}
The identity for expectations follows from linearity. If $T$ is nondecreasing, the quantile of the push-forward distribution is obtained by applying $T$ to the quantile of the original distribution at continuity levels, with the usual generalized-inverse convention. This gives the stated relation.
\end{proof}

The condition $T(x)\le x$ is natural for a loss-reduction intervention, but it is not automatic in real data. A measure may reduce severe claims but increase small reported claims, for example because more losses are detected or because deductibles change. In such cases the prevention map may cross the identity.

\subsection{Measuring prevention}\label{subsec:measuring-prevention}

A prevention evaluation should not rely on a single metric. The following quantities capture complementary aspects:
\[
  \Delta_{\mathrm{mean}}
  =\E[L_{\mathrm{before}}]-\E[L_{\mathrm{after}}],
\]
\[
  \Delta_{\VaR,\alpha}
  =\VaR_\alpha(L_{\mathrm{before}})-\VaR_\alpha(L_{\mathrm{after}}),
\]
\[
  \Delta_{\TVaR,\alpha}
  =\TVaR_\alpha(L_{\mathrm{before}})-\TVaR_\alpha(L_{\mathrm{after}}),
\]
\[
  \Delta_W=\Wtwo(\mu_{\mathrm{before}},\mu_{\mathrm{after}}).
\]
The first three are reductions in actuarial risk functionals. The last one measures the size of the distributional movement. A large $\Delta_W$ does not necessarily imply a large reduction in risk: the distribution may move in shape without much mean reduction. Conversely, a large mean reduction concentrated in a narrow part of the distribution may produce a moderate Wasserstein distance.

\begin{definition}[Prevention profile]\label{def:prevention-profile}
For scalar loss distributions, the prevention profile is the quantile-level function
\[
  \mathcal A(u)=Q_{\mathrm{before}}(u)-Q_{\mathrm{after}}(u),
  \qquad u\in(0,1).
\]
Positive values indicate avoided loss at quantile level $u$. The profile is tail-oriented if $\mathcal A(u)$ is increasing near $u=1$.
\end{definition}

\begin{example}[Two prevention measures with same mean effect]
Suppose two interventions reduce expected annual loss by the same amount. Measure A reduces many small claims slightly. Measure B leaves most small claims unchanged but reduces extreme losses. They may have the same $\Delta_{\mathrm{mean}}$, but Measure B may have much larger $\Delta_{\TVaR,\alpha}$ and a prevention profile concentrated near $u=1$. From a solvency or reinsurance perspective, the two measures are not equivalent.
\end{example}

\subsection{Counterfactual prevention and attribution}\label{subsec:counterfactual-prevention-attribution}

A prevention claim is always counterfactual: it compares observed or projected losses with losses that would have occurred without the intervention. Let
\[
  L^{0}\quad\text{and}\quad L^{1}
\]
be potential losses without and with prevention. The distributional effect is the pair
\[
  \Law(L^0),\qquad \Law(L^1).
\]
If the two potential losses can be coupled at the event level, one may study avoided losses
\[
  A=L^0-L^1.
\]
If only marginal laws are available, optimal transport can provide a cost-minimizing coupling, but it should not automatically be interpreted causally. It is a geometric matching of quantiles or scenarios, not evidence by itself.

\begin{remark}[Transport coupling versus causal coupling]\label{rem:transport-versus-causal-prevention-coupling}
The optimal coupling between pre- and post-prevention distributions minimizes a mathematical cost. A causal coupling pairs the same underlying event under two potential worlds. These couplings may coincide under monotone rank preservation, but they need not coincide in general.
\end{remark}

This distinction is important in climate-risk accounting. If a prevention project is evaluated by comparing distributions before and after, the Wasserstein map describes how one distribution can be moved into the other. To claim avoided losses, one also needs a credible baseline and evidence that the intervention caused the shift.

\subsection{Exposure changes and unbalanced transport}\label{subsec:exposure-unbalanced-transport}

Classical optimal transport assumes equal total mass. In actuarial climate applications, this assumption may fail. Exposure can be created, destroyed or relocated:
\begin{itemize}
  \item new properties are built in high-risk zones;
  \item properties are removed through buyout programs;
  \item insured values change through inflation or reconstruction cost;
  \item policyholders enter or leave the insurance market;
  \item deductibles and limits alter the insured portion of losses.
\end{itemize}

If $\mu$ and $\nu$ are finite measures with different total masses, exact marginal constraints are too rigid. Unbalanced optimal transport relaxes the constraints. A common formulation is
\[
  \inf_{\pi\ge0}
  \int c(\bx,\by)\,\pi(\dd \bx,\dd \by)
  +\lambda_1 D(\pi_1\mid \mu)
  +\lambda_2 D(\pi_2\mid \nu),
\]
where $\pi_1,\pi_2$ are the marginals of $\pi$, and $D$ is a divergence such as relative entropy or total variation. The parameters $\lambda_1,\lambda_2$ control the cost of creating or destroying mass. The residual-mass interpretation depends on the chosen divergence: with entropic penalties, for example, the fitted marginals need not be literal submeasures of the original measures.

\begin{actuarialbox}
Balanced transport asks: how do we move the same portfolio from one distribution to another? Unbalanced transport asks: how do we move, create and remove exposure mass? The second question is often closer to climate adaptation and insurance availability.
\end{actuarialbox}

\begin{example}[Managed retreat]
Consider a flood portfolio before and after a buyout program. Some exposed homes disappear from the insured portfolio. A balanced transport comparison would be forced to match their mass to other homes. An unbalanced formulation can instead pay a penalty for removing that mass, making the interpretation closer to the intervention.
\end{example}

\subsection{Parametric triggers and basis risk}
\label{sec:parametric-basis-risk-ot}

Traditional indemnity insurance pays according to the realized loss. Parametric insurance pays according to an index or trigger. For example, a catastrophe contract may pay when wind speed, rainfall, river height, ground acceleration or drought index crosses a pre-specified threshold. The advantage is operational: payments can be fast, transparent and less costly to administer. The difficulty is actuarial: the index payment need not match the policyholder's actual loss.

Let \(L\) denote the economic loss and let \(I\) denote the parametric index. A parametric contract is a measurable function
\[
  P = g(I),
\]
where \(g\) is the payout function. For example,
\[
  g(i)
  =
  C\,(i-u)_+\wedge M
\]
for a trigger level \(u\), a payout slope \(C\), and a maximum payment \(M\). The basis risk is the mismatch between the loss \(L\) and the payment \(P\). The signed residual
\[
  B = L-P
\]
is useful, but no single scalar residual summarizes basis risk. One usually needs to distinguish underpayment, overpayment, timing mismatch, spatial mismatch and tail mismatch.

\begin{definition}[Basis-risk law and shortfall law]\label{def:basis-risk-shortfall-law}
Let $L$ be the loss and $P=g(I)$ the parametric payment. The basis-risk law is the distribution of $B=L-P$. The shortfall law is the distribution of
\[
  S=(L-P)_+.
\]
A distributional basis-risk audit should report both the symmetric mismatch between $\Law(L)$ and $\Law(P)$ and the directional shortfall law $\Law(S)$.
\end{definition}

Optimal transport provides a useful way to compare the law of actual losses and the law of parametric payments:
\[
  W_p\bigl(\Law(L),\Law(P)\bigr).
\]
This distance is zero only when the two one-dimensional distributions coincide. In dimension one it compares quantiles:
\[
  W_p^p\bigl(\Law(L),\Law(P)\bigr)
  =
  \int_0^1
  \left|Q_L(t)-Q_P(t)\right|^p\,\dd t.
\]
Thus the discrepancy is not only a mean error. It records where the parametric contract under- or over-reproduces the loss distribution.

\begin{actuarialbox}
Parametric insurance can be audited as a comparison between two distributions: the distribution of actual losses and the distribution of index-based payments. Basis risk is not only the expectation of \(L-P\); it also includes the geometry and direction of the mismatch between \(\Law(L)\) and \(\Law(g(I))\).
\end{actuarialbox}

A useful decomposition separates calibration risk and structural basis risk. Suppose
the actuary can choose the payout function \(g\) in a class \(\mathcal G\). A
distributional calibration problem is
\[
  \inf_{g\in\mathcal G}
  W_p^p\bigl(\Law(L),\Law(g(I))\bigr).
\]
If \(\mathcal G\) contains only simple trigger functions, the infimum may remain
large. This remaining distance is not sampling error; it is the distributional cost of
using the index rather than the loss itself.

One may also focus on underpayment, which is often more important than symmetric
basis risk. Define the shortfall
\[
  S=(L-P)_+.
\]
A tail-sensitive diagnostic is
\[
  \TVaR_\alpha(S)
  =
  \frac{1}{1-\alpha}\int_\alpha^1 Q_S(t)\,\dd t.
\]
Alternatively, one can compare the upper tails of \(L\) and \(P\) through
\[
  W_{p,\alpha}^{\mathrm{tail}}\bigl(\Law(L),\Law(P)\bigr)^p
  =
  \frac{1}{1-\alpha}
  \int_\alpha^1
  \left|Q_L(t)-Q_P(t)\right|^p\,\dd t.
\]

\begin{remark}[Basis risk is directional]
The distance \(W_p(\Law(L),\Law(P))\) is symmetric, but policyholder welfare is
not. Underpayment and overpayment have different meanings. For contract design,
one should therefore report both a distributional distance and directional quantities
such as \(\E[(L-P)_+]\), \(\TVaR_\alpha((L-P)_+)\), and
\(\mathbb P(P<L)\).
\end{remark}

In spatial catastrophe insurance, the mismatch may be multivariate. If components are indexed by locations, write
\[
  \bL=(L_1,\ldots,L_d),\qquad \bP=(P_1,\ldots,P_d).
\]
A direct vector cost such as
\[
  c(\bl,\bp)=\sum_{j=1}^d w_j |l_j-p_j|^p
\]
measures amount mismatch at each location. If spatial misallocation matters, it is often better to represent losses and payments as measures over locations, for example
\[
  \lambda_L=\sum_{j=1}^d L_j\delta_{\bz_j},\qquad \lambda_P=\sum_{j=1}^d P_j\delta_{\bz_j},
\]
and to use a ground cost based on geographical or hazard distance between locations. This separates amount mismatch from the question of whether payment mass is allocated to the right places.

\begin{computebox}
A practical parametric-insurance diagnostic can report:
\[
  W_1(\Law(L),\Law(P)),\qquad
  W_{1,\alpha}^{\mathrm{tail}}(\Law(L),\Law(P)),\qquad
  \E[(L-P)_+],
  \qquad
  \TVaR_\alpha((L-P)_+).
\]
The first two are distributional comparisons. The last two are directional
policyholder shortfall measures.
\end{computebox}

\subsection{A small computational lab}\label{subsec:reserving-climate-computational-lab}

The following pseudo-code illustrates a simple one-dimensional comparison of reserve distributions. The same structure can be used for climate loss distributions.

\begin{lstlisting}[language=R]
set.seed(123)

n <- 5000
R1 <- rlnorm(n, meanlog = log(100), sdlog = 0.25)
R2 <- rlnorm(n, meanlog = log(102), sdlog = 0.40)

u <- ((1:n) - 0.5) / n
q1 <- sort(R1)
q2 <- sort(R2)

W1 <- mean(abs(q1 - q2))
W2 <- sqrt(mean((q1 - q2)^2))

emp_qleft <- function(x, alpha) {
  x <- sort(x)
  x[max(1, ceiling(alpha * length(x)))]
}

VaR <- function(x, alpha) emp_qleft(x, alpha)
TVaR <- function(x, alpha) mean(x[x >= VaR(x, alpha)])

alpha <- 0.995
out <- c(
  mean_1 = mean(R1),
  mean_2 = mean(R2),
  VaR_1  = VaR(R1, alpha),
  VaR_2  = VaR(R2, alpha),
  TVaR_1 = TVaR(R1, alpha),
  TVaR_2 = TVaR(R2, alpha),
  W1 = W1,
  W2 = W2
)
print(round(out, 2))

plot(u, q2 - q1, type = "l",
     xlab = "quantile level u",
     ylab = "q2(u) - q1(u)",
     main = "Quantile displacement between reserve laws")
abline(h = 0, lty = 2)
\end{lstlisting}

For a cash-flow distribution in $\R_+^T$, one may represent simulations as an $N\times T$ matrix and compute an empirical transport distance using an optimal transport package. Even without solving full multivariate OT, comparing one-dimensional marginals year by year is a useful diagnostic:
\[
  \Wtwo(\Law(R_t^{(1)}),\Law(R_t^{(2)})),
  \qquad t=1,\ldots,T.
\]
This does not capture dependence across future years, but it is easy to explain and often reveals whether model differences are concentrated in early or late payments.

\subsection{Practical workflow}\label{subsec:reserving-climate-workflow}

A distributional workflow for reserving and prevention can be organized as follows.

\paragraph{Step 1: choose the object.}
Decide whether the comparison concerns total reserves, cash-flow vectors, triangle cell mass, development patterns, annual aggregate climate losses or event-level losses.

\paragraph{Step 2: define the metric.}
Choose the cost function. For cash-flows, the cost should reflect discounting, capital relevance and timing. For spatial climate losses, the cost may combine geographic distance, hazard intensity and insured value.

\paragraph{Step 3: compute scalar summaries.}
Report changes in best estimate, $\VaR$, $\TVaR$, and possibly cost-of-capital quantities. These remain essential actuarial outputs.

\paragraph{Step 4: compute distributional summaries.}
Compute Wasserstein distances, barycenters, quantile displacement plots or scenario transport maps.

\paragraph{Step 5: separate diagnosis from attribution.}
A Wasserstein displacement diagnoses a distributional change. Attribution to a reserving assumption, inflation scenario or prevention measure requires additional modelling assumptions and evidence.

\begin{readingbox}
For reserving, this lecture connects naturally to stochastic claims reserving, GLM reserving, bootstrap reserving and Bayesian predictive reserving. For the transport side, the relevant tools are one-dimensional quantile transport, empirical Wasserstein distances, barycenters and entropic regularization. For climate prevention, the key distinction is between distributional comparison and causal avoided-loss attribution.
\end{readingbox}

\subsection{Exercises}\label{subsec:reserving-climate-exercises}

\begin{exercise}[Best estimate versus predictive law]\label{ex:11:1}
Construct two scalar reserve distributions with the same mean but different $\VaR_{0.995}$ and $\TVaR_{0.995}$. Compute their $\Wone$ and $\Wtwo$ distances using the quantile formula. Explain why the best estimate is insufficient.
\end{exercise}

\begin{exercise}[Aggregation contraction]\label{ex:11:2}
Prove the aggregation Lipschitz proposition for $p=2$ directly using the Cauchy--Schwarz inequality. Give an example showing that two cash-flow laws can have the same total-reserve law but different timing distributions.
\end{exercise}

\begin{exercise}[Reserve barycenter]\label{ex:11:3}
Let $\mu_1,\mu_2,\mu_3$ be three scalar predictive reserve laws with quantile functions $q_1,q_2,q_3$. Derive the $\Wtwo$-barycenter quantile for weights $(1/2,1/4,1/4)$. Compare the barycenter with the mixture $(1/2)\mu_1+(1/4)\mu_2+(1/4)\mu_3$.
\end{exercise}

\begin{exercise}[Inflation stress]\label{ex:11:4}
Let $\bR=(R_1,R_2,R_3)$ be a predictive cash-flow vector. Consider the inflation map $T_{\ba}(\br)=T_{\ba}(r_1,r_2,r_3)=(a_1r_1,a_2r_2,a_3r_3)$. Show that if $a_1=a_2=a_3=a$, then $\Wp{2}(\Law(A(\bR)),\Law(A(T_{\ba}(\bR))))=|a-1|\sqrt{\E[A(\bR)^2]}$. Discuss why the formula is more complicated when the $a_t$'s differ.
\end{exercise}

\begin{exercise}[Prevention profile]\label{ex:11:5}
Let $L$ be Pareto distributed and let $T(x)=\min(x,M)+\gamma(x-M)_+$ with $0\le\gamma\le1$. Interpret $T$ as a prevention map that reduces the excess above $M$. Derive the quantile function of $T(L)$ and plot the prevention profile $Q_\mu(u)-Q_{T\push\mu}(u)$.
\end{exercise}

\begin{exercise}[Balanced versus unbalanced comparison]\label{ex:11:6}
Give an actuarial example where exposure mass changes between two portfolios. Explain why balanced transport may give a misleading interpretation. Propose an unbalanced formulation by choosing a transport cost, a divergence term and a qualitative meaning for the penalty parameter.
\end{exercise}

\begin{exercise}[Mini-project]\label{ex:11:7}
Simulate two stochastic reserving models for future payments over five years. One model should increase total severity; the other should only delay payments. Compare them using: yearly means, total best estimate, total $\VaR$, year-by-year Wasserstein distances, and a Wasserstein distance between total reserve distributions. Write a short paragraph explaining which diagnostics detect level risk and which detect timing risk.
\end{exercise}

\newpage
\lecture{Fairness, Discrimination and Wasserstein Projections}
\label{lec:fairness-wasserstein-projections}

\subsection*{Learning objectives}

The previous lectures used optimal transport to compare risks, portfolios, reserving models, and climate scenarios. This lecture applies the same distributional geometry to fairness and discrimination in insurance. The central object is again a probability distribution: the distribution of scores, predicted premiums, residuals, losses, or decisions conditional on a sensitive attribute. Wasserstein distances provide an audit tool, by quantifying disparities between conditional laws, and a correction tool, by transporting group-specific distributions toward a common reference law. This lecture focuses on one-dimensional score distributions.
% Multi-sensitive and multivariate extensions are left in comments for v2.

After this lecture, the reader should be able to:
\begin{enumerate}[label=(\roman*)]
  \item distinguish individual risk classification, group disparity, proxy discrimination, and distributional fairness;
  \item formulate demographic parity and related constraints as equalities between conditional distributions;
  \item compute one-dimensional Wasserstein distances between group score distributions using quantile functions;
  \item define a Wasserstein barycenter of group distributions and interpret it as a fair reference distribution;
  \item construct monotone group-wise repair maps in dimension one;
  \item quantify the cost of fairness using Wasserstein displacement, changes in premiums, and changes in calibration;
  \item understand the tension between distributional parity and actuarial calibration;
  \item design empirical diagnostics and simulation exercises for insurance fairness audits.
\end{enumerate}

\begin{actuarialbox}
Fairness questions in insurance are not only questions about individual predictions. They are also questions about distributions: who receives which range of premiums, which groups are concentrated in which tail, and how much mass would have to be moved to remove a group disparity.
\end{actuarialbox}

\begin{actuarialbox}%[Distributional, not causal]
The transport maps in this lecture align conditional score distributions. They do not identify individual causal effects of group membership. A causal interpretation requires a structural model, an intervention design, or other identifying assumptions.
\end{actuarialbox}

\begin{readingbox}
For general algorithmic-fairness background, see \cite{dwork2012fairness,feldman2015certifying,hardt2016equality,chouldechova2017fair,kleinberg2017inherent,barocas2023fairness}. For counterfactual fairness, see \cite{kusner2017counterfactual}. Wasserstein and optimal-transport approaches to distributional fairness are developed in \cite{gordaliza2019obtaining,chzhen2020fair}. For insurance-specific discussions of discrimination and pricing, see \cite{lindholm2022discrimination,charpentier2024insurance}. The Wasserstein notions used here rely on the quantile and barycenter tools from Lectures~\ref{lec:continuous-ot} and~\ref{lec:wasserstein-barycenters}, while statistical uncertainty is discussed in Lecture~\ref{lec:statistical-ot}.
\end{readingbox}

\subsection{The actuarial setting}
\label{sec:fairness-actuarial-setting}

Let \(Y\) denote an insured outcome, such as claim frequency, claim severity, annual aggregate loss, or a binary claim indicator. Let \(\bX\) be the vector of rating variables available to the insurer and let \(S\) be a sensitive attribute, such as sex, age category, race, immigration status, or another legally relevant grouping variable. In many jurisdictions, some components of \(S\) may be legally restricted, socially contested, or acceptable only under specific conditions.

A predictive pricing model produces a score or premium
\[
  Z = \widehat m(\bX),
\]
where \(\widehat m\) is typically trained to approximate a risk quantity such as
\[
  m(\bx)=\E[Y\mid \bX=\bx].
\]
For a pure premium model, \(Z\) is often interpreted as an estimate of expected cost; when the score is a tariff or a technical premium, we write more specifically \(Z=\Premium(\bX)\). For a classification model, \(Z\) may be a probability of claim, a fraud score, or an underwriting score.

The actuarial argument for risk classification is conditional accuracy: under a cost-based pricing principle, different expected losses may lead to different premiums. Fairness concerns arise because this classification can reproduce, amplify, or hide group-level disparities. Even if \(S\) is excluded from the model, other variables may act as proxies for \(S\). The law of \(Z\) conditional on \(S=s\) can therefore differ substantially across groups.

\begin{definition}[Group-specific score distribution]\label{def:group-specific-score-distribution}
Let \(Z\) be a score or premium and let \(S\) take values in a finite set \(\mathcal S\). The group-specific score distribution is
\[
  \mu_s = \Law(Z\mid S=s), \qquad s\in\mathcal S.
\]
If \(p_s=\Pbb(S=s)\), the marginal score distribution is the mixture
\[
  \mu = \sum_{s\in\mathcal S} p_s \mu_s.
\]
\end{definition}

The objects \(\mu_s\) are the natural starting point for a distributional fairness audit. They describe whether the score distribution for one group is shifted upward, more dispersed, heavier tailed, or more concentrated in high-premium regions than the score distribution for another group.

\begin{example}[A proxy effect]
Suppose that \(S\in\{0,1\}\) is not used directly in the tariff, but a geographic variable \(G\) is strongly associated with \(S\). If \(Z=\widehat m(G,A,V)\), where \(A\) is age and \(V\) is vehicle value, then the conditional laws \(\Law(Z\mid S=0)\) and \(\Law(Z\mid S=1)\) may differ even though \(S\) never appears in the formula. Fairness through unawareness does not imply distributional fairness.
\end{example}

\begin{definition}[Proxy discrimination]
\label{def:proxy-discrimination}
A variable or score exhibits proxy discrimination with respect to a sensitive attribute \(S\) when it does not use \(S\) explicitly but remains strongly informative about \(S\), or reproduces a disparity associated with \(S\), through other variables. In a distributional audit, this is often visible through differences between the conditional laws \(\Law(Z\mid S=s)\).
\end{definition}

\subsection{From group averages to group distributions}
\label{sec:fairness-group-distributions}

A first audit may compare group averages,
\[
  \E[Z\mid S=s],\qquad s\in\mathcal S.
\]
This is useful but incomplete. Two groups can have the same average premium and very different distributions. One group may have many very low and very high premiums, while another group may be concentrated around the average. Conversely, two groups can have different averages but very similar distributions after a small location shift.

Distributional audits retain the full conditional law. For example, one may compare:
\begin{align*}
  \text{mean disparity:}\quad & \E[Z\mid S=s]-\E[Z\mid S=s'],\\
  \text{tail disparity:}\quad & \VaR_\alpha(Z\mid S=s)-\VaR_\alpha(Z\mid S=s'),\\
  \text{spread disparity:}\quad & \Var(Z\mid S=s)-\Var(Z\mid S=s'),\\
  \text{distributional disparity:}\quad & \Wp{p}(\mu_s,\mu_{s'}).
\end{align*}
The last quantity is especially informative because it is sensitive to the whole distribution, not only one functional; see Section~\ref{sec:wasserstein-quantile-formula} for the one-dimensional formula and Section~\ref{sec:wasserstein-barycenters} for barycenters.

\begin{definition}[Distributional demographic parity]\label{def:distributional-demographic-parity}
A score \(Z\) satisfies distributional demographic parity with respect to \(S\) if
\[
  \Law(Z\mid S=s)=\Law(Z\mid S=s')
\]
for all \(s,s'\in\mathcal S\). Equivalently, all group-specific score distributions \(\mu_s\) are identical.
\end{definition}

This condition is strong. In insurance, it should usually not be presented as the only definition of fairness. It is better understood as one benchmark: it asks what would be needed to remove all distributional information about \(S\) from \(Z\). The answer is useful even when the benchmark is not ultimately imposed.

\begin{proposition}[Distributional parity and independence]
\label{prop:distributional-parity-independence}
Assume that \(S\) is discrete and that \(\Pbb(S=s)>0\) for all \(s\in\mathcal S\). Then \(Z\) satisfies distributional demographic parity with respect to \(S\) if and only if \(Z\) is independent of \(S\).
\end{proposition}

\begin{proof}
If all conditional laws \(\Law(Z\mid S=s)\) are equal to a common law \(\bar\mu\), then for every measurable set \(B\),
\[
  \Pbb(Z\in B,S=s)=\Pbb(S=s)\bar\mu(B)=\Pbb(S=s)\Pbb(Z\in B),
\]
so \(Z\) and \(S\) are independent. The converse follows immediately from independence.
\end{proof}

\begin{remark}[Fairness constraints and fairness audits]
A fairness audit may compute a disparity without recommending that the disparity be set exactly to zero. This distinction is important. Wasserstein distances can quantify how different groups are treated by a score, while the final regulatory or ethical decision may involve additional considerations: cost-based pricing, social acceptability, legal constraints, adverse selection, transparency, and affordability.
\end{remark}

\subsection{Wasserstein disparity indices}
\label{sec:fairness-wasserstein-disparity}

Let \(\mu_s=\Law(Z\mid S=s)\) be group score distributions on \(\R\). A simple pairwise Wasserstein audit computes
\[
  D_{p}(s,s') = \Wp{p}(\mu_s,\mu_{s'}),
\]
or, when \(p=2\), the squared distance
\[
  D_{2}^2(s,s') = \Wtwo^2(\mu_s,\mu_{s'}).
\]
For one-dimensional distributions with finite second moments,
\[
  \Wtwo^2(\mu_s,\mu_{s'})
  = \int_0^1 \left\{ Q_s(u)-Q_{s'}(u)\right\}^2\dd u,
\]
where \(Q_s\) is the quantile function of \(\mu_s\).

Pairwise distances are useful when there are only two groups. With many groups, it is more convenient to compare each group with a common reference distribution. This leads to a barycentric disparity index.

\begin{definition}[Barycentric Wasserstein disparity]\label{def:barycentric-wasserstein-disparity}
Let \(p_s=\Pbb(S=s)\). A Wasserstein barycenter of the group distributions is any solution of
\[
  \bar\mu \in \argmin_{\nu\in\Probp{2}(\R)}
  \sum_{s\in\mathcal S} p_s \Wtwo^2(\mu_s,\nu).
\]
The total disparity around the barycenter is
\[
  D_{\mathrm{bar}}^2(Z,S)
  = \sum_{s\in\mathcal S} p_s \Wtwo^2(\mu_s,\bar\mu).
\]
The group contribution is
\[
  C_s = \Wtwo^2(\mu_s,\bar\mu).
\]
\end{definition}

In dimension one, the barycenter has an explicit quantile representation:
\[
  \bar Q(u) = \sum_{s\in\mathcal S} p_s Q_s(u),
  \qquad 0<u<1.
\]
Thus
\[
  D_{\mathrm{bar}}^2(Z,S)
  = \sum_s p_s\int_0^1\{Q_s(u)-\bar Q(u)\}^2\dd u.
\]
This is the weighted variance, across groups, of the group quantiles at each rank \(u\), integrated over \(u\).

\begin{proposition}[Quantile variance identity]\label{prop:quantile-variance-identity}
Suppose that \(Z\mid S=s\) has finite second moment for all \(s\). Let \(Q_s\) be the group quantile functions and let \(\bar Q=\sum_s p_s Q_s\). Then
\[
  D_{\mathrm{bar}}^2(Z,S)
  = \int_0^1 \sum_s p_s\{Q_s(u)-\bar Q(u)\}^2\dd u.
\]
Moreover, for two groups with weights \(p\) and \(1-p\),
\[
  D_{\mathrm{bar}}^2(Z,S)
  = p(1-p)\Wtwo^2(\mu_0,\mu_1).
\]
\end{proposition}

\begin{proof}
The first identity follows from the one-dimensional representation of the \(\Wtwo\) distance and the quantile formula for the barycenter. For two groups, \(\bar Q=pQ_1+(1-p)Q_0\), hence
\[
  p\{Q_1-\bar Q\}^2+(1-p)\{Q_0-\bar Q\}^2
  =p(1-p)(Q_1-Q_0)^2.
\]
Integrating over \(u\in(0,1)\) gives the result.
\end{proof}

\begin{actuarialbox}
The barycentric disparity index has a direct interpretation: it is the average squared movement required to bring each group score distribution to a common reference distribution. In one dimension, this movement is measured rank by rank.
\end{actuarialbox}

\subsection{Wasserstein projections and score repair}
\label{sec:fairness-wasserstein-repair}

The previous section described an audit. We now describe a correction. This uses the monotone rearrangement from Theorem~\ref{thm:monotone-rearrangement} and the barycenter construction from Definition~\ref{def:wasserstein-barycenter}. The goal is to transform \(Z\) into a new score \(Z^\star\) such that the conditional laws \(\Law(Z^\star\mid S=s)\) are identical across groups. Among all such transformations, the Wasserstein barycentric repair is the one obtained by transporting each group distribution to the barycenter.

Assume first that \(Z\mid S=s\) has a continuous distribution function \(F_s\). Let \(\bar F\) be the distribution function of the barycenter \(\bar\mu\), with quantile function \(\bar Q\). Define
\[
  T_s(z)=\bar Q(F_s(z)).
\]
Then, conditionally on \(S=s\),
\[
  Z^\star=T_s(Z)
\]
has distribution \(\bar\mu\). Therefore
\[
  \Law(Z^\star\mid S=s)=\bar\mu
\]
for all \(s\). This is a distributionally fair score. Any decision rule that depends only on \(Z^\star\) will then have the same conditional distribution across groups. This conclusion does not apply to rules that also use \(S\) or unrepaired covariates.

\begin{definition}[Barycentric repair]\label{def:barycentric-repair}
When the group conditional distribution functions are continuous, the barycentric Wasserstein repair of a one-dimensional score \(Z\) is the group-wise transformation
\[
  Z^\star = T_S(Z),
  \qquad T_s = \bar Q\circ F_s,
\]
where \(F_s\) is the group conditional distribution function of \(Z\) and \(\bar Q\) is the barycenter quantile function. With atoms, the same construction should be read through monotone transport plans, randomized ranks, or a documented empirical tie convention.
\end{definition}

The map \(T_s\) is monotone. It preserves ranks within each group: two policyholders in the same group keep their within-group ordering. This is often an attractive feature in insurance. The correction does not arbitrarily reshuffle policyholders within a group; it changes the scale on which group ranks are expressed.

\begin{proposition}[Optimality of monotone repair]\label{prop:optimality-monotone-repair}
Assume that \(\mu_s\) and \(\bar\mu\) belong to \(\Probp{2}(\R)\), and that either \(\mu_s\) is atomless or a randomized-rank convention is used. In dimension one, among all admissible transformations that send \(\mu_s\) to \(\bar\mu\), the monotone map \(T_s=\bar Q\circ F_s\) minimizes
\[
  \E\left[(T(Z)-Z)^2\mid S=s\right].
\]
The minimal value is \(\Wtwo^2(\mu_s,\bar\mu)\).
\end{proposition}

\begin{proof}
This is the one-dimensional optimal transport result for the squared cost. The increasing rearrangement transports \(\mu_s\) to \(\bar\mu\) and minimizes the expected squared displacement. The value of the minimum is the squared \(\Wtwo\) distance.
\end{proof}

\begin{remark}[Ties and empirical scores]
Insurance scores often have ties, for example because tariffs are rounded, binned, or table-based. Then \(F_s(Z)\) is not uniform without additional randomization. In practice one may use randomized ranks inside ties, mid-ranks, or a smoothed empirical distribution. The choice should be documented because it can affect individual corrections when many policyholders share the same tariff cell.
\end{remark}

\subsection{Premium balance and mean effects}
\label{sec:fairness-premium-balance}

Let \(Z\) be a premium and \(Z^\star\) its repaired version. Since all groups are mapped to the barycenter, the group mean after repair is
\[
  \E[Z^\star\mid S=s]=\int_0^1\bar Q(u)\dd u,
\]
which is the same for all groups. The marginal mean is preserved if the barycenter weights are the empirical group proportions:
\[
  \E[Z^\star]
  = \sum_s p_s\int_0^1\bar Q(u)\dd u
  = \int_0^1\sum_s p_s Q_s(u)\dd u
  = \sum_s p_s\E[Z\mid S=s]
  = \E[Z].
\]
Thus the barycentric repair is globally premium-balanced, at least before rounding, expenses, taxes, and other business constraints.

\begin{proposition}[Global mean preservation]\label{prop:global-mean-preservation}
Assume \(Z\mid S=s\) is integrable for all \(s\). If the barycenter uses weights \(p_s=\Pbb(S=s)\), then the one-dimensional barycentric repair satisfies
\[
  \E[Z^\star]=\E[Z].
\]
\end{proposition}

\begin{proof}
The proof follows from the quantile formula above and the identity \(\E[Z\mid S=s]=\int_0^1 Q_s(u)\dd u\) for integrable random variables.
\end{proof}

Mean preservation does not imply that all individual changes are small. Some policyholders may receive large decreases or increases. It is therefore useful to report the distribution of displacements
\[
  \Delta = Z^\star-Z,
\]
including its mean, standard deviation, quantiles, and tail values by group. In an actuarial context, one may also impose additional constraints, for example:
\begin{align*}
  \text{no global revenue change:}\quad & \E[Z^\star]=\E[Z],\\
  \text{limited individual change:}\quad & |Z^\star-Z|\leq b,\\
  \text{limited multiplicative change:}\quad & a\leq Z^\star/Z\leq A,\\
  \text{minimum premium:}\quad & Z^\star\geq z_{\min}.
\end{align*}
These constraints may break the simple barycentric formula. They lead to constrained projection problems, often still convex in discretized form.

\begin{example}[Two-group location shift]
Suppose
\[
  Z\mid S=0 \sim \mathcal N(100,10^2),\qquad
  Z\mid S=1 \sim \mathcal N(120,10^2),
\]
with group weights \(p_0=p_1=1/2\). The barycenter is \(\mathcal N(110,10^2)\). The repair decreases group 1 premiums by 10 and increases group 0 premiums by 10. This removes the distributional disparity while preserving the global average premium.
\end{example}

\begin{example}[Scale difference]
Suppose
\[
  Z\mid S=0 \sim \mathcal N(100,10^2),\qquad
  Z\mid S=1 \sim \mathcal N(100,30^2),
\]
again with equal weights. The means are equal, but the distributions are not. The barycenter has quantile function
\[
  \bar Q(u)=100+20\Phi^{-1}(u),
\]
so group 0 is spread out and group 1 is compressed. A mean-based audit would see no disparity; a distributional audit would.
\end{example}

\subsection{Calibration and the impossibility tension}
\label{sec:fairness-calibration-tension}

Distributional parity is not the only desirable property. In insurance, calibration is central. A score \(Z\) is calibrated, in a simplified regression sense, if
\[
  \E[Y\mid Z=z]=z.
\]
A group-calibrated score satisfies
\[
  \E[Y\mid Z=z,S=s]=z
\]
for all groups and relevant scores. Calibration says that a premium level should correspond to the same expected loss.

Distributional demographic parity and group calibration can be incompatible when groups have different base risks. The following proposition gives a simple version of this tension.

\begin{proposition}[A calibration and parity tension]\label{prop:calibration-parity-tension}
Assume that \(Z\) is integrable and satisfies group calibration,
\[
  \E[Y\mid Z,S]=Z.
\]
If \(Z\) also satisfies distributional demographic parity with respect to \(S\), then
\[
  \E[Y\mid S=s]
\]
is the same for all groups \(s\).
\end{proposition}

\begin{proof}
By iterated expectation,
\[
  \E[Y\mid S=s]
  = \E[\E[Y\mid Z,S=s]\mid S=s]
  = \E[Z\mid S=s].
\]
If \(Z\) has the same conditional distribution for all groups, then \(\E[Z\mid S=s]\) is the same for all groups. Hence the group expected losses must be equal.
\end{proof}

This result does not say that parity is wrong. It says that exact parity and exact group calibration cannot both hold unless group expected losses are equal. In practice, this creates a frontier: one can move toward parity at the cost of calibration, or preserve calibration at the cost of group disparities.

\begin{definition}[Calibration error after repair]\label{def:calibration-error-after-repair}
Let \(Z^\star\) be a repaired score. A group calibration diagnostic is
\[
  \mathrm{CalErr}_s
  = \E\left[\left(\E[Y\mid Z^\star,S=s]-Z^\star\right)^2\mid S=s\right].
\]
Empirically, this can be estimated by binning \(Z^\star\), fitting a calibration curve, or using a flexible regression of \(Y\) on \((Z^\star,S)\).
\end{definition}

\begin{actuarialbox}
A Wasserstein repair can remove a distributional disparity, but it may also change the actuarial meaning of the score. After any correction, one should check calibration, loss ratios, residual distributions, and the distribution of premium changes.
\end{actuarialbox}

\subsection{Partial repair and fairness frontiers}
\label{sec:fairness-partial-repair-frontiers}

Exact distributional parity may be too strong. A useful compromise is partial repair. The following formula is specific to the one-dimensional monotone repair path. Let \(T_s\) be the barycentric repair map and let \(\lambda\in[0,1]\). Define
\[
  Z^{(\lambda)} = (1-\lambda)Z + \lambda T_S(Z).
\]
For \(\lambda=0\), there is no correction. For \(\lambda=1\), there is full barycentric repair. Intermediate values trace a path from the original score to the distributionally fair score.

In one dimension, if each group is transformed by
\[
  T_{s,\lambda}(z)=(1-\lambda)z+\lambda T_s(z),
\]
then the group quantile functions along the path are
\[
  Q_{s,\lambda}(u)=(1-\lambda)Q_s(u)+\lambda\bar Q(u).
\]
Therefore the distance from group \(s\) to the barycenter shrinks linearly in \(\lambda\), and the squared distance shrinks quadratically:
\[
  \Wtwo^2(\mu_{s,\lambda},\bar\mu)
  =(1-\lambda)^2\Wtwo^2(\mu_s,\bar\mu).
\]
The total barycentric disparity is therefore
\[
  D_{\mathrm{bar}}^2(Z^{(\lambda)},S)
  =(1-\lambda)^2D_{\mathrm{bar}}^2(Z,S).
\]

\begin{proposition}[Quadratic decrease of disparity along the repair path]\label{prop:quadratic-decrease-repair-path}
Under the one-dimensional barycentric repair path above,
\[
  D_{\mathrm{bar}}^2(Z^{(\lambda)},S)
  =(1-\lambda)^2D_{\mathrm{bar}}^2(Z,S).
\]
\end{proposition}

\begin{proof}
The quantile of the repaired group distribution is \(Q_{s,\lambda}=(1-\lambda)Q_s+\lambda\bar Q\). Hence
\[
  Q_{s,\lambda}-\bar Q=(1-\lambda)(Q_s-\bar Q).
\]
Squaring, integrating, and summing over groups gives the result.
\end{proof}

This path is useful for drawing a fairness and accuracy or fairness and calibration frontier. For each \(\lambda\), one can compute:
\begin{align*}
  &\text{distributional disparity: }D_{\mathrm{bar}}(Z^{(\lambda)},S),\\
  &\text{predictive performance: }\E[\ell(Y,Z^{(\lambda)})],\\
  &\text{calibration error: }\mathrm{CalErr}_s(\lambda),\\
  &\text{business disruption: }\E[|Z^{(\lambda)}-Z|],\\
  &\text{maximum or tail premium change: }\VaR_\alpha(|Z^{(\lambda)}-Z|).
\end{align*}

\begin{remark}[Choice of loss function]
The loss function \(\ell\) should match the actuarial task. For frequency modeling one may use Poisson deviance; for binary outcomes, log-loss or Brier score; for pure premium regression, squared error, Gamma deviance, Tweedie deviance, or another scoring rule. The fairness correction should not be evaluated only by RMSE if the pricing model was trained under a different actuarial objective.
\end{remark}

\subsection{Residual distributions and loss-ratio diagnostics}
\label{sec:fairness-residual-loss-ratio}

Auditing the score distribution alone can be misleading. Suppose that one group has higher premiums because it has higher predicted risk, and another group has lower premiums because it has lower predicted risk. A distributional parity constraint would remove this premium difference, but it may generate systematic loss-ratio differences.

For this reason, a fairness audit should also examine residuals and loss ratios. Let
\[
  R = Y-Z,
  \qquad R^\star=Y-Z^\star.
\]
Group residual distributions
\[
  \Law(R\mid S=s), \qquad \Law(R^\star\mid S=s)
\]
show whether the tariff is systematically high or low for some groups. Similarly, if \(Z>0\), the loss ratio
\[
  LR=Y/Z
\]
or its aggregated version by group can be examined.

\begin{definition}[Distributional residual audit]\label{def:distributional-residual-audit}
A distributional residual audit compares the conditional laws
\[
  \Law(Y-Z\mid S=s)
\]
across groups, before and after any correction. A Wasserstein residual disparity can be defined as
\[
  D_{\mathrm{res}}^2
  = \sum_s p_s\Wtwo^2\left(\Law(Y-Z\mid S=s),\bar\nu\right),
\]
where \(\bar\nu\) is the Wasserstein barycenter of the residual distributions.
\end{definition}

Residual parity is not the same as score parity. A model may have equal score distributions but unequal residual distributions, or conversely unequal score distributions but well-balanced residuals. This distinction is important in insurance because fairness may concern both access to affordable premiums and cost-based adequacy.

\subsection{Empirical computation}
\label{sec:fairness-empirical-computation}

Suppose the data consist of observations
\[
  (z_i,s_i,y_i),\qquad i=1,\ldots,n.
\]
For each group \(s\), let \(z_{s,(1)}\leq\cdots\leq z_{s,(n_s)}\) be the ordered scores. The empirical quantile function is a step function. To compare groups with different sizes, one can evaluate all group quantile functions on a common grid
\[
  u_j=\frac{j-1/2}{M},\qquad j=1,\ldots,M.
\]
Then approximate
\[
  Q_s(u_j)\approx \widehat Q_s(u_j),
\]
compute
\[
  \widehat{\bar Q}(u_j)=\sum_s \widehat p_s\widehat Q_s(u_j),
\]
and estimate the barycentric disparity by
\[
  \widehat D_{\mathrm{bar}}^2
  = \frac{1}{M}\sum_{j=1}^M\sum_s \widehat p_s
  \left\{\widehat Q_s(u_j)-\widehat{\bar Q}(u_j)\right\}^2.
\]

The repaired score for an observation in group \(s\) can be obtained by replacing its within-group rank with the barycenter quantile at the same rank. For observation \(i\), define an empirical rank
\[
  \widehat U_i = \widehat F_{s_i}(z_i),
\]
or a mid-rank version. Then
\[
  \widehat z_i^\star = \widehat{\bar Q}(\widehat U_i).
\]

\begin{computebox}
In empirical work, use the same quantile grid for all groups. This makes Wasserstein distances, barycenters, partial repairs, and plots all consistent. It also avoids comparing empirical quantiles at incompatible ranks.
\end{computebox}

\subsection{A minimal R lab}
\label{sec:fairness-minimal-r-lab}

The following code illustrates a one-dimensional barycentric repair. It is the in-chapter counterpart of the dedicated Wasserstein fairness repair lab in the appendices, and it should be read together with the statistical cautions in Lecture~\ref{lec:statistical-ot}. It is intentionally simple. In a production audit, one would add sample splitting, bootstrap uncertainty, calibration curves, and careful treatment of ties.

\begin{lstlisting}[language=R]
set.seed(1)

n0 <- 800
n1 <- 600
s  <- c(rep(0, n0), rep(1, n1))
z  <- c(rlnorm(n0, log(600), .25),
        rlnorm(n1, log(760), .35))

# common quantile grid
M <- 1000
u <- (seq_len(M) - .5) / M

q_emp_left <- function(x, probs) {
  x <- sort(x)
  n <- length(x)
  x[pmax(1, ceiling(n * probs))]
}

q0 <- q_emp_left(z[s == 0], u)
q1 <- q_emp_left(z[s == 1], u)

p0 <- mean(s == 0)
p1 <- mean(s == 1)
qbar <- p0 * q0 + p1 * q1

# barycentric disparity
D2 <- mean(p0 * (q0 - qbar)^2 + p1 * (q1 - qbar)^2)
sqrt(D2)

# empirical ranks within groups
rank01 <- function(x) rank(x, ties.method = "average") / (length(x) + 1)
u_hat <- numeric(length(z))
u_hat[s == 0] <- rank01(z[s == 0])
u_hat[s == 1] <- rank01(z[s == 1])

# interpolation of barycenter quantile
z_star <- approx(u, qbar, xout = u_hat, rule = 2)$y

# partial repair
lambda <- .5
z_half <- (1 - lambda) * z + lambda * z_star

aggregate(cbind(original = z, repaired = z_star,
                half = z_half), by = list(group = s), mean)

# premium changes
delta <- z_star - z
aggregate(delta, by = list(group = s),
          function(x) quantile(x, c(.05, .5, .95)))
\end{lstlisting}

This code builds two group-specific lognormal score distributions, computes their quantile barycenter, repairs the scores by rank matching, and summarizes group mean changes. The same structure can be used with fitted GLM, GAM, random forest, or gradient boosting predictions.

\subsection{A small numerical illustration}
\label{sec:fairness-numerical-illustration}

Consider two groups with equal weights and scores
\[
  Z_0\sim \mathcal N(600,100^2),
  \qquad
  Z_1\sim \mathcal N(750,150^2).
\]
The barycenter is Gaussian with quantile function
\[
  \bar Q(u)=\frac{1}{2}\{600+100\Phi^{-1}(u)\}
          +\frac{1}{2}\{750+150\Phi^{-1}(u)\}
          =675+125\Phi^{-1}(u).
\]
Thus the repair maps are affine:
\[
  T_0(z)=675+125\frac{z-600}{100},
  \qquad
  T_1(z)=675+125\frac{z-750}{150}.
\]
The first group is shifted upward and made more dispersed. The second group is shifted downward and compressed. The squared \(\Wtwo\) distance between the two original distributions is
\[
  \Wtwo^2(\mu_0,\mu_1)=(600-750)^2+(100-150)^2=25000.
\]
With equal weights, the barycentric disparity is
\[
  D_{\mathrm{bar}}^2=\frac{1}{4}\Wtwo^2(\mu_0,\mu_1)=6250.
\]
A half repair, \(\lambda=1/2\), reduces the squared disparity to \(6250/4\).

\begin{remark}[Gaussian repairs are special]
The affine form above relies on both group distributions being Gaussian. In general, the repair map is nonlinear. For skewed insurance premiums, nonlinear maps are common: lower quantiles may be moved little, while upper-tail scores may be compressed or expanded substantially.
\end{remark}

\subsection{Fairness as projection}
\label{sec:fairness-as-projection}

The barycentric repair can be described as a projection onto a fairness constraint. Here projection means minimization of the chosen transport displacement, not an ordinary Euclidean projection. Let \(\mathcal F\) be the set of random variables \(\widetilde Z\) such that
\[
  \Law(\widetilde Z\mid S=s)
\]
is the same for all \(s\). Consider the optimization problem
\[
  \inf_{\widetilde Z\in\mathcal F}\E[(\widetilde Z-Z)^2].
\]
Under the restriction that each group is transformed by an optimal one-dimensional transport map, the solution is the barycentric repair. More generally, with discrete data, this becomes a finite-dimensional projection problem.

This projection view clarifies the meaning of the correction. We are not simply forcing all means to be equal. We are finding the closest score, in squared displacement, whose group conditional distributions coincide.

\begin{definition}[Wasserstein fairness projection]\label{def:wasserstein-fairness-projection}
A Wasserstein fairness projection of a score \(Z\) is a score \(Z^\star\) satisfying distributional demographic parity and minimizing a transportation cost from \(Z\) to \(Z^\star\), typically
\[
  \E[(Z^\star-Z)^2].
\]
In one dimension, with group-wise monotone maps, this projection is given by the barycentric repair.
\end{definition}

\begin{proposition}[One-dimensional barycentric repair as a projection]
\label{prop:barycentric-repair-projection}
Assume that each conditional law \(\mu_s=\Law(Z\mid S=s)\) belongs to \(\Probp{2}(\R)\). Among group-wise monotone transformations \(T_s\) such that \((T_s)_\#\mu_s\) is the same law for all \(s\), the minimum of
\[
  \sum_s p_s\int |T_s(z)-z|^2\,\mu_s(\dd z)
\]
is attained by choosing the common law to be the \(\Wtwo\) barycenter \(\bar\mu\) and \(T_s\) to be the monotone map from \(\mu_s\) to \(\bar\mu\).
\end{proposition}

\begin{proof}
For any common target law \(\nu\), the optimal monotone group-wise transformation has cost \(\Wtwo^2(\mu_s,\nu)\). Minimizing the weighted sum of these costs over \(\nu\) is exactly the barycenter problem from Definition~\ref{def:wasserstein-barycenter}.
\end{proof}

\subsection{Beyond demographic parity}
\label{sec:fairness-beyond-demographic-parity}

Distributional parity of scores is only one possible fairness target. Other targets compare conditional distributions after controlling for a legitimate risk variable. Let \(A\) be an admissible risk summary, such as a risk tier, exposure class, or predicted loss band. One may ask for
\[
  \Law(Z\mid S=s,A=a)=\Law(Z\mid S=s',A=a)
\]
for each value of \(a\). This is a conditional distributional parity requirement.

Similarly, one may compare residuals conditional on risk tiers:
\[
  \Law(Y-Z\mid S=s,A=a).
\]
This type of audit is often more actuarially meaningful than unconditional demographic parity, because it distinguishes disparities explained by accepted risk factors from disparities that remain within comparable risk classes.

\begin{definition}[Conditional Wasserstein disparity]\label{def:conditional-wasserstein-disparity}
Let \(A\) be a finite admissible stratification variable. A conditional Wasserstein disparity can be defined as
\[
  D_{\mathrm{cond}}^2
  = \sum_a \Pbb(A=a)
    \sum_s \Pbb(S=s\mid A=a)
    \Wtwo^2\bigl(\Law(Z\mid S=s,A=a),\bar\mu_a\bigr),
\]
where \(\bar\mu_a\) is the barycenter of the conditional group laws within stratum \(a\).
\end{definition}

The choice of \(A\) is normative and actuarial. If \(A\) is too rich, it may absorb the sensitive attribute through proxies. If \(A\) is too coarse, it may ignore legitimate risk heterogeneity. The audit should therefore report which variables are treated as admissible and why.

\subsection{Practical audit protocol}
\label{sec:fairness-practical-audit-protocol}

A practical Wasserstein fairness audit for an insurance score can be organized as follows.

\begin{enumerate}[label=\textbf{Step \arabic*.},leftmargin=2.8cm]
  \item \textbf{Define the object.} Decide whether the audit concerns a predicted frequency, predicted severity, pure premium, final tariff, underwriting score, residual, or loss ratio.
  \item \textbf{Define groups.} Choose the group variable used in the audit. Record sample sizes and missingness.
  \item \textbf{Estimate group laws.} Compute empirical distributions, quantile functions, and uncertainty bands.
  \item \textbf{Measure disparity.} Report pairwise distances, barycentric contributions, mean differences, tail differences, and plots.
  \item \textbf{Construct repairs.} If correction is considered, compute full and partial Wasserstein repairs.
  \item \textbf{Check actuarial effects.} Evaluate calibration, loss ratios, premium changes, predictive loss, and business constraints.
  \item \textbf{Document choices.} Explain the fairness target, the admissible variables, the treatment of ties, and uncertainty.
\end{enumerate}

\begin{actuarialbox}
The output of a fairness audit should not be a single number. A useful report contains at least four objects: group distributions, Wasserstein disparities, premium-change distributions under possible repairs, and calibration or loss-ratio diagnostics.
\end{actuarialbox}

\subsection{Limitations}
\label{sec:fairness-limitations}

Several limitations should be kept in mind.

First, the choice of sensitive variables and admissible controls is normative. Mathematics can measure a disparity conditional on a chosen set of variables, but it cannot decide by itself which variables are legitimate.

Second, distributional parity may conflict with actuarial calibration when groups have different expected losses. This does not invalidate the audit; it clarifies the trade-off.

Third, empirical Wasserstein distances can be noisy, especially in small groups and in the tails. Bootstrap uncertainty and sensitivity analysis are important.

Fourth, individual-level corrections can be legally and ethically sensitive. A monotone within-group map preserves ranks, but it still changes premiums partly as a function of group membership. Depending on the context, this may or may not be acceptable.

Fifth, a score distribution is only one part of the insurance process. Underwriting rules, marketing, claims handling, cancellation, deductibles, coverage limits, and payment plans can also produce disparate effects.

\subsection{Further reading}
\label{sec:fairness-further-reading}

\begin{readingbox}
For the optimal transport background behind Wasserstein distances, one-dimensional quantile formulas and barycenters, see \cite{villani2003topics,villani2008oldnew,peyre2019computational,figalli2021invitation}. For general algorithmic fairness, see \cite{dwork2012fairness,feldman2015certifying,hardt2016equality,chouldechova2017fair,kleinberg2017inherent,barocas2023fairness}. For counterfactual fairness, see \cite{kusner2017counterfactual}. For optimal-transport fairness repairs and fair regression by barycenters, see \cite{gordaliza2019obtaining,chzhen2020fair}. For actuarial discussions of insurance, discrimination and fairness, see \cite{lindholm2022discrimination,charpentier2024insurance}. The computational ideas in this lecture build on the discrete, statistical and regularized transport tools developed in previous lectures.
\end{readingbox}

\subsection{Exercises}
\label{sec:fairness-exercises}

\begin{exercise}[Two-group barycenter]\label{ex:12:1}
Let \(S\in\{0,1\}\), with \(\Pbb(S=1)=p\). Let \(Q_0\) and \(Q_1\) be the quantile functions of \(Z\mid S=0\) and \(Z\mid S=1\). Show that the \(\Wtwo\) barycenter has quantile function
\[
  \bar Q(u)=(1-p)Q_0(u)+pQ_1(u).
\]
Then show that
\[
  D_{\mathrm{bar}}^2=p(1-p)\int_0^1\{Q_1(u)-Q_0(u)\}^2\dd u.
\]
\end{exercise}

\begin{exercise}[Gaussian score repair]\label{ex:12:2}
Assume
\[
  Z\mid S=s\sim\mathcal N(m_s,\sigma_s^2),
  \qquad s=1,\ldots,K,
\]
with group weights \(p_s\). Derive the barycenter distribution and the group-specific affine repair map \(T_s\). Interpret separately the effect on location and dispersion.
\end{exercise}

\begin{exercise}[Calibration tension]\label{ex:12:3}
Suppose that \(Z\) satisfies group calibration,
\[
  \E[Y\mid Z,S]=Z.
\]
Prove that if \(\Law(Z\mid S=s)\) is identical for all \(s\), then \(\E[Y\mid S=s]\) is identical for all \(s\). Explain why this creates a tension in insurance pricing when group base risks differ.
\end{exercise}

\begin{exercise}[Partial repair path]\label{ex:12:4}
Let \(Z^{(\lambda)}=(1-\lambda)Z+\lambda T_S(Z)\), where \(T_S\) is the barycentric repair. Prove that the squared barycentric disparity decreases as \((1-\lambda)^2\). Then propose three quantities, other than Wasserstein disparity, that should be monitored along the path.
\end{exercise}

\begin{exercise}[Empirical implementation]\label{ex:12:5}
Given a data set with scores \(z_i\) and group labels \(s_i\), write pseudo-code to compute:
\begin{enumerate}[label=(\alph*)]
  \item empirical group quantile functions on a common grid;
  \item the empirical Wasserstein barycenter;
  \item the group contributions \(C_s\);
  \item the repaired scores \(z_i^\star\).
\end{enumerate}
Discuss how you would handle ties.
\end{exercise}

\begin{exercise}[Residual audit]\label{ex:12:6}
Let \(R=Y-Z\) and \(R^\star=Y-Z^\star\). Explain why score parity does not imply residual parity. Construct a simple example with two groups where \(\Law(Z\mid S=0)=\Law(Z\mid S=1)\), but \(\Law(R\mid S=0)\neq\Law(R\mid S=1)\).
\end{exercise}

\begin{exercise}[Conditional disparity]\label{ex:12:7}
Let \(A\) be an admissible binary risk class. Define a conditional Wasserstein disparity by comparing \(\Law(Z\mid S=s,A=a)\) across groups within each level of \(A\). Show how the unconditional disparity can be small while the conditional disparity is large, or conversely. Give an actuarial interpretation of both cases.
\end{exercise}

\newpage
\lecture{Reinsurance, Capital Allocation and Multi-Marginal Transport}
\label{lec:reinsurance-capital-mmot}

\subsection*{Learning objectives}

This lecture brings together several themes developed throughout the course: couplings, push-forwards, Wasserstein distances, barycenters, duality, computation, and actuarial interpretation. The guiding observation is that reinsurance and capital allocation are distributional operations. A treaty transforms a gross loss distribution into retained and ceded distributions. A dependence model transforms marginal line-of-business distributions into an aggregate loss distribution. A capital allocation rule transforms aggregate capital into line-level charges. Optimal transport provides a common language for all three transformations.

After this lecture, the reader should be able to:
\begin{enumerate}[label=(\roman*)]
  \item represent classical reinsurance treaties as measurable maps and push-forwards of loss distributions;
  \item compare retained and ceded distributions under different treaties using Wasserstein distances and tail functionals;
  \item formulate dependence uncertainty as an optimization problem over couplings with fixed marginals;
  \item recognize worst-case and best-case aggregate capital problems as multi-marginal optimal transport problems;
  \item write the discrete multi-marginal problem and its Kantorovich dual;
  \item interpret dual potentials as marginal shadow prices of capital or risk contribution;
  \item connect capital allocation, Euler allocation, and Shapley-style allocation to scenario-level decompositions;
  \item design empirical algorithms for reinsurance and capital allocation based on grids, Monte Carlo scenarios, Sinkhorn regularization, and scenario reduction;
  \item understand how dependence, treaty design, and allocation rules interact in insurance portfolios.
\end{enumerate}

\begin{actuarialbox}
Reinsurance is a distributional transformation. Capital aggregation is a coupling problem. Capital allocation is a disaggregation problem. Optimal transport links these three operations by asking what mass is moved, under which constraints, and at which cost.
\end{actuarialbox}

\begin{readingbox}
For reinsurance and risk sharing, see the classical actuarial treatments of reinsurance design and ruin theory, for instance \cite{Borch1960,Gerber1979,AlbrecherAsmussen2010,AlbrecherTeugelsSchoutens2017}. For dependence, comonotonicity and risk aggregation, see \cite{DhaeneDenuitGoovaertsKaasVyncke2002,DenuitDhaeneGoovaertsKaas2005,McNeilFreyEmbrechts2015,EmbrechtsPuccettiRueschendorf2013}. Multi-marginal optimal transport is reviewed in \cite{Pass2015}; computational and entropic viewpoints are discussed in \cite{peyre2019computational}. For capital allocation, see \cite{Shapley1953,AumannShapley1974,Denault2001,Tasche2008}. The martingale optimal transport section is connected to \cite{Hobson1998,BeiglbockHenryLaborderePenkner2013,GalichonHenryLabordereTouzi2014}.
\end{readingbox}

\paragraph{Notation used in this lecture.}
We write \(\bX=(X_1,\ldots,X_d)\) for line losses, \(\bx=(x_1,\ldots,x_d)\) for deterministic scenarios, and
\[
  \Sigma(\bx)=x_1+\cdots+x_d
\]
for the aggregation map. The notation \(Q_\mu\) denotes the generalized quantile function of a one-dimensional law \(\mu\), as in Lecture~\ref{lec:wasserstein-barycenters}. Push-forwards, couplings and Wasserstein distances are used in the sense of Definitions~\ref{def:push-forward}, \ref{def:coupling} and \ref{def:wasserstein-distance}. Risk functionals use the notation of Lecture~\ref{lec:risk-measures-robustness}.

\subsection{Why optimal transport appears in reinsurance}
\label{sec:reinsurance-why-ot}

Consider an insurer with gross portfolio loss \(L\). A reinsurance treaty decomposes \(L\) into a retained loss and a ceded loss,
\[
  L = L^{\mathrm{ret}} + L^{\mathrm{ced}}.
\]
In the simplest case, both components are deterministic functions of the gross loss:
\[
  L^{\mathrm{ret}} = r(L), \qquad L^{\mathrm{ced}} = c(L)=L-r(L).
\]
If \(\mu=\Law(L)\), then the retained and ceded distributions are the push-forwards
\[
  \mu^{\mathrm{ret}} = r_\#\mu, \qquad \mu^{\mathrm{ced}} = c_\#\mu.
\]
Thus a treaty is not only a contractual formula. It is also a map between probability distributions.

\begin{definition}[Treaty map and retained/ceded push-forwards]\label{def:treaty-push-forward}
Let \(L\) be a gross loss with law \(\mu\). A deterministic reinsurance treaty is represented by a measurable retained-loss map \(r\) and a ceded-loss map \(c\), usually satisfying \(c(x)=x-r(x)\). The retained and ceded laws are
\[
  \mu^{\mathrm{ret}}=r_\#\mu,
  \qquad
  \mu^{\mathrm{ced}}=c_\#\mu.
\]
When the treaty depends on an event vector \(\bZ\), the same definition applies with \(r\) and \(c\) defined on the event-state space and \(\mu=\Law(\bZ)\).
\end{definition}

This is already useful. Suppose two treaties \(r_0\) and \(r_1\) are proposed. They induce retained distributions
\[
  \mu^{\mathrm{ret}}_0=(r_0)_\#\mu,
  \qquad
  \mu^{\mathrm{ret}}_1=(r_1)_\#\mu.
\]
The quantity
\[
  \Wp{p}\bigl(\mu^{\mathrm{ret}}_0,\mu^{\mathrm{ret}}_1\bigr)
\]
measures how much the retained loss distribution is changed by switching treaty. This comparison is not limited to the mean ceded loss. It compares the whole retained distribution: moderate losses, tail losses, attachment regions, and exhaustion regions.

The same idea applies to portfolios with several lines of business. Let
\[
  \bX=(X_1,\ldots,X_d)
\]
denote line losses, with marginal laws \(\mu_1,\ldots,\mu_d\). The aggregate loss is
\[
  S=X_1+\cdots+X_d.
\]
The law of \(S\) is not determined by the marginal laws alone. It depends on the joint law \(\gamma\) of \(\bX\), that is, on the coupling
\[
  \gamma\in\Coup(\mu_1,
  \ldots,
  \mu_d).
\]

\begin{definition}[Multi-marginal coupling]\label{def:multi-marginal-coupling}
Let \(\mu_1,\ldots,\mu_d\) be probability measures on spaces \(\mathcal X_1,\ldots,\mathcal X_d\). A multi-marginal coupling of these laws is a probability measure \(\gamma\) on \(\mathcal X_1\times\cdots\times\mathcal X_d\) such that
\[
  (p_k)_\#\gamma=\mu_k,\qquad k=1,\ldots,d,
\]
where \(p_k\) is the projection onto the \(k\)-th coordinate. The set of all such couplings is denoted
\[
  \Coup(\mu_1,\ldots,\mu_d).
\]
For \(d=2\), this reduces to the two-marginal coupling set introduced in Definition~\ref{def:coupling}.
\end{definition}

Capital, reinsurance demand, and allocation can therefore vary substantially across admissible dependence structures.

\begin{actuarialbox}
In actuarial terms, a coupling is a dependence assumption. A multi-marginal transport problem is a dependence stress test under fixed marginal line-of-business distributions.
\end{actuarialbox}

\subsection{Treaties as maps}
\label{sec:treaties-as-maps}

We begin with one-dimensional treaties, where the gross loss \(L\geq 0\) has distribution \(\mu\). A treaty is represented by a retention function
\[
  r:[0,\infty)\to[0,\infty),
\]
with ceded amount \(c(x)=x-r(x)\). Usually one assumes
\[
  0\leq r(x)\leq x,
  \qquad
  r \text{ nondecreasing},
  \qquad
  c \text{ nondecreasing or layer-wise nondecreasing}.
\]
The monotonicity condition excludes perverse treaties in which a larger gross loss leads to a smaller retained amount.

\begin{definition}[Admissible retention function]\label{def:admissible-retention-function}
A retention function is called admissible, in this chapter, if \(r:[0,\infty)\to[0,\infty)\) is measurable, nondecreasing, and satisfies
\[
  0\le r(x)\le x,
  \qquad x\ge 0.
\]
The inequality \(r(x)\le x\) is equivalently the non-negativity of the ceded function \(c(x)=x-r(x)\). Additional contractual restrictions, such as limits, reinstatements or hours clauses, can be added by restricting the class of admissible maps.
\end{definition}

\subsubsection*{Quota-share treaties}

A quota-share treaty with retention proportion \(q\in[0,1]\) is
\[
  r_q(x)=q x,
  \qquad
  c_q(x)=(1-q)x.
\]
The retained distribution is
\[
  \mu_q^{\mathrm{ret}}=(x\mapsto qx)_\#\mu.
\]
If \(Q_L\) is the quantile function of \(L\), then the retained quantile function is
\[
  Q_{r_q(L)}(u)=q Q_L(u),\qquad 0<u<1.
\]
For \(p\geq 1\), the Wasserstein distance between two retained distributions with retentions \(q_0\) and \(q_1\) is therefore
\[
  \Wp{p}\bigl((r_{q_0})_\#\mu,(r_{q_1})_\#\mu\bigr)
  = |q_0-q_1|\,\Bigl(\E[|L|^p]\Bigr)^{1/p},
\]
whenever \(L\in L^p\). This gives a direct geometric interpretation of changing the quota-share retention.

\begin{proposition}[Quota-share scaling in Wasserstein distance]\label{prop:quota-share-wasserstein-scaling}
Let \(L\ge0\) have law \(\mu\in\Probp{p}(\R)\), and let \(r_q(x)=qx\) for \(q\in[0,1]\). Then
\[
  \Wp{p}\bigl((r_{q_0})_\#\mu,(r_{q_1})_\#\mu\bigr)
  = |q_0-q_1|\,\|L\|_{p}.
\]
\end{proposition}

\begin{proof}
Since \(r_q(L)=qL\) and the maps \(q_0x\) and \(q_1x\) are increasing functions of the same one-dimensional variable, the monotone quantile coupling is the coupling induced by the common rank of \(L\). Hence
\[
  \Wp{p}^p
  =\E\bigl[|q_0L-q_1L|^p\bigr]
  =|q_0-q_1|^p\E[|L|^p].
\]
\end{proof}

\subsubsection*{Stop-loss and excess-of-loss treaties}

A stop-loss treaty with priority \(a\geq0\) has retained loss
\[
  r_a(x)=\min(x,a),
\]
and ceded loss
\[
  c_a(x)=(x-a)_+.
\]
The retained distribution has an atom at \(a\) whenever \(\Pbb(L>a)>0\). Indeed,
\[
  \Pbb(r_a(L)\leq t)=
  \begin{cases}
    \Pbb(L\leq t), & 0\leq t<a,\\
    1, & t\geq a.
  \end{cases}
\]
The atom at the retention point is
\[
  \Pbb(r_a(L)=a)=\Pbb(L\geq a)
\]
when \(L\) has no atom at \(a\); more generally, the atom includes the mass of all losses mapped to \(a\).

For an excess-of-loss layer with attachment \(a\) and limit \(\ell\), the ceded amount is
\[
  c_{a,\ell}(x)=\min\{(x-a)_+,\ell\},
\]
and the retained amount is
\[
  r_{a,\ell}(x)=x-c_{a,\ell}(x).
\]
The layer creates atoms at \(0\) and \(\ell\) in the ceded distribution, corresponding respectively to losses below attachment and losses above exhaustion.

\begin{example}[Layer distribution]
Let \(C=c_{a,\ell}(L)\). For \(0<t<\ell\),
\[
  \Pbb(C\leq t)=\Pbb(L\leq a+t).
\]
Moreover,
\[
  \Pbb(C=0)=\Pbb(L\leq a),
  \qquad
  \Pbb(C=\ell)=\Pbb(L\geq a+\ell),
\]
up to endpoint conventions when \(L\) has atoms. Thus a smooth gross loss distribution may induce a ceded distribution with large atoms. This matters when comparing treaties by Wasserstein distances, because atoms represent mass concentrated at contract thresholds.
\end{example}

\subsubsection*{Reinstatements, aggregate covers, and non-smooth maps}

Real treaties often involve aggregate deductibles, annual aggregate limits, reinstatements, event definitions, or several layers. Even then, the basic representation remains useful. The treaty induces a measurable map from the vector of underlying losses to retained and ceded amounts.

For example, if \(N\) events generate the loss vector \(\bZ=(Z_1,\ldots,Z_N)\), an annual aggregate cover may depend on
\[
  L=Z_1+\cdots+Z_N.
\]
The ceded amount could be
\[
  c(L)=\min\{(L-A)_+,M\},
\]
where \(A\) is an aggregate deductible and \(M\) an annual aggregate limit. The retained law is again a push-forward of \(\Law(L)\).

When the treaty depends on the full event vector, say
\[
  c(\bZ),
\]
the ceded distribution is
\[
  c_\#\Law(\bZ).
\]
This emphasizes a key point: in the presence of event-level dependence, the treaty output depends on the joint law, not only on the marginal event distributions.

\subsection{Comparing treaties distributionally}
\label{sec:comparing-treaties-distributionally}

Classical treaty comparison often uses expected ceded loss, variance reduction, tail reduction, or impact on a solvency capital requirement. Optimal transport adds a complementary diagnostic: the distance between whole distributions.

Let \(r_0\) and \(r_1\) be two retention functions and define
\[
  R_0=r_0(L),
  \qquad
  R_1=r_1(L).
\]
A natural comparison is
\[
  \Wtwo^2(\Law(R_0),\Law(R_1)).
\]
In dimension one,
\[
  \Wtwo^2(\Law(R_0),\Law(R_1))
  =\int_0^1\left(Q_{R_0}(u)-Q_{R_1}(u)\right)^2\dd u.
\]
This makes the comparison easy to compute from simulated samples.

\begin{definition}[Distributional treaty distance]\label{def:distributional-treaty-distance}
For two treaties \(T_0,T_1\), with retained loss maps \(r_0,r_1\), define the retained distributional distance
\[
  D_p^{\mathrm{ret}}(T_0,T_1)
  =\Wp{p}\bigl((r_0)_\#\mu,(r_1)_\#\mu\bigr).
\]
Similarly, the ceded distributional distance is
\[
  D_p^{\mathrm{ced}}(T_0,T_1)
  =\Wp{p}\bigl((c_0)_\#\mu,(c_1)_\#\mu\bigr).
\]
\end{definition}

These two distances need not coincide. Two treaties may have similar retained distributions but quite different ceded distributions, especially when one treaty removes moderate losses and another removes rare large losses. The insurer and reinsurer may therefore view the same treaty switch differently.

\begin{proposition}[Map stability and policy-level upper bound]\label{prop:treaty-map-stability}
Let \(r\) be \(K\)-Lipschitz on \([0,\infty)\). For any two gross-loss laws \(\mu,\nu\in\Probp{p}(\R_+)\),
\[
  \Wp{p}(r_\#\mu,r_\#\nu)\le K\Wp{p}(\mu,\nu).
\]
Moreover, if two treaties \(r_0,r_1\) are applied to the same gross loss \(L\sim\mu\), then
\[
  \Wp{p}\bigl((r_0)_\#\mu,(r_1)_\#\mu\bigr)
  \le
  \left(\E\bigl[|r_0(L)-r_1(L)|^p\bigr]\right)^{1/p}.
\]
\end{proposition}

\begin{proof}
For the first inequality, push any coupling \(\gamma\in\Coup(\mu,\nu)\) through \((x,y)\mapsto(r(x),r(y))\). Its cost is bounded by \(K^p|x-y|^p\). Taking the infimum over \(\gamma\) gives the result. The second inequality uses the particular coupling \((r_0(L),r_1(L))\); the optimal Wasserstein coupling can only be cheaper.
\end{proof}

\begin{remark}[Paired versus distributional treaty change]\label{rem:paired-versus-distributional-treaty-change}
The second inequality separates a paired claim-level movement from an unpaired distributional movement. The paired movement is an upper bound on the Wasserstein movement between retained laws, and the inequality may be strict when different pointwise treaty effects produce similar retained distributions.
\end{remark}

\subsubsection*{A retention path}

Let \(\theta\in[0,1]\) index a family of treaties, with retention maps \(r_\theta\). For example, \(\theta\) may represent an attachment point, a limit, or a quota-share retention. The path
\[
  \theta\mapsto (r_\theta)_\#\mu
\]
is a path in Wasserstein space. Its length, curvature, or local derivative can be used to understand treaty sensitivity.

A simple local diagnostic is
\[
  \frac{\Wtwo((r_{\theta+h})_\#\mu,(r_\theta)_\#\mu)}{|h|}
\]
for small \(h\). Large values indicate that small contract changes produce large distributional changes. For excess-of-loss treaties, such sensitivity is often concentrated near the attachment and exhaustion points.

\begin{example}[Moving an attachment point]
Let \(c_a(x)=(x-a)_+\) and \(R_a=x-c_a(x)=\min(x,a)\). Increasing \(a\) by a small amount affects all losses above \(a\), but does not affect losses below \(a\). Thus the distributional derivative depends on the survival probability \(\Pbb(L>a)\). The impact of changing a low attachment point is therefore different from the impact of changing a high attachment point, even if the numerical change in attachment is the same.
\end{example}

\subsection{Premiums, risk measures, and treaty objectives}
\label{sec:reinsurance-premiums-risk-objectives}

A treaty is not selected only by comparing retained distributions. It also has a price. Let \(\pi(T)\) denote the reinsurance premium for treaty \(T\). A stylized insurer objective is
\[
  J(T)=\Risk\bigl(L^{\mathrm{ret}}_T\bigr)+\pi(T),
\]
where \(\Risk\) is a risk functional, for example \(\VaR_\alpha\), \(\TVaR_\alpha\), a distortion risk measure, or an economic capital functional.

If \(T\) is parameterized by \(\theta\), this becomes
\[
  J(\theta)=\Risk(r_\theta(L))+\pi(\theta).
\]
The first term is a function of the push-forward \((r_\theta)_\#\mu\). The second term may be based on expected ceded loss, a loading, a cost-of-capital principle, or market quotes.

\begin{definition}[Distributional benefit of a treaty]\label{def:distributional-treaty-benefit}
Relative to no reinsurance, the distributional benefit of treaty \(T\) may be summarized by
\[
  B_p(T)=\Wp{p}\bigl(\Law(L),\Law(L_T^{\mathrm{ret}})\bigr),
\]
or by tail reductions such as
\[
  \Delta \VaR_\alpha(T)
  =\VaR_\alpha(L)-\VaR_\alpha(L_T^{\mathrm{ret}}),
\]
\[
  \Delta \TVaR_\alpha(T)
  =\TVaR_\alpha(L)-\TVaR_\alpha(L_T^{\mathrm{ret}}).
\]
\end{definition}

The Wasserstein benefit and the tail benefit answer different questions. A treaty may have a modest global distributional effect but a large tail effect, or conversely a large global effect concentrated on frequent moderate claims but limited solvency benefit.

\begin{actuarialbox}
Expected ceded loss measures how much is transferred on average. A tail risk reduction measures how much capital is relieved. A Wasserstein distance measures how much the entire retained distribution is moved. These are related but not interchangeable.
\end{actuarialbox}

\subsection{Dependence uncertainty as a coupling problem}
\label{sec:dependence-uncertainty-coupling-problem}

Now consider \(d\) lines of business with line-loss vector \(\bX=(X_1,\ldots,X_d)\). Assume that the marginal laws \(\mu_1,\ldots,\mu_d\) are estimated separately. This is common in practice: frequency and severity models may be built by line, reserving models may produce line-level predictive distributions, and catastrophe models may output peril-level or region-level loss distributions.

The aggregate loss is
\[
  S=X_1+\cdots+X_d.
\]
But \(\Law(S)\) depends on the joint distribution \(\gamma\) of \(\bX\). The set of all admissible joint laws with given marginals is
\[
  \Coup(\mu_1,\ldots,\mu_d).
\]
For any coupling \(\gamma\), the aggregate distribution is
\[
  \Law_\gamma(S)=\Sigma_\#\gamma,
  \qquad
  \Sigma(\bx)=x_1+\cdots+x_d.
\]

A dependence stress test asks how large or small a risk functional can be when the marginals are fixed:
\[
  \overline \Risk
  =\sup_{\gamma\in\Coup(\mu_1,
  \ldots,
  \mu_d)}
  \Risk\bigl(\Sigma_\#\gamma\bigr),
\]
\[
  \underline \Risk
  =\inf_{\gamma\in\Coup(\mu_1,
  \ldots,
  \mu_d)}
  \Risk\bigl(\Sigma_\#\gamma\bigr).
\]
This formulation separates two tasks: marginal modeling and dependence stress testing.

\begin{example}[Two lines]
Let \(X_1\) be attritional motor loss and \(X_2\) be catastrophe property loss. Their marginal distributions may be estimated from different sources. The aggregate capital for \(S=X_1+X_2\) depends on whether large motor losses and large catastrophe losses tend to occur together. If the dependence structure is uncertain, optimizing over couplings gives a range of plausible aggregate capital values.
\end{example}

\subsection{Multi-marginal optimal transport}
\label{sec:multi-marginal-ot-reinsurance}

\begin{definition}[Multi-marginal optimal transport]\label{def:multi-marginal-ot}
Let \(\mu_1,\ldots,\mu_d\) be probability measures and let \(c:\mathcal X_1\times\cdots\times\mathcal X_d\to\R\cup\{+\infty\}\) be a measurable cost. The multi-marginal optimal transport problem is
\[
  \inf_{\gamma\in\Coup(\mu_1,\ldots,\mu_d)}
  \int c(\bx)\dd\gamma(\bx).
\]
\end{definition}
For actuarial aggregation, natural costs are functions of the sum:
\[
  c(\bx)=h(x_1+\cdots+x_d),
\]
where \(h\) may represent capital, penalty, utility loss, or a tail indicator.

If \(h\) is convex and increasing, then minimizing or maximizing
\[
  \E_\gamma[h(S)]
\]
over \(\gamma\in\Coup(\mu_1,\ldots,\mu_d)\) produces best-case and worst-case aggregation under fixed marginals. This is a dependence problem, not a marginal problem.

\begin{proposition}[Comonotonic extremality for convex aggregation]\label{prop:comonotonic-extremality-convex-aggregation}
Let \(X_k\sim\mu_k\) be integrable real-valued losses and let \(h:\R\to\R\) be convex and increasing, with finite expectations. Among all couplings of \(\mu_1,\ldots,\mu_d\), the comonotonic coupling
\[
  X_k^{\mathrm{com}}=Q_{\mu_k}(U),\qquad U\sim\mathrm{Unif}(0,1),
\]
maximizes \(\E[h(X_1+\cdots+X_d)]\). In particular, it is a conservative benchmark for convex increasing aggregate penalties.
\end{proposition}

\begin{proof}[Proof idea]
The function \((x_1,\ldots,x_d)\mapsto h(x_1+\cdots+x_d)\) is supermodular when \(h\) is convex. Classical rearrangement and supermodular-order results imply that the comonotonic coupling is maximal for expectations of supermodular functions among couplings with fixed marginals; see, for example, \cite{DhaeneDenuitGoovaertsKaasVyncke2002,DenuitDhaeneGoovaertsKaas2005,Rueschendorf1982}.
\end{proof}

\subsubsection*{Discrete formulation}
\label{subsec:mmot-discrete-formulation}

Suppose each marginal is supported on a grid:
\[
  \mu_k=\sum_{i=1}^{n_k} a_i^{(k)}\delta_{x_i^{(k)}}.
\]
A multi-marginal coupling is an array
\[
  \gamma_{i_1,
  \ldots,i_d}\geq0.
\]
The marginal constraints are
\[
  \sum_{i_1,
  \ldots,i_{k-1},i_{k+1},
  \ldots,i_d}
  \gamma_{i_1,
  \ldots,i_d}
  = a_{i_k}^{(k)},
  \qquad
  k=1,
  \ldots,d.
\]
The objective is
\[
  \sum_{i_1,
  \ldots,i_d}
  c\bigl(x_{i_1}^{(1)},
  \ldots,x_{i_d}^{(d)}\bigr)
  \gamma_{i_1,
  \ldots,i_d}.
\]
This is a linear program, but its dimension grows quickly with \(d\). For \(d=2\), the coupling is a matrix. For \(d=5\), it is a five-dimensional array. This curse of dimensionality is the main computational obstacle.

\subsubsection*{Dual formulation}
\label{subsec:mmot-dual-formulation}

The Kantorovich dual is
\[
  \sup_{\varphi_1,\ldots,\varphi_d}
  \sum_{k=1}^d \int \varphi_k(x)\dd\mu_k(x)
\]
subject to
\[
  \varphi_1(x_1)+\cdots+\varphi_d(x_d)\leq c(\bx)
\]
for all \(\bx=(x_1,\ldots,x_d)\). The functions \(\varphi_k\) are dual potentials.

In an actuarial interpretation, \(\varphi_k(x)\) is a marginal shadow price assigned to line \(k\) when that line realizes loss level \(x\). The inequality says that the sum of line-level shadow prices cannot exceed the aggregate scenario cost. At optimality, equality holds on the support of the optimal coupling.

\begin{actuarialbox}
The primal problem searches over dependence structures. The dual problem searches over line-level price functions that support the aggregate capital functional.
\end{actuarialbox}

\subsection{Worst-case capital over couplings}
\label{sec:worst-case-capital-couplings}

Let \(\Risk\) be a risk measure applied to aggregate loss \(S\). For expectations of transformed aggregate losses,
\[
  \Risk(S)=\E[h(S)],
\]
the worst-case problem is directly a multi-marginal OT problem with cost
\[
  c(\bx)=h(x_1+\cdots+x_d).
\]
For quantile-based capital, such as \(\VaR_\alpha\), the problem is more delicate because \(\VaR\) is nonlinear in the distribution. Nevertheless, it can often be reformulated or approximated using tail indicators, expected shortfall, or optimization over thresholds.

For example, expected shortfall, also called tail value-at-risk in actuarial notation, has the representation
\[
  \TVaR_\alpha(S)
  =\inf_{t\in\R}\left\{t+\frac{1}{1-\alpha}\E[(S-t)_+]\right\}.
\]
Thus, for a fixed threshold \(t\), the dependence-sensitive term is
\[
  \E_\gamma[(\Sigma(\bX)-t)_+],
\]
which is a multi-marginal OT objective. One can therefore study worst-case \(\TVaR\) by combining threshold optimization and coupling optimization.

\begin{remark}[A minimax caution for tail risk]\label{rem:tvar-minimax-caution}
Because \(\TVaR_\alpha(S)=\inf_t\{t+(1-\alpha)^{-1}\E[(S-t)_+]\}\), optimizing over couplings and over thresholds does not automatically commute:
\[
  \sup_\gamma \inf_t F(\gamma,t)
  \le
  \inf_t \sup_\gamma F(\gamma,t).
\]
Thus fixed-threshold multi-marginal OT problems give useful building blocks and upper bounds for worst-case \(\TVaR\); equality requires additional compactness, convexity and minimax conditions, or a more careful reformulation.
\end{remark}

\subsubsection*{Comonotonic upper bound}

In one-dimensional aggregation, the comonotonic coupling is obtained by taking a common uniform random variable \(U\sim\mathrm{Unif}(0,1)\) and setting
\[
  X_k=Q_k(U),
  \qquad k=1,
  \ldots,d,
\]
where \(Q_k\) is the quantile function of \(\mu_k\). The aggregate is
\[
  S^{\mathrm{com}}=Q_1(U)+\cdots+Q_d(U).
\]
For many convex increasing functionals of the aggregate, this coupling gives a conservative dependence structure. It represents perfect positive rank dependence: large losses occur together across lines.

The comonotonic model is actuarially important because it is transparent and easy to compute. It is not always the exact worst case for every functional and every constraint set, but it is a natural benchmark for stress testing.

\subsubsection*{Diversification as movement away from comonotonicity}

Diversification may be viewed as a movement from a concentrated, highly aligned coupling toward a less aligned coupling. If \(\gamma^{\mathrm{com}}\) is the comonotonic coupling and \(\gamma\) is a selected dependence model, one can compare them using a distance between joint laws, for example
\[
  \Wp{2}(\gamma,\gamma^{\mathrm{com}}),
\]
when both are distributions on \(\R^d\). A large distance indicates a substantial departure from perfect positive dependence.

Another diagnostic compares aggregate distributions:
\[
  \Wp{2}\bigl(\Sigma_\#\gamma,
  \Sigma_\#\gamma^{\mathrm{com}}\bigr).
\]
This measures the distributional diversification benefit at the aggregate level.

\begin{definition}[Distributional diversification benefit]\label{def:distributional-diversification-benefit}
Given a reference conservative coupling \(\gamma^{\mathrm{ref}}\), define the distributional diversification benefit of \(\gamma\) by
\[
  \mathrm{DDB}_p(\gamma;\gamma^{\mathrm{ref}})
  =\Wp{p}\bigl(\Sigma_\#\gamma,
  \Sigma_\#\gamma^{\mathrm{ref}}\bigr).
\]
If \(\gamma^{\mathrm{ref}}=\gamma^{\mathrm{com}}\), this compares the selected aggregate distribution to the comonotonic aggregate distribution.
\end{definition}

\subsection{Reinsurance under dependence uncertainty}
\label{sec:reinsurance-dependence-uncertainty}

Reinsurance effectiveness depends on dependence. Consider a treaty \(T\) applied to the aggregate loss
\[
  S=X_1+\cdots+X_d.
\]
The retained loss is
\[
  R_T=r_T(S).
\]
For a coupling \(\gamma\), its distribution is
\[
  \Law_\gamma(R_T)=(r_T\circ\Sigma)_\#\gamma.
\]
A robust assessment of the treaty considers
\[
  \sup_{\gamma\in\Coup(\mu_1,\ldots,\mu_d)}
  \Risk\bigl((r_T\circ\Sigma)_\#\gamma\bigr).
\]
This is the worst retained capital under fixed line marginals. Similarly,
\[
  \sup_{\gamma\in\Coup(\mu_1,\ldots,\mu_d)}
  \E_\gamma[c_T(\Sigma(\bX))]
\]
quantifies the worst-case expected ceded loss for the reinsurer.

\begin{example}[Aggregate excess-of-loss]
Let \(T\) be an aggregate excess-of-loss treaty with retained loss
\[
  r_T(s)=s-\min\{(s-a)_+,\ell\}.
\]
For fixed marginals, the expected ceded loss is
\[
  \E_\gamma\left[\min\{(\Sigma(\bX)-a)_+,\ell\}\right].
\]
This is a multi-marginal OT objective with cost
\[
  c(\bx)=\min\{(\Sigma(\bx)-a)_+,\ell\}.
\]
Thus dependence uncertainty directly affects the reinsurance price.
\end{example}

\begin{actuarialbox}
\textbf{Professional lens: cyber accumulation and dependence uncertainty.}

Cyber risk is an accumulation risk. Marginal loss models for individual firms are not enough, because many insureds may depend on the same cloud provider, software library, payment processor, authentication service, managed-service provider or security product. A single incident can therefore generate losses across many policies at once.

Let
\[
  \bL=(L_1,\ldots,L_d)
\]
be the vector of cyber losses across insured firms, sectors or coverage components.
The marginal laws \(\Law(L_j)\) may be reasonably calibrated from claims, scenarios or expert judgement. The aggregate loss
\[
  S=\sum_{j=1}^d L_j
\]
depends crucially on the coupling of these marginals. Transport language makes this explicit:
\[
  \gamma\in\Coup(\mu_1,\ldots,\mu_d)
\]
is not a technical detail; it is the accumulation assumption.

A cyber accumulation stress can therefore be written as an optimization over couplings:
\[
  \sup_{\gamma\in\Coup(\mu_1,\ldots,\mu_d)}
  \Risk\left(\sum_{j=1}^d L_j\right),
\]
possibly with constraints reflecting known common vulnerabilities, vendor concentration, sectoral dependence or geographical exposure. Reinsurance, limits, exclusions and aggregate covers then act on the transported aggregate distribution.
\end{actuarialbox}

\subsection{Capital allocation}
\label{sec:capital-allocation-transport}

Let \(S=X_1+\cdots+X_d\) and let aggregate capital be
\[
  K=\Risk(S).
\]
A capital allocation rule assigns amounts \(K_1,\ldots,K_d\) to lines, business units, or risk cells, with
\[
  K_1+\cdots+K_d=K.
\]

\begin{definition}[Capital allocation rule]\label{def:capital-allocation-rule}
Let \(S=X_1+\cdots+X_d\) and \(K=\Risk(S)\). A full capital allocation is a vector \((K_1,\ldots,K_d)\) such that \(\sum_{k=1}^d K_k=K\). It is called dependence-sensitive if \((K_1,\ldots,K_d)\) depends on the coupling \(\gamma=\Law(X_1,\ldots,X_d)\), not only on the marginal laws \(\mu_1,\ldots,\mu_d\).
\end{definition}

The allocation should reflect both marginal distributions and dependence. This dependence on the coupling is the main point for allocation.

\subsubsection*{Scenario-level allocation}

\begin{definition}[Scenario-level allocation]\label{def:scenario-level-allocation}
A scenario-level allocation is a family of functions
\[
  a_k:\R^d\to\R,
  \qquad k=1,
  \ldots,d,
\]
such that
\[
  a_1(\bx)+\cdots+a_d(\bx)=A(\bx),
\]
where \(A(\bx)\) is a scenario-level aggregate capital or loss quantity. For example, one may take
\[
  A(\bx)=\Sigma(\bx),
\]
or
\[
  A(\bx)=(\Sigma(\bx)-t)_+.
\]
Then line-level allocations are expectations under the selected coupling \(\gamma\):
\[
  K_k=\int a_k(\bx)\dd\gamma(\bx).
\]
\end{definition}

For tail-based allocation, a common choice is based on tail scenarios. If \(S=\Sigma(\bX)\) and \(t=\VaR_\alpha(S)\), then one may allocate expected shortfall / TVaR using
\[
  K_k=\E[X_k\mid S\geq t]
\]
or a centered variant. This depends crucially on the joint law of \(\bX\).

\begin{actuarialbox}
A capital allocation rule is not only a formula. It is a way of assigning scenario-level aggregate risk back to coordinates. The selected coupling determines which line-level losses appear in tail aggregate scenarios.
\end{actuarialbox}

\subsubsection*{Euler allocation}
\label{subsec:euler-allocation}

Suppose \(\Risk\) is positively homogeneous and differentiable in line weights. Define
\[
  S(\bw)=\sum_{k=1}^d w_k X_k,
\]
and
\[
  K(\bw)=\Risk(S(\bw)).
\]
\begin{definition}[Euler allocation]\label{def:euler-allocation}
The Euler allocation is
\[
  K_k = \left.\frac{\partial K(\bw)}{\partial w_k}\right|_{\bw=\1}.
\]
\end{definition}
If \(\Risk=\TVaR_\alpha\) and appropriate regularity conditions hold, this often leads to
\[
  K_k=\E[X_k\mid S\geq \VaR_\alpha(S)].
\]
This expression reveals the dependence issue: the allocation depends on the conditional distribution of \(X_k\) in aggregate tail scenarios.

\subsubsection*{Shapley-style allocation}
\label{subsec:shapley-allocation}

Another approach allocates capital by averaging marginal contributions across coalitions. For a coalition \(A\subset\{1,\ldots,d\}\), let
\[
  S_A=\sum_{k\in A}X_k,
\qquad
  v(A)=\Risk(S_A).
\]
\begin{definition}[Shapley allocation]\label{def:shapley-allocation}
The Shapley allocation to line \(k\) is
\[
  \phi_k
  =\sum_{A\subseteq N\setminus\{k\}}
  \frac{|A|!(d-|A|-1)!}{d!}
  \left(v(A\cup\{k\})-v(A)\right),
\]
where \(N=\{1,\ldots,d\}\).
\end{definition}

This allocation depends on the laws of all coalition sums \(S_A\), hence on the dependence structure within every coalition.

Optimal transport can be used to stress-test Shapley allocations by optimizing coalition risk measures over admissible couplings. It can also compare allocations under different dependence models.

\subsection{Transport-based allocation diagnostics}
\label{sec:transport-based-allocation-diagnostics}

Suppose two dependence models \(\gamma_0\) and \(\gamma_1\) have the same marginals but induce different capital allocations
\[
  \bK^{(0)}=(K_1^{(0)},\ldots,K_d^{(0)}),
  \qquad
  \bK^{(1)}=(K_1^{(1)},\ldots,K_d^{(1)}).
\]
A simple diagnostic is the Euclidean difference
\[
  \|\bK^{(0)}-\bK^{(1)}\|.
\]
But a distributional diagnostic is also possible. If allocation is scenario-level, it induces for each line a distribution of allocated scenario charges,
\[
  \nu_k^{(j)}=\Law_{\gamma_j}(a_k(\bX)),
  \qquad j=0,1.
\]
Then
\[
  \Wp{p}(\nu_k^{(0)},\nu_k^{(1)})
\]
measures how the whole distribution of allocated charges changes for line \(k\), not only its average.

Similarly, if an allocation rule produces a random vector of line charges
\[
  \bA^{(j)}=(A_1^{(j)},\ldots,A_d^{(j)}),
\]
then the joint allocation distributions can be compared by
\[
  \Wp{p}\bigl(\Law(\bA^{(0)}),\Law(\bA^{(1)})\bigr).
\]
This is useful when allocations are later used as internal prices, performance targets, or business-unit risk budgets.

\subsection{Optimal risk sharing and infimal convolution}
\label{sec:optimal-risk-sharing-infimal-convolution}

Capital allocation starts from a total risk and splits it across units. Optimal risk sharing asks the reverse question: how should a total loss be split among agents?

Let \(L\) be a total loss. A risk sharing rule is a vector of functions
\[
  f_1,
  \ldots,f_m
\]
such that
\[
  f_1(x)+\cdots+f_m(x)=x.
\]
Agent \(j\) receives loss \(f_j(L)\) and evaluates it using risk functional \(\Risk_j\). A stylized optimal risk sharing problem is
\[
  \inf_{f_1+
  \cdots+f_m=\Id}
  \sum_{j=1}^m \lambda_j \Risk_j(f_j(L)).
\]
This is a functional optimization problem over loss-sharing maps.

In convex analysis, a related object is the infimal convolution of risk functionals; see also the convex-analysis insert in Appendix~\ref{app:convex-analysis-ot}:
\begin{definition}[Infimal convolution of risk functionals]\label{def:infimal-convolution-risk-functionals}
The infimal convolution of risk functionals \(\Risk_1,\ldots,\Risk_m\) is
\[
  (\Risk_1\Box\cdots\Box\Risk_m)(L)
  =\inf\left\{\sum_{j=1}^m\Risk_j(L_j):
  L_1+
  \cdots+L_m=L\right\}.
\]
\end{definition}
This expression resembles a transport problem because it optimizes over decompositions of a total mass or total loss into components. In insurance, those components correspond to retained loss, ceded loss, retroceded loss, or capital-bearing entities.

\begin{example}[Insurer and reinsurer]
Let \(L_1\) be retained by the insurer and \(L_2\) ceded to a reinsurer, with \(L_1+L_2=L\). If the insurer evaluates \(L_1\) using \(\Risk_1\) and the reinsurer evaluates \(L_2\) using \(\Risk_2\), the efficient frontier of treaty design can be studied through
\[
  \inf_{L_1+L_2=L}
  \left\{\Risk_1(L_1)+\Risk_2(L_2)\right\}.
\]
Contractual constraints restrict the admissible decompositions, for example by requiring \(L_2=(L-a)_+\wedge \ell\).
\end{example}

\subsection{Multi-marginal transport for market clearing}
\label{sec:mmot-market-clearing}

A reinsurance market can also be viewed as matching ceded risks with reinsurer capacities. Suppose an insurer has ceded-loss scenarios \(C\), while reinsurers have capital or appetite profiles. A stylized market-clearing problem assigns pieces of ceded loss to reinsurers subject to capacity constraints.

Let \(Y_j\) be the loss assigned to reinsurer \(j\), with
\[
  Y_1+
  \cdots+Y_m=C.
\]
If each reinsurer has a cost \(\Risk_j(Y_j)\), market clearing may be represented by
\[
  \inf_{Y_1+
  \cdots+Y_m=C}
  \sum_{j=1}^m \Risk_j(Y_j).
\]
When scenario distributions are prescribed or when matching constraints are imposed, the problem becomes multi-marginal. The cost combines aggregate feasibility, risk appetite, and capital cost.

This abstract formulation is not meant to replace detailed reinsurance pricing. Rather, it gives a geometric language for understanding why reinsurance allocation, retrocession, and capital relief are intrinsically linked to transport-like constraints.

\subsection{Advanced topic: martingale optimal transport}
\label{sec:martingale-optimal-transport}

Classical optimal transport fixes two marginal distributions and optimizes over all couplings between them:
\[
  \Coup(\mu,\nu)
  =
  \{\gamma:\gamma_X=\mu,\;\gamma_Y=\nu\}.
\]
Martingale optimal transport adds a financial or actuarial consistency constraint. When \(X\) and \(Y\) represent values of a market-consistent quantity at two dates, one may require
\[
  \E_\gamma[Y\mid X]=X.
\]
\begin{definition}[Martingale coupling]\label{def:martingale-coupling}
A coupling \(\gamma\in\Coup(\mu,\nu)\) is a martingale coupling if, under \((X,Y)\sim\gamma\),
\[
  \E_\gamma[Y\mid X]=X.
\]
The set of martingale couplings is denoted
\[
  \Coup_{\mathrm M}(\mu,\nu)
  =
  \left\{
    \gamma\in\Coup(\mu,\nu):
    \E_\gamma[Y\mid X]=X
  \right\}.
\]
\end{definition}
The martingale optimal transport problem is then
\[
  \inf_{\gamma\in\Coup_{\mathrm M}(\mu,\nu)}
  \int c(x,y)\,\dd\gamma(x,y),
\]
or, depending on the application,
\[
  \sup_{\gamma\in\Coup_{\mathrm M}(\mu,\nu)}
  \int \Phi(x,y)\,\dd\gamma(x,y).
\]

\begin{actuarialbox}
Martingale optimal transport is optimal transport with a no-arbitrage-type constraint. It is useful when the marginals are calibrated at different dates but the coupling must be compatible with a martingale dynamics.
\end{actuarialbox}

The martingale constraint is restrictive. It requires that \(\mu\) and \(\nu\) be in convex order.

\begin{definition}[Convex order]\label{def:convex-order}
For integrable probability measures \(\mu\) and \(\nu\) on \(\R\), we write
\[
  \mu\preceq_{\mathrm{cx}}\nu
\]
if
\[
  \int \varphi\,\dd\mu
  \le
  \int \varphi\,\dd\nu
\]
for every convex function \(\varphi\) for which the integrals exist. Convex order also requires equality of means, obtained by testing both \(\varphi(x)=x\) and \(\varphi(x)=-x\).
\end{definition}

Intuitively, \(\nu\) must be more spread out than \(\mu\) while having the same mean. This is consistent with a martingale that accumulates uncertainty over time.

\begin{remark}[Why convex order appears]
If \(Y\) is a martingale successor of \(X\), then for any convex function
\(\varphi\),
\[
  \varphi(X)
  =
  \varphi(\E[Y\mid X])
  \le
  \E[\varphi(Y)\mid X]
\]
by Jensen's inequality. Taking expectations gives
\[
  \E[\varphi(X)]\le \E[\varphi(Y)].
\]
Thus convex order is necessary for the existence of a martingale coupling. In the one-dimensional integrable setting, it is also sufficient for existence of a martingale coupling. Outside that standard setting, the precise existence statement must be checked with the appropriate version of the theorem.
\end{remark}

Martingale optimal transport is most developed in mathematical finance, where it is used for model-independent bounds on derivative prices. The marginal laws at different maturities may be inferred from option prices, while the unknown coupling between maturities represents model uncertainty. The martingale constraint enforces absence of arbitrage.

In actuarial science, the same idea arises when liabilities contain financial guarantees or market-consistent components. Examples include:
\begin{itemize}
\item variable annuities with guarantees;
\item participating life insurance contracts;
\item equity-linked insurance;
\item pension guarantees;
\item asset-liability management under market-consistent valuation;
\item robust bounds for embedded options.
\end{itemize}

Suppose \(S_0\) and \(S_1\) are asset values at two dates under a risk-neutral
measure. A guarantee payoff may be
\[
  \Phi(S_0,S_1)=(K-S_1)_+
\]
or a path-dependent payoff involving both dates. If only the marginal laws of
\(S_0\) and \(S_1\) are trusted, a robust upper price is
\[
  \sup_{\gamma\in\Coup_{\mathrm M}(\mu_0,\mu_1)}
  \int \Phi(s_0,s_1)\,\dd\gamma(s_0,s_1).
\]
This is a model-independent bound over all martingale models consistent with the given marginals.

\begin{example}[Embedded guarantee]
Consider a simplified equity-linked insurance contract with payoff
\[
  \Phi(S_1)=\max(G,S_1)-S_1=(G-S_1)_+.
\]
If the marginal law of \(S_1\) is fixed, the expectation of \(\Phi(S_1)\) is fixed.
For more complex guarantees depending on intermediate values, such as ratchets or
cliquet-type benefits, the joint law across dates matters. Martingale optimal
transport gives robust bounds over all martingale couplings consistent with the
calibrated marginals.
\end{example}

There is a useful contrast with the dependence uncertainty problems discussed
earlier in this chapter. In a classical Fr\'echet problem, one fixes marginals and
optimizes over all couplings:
\[
  \gamma\in\Coup(\mu_1,\ldots,\mu_d).
\]
In a martingale problem, one fixes marginals across time and imposes dynamic
consistency:
\[
  \E[X_{t+1}\mid X_0,\ldots,X_t]=X_t.
\]
Thus martingale transport is not only about dependence; it is about admissible dependence under a time-consistency constraint.

\begin{computebox}
Classical OT asks:
\[
  \text{How should mass move from }\mu\text{ to }\nu?
\]
Martingale OT asks:
\[
  \text{How should mass move from }\mu\text{ to }\nu
  \text{ without creating predictable gains?}
\]
The second question is central when actuarial liabilities interact with financial markets.
\end{computebox}

The dual formulation also has an actuarial interpretation. A typical martingale OT dual writes a robust payoff bound as the cost of a semi-static hedge: static positions in instruments that determine the marginals, plus a dynamic trading strategy in the underlying. In schematic form,
\[
  \sup_{\gamma\in\Coup_{\mathrm M}(\mu,\nu)}
  \E_\gamma[\Phi(X,Y)]
  =
  \inf_{\varphi,\psi,h}
  \left\{
    \int\varphi\,\dd\mu+\int\psi\,\dd\nu
  \right\},
\]
subject to
\[
  \varphi(x)+\psi(y)+h(x)(y-x)\ge \Phi(x,y).
\]
The term \(h(x)(y-x)\) is the gain from dynamic trading. In actuarial language, it represents the part of the guarantee risk that can be hedged dynamically, while \(\varphi\) and \(\psi\) represent static calibration instruments.

\begin{remark}[Scope]
Martingale optimal transport is an advanced topic. It is not needed for ordinary loss modelling or distributional comparison. It becomes relevant when insurance liabilities are linked to tradable assets, when valuation is market-consistent, or when one wants model-independent bounds under no-arbitrage constraints.
\end{remark}

For multi-period problems, the admissible set contains laws of processes
\[
  (X_0,\ldots,X_T)
\]
such that
\[
  \E[X_{t+1}\mid X_0,\ldots,X_t]=X_t,
  \qquad t=0,\ldots,T-1.
\]
This connects martingale OT with adapted transport. Both impose temporal structure on couplings. The martingale constraint imposes financial consistency; adapted transport imposes non-anticipativity.

\subsection{Computational approaches}
\label{sec:reinsurance-computational-approaches}

The exact multi-marginal problem can be computationally expensive. In actuarial applications, one typically uses approximations adapted to the data structure.

\subsubsection*{Grid discretization}

For small \(d\), discretize each marginal on a grid and solve the linear program. This is transparent and useful for teaching or benchmarking, but it scales poorly.

\begin{computebox}
For \(d\) marginals with \(n\) grid points each, the full coupling array has \(n^d\) entries. Even \(n=50\) and \(d=4\) gives more than six million variables. This motivates structure, regularization, or simulation-based approximations.
\end{computebox}

\subsubsection*{Entropic regularization}

Add entropy to the objective:
\[
  \inf_{\gamma\in\Coup(\mu_1,
  \ldots,
  \mu_d)}
  \int c\dd\gamma+
  \eps\int \log\left(\frac{\dd\gamma}{\dd m}\right)\dd\gamma,
\]
where \(m\) is a reference product measure. This produces smoother couplings and enables generalized Sinkhorn scaling. The price is regularization bias, controlled by \(\eps\).

Entropic methods are especially useful when the goal is not an exact worst case but a stable dependence stress test.

\subsubsection*{Scenario optimization}

Another approach starts from simulated scenarios
\[
  \bx^{(1)},
  \ldots,
  \bx^{(N)}.
\]
One can reweight scenarios to match target marginals or stress constraints. If \(w_i\) is the weight assigned to scenario \(i\), a reweighted stress test solves
\[
  \max_{w_i\geq0}
  \sum_{i=1}^N w_i h\left(\sum_{k=1}^d x_k^{(i)}\right)
\]
subject to
\[
  \sum_{i=1}^N w_i=1
\]
and calibration constraints ensuring that the reweighted line-level distributions remain close to target marginals. This is a practical compromise between full coupling optimization and fixed scenario analysis.

\subsubsection*{Sliced and pairwise approximations}

For high-dimensional portfolios, one may use pairwise dependence diagnostics or random projections. For example, compare projected aggregate losses
\[
  \btheta^\top\bX
\]
for several directions \(\btheta\). Sliced Wasserstein distances can then summarize how joint dependence changes across models. This is less exact than full multi-marginal OT but often more scalable.

\subsection{Mini-lab: comparing aggregate dependence models}
\label{sec:lab-aggregate-dependence-models}

The following pseudo-code illustrates a simple simulation-based comparison of aggregate distributions under independent and comonotonic dependence. It extends the one-dimensional empirical Wasserstein computations of Lab~\ref{lab:empirical-wasserstein-noise} and the discrete coupling ideas from Lecture~\ref{lec:discrete-ot}. It is deliberately minimal and can be adapted to any marginal model.

\begin{lstlisting}[language=R]
set.seed(1)
N <- 20000
alpha <- 0.995

# Marginal quantile functions
Q1 <- function(u) qlnorm(u, meanlog = 10, sdlog = 0.6)
Q2 <- function(u) qlnorm(u, meanlog = 9.5, sdlog = 1.0)
Q3 <- function(u) qgamma(u, shape = 2, rate = 1/50000)

# Independent coupling
U_ind <- matrix(runif(3 * N), ncol = 3)
X_ind <- cbind(Q1(U_ind[,1]), Q2(U_ind[,2]), Q3(U_ind[,3]))
S_ind <- rowSums(X_ind)

# Comonotonic coupling
U <- runif(N)
X_com <- cbind(Q1(U), Q2(U), Q3(U))
S_com <- rowSums(X_com)

emp_quantile_left <- function(x, a){
  xs <- sort(x)
  xs[max(1, ceiling(a * length(xs)))]
}
VaR <- function(x, a) emp_quantile_left(x, a)
ES <- function(x, a){
  v <- VaR(x, a)
  mean(x[x >= v])
}

c(VaR_ind = VaR(S_ind, alpha),
  VaR_com = VaR(S_com, alpha),
  ES_ind  = ES(S_ind, alpha),
  ES_com  = ES(S_com, alpha))

# Wasserstein-2 distance between aggregate distributions
w2_empirical <- function(x, y){
  sqrt(mean((sort(x) - sort(y))^2))
}

w2_empirical(S_ind, S_com)
\end{lstlisting}

This lab can be extended by adding an aggregate excess-of-loss treaty:
\[
  c(s)=\min\{(s-a)_+,\ell\},
\]
then comparing retained and ceded distributions under the two couplings.

\subsection{A small worked example}
\label{sec:worked-example-layer-dependence}

Suppose two lines have marginal losses \(X_1\) and \(X_2\), each taking values in \(\{0,1,2\}\) with probabilities
\[
  \ba=(0.6,0.3,0.1),
  \qquad
  \bb=(0.7,0.2,0.1).
\]
A coupling is a matrix \(\gamma=(\gamma_{ij})\) satisfying
\[
  \sum_j\gamma_{ij}=a_i,
  \qquad
  \sum_i\gamma_{ij}=b_j,
  \qquad
  \gamma_{ij}\geq0.
\]
If the cost is
\[
  c(i,j)=h(x_i+y_j),
\]
with \(h(s)=(s-2)_+\), then the objective
\[
  \sum_{i,j}h(x_i+y_j)\gamma_{ij}
\]
is expected excess loss above threshold \(2\). Maximizing it over \(\gamma\) finds the dependence structure that creates the largest expected layer loss. Minimizing it finds the dependence structure that creates the smallest expected layer loss.

The dual problem is
\[
  \max_{\varphi,\psi}
  \sum_i a_i\varphi_i+
  \sum_j b_j\psi_j
\]
subject to
\[
  \varphi_i+
  \psi_j\leq h(x_i+y_j)
\]
for all \(i,j\). The dual variables can be interpreted as line-level marginal prices that support the layer loss cost.

\subsection{Interpretation for internal models}
\label{sec:internal-models-transport-chain}

Internal capital models often combine line-level distributions, dependence assumptions, reinsurance programs, and allocation methodologies. These components are sometimes calibrated separately. Transport notation makes the chain of operations explicit.

\begin{enumerate}[label=(\arabic*)]
  \item \textbf{Marginals.} Line-level or cell-level distributions define the constraints.
  \item \textbf{Dependence.} Couplings define admissible joint laws.
  \item \textbf{Aggregation.} The sum map pushes joint laws to aggregate laws.
  \item \textbf{Reinsurance.} Treaty maps push aggregate or event-level laws to retained and ceded laws.
  \item \textbf{Capital.} Risk functionals act on retained aggregate laws.
  \item \textbf{Allocation.} Scenario-level decompositions push capital back to lines.
\end{enumerate}

This sequence is a chain of maps and optimizations. Each step can be audited by asking:
\[
  \text{What is fixed? What is transported? What cost is minimized or maximized?}
\]

\subsection{Limitations}
\label{sec:reinsurance-transport-limitations}

The transport formulation is powerful but not automatic. Several limitations should be kept in mind.

First, optimizing over all couplings can be too conservative if some dependence information is available. In practice, one may constrain the coupling set by rank correlations, tail dependence coefficients, common factor structures, or expert judgment.

Second, finite samples make extreme tail estimates unstable. Worst-case coupling optimization can amplify this instability. Regularization and sensitivity analysis are therefore essential.

Third, contractual details matter. Real reinsurance treaties include reinstatement provisions, event definitions, hours clauses, exclusions, aggregate limits, and legal wording. A push-forward representation remains possible, but the map may be complex.

Fourth, capital allocation is partly normative. Different allocation rules answer different questions: marginal contribution, standalone risk, diversification benefit, equity among lines, or managerial incentives. Optimal transport clarifies the geometry but does not by itself decide which allocation principle is appropriate.

\subsection{Summary}
\label{sec:reinsurance-capital-summary}

This lecture has presented reinsurance, dependence stress testing, and capital allocation as transport problems. The main lessons are as follows.

\begin{itemize}
  \item A treaty maps gross losses to retained and ceded losses; therefore it pushes forward loss distributions.
  \item Comparing treaties by Wasserstein distances gives a distributional view beyond expected ceded loss and capital reduction.
  \item Aggregating line losses with fixed marginals but uncertain dependence is a coupling problem.
  \item Worst-case and best-case aggregate risk can often be expressed as multi-marginal optimal transport problems.
  \item Dual potentials can be interpreted as line-level shadow prices supporting aggregate capital costs.
  \item Capital allocation depends on the coupling because tail scenarios depend on the joint distribution.
  \item Computational practice requires discretization, regularization, scenario reduction, or approximate diagnostics.
\end{itemize}

\begin{readingbox}
For dependence and actuarial aggregation, see \cite{DenuitDhaeneGoovaertsKaas2005,McNeilFreyEmbrechts2015,EmbrechtsPuccettiRueschendorf2013}. For multi-marginal OT, see \cite{GangboSwiech1998,Pass2015}. For computational transport and entropic regularization, see Lectures~\ref{lec:entropic-sinkhorn} and \ref{lec:statistical-ot}. For reinsurance practice and optimal treaty design, see \cite{Borch1960,Gerber1979,AlbrecherTeugelsSchoutens2017}. For capital allocation, see \cite{Denault2001,Tasche2008,Shapley1953}.
\end{readingbox}

\subsection{Exercises}
\label{sec:reinsurance-capital-exercises}

\begin{exercise}[Push-forward of a quota-share treaty]\label{ex:13:1}
Let \(L\geq0\) have distribution \(\mu\) and finite second moment. For \(q\in[0,1]\), define \(R_q=qL\). Show that
\[
  \Wtwo(\Law(R_q),\Law(R_{q'}))
  =|q-q'|\sqrt{\E[L^2]}.
\]
Explain why this identity is special to one-dimensional scaling.
\end{exercise}

\begin{exercise}[Layer distribution]\label{ex:13:2}
Let \(C=\min\{(L-a)_+,\ell\}\), where \(L\) has continuous distribution function \(F\). Derive the distribution function of \(C\), identifying its atoms. Compute \(\E[C]\) in terms of the survival function \(\overline F(x)=1-F(x)\).
\end{exercise}

\begin{exercise}[Worst-case expected layer loss]\label{ex:13:3}
Let \(X\) and \(Y\) have known discrete marginal distributions supported on \(\{x_1,\ldots,x_m\}\) and \(\{y_1,\ldots,y_n\}\). Formulate as a linear program the maximization of
\[
  \E[(X+Y-a)_+]
\]
over all couplings of \(X\) and \(Y\). Write the corresponding dual problem and interpret the dual variables.
\end{exercise}

\begin{exercise}[Comonotonic aggregation]\label{ex:13:4}
Let \(X_k=Q_k(U)\), \(k=1,\ldots,d\), where \(U\sim\mathrm{Unif}(0,1)\). Show that
\[
  \VaR_\alpha\left(\sum_{k=1}^d X_k\right)
  =\sum_{k=1}^d \VaR_\alpha(X_k)
\]
when the marginal quantile functions are continuous at \(\alpha\). Discuss why this identity may fail under other dependence structures.
\end{exercise}

\begin{exercise}[Expected shortfall and multi-marginal transport]\label{ex:13:5}
Use the representation
\[
  \TVaR_\alpha(S)=\inf_{t\in\R}\left\{t+\frac{1}{1-\alpha}\E[(S-t)_+]\right\}
\]
to explain how worst-case expected shortfall over all couplings can be approximated by solving a collection of multi-marginal transport problems indexed by \(t\). What numerical difficulties arise in the tail?
\end{exercise}

\begin{exercise}[Euler allocation under two dependence models]\label{ex:13:6}
Simulate two marginal loss distributions. Construct an independent coupling and a comonotonic coupling. For each coupling, estimate
\[
  K_k=\E[X_k\mid S\geq \VaR_\alpha(S)].
\]
Compare the resulting allocations. Which line receives more capital under comonotonic dependence, and why?
\end{exercise}

\begin{exercise}[Treaty sensitivity]\label{ex:13:7}
Let \(R_a=\min(L,a)\). Using simulated losses, plot
\[
  a\mapsto \Wtwo(\Law(R_a),\Law(L))
\]
and
\[
  a\mapsto \VaR_\alpha(L)-\VaR_\alpha(R_a).
\]
Explain why these two curves need not have the same shape.
\end{exercise}

\begin{exercise}[Scenario reweighting]\label{ex:13:8}
Suppose that \(N\) joint scenarios \(\bx^{(i)}\in\R^d\) are available from an internal model. Formulate a convex optimization problem that chooses scenario weights \(w_i\) to maximize expected aggregate layer loss subject to constraints matching selected marginal means and variances. How would you add an entropy penalty to avoid concentrating all mass on a few scenarios?
\end{exercise}

\newpage
\lecture{Counterfactual Worlds and Transport}
\label{lec:counterfactual-worlds-transport}

\subsection*{Learning objectives}

This lecture introduces counterfactual reasoning as a bridge between optimal transport and actuarial applications. After this lecture, the reader should be able to:
\begin{enumerate}[label=(\roman*)]
  \item distinguish philosophical, causal, algorithmic and distributional meanings of counterfactuals;
  \item represent counterfactual worlds through couplings and push-forwards;
  \item explain why optimal transport selects one plausible counterfactual matching among many possible couplings;
  \item separate individual causal counterfactuals from distributional and actuarial counterfactuals;
  \item apply transport-based counterfactuals to recourse, catastrophe modelling, cyber risk, prevention and stochastic scenario generation;
  \item state precisely what is identified by marginals, what is selected by a cost, and what requires causal assumptions.
\end{enumerate}

\begin{readingbox}
Counterfactuals have several intellectual origins. The possible-world semantics is classically associated with \cite{lewis1973counterfactuals}. The causal view is developed through potential outcomes and structural causal models; see \cite{rubin1974estimating,holland1986statistics,pearl2009causality,imbens2015causal,hernan2020whatif,peters2017elements}. Counterfactual explanations and recourse in machine learning are discussed in \cite{wachter2017counterfactual,verma2020counterfactual,karimi2021algorithmic}. Transport-based counterfactual models are developed in \cite{delara2024transportcounterfactual}. Actuarial examples include counterfactual disaster risk analysis \cite{woo2018counterfactual}, cyberinsurance and causal attribution \cite{farahmand2020cyberinsurance}. For Schr\"odinger bridges and stochastic transport paths, see \cite{leonard2014survey,chen2021stochastic,peyre2019computational}. The optimal transport background is standard in \cite{villani2003topics,villani2008oldnew,figalli2021invitation,ambrosio2021lectures,peyre2019computational}.
\end{readingbox}

\subsection{Why counterfactual worlds in actuarial science?}
\label{subsec:cf-why-actuarial}

Actuaries reason counterfactually all the time. A technical premium is a prediction of what would happen under a contract. A reserve is a valuation of future payments under possible claims developments. A catastrophe model evaluates losses under events that have not occurred. A solvency capital requirement asks what would happen under an adverse scenario. A model audit asks how a score would change under another statistical environment. A prevention analysis asks what losses would have been observed without the protection measure, or what losses would be observed with a new one.

The word ``counterfactual'' therefore hides several questions:
\begin{itemize}
  \item what would have happened to the same policyholder under another tariff?
  \item what would the loss distribution have been under another climate, another mitigation policy, or another dependence structure?
  \item what would the portfolio have looked like if its covariate distribution had matched another market?
% To be restored in v2: fairness counterfactuals based on protected group transformations.
%  \item what would a score have been if the individual had belonged to another protected group, while remaining statistically plausible?
  \item what small perturbation of a historical event would have generated a catastrophe-level loss?
\end{itemize}

These questions are not identical. Some concern the same individual under another intervention. Some concern a whole population under another distribution. Some concern explanations of a black-box decision. Some concern stress scenarios. This lecture develops a common mathematical language for these questions: a counterfactual world is often represented by a probability measure, and a relation between the observed world and the counterfactual world is often represented by a coupling. The static constructions below point back to the couplings and Wasserstein maps of Lectures~\ref{lec:couplings}--\ref{lec:wasserstein-barycenters}; the stochastic-path version in Section~\ref{subsec:cf-schrodinger-bridges} points forward to dynamic scenario generation.

\begin{actuarialbox}
In actuarial science, the most stable and useful counterfactual object is often not an individual statement of the form ``this person's loss would have been $y$'', but a distributional statement of the form ``the loss distribution would have been $\nu$''. Optimal transport then supplies a principled way to connect the observed distribution $\mu$ and the counterfactual distribution $\nu$.
\end{actuarialbox}

\subsection{Three languages of counterfactuals}
\label{subsec:cf-three-languages}

Counterfactuals appear in philosophy, causal inference, machine learning and insurance practice. The same words are used, but the mathematical objects are different.

\subsubsection{Possible-world counterfactuals}

A possible-world counterfactual is a statement of the form
\[
  A \;\Box\!\!\to\; B,
\]
read as ``if $A$ had been the case, then $B$ would have been the case''. The Lewisian idea is that this statement is true if $B$ is true in the closest possible worlds in which $A$ is true. The key mathematical difficulty is the word ``closest''. It asks for a notion of similarity between worlds.

The connection with optimal transport is suggestive. Optimal transport also requires a notion of closeness, but it makes it explicit through a cost function $c(\bx,\by)$. In a possible-world semantics, closeness is a semantic primitive. In optimal transport, closeness is a modelling choice and can be audited.

\subsubsection{Causal counterfactuals}

In causal inference, counterfactuals are usually formalized through potential outcomes or structural causal models. For a binary treatment $T\in\{0,1\}$, the two potential outcomes are
\[
  Y_i(1),\qquad Y_i(0).
\]
The individual causal effect would be
\[
  Y_i(1)-Y_i(0),
\]
but only one of $Y_i(1)$ and $Y_i(0)$ is observed. This is the fundamental problem of causal inference.

A structural causal model gives more structure. It writes variables as functions of their parents and exogenous noises:
\[
  X_j=f_j\bigl(\bX_{\operatorname{pa}(j)},U_j\bigr),
  \qquad j=1,\ldots,d.
\]
A counterfactual is then computed by the abduction--action--prediction recipe: infer the latent noise compatible with the observation, modify the structural equations by an intervention, and propagate the resulting system.

\begin{definition}[Structural causal model]
\label{def:cf-structural-causal-model}
Let $G$ be a directed acyclic graph on variables $X_1,\ldots,X_d$. A structural causal model consists of exogenous random variables $U_1,
\ldots,U_d$, usually jointly independent, and measurable functions $f_j$ such that
\[
  X_j=f_j\bigl(\bX_{\operatorname{pa}(j)},U_j\bigr),
  \qquad j=1,\ldots,d.
\]
An intervention $do(X_k=x)$ replaces the equation for $X_k$ by the constant equation $X_k=x$, leaving the other structural equations unchanged.
\end{definition}

Causal counterfactuals are powerful because they aim at the same unit under another intervention. They are demanding because they require structural information that is usually not identified from the observational law alone.

\subsubsection{Algorithmic counterfactuals}

In machine learning, a counterfactual explanation is often a nearby point $\bx'$ that changes a decision. If $m:\mathcal X\to\R$ is a score and $y^\star$ is the desired decision, a typical formulation is
\[
  \bx^\star\in\argmin_{\bz\in\mathcal A(\bx)}
  \Bigl\{d(\bx,\bz)+\lambda\,\ell(m(\bz),y^\star)\Bigr\},
\]
where $\mathcal A(\bx)$ encodes feasibility and actionability constraints. In insurance, $\bx^\star$ may be interpreted as a recourse recommendation: what should change for a premium to decrease, for a policy to be accepted, or for a risk score to cross a threshold?

This is useful, but potentially misleading. A closest point in Euclidean distance may be out of distribution. It may recommend changing non-actionable features. It may alter a variable while leaving implausible values of correlated variables unchanged. Transport-based counterfactuals are one way to keep counterfactual profiles distributionally plausible.

\subsection{Observed and counterfactual distributions}
\label{subsec:cf-observed-counterfactual-distributions}

Let $\mathcal X$ be a Polish space. We write $\mu\in\Prob(\mathcal X)$ for the observed distribution and $\nu\in\Prob(\mathcal X)$ for a counterfactual distribution. The notation follows the measure-theoretic conventions introduced in Definition~\ref{def:push-forward} and Definition~\ref{def:coupling}. The two measures may correspond to two populations, two time periods, two climates, two legal regimes, two prevention states or two portfolios.

\begin{definition}[Distributional counterfactual]
\label{def:cf-distributional-counterfactual}
A distributional counterfactual for an observed law $\mu\in\Prob(\mathcal X)$ is an alternative law $\nu\in\Prob(\mathcal X)$ representing the population, portfolio or scenario distribution under another regime. It is a statement about a law, not yet about the same individual unit under another intervention.
\end{definition}

A distributional counterfactual is not yet a statement about individuals. It says what the alternative population law is. To make statements of the form ``this observed profile $\bx$ is matched to that counterfactual profile $\bx^\dagger$'', we need a joint law, that is, a coupling in the sense of Definition~\ref{def:coupling}.

\begin{definition}[Counterfactual coupling]
\label{def:cf-counterfactual-coupling}
Let $\mu,\nu\in\Prob(\mathcal X)$. A counterfactual coupling from $\mu$ to $\nu$ is an element
\[
  \gamma\in\Coup(\mu,\nu).
\]
If $(\bX,\bX^\dagger)\sim\gamma$, then $\bX\sim\mu$ is the observed profile and $\bX^\dagger\sim\nu$ is the counterfactual profile.
\end{definition}

The dagger notation is deliberate. It denotes a profile constructed by a selected distributional matching. It does not, by itself, denote a structural potential outcome.

\begin{proposition}[Couplings as distributional counterfactual models]
\label{prop:cf-couplings}
Let $\mu,\nu\in\Prob(\mathcal X)$. Any coupling $\gamma\in\Coup(\mu,\nu)$ defines a distributional counterfactual model by specifying the joint law of an observed profile $\bX\sim\mu$ and an alternative profile $\bX^\dagger\sim\nu$. Conversely, any joint law of $(\bX,\bX^\dagger)$ with marginals $\mu$ and $\nu$ is an element of $\Coup(\mu,\nu)$.
\end{proposition}

\begin{proof}
This is exactly Kantorovich's definition of a transference plan; see Chapter~1 of \cite{villani2003topics}, Chapter~4 of \cite{villani2008oldnew}, and the introductory chapters of \cite{figalli2021invitation}. Sketch of the proof: if $\gamma\in\Coup(\mu,\nu)$, then by definition its first marginal is $\mu$ and its second marginal is $\nu$. Thus a random vector $(\bX,\bX^\dagger)\sim\gamma$ has $\bX\sim\mu$ and $\bX^\dagger\sim\nu$. Conversely, the law of any pair $(\bX,\bX^\dagger)$ with these two marginals satisfies the two marginal constraints in the definition of $\Coup(\mu,\nu)$.
\end{proof}

\begin{remark}[A coupling is not a causal model]
A coupling specifies how two marginal distributions are matched. It does not specify why one world changes into the other. A causal reading requires assumptions that select the coupling as the structural relation between observed and counterfactual units.
\end{remark}

\subsection{Non-identification from marginals}
\label{subsec:cf-nonidentification}

The two marginals $\mu$ and $\nu$ do not identify the joint law $\gamma$. This is the first warning in any transport-based counterfactual analysis. If we know only the current distribution and the target distribution, we do not know which individual in the first distribution corresponds to which individual in the second.

\begin{proposition}[Non-identification of individual counterfactuals]
\label{prop:nonidentification-cf}
Let $\mu,\nu\in\Prob(\mathcal X)$. Unless additional assumptions select a unique element of $\Coup(\mu,\nu)$, the joint law of $(\bX,\bX^\dagger)$ is not identified by the two marginal laws $\mu$ and $\nu$. In particular, individual counterfactual quantities depending on the pair $(\bX,\bX^\dagger)$ are generally not identified by $\mu$ and $\nu$ alone.
\end{proposition}

\begin{proof}
This is the elementary non-uniqueness of Kantorovich couplings; compare the discussion of transport plans in \cite{villani2003topics,villani2008oldnew} and the discrete formulation in \cite{peyre2019computational}. Sketch of the proof: it is enough to give a two-point example. Let $\mu=\nu=\frac12\delta_0+\frac12\delta_1$. Both
\[
  \gamma_1=\frac12\delta_{(0,0)}+\frac12\delta_{(1,1)}
  \qquad\text{and}\qquad
  \gamma_2=\frac12\delta_{(0,1)}+\frac12\delta_{(1,0)}
\]
belong to $\Coup(\mu,\nu)$. Under $\gamma_1$, $X=X^\dagger$ almost surely, whereas under $\gamma_2$, $|X-X^\dagger|=1$ almost surely. The marginal laws are identical, but the implied individual counterfactual displacements are different.
\end{proof}

\begin{actuarialbox}
A before/after distribution is not an audit trail. The audit trail is the coupling: it tells which units, or which pieces of mass, moved from which observed states to which counterfactual states.
\end{actuarialbox}

\subsection{Transport maps as deterministic counterfactuals}
\label{subsec:cf-transport-maps}

A counterfactual coupling may be deterministic. If $T:\mathcal X\to\mathcal X$ satisfies
\[
  T_{\push}\mu=\nu,
\]
then $\bX^\dagger=T(\bX)$ has law $\nu$. The associated coupling is $(\Id,T)_{\push}\mu$. In this case every observed point is assigned a unique counterfactual point.

\begin{definition}[Transport-based counterfactual map]
\label{def:cf-transport-map}
Let $\mu,\nu\in\Prob(\mathcal X)$. A measurable map $T:\mathcal X\to\mathcal X$ is a transport-based counterfactual map from $\mu$ to $\nu$ if $T_{\push}\mu=\nu$. The counterfactual profile associated with $\bX\sim\mu$ is
\[
  \bX^\dagger=T(\bX).
\]
\end{definition}

\begin{proposition}[Maps are graph couplings]
\label{prop:cf-graph-couplings}
Let $T:\mathcal X\to\mathcal X$ be measurable and let $\mu\in\Prob(\mathcal X)$. Then
\[
  T_{\push}\mu=\nu
  \quad\Longleftrightarrow\quad
  (\Id,T)_{\push}\mu\in\Coup(\mu,\nu).
\]
Moreover, the coupling $(\Id,T)_{\push}\mu$ is concentrated on the graph of $T$.
\end{proposition}

\begin{proof}
This is the standard graph-plan construction of Monge transport; see Chapter~1 of \cite{villani2003topics}, Chapter~1 of \cite{figalli2021invitation}, and \cite{ambrosio2021lectures}. Sketch of the proof: for any measurable $A\subseteq\mathcal X$,
\[
  (\Id,T)_{\push}\mu(A\times\mathcal X)=\mu(A),
\]
so the first marginal is $\mu$. For any measurable $B\subseteq\mathcal X$,
\[
  (\Id,T)_{\push}\mu(\mathcal X\times B)=\mu(T^{-1}(B))=T_{\push}\mu(B).
\]
Hence the second marginal is $\nu$ exactly when $T_{\push}\mu=\nu$. The graph concentration follows because $(\bx,T(\bx))$ lies on the graph of $T$ for $\mu$-almost every $x$.
\end{proof}

Deterministic maps are attractive for explanation because they provide a single answer. But they can be too restrictive: discrete mass may need to be split, and in many applications a stochastic counterfactual rule is more realistic.

\subsection{Optimal transport as a selection rule}
\label{subsec:cf-ot-selection-rule}

The set $\Coup(\mu,\nu)$ contains many counterfactual matchings. Optimal transport selects one by minimizing an explicit cost. Let $c:\mathcal X\times\mathcal X\to\R\cup\{+\infty\}$ be a lower semicontinuous cost. The optimal transport problem is
\[
  \OT_c(\mu,\nu)
  =
  \inf_{\gamma\in\Coup(\mu,\nu)}
  \int_{\mathcal X\times\mathcal X} c(\bx,\bx^\dagger)\,\gamma(\dd \bx,\dd \bx^\dagger).
\]

In counterfactual language, $c(\bx,\bx^\dagger)$ is the cost of declaring $\bx^\dagger$ to be the alternative version of $\bx$. This cost may measure physical distance, financial effort, behavioural change, social plausibility, causal implausibility, legal inadmissibility or statistical displacement.

\begin{theorem}[Existence of an optimal counterfactual coupling]
\label{thm:existence-cf-ot}
Let $\mathcal X$ be Polish. Let $\mu,\nu\in\Prob(\mathcal X)$, and let $c:\mathcal X\times\mathcal X\to[0,+\infty]$ be lower semicontinuous. If there exists at least one coupling $\gamma\in\Coup(\mu,\nu)$ with finite cost, then there exists an optimal coupling $\gamma^\star\in\Coup(\mu,\nu)$ such that
\[
  \int c\,\dd\gamma^\star
  =
  \inf_{\gamma\in\Coup(\mu,\nu)}\int c\,\dd\gamma.
\]
\end{theorem}

\begin{proof}
This is the standard existence theorem for Kantorovich's problem; see Theorem~4.1 in \cite{villani2008oldnew}, Chapter~1 of \cite{villani2003topics}, and \cite{ambrosio2021lectures}. Sketch of the proof: take a minimizing sequence $(\gamma_n)\subset\Coup(\mu,\nu)$. The set of couplings is tight and closed, hence compact for weak convergence on Polish spaces. Extract a weakly convergent subsequence $\gamma_{n_k}\Rightarrow\gamma^\star$. Closedness of the marginal constraints gives $\gamma^\star\in\Coup(\mu,\nu)$. Lower semicontinuity of the integral functional $\gamma\mapsto\int c\,\dd\gamma$ gives
\[
  \int c\,\dd\gamma^\star
  \le
  \liminf_k \int c\,\dd\gamma_{n_k},
\]
which proves optimality.
\end{proof}

\begin{actuarialbox}
Optimal transport does not identify the true counterfactual world. It selects a counterfactual world by a declared criterion. This is a strength for governance: the cost function can be discussed, challenged and documented.
\end{actuarialbox}

\subsection{One-dimensional counterfactuals}
\label{subsec:cf-one-dimensional}

On the real line, the monotone transport gives an explicit deterministic counterfactual map. Let $\mu,\nu\in\Prob(\R)$, with distribution functions $F_\mu,F_\nu$ and quantile functions
\[
  Q_\mu=F_\mu^{-1},\qquad Q_\nu=F_\nu^{-1}.
\]
This notation is the same as in the Wasserstein and risk-measure chapters. If $X\sim\mu$, define
\[
  T(x)=Q_\nu(F_\mu(x))
\]
when $\mu$ is continuous. More generally, the increasing rearrangement is represented by
\[
  X=Q_\mu(U),
  \qquad
  X^\dagger=Q_\nu(U),
  \qquad U\sim\mathrm{Unif}(0,1).
\]

\begin{proposition}[Quantile counterfactual on the real line]
\label{prop:quantile-cf}
Let $\mu,\nu\in\Prob_p(\R)$ and let $U\sim\mathrm{Unif}(0,1)$. Then
\[
  (X,X^\dagger)=\bigl(Q_\mu(U),Q_\nu(U)\bigr)
\]
defines a coupling of $\mu$ and $\nu$. Moreover, for the cost $c(x,y)=|x-y|^p$, this coupling is optimal and
\[
  \Wp{p}^p(\mu,\nu)
  =
  \int_0^1\bigl|Q_\mu(u)-Q_\nu(u)\bigr|^p\,\dd u.
\]
\end{proposition}

\begin{proof}
This is the classical one-dimensional monotone rearrangement theorem; see Chapter~2 of \cite{villani2003topics}, Chapter~2 of \cite{figalli2021invitation}, and the computational account in \cite{peyre2019computational}. Sketch of the proof: the quantile transform gives $F_\mu^{-1}(U)\sim\mu$ and $F_\nu^{-1}(U)\sim\nu$, hence the pair is a coupling. Optimality follows from monotonicity: if two pairs are crossed, replacing $(x_1,y_2),(x_2,y_1)$ by $(x_1,y_1),(x_2,y_2)$, with $x_1\le x_2$ and $y_1\le y_2$, cannot increase the convex cost $|x-y|^p$. Iterating this uncrossing argument yields the increasing coupling, and integration over quantiles gives the formula.
\end{proof}

\begin{example}[Severity ranks]
Let $\mu$ be the distribution of baseline claim severities and $\nu$ the distribution under a stress scenario. The monotone counterfactual map sends each baseline severity quantile to the same stressed severity quantile. It preserves ranks. This is often interpretable, but it may be inappropriate if the stress changes the ordering of risks in a meaningful way.
\end{example}

\subsection{Gaussian counterfactuals}
\label{subsec:cf-gaussian}

In multivariate actuarial examples, Gaussian approximations often appear: log-loss factors after transformation, latent scores, reserving residuals or climate-index perturbations. For quadratic cost, the optimal map between non-degenerate Gaussian distributions is affine.

\begin{proposition}[Gaussian quadratic transport]
\label{prop:gaussian-cf}
Let $\mu=\mathcal N(\bm_0,\Sigma_0)$ and $\nu=\mathcal N(\bm_1,\Sigma_1)$ on $\R^d$, where $\Sigma_0$ and $\Sigma_1$ are positive definite. For the quadratic cost $c(\bx,\by)=\|\bx-\by\|^2$ (equivalently, for $W_2^2$), the optimal transport map from $\mu$ to $\nu$ is
\[
  T(\bx)=\bm_1+A(\bx-\bm_0),
\]
where
\[
  A=
  \Sigma_0^{-1/2}
  \bigl(\Sigma_0^{1/2}\Sigma_1\Sigma_0^{1/2}\bigr)^{1/2}
  \Sigma_0^{-1/2}.
\]
In particular, $T_{\push}\mu=\nu$, and
\[
  \Wtwo^2(\mu,\nu)
  =
  \|\bm_0-\bm_1\|^2
  +\operatorname{tr}
  \left(
  \Sigma_0+\Sigma_1
  -2(\Sigma_0^{1/2}\Sigma_1\Sigma_0^{1/2})^{1/2}
  \right).
\]
\end{proposition}

\begin{proof}
This is the standard Gaussian formula for quadratic optimal transport; see Chapter~2 of \cite{villani2003topics}, Chapter~7 of \cite{villani2008oldnew}, and the Gaussian examples in \cite{peyre2019computational}. Sketch of the proof: if $\bX\sim\mathcal N(\bm_0,\Sigma_0)$, then $T(\bX)$ is Gaussian with mean $\bm_1$ and covariance $A\Sigma_0A$. The displayed matrix $A$ is symmetric positive definite and satisfies $A\Sigma_0A=\Sigma_1$, hence $T_{\push}\mu=\nu$. Brenier's theorem identifies the quadratic optimal map as the gradient of a convex function; the affine map above has symmetric positive definite linear part, so it is such a gradient. Uniqueness of the Brenier map gives optimality.
\end{proof}

\iffalse
% To be restored in v2: fairness diagnostic example.
\begin{example}[A Gaussian fairness counterfactual]
Suppose that $X\mid S=0\sim\mathcal N(m_0,\Sigma_0)$ and $X\mid S=1\sim\mathcal N(m_1,\Sigma_1)$. The Gaussian transport map gives a plausible transformed profile
\[
  X^\dagger=m_1+A(X-m_0)
\]
with the distribution of group $S=1$. A counterfactual fairness diagnostic for a score $m(x,s)$ may compare $m(X,0)$ and $m(X^\dagger,1)$. This is a transport-based diagnostic, not automatically a causal effect of changing $S$.
\end{example}
\fi

\subsection{Individual, distributional and actuarial counterfactuals}
\label{subsec:cf-three-levels}

It is helpful to separate three levels.

\paragraph{Individual causal counterfactual.} For unit $i$, an expression such as $Y_i(1)-Y_i(0)$ compares two potential outcomes of the same unit. It requires a causal model or an identification design.

\paragraph{Distributional counterfactual.} For a population, a counterfactual law $\nu$ describes the distribution of outcomes in an alternative world. We may compare $\mu$ and $\nu$ by expectations, quantiles, Wasserstein distances or stochastic orders.

\paragraph{Actuarial counterfactual.} For a loss variable $L$, a risk functional $\Risk$ produces quantities such as
\[
  \Risk(L^\dagger)-\Risk(L),
  \qquad
  \Risk\in\{\E,\VaR_\alpha,\TVaR_\alpha,\ES_\alpha,\SCR\}.
\]
This is the natural scale for pricing, capital, reserving, prevention and reinsurance.

\begin{definition}[Transport-induced actuarial counterfactual]
\label{def:cf-actuarial-counterfactual}
Let $L=\ell(\bX)$ be a loss generated by an observed profile $\bX\sim\mu$. Let $\gamma\in\Coup(\mu,\nu)$ and $(\bX,\bX^\dagger)\sim\gamma$. Given a loss function $\ell^\dagger$ in the counterfactual world, define
\[
  L^\dagger=\ell^\dagger(\bX^\dagger).
\]
For a risk functional $\Risk$, the transport-induced actuarial counterfactual effect is
\[
  \Delta_\Risk(\gamma)
  =
  \Risk(L^\dagger)-\Risk(L).
\]
\end{definition}

\begin{proposition}[Dependence of pathwise effects on the coupling]
\label{prop:pathwise-depend-on-coupling}
Let $L=\ell(\bX)$ and $L^\dagger=\ell^\dagger(\bX^\dagger)$. Distributional quantities depending only on $\Law(L)$ and $\Law(L^\dagger)$, such as $\E[L^\dagger]-\E[L]$, are determined by the two marginal laws. Pathwise quantities depending on $(L,L^\dagger)$, such as $\E[|L^\dagger-L|]$ or $\Pbb(L^\dagger<L)$, depend on the coupling.
\end{proposition}

\begin{proof}
This is the elementary distinction between marginal functionals and joint functionals; it is the same distinction underlying the Fr{\'e}chet class in actuarial dependence theory and the set of couplings in \cite{villani2003topics,villani2008oldnew}. Sketch of the proof: if a functional can be written as $\int h(l)\,\Law(L)(\dd l)$ and $\int h^\dagger(l)\,\Law(L^\dagger)(\dd l)$, then it is determined by the marginal laws. In contrast, $\E[|L^\dagger-L|]=\int |l^\dagger-l|\,\eta(\dd l,\dd l^\dagger)$, where $\eta=\Law(L,L^\dagger)$. The joint law $\eta$ is induced by the coupling $\gamma$, and different couplings with the same marginals can give different values, as in Proposition~\ref{prop:nonidentification-cf}.
\end{proof}

\subsection{Counterfactual explanations and recourse in insurance}
\label{subsec:cf-recourse-insurance}

Counterfactual explanations are often individual and local. Given a model $m$, a policyholder profile $\bx$, and a target premium $p^\star$, one may solve
\[
  \bx^\star
  \in
  \argmin_{\bz\in\mathcal A(\bx)} d(\bx,\bz)
  \quad\text{subject to}\quad
  m(\bz)\le p^\star.
\]
The set $\mathcal A(\bx)$ is crucial. It encodes which changes are feasible, actionable, legal and meaningful. In insurance, the following distinction matters.

\begin{center}
\begin{tabular}{p{.28\textwidth}p{.58\textwidth}}
\toprule
Type of variable & Counterfactual recourse interpretation \\
\midrule
Contract choices & Deductible, limit, coverage, usage-based options; often actionable \\
Behavioural variables & Mileage, prevention, maintenance, security practices; potentially actionable but costly \\
Historical variables & Past claims, previous defaults, prior incidents; not directly changeable \\
Contextual variables & Territory, occupation, climate exposure; sometimes changeable, often socially constrained \\
Protected attributes & Sex, race, disability, age in some contexts; not legitimate recourse variables \\
Proxy variables & May be technically actionable but normatively problematic \\
\bottomrule
\end{tabular}
\end{center}

\begin{proposition}[Linear premium recourse]
\label{prop:linear-recourse}
Let $m(\bx)=\beta_0+\bbeta^\top \bx$ on $\R^d$. Suppose all coordinates are actionable and distance is Euclidean. If $m(\bx)>p^\star$ and $\bbeta\ne\bzero$, the closest point satisfying $m(\bz)\le p^\star$ is
\[
  \bz^\star
  =
  \bx-
  \frac{m(\bx)-p^\star}{\|\bbeta\|^2}\,\bbeta.
\]
If $m(\bx)\le p^\star$, then $\bz^\star=\bx$.
\end{proposition}

\begin{proof}
This is the Euclidean projection onto a closed half-space; see the projection theorem in Hilbert spaces in \cite{bauschke2017convex} and the convex optimization background in \cite{boyd2004convex}. Sketch of the proof: the feasible set is the half-space $H=\{\bz:\bbeta^\top \bz\le p^\star-\beta_0\}$. If $\bx\in H$, the projection is $\bx$. Otherwise, the projection lies on the boundary $\bbeta^\top \bz=p^\star-\beta_0$ and has the form $\bz=\bx-t\bbeta$. Enforcing the boundary condition gives $t=(m(\bx)-p^\star)/\|\bbeta\|^2$.
\end{proof}

\begin{actuarialbox}
A recourse recommendation is not neutral. Telling a policyholder to increase a deductible is different from telling them to move neighborhood, change occupation, or alter a proxy for a protected attribute. Counterfactual explanations need actuarial, legal and ethical constraints, not only a distance function.
\end{actuarialbox}

\subsection{Catastrophe and disaster counterfactuals}
\label{subsec:cf-catastrophe}

Catastrophe risk is naturally counterfactual. The observed historical record is only one realization. Many near misses did not enter loss databases because no large loss occurred. Counterfactual disaster analysis asks what perturbation of a historical event would have generated a severe loss, and how plausible that perturbation was.

A simple representation is the following. For each historical event $j$, define possible counterfactual losses
\[
  C_1^{(j)},C_2^{(j)},\ldots,C_K^{(j)}
\]
with probabilities
\[
  p_1^{(j)},p_2^{(j)},\ldots,p_K^{(j)}.
\]
The event generates a local counterfactual loss distribution
\[
  \nu_j=\sum_{k=1}^K p_k^{(j)}\delta_{C_k^{(j)}}.
\]
Aggregating across historical events produces a counterfactual catalogue. This is close to catastrophe modelling, but with a past event as the anchor and perturbations as the scenario generator.

\begin{definition}[Counterfactual catastrophe catalogue]
\label{def:cf-catastrophe-catalogue}
Let historical events be indexed by $j=1,\ldots,J$. For each event, let $\nu_j\in\Prob(\R_+)$ be a distribution of counterfactual losses generated by perturbing the event. Given event weights $w_j\ge0$, $\sum_jw_j=1$, the counterfactual catalogue distribution is
\[
  \nu_{\mathrm{cat}}
  =
  \sum_{j=1}^J w_j\nu_j.
\]
\end{definition}

\begin{proposition}[Risk metrics from a counterfactual catalogue]
\label{prop:cat-catalogue-risk}
Let $L^\dagger\sim\nu_{\mathrm{cat}}$. Then expected annual loss, value-at-risk and tail value-at-risk are computed from the single distribution $\nu_{\mathrm{cat}}$:
\[
  \mathrm{EAL}^\dagger=\E[L^\dagger],
  \qquad
  \VaR_\alpha^\dagger=F_{\nu_{\mathrm{cat}}}^{-1}(\alpha),
\]
and, when integrable,
\[
  \TVaR_\alpha^\dagger
  =
  \frac{1}{1-\alpha}
  \int_\alpha^1 F_{\nu_{\mathrm{cat}}}^{-1}(u)\,\dd u.
\]
\end{proposition}

\begin{proof}
This is the standard quantile representation of VaR and TVaR; see actuarial risk-measure references such as \cite{denuit2005actuarial} and the quantile notation in \cite{charpentier2024insurance}. Sketch of the proof: by definition, VaR is the generalized inverse of the distribution function. For integrable nonnegative losses, TVaR is the average of upper quantiles. Applying these definitions to the distribution $\nu_{\mathrm{cat}}$ gives the formulas.
\end{proof}

\subsection{Cyber counterfactuals and causal attribution}
\label{subsec:cf-cyber}

Cyber risk is particularly sensitive to counterfactual thinking because historical data are unstable. A vulnerability may be patched after an incident. Attackers adapt. Reporting changes. Legal liability depends on whether a loss would have happened under reasonable security practices.

Let $A$ denote a security action, such as multi-factor authentication, segmentation, patching, backup isolation or incident response. Let $L(A)$ be the loss under action $A$. A causal cyber question is
\[
  L(A=1)-L(A=0),
\]
where $A=1$ may mean that a control is implemented. A distributional cyber question is
\[
  \Law(L\mid A=1)\quad\text{versus}\quad \Law(L\mid A=0),
\]
possibly after controlling for firm size, sector and threat environment.

Transport can help describe how a loss distribution moves under a security intervention:
\[
  T_{\push}\Law(L^{\mathrm{without}})=\Law(L^{\mathrm{with}}).
\]
But legal attribution usually needs more than this distributional movement. It asks whether the intervention would have prevented this incident, not only whether it improves the portfolio distribution.

\begin{remark}[Cyberinsurance and non-stationarity]
Cyber counterfactuals often mix causal and distributional questions. Pricing requires a distribution under future threat conditions. Liability requires a causal statement about a specific incident. Governance requires documenting the assumptions connecting the two.
\end{remark}

\subsection{Prevention counterfactuals}
\label{subsec:cf-prevention}

Prevention is a central actuarial counterfactual. Let $L^{(0)}$ denote the loss without prevention and $L^{(1)}$ the loss with prevention. A prevention benefit can be expressed at several levels:
\[
  \Delta_{\E}
  =
  \E[L^{(0)}]-\E[L^{(1)}],
\]
\[
  \Delta_{\VaR_\alpha}
  =
  \VaR_\alpha(L^{(0)})-\VaR_\alpha(L^{(1)}),
\]
\[
  \Delta_{\TVaR_\alpha}
  =
  \TVaR_\alpha(L^{(0)})-\TVaR_\alpha(L^{(1)}).
\]
These quantities may tell different stories. A prevention action can reduce frequent small losses while leaving the tail unchanged; another can leave the mean almost unchanged while cutting the extreme tail.

\begin{definition}[Distributional prevention value]
\label{def:cf-prevention-value}
Let $\mu_0=\Law(L^{(0)})$ and $\mu_1=\Law(L^{(1)})$. For a risk functional $\Risk$, the distributional prevention value is
\[
  V_\Risk(\mu_0,\mu_1)
  =
  \Risk(\mu_0)-\Risk(\mu_1),
\]
where $\Risk(\mu)$ denotes the value of $\Risk(L)$ for $L\sim\mu$.
\end{definition}

\begin{proposition}[Mean prevention value is marginal]
\label{prop:mean-prevention-marginal}
Let $(L^{(0)},L^{(1)})$ be any coupling of $\mu_0$ and $\mu_1$, with both means finite. Then
\[
  \E[L^{(0)}-L^{(1)}]
  =
  \int l\,\mu_0(\dd l)-\int l\,\mu_1(\dd l),
\]
so the mean prevention value does not depend on the coupling. By contrast, $\Pbb(L^{(1)}<L^{(0)})$ and $\E[(L^{(0)}-L^{(1)})_+]$ generally depend on the coupling.
\end{proposition}

\begin{proof}
This is the linearity of expectation under couplings; compare the marginal-versus-joint distinction in \cite{villani2008oldnew} and actuarial dependence discussions in \cite{denuit2005actuarial}. Sketch of the proof: for any coupling $\eta\in\Coup(\mu_0,\mu_1)$,
\[
  \int(l_0-l_1)\,\eta(\dd l_0,\dd l_1)
  =
  \int l_0\,\mu_0(\dd l_0)-\int l_1\,\mu_1(\dd l_1),
\]
because the marginals of $\eta$ are $\mu_0$ and $\mu_1$. The final claim follows because nonlinear functions of the pair depend on its joint law, as shown by the two-point example in Proposition~\ref{prop:nonidentification-cf}.
\end{proof}

\begin{actuarialbox}
For prevention accounting, it is often safer to report marginal distributional benefits such as $\Delta_\E$, $\Delta_{\VaR}$ and $\Delta_{\TVaR}$. Statements such as ``this specific loss was avoided'' require a stronger coupling or causal model.
\end{actuarialbox}

\subsection{Sequential and graph-aligned counterfactuals}
\label{subsec:cf-sequential-graph-aligned}

A global transport map can be difficult to interpret when covariates have an internal structure. Suppose the variables satisfy a graph
\[
  X_1\longrightarrow X_2\longrightarrow X_3\longrightarrow Y.
\]
If we transport $(X_1,X_2,X_3)$ as a single vector, we may align the joint distribution while obscuring how each variable changes conditionally on its parents. A graph-aligned counterfactual transports variables sequentially.

Let $G$ be a directed acyclic graph on $\bX=(X_1,\ldots,X_d)$. For two groups $A=0$ and $A=1$, write the conditional factorization
\[
  P(\bX\mid A=a)
  =
  \prod_{j=1}^d
  P\bigl(X_j\mid X_{\operatorname{pa}(j)},A=a\bigr)
\]
for a topological order. A sequential transport constructs $\bX^\dagger$ one coordinate at a time:
\[
  X_1^\dagger=T_1(X_1),
\]
\[
  X_j^\dagger
  =
  T_j\bigl(X_j\mid X_{\operatorname{pa}(j)}^\dagger\bigr),
  \qquad j=2,\ldots,d,
\]
where each $T_j$ transports a conditional law in group $0$ toward a conditional law in group $1$.

\begin{definition}[Graph-aligned transport counterfactual]
\label{def:cf-graph-aligned-transport}
Let $G$ be a DAG on $\bX=(X_1,\ldots,X_d)$. A graph-aligned transport counterfactual from group $0$ to group $1$ is a sequential construction of $X^\dagger$ such that, for each node $j$, the conditional law of $X_j^\dagger$ given its already transported parents is aligned with the target conditional law
\[
  \Law(X_j\mid X_{\operatorname{pa}(j)},A=1).
\]
\end{definition}

\begin{proposition}[Sequential factorization preserves the target law]
\label{prop:sequential-target-law}
Assume that, in a topological order, each step of a sequential construction samples or transports $X_j^\dagger$ from the target conditional kernel
\[
  P_1(\dd x_j\mid x_{\operatorname{pa}(j)}^\dagger)
  =
  \Law(X_j\in\dd x_j\mid X_{\operatorname{pa}(j)}=x_{\operatorname{pa}(j)}^\dagger,A=1).
\]
Then $\bX^\dagger\sim\Law(\bX\mid A=1)$.
\end{proposition}

\begin{proof}
This is the standard chain-rule construction of a joint distribution from conditional kernels on a DAG; compare the disintegration theorem in \cite{villani2008oldnew}, the causal factorization used in \cite{pearl2009causality}, and graph-aligned transport constructions such as \cite{fernandesmachado2026sequential}. Sketch of the proof: in a topological order, the joint law generated by the sequential kernels is
\[
  \prod_{j=1}^d P_1(\dd x_j\mid x_{\operatorname{pa}(j)}),
\]
which is precisely the conditional factorization of $\Law(\bX\mid A=1)$ under the DAG. Therefore the generated vector has the target group law.
\end{proof}

\begin{remark}[Sequential transport is still not automatically causal]
The graph may encode a statistical ordering, a causal ordering, or an audit ordering. Only in the causal case, with appropriate structural assumptions, should the sequentially transported profile be interpreted as a causal counterfactual.
\end{remark}

\subsection{What the cost means}
\label{subsec:cf-cost-meaning}

The cost function is where mathematical transport becomes modelling. In actuarial counterfactuals, common choices include:
\begin{itemize}
  \item \textbf{geometric costs}, such as $\|\bx-\by\|^2$, for numerical covariates;
  \item \textbf{rank costs}, based on quantiles, for one-dimensional scores and losses;
  \item \textbf{economic costs}, such as the financial effort required to change a deductible, install protection, or reduce exposure;
  \item \textbf{plausibility costs}, penalizing moves outside the support of the data;
  \item \textbf{legal costs}, assigning infinite cost to prohibited changes;
  \item \textbf{causal costs}, penalizing changes that violate a known structural ordering.
\end{itemize}

A cost is therefore not only a numerical device. It encodes a theory of what changes are small, plausible, admissible or actionable.

\begin{definition}[Actionability-constrained transport cost]
\label{def:cf-actionability-cost}
Let $\mathcal A\subseteq\mathcal X\times\mathcal X$ be the set of admissible counterfactual moves. Given a base cost $c_0$, define
\[
  c_{\mathcal A}(\bx,\by)
  =
  \begin{cases}
    c_0(\bx,\by), & (\bx,\by)\in\mathcal A,\\
    +\infty, & (\bx,\by)\notin\mathcal A.
  \end{cases}
\]
The corresponding optimal transport problem selects the cheapest admissible counterfactual coupling.
\end{definition}

\begin{proposition}[Hard constraints as infinite costs]
\label{prop:hard-constraints-infinite-costs}
Let $c_{\mathcal A}$ be defined as above. A coupling $\gamma\in\Coup(\mu,\nu)$ has finite $c_{\mathcal A}$-cost only if $\gamma(\mathcal A)=1$. Conversely, if $\gamma(\mathcal A)=1$ and $\int c_0\,\dd\gamma<\infty$, then $\gamma$ has finite $c_{\mathcal A}$-cost.
\end{proposition}

\begin{proof}
This is the standard way of encoding constraints in variational problems and optimal transport; see the constrained formulations in \cite{villani2008oldnew} and the optimization background in \cite{boyd2004convex}. Sketch of the proof: if $\gamma(\mathcal A^c)>0$, then $c_{\mathcal A}=+\infty$ on a set of positive $\gamma$-mass, so $\int c_{\mathcal A}\,\dd\gamma=+\infty$. If $\gamma(\mathcal A)=1$, then $c_{\mathcal A}=c_0$ $\gamma$-almost surely, so the two integrals coincide.
\end{proof}

\subsection{Mini-lab: transport counterfactuals in R}
\label{subsec:cf-mini-lab}

The following code illustrates three counterfactual matchings between two score distributions: rank-preserving, independent and random permutation. The marginals are the same; the implied individual displacements differ.

\begin{lstlisting}[language=R]
set.seed(123)
n <- 2000

# observed and counterfactual score distributions
x <- rnorm(n, mean = 0.0, sd = 1.0)
y <- rnorm(n, mean = 0.8, sd = 1.4)

# 1. rank-preserving / monotone transport
x_ord <- sort(x)
y_ord <- sort(y)
disp_rank <- y_ord - x_ord

# 2. independent coupling
x_ind <- sample(x, n, replace = FALSE)
y_ind <- sample(y, n, replace = FALSE)
disp_ind <- y_ind - x_ind

# 3. reverse-rank coupling
x_rev <- sort(x)
y_rev <- sort(y, decreasing = TRUE)
disp_reverse <- y_rev - x_rev

out <- rbind(
  rank_preserving = c(mean = mean(disp_rank),
                      mean_abs = mean(abs(disp_rank)),
                      q95_abs = quantile(abs(disp_rank), .95)),
  independent = c(mean = mean(disp_ind),
                  mean_abs = mean(abs(disp_ind)),
                  q95_abs = quantile(abs(disp_ind), .95)),
  reverse_rank = c(mean = mean(disp_reverse),
                   mean_abs = mean(abs(disp_reverse)),
                   q95_abs = quantile(abs(disp_reverse), .95))
)
round(out, 3)

# all three have approximately the same marginal means difference,
# but the pathwise displacement metrics differ.
\end{lstlisting}

\begin{computebox}
This lab is deliberately simple. It shows the distinction between marginal effects and coupling-dependent effects. The mean displacement is approximately the difference between marginal means. The mean absolute displacement depends heavily on the coupling.
\end{computebox}

\subsection{Schr\"odinger bridges and stochastic counterfactual paths}
\label{subsec:cf-schrodinger-bridges}

Optimal transport gives a static counterfactual rule. Given an observed law \(\mu_0\) and a target law \(\mu_1\), one selects a coupling
\[
  \gamma^\star \in \Coup(\mu_0,\mu_1)
\]
that minimizes a cost. If the optimal plan is induced by a map \(T\), the counterfactual version of an observation \(\bx\) is \(T(\bx)\). This is a powerful construction, but it is static: it says where mass starts and where it ends, not how it travels from one distribution to the other.

In many actuarial problems, the path matters. A loss distribution may evolve over several years under climate change. A portfolio may drift gradually as new business enters and older contracts lapse. A prevention policy may reduce risk progressively. A model intervention may be phased in over renewal cycles. In such cases, a static counterfactual endpoint is not enough. We may want a stochastic path
\[
  (X_t)_{0\le t\le 1}
\]
such that
\[
  X_0 \sim \mu_0,
  \qquad
  X_1 \sim \mu_1,
\]
while the whole process remains as close as possible to a reference dynamics.

This is the idea behind Schr\"odinger bridges; it connects the present lecture with dynamic OT in Lecture~\ref{lec:dynamic-ot} and entropic regularization in Lecture~\ref{lec:entropic-sinkhorn}.

\begin{definition}[Reference dynamics]
\label{def:cf-reference-dynamics}
Let \(R\) be a probability measure on path space
\[
  \Omega = C([0,1],\R^d)
\]
or, more generally, on a measurable space of trajectories. We write \(X_t(\omega)=\omega(t)\) for the canonical process. The law \(R\) represents a reference dynamics, for example an uncontrolled diffusion, a baseline portfolio evolution, or a prior scenario generator.
\end{definition}

\begin{definition}[Schr\"odinger bridge problem]
\label{def:cf-schrodinger-bridge}
Let \(\mu_0,\mu_1\in\Prob(\R^d)\). The Schr\"odinger bridge from \(\mu_0\) to \(\mu_1\), relative to the reference path law \(R\), is the solution of
\[
  \inf_{P}
  \left\{
    \KL(P\,\|\,R)
    :
    P\circ X_0^{-1}=\mu_0,\;
    P\circ X_1^{-1}=\mu_1
  \right\},
\]
where \(P\) ranges over probability measures on path space and \(\KL(P\,\|\,R)\) is the relative entropy of \(P\) with respect to \(R\).
\end{definition}

Thus the Schr\"odinger bridge is the least informative change of dynamics that forces the prescribed initial and terminal laws. It does not choose the shortest path in a deterministic sense. It chooses the path law closest to a reference stochastic model, under marginal constraints.

\begin{actuarialbox}
Optimal transport asks: what is the cheapest way to move the distribution \(\mu_0\) into \(\mu_1\)? A Schr\"odinger bridge asks: what is the least surprising stochastic evolution, relative to a baseline model, that starts from \(\mu_0\) and ends at \(\mu_1\)?
\end{actuarialbox}

\subsubsection{Static entropic transport as a two-time bridge}
\label{subsubsec:cf-static-entropic-bridge}

The link with entropic optimal transport appears when one looks only at the joint law of \((X_0,X_1)\). Let
\[
  R_{01} = R\circ (X_0,X_1)^{-1}
\]
be the two-time marginal of the reference dynamics. Then any path law \(P\) has a two-time marginal
\[
  \gamma = P\circ (X_0,X_1)^{-1}.
\]
The static two-time projection of the Schr\"odinger bridge solves the entropy projection problem
\[
  \inf_{\gamma\in\Coup(\mu_0,\mu_1)}
  \KL(\gamma\,\|\,R_{01}).
\]
If the reference two-time law has the Gibbs form
\[
  R_{01}(\dd \bx,\dd \by)
  \propto
  \exp\left(-\frac{c(\bx,\by)}{\varepsilon}\right)\,m(\dd \bx,\dd \by),
\]
for a base measure \(m\), then this is equivalent, up to constants depending on the endpoints and the base measure, to an entropic optimal transport problem of the form
\[
  \inf_{\gamma\in\Coup(\mu_0,\mu_1)}
  \left\{
    \int c(\bx,\by)\,\gamma(\dd\bx,\dd\by)
    +
    \varepsilon \KL(\gamma\,\|\,m)
  \right\}.
\]
A common discrete choice is \(m=\mu_0\otimes\mu_1\), but the reference law should be stated explicitly.

For Brownian reference dynamics on \(\R^d\), the transition density is Gaussian:
\[
  r_\varepsilon(\bx,\by)
  =
  (2\pi\varepsilon)^{-d/2}
  \exp\left(-\frac{\|\bx-\by\|^2}{2\varepsilon}\right).
\]
The negative log-density is proportional to the quadratic transport cost. Thus the Schr\"odinger bridge can be viewed as a stochastic, entropy-regularized version of quadratic optimal transport.

\begin{remark}[Zero-noise limit]
As the noise parameter \(\varepsilon\) tends to zero, the Schr\"odinger bridge concentrates around optimal transport trajectories. In this sense, optimal transport is the small-noise limit of Schr\"odinger bridge problems. Conversely, for positive \(\varepsilon\), the bridge keeps stochastic dispersion around the deterministic OT path.
\end{remark}

\subsubsection{The dynamic formulation}
\label{subsubsec:cf-sb-dynamic-formulation}

For a Brownian reference process, the Schr\"odinger bridge has a controlled diffusion representation. Consider the reference dynamics
\[
  \dd X_t = \sqrt{\varepsilon}\,\dd B_t,
\]
and allow a progressively measurable drift \(v_t\). The controlled process is
\[
  \dd X_t = v_t(X_t)\,\dd t + \sqrt{\varepsilon}\,\dd B_t.
\]
If \(\rho_t\) denotes the density of \(X_t\), the pair \((\rho_t,v_t)\) must satisfy the Fokker--Planck equation
\[
  \partial_t\rho_t
  +
  \nabla\cdot(\rho_t v_t)
  =
  \frac{\varepsilon}{2}\Delta\rho_t,
\]
with boundary conditions
\[
  \rho_0=\mu_0,
  \qquad
  \rho_1=\mu_1.
\]
With the convention above, Girsanov's formula gives the following control cost. The dynamic Schr\"odinger problem can be written formally as
\[
  \inf_{\rho,v}
  \left\{
    \frac{1}{2\varepsilon}
    \int_0^1\int_{\R^d}
    \|v_t(\bx)\|^2 \rho_t(\bx)\,\dd \bx\,\dd t
  \right\}
\]
subject to the Fokker--Planck equation and the endpoint constraints.

This should be compared with the Benamou--Brenier formulation of quadratic optimal transport:
\[
  W_2^2(\mu_0,\mu_1)
  =
  \inf_{\rho,v}
  \left\{
    \int_0^1\int_{\R^d}
    \|v_t(\bx)\|^2\rho_t(\bx)\,\dd \bx\,\dd t
  \right\},
\]
subject to the deterministic continuity equation
\[
  \partial_t\rho_t+\nabla\cdot(\rho_t v_t)=0.
\]
The only difference is the diffusion term
\[
  \frac{\varepsilon}{2}\Delta\rho_t.
\]
Thus Schr\"odinger bridges are to stochastic paths what Benamou--Brenier transport is to deterministic displacement paths.

\begin{actuarialbox}
Benamou--Brenier gives a least-action deterministic trajectory of distributions. The Schr\"odinger bridge gives a least-action stochastic trajectory, allowing random fluctuations around the transported path. This distinction is important when actuarial trajectories are not deterministic: claim processes, climate losses, lapses, portfolio renewal and cyber events all include genuine noise.
\end{actuarialbox}

\subsubsection{Counterfactual interpretation}
\label{subsubsec:cf-sb-counterfactual-interpretation}

Suppose that \(\mu_0\) is the observed baseline distribution and \(\mu_1\) is a counterfactual target distribution. A static transport map answers:
\[
  \text{if the observed state was }\bx,\text{ what target state }T(\bx)
  \text{ should correspond to it?}
\]
A Schr\"odinger bridge answers a different question:
\[
  \text{what stochastic evolution from baseline to target is closest to the
  reference dynamics?}
\]

This is useful when the counterfactual is not a one-shot change but a process. Examples include:

\begin{itemize}
\item the counterfactual path of annual losses under a prevention policy;
\item the path of portfolio composition under a different underwriting strategy;
\item the evolution of claim severity under a climate adaptation scenario;
\item the trajectory of scores under a phased model adjustment;
\item the cyber-loss process under a different security investment trajectory.
\end{itemize}

The bridge therefore provides a stochastic counterfactual path, not merely a counterfactual endpoint.

\begin{definition}[Stochastic counterfactual path]
\label{def:cf-stochastic-counterfactual-path}
Let \(R\) be a baseline path law and let \(\mu_0,\mu_1\) be prescribed endpoint distributions. A stochastic counterfactual path is a path law \(P^\star\) solving
\[
  P^\star
  \in
  \argmin_P
  \left\{
    \KL(P\,\|\,R)
    :
    P_0=\mu_0,\;P_1=\mu_1
  \right\}.
\]
The intermediate laws
\[
  \mu_t^\star = P^\star\circ X_t^{-1},
  \qquad 0<t<1,
\]
are the counterfactual trajectory of distributions.
\end{definition}

\subsubsection{Why entropy is the right penalty}
\label{subsubsec:cf-sb-entropy}

Relative entropy measures how much the counterfactual dynamics departs from the baseline dynamics. If \(P\ll R\), then
\[
  \KL(P\,\|\,R)
  =
  \E_P\left[
    \log\frac{\dd P}{\dd R}
  \right].
\]
A small value means that the counterfactual path law is difficult to distinguish from the reference path law. A large value means that the counterfactual dynamics requires a strong modification of the baseline model.

This gives a useful modelling principle:

\[
  \text{counterfactual path}
  =
  \text{baseline dynamics}
  +
  \text{minimal information update to match endpoints}.
\]

In actuarial language, the endpoint constraints encode the counterfactual facts we want to impose. The entropy objective keeps everything else as close as possible to the reference model.

\subsubsection{Example: prevention as a stochastic bridge}
\label{subsubsec:cf-sb-prevention}

Let \(L_t\) denote the annual loss distribution over a time interval \(t\in[0,1]\). Suppose \(\mu_0\) is the current loss distribution and \(\mu_1^{\mathrm{prev}}\) is a target loss distribution after a prevention policy. A deterministic OT map gives a direct transformation
\[
  L_1^{\mathrm{prev}} = T(L_0).
\]
A Schr\"odinger bridge instead constructs a path law
\[
  P^\star
  =
  \argmin_P
  \left\{
    \KL(P\,\|\,R)
    :
    P_0=\mu_0,\;
    P_1=\mu_1^{\mathrm{prev}}
  \right\},
\]
where \(R\) is a baseline stochastic loss evolution.

The intermediate laws
\[
  \mu_t^\star = \Law_{P^\star}(L_t)
\]
describe how the loss distribution changes during the prevention period. This matters when benefits are not instantaneous. For instance, flood adaptation may gradually reduce moderate losses before affecting tail losses; fire prevention may reduce frequency before reducing severity; building-code changes may affect new exposures before old exposures.

\begin{actuarialbox}
A prevention policy is often evaluated by comparing a baseline loss law and a post-prevention loss law. Schr\"odinger bridges add a dynamic layer: they describe a plausible stochastic path between the two, constrained by the baseline dynamics.
\end{actuarialbox}

\iffalse
% To be restored in v2: dynamic fairness repair.
\subsubsection{Example: dynamic fairness repair}
\label{subsubsec:cf-sb-dynamic-fairness}

Let \(Z_t\) denote a score or premium at renewal cycle \(t\), and let \(S\in\{0,1\}\) be a protected group. A static Wasserstein fairness repair maps group-specific distributions
\[
  \mu_s = \Law(Z\mid S=s)
\]
to a common target distribution \(\bar\mu\).

In practice, however, an insurer may not want to impose the full correction at once. Abrupt premium changes may be commercially, legally or politically difficult. A stochastic bridge gives a gradual correction path. For each group \(s\), one may define
\[
  P_s^\star
  =
  \argmin_P
  \left\{
    \KL(P\,\|\,R_s)
    :
    P_0=\mu_s,\;
    P_1=\bar\mu
  \right\},
\]
where \(R_s\) is a reference dynamics for renewal-to-renewal premium changes in group \(s\). The intermediate laws
\[
  \mu_{s,t}^\star = P_s^\star\circ Z_t^{-1}
\]
describe a phased fairness repair.

This differs from the displacement interpolation
\[
  \mu_{s,t}=((1-t)\Id+tT_s)_\#\mu_s.
\]
The displacement interpolation gives a deterministic path. The Schr\"odinger bridge gives a stochastic path that remains close to historical premium dynamics.

\fi

\subsubsection{Example: cyber counterfactuals}
\label{subsubsec:cf-sb-cyber}

Cyber risk is dynamic and path-dependent. The loss distribution after a breach depends on detection time, containment, propagation, backup restoration, liability notification and regulatory response. A static counterfactual such as ``what if the firm had invested in better security?'' hides this temporal structure.

Let \(R\) be a baseline incident-response dynamics calibrated from scenarios or expert judgement. Let \(\mu_0\) be the distribution of initial cyber states, and let \(\mu_1^{\mathrm{secure}}\) be a target distribution of terminal losses under a stronger security posture. The Schr\"odinger bridge
\[
  \argmin_P
  \left\{
    \KL(P\,\|\,R):
    P_0=\mu_0,\;
    P_1=\mu_1^{\mathrm{secure}}
  \right\}
\]
gives the least informative modification of the incident path that achieves the target terminal law.

This can be read as a stochastic counterfactual scenario tool: rather than claiming that a specific breach would have produced a specific alternative loss, the bridge produces a distribution over alternative incident paths.

\subsubsection{Discrete-time formulation}
\label{subsubsec:cf-sb-discrete-time}

For actuarial applications, a discrete-time version is often more practical. Let
\[
  X_0,X_1,\ldots,X_T
\]
be a Markov chain with reference transition kernels
\[
  R_t(\bx_{t-1},\dd \bx_t).
\]
The reference path law is
\[
  R(\dd \bx_{0:T})
  =
  \mu_0^R(\dd \bx_0)
  \prod_{t=1}^T R_t(\bx_{t-1},\dd \bx_t).
\]
Given endpoint laws \(\mu_0\) and \(\mu_T\), the discrete Schr\"odinger bridge solves
\[
  \inf_P
  \left\{
    \KL(P\,\|\,R):
    P_0=\mu_0,\;P_T=\mu_T
  \right\}.
\]

Under standard positivity assumptions, the solution has the form
\[
  \dd P^\star(\bx_{0:T})
  =
  f_0(\bx_0)g_T(\bx_T)\,\dd R(\bx_{0:T}),
\]
where \(f_0\) and \(g_T\) are nonnegative scaling functions chosen so that the endpoint constraints hold. This is the path-space analogue of Sinkhorn scaling.

\begin{computebox}
In a finite state space, the Schr\"odinger bridge can be computed by iterative scaling. One alternately enforces the initial marginal and the terminal marginal, exactly as Sinkhorn alternately enforces row and column sums in entropic optimal transport.
\end{computebox}

\subsubsection{Finite-state actuarial example}
\label{subsubsec:cf-sb-finite-state-example}

Suppose a portfolio has three risk states:
\[
  \mathcal X=\{\text{low},\text{medium},\text{high}\}.
\]
The baseline transition matrix over one year is
\[
  K =
  \begin{pmatrix}
  0.85 & 0.12 & 0.03\\
  0.10 & 0.78 & 0.12\\
  0.04 & 0.16 & 0.80
  \end{pmatrix}.
\]
Let \(\mu_0\) be the current portfolio composition and let \(\mu_1^{\mathrm{cf}}\) be a counterfactual target composition after a prevention or underwriting policy. A static coupling describes how much mass moves from each current risk state to each terminal risk state. A Schr\"odinger bridge instead modifies the transition dynamics as little as possible, in relative entropy, while matching the target composition.

For one period, this becomes the entropic coupling problem
\[
  \min_{\gamma\in\Coup(\mu_0,\mu_1^{\mathrm{cf}})}
  \KL(\gamma\,\|\,\mu_0\otimes K),
\]
where
\[
  (\mu_0\otimes K)(i,j)=\mu_0(i)K_{ij}.
\]
The solution has the scaling form
\[
  \gamma_{ij}^\star
  =
  a_i\,\mu_0(i)K_{ij}\,b_j,
\]
with positive scaling vectors \(\ba\) and \(\bb\) chosen to match the two marginals.

\begin{proposition}[One-period Schr\"odinger bridge as KL projection]
\label{prop:cf-one-period-bridge-kl}
Let \(Q\) be a strictly positive prior joint law on a finite space \(\mathcal X\times\mathcal X\). Let \(\mu_0\) and \(\mu_1\) be endpoint laws with strictly positive masses. Then the problem
\[
  \min_{\gamma\in\Coup(\mu_0,\mu_1)}\KL(\gamma\,\|\,Q)
\]
has a unique solution. It has the multiplicative scaling form
\[
  \gamma_{ij}^\star=a_i Q_{ij} b_j,
\]
where the positive scaling vectors \(\ba=(a_i)\) and \(\bb=(b_j)\) are chosen so that \(\gamma^\star\) has row sums \(\mu_0\) and column sums \(\mu_1\).
\end{proposition}

\begin{proof}
The feasible set is a compact convex polytope and the relative entropy is strictly convex on the positive orthant. Since \(Q_{ij}>0\) and the marginals are strictly positive, the optimum is unique. The Lagrangian for the marginal constraints is
\[
  \sum_{i,j}\gamma_{ij}\log\frac{\gamma_{ij}}{Q_{ij}}+\sum_i\alpha_i\left(\mu_0(i)-\sum_j\gamma_{ij}\right)+\sum_j\beta_j\left(\mu_1(j)-\sum_i\gamma_{ij}\right).
\]
The first-order condition gives \(\log(\gamma_{ij}/Q_{ij})=\alpha_i+\beta_j-1\). Setting \(a_i=\exp(\alpha_i-1/2)\) and \(b_j=\exp(\beta_j-1/2)\) gives the scaling form. The scaling factors are determined by the row and column constraints, up to the usual common multiplicative normalization.
\end{proof}

\subsubsection{A minimal R lab}
\label{subsubsec:cf-sb-r-lab}

The following code computes the one-period finite-state bridge. It is just Sinkhorn scaling applied to a prior joint law.

\begin{lstlisting}[language=R]
# One-period finite-state Schrodinger bridge

mu0 <- c(low = 0.50, medium = 0.35, high = 0.15)

K <- matrix(c(
  0.85, 0.12, 0.03,
  0.10, 0.78, 0.12,
  0.04, 0.16, 0.80
), nrow = 3, byrow = TRUE)

colnames(K) <- rownames(K) <- names(mu0)

# Counterfactual terminal distribution:
# fewer high-risk policies after prevention
mu1 <- c(low = 0.58, medium = 0.34, high = 0.08)

# Prior joint law under baseline dynamics
Q <- diag(mu0) %*% K

sinkhorn_bridge <- function(Q, mu0, mu1, n_iter = 1000, tol = 1e-12) {
  A <- Q
  for (it in 1:n_iter) {
    A <- A * (mu0 / rowSums(A))
    A <- t(t(A) * (mu1 / colSums(A)))
    err <- max(abs(rowSums(A) - mu0), abs(colSums(A) - mu1))
    if (err < tol) break
  }
  A
}

Gamma <- sinkhorn_bridge(Q, mu0, mu1)

Gamma
rowSums(Gamma)
colSums(Gamma)

# Baseline and counterfactual transition matrices
K_baseline <- K
K_cf <- Gamma / mu0

K_baseline
K_cf
\end{lstlisting}

The matrix \(K_{\mathrm{cf}}\) is the counterfactual transition matrix closest to the baseline transition matrix \(K\), subject to the target terminal composition. It can be interpreted as the least disruptive transition adjustment achieving the counterfactual scenario.

\subsubsection{Interpretation for actuarial governance}
\label{subsubsec:cf-sb-governance}

Schr\"odinger bridges are useful because they separate three ingredients:

\begin{enumerate}
\item the baseline dynamics \(R\);
\item the counterfactual endpoint constraints \(\mu_0,\mu_1\);
\item the entropy-minimizing adjustment from \(R\) to \(P^\star\).
\end{enumerate}

This separation is valuable in governance. The baseline dynamics can be documented as an internal model or scenario generator. The endpoint constraints can be justified by a prevention target, regulatory constraint, stress scenario or policy intervention. The entropy objective provides a transparent rule for selecting the least informative path law compatible with these constraints.

\begin{actuarialbox}
A Schr\"odinger bridge is not a causal model by itself. It is a principled selection rule for stochastic counterfactual paths. Causal interpretation still requires assumptions about the intervention, the structural mechanism and the meaning of the endpoint constraints.
\end{actuarialbox}

\subsubsection{Limitations}
\label{subsubsec:cf-sb-limitations}

Schr\"odinger bridges should be used with care.

\begin{itemize}
\item They depend on the reference dynamics \(R\). A poorly chosen baseline
process leads to a poorly justified counterfactual path.

\item They enforce marginal constraints, but do not identify structural causal
effects without additional assumptions.

\item The entropy criterion favours minimal information change, not necessarily
economic realism or legal admissibility.

\item In high dimension, estimation of path laws can be difficult. Practical
implementations usually require Markov structure, finite-state approximations, Gaussian approximations or entropic numerical methods.

\item Endpoint constraints may be too weak. Many different path-level
phenomena can share the same initial and terminal laws.
\end{itemize}

Thus Schr\"odinger bridges should be presented as stochastic counterfactual scenario generators, not as automatic causal truth machines.

\subsubsection{What to remember}
\label{subsubsec:cf-sb-remember}

\begin{itemize}
\item Optimal transport gives static counterfactual couplings or maps.
\item Benamou--Brenier transport gives deterministic distributional paths.
\item Schr\"odinger bridges give stochastic distributional paths.
\item The bridge is selected by minimizing relative entropy with respect to a
reference dynamics.
\item In finite state spaces, the bridge is computed by Sinkhorn-type scaling.
\item For actuarial science, this is useful for prevention, climate transition,
portfolio drift, phased model adjustment, cyber incident paths and dynamic scenario generation.
\end{itemize}

\subsection{A small dictionary for actuarial use}
\label{subsec:cf-dictionary}

\begin{center}
\begin{tabular}{p{.34\textwidth}p{.52\textwidth}}
\toprule
Counterfactual object & Actuarial interpretation \\
\midrule
$\nu$ & Alternative distribution: target group, future climate, post-prevention losses \\
$\gamma\in\Coup(\mu,\nu)$ & Matching between observed and alternative worlds \\
$T_{\push}\mu=\nu$ & Deterministic transformation of a portfolio or loss distribution \\
$\int c\,\dd\gamma$ & Total plausibility, effort or displacement cost \\
Quantile transport & Rank-preserving counterfactual for scores, premiums or losses \\
Gaussian transport & Linear distributional counterfactual for multivariate normal approximations \\
Recourse point $\bx^\star$ & Individual explanation or recommended action \\
$\Risk(L^\dagger)-\Risk(L)$ & Actuarial value of a counterfactual scenario \\
Sequential transport & Graph-aligned mediator or covariate counterfactual \\
\bottomrule
\end{tabular}
\end{center}

\subsection{What to remember}
\label{subsec:cf-remember}

The main lesson of this lecture is not that optimal transport solves causality. It does not. The lesson is that many actuarial counterfactuals are distributional, and optimal transport provides a rigorous way to construct and audit distributional counterfactual worlds.

\begin{itemize}
  \item Marginals describe the observed and alternative worlds.
  \item Couplings describe how those worlds are matched.
  \item Costs select which matching is preferred.
  \item Maps provide deterministic counterfactual profiles.
  \item Causal interpretation requires additional structural assumptions.
\end{itemize}

\subsection{Exercises}
\label{subsec:cf-exercises}

\begin{exercise}[Two-point counterfactual worlds]\label{ex:14:1}
Let $\mu=\mathrm{Bernoulli}(p)$ and $\nu=\mathrm{Bernoulli}(q)$. Write all couplings as $2\times2$ matrices. Determine the admissible interval for $a=\Pbb(X=1,X^\dagger=1)$. Compute $\E[|X-X^\dagger|]$ as a function of $a$.
\end{exercise}

\begin{exercise}[Non-identification]\label{ex:14:2}
Let $\mu=\nu=\frac12\delta_0+\frac12\delta_1$. Construct two couplings with the same marginals but different values of $\Pbb(X=X^\dagger)$, $\E[|X-X^\dagger|]$ and $\E[(X-X^\dagger)^2]$. Explain the implication for individual counterfactual interpretation.
\end{exercise}

\begin{exercise}[Premium recourse under a linear tariff]\label{ex:14:3}
Let $m(\bx)=\beta_0+\bbeta^\top \bx$ and suppose $m(\bx)>p^\star$. Derive the closest Euclidean counterfactual $\bz^\star$ such that $m(\bz^\star)\le p^\star$. Repeat the calculation when only a subset of coordinates $J\subset\{1,\ldots,d\}$ is actionable.
\end{exercise}

\begin{exercise}[Quantile transport between normal laws]\label{ex:14:4}
Let $\mu=\mathcal N(m_0,\sigma_0^2)$ and $\nu=\mathcal N(m_1,\sigma_1^2)$ on $\R$. Show that the monotone transport map is
\[
  T(x)=m_1+\frac{\sigma_1}{\sigma_0}(x-m_0).
\]
Compute $\Wtwo^2(\mu,\nu)$.
\end{exercise}

\iffalse
% To be restored in v2: fairness counterfactual exercise.
\begin{exercise}[Transport counterfactual fairness]\label{ex:14:5}
Let $X\mid S=0\sim\mathcal N(0,1)$ and $X\mid S=1\sim\mathcal N(1,4)$. Let $m(x,s)=x+\alpha s$. Use the monotone transport from group $0$ to group $1$ to compute
\[
  \E\left[(m(X,0)-m(T(X),1))^2\mid S=0\right]
\]
as a function of $\alpha$.
\end{exercise}

\fi

\begin{exercise}[Counterfactual catastrophe catalogue]\label{ex:14:6}
A historical event has three counterfactual losses $\bC=(10,50,200)$ with probabilities $\bp=(.8,.15,.05)$. Another event has counterfactual losses $(5,30,300)$ with probabilities $(.7,.25,.05)$. Give equal weight to the two events. Compute the catalogue mean, $\VaR_{.95}$ and $\TVaR_{.95}$.
\end{exercise}

\begin{exercise}[Prevention and coupling]\label{ex:14:7}
Let $L^{(0)}\sim\frac12\delta_0+\frac12\delta_{100}$ and $L^{(1)}\sim\frac12\delta_0+\frac12\delta_{80}$. Compare the comonotone coupling and the countermonotone coupling. Show that $\E[L^{(0)}-L^{(1)}]$ is the same under both, but $\Pbb(L^{(1)}<L^{(0)})$ differs.
\end{exercise}

\begin{exercise}[Hard actionability constraints]\label{ex:14:8}
Let $\bx=(x_A,x_N)$, where $x_A$ is actionable and $x_N$ is non-actionable. Define the admissible set $\mathcal A=\{((x_A,x_N),(y_A,y_N)):y_N=x_N\}$. Write the corresponding infinite cost $c_{\mathcal A}$. Explain what happens if the target distribution $\nu$ has a different marginal distribution of $x_N$ than $\mu$.
\end{exercise}

\begin{exercise}[Sequential counterfactuals]\label{ex:14:9}
Suppose $X_1\to X_2$, with
\[
  X_1\mid A=a\sim\mathcal N(m_a,s_a^2),
  \qquad
  X_2\mid X_1=x,A=a\sim\mathcal N(\alpha_a+\beta_a x,\sigma_a^2).
\]
Construct a sequential Gaussian transport from $A=0$ to $A=1$. Compare it with a global Gaussian transport on $(X_1,X_2)$.
\end{exercise}

\newpage
\lecture{Dynamic Optimal Transport and Actuarial Trajectories}
\label{lec:dynamic-ot-actuarial-trajectories}

\subsection*{Learning objectives}

The previous lectures treated optimal transport mostly as a static problem: compare two distributions, compute a distance, find a coupling, or construct a barycenter. Many actuarial questions, however, are dynamic. A portfolio ages, claim inflation evolves over calendar time, a climate scenario modifies loss distributions gradually, a prevention policy changes losses over many years, and a reserving model produces a future cash-flow trajectory rather than a single scalar reserve.

This lecture introduces the dynamic viewpoint on optimal transport. The main idea is to replace a single coupling between two distributions by a path of distributions and a velocity field that moves probability mass through time.

After this lecture, the reader should be able to:
\begin{enumerate}[label=(\roman*)]
  \item distinguish a static transport plan from a dynamic transport trajectory;
  \item describe a time-indexed family of actuarial laws \((\mu_t)_{0\leq t\leq 1}\);
  \item state the continuity equation for moving probability measures;
  \item interpret a velocity field as the infinitesimal movement of risk mass;
  \item state the Benamou--Brenier dynamic formulation of \(\Wtwo\);
  \item compute one-dimensional Wasserstein geodesics through quantile interpolation;
  \item write the Gaussian \(\Wtwo\)-geodesic and interpret mean and covariance movements;
  \item define dynamic drift, action, path length and kinetic energy for actuarial trajectories;
  \item formulate discrete-time dynamic transport problems for yearly portfolios, reserving scenarios and climate trajectories;
  \item connect dynamic OT to climate adaptation, prevention accounting and scenario generation.
\end{enumerate}

\begin{readingbox}
The dynamic formulation of optimal transport goes back to the fluid-dynamical formulation of \cite{benamou2000computational}; see also \cite{villani2003topics,villani2008oldnew,ambrosio2005gradient,ambrosio2021lectures,santambrogio2015optimal,figalli2021invitation}. Computational aspects, including dynamic viewpoints, entropic variants and Sinkhorn-type algorithms, are discussed in \cite{peyre2019computational}. Statistical aspects of Wasserstein geometry are treated in \cite{chewi2025statistical}. Adapted and causal versions of transport for stochastic processes are related to \cite{lassalle2018causal,backhoff2020adapted}. Schr\"odinger bridges and stochastic transport paths are surveyed in \cite{leonard2014survey,chen2021stochastic}. The actuarial interpretation of moving loss, reserve, cash-flow and scenario distributions builds on the distributional viewpoint developed throughout these notes and on the insurance examples in \cite{charpentier2024insurance,wuthrich2023statistical,denuit2005actuarial}.
\end{readingbox}

\begin{actuarialbox}
Static OT compares two actuarial distributions. Dynamic OT describes how one distribution can move into another through time. It is therefore a useful language for portfolio drift, climate transition, prevention effects, reserving cash-flow evolution and stochastic scenario generation.
\end{actuarialbox}

\begin{actuarialbox}%[Distributional, not causal]
A dynamic transport path is a distributional construction. It should not be read as an identified causal trajectory for an individual policyholder, claim or asset unless a causal model or intervention design has been added separately.
\end{actuarialbox}

\subsection{From static comparison to dynamic movement}
\label{sec:dynamic-from-static}

Let \(\mu_0\) and \(\mu_1\) be two probability measures on \(\R^d\). In the static formulation of Lectures~\ref{lec:discrete-ot}--\ref{lec:wasserstein-barycenters}, an optimal coupling
\[
  \gamma^\star\in\Coup(\mu_0,\mu_1)
\]
solves
\[
  \inf_{\gamma\in\Coup(\mu_0,\mu_1)}
  \int_{\R^d\times\R^d}\|\bx-\by\|^2\,\gamma(\dd \bx,\dd \by).
\]
This tells us which source states \(\bx\) are matched to which target states \(\by\). It does not, by itself, describe the intermediate states between \(\bx\) and \(\by\).

The dynamic viewpoint asks for a curve
\[
  (\mu_t)_{0\leq t\leq 1}
\]
with endpoints
\[
  \mu_{t=0}=\mu_0,\qquad \mu_{t=1}=\mu_1,
\]
and for a velocity field \(\bv_t(\bx)\) such that probability mass moves according to this velocity. Instead of paying directly for the jump \(\bx\to \by\), one pays kinetic energy along the path.

This distinction matters in insurance. Suppose \(\mu_0\) is the current loss distribution and \(\mu_1\) is a stressed climate loss distribution. A static Wasserstein distance answers:
\[
  \text{how far are the two loss distributions?}
\]
A dynamic trajectory asks:
\[
  \text{through which intermediate loss distributions could the portfolio move?}
\]
For adaptation, prevention and long-term solvency, the second question is often closer to the actuarial problem.

\begin{example}[Three actuarial trajectories]
\leavevmode
\begin{enumerate}[label=(\alph*)]
  \item \textbf{Portfolio ageing.} Let \(\bX_t\) be the vector of covariates of a representative policyholder at time \(t\). Then \(\mu_t=\Law(\bX_t)\) is a path of portfolio distributions.
  \item \textbf{Loss inflation.} Let \(L_0\sim\mu_0\) and let \(L_t=(1+i_t)L_0\), where \(i_t\) is cumulative inflation. Then \(\mu_t=\Law(L_t)\) is a deterministic push-forward path.
  % Fairness post-processing item to be restored in v2.
\end{enumerate}
\end{example}

\subsection{Curves of probability measures}
\label{sec:dynamic-curves-of-measures}

We work on \(\R^d\), equipped with the Euclidean norm. A curve of probability measures is a map
\[
  t\in[0,1]\longmapsto \mu_t\in\Prob(\R^d).
\]
For dynamic \(\Wtwo\)-geometry, one usually assumes \(\mu_t\in\Probp{2}(\R^d)\), so that second moments are finite.

\begin{definition}[Actuarial distributional trajectory]
\label{def:actuarial-distributional-trajectory}
An actuarial distributional trajectory is a time-indexed family
\[
  t\longmapsto \mu_t
\]
where each \(\mu_t\) is the law of the same type of actuarial object at time \(t\): losses, covariates, reserves, cash-flow vectors, mortality surfaces or exposure-weighted measures. The time scale must be specified: calendar time, development time, underwriting year, climate horizon, valuation date or intervention strength.
\end{definition}

\begin{definition}[Wasserstein absolutely continuous curve]
\label{def:wasserstein-ac-curve}
A curve \((\mu_t)_{0\leq t\leq 1}\subset\Probp{2}(\R^d)\) is absolutely continuous for \(\Wtwo\) if there exists an integrable function \(m\in L^1(0,1)\) such that, for all \(0\leq s<t\leq 1\),
\[
  \Wtwo(\mu_s,\mu_t)\leq \int_s^t m(r)\,\dd r.
\]
\end{definition}

The smallest admissible instantaneous bound, when it exists, is the metric derivative
\[
  |\dot\mu_t| = \lim_{h\to 0}\frac{\Wtwo(\mu_{t+h},\mu_t)}{|h|}.
\]
This is the speed of the distributional trajectory.

\begin{example}[Deterministic push-forward path]
Let \(\bX_0\sim\mu_0\), and define
\[
  \bX_t = T_t(\bX_0),
\]
where \((T_t)_{0\leq t\leq 1}\) is a family of measurable maps. Then
\[
  \mu_t=(T_t)_\#\mu_0.
\]
If \(T_t(\bx)=(1-t)\bx+tT(\bx)\), then
\[
  \mu_t=((1-t)\Id+tT)_\#\mu_0.
\]
This is the displacement interpolation associated with the map \(T\), formalized in Definition~\ref{def:wasserstein-geodesic} and Proposition~\ref{prop:displacement-interpolation-geodesic}.
\end{example}

\begin{example}[Calendar-year severity inflation]
Let \(L_0\sim\mu_0\), and let cumulative inflation be represented by a differentiable function \(a_t>0\), with \(a_0=1\). Set
\[
  L_t=a_tL_0,
  \qquad \mu_t=(x\mapsto a_tx)_\#\mu_0.
\]
Then the trajectory is deterministic. A loss amount \(x_t\) at time \(t\) evolves according to
\[
  \dot x_t = \frac{\dot a_t}{a_t}x_t.
\]
The velocity field is therefore
\[
  v_t(x)=\frac{\dot a_t}{a_t}x.
\]
This is a simple actuarial velocity: large claims move faster in absolute monetary units under proportional inflation.
\end{example}

\subsection{The continuity equation}
\label{sec:dynamic-continuity-equation}

A velocity field \(\bv_t(\bx)\in\R^d\) describes how mass at location \(\bx\) moves at time \(t\). The evolution of the probability law \(\mu_t\) is governed by the continuity equation
\[
  \partial_t\mu_t + \nabla\cdot(\bv_t\mu_t)=0.
\]
This equation is understood weakly, because \(\mu_t\) may not have a density.

\begin{definition}[Continuity equation]
\label{def:continuity-equation}
Let \((\mu_t)_{0\leq t\leq 1}\subset\Prob(\R^d)\) and let \((\bv_t)_{0\leq t\leq 1}\) be a Borel vector field such that
\[
  \int_0^1\int_{\R^d}\|\bv_t(\bx)\|\,\mu_t(\dd \bx)\,\dd t<\infty.
\]
We say that \((\mu_t,\bv_t)\) solves the continuity equation if, for every smooth compactly supported test function \(\varphi\in C_c^\infty((0,1)\times\R^d)\),
\[
  \int_0^1\int_{\R^d}
  \left\{\partial_t\varphi(t,\bx)+\nabla_{\bx}\varphi(t,\bx)\cdot \bv_t(\bx)\right\}
  \mu_t(\dd \bx)\,\dd t=0.
\]
\end{definition}

If \(\mu_t\) has density \(\rho_t(\bx)\), the equation becomes the usual PDE
\[
  \partial_t\rho_t(\bx)+\nabla\cdot\{\rho_t(\bx)\bv_t(\bx)\}=0.
\]
It says that probability mass is conserved as it flows.

\begin{proposition}[Push-forward flows solve the continuity equation]
\label{prop:push-forward-flows-continuity}
Let \(\bX_t\) solve the ordinary differential equation
\[
  \dot\bX_t=\bv_t(\bX_t),\qquad \bX_0\sim\mu_0,
\]
where the flow is sufficiently regular. Let \(\mu_t=\Law(\bX_t)\). Then \((\mu_t,\bv_t)\) solves the continuity equation.
\end{proposition}

\begin{proof}
This is the standard weak form of mass conservation for a flow; see the dynamic formulation in \cite{villani2008oldnew,ambrosio2021lectures,figalli2021invitation}. Sketch of the proof: for a smooth test function \(\phi\), the chain rule gives
\[
  \frac{\dd}{\dd t}\phi(\bX_t)=\nabla\phi(\bX_t)\cdot \bv_t(\bX_t).
\]
Taking expectations yields
\[
  \frac{\dd}{\dd t}\int \phi(\bx)\,\mu_t(\dd \bx)
  =\int \nabla\phi(x)\cdot \bv_t(\bx)\,\mu_t(\dd \bx).
\]
This is the weak form of
\(\partial_t\mu_t+\nabla\cdot(\bv_t\mu_t)=0\). A time-dependent test function gives the displayed distributional formulation.
\end{proof}

\begin{actuarialbox}
The continuity equation says that risk mass is not created or destroyed along the trajectory. This is appropriate for balanced transport: the portfolio distribution changes shape, but total probability remains one. When exposure enters or leaves the portfolio, one needs unbalanced transport or a source term.
\end{actuarialbox}

\subsection{The dynamic action}
\label{sec:dynamic-action}

Given a curve \((\mu_t,\bv_t)\), define its kinetic action by
\[
  \mathcal A(\mu,\bv)=\int_0^1\int_{\R^d}\|\bv_t(\bx)\|^2\,\mu_t(\dd \bx)\,\dd t.
\]
This is the expected squared speed accumulated over time. It is the dynamic analogue of the static quadratic transport cost.

\begin{definition}[Admissible dynamic transport]
\label{def:admissible-dynamic-transport}
A pair \((\mu_t,\bv_t)_{0\leq t\leq 1}\) is dynamically admissible between \(\mu_0\) and \(\mu_1\) if:
\begin{enumerate}[label=(\roman*)]
  \item \(\mu_t\in\Probp{2}(\R^d)\) for each \(t\);
  \item \((\mu_t,\bv_t)\) solves the continuity equation;
  \item the endpoint constraints hold: \(\mu_{t=0}=\mu_0\) and \(\mu_{t=1}=\mu_1\);
  \item \(\mathcal A(\mu,\bv)<\infty\).
\end{enumerate}
\end{definition}

The central theorem is that, for quadratic cost, the minimum dynamic action equals the static squared Wasserstein distance.

\begin{theorem}[Benamou--Brenier formula]
\label{thm:benamou-brenier}
Let \(\mu_0,\mu_1\in\Probp{2}(\R^d)\). Then
\[
  \Wtwo^2(\mu_0,\mu_1)
  =
  \inf_{(\mu_t,\bv_t)}
  \int_0^1\int_{\R^d}\|\bv_t(\bx)\|^2\,\mu_t(\dd \bx)\,\dd t,
\]
where the infimum is over all dynamically admissible pairs \((\mu_t,\bv_t)\) joining \(\mu_0\) to \(\mu_1\).
\end{theorem}

\begin{proof}
This is the classical dynamic formulation of quadratic optimal transport; see \cite{villani2008oldnew,ambrosio2021lectures,figalli2021invitation,peyre2019computational}. Sketch of the proof: first assume an optimal Monge map \(T\) exists. Define
\[
  \bX_t=(1-t)\bX_0+tT(\bX_0),\qquad \bX_0\sim\mu_0.
\]
Then \(\mu_t=\Law(\bX_t)\) and the particle velocity is constant along each trajectory:
\[
  \dot\bX_t=T(\bX_0)-\bX_0.
\]
The action is
\[
  \int_0^1\E\|T(\bX_0)-\bX_0\|^2\,\dd t
  =\E\|T(\bX_0)-\bX_0\|^2
  =\Wtwo^2(\mu_0,\mu_1).
\]
This proves that the dynamic infimum is at most \(\Wtwo^2\). Conversely, for any admissible curve, a particle representation or superposition principle gives trajectories whose expected squared length is bounded by the action. By Cauchy--Schwarz, the squared endpoint displacement is bounded by the time-integrated squared velocity. Taking endpoint laws yields a coupling of \(\mu_0\) and \(\mu_1\) with cost at most the action. Hence \(\Wtwo^2\) is at most any admissible action. The two inequalities give equality. The general case follows by approximation and the superposition principle.
\end{proof}

\begin{remark}[Why quadratic cost?]
The kinetic-energy form is naturally quadratic. There are dynamic formulations for other costs, but the cleanest Riemannian-type geometry appears for \(\Wtwo\). This is why \(\Wtwo\) is central in Wasserstein geodesics, barycenters and gradient flows.
\end{remark}

\begin{actuarialbox}
The Benamou--Brenier formula gives a physical interpretation of \(\Wtwo^2\): it is the least kinetic energy required to move a portfolio, loss or cash-flow distribution from its initial law to its final law.
\end{actuarialbox}

\subsection{Displacement interpolation as the least-action path}
\label{sec:dynamic-displacement-interpolation}

Assume that the optimal transport from \(\mu_0\) to \(\mu_1\) is induced by a map \(T\). The least-action path is
\[
  \mu_t=((1-t)\Id+tT)_\#\mu_0.
\]
This is the \(\Wtwo\)-geodesic between \(\mu_0\) and \(\mu_1\).

\begin{definition}[Wasserstein geodesic]
\label{def:wasserstein-geodesic}
A curve \((\mu_t)_{0\leq t\leq 1}\) is a constant-speed \(\Wtwo\)-geodesic from \(\mu_0\) to \(\mu_1\) if
\[
  \Wtwo(\mu_s,\mu_t)=|t-s|\,\Wtwo(\mu_0,\mu_1)
\]
for all \(s,t\in[0,1]\).
\end{definition}

\begin{proposition}[Displacement interpolation is a geodesic]
\label{prop:displacement-interpolation-geodesic}
Suppose \(T\) is an optimal \(\Wtwo\)-transport map from \(\mu_0\) to \(\mu_1\). Define
\[
  \mu_t=((1-t)\Id+tT)_\#\mu_0.
\]
Then \((\mu_t)_{0\leq t\leq 1}\) is a constant-speed \(\Wtwo\)-geodesic.
\end{proposition}

\begin{proof}
This is the standard displacement interpolation result in quadratic optimal transport; see \cite{villani2003topics,villani2008oldnew,figalli2021invitation}. Sketch of the proof: let
\[
  T_t(\bx)=(1-t)\bx+tT(\bx).
\]
For \(0\leq s<t\leq 1\), the coupling induced by \((T_s(\bX),T_t(\bX))\), \(\bX\sim\mu_0\), gives
\[
  \Wtwo^2(\mu_s,\mu_t)
  \leq \E\|T_t(\bX)-T_s(\bX)\|^2
  =(t-s)^2\E\|T(\bX)-\bX\|^2.
\]
Since \(T\) is optimal, the last expectation is \(\Wtwo^2(\mu_0,\mu_1)\). The reverse inequality follows from the triangle inequality:
\[
  \Wtwo(\mu_0,\mu_1)
  \leq \Wtwo(\mu_0,\mu_s)+\Wtwo(\mu_s,\mu_t)+\Wtwo(\mu_t,\mu_1),
\]
and the analogous upper bounds on the other segments. Equality must hold, so
\[
  \Wtwo(\mu_s,\mu_t)=|t-s|\Wtwo(\mu_0,\mu_1).
\]
\end{proof}

\iffalse
% To be restored in v2 with the premium distribution and demutualization material.
\begin{example}[Linear premium migration]
Let \(P_0\) be the old premium and let \(P_1=T(P_0)\) be a new premium obtained by an optimal monotone map. The path
\[
  P_t=(1-t)P_0+tP_1
\]
corresponds to phasing in the tariff change. If \(t=0.25\), only one quarter of the distributional move has been implemented. This is a displacement path, not a mixture of old and new portfolios.
\end{example}
\fi

\subsection{One-dimensional dynamic transport}
\label{sec:dynamic-one-dimensional-transport}

Dimension one is again explicit. Let \(Q_0=F^{-1}_{\mu_0}\) and \(Q_1=F^{-1}_{\mu_1}\) be the quantile functions. The \(\Wtwo\)-geodesic has quantile function
\[
  Q_t(u)=(1-t)Q_0(u)+tQ_1(u),\qquad 0<u<1.
\]
Thus each rank \(u\) moves linearly from its initial quantile to its final quantile. Throughout this section, quantiles use the lower generalized inverse convention; with atoms or ties, the path should be read through quantile functions or randomized ranks rather than through a pointwise unique transport map.

\begin{proposition}[One-dimensional geodesic]
\label{prop:one-dimensional-geodesic}
Let \(\mu_0,\mu_1\in\Probp{2}(\R)\). Define \(\mu_t\) by the quantile function
\[
  F^{-1}_{\mu_t}(u)=(1-t)F^{-1}_{\mu_0}(u)+tF^{-1}_{\mu_1}(u).
\]
Then \((\mu_t)_{0\leq t\leq 1}\) is a constant-speed \(\Wtwo\)-geodesic.
\end{proposition}

\begin{proof}
This follows from the one-dimensional quantile embedding of \((\Probp{2}(\R),\Wtwo)\) into \(L^2(0,1)\); see \cite{villani2003topics,figalli2021invitation,chewi2025statistical}. Sketch of the proof: in dimension one,
\[
  \Wtwo^2(\mu_s,\mu_t)=\int_0^1 |Q_s(u)-Q_t(u)|^2\,\dd u.
\]
Since \(Q_t=(1-t)Q_0+tQ_1\),
\[
  Q_t-Q_s=(t-s)(Q_1-Q_0),
\]
so
\[
  \Wtwo^2(\mu_s,\mu_t)=(t-s)^2\int_0^1 |Q_1(u)-Q_0(u)|^2\,\dd u
  =(t-s)^2\Wtwo^2(\mu_0,\mu_1).
\]
\end{proof}

\begin{example}[Climate severity trajectory]
Suppose \(\mu_0\) is the baseline annual loss distribution and \(\mu_1\) is a 2050 climate-stressed distribution. The quantile path
\[
  Q_t(u)=(1-t)Q_0(u)+tQ_1(u)
\]
creates an interpretable family of intermediate scenarios. The 99.5\% capital quantile along the path is simply
\[
  \VaR_{0.995}(\mu_t)=(1-t)\VaR_{0.995}(\mu_0)+t\VaR_{0.995}(\mu_1),
\]
provided the same quantile convention is used. This is a rank-preserving stress trajectory.
\end{example}

\begin{proposition}[Derivative of tail averages along a one-dimensional geodesic]
\label{prop:tvar-derivative-geodesic}
Let \(Q_t=(1-t)Q_0+tQ_1\), and assume the relevant integrals are finite. Then for \(0<\alpha<1\),
\[
  \frac{\dd}{\dd t}\TVaR_\alpha(\mu_t)
  =\frac{1}{1-\alpha}\int_\alpha^1 \{Q_1(u)-Q_0(u)\}\,\dd u.
\]
\end{proposition}

\begin{proof}
Using the integrated-quantile representation,
\[
  \TVaR_\alpha(\mu_t)=\frac{1}{1-\alpha}\int_\alpha^1 Q_t(u)\,\dd u.
\]
Since \(Q_t\) is affine in \(t\), differentiation under the integral sign gives the result.
\end{proof}

\begin{actuarialbox}
Along a one-dimensional Wasserstein geodesic, VaR and TVaR evolve linearly in quantile coordinates. This is often the simplest way to build interpretable capital trajectories between baseline and stressed distributions.
\end{actuarialbox}

\subsection{The Gaussian trajectory}
\label{sec:dynamic-gaussian-trajectory}

Let
\[
  \mu_0=\mathcal N(\bm_0,\Sigma_0),\qquad \mu_1=\mathcal N(\bm_1,\Sigma_1),
\]
where \(\Sigma_0\) and \(\Sigma_1\) are positive definite. The optimal \(\Wtwo\)-map from \(\mu_0\) to \(\mu_1\) is affine:
\[
  T(\bx)=\bm_1+A(\bx-\bm_0),
\]
with
\[
  A=\Sigma_0^{-1/2}\left(\Sigma_0^{1/2}\Sigma_1\Sigma_0^{1/2}\right)^{1/2}\Sigma_0^{-1/2}.
\]
The displacement interpolation is
\[
  \bX_t=(1-t)\bX_0+tT(\bX_0),\qquad \bX_0\sim \mathcal N(\bm_0,\Sigma_0).
\]
Therefore \(\mu_t=\mathcal N(\bm_t,\Sigma_t)\), where
\[
  \bm_t=(1-t)\bm_0+t\bm_1,
\]
and
\[
  \Sigma_t=((1-t)I+tA)\Sigma_0((1-t)I+tA).
\]

\begin{proposition}[Gaussian \(\Wtwo\)-geodesic]
\label{prop:gaussian-w2-geodesic}
The curve \(\mu_t=\mathcal N(\bm_t,\Sigma_t)\) defined above is the \(\Wtwo\)-geodesic between \(\mathcal N(\bm_0,\Sigma_0)\) and \(\mathcal N(\bm_1,\Sigma_1)\).
\end{proposition}

\begin{proof}
This is the standard Gaussian formula for quadratic optimal transport; see \cite{villani2008oldnew,chewi2025statistical}. Sketch of the proof: the optimal map between two non-degenerate Gaussian laws is affine and given by \(T(\bx)=\bm_1+A(\bx-\bm_0)\). Applying the interpolation map \((1-t)\Id+tT\) to a Gaussian random vector preserves Gaussianity. Its mean and covariance are exactly the displayed \(\bm_t\) and \(\Sigma_t\).
\end{proof}

\begin{example}[Dynamic capital model comparison]
Suppose a multivariate capital model represents line-of-business losses by a Gaussian vector. A stress scenario changes both the mean vector and the covariance matrix. The Wasserstein geodesic decomposes the movement into:
\begin{enumerate}[label=(\roman*)]
  \item a linear movement of expected line losses;
  \item a nonlinear movement of the covariance matrix;
  \item an induced change in diversification along the path.
\end{enumerate}
This can be more informative than jumping directly from baseline SCR to stressed SCR.
\end{example}

\subsection{Actuarial speed, path length and action}
\label{sec:dynamic-speed-length-action}

A dynamic path \((\mu_t)\) can be summarized by several quantities.

\begin{definition}[Distributional speed]
\label{def:distributional-speed}
If the metric derivative exists, the instantaneous \(\Wtwo\)-speed is
\[
  s(t)=|\dot\mu_t|.
\]
For a smooth velocity representation, the speed satisfies
\[
  s(t)^2\leq \int\|\bv_t(\bx)\|^2\,\mu_t(\dd \bx),
\]
with equality for the minimal velocity field.
\end{definition}

\begin{definition}[Path length and action]
\label{def:path-length-action}
The \(\Wtwo\)-length of a curve is
\[
  \mathcal L(\mu)=\int_0^1 |\dot\mu_t|\,\dd t,
\]
and its quadratic action is
\[
  \mathcal A(\mu)=\int_0^1 |\dot\mu_t|^2\,\dd t,
\]
when these quantities are defined.
\end{definition}

The path length measures cumulative distributional movement. The action penalizes rapid movement more strongly.

\begin{proposition}[Direct displacement is no larger than path length]
\label{prop:distance-bounded-by-path-length}
For any absolutely continuous path \((\mu_t)_{0\leq t\leq 1}\),
\[
  \Wtwo(\mu_0,\mu_1)\leq \mathcal L(\mu).
\]
If the curve is a constant-speed geodesic, equality holds.
\end{proposition}

\begin{proof}
This is a direct consequence of the definition of metric derivative in a metric space. For a partition \(0=t_0<t_1<\cdots<t_n=1\), the triangle inequality gives
\[
  \Wtwo(\mu_0,\mu_1)\leq \sum_{k=0}^{n-1}\Wtwo(\mu_{t_k},\mu_{t_{k+1}}).
\]
Taking the supremum over partitions gives the metric length, which equals \(\int_0^1|\dot\mu_t|\,\dd t\) for absolutely continuous curves. For a constant-speed geodesic, the distance between endpoints equals the length by definition.
\end{proof}

\begin{actuarialbox}
A portfolio may end close to where it started but experience large intermediate movement. The endpoint distance \(\Wtwo(\mu_0,\mu_1)\) misses this churn. The path length \(\mathcal L(\mu)\) captures cumulative distributional movement.
\end{actuarialbox}

\subsection{Discrete-time actuarial trajectories}
\label{sec:dynamic-discrete-time-trajectories}

In practice, we often observe distributions at discrete dates:
\[
  \mu_0,\mu_1,\ldots,\mu_T.
\]
For example, \(\mu_t\) may be the distribution of claim severities in calendar year \(t\), the distribution of covariates in a yearly portfolio, or the distribution of reserve cash-flow scenarios at successive valuations.

A simple dynamic diagnostic is the sequence of one-step Wasserstein distances
\[
  D_t=\Wtwo(\mu_t,\mu_{t+1}),\qquad t=0,
  \ldots,T-1.
\]
The cumulative distributional movement is
\[
  \mathcal L_T=\sum_{t=0}^{T-1}\Wtwo(\mu_t,\mu_{t+1}),
\]
and the discrete quadratic action is
\[
  \mathcal A_T=\sum_{t=0}^{T-1}\Wtwo^2(\mu_t,\mu_{t+1}).
\]
If the time intervals have different lengths \(\Delta t_t\), one uses
\[
  \mathcal A_T=\sum_{t=0}^{T-1}\frac{\Wtwo^2(\mu_t,\mu_{t+1})}{\Delta t_t}.
\]

\begin{example}[Portfolio drift dashboard]
Let \(\mu_t=\Law(\bX_t)\), where \(\bX_t\) is the vector of policyholder characteristics in year \(t\). The sequence
\[
  \Wtwo(\mu_t,\mu_{t+1})
\]
provides a distributional drift dashboard. A large value identifies a year where the portfolio composition changed substantially. The cumulative length records total portfolio churn.
\end{example}

\iffalse
% To be restored in v2 with the premium distribution and demutualization material.
\begin{example}[Tariff phasing]
Suppose an insurer moves from tariff \(\pi_0\) to tariff \(\pi_1\) over four quarters. Let \(\mu_t\) be the distribution of premiums after quarter \(t\). A smooth phase-in should have relatively stable one-step distances. A sudden jump in \(\Wtwo(\mu_t,\mu_{t+1})\) indicates a concentrated distributional shock to the book.
\end{example}
\fi

\begin{proposition}[Discrete triangle bound]
\label{prop:discrete-triangle-bound}
For any sequence \(\mu_0,\ldots,\mu_T\in\Probp{2}(\R^d)\),
\[
  \Wtwo(\mu_0,\mu_T)\leq \sum_{t=0}^{T-1}\Wtwo(\mu_t,\mu_{t+1}).
\]
\end{proposition}

\begin{proof}
Apply the triangle inequality repeatedly.
\end{proof}

\begin{proposition}[Stability of Lipschitz actuarial indicators along a path]
\label{prop:lipschitz-indicator-dynamic-bound}
Let \(f:\R^d\to\R\) be \(K\)-Lipschitz. For any sequence \(\mu_0,\ldots,\mu_T\in\Probp{1}(\R^d)\),
\[
  \left|\int f\,\dd\mu_T-\int f\,\dd\mu_0\right|
  \le K\sum_{t=0}^{T-1}\Wone(\mu_t,\mu_{t+1}).
\]
\end{proposition}

\begin{proof}
By the Kantorovich--Rubinstein duality,
\(|\int f\,\dd\mu_{t+1}-\int f\,\dd\mu_t|\le K\Wone(\mu_t,\mu_{t+1})\).
Summing over \(t\) gives the result.
\end{proof}

\subsection{Dynamic couplings over several dates}
\label{sec:dynamic-multitime-couplings}

The pairwise distances \(\Wtwo(\mu_t,\mu_{t+1})\) compare adjacent dates. Sometimes one wants a joint scenario path across all dates. This requires a multi-time coupling.

\begin{definition}[Multi-time coupling]
\label{def:multi-time-coupling}
Given probability measures \(\mu_0,\ldots,\mu_T\) on \(\R^d\), a multi-time coupling is a probability measure
\[
  \Gamma\in\Coup(\mu_0,\ldots,\mu_T)\subset\Prob((\R^d)^{T+1})
\]
such that the \(t\)-th marginal of \(\Gamma\) is \(\mu_t\) for each \(t=0,\ldots,T\).
\end{definition}

A path sampled from \(\Gamma\) is
\[
  (\bX_0,\bX_1,\ldots,\bX_T),\qquad \bX_t\sim\mu_t.
\]
The natural dynamic cost is
\[
  \E_\Gamma\left[\sum_{t=0}^{T-1}\|\bX_{t+1}-\bX_t\|^2\right].
\]
This leads to the discrete-time dynamic OT problem
\[
  \inf_{\Gamma:\, (\bX_t)\sim\mu_t}
  \E_\Gamma\left[\sum_{t=0}^{T-1}\|\bX_{t+1}-\bX_t\|^2\right].
\]

\begin{proposition}[Pairwise Markov construction]
\label{prop:pairwise-markov-gluing}
Suppose, for each \(t\), \(\gamma_t\in\Coup(\mu_t,\mu_{t+1})\). Under standard regularity assumptions, the pairwise couplings can be glued into a multi-time law \(\Gamma\) such that \((\bX_t,\bX_{t+1})\sim\gamma_t\) for each \(t\). If each \(\gamma_t\) is optimal, then
\[
  \E_\Gamma\left[\sum_{t=0}^{T-1}\|\bX_{t+1}-\bX_t\|^2\right]
  =\sum_{t=0}^{T-1}\Wtwo^2(\mu_t,\mu_{t+1}).
\]
\end{proposition}

\begin{proof}
This is an application of the gluing lemma for probability measures; see \cite{villani2008oldnew,ambrosio2021lectures}. Sketch of the proof: disintegrate each pairwise coupling as
\[
  \gamma_t(\dd \bx_t,\dd \bx_{t+1})=\mu_t(\dd \bx_t)K_t(\bx_t,\dd \bx_{t+1}).
\]
Starting from \(\bX_0\sim\mu_0\), generate \(\bX_{t+1}\) from \(K_t(\bX_t,\cdot)\). This defines a Markov law \(\Gamma\) with the desired pairwise marginals. The cost identity follows by summing expectations.
\end{proof}

\begin{remark}[Pairwise optimal is not always globally meaningful]
Pairwise optimal couplings minimize adjacent costs. They need not preserve individual identity, contractual continuity or causal consistency across all dates. For policyholder-level modelling, additional constraints may be necessary: same policyholder tracking, no impossible jumps in covariates, monotonic ageing, renewal behaviour or lapse.
\end{remark}

\begin{actuarialbox}
A multi-time coupling is a scenario generator. It turns marginal distributions at several dates into full sample paths. This is useful for climate scenarios, reserve cash-flow paths, portfolio migration and staged interventions.
\end{actuarialbox}

\subsection{Adapted transport and non-anticipative actuarial paths}
\label{sec:adapted-transport-actuarial}

The dynamic transport constructions introduced above compare trajectories of
distributions. In many actuarial applications, however, trajectories carry
information over time. A coupling between two stochastic processes should not use
future information when matching the present. This leads to the idea of adapted,
or non-anticipative, transport.

Let
\[
  X=(X_0,\ldots,X_T),
  \qquad
  Y=(Y_0,\ldots,Y_T)
\]
be two stochastic processes with laws \(\mu\) and \(\nu\) on a path space
\(\mathcal X^{T+1}\). A usual coupling \(\lambda\in\Coup(\mu,\nu)\) is a joint law of
\((\bX,\bY)\) with the prescribed marginals. It may, however, match \(\bX_t\) using
information from \(\bY_{t+1},\ldots,\bY_T\), or match \(\bY_t\) using future values of
\(\bX\). This is not acceptable when the coupling is meant to represent a feasible
scenario transformation over time.

An adapted coupling imposes a causality constraint.

\begin{definition}[Causal and bicausal couplings]
\label{def:causal-bicausal-coupling}
A coupling \(\lambda\in\Coup(\mu,\nu)\) is causal from \(\bX\) to \(\bY\) if, for each date \(t\), the conditional law of \(\bY_{0:t}\) given the whole path \(\bX_{0:T}\) depends only on the past and present \(\bX_{0:t}\):
\[
  \Law_\lambda(\bY_{0:t}\mid \bX_{0:T})
  =
  \Law_\lambda(\bY_{0:t}\mid \bX_{0:t}),
  \qquad t=0,\ldots,T.
\]
It is bicausal if the same condition holds in both directions, from \(\bX\) to \(\bY\) and from \(\bY\) to \(\bX\). We write \(\Coup_{\mathrm{ad}}(\mu,\nu)\) or \(\Coup_{\mathrm{bc}}(\mu,\nu)\) for the corresponding adapted or bicausal feasible sets, depending on the convention used.
\end{definition}

The corresponding adapted transport problem is
\[
  \inf_{\lambda\in\Coup_{\mathrm{ad}}(\mu,\nu)}
  \E_\lambda\left[
    \sum_{t=0}^T c_t(\bX_t,\bY_t)
  \right].
\]

\begin{actuarialbox}
Ordinary transport compares distributions. Adapted transport compares stochastic
processes without allowing the coupling to look into the future. This is the right
notion when trajectories represent reserving paths, cash-flow projections, epidemic
states, mortality scenarios or portfolio evolution.
\end{actuarialbox}

\subsubsection{Why non-anticipativity matters}

Consider two reserving scenarios
\[
  R^0=(R^0_0,\ldots,R^0_T),
  \qquad
  R^1=(R^1_0,\ldots,R^1_T),
\]
where \(R^k_t\) is the outstanding liability at valuation date \(t\). A static
coupling of the full paths may match an early reserve under one model with a path
whose later development is already known under the other model. Such a coupling
can be useful for comparing laws of paths, but it cannot be interpreted as a feasible
online transformation.

For model monitoring, the actuary often wants a diagnostic of the following kind:
given the information available at time \(t\), how much must the reserve process
move to resemble the target process? This is a non-anticipative question.

Similarly, for mortality and longevity projections, one may compare two stochastic
mortality surfaces
\[
  q^0_{x,t},
  \qquad
  q^1_{x,t}.
\]
A pathwise coupling should not match the mortality rate at date \(t\) using future
realizations of the mortality trend. The same issue arises in climate trajectories,
cyber accumulation scenarios and epidemic state processes.

\subsubsection{A discrete-time formulation}

Let \(\mathcal F_t^{\bX}=\sigma(\bX_0,\ldots,\bX_t)\) and
\(\mathcal F_t^{\bY}=\sigma(\bY_0,\ldots,\bY_t)\). A bicausal coupling may be described
by the condition that, under \(\lambda\),
\[
  \Law(\bY_{t+1:T}\mid \bX_{0:T},\bY_{0:t})
  =
  \Law(\bY_{t+1:T}\mid \bX_{0:t},\bY_{0:t}),
\]
and symmetrically with \(\bX\) and \(\bY\) exchanged. Equivalently, the matching rule
at time \(t\) cannot depend on future observations.

For a finite state space, this becomes a dynamic programming problem. Suppose
\(X_t,Y_t\in E\), with \(E\) finite. A path coupling is a probability mass function
\[
  \lambda(x_{0:T},y_{0:T}).
\]
The marginal constraints impose
\[
  \sum_{y_{0:T}}\lambda(x_{0:T},y_{0:T})=\mu(x_{0:T}),
  \qquad
  \sum_{x_{0:T}}\lambda(x_{0:T},y_{0:T})=\nu(y_{0:T}).
\]
The adapted constraints restrict how the conditional kernels factor over time. A
typical bicausal factorization is
\[
  \lambda(\dd x_{0:T},\dd y_{0:T})
  =
  \lambda_0(\dd x_0,\dd y_0)
  \prod_{t=0}^{T-1}
  K_t(\dd x_{t+1},\dd y_{t+1}\mid x_{0:t},y_{0:t}),
\]
with kernels compatible with the marginal transition laws.

\subsubsection{Actuarial path costs}

The choice of path cost should reflect the actuarial object being compared. For
cash-flow vectors, one may use
\[
  c(\bX,\bY)
  =
  \sum_{t=0}^T v_t |X_t-Y_t|^p,
\]
where \(v_t\) are discount factors or reporting weights. For reserving paths, changes
near the valuation date may matter more:
\[
  c(X,Y)
  =
  \sum_{t=0}^T w_t |R^{\bX}_t-R^{\bY}_t|^p,
  \qquad
  w_0\ge w_1\ge\cdots\ge w_T.
\]
For epidemic or health-state trajectories, the cost may be a state-transition cost:
\[
  c(X,Y)
  =
  \sum_{t=0}^T \kappa_t(\bX_t,\bY_t),
\]
where \(\kappa_t\) penalizes mismatches between health states such as susceptible,
infected, hospitalized, disabled or recovered.

\begin{computebox}
For actuarial monitoring, adapted transport should be reported together with the
ordinary path Wasserstein distance. A large difference between the two indicates
that the apparent similarity of two path distributions relies on anticipative
matching.
\end{computebox}

\subsubsection{Connection with martingale and causal transport}

Adapted transport is close in spirit to martingale optimal transport, but the
constraints are different. Martingale optimal transport imposes
\[
  \E[\bY_{t+1}\mid \bY_{0:t}]=\bY_t,
\]
or a similar financial consistency constraint. Adapted transport imposes
non-anticipativity of the coupling. In insurance applications, both may be relevant:
market-consistent guarantees require martingale constraints, while cash-flow,
mortality and reserving paths require adaptedness.

\begin{remark}[Governance interpretation]
A non-anticipative transport diagnostic is useful for governance because it respects
the information structure of actuarial work. At valuation date \(t\), the actuary has
observed the past, not the future. A path comparison that uses future information
may be mathematically valid, but it should not be interpreted as an implementable
scenario transformation.
\end{remark}

\subsection{Dynamic OT for reserving cash-flow trajectories}
\label{sec:dynamic-reserving-cashflow-trajectories}

Reserving is naturally dynamic. A stochastic reserving model produces a random vector
\[
  \bR=(R_1,\ldots,R_H),
\]
where \(R_h\) is the future payment in development or calendar period \(h\). A model gives a law
\[
  \mu=\Law(\bR)\in\Prob(\R^H).
\]
Comparing two reserving models can be done statically through \(\Wtwo(\mu_0,\mu_1)\), as in Lecture~\ref{lec:reserving-climate-prevention}. But there is also a dynamic viewpoint across valuation dates. At valuation date \(t\), the predictive law of future cash-flows is
\[
  \mu_t=\Law_t(R_{t+1},\ldots,R_{t+H}).
\]
As claims develop, \(\mu_t\) should evolve. A large distributional speed \(\Wtwo(\mu_t,\mu_{t+1})\) may indicate:
\begin{enumerate}[label=(\roman*)]
  \item new information from reported claims;
  \item changes in inflation assumptions;
  \item calendar-year effects;
  \item model instability;
  \item large payments or case reserve revisions.
\end{enumerate}

\begin{example}[Reserve revaluation shock]
Let \(\mu_t\) be the predictive distribution of total unpaid claims at quarter \(t\). If
\[
  \Wtwo(\mu_t,\mu_{t+1})
\]
is much larger than usual, the reserve distribution has moved significantly. This may happen even if the best estimate changes only moderately, because the tail or dispersion of the reserve distribution may have shifted.
\end{example}

\subsection{Dynamic OT for climate and prevention}
\label{sec:dynamic-climate-prevention}

Climate risk is not a one-step perturbation. A baseline climate distribution, an intermediate transition distribution and a long-run stressed distribution should be linked by a trajectory. Such a trajectory is a distributional scenario device. It is not, by itself, an identification argument for the causal effect of a prevention policy on an individual loss.

Let \(\mu_t\) denote the annual aggregate loss distribution at climate time \(t\), where \(t=0\) is current climate and \(t=1\) is a long-run scenario. A dynamic OT path can be used to define:
\begin{enumerate}[label=(\roman*)]
  \item a smooth climate stress trajectory;
  \item a prevention trajectory from no intervention to intervention;
  \item a sequence of adaptation benefits over time;
  \item a cumulative prevented-loss measure.
\end{enumerate}

If \(L_t\sim\mu_t\), the expected annual loss is
\[
  \E[L_t]=\int x\,\mu_t(\dd \bx).
\]
\begin{definition}[Dynamic prevention profile]
\label{def:dynamic-prevention-profile}
Let \((\mu_t^{0})_{0\le t\le 1}\) be a baseline loss trajectory and \((\mu_t^{1})_{0\le t\le 1}\) a post-prevention trajectory. A dynamic prevention profile is the family of distributional differences
\[
  t\longmapsto \left(\mu_t^{0},\mu_t^{1},\gamma_t,\Delta_t^{\mathrm{mean}},\Delta_t^{\mathrm{tail}}\right),
\]
where \(\gamma_t\in\Coup(\mu_t^{0},\mu_t^{1})\) is a chosen comparison coupling,
\(\Delta_t^{\mathrm{mean}}=\E[L_t^{0}]-\E[L_t^{1}]\), and \(\Delta_t^{\mathrm{tail}}\) denotes a tail functional difference such as \(\TVaR_\alpha(\mu_t^{0})-\TVaR_\alpha(\mu_t^{1})\).
\end{definition}

A prevention effect may be summarized by
\[
  \Delta_t=\E[L_t^{\mathrm{baseline}}]-\E[L_t^{\mathrm{prevention}}],
\]
but dynamic OT also measures how the full distributions move:
\[
  \Wtwo(\mu_t^{\mathrm{baseline}},\mu_t^{\mathrm{prevention}}).
\]

\begin{example}[Tail-focused prevention]
Two prevention measures may reduce expected annual loss by the same amount. The first reduces frequent small claims; the second reduces rare large claims. Their expected benefits may match, but their dynamic quantile paths differ. In one dimension, this is visible through
\[
  Q_t^{\mathrm{baseline}}(u)-Q_t^{\mathrm{prevention}}(u),\qquad u\in(0,1).
\]
A tail-focused prevention measure has larger differences near \(u=1\).
\end{example}

\begin{actuarialbox}
Prevention accounting should not only ask how much expected loss was avoided. It should also ask which parts of the loss distribution were moved, when they were moved, and how persistent the movement is. Dynamic OT provides a language for this comparison.
\end{actuarialbox}

\subsection{When total mass changes: toward unbalanced dynamics}
\label{sec:dynamic-unbalanced-source-term}

The continuity equation conserves total mass. Probability measures always have mass one, but actuarial exposure may change. A portfolio can grow, shrink, lapse or acquire new business. A climate model can change the number of exposed assets. A reserving triangle can gain or lose observed cells as time passes.

To model changing exposure explicitly, one introduces a source term:
\[
  \partial_t\eta_t+\nabla\cdot(\bv_t\eta_t)=\sigma_t,
\]
where \(\eta_t\) is now a finite exposure measure and \(\sigma_t\) creates or destroys mass.

\begin{definition}[Source-term dynamics]
\label{def:source-term-dynamics}
A source-term dynamic transport model is a triple \((\eta_t,\bv_t,\sigma_t)\) where \(\eta_t\) is a finite measure, \(\bv_t\) is a velocity field, and \(\sigma_t\) is a signed measure satisfying
\[
  \partial_t\eta_t+\nabla\cdot(\bv_t\eta_t)=\sigma_t.
\]
Positive mass in \(\sigma_t\) represents new exposure, while negative mass represents lapse, exit, death, settlement or other disappearance of exposure.
\end{definition}

In probability-normalized applications, the source term is hidden by renormalization. In exposure-weighted applications, it is essential.

\begin{example}[Lapse and new business]
Let \(\eta_t\) be an exposure measure rather than a probability measure. If policies lapse, some mass disappears. If new business enters, new mass appears. Balanced OT can compare normalized distributions \(\mu_t=\eta_t/\eta_t(\R^d)\), but it cannot distinguish portfolio composition drift from total exposure growth. Unbalanced transport is needed for that distinction.
\end{example}

\begin{remark}[Practical rule]
If the actuarial question is about composition, normalize to probability measures and use balanced OT. If the actuarial question is about both composition and volume, keep exposure weights and use unbalanced OT or a source term.
\end{remark}

\subsection{Epidemic trajectories and state-dependent insurance risk}
\label{sec:epidemic-dynamic-ot}

Epidemic risk creates time-varying heterogeneity in an insured population. The
risk profile of an individual depends on an evolving health state, and the aggregate
claim process depends on the changing composition of the portfolio. This is not
well represented by a single static loss distribution.

Let the health state space be
\[
  E=\{\mathrm{Sus},\mathrm{Inf},\mathrm{Rec}\},
\]
where \(\mathrm{Sus}\) denotes susceptible, \(\mathrm{Inf}\) infected, and \(\mathrm{Rec}\) recovered or removed.
At time \(t\), let
\[
  \mu_t\in\Prob(E)
\]
be the distribution of the insured population across states. In a more detailed
model, \(E\) may include hospitalization, disability, long-term complications,
death, return to work, or vaccination status.

The insurance loss at time \(t\) is state-dependent. If \(c_t(e)\) denotes the expected
claim cost for an insured in state \(e\), then the expected portfolio cost is
\[
  \E[C_t]
  =
  \sum_{e\in E} c_t(e)\mu_t(e).
\]
The full distribution of claims depends not only on the marginal state distribution
\(\mu_t\), but also on the dependence between individuals, common shocks and
transition dynamics.

Dynamic optimal transport provides a language for comparing epidemic scenarios.
Let
\[
  (\mu_t^0)_{t=0}^T
  \quad\text{and}\quad
  (\mu_t^1)_{t=0}^T
\]
be two trajectories of population state distributions, for example a baseline
scenario and a severe epidemic scenario. A simple trajectory discrepancy is
\[
  \sum_{t=0}^T
  w_t W_p^p(\mu_t^0,\mu_t^1).
\]
This compares state distributions date by date. A more dynamic comparison also
penalizes the movement between states over time.

\begin{actuarialbox}
Epidemic insurance risk is distributional and dynamic. The relevant object is not
only a claim severity law, but a trajectory of population compositions:
\[
  t\mapsto \mu_t.
\]
Transport compares how much mass must move between health states, calendar
dates or cost states to transform one epidemic scenario into another.
\end{actuarialbox}

Suppose that the state distribution evolves according to a transition matrix
\(K_t\):
\[
  \mu_{t+1}=\mu_t K_t.
\]
In an SIR-type model, the transition probabilities may depend on the current
infected mass. For example, in discrete time,
\[
  \Pbb(\mathrm{Sus}\to \mathrm{Inf}\mid \mu_t)=\beta \mu_t(\mathrm{Inf}),
  \qquad
  \Pbb(\mathrm{Inf}\to \mathrm{Rec})=\gamma.
\]
Thus the transition kernel is endogenous. Two epidemic scenarios can differ because
they have different initial states, different transition intensities, different
intervention policies, or different reporting conventions.

A transport-based stress can be described at the level of trajectories:
\[
  \mu_0
  \longrightarrow
  \mu_1
  \longrightarrow
  \cdots
  \longrightarrow
  \mu_T.
\]
If the state space is finite, a dynamic coupling consists of a joint law over paths
\[
  (E_0,\ldots,E_T).
\]
The cost of moving from one path to another can encode actuarial meaning:
\[
  c(e_{0:T},f_{0:T})
  =
  \sum_{t=0}^T \kappa_t(e_t,f_t)
  +
  \lambda
  \sum_{t=0}^{T-1}
  \1_{\{e_t\ne e_{t+1}\}},
\]
where the first term compares health states and the second term penalizes transition
events.

\begin{remark}[Static versus dynamic comparison]
A date-by-date comparison of \(\mu_t^0\) and \(\mu_t^1\) may miss the timing of
the epidemic wave. Two scenarios can have similar total infection counts but very
different peaks, durations and hospitalization timing. A path-level transport cost
can penalize these differences directly.
\end{remark}

Insurance cash flows are obtained by pushing the state trajectory through a cost
map. Let
\[
  h_t:E\to\R_+
\]
be the claim cost at time \(t\). The cumulative discounted claim is
\[
  L
  =
  \sum_{t=0}^T v_t h_t(E_t).
\]
Therefore
\[
  \Law(L)
  =
  \left(
    e_{0:T}\mapsto \sum_{t=0}^T v_t h_t(e_t)
  \right)_\#\Law(E_{0:T}).
\]
This connects epidemic dynamics to the push-forward viewpoint used throughout
these notes.

The same framework applies to multiple insurance lines:
\[
  L^{\mathrm{health}},
  \qquad
  L^{\mathrm{life}},
  \qquad
  L^{\mathrm{disability}},
  \qquad
  L^{\mathrm{business\ interruption}}.
\]
An epidemic can increase claims in some lines while reducing or delaying claims in
others. The aggregate effect depends on the joint law of the line-specific losses.

\begin{computebox}
A practical epidemic stress test can report:
\[
  \sum_{t=0}^T W_1(\mu_t^{\mathrm{base}},\mu_t^{\mathrm{stress}}),
  \qquad
  \max_t \mu_t^{\mathrm{stress}}(\mathrm{Inf}),
  \qquad
  \Law\left(\sum_{t=0}^T v_t h_t(E_t)\right),
\]
and a capital measure of the resulting aggregate loss. The Wasserstein term
measures population-state displacement; the loss distribution measures financial
impact.
\end{computebox}

Finally, epidemic trajectories are a natural example where balanced transport may
be insufficient. Entry, exit, death, lapse, delayed reporting and changes in exposure
can alter the total insured mass. In such cases, unbalanced transport or source-term
dynamics are more appropriate:
\[
  \partial_t \rho_t+\nabla\cdot(\rho_t \bv_t)=\sigma_t,
\]
where \(\sigma_t\) represents creation or destruction of exposure mass, as in Definition~\ref{def:source-term-dynamics}.

\subsection{Entropic dynamics and stochastic paths}
\label{sec:dynamic-entropic-stochastic-paths}

Exact dynamic OT produces least-action paths. Entropic regularization produces smoother, more diffuse path measures. This is connected to the Schr\"odinger bridge viewpoint: among stochastic processes matching prescribed endpoint distributions, choose the one closest to a reference diffusion or random walk in relative entropy.

For these notes, the key idea is simple. Static OT selects a cheapest endpoint coupling. Dynamic entropic OT selects a stochastic evolution that balances:
\begin{enumerate}[label=(\roman*)]
  \item endpoint constraints;
  \item movement cost;
  \item closeness to a reference stochastic dynamics;
  \item computational smoothness.
\end{enumerate}

\begin{example}[Stochastic climate scenario paths]
A deterministic displacement path may be too rigid for climate risk. A stochastic path allows individual scenarios to diffuse around a trend while still matching endpoint distributions. Entropic dynamic OT provides a principled way to construct such paths.
\end{example}

\begin{computebox}
Dynamic entropic transport is computationally more involved than pairwise Sinkhorn, but the intuition is the same: exact deterministic motion is replaced by a smoother probabilistic motion. For many actuarial dashboards, a practical first step is to compute pairwise entropic couplings between consecutive dates and glue them into Markov scenario paths.
\end{computebox}

\subsection{Transport flows and synthetic scenario generation}
\label{sec:transport-generative-scenarios}

Optimal transport gives deterministic maps and geodesic interpolations between
probability distributions. Modern generative modelling extends this idea: instead
of only computing a map between two known distributions, one learns a mechanism
that transforms a simple latent distribution into a complex target distribution.

Let \(\bZ_0\sim\mu_0\) be a simple random vector, for example a standard Gaussian,
and let \(\bZ_1\sim\mu_1\) represent actuarial scenarios: claim amounts, catastrophe
losses, reserve cash-flows, cyber accumulation losses, mortality paths or economic
risk factors. A generative transport model seeks a transformation
\[
  T_{\btheta}:\R^d\to\R^d
\]
such that
\[
  (T_{\btheta})_\#\mu_0\approx \mu_1.
\]
Once learned, the map can generate synthetic scenarios by sampling \(\bZ_0\) and computing \(T_{\btheta}(\bZ_0)\).

\begin{actuarialbox}
Synthetic scenario generation is a transport problem: move a tractable latent law
to a realistic actuarial scenario law. The actuarial difficulty is not only to generate
samples, but to preserve tails, dependence, constraints and interpretability.
\end{actuarialbox}

\subsubsection{Deterministic flows}

A continuous deterministic flow is defined by an ordinary differential equation
\[
  \frac{\dd \bZ_t}{\dd t}=\bv_t(\bZ_t),
  \qquad
  t\in[0,1],
\]
with initial law \(\bZ_0\sim\mu_0\). If \(\rho_t=\Law(\bZ_t)\), then \(\rho_t\) solves
the continuity equation
\[
  \partial_t\rho_t+\nabla\cdot(\rho_t \bv_t)=0.
\]
Learning the vector field \(\bv_t\) is one way to learn a generative transport from
\(\mu_0\) to \(\mu_1\). This viewpoint connects continuous normalizing flows, flow matching and dynamic optimal transport; see, for example, \cite{chen2018neural,rezende2015variational,dinh2017density,lipman2023flow}.

For actuarial applications, the vector field may be constrained to respect
properties of the scenario:
\[
  \bZ_t\in\R_+^d,
  \qquad
  \sum_{j=1}^d \bZ_{t,j}\le M,
  \qquad
  \bZ_{t,1}\le \bZ_{t,2}\le\cdots\le \bZ_{t,T},
\]
depending on whether the scenario represents non-negative losses, limited
coverages, or cumulative paid losses.

\subsubsection{Stochastic flows and diffusion models}

A stochastic generative flow is defined by a stochastic differential equation
\[
  \dd \bZ_t=\bb_t(\bZ_t)\,\dd t+\sigma_t\,\dd \bW_t.
\]
Diffusion-based generative models use noise to move between data and latent
distributions; see \cite{song2021score,ho2020denoising}. In a forward diffusion, the data distribution is gradually noised until
it becomes close to a simple reference distribution. A reverse process is then learned
to transform noise back into data-like samples.

The deterministic and stochastic viewpoints are closely related. Deterministic
flows provide direct scenario maps. Diffusions provide stochastic scenario
generation and may better represent uncertainty, multimodality and model
ambiguity.

\begin{remark}[Map versus generator]
A deterministic map sends the same latent point to the same scenario. A diffusion
generator adds stochasticity along the path. For actuarial stress testing, the first is
useful for traceability; the second is useful for rich scenario generation.
\end{remark}

\subsubsection{Scenario generation tasks}

Transport-based generators are useful for several actuarial tasks.

\paragraph{Catastrophe and climate scenarios.}
A generator can produce spatial loss fields, rainfall maps, wind fields or portfolio
loss vectors. The transport structure can help align generated scenarios with a
baseline distribution, a climate-stressed distribution or an event catalogue.

\paragraph{Reserving cash-flow paths.}
A cash-flow scenario is a vector
\[
  C=(C_1,\ldots,C_T).
\]
A generative model can learn the joint law of paid or incurred development paths,
including timing and tail dependence. Transport diagnostics can then compare
synthetic and historical cash-flow distributions.

\paragraph{Mortality and longevity paths.}
A mortality scenario may be a surface
\[
  \bq=(q_{x,t})_{x,t}.
\]
A generator can produce stochastic mortality improvements while preserving
smoothness over ages, cohort effects and plausible long-run trends.

\paragraph{Cyber accumulation.}
Cyber scenarios require dependence across firms, vendors and coverages. A
transport generator can learn or impose common shock structures, vendor
concentration effects and heavy-tailed accumulation losses.

\subsubsection{Validation by transport diagnostics}

Generated scenarios should be validated distributionally. If \(\widehat\mu_{\mathrm{real}}\)
is the empirical law of observed or benchmark scenarios and
\(\widehat\mu_{\mathrm{gen}}\) is the law of generated scenarios, one may report
\[
  W_p(\widehat\mu_{\mathrm{real}},\widehat\mu_{\mathrm{gen}}),
\]
or sliced Wasserstein distances in high dimension:
\[
  SW_p(\widehat\mu_{\mathrm{real}},\widehat\mu_{\mathrm{gen}}).
\]
For actuarial work, these global diagnostics should be complemented by tail and
business metrics:
\[
  \VaR_\alpha,\qquad
  \TVaR_\alpha,\qquad
  \E[(L-d)_+\wedge \ell],
  \qquad
  \Corr(L_i,L_j).
\]

\begin{computebox}
A generative actuarial model should not be validated only by visual realism or
average error. A minimum validation report should include Wasserstein or sliced
Wasserstein distance, marginal calibration, tail calibration, dependence diagnostics
and layer-specific errors.
\end{computebox}

\subsubsection{Stress generation by interpolation}

A simple use of transport flows is to interpolate between a baseline law \(\mu_0\)
and a stressed law \(\mu_1\). If \(T\) transports \(\mu_0\) to \(\mu_1\), the
displacement interpolation is
\[
  \bZ_t=(1-t)\bZ_0+tT(\bZ_0),
  \qquad t\in[0,1].
\]
Then
\[
  \mu_t=\Law(\bZ_t)
\]
is a path of scenarios from baseline to stress. This gives a transparent stress ladder:
\[
  \mu_0,\mu_{0.25},\mu_{0.5},\mu_{0.75},\mu_1.
\]
For more complex settings, a learned flow can replace the explicit map \(T\).

\begin{remark}[Synthetic does not mean plausible]
A generated scenario is not automatically a plausible actuarial scenario. It must
satisfy accounting identities, policy conditions, exposure constraints, monotonicity
constraints, tail behavior and expert review. Transport gives a geometry for
generation and validation, but governance remains essential.
\end{remark}

\subsubsection{Connection with Schr\"odinger bridges}

Schr\"odinger bridges provide a stochastic transport between two endpoint
distributions. They can be viewed as the most likely stochastic evolution between
\(\mu_0\) and \(\mu_1\) relative to a reference diffusion. In scenario generation,
this is useful when one wants both endpoint calibration and stochastic path
variability.

For example, suppose \(\mu_0\) is a current reserve distribution and \(\mu_1\) is a
one-year stressed reserve distribution. A deterministic map gives a single
transformation of each scenario. A Schr\"odinger bridge gives a distribution over
intermediate paths, allowing stochastic transition scenarios between the endpoints.

\begin{actuarialbox}
Transport flows, diffusion models and Schr\"odinger bridges are different ways of
turning distributional information into scenarios. The actuarial choice depends on
the objective: traceable stress maps, rich synthetic generation, or stochastic paths
between calibrated endpoints.
\end{actuarialbox}

\subsection{A mini-lab: dynamic severity stress in R}
\label{sec:lab-dynamic-severity-stress}

The following lab, which complements the statistical labs in Lecture~\ref{lec:statistical-ot}, constructs a one-dimensional Wasserstein geodesic between a baseline severity distribution and a stressed severity distribution. It then tracks mean, VaR and TVaR along the path.

\begin{lstlisting}[language=R]
set.seed(123)

n <- 50000
u <- (1:n) / (n + 1)

# Baseline and stressed severity quantile functions
q0 <- function(u) qlnorm(u, meanlog = 8.0, sdlog = 0.75)
q1 <- function(u) qlnorm(u, meanlog = 8.25, sdlog = 1.00)

Q0 <- q0(u)
Q1 <- q1(u)

# Wasserstein-2 distance between endpoints
W2 <- sqrt(mean((Q1 - Q0)^2))
W2

# Integrated tail functional
TVaR_from_quantiles <- function(Q, u, alpha = 0.99) {
  mean(Q[u >= alpha])
}

VaR_from_quantiles <- function(Q, u, alpha = 0.995) {
  Q[which.min(abs(u - alpha))]
}

# Dynamic path indexed by t
grid_t <- seq(0, 1, by = 0.05)
out <- data.frame(
  t = grid_t,
  mean = NA_real_,
  VaR995 = NA_real_,
  TVaR99 = NA_real_,
  W2_from_start = NA_real_,
  W2_to_target = NA_real_
)

for (k in seq_along(grid_t)) {
  t <- grid_t[k]
  Qt <- (1 - t) * Q0 + t * Q1
  out$mean[k] <- mean(Qt)
  out$VaR995[k] <- VaR_from_quantiles(Qt, u, alpha = 0.995)
  out$TVaR99[k] <- TVaR_from_quantiles(Qt, u, alpha = 0.99)
  out$W2_from_start[k] <- sqrt(mean((Qt - Q0)^2))
  out$W2_to_target[k] <- sqrt(mean((Qt - Q1)^2))
}

print(out)

par(mfrow = c(1, 3))
plot(out$t, out$mean, type = "l", lwd = 2,
     xlab = "stress time t", ylab = "mean loss")
plot(out$t, out$VaR995, type = "l", lwd = 2,
     xlab = "stress time t", ylab = "VaR 99.5%")
plot(out$t, out$W2_from_start, type = "l", lwd = 2,
     xlab = "stress time t", ylab = "W2 from baseline")
\end{lstlisting}

Along this geodesic, \(\Wtwo(\mu_0,\mu_t)=t\Wtwo(\mu_0,\mu_1)\). The mean, VaR and TVaR evolve linearly in quantile coordinates, but their monetary interpretation depends on the tail region under consideration.

\subsection{Practical workflow}
\label{sec:dynamic-practical-workflow}

A practical dynamic OT workflow for actuarial applications can be organized as follows.

\begin{enumerate}[label=\textbf{Step \arabic*.}]
  \item \textbf{Define the moving object.} Decide whether \(\mu_t\) represents losses, covariates, reserves, cash-flow vectors, climate scenarios or exposure-weighted measures.
  \item \textbf{Choose the time scale.} Calendar year, underwriting year, development period, quarterly monitoring or climate horizon.
  \item \textbf{Choose the ground metric.} Monetary distance, standardized covariate distance, geographic distance, cash-flow vector distance, or a mixed actuarial cost.
  \item \textbf{Compute one-step distances.} Estimate \(\Wtwo(\mu_t,\mu_{t+1})\) or a regularized approximation.
  \item \textbf{Compute cumulative movement.} Track path length, action and direct distance from baseline.
  \item \textbf{Interpret velocity.} Identify which regions of the distribution are moving: low losses, tail losses, high-exposure regions or long-tail reserves.
  \item \textbf{Stress or intervene.} Construct geodesic paths to target distributions for scenario analysis or prevention.
  \item \textbf{Report uncertainty.} Use bootstrap, sensitivity to ground cost, and alternative radii or regularization parameters.
\end{enumerate}

\subsection{A small dictionary for dynamic actuarial OT}
\label{sec:dynamic-small-dictionary}

\begin{center}
\begin{tabular}{ll}
\toprule
Mathematical object & Actuarial interpretation\\
\midrule
\((\mu_t)_{t\in[0,1]}\) & Path of loss, reserve, cash-flow or scenario laws\\
\(\bv_t\) & Instantaneous velocity of risk mass\\
Continuity equation & Conservation of normalized portfolio mass\\
Benamou--Brenier action & Least kinetic energy of distributional movement\\
\(\Wtwo\)-geodesic & Shortest path between two actuarial laws\\
Path length & Cumulative distributional movement or churn\\
Multi-time coupling \(\Gamma\) & Joint scenario law across dates\\
Adapted coupling & Non-anticipative comparison of stochastic processes\\
Source term \(\sigma_t\) & Entry, lapse, settlement, death or exposure growth\\
Schr\"odinger bridge & Stochastic path calibrated to endpoint laws\\
Transport generator & Map or kernel from latent scenarios to actuarial scenarios\\
\bottomrule
\end{tabular}
\end{center}

\subsection{Summary}
\label{sec:dynamic-summary}

Dynamic optimal transport turns Wasserstein distance into a theory of trajectories. A path of probability measures \((\mu_t)\) evolves under a velocity field \(\bv_t\) satisfying the continuity equation. The Benamou--Brenier formula states that the squared \(\Wtwo\) distance is the least kinetic action needed to move \(\mu_0\) into \(\mu_1\). In one dimension, the geodesic is obtained by linear interpolation of quantile functions. For Gaussian laws, the geodesic has explicit mean and covariance formulas.

For actuarial science, dynamic OT provides a language for portfolio drift, reserving revaluations, climate transition, prevention accounting and scenario generation. Static OT tells us how far two distributions are. Dynamic OT tells us how they move.

\subsection{Exercises}
\label{sec:dynamic-exercises}

\begin{exercise}[Inflation velocity]\label{ex:15:1}
Let \(L_t=a_tL_0\), where \(a_t>0\) is differentiable and \(L_0\in L^2\). Show that \(\mu_t=\Law(L_t)\) solves the continuity equation with velocity
\[
  v_t(x)=\frac{\dot a_t}{a_t}x.
\]
Compute the kinetic action \(\int_0^1\E[v_t(L_t)^2]\,\dd t\).
\end{exercise}

\begin{exercise}[Quantile geodesic]\label{ex:15:2}
Let \(\mu_0,\mu_1\in\Probp{2}(\R)\) have quantile functions \(Q_0,Q_1\). Define \(Q_t=(1-t)Q_0+tQ_1\). Prove that
\[
  \Wtwo(\mu_s,\mu_t)=|t-s|\Wtwo(\mu_0,\mu_1).
\]
\end{exercise}

\begin{exercise}[Uniform-to-uniform trajectory]\label{ex:15:3}
Let \(\mu_0=U[0,1]\) and \(\mu_1=U[0,2]\). Compute the quantile geodesic \(\mu_t\). What is \(\mu_t\)? Compute \(\Wtwo(\mu_0,\mu_t)\) and \(\Wtwo(\mu_t,\mu_1)\).
\end{exercise}

\begin{exercise}[Dirac-to-two-point target]\label{ex:15:4}
Let \(\mu_0=\delta_0\) and \(\mu_1=(\delta_{-1}+\delta_1)/2\). Explain why no Monge map from \(\mu_0\) to \(\mu_1\) exists. Construct a dynamic interpolation induced by the optimal Kantorovich coupling. What are the intermediate measures \(\mu_t\)?
\end{exercise}

\begin{exercise}[TVaR along a geodesic]\label{ex:15:5}
Let \(Q_t=(1-t)Q_0+tQ_1\). Prove that \(\TVaR_\alpha(\mu_t)\) is affine in \(t\). Interpret this result for a climate stress trajectory.
\end{exercise}

\begin{exercise}[Gaussian covariance path]\label{ex:15:6}
Let \(\mu_0=N(0,\sigma_0^2)\) and \(\mu_1=N(0,\sigma_1^2)\). Derive the \(\Wtwo\)-geodesic using quantiles. Show that \(\mu_t=N(0,\sigma_t^2)\), where \(\sigma_t=(1-t)\sigma_0+t\sigma_1\). Compare this with linear interpolation of variances.
\end{exercise}

\begin{exercise}[Path length and churn]\label{ex:15:7}
Construct three one-dimensional distributions \(\mu_0,\mu_1,\mu_2\) such that \(\mu_2=\mu_0\) but
\[
  \Wtwo(\mu_0,\mu_1)+\Wtwo(\mu_1,\mu_2)>0.
\]
Interpret this as portfolio churn that is invisible from endpoint comparison alone.
\end{exercise}

\iffalse
% To be restored in v2 with the premium distribution and demutualization material.
\begin{exercise}[Tariff phase-in]\label{ex:15:8}
Let \(P_0\) be an old premium and \(P_1=T(P_0)\) a new premium obtained by monotone quantile transport. Define \(P_t=(1-t)P_0+tP_1\). Show that the premium distribution at time \(t\) lies on a \(\Wtwo\)-geodesic. Discuss why this may be preferable to randomly assigning a fraction \(t\) of policies to the new tariff and leaving the others under the old tariff.
\end{exercise}
\fi

\begin{exercise}[Discrete dynamic action]\label{ex:15:9}
Let \(\mu_0,\ldots,\mu_T\in\Probp{2}(\R^d)\). Prove that
\[
  \Wtwo(\mu_0,\mu_T)^2\leq T\sum_{t=0}^{T-1}\Wtwo^2(\mu_t,\mu_{t+1}).
\]
What inequality did you use? What does the bound say about a sequence of small yearly drifts?
\end{exercise}

\iffalse
% To be restored in v2 with the fairness material.
\begin{exercise}[Fairness repair path]\label{ex:15:10}
Let \(\mu_s\) be a group-specific score distribution and \(\bar\mu\) a target distribution on \(\R\). Let \(Q_s\) and \(\bar Q\) be their quantile functions. Define
\[
  Q_{s,t}=(1-t)Q_s+t\bar Q.
\]
Show that the squared distance to the target decreases as
\[
  \Wtwo^2(\mu_{s,t},\bar\mu)=(1-t)^2\Wtwo^2(\mu_s,\bar\mu).
\]
How could this be used to build a fairness dashboard?
\end{exercise}
\fi

\begin{exercise}[Exposure growth]\label{ex:15:11}
Suppose an insurer's exposure measure doubles between two dates, but the normalized portfolio composition remains unchanged. What does balanced OT between normalized laws report? What information is lost? Propose an exposure-weighted alternative.
\end{exercise}

\begin{exercise}[Mini-lab extension]\label{ex:15:12}
Modify the severity-stress R lab by replacing the target lognormal distribution with a Pareto-type tail. Compare the paths of the mean, \(\VaR_{0.99}\), \(\VaR_{0.995}\) and \(\TVaR_{0.99}\). Which quantity is most sensitive to tail thickening?
\end{exercise}

\newpage
\lecture{Unbalanced Transport, Exposure and Portfolio Growth}
\label{lec:unbalanced-ot-exposure-growth}

\begin{readingbox}
Balanced optimal transport compares probability measures with the same total mass. This is the right model when we compare two normalized loss distributions, two severity distributions, or two reserving scenario distributions after rescaling to total mass one. Many actuarial objects, however, are finite measures rather than probability measures. A portfolio grows or shrinks; exposures enter and leave; claims are created by frequency variation; insured values increase; climate change changes both the shape and the mass of loss distributions. In such cases, forcing all mass to be matched is artificial.

This lecture introduces unbalanced optimal transport. The key idea is to relax the marginal constraints of Kantorovich transport. Instead of requiring the first and second marginals of a plan to be exactly equal to the source and target measures, we penalize deviations from them. This allows mass to be transported, created or destroyed. The mathematical references are the unbalanced and entropic transport chapters in \cite{peyre2019computational}, together with the measure-theoretic and dynamic background in \cite{villani2003topics,villani2008oldnew,ambrosio2021lectures}.
For the modern theory of unbalanced, partial and Wasserstein--Fisher--Rao transport, see
\cite{figalli2010optimal,chizat2018unbalanced,liero2018optimal,kondratyev2016new,piccoli2014generalized}.
The computational generalized Sinkhorn viewpoint is closely related to \cite{frogner2015learning,chizat2018scaling}.
The dynamic source-term formulation should be read together with Lecture~\ref{lec:dynamic-ot-actuarial-trajectories},
while the numerical implementation below extends the entropic ideas of Lecture~\ref{lec:entropic-sinkhorn}.
\end{readingbox}

\subsection{Learning objectives}\label{subsec:uot-learning-objectives}

After this lecture, the reader should be able to:
\begin{enumerate}[label=(\roman*)]
\item distinguish probability measures from finite exposure measures;
\item explain why balanced OT can be misleading when total exposure changes;
\item formulate static unbalanced optimal transport with marginal penalties;
\item interpret the Kullback--Leibler marginal penalty for finite measures;
\item derive and implement generalized Sinkhorn iterations for entropic unbalanced OT;
\item separate shape displacement, transported mass, created mass and destroyed mass;
\item connect unbalanced OT to lapse, new business, portfolio growth, claim frequency and climate exposure;
\item understand the dynamic continuity equation with a source term;
\item use unbalanced transport as an actuarial diagnostic rather than only as a distance.
\end{enumerate}

\begin{actuarialbox}
Balanced OT answers: how should one unit of mass be rearranged into one unit of mass? Unbalanced OT answers: how much of the source mass should be transported, how much should disappear, and how much target mass should be newly created? This is often the actuarial question when portfolios, exposures or claim counts change.
\end{actuarialbox}

\begin{actuarialbox}%[Distributional, not causal]
An unbalanced plan is a distributional diagnostic. It can suggest which parts of a source measure are plausibly matched to a target measure and which parts are better read as disappearing or newly created mass. It does not identify a causal migration mechanism for a specific policy, exposure or claim without additional assumptions.
\end{actuarialbox}

\subsection{Why balanced transport is sometimes the wrong model}\label{subsec:uot-why-unbalanced}

Let $\mu$ and $\nu$ be probability measures. Balanced optimal transport solves
\[
\inf_{\gamma\in\Coup(\mu,\nu)} \int c(\bx,\by)\,\gamma(\dd \bx,\dd \by),
\]
where the two marginal constraints are
\[
(p_1)_\#\gamma=\mu,\qquad (p_2)_\#\gamma=\nu.
\]
This is appropriate when the total amount of mass is the same and must be conserved.

But many actuarial objects are not naturally normalized to mass one. For instance:
\begin{itemize}
\item a portfolio exposure measure may have total mass equal to the number of vehicle-years;
\item a territorial risk measure may have mass equal to total insured value in each region;
\item a claim measure may have mass equal to expected claim count rather than one;
\item a loss measure may have mass equal to total expected loss volume;
\item a reserving cash-flow measure may have mass equal to expected future payments;
\item a climate exposure measure may gain mass because more assets enter high-risk regions.
\end{itemize}

\begin{figure}
    \centering
    \includegraphics[width=0.95\linewidth]{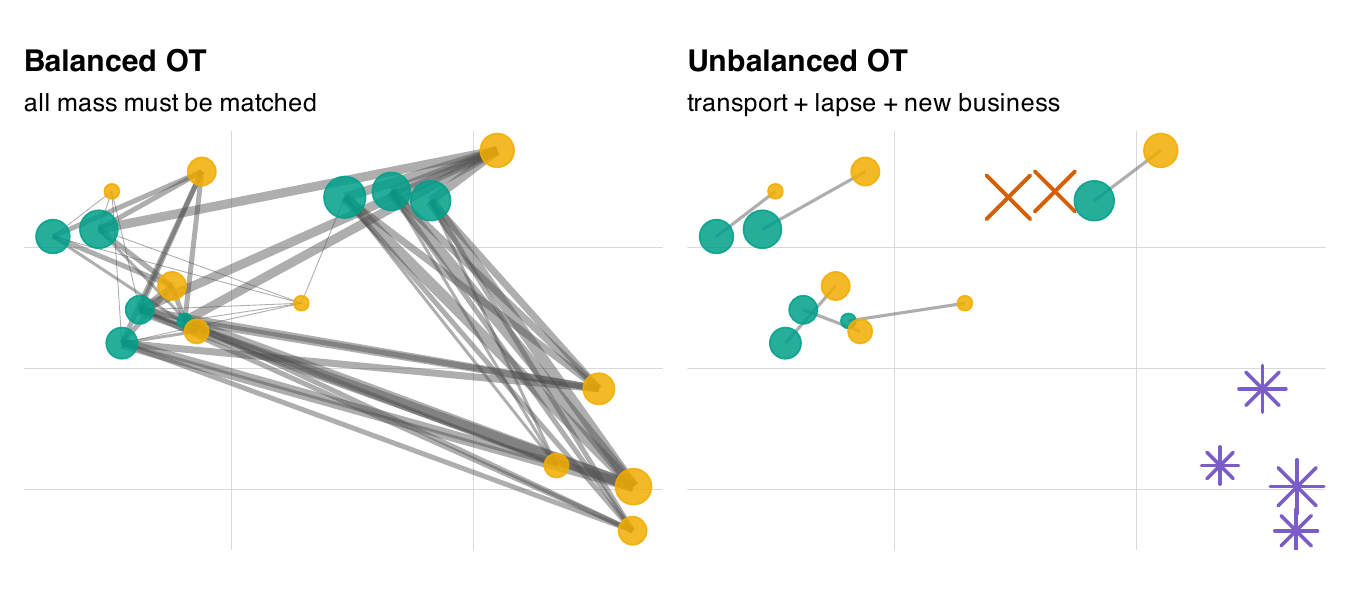}
    \caption{{\bf Balanced and unbalanced transport for exposure changes}. Balanced transport forces all source mass to be matched to target mass. Unbalanced transport separates three effects: transported exposure, disappearing exposure such as lapses, and newly created exposure such as new business.}
    \label{fig:balanced-unbalanced-exposure}
\end{figure}

If the total mass changes from $M_0$ to $M_1$, a balanced comparison has two common workarounds.
The first is to normalize both measures into probability measures and compare only their shapes.
The second is to compare total masses separately, for example through $|M_1-M_0|$. Both are useful,
but they miss the interaction between shape change and mass change. Figure~\ref{fig:balanced-unbalanced-exposure}
illustrates the difference between forcing all mass to move and allowing some mass to disappear or be created.

\begin{example}[A growing motor portfolio]
Suppose that in year 0 an insurer has $100{,}000$ vehicle-years distributed over territories, and
in year 1 it has $130{,}000$ vehicle-years. If we normalize both territorial exposure measures to
probability measures, we ignore the fact that the book grew by $30\%$. If we compare total exposure
only, we ignore where the growth occurred. Unbalanced OT combines both questions: mass can be
matched geographically where the same type of exposure persists, but additional mass must be
created in the territories where the portfolio grows.
\end{example}

\begin{example}[Frequency versus severity]
A severity distribution is often normalized: conditional on a claim occurring, the severity has
law $\mu$. A claim-cost measure may instead be
\[
\alpha = \lambda\,\mu,
\]
where $\lambda$ is expected claim frequency. If frequency changes from $\lambda_0$ to $\lambda_1$ and
severity changes from $\mu_0$ to $\mu_1$, then comparing only $\mu_0$ and $\mu_1$ ignores frequency.
Unbalanced OT can compare the finite measures $\alpha_0=\lambda_0\mu_0$ and
$\alpha_1=\lambda_1\mu_1$ directly.
\end{example}

\subsection{Finite measures as actuarial objects}\label{subsec:uot-finite-measures}

Let $\mathcal X$ be a Polish space. We denote by $\mathcal M_+(\mathcal X)$ the set of finite nonnegative Borel measures on $\mathcal X$. For $\alpha\in\mathcal M_+(\mathcal X)$, its total mass is
\[
|\alpha| = \alpha(\mathcal X).
\]
When $|\alpha|=1$, $\alpha$ is a probability measure. In unbalanced transport, we allow
$|\alpha|\ne |\beta|$.

\begin{definition}[Finite positive measure]\label{def:finite-positive-measure}
A finite positive measure on $\mathcal X$ is a nonnegative Borel measure $\alpha$ such that
$|\alpha|=\alpha(\mathcal X)<\infty$. Its normalized shape, when $|\alpha|>0$, is the probability
measure $\alpha/|\alpha|$. In this lecture, $|\alpha|$ carries actuarial information and should not
be silently normalized away.
\end{definition}

A finite exposure measure may be written as
\[
\alpha = \sum_{i=1}^n a_i\delta_{\bx_i},
\]
where $a_i\ge 0$ is the exposure volume in cell $i$ and $\bx_i$ describes the cell. The total mass
$\sum_i a_i$ is not a nuisance: it is part of the actuarial object.

\begin{definition}[Exposure measure]\label{def:exposure-measure}
An exposure measure on a state space $\mathcal X$ is a finite measure $\alpha\in\mathcal M_+(\mathcal X)$
whose mass $\alpha(A)$ represents the amount of exposure in states $A\subseteq\mathcal X$. The state
$\bx$ may encode territory, age, vehicle type, insured value, coverage, climate zone or a vector of
rating variables.
\end{definition}

\begin{definition}[Claim-intensity measure]\label{def:claim-intensity-measure}
Let $\mu$ be a probability distribution of claim amounts and let $\lambda>0$ be an expected number
of claims. The finite measure
\[
\alpha=\lambda\mu
\]
is called a claim-intensity measure. Its total mass is the expected claim count. Its normalized
shape is the conditional severity distribution.
\end{definition}

\begin{actuarialbox}
A probability law answers a conditional question: what does a randomly selected claim, policy or
scenario look like? A finite measure answers an unconditional portfolio question: how much mass
is there, and where is it located? This distinction will reappear in the reserving volume application in Section~\ref{subsec:uot-reserving-volume}.
\end{actuarialbox}

\subsection{Three ways to compare finite measures}\label{subsec:uot-three-comparisons}

Let $\alpha,\beta\in\mathcal M_+(\mathcal X)$ and write
\[
M_\alpha=|\alpha|,\qquad M_\beta=|\beta|.
\]
If both masses are positive, define normalized shapes
\[
\bar\alpha=\frac{\alpha}{M_\alpha},\qquad \bar\beta=\frac{\beta}{M_\beta}.
\]
There are at least three approaches.

\paragraph{Shape-only comparison.}
Compute $W_p(\bar\alpha,\bar\beta)$. This compares the distribution of mass but ignores the total
change $M_\beta-M_\alpha$.

\paragraph{Mass plus shape comparison.}
Report both $|M_\beta-M_\alpha|$ and $W_p(\bar\alpha,\bar\beta)$. This is transparent and often
sufficient for a first report, but it does not say which part of the mass change is spatially or
actuarially aligned with the shape change.

\paragraph{Unbalanced comparison.}
Optimize over partial transport plans and penalize deviations from the original marginals. This
lets the model decide how much mass should be transported, created or destroyed.

\begin{example}[A localized exposure increase]
Suppose that total exposure grows by $10\%$, all in flood-prone coastal cells. A mass-plus-shape
summary reports growth and distributional drift, but it does not identify how much of the year-1
coastal exposure corresponds to transported year-0 mass and how much should be treated as new
business. An unbalanced plan gives such a distributional attribution.
\end{example}

\subsection{Partial transport as a first unbalanced idea}\label{subsec:uot-partial-transport}

Before introducing marginal penalties, it is useful to consider partial transport. Suppose
$\alpha,\beta\in\mathcal M_+(\mathcal X)$ and choose a mass level $m\le |\alpha|\wedge |\beta|$. A partial
transport plan of mass $m$ is a measure $\gamma\in\mathcal M_+(\mathcal X\times\mathcal X)$ such that
\[
(p_1)_\#\gamma\le \alpha,
\qquad
(p_2)_\#\gamma\le \beta,
\qquad
|\gamma|=m.
\]
The partial transport problem is
\[
\inf_{\gamma}\left\{\int c(\bx,\by)\,\gamma(\dd \bx,\dd \by):
(p_1)_\#\gamma\le \alpha,\ (p_2)_\#\gamma\le \beta,
|\gamma|=m\right\}.
\]
This transports only $m$ units of mass. Unused source mass is left behind; unmatched target mass
is not explained by transport.

\begin{remark}[Partial transport versus unbalanced transport]\label{rem:partial-vs-unbalanced}
Partial transport fixes the transported mass $m$ externally. Unbalanced transport typically lets
the transported mass be chosen endogenously through a trade-off: transporting mass has a cost,
while destroying or creating mass has a penalty. Partial transport is useful when the analyst has
an externally meaningful retained mass, whereas KL-penalized unbalanced OT is useful when the
retained mass itself is part of the diagnostic.
\end{remark}

\subsection{Relative entropy for finite measures}\label{subsec:uot-finite-kl}

A standard way to penalize marginal deviations is through a divergence between finite measures.
The most common choice in computational unbalanced OT is the finite-measure Kullback--Leibler
divergence.

\begin{definition}[Finite-measure KL divergence]\label{def:finite-measure-kl}
Let $\rho,\alpha\in\mathcal M_+(\mathcal X)$. If $\rho\ll\alpha$ and $f=\dd\rho/\dd\alpha$, define
\[
\KL(\rho\,|\,\alpha)
=\int_{\mathcal X}\bigl(f\log f-f+1\bigr)\,\alpha(\dd \bx).
\]
If $\rho$ is not absolutely continuous with respect to $\alpha$, set
\[
\KL(\rho\,|\,\alpha)=+\infty.
\]
\end{definition}

For probability measures, this coincides with the usual relative entropy, because both total masses
are equal to one. For finite measures, the terms $-f+1$ are essential: they make
\[\KL(\alpha\,|\,\alpha)=0\]
and allow different total masses.

If $\rho=q\delta_{\bx}$ and $\alpha=m\delta_{\bx}$, then
\[
\KL(q\delta_{\bx}\,|\,m\delta_{\bx})
= q\log\left(\frac qm\right)-q+m,
\]
with the convention $0\log 0=0$.

\begin{proposition}[Basic properties of finite KL]\label{prop:finite-kl-basic-properties}
Let $\rho,\alpha\in\mathcal M_+(\mathcal X)$. Then
\[
\KL(\rho\,|\,\alpha)\ge 0,
\]
with equality if and only if $\rho=\alpha$.
\end{proposition}

\begin{proof}
This is the standard convexity argument for relative entropy; see the entropy regularization
and unbalanced transport discussion in \cite{peyre2019computational}. Sketch: the function
$h(t)=t\log t-t+1$ is convex and satisfies $h(t)\ge 0$, with equality only at $t=1$. Therefore
\[\int h(f)\,\dd\alpha\ge 0.\]
Equality implies $f=1$ $\alpha$-almost surely, hence $\rho=\alpha$.
\end{proof}

\begin{remark}[The cost of deleting all mass]
If $\rho=0$, then $f=0$ and
\[
\KL(0\,|\,\alpha)=\alpha(\mathcal X)=|\alpha|.
\]
Thus deleting one unit of source mass costs one unit of KL penalty before multiplication by the
penalty parameter. This gives a simple interpretation of the penalty scale.
\end{remark}

\subsection{Static unbalanced optimal transport}\label{subsec:uot-static}

Let $\alpha\in\mathcal M_+(\mathcal X)$ and $\beta\in\mathcal M_+(\mathcal Y)$. Let $c:\mathcal X\times\mathcal Y\to[0,\infty]$ be a lower semicontinuous cost. A transport plan is now any finite measure
$\gamma\in\mathcal M_+(\mathcal X\times\mathcal Y)$, not necessarily with prescribed marginals.
Its marginals are
\[
\gamma_1=(p_1)_\#\gamma,
\qquad
\gamma_2=(p_2)_\#\gamma.
\]
Unbalanced OT penalizes the fact that $\gamma_1$ may differ from $\alpha$ and $\gamma_2$ may differ
from $\beta$.

\begin{definition}[KL-penalized unbalanced OT]\label{def:kl-penalized-uot}
For parameters $\tau_1,\tau_2>0$, define
\[
\mathrm{UOT}_{c,\tau_1,\tau_2}(\alpha,\beta)
=\inf_{\gamma\in\mathcal M_+(\mathcal X\times\mathcal Y)}
\left\{
\int c(\bx,\by)\,\gamma(\dd \bx,\dd \by)
+\tau_1\KL(\gamma_1\,|\,\alpha)
+\tau_2\KL(\gamma_2\,|\,\beta)
\right\}.
\]
\end{definition}

\begin{proposition}[Existence, standard KL-penalized form]\label{prop:static-uot-existence}
Assume that $\mathcal X$ and $\mathcal Y$ are Polish spaces, $c$ is lower semicontinuous and
nonnegative, and $\alpha,\beta$ are finite positive measures. Under the usual tightness or coercivity
conditions ensuring that transported mass cannot escape to infinity at finite cost, the problem in
Definition~\ref{def:kl-penalized-uot} admits an optimal finite plan.
\end{proposition}

\begin{remark}[Unbalanced value versus unbalanced metric]\label{rem:uot-value-not-metric}
The KL-penalized value $\mathrm{UOT}_{c,\tau_1,\tau_2}$ is an optimization criterion, not automatically
a metric. It may be asymmetric when $\tau_1\ne\tau_2$, and even in symmetric cases it does not have
the same status as $W_p$. Metric unbalanced geometries require additional structure, as in
Hellinger--Kantorovich or Wasserstein--Fisher--Rao distances. This distinction is useful when the
quantity is used as a governance diagnostic rather than as a mathematical distance.
\end{remark}

The parameters $\tau_1$ and $\tau_2$ tune the cost of violating the source and target marginals.
Large $\tau$ makes marginal deviations expensive; small $\tau$ allows more deletion and creation.

\begin{actuarialbox}
The plan $\gamma$ represents mass that is explained by transport. The residual measure
$\alpha-\gamma_1$, when meaningful, represents source exposure that disappears, lapses or is not
matched. The residual $\beta-\gamma_2$ represents new target exposure or newly created mass. With
KL penalties, residuals need not be simple positive differences everywhere, but the same intuition
remains: the marginals of $\gamma$ indicate the mass actually transported.
\end{actuarialbox}

\begin{definition}[Transported, unused and unexplained mass]\label{def:transported-unused-unexplained-mass}
For an unbalanced plan $\gamma$, the transported mass is $|\gamma|$. Its source usage and target
explanation marginals are $\gamma_1=(p_1)_\#\gamma$ and $\gamma_2=(p_2)_\#\gamma$. When
$\gamma_1\le\alpha$ and $\gamma_2\le\beta$, the unused source mass and unexplained target mass are
$\alpha-\gamma_1$ and $\beta-\gamma_2$. For KL penalties, these signed residual interpretations may
fail locally, but $\gamma_1$ and $\gamma_2$ still identify where the plan uses source mass and explains
target mass.
\end{definition}

\begin{proposition}[Balanced OT as a limiting case]\label{prop:uot-balanced-limit}
Assume $|\alpha|=|\beta|$ and suppose the balanced problem has finite value. As
$\tau_1,\tau_2\to\infty$, unbalanced transport increasingly enforces
\[
\gamma_1=\alpha,
\qquad
\gamma_2=\beta,
\]
so that the limiting problem is the balanced Kantorovich problem.
\end{proposition}

\begin{proof}
This is the usual penalty-method interpretation of constrained optimization, as developed in
unbalanced OT references such as \cite{peyre2019computational}. Sketch: if a sequence of plans
has objective values bounded independently of $\tau_1,\tau_2$, then the terms
$\KL(\gamma_1\,|\,\alpha)$ and $\KL(\gamma_2\,|\,\beta)$ must tend to zero as the penalties diverge.
By the non-negativity and definiteness of KL, this forces $\gamma_1\to\alpha$ and
$\gamma_2\to\beta$ in the appropriate weak sense, under tightness assumptions. The remaining term
is then the balanced transport cost.
\end{proof}

\begin{remark}[Why two penalties?]
There is no reason the cost of deleting old mass should equal the cost of creating new mass.
In portfolio management, lapse and acquisition may have different economic meanings. In climate
applications, disappearing exposure may be an adaptation success, while newly created exposure in
a high-risk zone may be a concern. Therefore $\tau_1$ and $\tau_2$ need not be equal.
\end{remark}

\subsection{A one-point calculation}\label{subsec:uot-one-point}

The simplest example explains the trade-off.

\begin{proposition}[One-point KL--UOT solution]\label{prop:one-point-kl-uot}
Let
\[
\alpha=m\delta_{\bx},
\qquad
\beta=n\delta_{\by},
\]
with $m,n>0$ and cost $c(\bx,\by)=c_0\ge 0$. Any plan is of the form
\[
\gamma=q\delta_{(\bx,\by)},\qquad q\ge 0.
\]
For the KL-penalized unbalanced objective of Definition~\ref{def:kl-penalized-uot}, the unique
minimizer is
\[
q^*=
\left(m^{\tau_1}n^{\tau_2}\exp(-c_0)\right)^{1/(\tau_1+\tau_2)}
=m^{\tau_1/(\tau_1+\tau_2)}n^{\tau_2/(\tau_1+\tau_2)}
\exp\left(-\frac{c_0}{\tau_1+\tau_2}\right).
\]
\end{proposition}

\begin{proof}
The objective is
\[
J(q)=q c_0+\tau_1\left\{q\log\frac qm-q+m\right\}
+\tau_2\left\{q\log\frac qn-q+n\right\}.
\]
For $q>0$, the first-order condition is
\[
c_0+\tau_1\log\frac qm+\tau_2\log\frac qn=0.
\]
Solving gives the displayed expression. Strict convexity in $q$ gives uniqueness.
\end{proof}

\begin{remark}[Interpretation]
If the transport cost $c_0$ is small, the transported mass is close to a weighted geometric average
of $m$ and $n$. If $c_0$ is large, less mass is transported: it is cheaper to delete source mass and
create target mass than to force a costly match. This is exactly the actuarial trade-off in a
portfolio where some new exposure is not plausibly a migration of old exposure.
\end{remark}

\subsection{Discrete unbalanced transport}\label{subsec:uot-discrete}

Let
\[
\alpha=\sum_{i=1}^n a_i\delta_{\bx_i},
\qquad
\beta=\sum_{j=1}^m b_j\delta_{\by_j},
\]
where $a_i,b_j\ge 0$, but the sums need not be one and need not be equal. Write $\ba=(a_1,\ldots,a_n)$ and $\bb=(b_1,\ldots,b_m)$, and let $\bC_{ij}=c(\bx_i,\by_j)$. A finite plan is a nonnegative matrix $\Gamma=(\gamma_{ij})\in\R_+^{n\times m}$. Its row and column
masses are
\[
r_i(\Gamma)=\sum_{j=1}^m\gamma_{ij},
\qquad
s_j(\Gamma)=\sum_{i=1}^n\gamma_{ij}.
\]
We write \(\br(\Gamma)=(r_1(\Gamma),\ldots,r_n(\Gamma))\) and \(\bs(\Gamma)=(s_1(\Gamma),\ldots,s_m(\Gamma))\).
\begin{definition}[Discrete KL-penalized unbalanced OT]\label{def:discrete-kl-uot}
The discrete KL-penalized problem is
\[
\min_{\Gamma\ge 0}
\left\{
\sum_{i,j}\bC_{ij}\gamma_{ij}
+\tau_1\sum_i\left(r_i\log\frac{r_i}{a_i}-r_i+a_i\right)
+\tau_2\sum_j\left(s_j\log\frac{s_j}{b_j}-s_j+b_j\right)
\right\},
\]
with the usual conventions when $r_i=0$ or $s_j=0$. If $a_i=0$, then finite KL forces $r_i=0$;
if $b_j=0$, it forces $s_j=0$.
\end{definition}

\begin{proposition}[Existence in the finite case]\label{prop:finite-discrete-uot-existence}
If all entries of $\bC$ are finite and nonnegative and $\ba,\bb$ are finite nonnegative vectors, the discrete unbalanced problem admits a minimizer.
\end{proposition}

\begin{proof}
This is a finite-dimensional convex optimization argument. Sketch: the objective is lower semicontinuous and convex on $\R_+^{n\times m}$. Because the KL terms grow superlinearly in the row and column masses, level sets are bounded as soon as the zero-mass rows and columns forced by finite KL are removed. A minimizing sequence therefore has a convergent subsequence, and lower semicontinuity gives a minimizer.
\end{proof}

\begin{remark}[Strict convexity and numerical stability]\label{rem:uot-strict-convexity}
Without entropic regularization, the KL marginal penalties regularize the marginal masses but do
not necessarily make the whole matrix $\Gamma$ strictly identifiable when several routes have the
same cost and the same row or column masses. The entropic term in
Section~\ref{subsec:uot-entropic-sinkhorn} removes this degeneracy and gives a smoother plan.
\end{remark}

\begin{remark}[No equality constraints]\label{rem:uot-no-equality-constraints}
The balanced transportation polytope is replaced by the full nonnegative orthant. The marginal
constraints are no longer hard constraints; they enter through the penalty. This is one reason why
unbalanced OT is flexible in actuarial applications with exposure changes.
\end{remark}

\subsection{Entropic unbalanced OT and generalized Sinkhorn}\label{subsec:uot-entropic-sinkhorn}

For computation, one often adds an entropy penalty to the plan itself.

\begin{definition}[Entropic unbalanced OT]\label{def:entropic-unbalanced-ot}
For $\varepsilon>0$, the entropically regularized discrete unbalanced problem is
\[
\min_{\Gamma\ge 0}\left\{
\langle \bC,\Gamma\rangle
+\varepsilon\sum_{i,j}\gamma_{ij}(\log\gamma_{ij}-1)
+\tau_1\KL(\Gamma\mathbf 1_m\,|\,\ba)
+\tau_2\KL(\Gamma^\top\mathbf 1_n\,|\,\bb)
\right\},
\]
where the KL divergences between vectors are understood componentwise.
\end{definition}

Set
\[
\bK_{ij}=\exp\left(-\frac{\bC_{ij}}{\varepsilon}\right).
\]
The minimizer has a scaling form
\[
\Gamma=\operatorname{diag}(\bu)\bK\operatorname{diag}(\bv)
\]
with nonnegative scaling vectors \(\bu\) and \(\bv\), positive on active rows and columns. Unlike balanced Sinkhorn, the updates are damped by
exponents.

\begin{theorem}[Generalized Sinkhorn updates]\label{thm:generalized-sinkhorn-uot}
For entropic unbalanced OT with KL marginal penalties, define
\[
\theta_1=\frac{\tau_1}{\tau_1+\varepsilon},
\qquad
\theta_2=\frac{\tau_2}{\tau_2+\varepsilon}.
\]
A standard generalized Sinkhorn iteration is
\[
\bu\leftarrow \left(\frac{\ba}{\bK\bv}\right)^{\theta_1},
\qquad
\bv\leftarrow \left(\frac{\bb}{\bK^\top \bu}\right)^{\theta_2},
\]
where powers and divisions are componentwise. The resulting plan is
\[
\Gamma=\operatorname{diag}(\bu)\bK\operatorname{diag}(\bv).
\]
Rows or columns with zero prescribed mass should be removed before the iteration, or handled with the corresponding scaling factor fixed at zero.
\end{theorem}

\begin{proof}
This is the first-order condition for the entropic unbalanced problem; see the computational
presentation of unbalanced Sinkhorn in \cite{peyre2019computational}. Sketch: differentiating the
regularized objective yields equations in the row sums, column sums and entries of $\Gamma$.
Writing $\Gamma=\operatorname{diag}(\bu)\bK\operatorname{diag}(\bv)$ reduces these equations to two nonlinear scaling relations. Solving each relation alternately gives the displayed updates. When
$\tau_1,\tau_2\to\infty$, the exponents tend to one and the balanced Sinkhorn updates are recovered.
\end{proof}

\begin{remark}[Log-domain implementation]\label{rem:uot-log-domain}
For small $\varepsilon$ or large costs, the kernel $\bK_{ij}=\exp(-\bC_{ij}/\varepsilon)$ may underflow.
The same warning as in Lecture~\ref{lec:entropic-sinkhorn} applies: production code should use
log-domain updates or stabilized scaling rather than dense exponentiation of $-\bC/\varepsilon$.
\end{remark}

\begin{computebox}
The parameter $\varepsilon$ smooths the coupling. The parameters $\tau_1$ and $\tau_2$ soften the
marginal constraints. These are different roles. A small $\varepsilon$ with small $\tau$ gives a sharp
but highly unbalanced matching. A large $\varepsilon$ with large $\tau$ gives a smooth approximation
to balanced OT.
\end{computebox}

\subsection{Interpreting the parameters}\label{subsec:uot-parameters}

The three main parameters in entropic unbalanced OT are:
\begin{itemize}
\item $\varepsilon$: smoothing of the transport plan;
\item $\tau_1$: penalty for not using source mass;
\item $\tau_2$: penalty for not explaining target mass.
\end{itemize}

The ratios
\[
\theta_k=\frac{\tau_k}{\tau_k+\varepsilon}
\]
are informative. If $\tau_k\gg\varepsilon$, then $\theta_k\approx 1$, and the corresponding marginal
is nearly enforced. If $\tau_k\ll\varepsilon$, then $\theta_k\approx 0$, and the corresponding
marginal is weakly enforced.

\begin{example}[Acquisition versus lapse]
In a portfolio-growth comparison from year 0 to year 1, $\tau_1$ may represent the cost of declaring
old exposure unmatched, while $\tau_2$ represents the cost of declaring year-1 exposure new. If
new-business acquisition is expected and cheap to explain, one may choose a smaller $\tau_2$. If
unexpected lapse is more concerning, one may choose a larger $\tau_1$ to force source mass to be
matched whenever plausible.
\end{example}

\begin{example}[Climate exposure]
If exposure in a coastal region grows because of construction, a low target penalty $\tau_2$ allows
the model to say: this is genuinely new exposure. If $\tau_2$ is high, the model is forced to match
that new exposure to some old exposure, possibly producing a misleading migration story.
\end{example}

\subsection{Diagnostics from an unbalanced plan}\label{subsec:uot-diagnostics}

Given a plan $\gamma$, recall from Definition~\ref{def:transported-unused-unexplained-mass} that the transported mass is
\[
|\gamma|=\gamma(\mathcal X\times\mathcal Y).
\]
The source marginal $\gamma_1$ indicates which part of $\alpha$ is used. The target marginal
$\gamma_2$ indicates which part of $\beta$ is explained by transported mass.

In a discrete setting, one may report:
\begin{align*}
\text{source mass} &= \sum_i a_i,\\
\text{target mass} &= \sum_j b_j,\\
\text{transported mass} &= \sum_{i,j}\gamma_{ij},\\
\text{average transport cost} &= \frac{\sum_{i,j}\bC_{ij}\gamma_{ij}}{\sum_{i,j}\gamma_{ij}},\\
\text{source usage ratio} &= \frac{\sum_{i,j}\gamma_{ij}}{\sum_i a_i},\\
\text{target explanation ratio} &= \frac{\sum_{i,j}\gamma_{ij}}{\sum_j b_j}.
\end{align*}

When the KL penalty is used, the difference $\ba-\br(\Gamma)$ need not be nonnegative componentwise,
because a row marginal may locally exceed $a_i$ if that is beneficial. Still, the ratios above remain
useful summaries of how much mass is explained through transport.

\begin{actuarialbox}
For governance, the transport value alone is rarely enough. A useful unbalanced OT report should include at least: total source mass, total target mass, transported mass, transport cost,
source usage, target explanation, largest disappearing cells and largest newly created cells.
\end{actuarialbox}

\subsection{Balanced normalization versus unbalanced comparison}\label{subsec:uot-normalization-vs-unbalanced}

Suppose $\alpha=M\mu$ and $\beta=N\nu$ with $M,N>0$ and $\mu,\nu\in\Prob(\mathcal X)$. A normalized
comparison computes $W_p(\mu,\nu)$. This is invariant to multiplying $M$ or $N$. Therefore it cannot
detect growth.

Unbalanced comparison is not invariant to total mass. This is a feature, not a bug, when total
exposure is meaningful. But it also means that one must be careful when interpreting values
across portfolios of different sizes.

\begin{remark}[Scale matters]
If all exposure weights are multiplied by $10$, the unbalanced objective typically scales by $10$.
Thus, for comparison across books, it is often useful to report both the total value and a per-unit
value, such as value divided by average exposure mass. The denominator should be stated explicitly:
source mass, target mass, average mass $(M+N)/2$ and insured value lead to different
per-unit interpretations.
\end{remark}

\begin{example}[Two portfolios with the same shape]
Let $\alpha=M\mu$ and $\beta=N\mu$ have the same shape but different total masses. Balanced OT
between normalized shapes gives zero. Unbalanced OT gives a positive value unless $M=N$, because
mass must be created or destroyed. This is appropriate if the question is portfolio growth, but not
if the question is only geographic composition.
\end{example}

\subsection{Unbalanced dynamic transport}\label{subsec:uot-dynamic}

The dynamic formulation of balanced $W_2$ uses the continuity equation
\[
\partial_t\rho_t+\nabla\cdot(\rho_t v_t)=0.
\]
This expresses conservation of mass along a trajectory. For unbalanced transport, we add a
source term:
\[
\partial_t\rho_t+\nabla\cdot(\rho_t v_t)=\zeta_t\rho_t.
\]
Here $v_t$ moves mass, while $\zeta_t$ creates or destroys mass locally. Positive $\zeta_t$ means
growth; negative $\zeta_t$ means decay.

\begin{definition}[Unbalanced dynamic trajectory]\label{def:unbalanced-dynamic-trajectory}
An unbalanced dynamic trajectory between finite measures $\alpha$ and $\beta$ is a curve of finite
measures $(\rho_t)_{0\le t\le 1}$, a velocity field $v_t$ and a growth-rate field $\zeta_t$ satisfying
\[
\partial_t\rho_t+\nabla\cdot(\rho_t v_t)=\zeta_t\rho_t,
\qquad \rho_0=\alpha,\quad \rho_1=\beta.
\]
\end{definition}

A common dynamic action has the form
\[
\int_0^1\int_{\mathcal X}
\left(\|v_t(\bx)\|^2+\kappa^2\zeta_t(\bx)^2\right)\rho_t(\dd \bx)\,\dd t,
\]
subject to the source continuity equation and endpoint constraints
\[
\rho_0=\alpha,
\qquad
\rho_1=\beta.
\]
The parameter $\kappa>0$ controls the cost of mass variation relative to spatial motion. This is
one route to Hellinger--Kantorovich or Wasserstein--Fisher--Rao type geometries.

\begin{remark}[Dynamic meaning]
Balanced dynamic OT says: the portfolio mass moves through state space. Unbalanced dynamic OT
says: the portfolio mass moves, grows and shrinks. This is much closer to actual insurance
portfolios, where new business, lapse, renewal, cancellation and exposure inflation occur at the
same time as changes in the covariate distribution.
\end{remark}

\begin{actuarialbox}
For actuarial trajectories, the source term $\zeta_t$ is the key new object. It can represent lapse,
new business, migration into coverage, migration out of coverage, claim frequency growth, insured
value growth, or exposure reduction due to prevention.
\end{actuarialbox}

\subsection{Application: portfolio growth and new business}\label{subsec:uot-portfolio-growth}

Let $\bX_t$ denote policyholder characteristics at time $t$, and let $E_t$ be exposure volume. A
probability law $\mu_t=\Law(\bX_t)$ describes the composition of the portfolio conditional on being
in the book. A finite measure
\[
\alpha_t=E_t\mu_t
\]
describes both composition and size.

Balanced OT between $\mu_0$ and $\mu_1$ measures composition drift. Unbalanced OT between
$\alpha_0$ and $\alpha_1$ measures exposure drift. The latter can distinguish:
\begin{itemize}
\item stable exposure transported from old to similar new states;
\item lapsed exposure not matched to the target;
\item new exposure not explained by the source;
\item exposure that moved to different risk states.
\end{itemize}

\begin{example}[Motor insurer expanding into a new territory]
Suppose a motor insurer enters a new province. The normalized portfolio distribution changes,
but a balanced transport map may misleadingly match old policies from existing provinces to new
policies in the new province. Unbalanced OT can instead declare part of the target mass as newly
created, which is the correct business interpretation.
\end{example}

\subsection{Application: lapse and retention}\label{subsec:uot-lapse-retention}

Consider a renewal portfolio. Let $\alpha_0$ be the exposure measure at the start of the year and
$\alpha_1$ the exposure measure among policies renewed at the next year, possibly including new
business. An unbalanced plan can be used to quantify retention by region of state space.

For each source cell $i$, define the transported row mass
\[
r_i=\sum_j\gamma_{ij}.
\]
The ratio
\[
\ell_i=1-\frac{r_i}{a_i}
\]
can be read as an unbalanced-transport lapse indicator, when $r_i\le a_i$. With KL penalties,
local ratios may need smoothing or truncation, but the diagnostic remains meaningful.

\begin{example}[Lapse after a premium increase]
Let $\bx_i$ include old premium, age, territory and claim history. If high-premium cells have low
row usage ratios $r_i/a_i$, the unbalanced plan suggests that these cells are not well matched to
the renewal portfolio. This can be interpreted as a distributional lapse signal.
\end{example}

\subsection{Application: exposure growth and climate risk}\label{subsec:uot-climate-exposure}

Climate risk is not only a change in hazard intensity. It is also a change in exposure: more assets
may be built in high-risk zones, property values may increase, and adaptation may remove or
protect some exposure.

Let $\bx$ encode location and vulnerability. Let
\[
\alpha_0(\dd \bx)=\text{insured value at baseline in state }\bx,
\]
and
\[
\alpha_1(\dd \bx)=\text{insured value under a future or stressed scenario}.
\]
Unbalanced OT separates three mechanisms:
\begin{enumerate}[label=(\roman*)]
\item existing exposure moves in risk space, for example because vulnerability or hazard changes;
\item exposure disappears, for example because of withdrawal, relocation or protection;
\item exposure appears, for example because of new construction or portfolio growth.
\end{enumerate}

\begin{actuarialbox}
A prevention policy can reduce loss by moving mass to safer states, by shrinking vulnerable mass,
or both. Balanced OT can describe the first mechanism. Unbalanced OT is needed for the second.
\end{actuarialbox}

\subsection{Application: claim frequency and severity jointly}\label{subsec:uot-frequency-severity}

Suppose losses are generated by a frequency-severity model. Let $\lambda_t$ be expected claim
frequency at time $t$, and let $\mu_t$ be the conditional severity distribution. The finite measure
\[
\alpha_t=\lambda_t\mu_t
\]
encodes the unconditional claim-intensity distribution.

If $\lambda_1>\lambda_0$ but $\mu_1=\mu_0$, then the severity shape is unchanged but the expected
number of claims has increased. If $\lambda_1=\lambda_0$ but $\mu_1$ has a heavier tail, then
frequency is unchanged but severity shape has changed. If both change, unbalanced OT compares
both effects in one object.

\begin{example}[Frequency growth without severity inflation]
Let $\alpha_0=10\mu$ and $\alpha_1=12\mu$. Balanced OT between normalized measures gives zero.
Unbalanced OT gives a positive creation cost corresponding to two additional expected claims. This
is the correct conclusion if the actuarial question concerns aggregate expected claim volume.
\end{example}

\iffalse
% To be restored in v2. This section relies on the paper on fairness with several sensitive variables and group-mass effects.
\subsection{Application: fairness with unequal group masses}\label{subsec:uot-fairness-unequal-masses}

In many fairness audits, one compares conditional score distributions
\[
\mu_s=\Law(Z\mid S=s),
\]
which are probability measures. This is appropriate when the goal is to compare shapes across
groups independently of group size.

But sometimes group size matters. Let
\[
\alpha_s = n_s\mu_s,
\]
where $n_s$ is the number of policies in group $s$. A finite-measure comparison incorporates both
the score distribution and the group exposure. This can be relevant when a fairness intervention
changes the composition of the insured population, or when accessibility differs by group.

\begin{remark}[Conditional fairness versus portfolio fairness]
Conditional fairness asks: are the score distributions equal after conditioning on group membership?
Portfolio fairness asks: how are total premium, total exposure and total burden distributed across
groups? Balanced OT is natural for the first; unbalanced OT can be useful for the second. This should be
read together with the conditional distributional parity framework in
Lecture~\ref{lec:fairness-wasserstein-projections}.
\end{remark}

\fi

\subsection{Application: reserving cash-flow volume}\label{subsec:uot-reserving-volume}

A predictive reserving model may produce a probability law of future cash-flow vectors, but it may
also produce a finite payment measure across development years. For instance,
\[
\alpha=\sum_{t=1}^T m_t\delta_t,
\]
where $m_t$ is expected payment volume in future year $t$. Comparing two reserving models through
balanced OT after normalization compares the timing shape of payments. Comparing finite measures
through unbalanced OT also accounts for differences in total reserve volume. This complements the
cash-flow distribution comparison in Lecture~\ref{lec:reserving-climate-prevention}.

\begin{example}[Same timing, different best estimate]
Two reserving models may have the same payment pattern but different total best-estimate reserve.
Balanced OT between normalized payment patterns is zero. Unbalanced OT is positive and reflects
the difference in total future payment mass.
\end{example}

\subsection{Cost engineering in unbalanced OT}\label{subsec:uot-cost-engineering}

The ground cost $c(\bx,\by)$ and the marginal penalties $\tau_1,\tau_2$ must be chosen together.
They answer different questions:
\begin{itemize}
\item $c(\bx,\by)$: how expensive is it to interpret target state $\by$ as a transformation of source state $\bx$?
\item $\tau_1$: how expensive is it to leave source mass unexplained?
\item $\tau_2$: how expensive is it to declare target mass newly created?
\end{itemize}

A useful calibration principle is to compare the cost of transporting one unit from $\bx$ to $\by$ with
the cost of deleting it and recreating it. In the one-point calculation, the relevant comparison is
between $c_0$ and $\tau_1+\tau_2$. If
\[
c(\bx,\by)\gg \tau_1+\tau_2,
\]
then matching $\bx$ to $\by$ is strongly discouraged; the model prefers destruction and creation. If
\[
c(\bx,\by)\ll \tau_1+\tau_2,
\]
then transport is encouraged.

\begin{actuarialbox}
The penalty parameters define the boundary between ``migration'' and ``new/lost exposure''. They
are therefore governance parameters, not merely numerical tuning constants.
\end{actuarialbox}

\subsection{Mini-lab: generalized Sinkhorn in R}\label{subsec:uot-mini-lab-generalized-sinkhorn}

The following code implements a simple generalized Sinkhorn algorithm for unbalanced entropic
transport. It is deliberately written in transparent form.

\begin{lstlisting}[language=R]
unbalanced_sinkhorn <- function(a, b, C, epsilon = 1, tau1 = 10, tau2 = 10,
                                n_iter = 1000, tol = 1e-10) {
  a <- as.numeric(a)
  b <- as.numeric(b)
  C <- as.matrix(C)

  n <- length(a)
  m <- length(b)

  if (epsilon <= 0 || tau1 <= 0 || tau2 <= 0) stop("epsilon, tau1 and tau2 must be positive")
  if (any(a < 0) || any(b < 0)) stop("a and b must be nonnegative")

  K <- exp(-C / epsilon)
  u <- rep(1, n)
  v <- rep(1, m)

  theta1 <- tau1 / (tau1 + epsilon)
  theta2 <- tau2 / (tau2 + epsilon)

  tiny <- 1e-300

  for (it in seq_len(n_iter)) {
    u_old <- u
    v_old <- v

    Kv <- as.numeric(K %*% v)
    u <- (a / pmax(Kv, tiny))^theta1

    Ktu <- as.numeric(t(K) %*% u)
    v <- (b / pmax(Ktu, tiny))^theta2

    err <- max(abs(log(pmax(u, tiny)) - log(pmax(u_old, tiny))),
               abs(log(pmax(v, tiny)) - log(pmax(v_old, tiny))))
    if (err < tol) break
  }

  Gamma <- diag(u, nrow = n) %*% K %*% diag(v, nrow = m)
  row_mass <- rowSums(Gamma)
  col_mass <- colSums(Gamma)

  list(plan = Gamma,
       row_mass = row_mass,
       col_mass = col_mass,
       transported_mass = sum(Gamma),
       transport_cost = sum(C * Gamma),
       iterations = it,
       theta = c(theta1, theta2))
}

# Example: exposure by territory-risk score
x <- c(0, 1, 2, 3, 4)
y <- c(0, 1, 2, 3, 4)

# Year 0 exposure: concentrated in low-risk cells
# Year 1 exposure: larger total mass, shifted toward higher-risk cells
a <- c(40, 35, 20, 5, 0)
b <- c(30, 30, 25, 20, 15)

C <- outer(x, y, function(u, v) (u - v)^2)

fit <- unbalanced_sinkhorn(a, b, C, epsilon = 0.5, tau1 = 5, tau2 = 5)
fit$transported_mass
fit$transport_cost
round(fit$plan, 2)
round(fit$row_mass, 2)
round(fit$col_mass, 2)

# Diagnostics
source_usage <- fit$row_mass / pmax(a, 1e-12)
target_explained <- fit$col_mass / pmax(b, 1e-12)

data.frame(cell = x,
           source_exposure = a,
           used_source_mass = round(fit$row_mass, 2),
           source_usage = round(source_usage, 3))

data.frame(cell = y,
           target_exposure = b,
           explained_target_mass = round(fit$col_mass, 2),
           target_explained = round(target_explained, 3))
\end{lstlisting}

The target has more total mass than the source. Balanced OT cannot compare $a$ and $b$ directly
without normalization. The unbalanced plan transports a substantial part of the old exposure and
creates additional target mass where needed. In a full report, the diagnostics in
Section~\ref{subsec:uot-diagnostics} should be displayed together with the plan.

\subsection{Mini-lab: sensitivity to the marginal penalty}\label{subsec:uot-mini-lab-penalty-sensitivity}

The next code varies $\tau=\tau_1=\tau_2$. Small values of $\tau$ allow more mass creation and
destruction. Large values force more mass to be transported.

\begin{lstlisting}[language=R]
taus <- c(0.5, 1, 2, 5, 10, 50)
summary_tau <- lapply(taus, function(tt) {
  fit <- unbalanced_sinkhorn(a, b, C, epsilon = 0.5, tau1 = tt, tau2 = tt)
  data.frame(tau = tt,
             transported_mass = fit$transported_mass,
             transport_cost = fit$transport_cost,
             source_mass = sum(a),
             target_mass = sum(b),
             source_usage = fit$transported_mass / sum(a),
             target_explained = fit$transported_mass / sum(b))
})
summary_tau <- do.call(rbind, summary_tau)
summary_tau

plot(summary_tau$tau, summary_tau$transported_mass, type = "b", log = "x",
     xlab = "marginal penalty tau",
     ylab = "transported mass")
abline(h = sum(a), lty = 2)
abline(h = sum(b), lty = 3)
\end{lstlisting}

The plot typically shows transported mass increasing with $\tau$. When marginal deviations are
cheap, the algorithm avoids costly long-distance moves. When marginal deviations are expensive,
it transports more mass even if the ground cost is higher.

\subsection{Practical workflow}\label{subsec:uot-workflow}

A practical unbalanced OT workflow for actuarial portfolios can be organized as follows.

\paragraph{Step 1: choose the mass.}
Decide what the measure represents: policy count, vehicle-years, insured value, expected claim
count, expected loss amount or reserve cash-flow volume.

\paragraph{Step 2: choose the state space.}
Decide what the locations $\bx$ represent: geography, rating-cell centroid, severity,
claim-count intensity, vulnerability index, cash-flow time, or multivariate covariates.

\paragraph{Step 3: choose the ground cost.}
Encode what counts as a plausible migration. Moving from one adjacent risk class may be cheap;
moving across territories or across product types may be expensive.

\paragraph{Step 4: choose marginal penalties.}
Calibrate $\tau_1$ and $\tau_2$ to the cost of declaring mass disappeared or newly created. Use
sensitivity analysis rather than a single arbitrary value.

\paragraph{Step 5: compute the plan.}
Use generalized Sinkhorn for large problems, exact convex optimization for small problems, or
problem-specific approximations for one-dimensional or gridded settings.

\paragraph{Step 6: report diagnostics.}
Report transported mass, cost per transported unit, source usage, target explanation, disappearing
cells and newly created cells.

\paragraph{Step 7: interpret actuarially.}
Translate mass creation/destruction into business concepts: lapse, new business, exposure growth,
portfolio withdrawal, frequency change, reserve strengthening, prevention or adaptation.

\subsection{Limitations and cautions}\label{subsec:uot-limitations}

Unbalanced OT is useful, but it can be misused.

\paragraph{Penalty calibration is normative.}
The choice of $\tau_1$ and $\tau_2$ determines whether a difference is interpreted as migration or
creation/destruction. This is not purely statistical.

\paragraph{Mass has units.}
A measure of policy count, a measure of insured value and a measure of expected loss are different
objects. Their unbalanced OT values are not directly comparable unless units are made explicit.

\paragraph{KL is asymmetric.}
The penalty $\KL(\gamma_1\,|\,\alpha)$ is not symmetric. The two penalties have directional meanings:
source usage and target explanation. This is often useful, but it should be documented.

\paragraph{Entropic smoothing can hide sharp changes.}
A large $\varepsilon$ spreads mass over many matches. This may improve numerical stability, but it
can blur localized migration, lapse or new-business signals.

\paragraph{Unbalanced transport is not causality.}
A plan can suggest a plausible matching between exposures in two periods, but it does not prove
that a specific policy migrated causally from one state to another. As in the counterfactual chapter,
causal interpretation requires additional design or structural assumptions.

\subsection{Summary}\label{subsec:uot-summary}

Balanced OT is a geometry for probability measures with conserved mass. Unbalanced OT extends
this geometry to finite measures, allowing mass to be transported, created or destroyed. The most
common static formulation penalizes deviations from the source and target marginals using
finite-measure KL divergences. Entropic regularization leads to generalized Sinkhorn iterations,
which are scalable and easy to implement.

For actuarial science, unbalanced OT is especially natural for exposure analysis. It can compare
portfolios with different sizes, distinguish composition drift from growth, analyze lapse and new
business, compare claim-intensity measures, quantify climate exposure growth, and compare
reserving payment volumes. The output should not be reduced to a single number: transported
mass, source usage and target explanation are often as important as the transport cost itself.

\subsection{Exercises}\label{subsec:uot-exercises}

\begin{exercise}[Finite-measure KL]\label{ex:16:1}
Let $\rho=q\delta_{\bx}$ and $\alpha=m\delta_{\bx}$ with $q,m\ge 0$ and $m>0$. Derive
\[
\KL(\rho\,|\,\alpha)=q\log(q/m)-q+m.
\]
Show that this is nonnegative and minimized at $q=m$.
\end{exercise}

\begin{exercise}[One-point unbalanced OT]\label{ex:16:2}
Let $\alpha=m\delta_{\bx}$, $\beta=n\delta_{\by}$ and $c(\bx,\by)=c_0$. Derive the minimizer $q^*$ of the
one-point problem. Study the limits $c_0\to 0$, $c_0\to\infty$, $\tau_1\to\infty$ and
$\tau_2\to\infty$.
\end{exercise}

\begin{exercise}[Balanced limit]\label{ex:16:3}
Assume $|\alpha|=|\beta|$ and consider KL-penalized UOT with $\tau_1=\tau_2=\tau$. Give a formal
argument showing that, as $\tau\to\infty$, any sequence of nearly optimal plans must have marginals
close to $\alpha$ and $\beta$.
\end{exercise}

\begin{exercise}[Normalization can hide growth]\label{ex:16:4}
Let $\alpha=100\delta_0$ and $\beta=120\delta_0$. What is the Wasserstein distance between the
normalized probability measures? What does unbalanced OT detect that normalization removes?
\end{exercise}

\begin{exercise}[Two-cell exposure]\label{ex:16:5}
Let
\[
\alpha=50\delta_0+50\delta_1,
\qquad
\beta=30\delta_0+90\delta_1.
\]
Use squared distance cost on $\{0,1\}$. Compare: (i) normalized balanced OT, (ii) total mass change,
(iii) a qualitative unbalanced interpretation. Which part of the change is shape drift and which
part is mass growth?
\end{exercise}

\begin{exercise}[Generalized Sinkhorn]\label{ex:16:6}
Starting from the entropic unbalanced objective, derive the scaling equations for
$\Gamma=\operatorname{diag}(\bu)\bK\operatorname{diag}(\bv)$. Show why the exponents are
$\tau_k/(\tau_k+\varepsilon)$.
\end{exercise}

\begin{exercise}[Penalty sensitivity]\label{ex:16:7}
Run the mini-lab with $\tau\in\{0.1,0.5,1,5,20,100\}$. Plot transported mass and transport cost as
functions of $\tau$. Explain the actuarial meaning of the curve.
\end{exercise}

\begin{exercise}[Asymmetric penalties]\label{ex:16:8}
Modify the mini-lab with $\tau_1\ne\tau_2$. Interpret the cases $\tau_1\gg\tau_2$ and
$\tau_2\gg\tau_1$ in terms of lapse versus new business.
\end{exercise}

\begin{exercise}[Claim-intensity measures]\label{ex:16:9}
Let $\mu_0=\mu_1=\mathrm{Exp}(1)$, $\lambda_0=10$ and $\lambda_1=15$. Explain why balanced OT
between $\mu_0$ and $\mu_1$ is zero. What is the actuarial meaning of comparing
$\lambda_0\mu_0$ and $\lambda_1\mu_1$ by unbalanced OT?
\end{exercise}

\begin{exercise}[Climate exposure]\label{ex:16:10}
Construct a three-zone example with baseline insured values and future insured values. Choose a
cost matrix based on distances between zones or risk scores. Compute an unbalanced plan and
identify which future exposure is best interpreted as transported exposure and which is best
interpreted as new exposure.
\end{exercise}

\iffalse
% To be restored in v2. This exercise belongs with the commented fairness material.
\begin{exercise}[Fairness with group masses]\label{ex:16:11}
Let $Z$ be a premium score and $S\in\{0,1\}$ a group variable. Explain the difference between
comparing $\Law(Z\mid S=0)$ and $\Law(Z\mid S=1)$ by balanced OT, and comparing
$n_0\Law(Z\mid S=0)$ and $n_1\Law(Z\mid S=1)$ by unbalanced OT. Which question is each comparison
answering?
\end{exercise}

\fi

\begin{exercise}[Dynamic source term]\label{ex:16:12}
Consider the one-dimensional equation
\[
\partial_t\rho_t+\partial_x(\rho_t v_t)=\zeta_t\rho_t.
\]
Assume $v_t=0$ and $\zeta_t=\zeta$ is constant. Solve for the total mass $|\rho_t|$ as a function of
$t$. Interpret the result as portfolio growth or shrinkage.
\end{exercise}

\newpage
\lecture{Life, Mortality and Longevity Transport}
\label{lec:life-mortality-longevity}

\begin{readingbox}
Life insurance mathematics starts from the distribution of a future lifetime. A life table, a mortality projection, an annuity portfolio or a longevity stress is therefore already a probability measure, or a family of probability measures indexed by age, calendar time and cohort. This lecture uses optimal transport to compare and stress such laws. Classical references on life contingencies include \cite{bowers1997actuarial,gerber1990life,dickson2009actuarial,devylder1997life}. For multi-state life insurance and long-term care models, see \cite{haberman1999actuarial,pitacco1999multistate,andersen1993statistical}. For stochastic mortality, longevity dynamics and annuity applications, see \cite{pitacco2009longevity,lee1992modeling,brouhns2002poisson,cairns2006pricing,renshaw2006cohort,plat2009stochastic,cairns2009quantitative}. The optimal-transport tools used below are those introduced in Lectures~\ref{lec:wasserstein-barycenters}, \ref{lec:risk-measures-robustness}, \ref{lec:dynamic-ot-actuarial-trajectories} and~\ref{lec:unbalanced-ot-exposure-growth}.
\end{readingbox}

\subsection{Learning objectives}
\label{subsec:life-learning-objectives}

After this lecture, the reader should be able to:
\begin{enumerate}[label=(\roman*)]
\item represent a life table as a probability measure on age at death or residual lifetime;
\item compute Wasserstein distances between residual lifetime laws;
\item interpret quantile transport as a rank-preserving longevity stress;
\item write annuity and death-benefit present values as push-forwards of lifetime laws;
\item use Wasserstein barycenters to combine mortality scenarios;
\item compare dynamic mortality trajectories and cohort-specific laws;
\item formulate longevity capital and prudent pricing with Wasserstein ambiguity sets;
\item connect tail-aware transport with old-age mortality and extreme longevity;
\item use multi-state and joint-life models as transport problems on richer state spaces;
\item understand how quantile sketches can support distributed mortality or longevity audits.
\end{enumerate}

\begin{actuarialbox}
In life insurance, the basic distributional object is often not a loss severity law but a lifetime law. Optimal transport compares how lifetime mass moves across ages, cohorts, health states, mortality scenarios or life-contingent present values.
\end{actuarialbox}

\begin{actuarialbox}%[Distributional, not causal]
A transport map between two lifetime laws is a distributional alignment. It is not, by itself, a statement about how long a particular person would have lived under another mortality regime. Individual counterfactual claims require a structural model or an intervention design in addition to the coupling.
\end{actuarialbox}

\subsection{Future lifetime as a probability measure}
\label{subsec:future-lifetime-measure}

Let \(T_x\) denote the future lifetime of an individual aged \(x\). In continuous time, \(T_x\) is a non-negative random variable with survival function
\[
  {}_tp_x=\Pbb(T_x>t),\qquad t\ge 0.
\]
Its distribution is a probability measure on \(\R_+\), denoted
\[
  \mu_x=\Law(T_x).
\]
In discrete time, let \(K_x\) be the curtate future lifetime,
\[
  K_x=\lfloor T_x\rfloor.
\]
Then \(K_x\) is a probability measure on \(\N\). If \(q_{x+k}\) is the one-year death probability at age \(x+k\), then, for an open-ended table,
\[
  \Pbb(K_x=k)={}_kp_x q_{x+k},\qquad k=0,1,2,\ldots,
\]
where
\[
  {}_kp_x=\prod_{j=0}^{k-1}(1-q_{x+j}),
  \qquad {}_0p_x=1.
\]
For a table closed at terminal age \(\omega\), one usually sets \(q_\omega=1\), or equivalently puts the remaining survival mass at \(K_x=\omega-x\). Thus a life table induces a collection of probability measures
\[
  x\longmapsto \mu_x.
\]

\begin{definition}[Life table as a lifetime measure]
\label{def:life-table-lifetime-measure}
Fix an initial age \(x\) and a terminal age \(\omega\). A discrete life table
\(\{q_y:x\le y\le \omega\}\), closed at \(\omega\), induces the probability measure
\[
  \mu_x^{\mathrm{curt}}
  =\sum_{k=0}^{\omega-x} p_{x,k}\,\delta_k,
\]
where
\[
  p_{x,k}={}_kp_x q_{x+k},\quad 0\le k<\omega-x,
  \qquad
  p_{x,\omega-x}={}_{\omega-x}p_x.
\]
The latter convention places at the terminal curtate age all mass surviving to \(\omega\).
\end{definition}

\begin{definition}[Residual lifetime law]
\label{def:residual-lifetime-law}
For a life aged \(x\), the residual lifetime law is
\[
  \mu_x=\Law(T_x)\in\Probp{1}(\R_+)
\]
when the mean residual lifetime is finite. In discrete time, the curtate residual lifetime law is
\[
  \mu_x^{\mathrm{curt}}=\Law(K_x)\in\mathcal P(\N).
\]
When \(p\)-th moments are finite, we view \(\mu_x\) as an element of \(\Probp{p}(\R_+)\).
\end{definition}

This viewpoint separates two questions. The actuarial table gives death probabilities. Optimal
transport compares the resulting lifetime distributions. The same distinction already appeared
for severity distributions in Lecture~\ref{lec:wasserstein-barycenters}; here the state space is
measured in years rather than in monetary units.

\begin{example}[Two mortality tables]
\label{ex:two-mortality-tables-w1}
Let \(\mu_x^0\) and \(\mu_x^1\) be residual lifetime laws at age \(x\) under two mortality
tables. The quantity
\[
  \Wone(\mu_x^0,\mu_x^1)
\]
measures the average number of residual years by which lifetime mass must be moved to
transform one table into the other. It is expressed in years and is therefore directly
interpretable.
\end{example}

\subsection{Transport between age-at-death distributions}
\label{subsec:age-at-death-transport}

Instead of residual lifetime \(T_x\), one may work with age at death
\[
  A_x=x+T_x.
\]
If \(\alpha_x=\Law(A_x)\), then \(\alpha_x\) is the push-forward of \(\mu_x\) by the translation map
\[
  \tau_x:t\longmapsto x+t.
\]
Thus
\[
  \alpha_x=(\tau_x)_\#\mu_x.
\]
For two populations or two mortality scenarios, \(a\in\{0,1\}\), write
\[
  \alpha_x^a=\Law(A_x^a),\qquad \mu_x^a=\Law(T_x^a).
\]

\begin{proposition}[Residual lifetime and age-at-death distances]
\label{prop:age-at-death-translation-invariance}
Let \(\mu_x^0,\mu_x^1\in\Probp{p}(\R_+)\), and let
\(\alpha_x^a=(\tau_x)_\#\mu_x^a\), where \(\tau_x(t)=x+t\). Then
\[
  \Wp{p}(\alpha_x^0,\alpha_x^1)=\Wp{p}(\mu_x^0,\mu_x^1).
\]
\end{proposition}

\begin{proof}
Translations by the same vector are isometries for the ground distance on \(\R\). If
\(\gamma\in\Coup(\mu_x^0,\mu_x^1)\), then \((\tau_x,\tau_x)_\#\gamma\in\Coup(\alpha_x^0,\alpha_x^1)\) and
\(|(x+s)-(x+t)|=|s-t|\). Taking infima in both directions gives the result.
\end{proof}

The choice between age at death and residual lifetime is therefore mostly interpretive. For a
single starting age, both contain the same geometry. For comparisons across several starting
ages, residual lifetimes are often easier to compare because all measures live on the same support
\(\R_+\).

In one dimension, the optimal transport map is the quantile map. We use the lower generalized inverse convention throughout this lecture:
\[
  Q_x^a(u)=Q_{\mu_x^a}(u)=F_{\mu_x^a}^{-1}(u),\qquad u\in(0,1).
\]
Then
\[
  \Wp{p}^p(\mu_x^0,\mu_x^1)
  =
  \int_0^1 |Q_x^0(u)-Q_x^1(u)|^p\,\dd u.
\]
The corresponding rank-preserving lifetime stress is
\[
  \mathsf T_x(t)=Q_x^1\left(F_{\mu_x^0}(t)\right),
\]
when the source law is continuous. With atoms, \(F_{\mu_x^0}(T_x)\) is not uniform; one should use the lower inverse together with a deterministic tie convention, or randomized ranks inside atoms. A life whose baseline lifetime lies at percentile \(u\) is moved to the target lifetime at the same percentile.

\begin{actuarialbox}
A quantile longevity stress preserves mortality rank. It does not say that an individual
would causally live exactly \(\mathsf T_x(t)\) years. It says that the baseline lifetime distribution is
transformed into the stressed lifetime distribution by matching percentiles. This is the same
interpretive caution as in Lectures~\ref{lec:couplings} and~\ref{lec:counterfactual-worlds-transport}.
\end{actuarialbox}

\subsection{Longevity stresses by quantile maps}
\label{subsec:longevity-quantile-stress}

A longevity stress often shifts mass toward older ages. Let \(\mu_x^0\) be a baseline lifetime law and \(\mu_x^1\) a stressed law with lower mortality. In one dimension, the monotone transport map
\[
  \mathsf T_x(t)=Q_x^1(F_{\mu_x^0}(t))
\]
gives a transparent scenario map.

\begin{definition}[Longevity quantile displacement]
\label{def:longevity-quantile-displacement}
For two residual lifetime laws \(\mu_x^0\) and \(\mu_x^1\), the quantile displacement curve is
\[
  \Delta_x(u)=Q_x^1(u)-Q_x^0(u),\qquad u\in(0,1).
\]
Positive values of \(\Delta_x(u)\) mean that the stressed scenario increases residual lifetime at
rank \(u\).
\end{definition}

If \(\mathsf T_x(t)>t\) for most \(t\), the stress increases residual lifetimes. If the increase is
concentrated in the upper tail, then the stress mainly affects annuity tail risk and old-age
capital. If it affects the body of the distribution, it changes expected annuity costs more broadly.
The curve \(\Delta_x\) is therefore often more informative than a single scalar distance.

\begin{proposition}[Expected lifetime stability]
\label{prop:expected-lifetime-stability}
Let \(\mu\) and \(\nu\) be two residual lifetime laws with finite first moments. Then
\[
  \left|\int t\,\dd\mu(t)-\int t\,\dd\nu(t)\right|\le \Wone(\mu,\nu).
\]
\end{proposition}

\begin{proof}
The function \(t\mapsto t\) is 1-Lipschitz on \(\R_+\). The Kantorovich--Rubinstein dual
for \(\Wone\) gives the inequality.
\end{proof}

\begin{remark}[Mean lifetime is not enough]
\label{rem:mean-lifetime-not-enough}
Two mortality stresses may have the same effect on expected lifetime but different effects on
old-age survival probabilities. Wasserstein distances and quantile displacement curves show
where the lifetime distribution moves.
\end{remark}

\subsection{Annuities as push-forwards of lifetime laws}
\label{subsec:annuities-pushforward-lifetime}

Let \(v=(1+i)^{-1}\) be a discount factor. In discrete time, the present value of a life annuity-due
paying one unit at the beginning of each year while alive is
\[
  g_{\mathrm{ann}}(k)=\sum_{j=0}^{k} v^j=\frac{1-v^{k+1}}{1-v},
  \qquad k\in\N.
\]
Thus the random present value is a measurable function of the curtate future lifetime:
\[
  A_x=g_{\mathrm{ann}}(K_x).
\]
Its law is the push-forward
\[
  \Law(A_x)=(g_{\mathrm{ann}})_\#\mu_x^{\mathrm{curt}}.
\]
Similarly, the present value of a death benefit of one paid at the end of the year of death is
\[
  B_x=g_{\mathrm{death}}(K_x)=v^{K_x+1},
\]
so
\[
  \Law(B_x)=(g_{\mathrm{death}})_\#\mu_x^{\mathrm{curt}}.
\]

\begin{definition}[Life-contingent present-value law]
\label{def:life-contingent-pv-law}
Let \(\mu\) be a lifetime law and let \(g\) be a measurable present-value map. The associated
life-contingent present-value law is
\[
  \lambda_g=g_\#\mu.
\]
For a single-life annuity, \(g=g_{\mathrm{ann}}\); for a death benefit, \(g=g_{\mathrm{death}}\).
\end{definition}

\begin{actuarialbox}
Life-contingent present values are push-forwards of lifetime laws. Mortality transport becomes
financial transport once the lifetime is mapped through annuity or insurance present-value
functions.
\end{actuarialbox}

This gives a direct stability bound. The annuity present-value map satisfies, for \(k,\ell\in\N\),
\[
  |g_{\mathrm{ann}}(k)-g_{\mathrm{ann}}(\ell)|
  \le |k-\ell|.
\]
Indeed, the difference is a sum of at most \(|k-\ell|\) terms, each bounded by one. Therefore
\[
  \Wone\left((g_{\mathrm{ann}})_\#\mu,(g_{\mathrm{ann}})_\#\nu\right)
  \le \Wone(\mu,\nu).
\]

\begin{proposition}[Push-forward stability of life-contingent values]
\label{prop:life-contingent-pushforward-stability}
Let \(g:\R_+\to\R\) be \(L_g\)-Lipschitz and let \(\mu,\nu\in\Probp{p}(\R_+)\). Then
\[
  \Wp{p}(g_\#\mu,g_\#\nu)\le L_g\,\Wp{p}(\mu,\nu).
\]
In particular, the discrete annuity-due map \(g_{\mathrm{ann}}\) is 1-Lipschitz on \(\N\) with the
usual distance.
\end{proposition}

\begin{proof}
If \(\gamma\in\Coup(\mu,\nu)\), then \((g,g)_\#\gamma\in\Coup(g_\#\mu,g_\#\nu)\). Hence
\[
  \int |a-b|^p\,\dd (g,g)_\#\gamma(a,b)
  =\int |g(s)-g(t)|^p\,\dd\gamma(s,t)
  \le L_g^p\int |s-t|^p\,\dd\gamma(s,t).
\]
Taking infima over \(\gamma\) gives the result.
\end{proof}

For death benefits, the map \(k\mapsto v^{k+1}\) is decreasing. Longer lifetimes decrease the
present value of death benefits but increase annuity present values. Thus the same longevity
stress has opposite financial effects for life annuities and death insurance.

\begin{proposition}[Opposite monotonicity of annuity and death-benefit values]
\label{prop:annuity-death-benefit-opposite-monotonicity}
Assume \(K^1\) first-order stochastically dominates \(K^0\). Then
\[
  \Ebb[g_{\mathrm{ann}}(K^1)]\ge \Ebb[g_{\mathrm{ann}}(K^0)],
  \qquad
  \Ebb[g_{\mathrm{death}}(K^1)]\le \Ebb[g_{\mathrm{death}}(K^0)].
\]
\end{proposition}

\begin{proof}
The map \(g_{\mathrm{ann}}\) is increasing and \(g_{\mathrm{death}}\) is decreasing. The claim follows from the
standard characterization of first-order stochastic dominance by expectations of increasing
functions, applied to \(g_{\mathrm{ann}}\) and to \(-g_{\mathrm{death}}\).
\end{proof}

\begin{example}[Annuity versus death benefit]
\label{ex:annuity-versus-death-benefit}
A stress that transports lifetime mass to older ages increases the distribution of annuity
present values but decreases the distribution of immediate death-benefit present values. The
same \(\Wp{p}(\mu_x^0,\mu_x^1)\) therefore has different business meanings depending on the
contract map applied after the lifetime law.
\end{example}

\subsection{Barycenters of mortality scenarios}
\label{subsec:mortality-scenario-barycenters}

Suppose \(M\) mortality models produce residual lifetime laws
\[
  \mu_x^{(1)},\ldots,\mu_x^{(M)}
\]
for a fixed age \(x\). A Wasserstein barycenter is a consensus lifetime distribution
\[
  \bar\mu_x\in
  \operatorname*{argmin}_{\mu\in\Probp{p}(\R_+)}
  \sum_{m=1}^M \omega_m \Wp{p}^p(\mu,\mu_x^{(m)}),
\]
where \(\omega_m\ge 0\) and \(\sum_m\omega_m=1\). This is the same Fr\'echet-mean construction
as in Lecture~\ref{lec:wasserstein-barycenters}, now applied to mortality scenarios.

\begin{definition}[Mortality-scenario barycenter]
\label{def:mortality-scenario-barycenter}
A mortality-scenario barycenter at age \(x\) is any minimizer
\[
  \bar\mu_x\in
  \operatorname*{argmin}_{\mu\in\Probp{p}(\R_+)}
  \sum_{m=1}^M \omega_m \Wp{p}^p(\mu,\mu_x^{(m)}).
\]
It represents a central residual lifetime law in Wasserstein geometry, not a mixture of mortality
models.
\end{definition}

\begin{proposition}[One-dimensional \(\Wtwo\)-barycenter of lifetime laws]
\label{prop:one-dimensional-lifetime-barycenter}
Assume \(\mu_x^{(1)},\ldots,\mu_x^{(M)}\in\Probp{2}(\R_+)\). In one dimension, a \(\Wtwo\)-barycenter
has quantile function
\[
  Q_{\bar\mu_x}(u)=\sum_{m=1}^M \omega_m Q_x^{(m)}(u),
  \qquad u\in(0,1).
\]
\end{proposition}

\begin{proof}
The one-dimensional \(\Wtwo\) objective is
\[
  \sum_{m=1}^M\omega_m\int_0^1 |Q(u)-Q_x^{(m)}(u)|^2\,\dd u.
\]
For each \(u\), the pointwise minimizer over real values is the weighted average
\(\sum_m\omega_mQ_x^{(m)}(u)\), which is nondecreasing because it is a weighted average of
nondecreasing functions. It is therefore a quantile function.
\end{proof}

\begin{remark}[Barycenter versus mixture]
\label{rem:mortality-barycenter-versus-mixture}
A mixture mortality law first selects a model and then samples a lifetime. A Wasserstein
barycenter averages quantiles. The two operations are different. Mixtures preserve model
heterogeneity; barycenters produce a central scenario in the geometry of lifetimes.
\end{remark}

For a portfolio with ages \(x_1,\ldots,x_n\), one can compute age-specific barycenters
\(\bar\mu_{x_i}\). Alternatively, one may first form the portfolio liability law under each
mortality model and then take a barycenter of the resulting liability distributions. These two
operations need not commute because aggregation and Wasserstein barycentering are nonlinear.

\subsection{Mortality surfaces and dynamic trajectories}
\label{subsec:mortality-surfaces-dynamic-trajectories}

Mortality modelling often describes a surface
\[
  q_{x,t},
\]
where \(x\) is age and \(t\) is calendar time. A stochastic mortality model produces random
future surfaces or paths
\[
  (q_{x,t+s})_{x,s\ge 0}.
\]
Each future path induces a law of lifetimes for each starting age and calendar date. Thus a
mortality model is not only a table but a distribution over future lifetime laws.

Lee--Carter type models describe log mortality rates by a low-dimensional time factor,
whereas Cairns--Blake--Dowd type models focus on older ages and annuity-relevant dynamics.
Cohort extensions add dependence along birth cohorts. From the transport viewpoint, all these
models define trajectories in a space of probability measures.

Let \(\mu_{x,t}\) denote the residual lifetime law of a life aged \(x\) at calendar time \(t\).
A dynamic mortality comparison between two scenarios \(a\in\{0,1\}\) may use
\[
  \sum_{s=0}^T w_s \Wp{p}^p\left(\mu_{x,t+s}^0,\mu_{x,t+s}^1\right).
\]
This compares the projected residual lifetime law date by date. A stronger comparison works
on whole paths of mortality rates, survival curves or annuity present values. This is the same
static-versus-dynamic distinction as in Lecture~\ref{lec:dynamic-ot-actuarial-trajectories}.

\begin{definition}[Mortality trajectory in Wasserstein space]
\label{def:mortality-wasserstein-trajectory}
For a fixed starting age \(x\), a mortality trajectory is a curve
\[
  t\longmapsto \mu_{x,t}\in\Probp{p}(\R_+),
\]
where \(\mu_{x,t}\) is the residual lifetime law at calendar date \(t\). Its discrete path length over
\(t_0,\ldots,t_T\) is
\[
  \sum_{s=0}^{T-1}\Wp{p}(\mu_{x,t_s},\mu_{x,t_{s+1}}).
\]
\end{definition}

\begin{actuarialbox}
Longevity risk is dynamic. A single distance between two tables may miss the timing of
mortality improvements. Dynamic OT compares the path of lifetime laws, not only the endpoint.
\end{actuarialbox}

\subsection{Cohorts and transport along diagonals}
\label{subsec:cohort-transport-diagonals}

A cohort is identified by year of birth \(b=t-x\). Mortality for a cohort evolves along the
diagonal
\[
  (x+s,t+s),\qquad s=0,1,2,\ldots.
\]
Thus the residual lifetime law at age \(x\) and time \(t\) depends on the future mortality
rates along this diagonal.

Let \(\mu_{x,t}^{b}\) denote a cohort-specific lifetime law. Comparing two cohorts \(b_0\) and
\(b_1\) may be expressed as
\[
  \Wp{p}\left(\mu_{x,t}^{b_0},\mu_{x,t}^{b_1}\right),
\]
or, for several starting ages,
\[
  \sum_x \omega_x \Wp{p}^p\left(\mu_{x,t}^{b_0},\mu_{x,t}^{b_1}\right).
\]
This gives a distributional measure of cohort improvement. Quantile displacement curves can
show whether differences occur around median lifetime or in the old-age tail.

\begin{remark}[Cohort effects are not only mean shifts]
\label{rem:cohort-effects-not-only-mean-shifts}
A cohort effect may increase survival mostly at advanced ages, or it may shift the whole
lifetime distribution. Wasserstein comparisons reveal the shape of the cohort displacement.
\end{remark}

\subsection{Annuity portfolios and liability laws}
\label{subsec:annuity-portfolios-liability-laws}

Consider a portfolio of annuitants \(i=1,\ldots,n\), with ages \(x_i\), annual payments \(b_i\) and future curtate lifetimes \(K_{x_i}\). Write \(\bb=(b_1,\ldots,b_n)\) and \(\bK=(K_{x_1},\ldots,K_{x_n})\). Conditional on a mortality scenario \(\Theta\), the portfolio present value is
\[
  L(\Theta)=\sum_{i=1}^n b_i\sum_{k=0}^{K_{x_i}} v^k.
\]
The law of \(L\) depends on both idiosyncratic lifetime risk and systematic longevity risk.
For large annuity portfolios, idiosyncratic risk diversifies, but systematic longevity risk
remains.

\begin{definition}[Annuity portfolio liability law]
\label{def:annuity-portfolio-liability-law}
Let \(\Gamma\) be a joint law of lifetimes \(\bK=(K_{x_1},\ldots,K_{x_n})\). The portfolio present-value map is
\[
  G(\bk)=\sum_{i=1}^n b_i\sum_{j=0}^{k_i}v^j,
  \qquad \bk=(k_1,\ldots,k_n).
\]
The associated liability law is
\[
  \lambda=G_\#\Gamma=\Law\{G(\bK)\}.
\]
\end{definition}

A mortality model therefore induces a liability law \(\lambda=\Law(L)\). Comparing two mortality
models can be done either before or after aggregation:
\[
  \Wp{p}(\mu_{x}^{0},\mu_{x}^{1})
  \quad\text{for each age }x,
\]
or
\[
  \Wp{p}(\lambda^0,
  \lambda^1)
  \quad\text{for the portfolio liability law.}
\]
Both are useful. Age-specific distances diagnose where mortality changes. Liability distances
measure financial impact.

\begin{proposition}[Lipschitz stability of annuity portfolio liabilities]
\label{prop:annuity-portfolio-lipschitz-stability}
Equip \(\N^n\) with the weighted \(\ell^1\) distance
\[
  d_{\bb}(\bk,\bl)=\sum_{i=1}^n b_i|k_i-\ell_i|.
\]
The annuity portfolio map \(G\) in Definition~\ref{def:annuity-portfolio-liability-law} is 1-Lipschitz from \((\N^n,d_{\bb})\) to \(\R\). Consequently, for any two joint lifetime laws \(\Gamma^0,\Gamma^1\),
\[
  \Wone(G_\#\Gamma^0,G_\#\Gamma^1)
  \le \Wone^{d_{\bb}}(\Gamma^0,\Gamma^1),
\]
where the right-hand side is \(\Wone\) computed with ground distance \(d_{\bb}\).
\end{proposition}

\begin{proof}
For each life \(i\), the map \(k_i\mapsto b_i\sum_{j=0}^{k_i}v^j\) is \(b_i\)-Lipschitz on
\(\N\). Summing over \(i\) gives the stated Lipschitz bound. The Wasserstein inequality is then
Proposition~\ref{prop:life-contingent-pushforward-stability} with \(p=1\).
\end{proof}

\begin{example}[Portfolio-level distance]
\label{ex:portfolio-level-longevity-distance}
Two mortality projections may differ substantially at young ages but have little effect on an
old annuity portfolio. Conversely, small differences after age 90 may matter greatly for a
portfolio concentrated at retirement ages. Liability push-forwards translate demographic
transport into financial transport.
\end{example}

\subsection{Prudent pricing and longevity capital}
\label{subsec:prudent-pricing-longevity-capital}

Let \(\widehat\mu_x\) be a baseline estimated residual lifetime law. A Wasserstein ambiguity
set is
\[
  \mathcal B_p(\widehat\mu_x,r)
  =
  \{\nu\in\Probp{p}(\R_+): \Wp{p}(\nu,\widehat\mu_x)\le r\}.
\]
A prudent annuity price may be written as
\[
  \Premium_r^{\mathrm{ann}}
  =
  \sup_{\nu\in\mathcal B_p(\widehat\mu_x,r)}
  \Ebb_\nu[g_{\mathrm{ann}}(T)].
\]
Similarly, a prudent capital diagnostic may use
\[
  \mathrm{Cap}_r
  =
  \sup_{\nu\in\mathcal B_p(\widehat\mu_x,r)}
  \Risk\left((g_{\mathrm{ann}})_\#\nu\right),
\]
where \(\Risk\) is a risk measure such as \(\VaR\), \(\TVaR\) or a spectral risk measure. This is
the longevity version of the Wasserstein-DRO construction in Lecture~\ref{lec:risk-measures-robustness}.

\begin{proposition}[Simple robust annuity-price bound]
\label{prop:robust-annuity-price-bound}
Assume \(g:\R_+\to\R\) is \(L_g\)-Lipschitz and \(\widehat\mu\in\Probp{1}(\R_+)\). Then
\[
  \sup_{\nu:\,\Wone(\nu,\widehat\mu)\le r}\Ebb_\nu[g(T)]
  \le
  \Ebb_{\widehat\mu}[g(T)]+L_g r.
\]
If \(g(t)=t\), the bound is sharp whenever the ambiguity set allows translation by \(r\) to the
right.
\end{proposition}

\begin{proof}
For every feasible \(\nu\), the Kantorovich--Rubinstein dual implies
\[
  \Ebb_\nu[g(T)]-\Ebb_{\widehat\mu}[g(T)]\le L_g\Wone(\nu,\widehat\mu)\le L_g r.
\]
Taking the supremum over \(\nu\) gives the result.
\end{proof}

\begin{actuarialbox}
A Wasserstein radius in longevity modelling represents uncertainty about the lifetime
law. It can be calibrated from sampling error, model disagreement, historical backtesting,
expert stress or regulatory prudence. The radius is a modelling decision and should be
reported.
\end{actuarialbox}

For a death benefit, the prudent direction may be opposite. A mortality improvement increases
annuity values but may reduce death-benefit present values. A robust framework should
therefore be contract-specific.

\subsection{Old-age mortality and tail-aware transport}
\label{subsec:old-age-tail-aware-transport}

For annuities, old-age survival matters disproportionately. A global Wasserstein distance
between lifetime laws may hide differences after age 95 or 100. Tail-aware comparisons are
therefore useful; they parallel the tail-aware severity tools of Lecture~\ref{lec:risk-measures-robustness}.

\begin{definition}[Tail-aware lifetime Wasserstein discrepancy]
\label{def:tail-aware-lifetime-wasserstein}
For \(\alpha\in(0,1)\), define the upper-tail lifetime discrepancy
\[
  \Wp{p,\alpha}^{\mathrm{tail}}(\mu,\nu)^p
  =
  \frac{1}{1-\alpha}
  \int_\alpha^1 |Q_\mu(u)-Q_\nu(u)|^p\,\dd u.
\]
It is the \(L^p\)-distance between quantiles restricted to the longest \(1-\alpha\) fraction of
lifetimes.
\end{definition}

This focuses on the longest lifetimes. It is especially relevant for extreme longevity,
late-life annuity cash-flows and longevity reinsurance. One may also compare survival
probabilities at advanced ages,
\[
  {}_tp_x=\Pbb(T_x>t),
\]
and report tail displacement curves
\[
  u\longmapsto Q_\nu(u)-Q_\mu(u),
  \qquad u\in[\alpha,1).
\]

\begin{remark}[Extreme longevity]
\label{rem:extreme-longevity-tail-transport}
Extreme-value thinking is not limited to non-life severity. In longevity risk, the upper tail
of the lifetime distribution corresponds to very old ages. Tail-aware transport makes explicit
whether a mortality improvement affects ordinary retirement ages or extreme survival.
\end{remark}

\subsection{Multi-state health and long-term care}
\label{subsec:multistate-health-ltc}

Life insurance and pensions often require more than alive/dead states. Health, disability and
long-term care risks can be represented by a finite state space
\[
  E=\{\text{healthy},\text{disabled},\text{long-term care},\text{dead}\}.
\]
A trajectory is the vector
\[
  \bZ=(Z_0,Z_1,\ldots,Z_T)\in E^{T+1}.
\]
A multi-state model defines a law on paths \(E^{T+1}\). Payments are push-forwards of this path law. For example,
\[
  L=\sum_{t=0}^T v^t h_t(Z_t),
\]
where \(h_t\) is the state-dependent benefit at time \(t\).

\begin{definition}[Multi-state life path law]
\label{def:multi-state-life-path-law}
Let \(E\) be a finite health-state space. A multi-state life path law over horizon \(T\) is a
probability measure
\[
  \Xi\in\mathcal P(E^{T+1})
\]
on paths \(\bz_{0:T}=(z_0,\ldots,z_T)\). A benefit schedule \(h_t:E\to\R_+\) induces the present-value push-forward
\[
  \Law(L)=G_\#\Xi,
  \qquad
  G(\bz_{0:T})=\sum_{t=0}^T v^t h_t(z_t).
\]
\end{definition}

Transport between multi-state models should use a state cost \(\kappa(e,f)\). For instance,
matching healthy to disabled should cost less than matching healthy to dead, depending on the
application. A path cost may be
\[
  c(\bz_{0:T},\bz'_{0:T})=
  \sum_{t=0}^T w_t\kappa(z_t,z'_t).
\]
This turns multi-state actuarial comparison into a dynamic transport problem on path space.

\begin{computebox}
For multi-state life and health models, the modelling work lies in the state cost. A good cost
should reflect benefit levels, transition severity, recovery possibilities and timing. This is an
instance of the cost-engineering principles collected in Appendix~\ref{app:cost-engineering}.
\end{computebox}

\subsection{Joint lives and dependence}
\label{subsec:joint-lives-dependence}

Many life products depend on more than one lifetime. Let \(T_x\) and \(T_y\) be future
lifetimes of two lives. A joint-life annuity may pay while both are alive; a last-survivor annuity
may pay until the second death. The relevant variables are
\[
  \min(T_x,T_y),\qquad \max(T_x,T_y).
\]
The marginal lifetime laws do not determine these distributions. The dependence coupling
matters.

\begin{definition}[Joint-life coupling]
\label{def:joint-life-coupling}
Let \(\mu_x\) and \(\mu_y\) be marginal residual lifetime laws. A joint-life model is a coupling
\[
  \gamma\in\Coup(\mu_x,\mu_y).
\]
For a last-survivor present value \(g(\max(T_x,T_y))\), the law of the benefit is the push-forward
of \(\gamma\) by the map \((s,t)\mapsto g(\max(s,t))\).
\end{definition}

Worst-case or prudent values can be formulated as optimization over couplings:
\[
  \sup_{\gamma\in\Coup(\mu_x,\mu_y)}
  \Ebb_\gamma\left[g\bigl(\max(T_x,T_y)\bigr)\right].
\]
This is a dependence-uncertainty problem analogous to the non-life aggregation problems of
Lecture~\ref{lec:reinsurance-capital-mmot}.

\begin{proposition}[Marginal lifetime laws do not identify survivor benefits]
\label{prop:joint-life-nonidentification}
The law of \(\max(T_x,T_y)\), and therefore the law of a last-survivor benefit, is generally not
determined by the marginal laws \(\mu_x\) and \(\mu_y\).
\end{proposition}

\begin{proof}
Take \(T_x\) and \(T_y\) with the same Bernoulli law on \(\{0,1\}\). Under the comonotonic
coupling, \(T_x=T_y\), hence \(\max(T_x,T_y)=T_x\). Under the countermonotonic coupling,
\(T_y=1-T_x\), hence \(\max(T_x,T_y)=1\) almost surely. The marginal laws are the same, but
the last-survivor distribution differs.
\end{proof}

\begin{actuarialbox}
Joint-life products are coupling problems. Marginal mortality tables determine individual
lifetime distributions, but not the dependence between lives. Transport provides the feasible
set and a language for prudent dependence analysis.
\end{actuarialbox}

For more than two lives, one obtains a multi-marginal problem. If \(\bT=(T_1,\ldots,T_d)\) has fixed marginal lifetime laws \(\mu_1,\ldots,\mu_d\), admissible joint models are
\[
  \gamma\in\Coup(\mu_1,\ldots,\mu_d).
\]
The aggregate liability may depend on all lifetimes, for example through survivor pensions,
group annuities or spouse benefits.

\subsection{Federated quantile sketches for mortality comparison}
\label{subsec:federated-mortality-quantile-sketches}

Mortality and longevity data may be distributed across insurers, countries or pension schemes.
Privacy, governance or competition may prevent sharing individual records. One can still share
summary information such as quantile sketches of residual lifetime distributions and annuity
present values. This mirrors the one-dimensional statistical-OT shortcuts from Lecture~\ref{lec:statistical-ot}; the full longevity lab uses the same quantile representation in Lab~\ref{lab:longevity-transport}.

\begin{definition}[Mortality quantile sketch]
\label{def:mortality-quantile-sketch}
A mortality quantile sketch for a one-dimensional law \(\mu\) is a finite collection
\[
  \{(u_k,\widehat Q(u_k)):k=1,\ldots,K\},
\]
where \(0<u_1<\cdots<u_K<1\) and \(\widehat Q(u_k)\) approximates \(Q_\mu(u_k)\). It is a
compressed representation of the residual lifetime, age-at-death or present-value distribution.
\end{definition}

Suppose institution \(m\) holds a residual lifetime sample and reports approximate quantiles
\[
  \widehat Q_m(u_1),\ldots,\widehat Q_m(u_K).
\]
A central analyst can approximate one-dimensional Wasserstein distances by
\[
  \widehat{\Wp{p}^p}(m,m')
  =
  \sum_{k=1}^K \omega_k
  \left|\widehat Q_m(u_k)-\widehat Q_{m'}(u_k)\right|^p.
\]
The same idea applies to subportfolio-specific longevity distributions, annuity present values or country-specific mortality laws.

\begin{remark}[What sketches do and do not contain]
\label{rem:mortality-sketches-limitations}
Quantile sketches are well suited for one-dimensional lifetime laws and present-value laws.
They do not fully describe multivariate dependence, joint-life dependence or path dependence.
For those objects, richer summaries are required.
\end{remark}

\subsection{Mini-lab: longevity transport in R}
\label{subsec:mini-lab-longevity-transport}

This lab builds synthetic mortality tables, converts them into residual lifetime laws, computes
Wasserstein distances, and maps lifetimes through annuity present values. A fuller version can
be used as Lab~\ref{lab:longevity-transport}.

\begin{lstlisting}[language=R]
# Synthetic longevity transport lab
set.seed(123)

ages <- 60:120
omega <- max(ages)

# Gompertz-Makeham style one-year death probabilities
qx_table <- function(ages, A = 0.0005, B = 0.00003, c = 1.095) {
  q <- A + B * c^ages
  pmin(q, 0.999)
}

q_base   <- qx_table(ages, A = 0.0005, B = 0.000030, c = 1.095)
q_longev <- qx_table(ages, A = 0.0004, B = 0.000022, c = 1.090)

curtate_pmf <- function(qx) {
  n <- length(qx)
  surv <- cumprod(c(1, 1 - qx[-n]))
  pmf <- surv * qx
  pmf[n] <- pmf[n] + prod(1 - qx)  # close the table at omega
  pmf / sum(pmf)
}

pmf_base   <- curtate_pmf(q_base)
pmf_longev <- curtate_pmf(q_longev)

support <- 0:(length(ages) - 1)

# Lower generalized inverse for a discrete law on a fixed support
q_discrete <- function(pmf, probs, support) {
  cdf <- cumsum(pmf)
  vapply(probs, function(u) support[which(cdf >= u)[1]], numeric(1))
}

u <- seq(0.001, 0.999, length.out = 2000)
Q_base   <- q_discrete(pmf_base,   u, support)
Q_longev <- q_discrete(pmf_longev, u, support)

W1 <- mean(abs(Q_base - Q_longev))
W2 <- sqrt(mean((Q_base - Q_longev)^2))

life_exp_base   <- sum(support * pmf_base)
life_exp_longev <- sum(support * pmf_longev)

# Annuity and death benefit present values
v <- 1 / 1.03
annuity_pv <- function(k) (1 - v^(k + 1)) / (1 - v)
death_pv   <- function(k) v^(k + 1)

A_base   <- annuity_pv(Q_base)
A_longev <- annuity_pv(Q_longev)
D_base   <- death_pv(Q_base)
D_longev <- death_pv(Q_longev)

ans <- rbind(
  lifetime = c(W1 = W1, W2 = W2,
               mean_base = life_exp_base,
               mean_longev = life_exp_longev),
  annuity_pv = c(W1 = mean(abs(A_base - A_longev)),
                 W2 = sqrt(mean((A_base - A_longev)^2)),
                 mean_base = mean(A_base),
                 mean_longev = mean(A_longev)),
  death_pv = c(W1 = mean(abs(D_base - D_longev)),
               W2 = sqrt(mean((D_base - D_longev)^2)),
               mean_base = mean(D_base),
               mean_longev = mean(D_longev))
)
round(ans, 4)

# Quantile displacement curve
plot(u, Q_longev - Q_base, type = "l",
     xlab = "lifetime percentile",
     ylab = "longevity stress in years",
     main = "Quantile longevity displacement")
abline(h = 0, lty = 2)
\end{lstlisting}

The output separates three layers: movement in residual lifetimes, movement in annuity present
values, and movement in death-benefit present values. The same longevity stress moves these
financial objects in different directions.

\subsection{Practical workflow}
\label{subsec:longevity-practical-workflow}

A transport workflow for longevity risk can be organized as follows:
\begin{enumerate}
\item choose the object: residual lifetime, age at death, annuity present value, death-benefit
      present value, health-state path or portfolio liability;
\item construct the probability law under each mortality table or stochastic mortality model;
\item choose the cost: years of lifetime, discounted cash-flow distance, state-transition cost or
      portfolio present-value distance;
\item compute Wasserstein distances, quantile displacement curves and tail-aware diagnostics;
\item map mortality differences through contract cash-flow functions;
\item report both demographic and financial distances;
\item perform sensitivity analysis with respect to mortality scenario, discount rate, tail focus
      and dependence assumptions.
\end{enumerate}

\subsection{What to remember}
\label{subsec:longevity-what-to-remember}

Life, mortality and longevity problems are naturally distributional. A mortality table induces
residual lifetime laws. A contract maps those lifetime laws into financial present-value laws.
Optimal transport compares mortality scenarios, builds longevity stresses, averages mortality
models through barycenters and defines prudent ambiguity sets. The actuarial interpretation
comes from the chosen state space: years of life, age at death, health states, cash-flow vectors
or portfolio liabilities.

\subsection{Exercises}
\label{subsec:longevity-exercises}

\begin{exercise}[Life table as a measure]
\label{ex:17:1}
Given one-year death probabilities \(q_x,q_{x+1},\ldots,q_\omega\), derive the probability mass
function of the curtate future lifetime \(K_x\). Check that the probabilities sum to one when
the table is closed at \(\omega\).
\end{exercise}

\begin{exercise}[Residual lifetime Wasserstein distance]
\label{ex:17:2}
Let \(K_x^0\) and \(K_x^1\) have discrete distributions on \(\{0,1,2\}\):
\[
  \mu^0=(0.3,0.4,0.3),\qquad \mu^1=(0.2,0.3,0.5).
\]
Compute \(\Wone(\mu^0,\mu^1)\) using the quantile formula.
\end{exercise}

\begin{exercise}[Quantile longevity stress]
\label{ex:17:3}
Let \(T_x^0\sim\mathrm{Exp}(\lambda_0)\) and \(T_x^1\sim\mathrm{Exp}(\lambda_1)\), with
\(\lambda_1<\lambda_0\). Derive the monotone transport map from \(T_x^0\) to \(T_x^1\).
Interpret it as a longevity stress.
\end{exercise}

\begin{exercise}[Annuity push-forward]
\label{ex:17:4}
Let \(K_x\) be curtate future lifetime and let
\[
  g(k)=\sum_{j=0}^{k}v^j.
\]
Write explicitly the distribution of \(g(K_x)\) in terms of the distribution of \(K_x\). Explain
why this is a push-forward.
\end{exercise}

\begin{exercise}[Opposite effects]
\label{ex:17:5}
Show that a first-order stochastic increase in lifetime increases the present value of a life
annuity but decreases the present value of a death benefit paid at the end of the year of death.
\end{exercise}

\begin{exercise}[Wasserstein barycenter]
\label{ex:17:6}
Let two residual lifetime laws have quantile functions \(Q_0\) and \(Q_1\). For weights
\(1-\lambda\) and \(\lambda\), derive the one-dimensional \(\Wtwo\) barycenter quantile function.
Interpret the result as a blended mortality scenario.
\end{exercise}

\begin{exercise}[Liability push-forward]
\label{ex:17:7}
Consider a portfolio of independent annuitants with payments \(b_i\) and lifetime vector \(\bK=(K_1,\ldots,K_n)\). Write the portfolio present value as a function of \(\bK\). Explain how a joint law of lifetimes is pushed forward to a liability law.
\end{exercise}

\begin{exercise}[Joint-life dependence]
\label{ex:17:8}
Let \(T_x\) and \(T_y\) have fixed marginal lifetime laws. Explain why the law of
\(\max(T_x,T_y)\) is not determined by the marginals. Compare the independent and
comonotonic couplings qualitatively for a last-survivor annuity.
\end{exercise}

\begin{exercise}[Tail-aware longevity]
\label{ex:17:9}
Define \(\Wp{1,0.95}^{\mathrm{tail}}\) between two residual lifetime laws. Explain why this
quantity may be more relevant than global \(\Wone\) for an old annuity portfolio.
\end{exercise}

\begin{exercise}[Multi-state transport]
\label{ex:17:10}
Consider the state space \(E=\{H,D,C,\Delta\}\), representing healthy, disabled, long-term
care and dead. Propose a cost matrix \(\kappa(e,f)\) for comparing two health-state paths.
Justify the ordering of costs.
\end{exercise}

\begin{exercise}[Prudent annuity price]
\label{ex:17:11}
Let \(\widehat\mu\) be a residual lifetime law and let \(g\) be a bounded Lipschitz annuity present-value function. Write the Wasserstein-robust prudent price over \(\mathcal B_1(\widehat\mu,r)\). Use the Lipschitz property of \(g\) to obtain a simple upper bound.
\end{exercise}

\begin{exercise}[Federated quantile sketches]
\label{ex:17:12}
Suppose three pension funds report lifetime quantiles at probabilities
\(0.05,0.10,\ldots,0.95\). Describe how to approximate pairwise \(\Wone\) distances. What
information about joint lives or path dependence is lost in such sketches?
\end{exercise}

\newpage
\printbibliography

\appendix

\newpage

\lecture{Mathematical Toolbox for Optimal Transport}
\label{app:math-toolbox}

This appendix collects the mathematical notions used repeatedly throughout the notes. It is a reference chapter on push-forwards, couplings, dual potentials, convexity, projections, Wasserstein geodesics, barycenters, dynamic formulations, entropic regularization, unbalanced transport and Schr\"odinger bridges. The aim is not to give a complete treatment of optimal transport, but to fix notation and to state the assumptions behind the objects used in the actuarial chapters.

\begin{readingbox}
Standard references for optimal transport include \cite{villani2003topics,villani2008oldnew,santambrogio2015optimal,ambrosio2008gradient,peyre2019computational,figalli2021invitation}. For convex analysis, see \cite{rockafellar1970convex,rockafellar1998variational}. For computational and entropic OT, see \cite{cuturi2013sinkhorn,benamou2015iterative,peyre2019computational}. For unbalanced transport and Schr\"odinger bridges, see \cite{chizat2018unbalanced,liero2018optimal,leonard2014survey,chen2016relation}.
\end{readingbox}

Throughout the appendix, \((\mathcal X,d_{\mathcal X})\) and \((\mathcal Y,d_{\mathcal Y})\) denote Polish spaces, usually subsets of Euclidean spaces in actuarial applications. We write \(\Prob(\mathcal X)\) for the set of probability measures on \(\mathcal X\). For \(p\ge 1\),
\[
  \Probp{p}(\mathcal X)
  =
  \left\{\mu\in\Prob(\mathcal X):
  \int_{\mathcal X} d_{\mathcal X}(x,x_0)^p\,\mu(\dd x)<\infty
  \text{ for some }x_0\in\mathcal X\right\}.
\]
The choice of \(x_0\) is immaterial. Finite positive measures are denoted by \(\mathcal M_+(\mathcal X)\). Push-forwards are written \(T_{\push}\mu\), and couplings are written \(\Coup(\mu,\nu)\), consistently with the main lectures.

\begin{actuarialbox}%[Quantile convention]
Throughout the notes, one-dimensional quantiles are lower generalized inverses, \(Q_\mu(u)=\inf\{x:F_\mu(x)\ge u\}\). With atoms or ties, formulas involving \(F_\mu(X)\) should be read either through the quantile coupling \((Q_\mu(U),Q_\nu(U))\) or through randomized ranks. This convention matters for empirical VaR, TVaR, one-dimensional barycenters and rank-preserving maps.
\end{actuarialbox}

\subsection{Measures, random variables and push-forwards}
\label{app:toolbox-pushforward}

A probability measure is the distributional object behind a random variable. In these notes, losses, premiums, reserves, capital scenarios and climate losses are often treated as probability measures rather than only as scalar summaries.

Let \(X:\Omega\to\mathcal X\) be a random variable defined on a probability space \((\Omega,\mathcal F,\Pbb)\). Its law is the push-forward of \(\Pbb\) by \(X\):
\[
  \Law(X)=X_{\push}\Pbb.
\]
Equivalently, for any measurable \(A\subseteq\mathcal X\),
\[
  \Law(X)(A)=\Pbb(X\in A).
\]

\begin{definition}[Moment space]
\label{def:app-moment-space}
Let \((\mathcal X,d)\) be a metric space and \(p\ge1\). The Wasserstein moment space \(\Probp{p}(\mathcal X)\) is the set of probability measures \(\mu\in\Prob(\mathcal X)\) with finite \(p\)-th moment, namely
\[
  \int d(x,x_0)^p\,\mu(\dd x)<\infty
\]
for some, hence every, \(x_0\in\mathcal X\).
\end{definition}

\begin{definition}[Push-forward]
\label{def:app-push-forward}
Let \(T:\mathcal X\to\mathcal Y\) be measurable and let \(\mu\in\Prob(\mathcal X)\). The push-forward of \(\mu\) by \(T\), denoted \(T_{\push}\mu\), is the probability measure on \(\mathcal Y\) defined by
\[
  T_{\push}\mu(B)=\mu(T^{-1}(B)),
  \qquad B\in\mathcal B(\mathcal Y).
\]
Equivalently, if \(X\sim\mu\), then \(T(X)\sim T_{\push}\mu\).
\end{definition}

\begin{proposition}[Change of variables for push-forwards]
\label{prop:app-pushforward-cov}
Let \(T:\mathcal X\to\mathcal Y\) be measurable and let \(\mu\in\Prob(\mathcal X)\). For every nonnegative measurable function \(h:\mathcal Y\to[0,\infty]\),
\[
  \int_{\mathcal Y} h(y)\,(T_{\push}\mu)(\dd y)
  =
  \int_{\mathcal X} h(T(x))\,\mu(\dd x).
\]
The same identity holds for any integrable real-valued \(h\).
\end{proposition}

\begin{proof}
First prove the identity for indicators \(h=\1_B\), where it is exactly the definition of \(T_{\push}\mu\). Extend by linearity to nonnegative simple functions, by monotone convergence to nonnegative measurable functions, and by decomposition into positive and negative parts to integrable functions.
\end{proof}

\paragraph{Discrete case.}
If \(\mu=\sum_{i=1}^n a_i\delta_{x_i}\), then
\[
  T_{\push}\mu=\sum_{i=1}^n a_i\delta_{T(x_i)}.
\]
A deterministic map moves locations while preserving their masses.

\paragraph{Smooth density case.}
If \(\mathcal X=\mathcal Y=\R^d\), \(T\) is a diffeomorphism, and \(\mu\) and \(\nu=T_{\push}\mu\) have densities \(\rho_\mu\) and \(\rho_\nu\), then
\[
  \rho_\mu(x)=\rho_\nu(T(x))\,|\det DT(x)|.
\]
This formula is useful for intuition, but most actuarial applications in the notes allow atoms, discrete grids or empirical measures, and therefore do not rely on smoothness.

\paragraph{Pull-back versus push-forward.}
The push-forward acts on measures. The pull-back acts on functions. If \(h:\mathcal Y\to\R\), then \(T^*h=h\circ T\), and
\[
  \int_{\mathcal Y} h\,\dd(T_{\push}\mu)
  =
  \int_{\mathcal X} T^*h\,\dd\mu.
\]

\begin{actuarialbox}
If \(\Premium:\mathcal X\to\R_+\) is a tariff rule and \(\bX\sim\mu_X\), the induced premium distribution is \(\Law(\Premium(\bX))=\Premium_{\push}\mu_X\). If \(T(l)=(l-d)_+\wedge u\) is a deductible-limit transformation, the payment law is \(T_{\push}\Law(L)\). The same notation covers stress maps, recalibrations and reserving aggregations.
% To be restored in v2: The same notation also covers fairness repairs.
\end{actuarialbox}

\subsection{Couplings, marginal projections and stochastic kernels}
\label{app:toolbox-couplings}

Optimal transport is optimization over joint distributions with prescribed marginals.

\begin{definition}[Coupling]
\label{def:app-coupling}
Let \(\mu\in\Prob(\mathcal X)\) and \(\nu\in\Prob(\mathcal Y)\). A coupling between \(\mu\) and \(\nu\) is a probability measure \(\gamma\in\Prob(\mathcal X\times\mathcal Y)\) such that
\[
  (p_1)_{\push}\gamma=\mu,
  \qquad
  (p_2)_{\push}\gamma=\nu,
\]
where \(p_1(x,y)=x\) and \(p_2(x,y)=y\). The set of all such couplings is denoted by
\[
  \Coup(\mu,\nu).
\]
\end{definition}

\begin{definition}[Product coupling]
\label{def:app-product-coupling}
The product coupling of \(\mu\) and \(\nu\) is the product measure \(\mu\otimes\nu\in\Coup(\mu,\nu)\). If \((X,Y)\sim\mu\otimes\nu\), then \(X\) and \(Y\) are independent.
\end{definition}

\begin{proposition}[Non-emptiness and convexity]
\label{prop:app-couplings-nonempty-convex}
For any \(\mu\in\Prob(\mathcal X)\) and \(\nu\in\Prob(\mathcal Y)\), the set \(\Coup(\mu,\nu)\) is non-empty and convex. If \(\mathcal X\) and \(\mathcal Y\) are Polish, then \(\Coup(\mu,\nu)\) is compact for weak convergence.
\end{proposition}

\begin{proof}
Non-emptiness follows from \(\mu\otimes\nu\). Convexity follows because marginalization is linear. Compactness follows from tightness of the fixed marginals and Prokhorov's theorem; closedness follows because weak convergence preserves marginals when tested against bounded continuous functions.
\end{proof}

\paragraph{Discrete case.}
For \(\mu=\sum_{i=1}^n a_i\delta_{x_i}\) and \(\nu=\sum_{j=1}^m b_j\delta_{y_j}\), a coupling is a nonnegative matrix \(\Gamma=(\gamma_{ij})\in\R_+^{n\times m}\) satisfying
\[
  \sum_{j=1}^m\gamma_{ij}=a_i,
  \qquad
  \sum_{i=1}^n\gamma_{ij}=b_j.
\]
This is the transportation polytope introduced in Definition~\ref{def:transportation-polytope}.

\begin{definition}[Markov kernel and disintegration]
\label{def:app-markov-kernel}
A Markov kernel from \(\mathcal X\) to \(\mathcal Y\) is a map \(K:\mathcal X\times\mathcal B(\mathcal Y)\to[0,1]\) such that \(K(x,\cdot)\) is a probability measure for every \(x\), and \(x\mapsto K(x,B)\) is measurable for every measurable \(B\). A coupling \(\gamma\in\Coup(\mu,\nu)\) admits a disintegration
\[
  \gamma(\dd x,\dd y)=\mu(\dd x)K(x,\dd y)
\]
under the usual standard-Borel assumptions.
\end{definition}

A deterministic map \(T\) corresponds to the special kernel \(K(x,\dd y)=\delta_{T(x)}(\dd y)\). Thus Monge maps are deterministic couplings, whereas Kantorovich plans allow randomization or mass splitting.

\begin{definition}[Graph coupling]
\label{def:app-graph-coupling}
If \(T:\mathcal X\to\mathcal Y\) is measurable and \(T_{\push}\mu=\nu\), the graph coupling induced by \(T\) is
\[
  (\Id,T)_{\push}\mu\in\Coup(\mu,\nu).
\]
It is concentrated on \(\{(x,T(x)):x\in\mathcal X\}\).
\end{definition}

\begin{actuarialbox}
A coupling is a dependence model with fixed marginals. It is also an audit trail: the kernel \(K(x,\dd y)\) records how mass starting at actuarial state \(x\) is redistributed to target states \(y\). This interpretation is used in portfolio drift, reinsurance dependence uncertainty and scenario design.
% To be restored in v2: This interpretation is also used in fairness repair and counterfactual scenario design.
\end{actuarialbox}

\begin{actuarialbox}%[Distributional, not causal]
A coupling aligns two marginal laws. It does not by itself identify the causal effect of an intervention on an individual unit. Such an interpretation requires additional structure, such as an intervention design, a structural causal model, ignorability assumptions or rank preservation justified outside the transport problem.
\end{actuarialbox}

\subsection{Monge maps, Kantorovich plans and optimal transport}
\label{app:toolbox-monge-kantorovich}

Let \(c:\mathcal X\times\mathcal Y\to\R\cup\{+\infty\}\) be a lower semicontinuous cost.

\begin{definition}[Monge problem]
\label{def:app-monge-problem}
The Monge problem from \(\mu\) to \(\nu\) is
\[
  M_c(\mu,\nu)
  =
  \inf_{T:T_{\push}\mu=\nu}
  \int_{\mathcal X} c(x,T(x))\,\mu(\dd x),
\]
with value \(+\infty\) if no admissible map exists.
\end{definition}

This formulation is deterministic: each source state \(x\) is sent to one target state \(T(x)\). It may fail to have feasible maps, especially with atoms.

\begin{definition}[Kantorovich problem]
\label{def:app-kantorovich-problem}
The Kantorovich relaxation is
\[
  \OT_c(\mu,\nu)
  =
  \inf_{\gamma\in\Coup(\mu,\nu)}
  \int_{\mathcal X\times\mathcal Y} c(x,y)\,\gamma(\dd x,\dd y).
\]
An optimizer \(\gamma^\star\), when it exists, is called an optimal transport plan.
\end{definition}

The Kantorovich problem is linear in \(\gamma\). In finite spaces, it is a linear program. This is the source of both its computational tractability and its dual interpretation.

\begin{proposition}[Existence of optimal plans]
\label{prop:app-ot-existence}
Let \(\mathcal X\) and \(\mathcal Y\) be Polish spaces and let \(c:\mathcal X\times\mathcal Y\to[0,+\infty]\) be lower semicontinuous. Then the Kantorovich problem admits an optimal plan, possibly with infinite value. If at least one coupling has finite cost, the optimal value is finite and attained.
\end{proposition}

\begin{proof}
The set \(\Coup(\mu,\nu)\) is compact for weak convergence. The functional \(\gamma\mapsto\int c\,\dd\gamma\) is lower semicontinuous by the portmanteau theorem. A minimizing sequence therefore has a weakly convergent subsequence, and the limit is optimal.
\end{proof}

\paragraph{Cost as modelling choice.}
The cost \(c\) decides which changes are small or large. For claim amounts, \(c(x,y)=|x-y|^p\) measures monetary displacement. For premium changes, a relative cost such as \(|\log x-\log y|^p\) may be more meaningful. For cash-flow vectors, \(c\) may combine amount, timing and discounting. For geographical exposures, \(c\) may be based on physical distance, hazard distance or administrative distance. Appendix~\ref{app:cost-engineering} is devoted to these choices.

\subsection{Wasserstein distances and convergence}
\label{app:toolbox-wasserstein}

When \(\mathcal X=\mathcal Y\) and \(c(x,y)=d(x,y)^p\), optimal transport defines the \(p\)-Wasserstein distance.

\begin{definition}[Wasserstein distance]
\label{def:app-wasserstein-distance}
Let \((\mathcal X,d)\) be a metric space and let \(p\ge1\). For \(\mu,\nu\in\Probp{p}(\mathcal X)\),
\[
  \Wp{p}(\mu,\nu)
  =
  \left(
  \inf_{\gamma\in\Coup(\mu,\nu)}
  \int_{\mathcal X\times\mathcal X} d(x,y)^p\,\gamma(\dd x,\dd y)
  \right)^{1/p}.
\]
\end{definition}

The distance \(\Wp{p}\) lifts the ground geometry \(d\) from points to probability measures. For Dirac masses,
\[
  \Wp{p}(\delta_x,\delta_y)=d(x,y).
\]
Thus Wasserstein distance recognizes that moving a loss from \(1000\) to \(1001\) is small, while moving it from \(1000\) to \(100000\) is large.

\begin{proposition}[Metric and convergence]
\label{prop:app-wasserstein-convergence}
For \(p\ge1\), \(\Wp{p}\) defines a metric on \(\Probp{p}(\mathcal X)\). On Polish spaces,
\[
  \Wp{p}(\mu_n,\mu)\to0
\]
if and only if \(\mu_n\Rightarrow\mu\) weakly and the \(p\)-th moments converge.
\end{proposition}

This proposition is the background for the statistical discussion in Lecture~\ref{lec:statistical-ot}: empirical Wasserstein distances estimate population distances, but the rate depends strongly on dimension and moment assumptions.

\subsection{Quantile maps and monotone rearrangement}
\label{app:toolbox-quantile-map}

In dimension one, the optimal transport map has an explicit form. For a distribution function \(F\), define the generalized inverse
\[
  Q_\mu(u)=F_\mu^{-1}(u)=\inf\{x\in\R:F_\mu(x)\ge u\},
  \qquad u\in(0,1).
\]
The notation \(Q_\mu\) is used throughout the notes.

\begin{definition}[Quantile map]
\label{def:app-quantile-map}
Let \(\mu,\nu\in\Prob(\R)\). A rank-preserving quantile map from \(\mu\) to \(\nu\) is
\[
  T(x)=Q_\nu(F_\mu(x))
\]
when \(F_\mu\) is continuous. In the presence of atoms, one may use a randomized rank
\[
  U=F_\mu(X-)+V\{F_\mu(X)-F_\mu(X-)\},
  \qquad V\sim\operatorname{Unif}(0,1),
\]
and map \(X\) to \(Q_\nu(U)\).
\end{definition}

\begin{theorem}[Monotone rearrangement]
\label{thm:app-monotone-rearrangement}
Let \(\mu,\nu\in\Probp{p}(\R)\). For \(p\ge1\),
\[
  \Wp{p}^p(\mu,\nu)
  =
  \int_0^1 |Q_\mu(u)-Q_\nu(u)|^p\,\dd u.
\]
The coupling \((Q_\mu(U),Q_\nu(U))\), with \(U\sim\operatorname{Unif}(0,1)\), is optimal. More generally, it is optimal for costs \(c(x,y)=h(x-y)\) with \(h\) convex.
\end{theorem}

\begin{actuarialbox}
Quantile transport preserves ranks. A claim, premium or other one-dimensional actuarial quantity at percentile \(u\) in the source distribution is sent to the value at percentile \(u\) in the target distribution. This is used for severity stresses, premium transformations, longevity stresses and one-dimensional barycenters.
% To be restored in v2: The same rank-preserving construction is also used for fairness repairs.
\end{actuarialbox}

\subsection{Dual potentials and the \texorpdfstring{\(c\)}{c}-transform}
\label{app:toolbox-potentials}

Duality gives a second view of optimal transport. It introduces potentials, which can be interpreted as shadow prices, sensitivity functions or certificates of optimality.

\begin{definition}[Kantorovich potentials]
\label{def:app-kantorovich-potentials}
Under the usual assumptions ensuring strong duality,
\[
  \OT_c(\mu,\nu)
  =
  \sup_{\varphi,\psi}
  \left\{
  \int \varphi\,\dd\mu+
  \int \psi\,\dd\nu:
  \varphi(x)+\psi(y)\le c(x,y)
  \right\}.
\]
A feasible pair \((\varphi,\psi)\) is called a pair of Kantorovich potentials. If it attains the supremum, it is called a pair of optimal Kantorovich potentials.
\end{definition}

In the discrete problem with cost matrix \(\bC=(C_{ij})\), the dual is
\[
  \max_{\bu,\bv}
  \left\{
  \sum_i a_i \bu_i+
  \sum_j b_j \bv_j:
  \bu_i+\bv_j\le \bC_{ij}\right\}.
\]

\begin{definition}[\(c\)-transform]
\label{def:app-c-transform}
For a function \(\varphi:\mathcal X\to\R\cup\{-\infty\}\), its \(c\)-transform is
\[
  \varphi^c(y)
  =
  \inf_{x\in\mathcal X}\{c(x,y)-\varphi(x)\}.
\]
Similarly, for \(\psi:\mathcal Y\to\R\cup\{-\infty\}\), define
\[
  \psi^{\bar c}(x)
  =
  \inf_{y\in\mathcal Y}\{c(x,y)-\psi(y)\}.
\]
\end{definition}

At the optimum, one can often take \(\psi=\varphi^c\) and \(\varphi=\psi^{\bar c}\). The \(c\)-transform is the analogue, for a general transport cost, of the Legendre--Fenchel transform.

\begin{definition}[Contact set]
\label{def:app-contact-set}
For a feasible pair \((\varphi,\psi)\), the contact set is
\[
  \mathcal C_{\varphi,\psi}
  =
  \{(x,y):\varphi(x)+\psi(y)=c(x,y)\}.
\]
\end{definition}

\begin{proposition}[Complementary slackness]
\label{prop:app-complementary-slackness}
If \(\gamma^\star\) is primal optimal and \((\varphi^\star,\psi^\star)\) is dual optimal, then
\[
  \operatorname{supp}(\gamma^\star)
  \subseteq
  \mathcal C_{\varphi^\star,\psi^\star}.
\]
In finite form,
\[
  \gamma^\star_{ij}>0
  \quad\Longrightarrow\quad
  \bu_i^\star+\bv_j^\star=\bC_{ij}.
\]
\end{proposition}

\begin{proof}
The slack \(c(x,y)-\varphi^\star(x)-\psi^\star(y)\) is nonnegative. Equality of primal and dual values implies that its integral under \(\gamma^\star\) is zero, hence it is zero \(\gamma^\star\)-almost surely.
\end{proof}

\begin{definition}[\(c\)-cyclical monotonicity]
\label{def:app-cyclical-monotonicity}
A set \(S\subseteq\mathcal X\times\mathcal Y\) is \(c\)-cyclically monotone if, for every finite family \((x_1,y_1),\ldots,(x_N,y_N)\in S\) and every permutation \(\sigma\) of \(\{1,\ldots,N\}\),
\[
  \sum_{k=1}^N c(x_k,y_k)
  \le
  \sum_{k=1}^N c(x_k,y_{\sigma(k)}).
\]
\end{definition}

Under standard assumptions, optimal plans are concentrated on \(c\)-cyclically monotone sets. In one dimension with \(c(x,y)=|x-y|^p\), this reduces to no-crossing monotone matching.

\begin{remark}[Potentials versus Brenier potentials]
\label{rem:app-potentials-warning}
The word ``potential'' is used in two related but distinct ways. Kantorovich potentials are dual variables. Brenier potentials are convex functions whose gradients define optimal maps for quadratic transport. They are related through convex conjugacy, but they are not the same object.
\end{remark}

\subsection{Convex analysis for quadratic optimal transport}
\label{app:convex-analysis-ot}

Quadratic optimal transport relies on elementary convex analysis. This subsection is placed after duality because the link becomes clearest once Kantorovich potentials and contact sets have been introduced.

\begin{definition}[Convex function and subdifferential]
\label{def:app-convex-function-subdifferential}
Let \(C\subseteq\R^d\) be convex. A function \(u:C\to\R\cup\{+\infty\}\) is convex if
\[
  u((1-t)\bx+t\by)\le (1-t)u(\bx)+tu(\by),
  \qquad \bx,\by\in C,
  \quad t\in[0,1].
\]
For a proper convex function \(u:\R^d\to\R\cup\{+\infty\}\), the subdifferential at \(\bx\) is
\[
  \partial u(\bx)
  =
  \{\by\in\R^d:u(\bz)\ge u(\bx)+\by\cdot(\bz-\bx)\text{ for all }\bz\in\R^d\}.
\]
If \(u\) is differentiable at \(\bx\), then \(\partial u(\bx)=\{\nabla u(\bx)\}\).
\end{definition}

\begin{definition}[Legendre--Fenchel transform]
\label{def:app-legendre-fenchel-transform}
The convex conjugate of a proper function \(u:\R^d\to\R\cup\{+\infty\}\) is
\[
  u^*(\by)=\sup_{\bx\in\R^d}\{\bx\cdot\by-u(\bx)\}.
\]
The Fenchel inequality is
\[
  u(\bx)+u^*(\by)\ge \bx\cdot\by.
\]
For proper closed convex functions, equality holds if and only if \(\by\in\partial u(\bx)\), equivalently \(\bx\in\partial u^*(\by)\).
\end{definition}

\begin{proposition}[Quadratic OT duality and Fenchel equality]
\label{prop:app-quadratic-ot-fenchel}
Let \(c(\bx,\by)=\frac12\|\bx-\by\|^2\). If \((\varphi,\psi)\) is dual feasible for the Kantorovich problem, and
\[
  u(\bx)=\frac12\|\bx\|^2-\varphi(\bx),
  \qquad
  v(\by)=\frac12\|\by\|^2-\psi(\by),
\]
then \(u(\bx)+v(\by)\ge \bx\cdot\by\). On the contact set
\[
  \varphi(\bx)+\psi(\by)=\frac12\|\bx-\by\|^2,
\]
one has equality in the Fenchel inequality. Thus, when \(v=u^*\), the contact set is contained in the graph of \(\partial u\).
\end{proposition}

\begin{proof}
The dual inequality is
\[
  \varphi(\bx)+\psi(\by)\le \frac12\|\bx-\by\|^2
  =\frac12\|\bx\|^2+\frac12\|\by\|^2-\bx\cdot\by.
\]
Rearranging gives \(u(\bx)+v(\by)\ge \bx\cdot\by\). Equality is exactly the contact condition. If \(v=u^*\), equality in Fenchel's inequality is equivalent to \(\by\in\partial u(\bx)\).
\end{proof}

\begin{theorem}[Brenier theorem, informal statement]
\label{thm:app-brenier}
Let \(\mu,\nu\in\Probp{2}(\R^d)\), and assume that \(\mu\) is absolutely continuous with respect to Lebesgue measure. For the quadratic cost \(c(\bx,\by)=\frac12\|\bx-\by\|^2\), there exists a convex function \(\Phi:\R^d\to\R\cup\{+\infty\}\) such that
\[
  T(\bx)=\nabla\Phi(\bx),
  \qquad
  T_{\push}\mu=\nu,
\]
and \(T\) is the optimal transport map from \(\mu\) to \(\nu\). The map is unique \(\mu\)-almost surely.
\end{theorem}

If \(\phi\) is a Kantorovich potential for the cost \(\frac12\|\bx-\by\|^2\), then one may write
\[
  \Phi(\bx)=\frac12\|\bx\|^2-\phi(\bx),
  \qquad
  T(\bx)=\bx-\nabla\phi(\bx)=\nabla\Phi(\bx),
\]
where differentiability holds.

\begin{definition}[Monge--Amp\`ere equation, formal]
\label{def:app-monge-ampere}
In the smooth positive-density setting, if \(T=\nabla\Phi\) pushes \(\rho_0(\bx)\,\dd\bx\) to \(\rho_1(\by)\,\dd\by\), then formally
\[
  \rho_0(\bx)
  =
  \rho_1(\nabla\Phi(\bx))\det(D^2\Phi(\bx)).
\]
This nonlinear PDE is the Monge--Amp\`ere equation associated with quadratic optimal transport.
\end{definition}

\begin{actuarialbox}
For quadratic OT, convex analysis explains why the optimal map is not an arbitrary regression-like transformation. Under Brenier's hypotheses, it is a gradient of a convex function. This imposes a monotonicity structure on multivariate stresses and scenario maps. In empirical actuarial data, however, atoms, ties and categorical covariates often lead to non-unique plans, mass splitting or barycentric summaries rather than true deterministic maps.
% To be restored in v2: The same caution applies to multivariate fairness repairs.
\end{actuarialbox}

\subsection{Projections in optimal transport}
\label{app:toolbox-projections}

The word ``projection'' appears in several different ways in the notes. The distinctions below avoid a common source of confusion.

\begin{definition}[Marginal projection]
\label{def:app-marginal-projection}
For \(\gamma\in\Prob(\mathcal X\times\mathcal Y)\), the marginal projections are \((p_1)_{\push}\gamma\) and \((p_2)_{\push}\gamma\). This is marginalization, not orthogonal projection.
\end{definition}

\begin{definition}[Barycentric projection of a plan]
\label{def:app-barycentric-projection}
Assume that \(\mathcal Y\subseteq\R^d\), \(\gamma\in\Coup(\mu,\nu)\), and \(\gamma(\dd\bx,\dd\by)=\mu(\dd\bx)K(\bx,\dd\by)\). The barycentric projection of \(\gamma\) is
\[
  \bar T_\gamma(\bx)=\int_{\mathcal Y}\by\,K(\bx,\dd\by),
\]
whenever the integral is well-defined. In the discrete case,
\[
  \bar T_\Gamma(\bx_i)
  =
  \frac{1}{a_i}\sum_{j=1}^m\gamma_{ij}\by_j,
  \qquad a_i=\sum_{j=1}^m\gamma_{ij}.
\]
\end{definition}

The barycentric projection is a deterministic summary of a plan. It generally does not push \(\mu\) exactly to \(\nu\). It loses conditional dispersion.

\begin{definition}[Metric projection in Wasserstein space]
\label{def:app-wasserstein-projection}
Let \(\mathcal C\subseteq\Probp{p}(\mathcal X)\). A Wasserstein projection of \(\mu\) onto \(\mathcal C\) is any solution of
\[
  \Pi_{\mathcal C}^{\Wp{p}}(\mu)
  \in
  \argmin_{\eta\in\mathcal C}\Wp{p}^p(\mu,\eta).
\]
The projection may not be unique.
\end{definition}

\begin{definition}[Bregman divergence]
\label{def:app-bregman-divergence}
Let \(\Phi\) be a differentiable convex function on a convex set \(D\). The Bregman divergence generated by \(\Phi\) is
\[
  D_\Phi(\bx,\by)=\Phi(\bx)-\Phi(\by)-\langle\nabla\Phi(\by),\bx-\by\rangle.
\]
It is nonnegative by convexity of \(\Phi\), but it is generally not symmetric and does not satisfy the triangle inequality.
\end{definition}

For the entropy generator \(\phi(t)=t\log t-t\), the associated Bregman divergence is
\[
  D_\phi(s,t)=s\log\frac{s}{t}-s+t.
\]
Summing over entries gives the generalized Kullback--Leibler divergence
\[
  \KL(P\mid Q)=\sum_{i,j}\left\{p_{ij}\log\frac{p_{ij}}{q_{ij}}-p_{ij}+q_{ij}\right\}.
\]
Sinkhorn iterations can be interpreted as alternating KL/Bregman projections onto marginal constraint sets; this is different from Wasserstein metric projection and from Euclidean projection.

\subsection{Displacement interpolation and Wasserstein geodesics}
\label{app:toolbox-geodesics}

A Wasserstein distance turns probability measures into points in a metric space. It is therefore natural to ask what a shortest path between two probability measures is.

\begin{definition}[Displacement interpolation]
\label{def:app-displacement-interpolation}
Let \(\gamma^\star\in\Coup(\mu_0,\mu_1)\) be an optimal plan for \(\Wtwo\) on \(\R^d\). For \(t\in[0,1]\), define
\[
  e_t(\bx,\by)=(1-t)\bx+t\by,
  \qquad
  \mu_t=(e_t)_{\push}\gamma^\star.
\]
The curve \((\mu_t)_{t\in[0,1]}\) is a displacement interpolation between \(\mu_0\) and \(\mu_1\).
\end{definition}

If \(\gamma^\star=(\Id,T)_{\push}\mu_0\), then
\[
  \mu_t=((1-t)\Id+tT)_{\push}\mu_0.
\]
In the discrete case,
\[
  \mu_t=\sum_{i,j}\gamma^\star_{ij}\delta_{(1-t)\bx_i+t\by_j}.
\]

\begin{definition}[Constant-speed Wasserstein geodesic]
\label{def:app-wasserstein-geodesic}
A curve \((\mu_t)_{t\in[0,1]}\subset\Probp{2}(\R^d)\) is a constant-speed \(\Wtwo\)-geodesic if
\[
  \Wtwo(\mu_s,\mu_t)=|t-s|\Wtwo(\mu_0,\mu_1),
  \qquad 0\le s,t\le1.
\]
\end{definition}

Displacement interpolations built from optimal plans are \(\Wtwo\)-geodesics \cite{mccann1997convexity}. They should not be confused with linear mixtures \((1-t)\mu_0+t\mu_1\): mixtures average mass vertically, while displacement interpolations move mass through the state space.

\subsection{Barycenters and Fr\'echet means}
\label{app:toolbox-barycenters}

Barycenters generalize averages from Euclidean points to probability measures.

\begin{definition}[Wasserstein barycenter]
\label{def:app-wasserstein-barycenter}
Let \(\mu_1,\ldots,\mu_K\in\Probp{p}(\mathcal X)\), and let \(\lambda=(\lambda_1,\ldots,\lambda_K)\) be weights in the simplex. A \(p\)-Wasserstein barycenter is any solution of
\[
  \bar\mu
  \in
  \argmin_{\mu\in\Probp{p}(\mathcal X)}
  \sum_{k=1}^K\lambda_k\Wp{p}^p(\mu,\mu_k).
\]
\end{definition}

For one-dimensional distributions, the quadratic Wasserstein barycenter satisfies
\[
  Q_{\bar\mu}(u)=\sum_{k=1}^K\lambda_k Q_{\mu_k}(u),
  \qquad u\in(0,1).
\]
This formula is central for one-dimensional severity, premium, output and longevity applications.

\paragraph{Mixture versus barycenter.}
The mixture \(\sum_k\lambda_k\mu_k\) keeps the original components and averages probabilities. The Wasserstein barycenter averages locations through transport. This distinction matters when constructing consensus reserve distributions, mortality-scenario summaries or climate benchmarks.
% To be restored in v2: It also matters when constructing fairness targets.

\subsection{Dynamic transport, velocities and action}
\label{app:toolbox-dynamic-transport}

The static Kantorovich problem compares two endpoint distributions. Dynamic optimal transport represents the movement between them through a time-indexed path.

Let \((\mu_t)_{t\in[0,1]}\) be a curve of probability measures on \(\R^d\), and let \((\bv_t)_{t\in[0,1]}\) be a velocity field. The continuity equation is
\[
  \partial_t\mu_t+\nabla\cdot(\mu_t \bv_t)=0,
\]
understood in the weak sense. It expresses conservation of mass along the path.

\begin{theorem}[Benamou--Brenier formula, informal statement]
\label{thm:app-benamou-brenier}
For \(\mu_0,\mu_1\in\Probp{2}(\R^d)\) \cite{benamou2000computational},
\[
  \Wtwo^2(\mu_0,\mu_1)
  =
  \inf_{(\mu_t,\bv_t)}
  \int_0^1\int_{\R^d}\|\bv_t(\bx)\|^2\,\mu_t(\dd\bx)\,\dd t,
\]
where the infimum is taken over curves satisfying the continuity equation with endpoints \(\mu_0\) and \(\mu_1\).
\end{theorem}

The integral is the kinetic action of the path. This formulation gives mathematical meaning to distributional speed, actuarial trajectory and least-action path.

\paragraph{Adapted transport.}
When distributions live on paths, a coupling may use future information unless non-anticipativity is imposed. Adapted, causal or bicausal transport restricts couplings so that matching up to time \(t\) depends only on information up to time \(t\). This distinction matters in reserving, sequential pricing and scenario generation; see Lecture~\ref{lec:dynamic-ot-actuarial-trajectories}.
% To be restored in v2: dynamic fairness.

\paragraph{Mass variation.}
If total mass changes, the continuity equation is replaced by a source-term equation
\[
  \partial_t\eta_t+\nabla\cdot(\eta_t \bv_t)=\sigma_t,
\]
where \(\eta_t\) is a finite positive measure and \(\sigma_t\) records creation or removal of mass. This is the dynamic counterpart of unbalanced transport.

\subsection{Entropic regularization and Sinkhorn potentials}
\label{app:toolbox-entropic}

Entropic regularization smooths the transport problem and leads to scalable algorithms. In the finite case, let \(\ba\in\Delta_n\), \(\bb\in\Delta_m\), and let \(\bC\in\R^{n\times m}\) be a cost matrix. For \(\eps>0\), the entropic problem is
\[
  \OT_{\eps,\bC}(\ba,\bb)
  =
  \min_{\Gamma\in U(\ba,\bb)}
  \left\{
  \langle \bC,\Gamma\rangle
  +
  \eps\sum_{i,j}\Gamma_{ij}(\log\Gamma_{ij}-1)
  \right\}.
\]
The solution has the scaling form
\[
  \Gamma^\eps=\operatorname{diag}(\bu)\bK\operatorname{diag}(\bv),
  \qquad
  \bK_{ij}=\exp(-\bC_{ij}/\eps),
\]
where \(\bu\in\R_+^n\) and \(\bv\in\R_+^m\) enforce the two marginal constraints.

\begin{proposition}[Entropic transport as KL projection]
\label{prop:app-eot-kl-projection}
Let \(\bK_{ij}=\exp(-\bC_{ij}/\eps)\). The entropic optimal transport plan is the solution of
\[
  \min_{\Gamma\in U(\ba,\bb)}\KL(\Gamma\mid \bK).
\]
Equivalently, it is the KL/Bregman projection of the Gibbs kernel \(\bK\) onto the marginal constraint set \(U(\ba,\bb)\).
\end{proposition}

\begin{proof}
Using the generalized KL divergence,
\[
  \eps\KL(\Gamma\mid \bK)
  =
  \langle \bC,\Gamma\rangle
  +
  \eps\sum_{i,j}\Gamma_{ij}(\log\Gamma_{ij}-1)
  +
  \eps\sum_{i,j}\bK_{ij}.
\]
The final term does not depend on \(\Gamma\).
\end{proof}

The logarithmic scalings
\[
  f_i^\eps=\eps\log \bu_i,
  \qquad
  g_j^\eps=\eps\log \bv_j
\]
are entropic dual potentials, up to additive constants. They are smooth analogues of Kantorovich potentials.

\begin{definition}[Log-sum-exp and soft minimum]
\label{def:app-logsumexp-softmin}
For \(\bz=(z_1,\ldots,z_m)\in\R^m\), define
\[
  \operatorname{LSE}_\eps(\bz)
  =\eps\log\sum_{j=1}^m\exp(z_j/\eps),
\]
and
\[
  \operatorname{softmin}_\eps(\bz)
  =-\eps\log\sum_{j=1}^m\exp(-z_j/\eps).
\]
As \(\eps\downarrow0\), \(\operatorname{softmin}_\eps(\bz)\to\min_j z_j\).
\end{definition}

The log-sum-exp function replaces the hard \(c\)-transform by a smooth \(c\)-transform. This is why the entropic dual is differentiable.

\begin{proposition}[Envelope derivative]
\label{prop:app-eot-envelope}
Assume the entropic optimal plan \(\Gamma_\eps(\bC)\) is unique. Then
\[
  \frac{\partial}{\partial \bC_{ij}}\OT_{\eps,\bC}(\ba,\bb)
  =
  \Gamma_{\eps,ij}(\bC).
\]
Consequently, if \(\bC=\bC(\theta)\) is differentiable, then
\[
  \frac{\dd}{\dd\theta}\OT_{\eps,\bC(\theta)}(\ba,\bb)
  =
  \sum_{i,j}\Gamma_{\eps,ij}(\bC(\theta))
  \frac{\partial \bC_{ij}(\theta)}{\partial\theta}.
\]
\end{proposition}

\paragraph{Bias and Sinkhorn divergence.}
Small \(\eps\) gives a plan close to a sparse OT plan, but can be numerically unstable. Large \(\eps\) gives a smoother plan but introduces entropic bias. The Sinkhorn divergence corrects part of this bias by subtracting self-transport terms, with the appropriate self-cost matrices on the source and target supports.

\subsection{Unbalanced transport and mass variation}
\label{app:toolbox-unbalanced}

Balanced OT compares measures with the same total mass and forces all mass to be matched. Unbalanced OT compares finite positive measures and allows mass to be created, destroyed or left unmatched at a penalty. This is natural for exposure growth, lapse, new business, changing portfolio volume, claim count variation and climate exposure changes.

Let \(\mu\in\mathcal M_+(\mathcal X)\) and \(\nu\in\mathcal M_+(\mathcal Y)\) be finite positive measures, not necessarily of the same total mass. A generic unbalanced formulation is
\[
  \inf_{\gamma\ge0}
  \left\{
  \int c(x,y)\,\gamma(\dd x,\dd y)
  +
  \tau_1 D((p_1)_{\push}\gamma\mid\mu)
  +
  \tau_2 D((p_2)_{\push}\gamma\mid\nu)
  \right\},
\]
where \(D\) is a divergence between finite positive measures, often a relative entropy, and \(\tau_1,\tau_2>0\) control the cost of changing mass.

\begin{remark}[Unbalanced OT is not automatically a distance]
\label{rem:app-uot-not-distance}
The KL-penalized value above is a useful comparison criterion, but it is not automatically a metric. Metric structures such as Hellinger--Kantorovich or Wasserstein--Fisher--Rao require specific dynamic or static formulations.
\end{remark}

The plan \(\gamma\) describes the mass actually transported. The deviations between its marginals and \(\mu,\nu\) describe source mass that is removed and target mass that is created. In actuarial terms, these terms may represent lapses, new policies, portfolio expansion, claim-count changes or unmatched exposure.

\subsection{Schr\"odinger bridges and stochastic transport paths}
\label{app:toolbox-schrodinger}

A Schr\"odinger bridge is an entropic dynamic analogue of optimal transport. Instead of choosing a least-cost deterministic displacement between two endpoint laws, one chooses the most likely stochastic path measure, relative to a reference process, subject to endpoint constraints.

\begin{definition}[Schr\"odinger bridge]
\label{def:app-schrodinger-bridge}
Let \(\Pbb\) be a reference path measure on trajectories \((\bX_t)_{t\in[0,1]}\). Given endpoint laws \(\mu_0\) and \(\mu_1\), the problem is
\[ \highlight{GoldenRod}{
  \min_{\Qbb}
  \KL(\Qbb\mid\Pbb)
  \quad\text{subject to}\quad
  (\bX_0)_{\push}\Qbb=\mu_0,
  \text{ and }
  (\bX_1)_{\push}\Qbb=\mu_1}
\]
\end{definition}

The static two-time marginal of this problem is first a KL projection onto the set of couplings with marginals \(\mu_0\) and \(\mu_1\). When the two-time reference law has a Gibbs form, this projection becomes an entropic OT problem. The resulting bridge produces a stochastic transition from \(\mu_0\) to \(\mu_1\), unlike displacement interpolation, which gives a deterministic least-action path once an optimal plan has been chosen.

\begin{actuarialbox}
Schr\"odinger bridges are useful when a stochastic path is more realistic than a deterministic map: climate trajectories, reserving scenario evolution, epidemic risk and cyber incidents.
% To be restored in v2: uncertain counterfactual paths.
\end{actuarialbox}

\subsection{A compact glossary}
\label{app:toolbox-glossary}

\begin{table}[h!]
\centering
\small
\begin{tabular}{p{0.25\linewidth}p{0.43\linewidth}p{0.24\linewidth}}
\toprule
Term & Meaning in the notes & Caution \\
\midrule
Probability measure & Distribution of losses, premiums, reserves or scenarios & Not only a scalar summary \\
Push-forward & Law after applying a map \(T\) & Acts on measures \\
Pull-back & Composition of a function with \(T\) & Acts on functions \\
Coupling & Joint law with prescribed marginals & Not a causal model without further assumptions \\
Product coupling & Independent coupling \(\mu\otimes\nu\) & Only one feasible coupling \\
Transport map & Deterministic push-forward \(T_{\push}\mu=\nu\) & May not exist \\
Transport plan & Coupling optimized for a cost & May split mass \\
Kernel & Conditional law \(K(x,\dd y)\) & Generalizes deterministic maps \\
Ground cost & Local cost \(c(x,y)\) & Encodes modelling choices \\
Kantorovich potential & Dual variable \(\varphi,\psi\) & Shadow price / certificate \\
Brenier potential & Convex function \(\Phi\) with \(T=\nabla\Phi\) & Specific to quadratic OT \\
Contact set & Where \(\varphi(x)+\psi(y)=c(x,y)\) & Contains optimal support \\
Barycentric projection & Conditional mean of a plan & Usually not an exact push-forward \\
Wasserstein projection & Closest feasible law in \(\Wp{p}\) & May be non-unique \\
Bregman projection & Projection for a Bregman divergence & Not a metric projection \\
Geodesic & Shortest path in Wasserstein space & Not a density mixture \\
Displacement interpolation & Mass moved along optimal pairings & May depend on the chosen plan \\
Barycenter & Fr\'echet mean in Wasserstein space & Not a mixture \\
Sinkhorn scaling & Vectors \(\bu,\bv\) enforcing marginals & Depends on \(\eps\) \\
Entropic potential & \(\eps\log \bu,\eps\log \bv\) & Smooth dual variable \\
Unbalanced OT & Transport with relaxed marginals & Allows mass variation \\
Schr\"odinger bridge & Entropic stochastic path matching endpoints & Depends on reference dynamics \\
\bottomrule
\end{tabular}
\caption{Glossary of mathematical terms used repeatedly in the notes.}
\label{tab:app-ot-glossary}
\end{table}

\newpage
\lecture{Cost Engineering for Actuarial Optimal Transport}\label{app:cost}\label{app:cost-engineering}

\begin{readingbox}
This appendix is about a modelling choice that is sometimes hidden behind
optimal transport notation: the ground cost. General references on optimal
transport explain how a cost selects a coupling; see \cite{villani2008oldnew,
peyre2019computational,galichon2016economics}. In actuarial applications, the
cost is not only geometric. It encodes money, time, regulation, actionability,
contractual constraints and governance. Choosing it is therefore an actuarial modelling task. For compositional or categorical states, the ground geometry may be the geometry of a simplex rather than a Euclidean coordinate system; see
\cite{aitchison1982statistical,egozcue2003isometric,palwong2020multiplicative,baxendalewong2022random,fernandesmachado2025categoricalsimplex}.
\end{readingbox}

\subsection{Why cost engineering matters}\label{app:cost:why}

An optimal transport problem has the form
\[
  \inf_{\gamma\in\Coup(\mu,\nu)}
  \int_{\mathcal X\times\mathcal Y} c(\bx,\by)\,\dd\gamma(\bx,\by).
\]
The two marginals \(\mu\) and \(\nu\) describe the distributions to be matched.
The coupling \(\gamma\) describes how mass is paired. The cost \(c(\bx,\by)\) says which pairings are cheap, expensive, allowed or forbidden.

In a mathematical introduction, one often takes
\[
  c(\bx,\by)=\|\bx-\by\|^p.
\]
This is a good default when the state space is Euclidean and all coordinates have comparable meaning. But actuarial objects are rarely that simple. A state
may contain monetary losses, premium levels, categorical rating variables, territory codes, legal contract terms, policyholder scores, payment calendars and exposure measures. A single Euclidean norm can then be misleading.

\begin{actuarialbox}
Cost engineering is the step where the actuary translates the question
``how different are these two distributions?'' into the question ``what is the
actuarial cost of matching state \(x\) with state \(y\)?'' The answer depends
on the application.
\end{actuarialbox}

There are three distinct roles for the cost.

\begin{enumerate}[label=(\roman*)]
\item \textbf{Geometry.} The cost defines what it means for two risks, two
claims, two policies, two territories or two scenarios to be close.

\item \textbf{Governance.} The cost can encode constraints: legally fixed variables should not be changed, illegal contract transformations are forbidden, large contractual jumps are penalized, and non-actionable covariates receive infinite or very large cost.

\item \textbf{Economics.} The cost determines the unit of the Wasserstein value:
dollars, squared dollars, log-points, kilometres, duration-weighted cash-flow
movement, or a dimensionless normalized index.

\end{enumerate}

\begin{table}[h]
\centering
\small
\begin{tabular}{p{0.23\textwidth}p{0.30\textwidth}p{0.35\textwidth}}
\toprule
Question & Typical cost scale & Main diagnostic \\
\midrule
Loss severity drift & monetary or log-monetary displacement & body/tail contribution \\
% To be restored in v2.
% Tariff disruption & absolute or relative premium movement & paired versus distributional change \\
% To be restored in v2.
% Fairness repair & score, logit, rank or constrained cost & group-wise repair size and calibration loss \\
Categorical probabilities & simplex or log-ratio geometry & boundary behaviour and category substitution \\
Cash-flow comparison & discounted, cumulative or duration-aware movement & level-shape decomposition \\
Portfolio alignment with different features & Gromov--Wasserstein or fused cost & quality of internal-distance matrices \\
\bottomrule
\end{tabular}
\caption{Cost engineering as a modelling decision. The same marginal laws may lead to different actuarial conclusions depending on the question encoded in the ground cost.}
\label{tab:app:cost-engineering-map}
\end{table}

The same pair of distributions can lead to very different transport plans under different costs. This is not a defect. It reflects different actuarial questions.

\subsection{A general template}\label{app:cost:template}

Let a state be written as
\[
  \bx=(x^{\mathrm{loss}},x^{\mathrm{prem}},x^{\mathrm{cov}},x^{\mathrm{geo}},
        x^{\mathrm{score}},x^{\mathrm{contract}},x^{\mathrm{cf}}),
\]
where the components may represent losses, premiums, covariates, geography, scores, contract terms and cash-flows. A general composite cost can be written as
\[
  c(\bx,\by)
  =
  \sum_{k=1}^K \lambda_k c_k(x_k,y_k)
  +
  \iota_{\mathcal A}(\bx,\by),
\]
where \(c_k\) are component costs, \(\lambda_k\ge 0\) are weights, and \(\iota_{\mathcal A}\) is a feasibility penalty:
\[
  \iota_{\mathcal A}(\bx,\by)
  =
  \begin{cases}
  0, & (\bx,\by)\in\mathcal A,\\
  +\infty, & (\bx,\by)\notin\mathcal A.
  \end{cases}
\]
The set \(\mathcal A\) contains admissible transformations. For example, it may forbid changing a legally fixed or non-actionable attribute, matching policies across incompatible contract types, or moving a claim scenario to an impossible geographical state.

\begin{definition}[Composite actuarial cost]\label{def:app:composite-actuarial-cost}
A composite actuarial cost is a cost of the form
\[
  c(\bx,\by)=\sum_{k=1}^K \lambda_k c_k(x_k,y_k)+\iota_{\mathcal A}(\bx,\by),
\]
where each component cost \(c_k\) measures movement along one actuarial coordinate, \(\lambda_k\ge 0\) is the exchange rate between that coordinate and other coordinates, and \(\mathcal A\) is the set of admissible pairs. The weights and the admissibility set are part of the model specification.
\end{definition}

\begin{definition}[Actuarial ground cost]\label{def:app:actuarial-ground-cost}
An actuarial ground cost is a measurable function
\[
  c:\mathcal X\times\mathcal Y\to[0,+\infty]
\]
chosen to encode the actuarial meaning of moving mass from state \(\bx\) to state \(\by\). It may be asymmetric, may take the value \(+\infty\), and need not be a metric.
\end{definition}

The cost does not always need to be symmetric. Increasing an insured amount, a coverage limit or a contribution by 20\% may have a different commercial or regulatory cost from decreasing it by 20\%. Moving a policyholder from a high-risk to a low-risk class may have a different interpretation from the reverse. Creating exposure in a region may not be the same as removing exposure from that region.

\begin{remark}[Metric versus cost]
A Wasserstein distance requires a metric-like ground cost \(d(x,y)^p\) if one wants metric properties. Many actuarial applications require only an optimal
transport \emph{cost}. For scenario design, stress testing or contract comparison, asymmetry and infinite penalties can be useful.
\end{remark}

\subsection{Loss amounts}\label{app:cost:loss-amounts}

Let \(x,y\ge 0\) denote two loss amounts or severity levels. Several costs are natural.

\paragraph{Absolute monetary displacement.}
\[
  c_1(x,y)=|x-y|.
\]
This cost is measured in monetary units. It is appropriate when the question is average claim amount displacement. It is robust and directly interpretable.

\paragraph{Squared monetary displacement.}
\[
  c_2(x,y)=(x-y)^2.
\]
This cost penalizes large movements more strongly. It is useful when tail mismatches should dominate the comparison, and it leads to the \(W_2\) geometry.

\paragraph{Relative or log-scale displacement.}
For heavy-tailed severities, absolute displacement may be dominated by a few large claims. A log-scale cost is
\[
  c_{\log}(x,y)
  =
  \left(\log(x+a)-\log(y+a)\right)^2,
\]
where \(a>0\) avoids singularity at zero. This cost compares multiplicative changes. Moving from \(100\) to \(200\) is then similar to moving from
\(10,000\) to \(20,000\).

\paragraph{Tail-weighted displacement.}
If one wants to emphasize large losses, one may use
\[
  c_{\mathrm{tail}}(x,y)
  =
  w(x,y)|x-y|^p,
\]
with, for instance,
\[
  w(x,y)=1+\eta\,\1\{x>u_0 \text{ or } y>u_0\},
\]
where \(u_0\) is a high loss threshold and \(\eta>0\). This makes movements
in the tail more expensive than movements in the body of the distribution.

\paragraph{Capped displacement.}
Sometimes extremely large claim differences should not completely dominate the
comparison. A capped cost is
\[
  c_{\mathrm{cap}}(x,y)=\min\{|x-y|^p,M\},
\]
where \(M>0\). This gives a bounded stress-diagnostic cost, but the resulting
transport value is no longer a standard \(W_p^p\) distance.

\begin{example}[Inflation versus tail deformation]
Suppose \(Y=(1+i)X\). Under \(|x-y|\), the cost is proportional to the average
monetary inflation effect. Under the log cost, the same transformation has
constant displacement \(\log(1+i)\) for all positive losses. The first cost
emphasizes large claims; the second emphasizes proportional change.
\end{example}

\begin{actuarialbox}
For losses, the first modelling decision is scale. Absolute costs answer a
monetary question. Log costs answer a multiplicative question. Tail-weighted
costs answer a capital or stress question.
\end{actuarialbox}

\subsection{Mixed actuarial covariates}\label{app:cost:mixed-covariates}

Policyholder states often contain continuous, ordinal, binary and categorical variables. Let
\[
  \bx=(x^{\mathrm{cont}},x^{\mathrm{ord}},x^{\mathrm{cat}},x^{\mathrm{bin}}).
\]
A useful cost is a weighted sum of normalized component distances:
\[
  c(\bx,\by)
  =
  \sum_{j\in\mathcal C} \lambda_j
  \left(\frac{x_j-y_j}{s_j}\right)^2
  +
  \sum_{j\in\mathcal O} \lambda_j d_j(x_j,y_j)^2
  +
  \sum_{j\in\mathcal K} \lambda_j \1\{x_j\ne y_j\}.
\]
Here \(\mathcal C\), \(\mathcal O\), and \(\mathcal K\) index continuous,
ordinal and categorical variables. The constants \(s_j\) are scale parameters,
such as robust standard deviations or actuarial materiality units.

\paragraph{Continuous variables.}
For variables such as age, insured value or claim history, one may use
\[
  c_j(x_j,y_j)=\left(\frac{x_j-y_j}{s_j}\right)^2.
\]
The scale \(s_j\) should not be chosen mechanically. It can be a standard
deviation, an interquartile range, or a business unit such as 10 years of age
or 10,000 dollars of insured value.

\paragraph{Ordinal variables.}
For bonus-malus level, credit-score band or risk tier, use the distance between
ordered levels:
\[
  c_j(x_j,y_j)=\left(\mathrm{rank}(x_j)-\mathrm{rank}(y_j)\right)^2.
\]

\paragraph{Categorical variables.}
For vehicle body type, occupation class or coverage type, a simple cost is
\[
  c_j(x_j,y_j)=\1\{x_j\ne y_j\}.
\]
A richer version uses a substitution matrix \(D_j(a,b)\):
\[
  c_j(x_j,y_j)=D_j(x_j,y_j).
\]
For example, moving between two similar vehicle classes may be cheaper than
moving from a private passenger car to a commercial truck.

\paragraph{Actionability and fixed variables.}
Some covariates should not be changed by a counterfactual construction or by a constrained matching rule. If \(A\) is a non-actionable, legally fixed or otherwise preserved variable, impose
\[
  c_A(x_A,y_A)
  =
  \begin{cases}
  0, & x_A=y_A,\\
  +\infty, & x_A\ne y_A.
  \end{cases}
\]
This forces transport to preserve that variable.

\begin{actuarialbox}
Mixed-variable costs are never purely statistical. They encode what the insurer
regards as comparable, actionable, legally admissible and economically material.
The cost matrix should therefore be documented like a model assumption.
\end{actuarialbox}

\subsection{Simplex-valued states and categorical probabilities}\label{app:cost:simplex}

Categorical variables often enter actuarial models through predicted probability
vectors. For instance, a policyholder may have probabilities over vehicle body
classes, occupation groups, coverage choices or claim-type categories. These
vectors live in a simplex rather than in an unconstrained Euclidean space.

For \(d\ge 2\), write the open probability simplex as
\[
  \Delta_d^\circ
  =
  \left\{\bx\in(0,1)^d:\sum_{r=1}^d x_r=1\right\}.
\]
The closed simplex \(\Delta_d\) allows zero components. The distinction matters:
log-ratio and Dirichlet-transport costs are naturally defined in the interior,
whereas empirical category probabilities may lie on or near the boundary.

\begin{definition}[Simplex-valued actuarial state]\label{def:app:simplex-valued-state}
A simplex-valued actuarial state is a vector
\(\bx=(x_1,\ldots,x_d)\in\Delta_d\) whose components represent probabilities or
shares over mutually exclusive categories. Examples include class-probability
outputs from a multinomial classifier, exposure shares by peril, payment-pattern
shares over development years, or territorial composition vectors.
\end{definition}

\paragraph{Why not use only Euclidean cost?}
The Euclidean cost \(\|x-y\|^p\) is easy to compute, but it treats the simplex as
an ordinary subset of \(\R^d\). This may be misleading because compositional
information is relative. Moving a component from \(.01\) to \(.02\) is small in
absolute terms but doubles the corresponding share; moving from \(.40\) to
\(.41\) has the same Euclidean size but a very different relative meaning. This
is the central message of compositional data analysis \cite{aitchison1982statistical,pawlowskyglahn2001geometric,egozcue2003isometric}.

\paragraph{Closure and log-ratio coordinates.}
For \(\bz\in\R_+^d\setminus\{0\}\), define the closure operation
\[
  \mathcal C(\bz)=\frac{\bz}{\sum_{r=1}^d z_r}\in\Delta_d.
\]
Log-ratio transforms move the simplex into a Euclidean space. Common choices are
\[
  \operatorname{alr}(\bx)
  =
  \left(\log\frac{x_1}{x_d},\ldots,\log\frac{x_{d-1}}{x_d}\right)
  \in\R^{d-1},
\]
\[
  \operatorname{clr}(\bx)_r
  =
  \log\frac{x_r}{g(\bx)},
  \qquad
  g(\bx)=\left(\prod_{r=1}^d x_r\right)^{1/d},
\]
and the isometric log-ratio transform \(\operatorname{ilr}\), which chooses an
orthonormal basis of the clr hyperplane. One can then compute a Euclidean OT
problem in transformed coordinates and map the result back to the simplex.

\begin{definition}[Log-ratio transport cost]\label{def:app:log-ratio-cost}
Let \(h:\Delta_d^\circ\to\R^{d-1}\) be a log-ratio coordinate map, for example
\(\operatorname{alr}\) or \(\operatorname{ilr}\). The induced log-ratio cost is
\[
  c_h(\bx,\by)=\|h(\bx)-h(\by)\|^p,
  \qquad \bx,\by\in\Delta_d^\circ.
\]
This cost compares categorical probability vectors through relative proportions rather than through raw coordinate differences.
\end{definition}

\paragraph{Aitchison geometry.}
The Aitchison inner product on \(\Delta_d^\circ\) may be written as
\[
  \langle \bx,\by\rangle_A
  =
  \frac1d\sum_{r<s}
  \log\frac{x_r}{x_s}\log\frac{y_r}{y_s},
\]
with associated Aitchison distance \(d_A(\bx,\by)=\|\operatorname{clr}(\bx)-\operatorname{clr}(\by)\|\), up to the chosen normalization. This is useful when category probabilities should be compared by ratios rather than by absolute gaps.

\paragraph{Dirichlet or multiplicative transport cost.}
An alternative is to work directly on the simplex. Following the multiplicative
transport geometry of \cite{palwong2020multiplicative} and related Dirichlet
transport ideas \cite{baxendalewong2022random,pistoneshoaib2024unified}, one
can use the cost
\[
  c_D(\bx,\by)
  =
  \log\left(\frac1d\sum_{r=1}^d\frac{y_r}{x_r}\right)
  -
  \frac1d\sum_{r=1}^d \log\left(\frac{y_r}{x_r}\right),
  \text{ where } \bx,\by\in\Delta_d^\circ .
\]
By Jensen's inequality, \(c_D(\bx,\by)\ge 0\), with equality when \(\by/\bx\) is constant
componentwise, hence, on the simplex, when \(\bx=\by\). This cost penalizes relative
composition changes and is asymmetric in general.

\begin{definition}[Simplex transport cost]\label{def:app:simplex-transport-cost}
A simplex transport cost is a ground cost on \(\Delta_d\) or \(\Delta_d^\circ\)
chosen to respect the compositional meaning of category probabilities. Typical
examples are Euclidean costs, log-ratio costs, Aitchison costs, Hellinger costs, Jensen--Shannon costs, KL-type costs and Dirichlet transport costs.
\end{definition}

\begin{remark}[Boundary handling]\label{rem:app:simplex-boundary}
Costs involving logarithms require strictly positive components. In empirical work, one must specify a boundary convention: add a small pseudo-count, group
rare categories, trim observations too close to the boundary, or use a bounded
alternative such as Hellinger or Jensen--Shannon cost. This is not a numerical
detail; it changes the meaning of categorical counterfactuals and rare-class transport.
\end{remark}

\paragraph{From categories to the simplex.}
One practical route is to replace a categorical variable \(C\in\{1,\ldots,d\}\)
by a predicted probability vector
\[
  \widehat p(\bx)=
  \left(\widehat\Pbb(C=1\mid \bx),\ldots,
        \widehat\Pbb(C=d\mid \bx)\right)\in\Delta_d.
\]
A transport problem can then be run on \(\widehat p(\bX)\) rather than on arbitrary
integer labels. This avoids imposing an artificial order on the categories and is
particularly useful for counterfactual and categorical-stability analyses of categorical variables; see the methodology and ternary illustrations in
\cite{fernandesmachado2025categoricalsimplex}.

\begin{actuarialbox}
For categorical variables, the cost should not depend on arbitrary label names.
A simplex representation lets the actuary compare uncertainty over categories:
occupation mixes, vehicle-body probabilities, claim-type probabilities, payment
shares or territorial compositions. The modelling choice then becomes the
geometry of the simplex.
\end{actuarialbox}

\paragraph{Simplex cost matrix in R.}
The following small helper computes three cost matrices for simplex-valued data:
Euclidean, Aitchison/clr, and Dirichlet transport. The parameter \texttt{eps}
implements a pseudo-count for boundary stability.

\begin{lstlisting}[language=R]
closure <- function(Z, eps = 1e-8) {
  Z <- pmax(as.matrix(Z), eps)
  Z / rowSums(Z)
}

clr <- function(Z, eps = 1e-8) {
  Z <- closure(Z, eps)
  logZ <- log(Z)
  sweep(logZ, 1, rowMeans(logZ), "-")
}

simplex_cost <- function(X, Y, type = c("euclidean", "aitchison", "dirichlet"),
                         eps = 1e-8) {
  type <- match.arg(type)
  X <- closure(X, eps)
  Y <- closure(Y, eps)

  if (type == "euclidean") {
    D <- as.matrix(dist(rbind(X, Y)))
    return(D[seq_len(nrow(X)), nrow(X) + seq_len(nrow(Y))]^2)
  }

  if (type == "aitchison") {
    Xc <- clr(X, eps)
    Yc <- clr(Y, eps)
    D <- as.matrix(dist(rbind(Xc, Yc)))
    return(D[seq_len(nrow(X)), nrow(X) + seq_len(nrow(Y))]^2)
  }

  # Dirichlet / multiplicative cost
  outer(seq_len(nrow(X)), seq_len(nrow(Y)), Vectorize(function(i, j) {
    r <- Y[j, ] / X[i, ]
    log(mean(r)) - mean(log(r))
  }))
}
\end{lstlisting}

\subsection{Geographical costs}\label{app:cost:geography}

Geography appears in property insurance, climate risk, motor insurance, health
access, catastrophe modelling and territorial pricing. The choice of geographical
cost should reflect the mechanism being modelled.

\paragraph{Euclidean or great-circle distance.}
If locations are coordinates \(\bs,\bt\in\R^2\), one may take
\[
  c_{\mathrm{geo}}(\bs,\bt)=\|\bs-\bt\|^p.
\]
On the sphere, replace Euclidean distance by great-circle distance. This is
appropriate when physical distance is the relevant notion of closeness.

\paragraph{Administrative adjacency.}
If locations are postal codes, municipalities or rating territories, physical
distance may be less relevant than adjacency in a graph. Let \(G=(V,E)\) be a
graph of territories. Define
\[
  c_{\mathrm{graph}}(i,j)=\mathrm{dist}_G(i,j)^p,
\]
where \(\mathrm{dist}_G\) is shortest-path distance. This cost respects the
territorial structure used by the insurer.

\paragraph{Hazard-based distance.}
Two territories can be geographically far but actuarially similar, or close but
hazard-wise different. Let \(h(s)\in\R^q\) be a vector of hazard features:
flood depth, wildfire score, wind speed, earthquake intensity, heat exposure.
Then
\[
  c_{\mathrm{haz}}(\bs,\bt)=\|h(\bs)-h(\bt)\|^2.
\]
This compares territories by risk profile rather than by location.

\paragraph{Hybrid distance.}
A hybrid geographical cost is
\[
  c(\bs,\bt)
  =
  \lambda_{\mathrm{km}} d_{\mathrm{km}}(\bs,\bt)^2
  +
  \lambda_{\mathrm{haz}}\|h(\bs)-h(\bt)\|^2
  +
  \lambda_{\mathrm{admin}} \1\{\mathrm{region}(\bs)\ne\mathrm{region}(\bt)\}.
\]
This is useful when both physical proximity and administrative boundaries
matter.

\begin{example}[Climate scenario matching]
Let \(\mu\) be the distribution of insured exposure by location today and
\(\nu\) the exposure distribution under a future climate scenario. A geographical
OT problem can match current and future exposure masses. If the cost is based
on hazard vectors, the transport plan answers: which current risk profiles are
closest to future risk profiles?
\end{example}

\subsection{When spaces do not align: Gromov--Wasserstein costs}
\label{sec:gromov-wasserstein-actuarial}

The transport problems considered so far compare measures living on the same
space, or at least on spaces that can be embedded into a common ground-cost
framework. This is often reasonable: two loss distributions live on \(\R_+\), two premium distributions live on \(\R_+\), two cash-flow vectors live on \(\R^T\), and two model-output distributions live on the same output scale.

In practice, however, actuarial objects are sometimes defined on different spaces.
Two portfolios may use different rating variables. Two countries may have different
geographical partitions. Two insurers may encode coverages differently. Two cyber
portfolios may involve different dependency networks. In such cases, it is not
obvious how to define a pointwise cost \(c(x,y)\), because \(x\) and \(y\) do not
belong to the same type of space.

Gromov--Wasserstein transport addresses this problem by comparing internal
geometries rather than points directly; see \cite{memoli2011gromov,peyre2016gromov,vayer2019optimal}. Let \((X,d_X,\mu)\) and \((Y,d_Y,\nu)\) be
two metric measure spaces. Instead of specifying a cost between \(x\in X\) and
\(y\in Y\), one compares pairwise distances:
\[
  d_X(x,x')
  \quad\text{and}\quad
  d_Y(y,y').
\]
A Gromov--Wasserstein discrepancy is defined by
\[
\highlight{GoldenRod}{  \GW_p^p(\mu,\nu)
  =
  \inf_{\gamma\in\Coup(\mu,\nu)}
  \iint
  \left|
    d_X(x,x')-d_Y(y,y')
  \right|^p
  \,\dd\gamma(x,y)\,\dd\gamma(x',y')}
\]
The coupling \(\gamma\) does not match points because their coordinates are
directly comparable. It matches points because their relational structure is
similar.

\begin{actuarialbox}
Ordinary Wasserstein transport asks whether two distributions place mass in
similar locations. Gromov--Wasserstein transport asks whether two portfolios have
similar internal structures. This is useful when the variables, territories, coverages
or networks are not directly aligned.
\end{actuarialbox}

\subsubsection{Examples in actuarial modelling}\label{app:cost:gw-examples}

\paragraph{Different geographical partitions.}
Suppose one catastrophe model uses postal codes and another uses municipalities.
There is no one-to-one correspondence between cells. Still, each space has an
internal distance, for example travel distance, exposure-weighted distance, or
hazard correlation distance. Gromov--Wasserstein compares the two spatial
exposure distributions through these internal distances.

\paragraph{Different rating structures.}
One insurer may classify motor policies by age, vehicle type and region, while
another uses telematics clusters and claims-history summaries. A direct cost between classes is arbitrary. But within each portfolio, one can define a distance between classes based on claim frequency, severity, lapse, exposure level or reserve response. A
Gromov--Wasserstein coupling then matches classes with similar relational roles.

\paragraph{Cyber dependency networks.}
Cyber risk is often structured by networks: firms share cloud providers, software
dependencies, managed-service providers or authentication systems. Two cyber
portfolios may not contain the same firms, but they may have similar dependency
graphs. A graph distance or diffusion distance can be used as \(d_X\) and \(d_Y\),
and Gromov--Wasserstein compares the network structures.

\paragraph{Contract and guarantee structures.}
Life or P\&C contracts may differ across markets. One space may contain deductibles,
limits and exclusions; another may contain riders and embedded guarantees. If an
internal distance between contracts can be defined through cash-flow behavior or
risk response, then Gromov--Wasserstein provides a way to compare portfolios
without requiring identical contract features.

\subsubsection{Discrete formulation}\label{app:cost:gw-discrete}

For finite spaces \(X=\{x_1,\ldots,x_n\}\) and \(Y=\{y_1,\ldots,y_m\}\), let
\[
  \bD^X_{ii'}=d_X(x_i,x_{i'}),
  \qquad
  \bD^Y_{jj'}=d_Y(y_j,y_{j'}).
\]
Let \(\ba\in\R_+^n\) and \(\bb\in\R_+^m\) be probability weights. A coupling is a matrix \(\bP\in\R_+^{n\times m}\) such that
\[
  \bP\1_m=\ba,
  \qquad
  \bP^\top\1_n=\bb.
\]
The discrete Gromov--Wasserstein objective is
\[
  \inf_{\bP}
  \sum_{i,i'=1}^n
  \sum_{j,j'=1}^m
  \left|
    \bD^X_{ii'}-\bD^Y_{jj'}
  \right|^p
  \bP_{ij}\bP_{i'j'}.
\]
This is no longer a linear program in \(\bP\). The objective is quadratic in the coupling, which makes the problem computationally more difficult than classical
discrete OT.

\begin{computebox}
In a finite actuarial application, the key inputs for Gromov--Wasserstein are not
only portfolio weights \(\ba,\bb\). They are the two internal distance matrices \(\bD^X\) and \(\bD^Y\). Most of the actuarial modelling work lies in constructing and
validating these matrices.
\end{computebox}

\subsubsection{Fused Gromov--Wasserstein}\label{app:cost:fused-gw}

Sometimes part of the structure is aligned and part is not. For example, two
portfolios may both have claim frequency and exposure level, but use different geographical or contractual features. One can then combine an ordinary transport
cost with a Gromov--Wasserstein structural term. A fused objective has the form
\[
  \inf_{\gamma\in\Coup(\mu,\nu)}
  \int c(x,y)\,\dd\gamma(x,y)
  +
  \lambda
  \iint
  |d_X(x,x')-d_Y(y,y')|^p
  \,\dd\gamma(x,y)\,\dd\gamma(x',y'),
\]
where \(c(x,y)\) compares aligned features and the second term compares internal
structures.

\begin{remark}[Cost engineering warning]
Gromov--Wasserstein does not remove modelling choices. It moves them from a
cross-space cost \(c(x,y)\) to within-space distances \(d_X\) and \(d_Y\). These
distances must be justified actuarially: geographical proximity, similarity of loss
behavior, similarity of contract response, or network dependence.
\end{remark}

\subsubsection{When to use it}\label{app:cost:gw-when-use}

Gromov--Wasserstein is useful when:
\begin{itemize}
\item the two portfolios have no common covariate representation;
\item the relevant object is a graph, network or relational structure;
\item geography or coverage categories differ across countries;
\item contract features are not directly comparable;
\item one wants to align clusters rather than individual observations.
\end{itemize}
It is less appropriate when the variables already live in a common, meaningful
metric space. In that case, ordinary Wasserstein transport is simpler, more stable
and easier to interpret.

\begin{actuarialbox}
Use Wasserstein when locations are comparable. Use Gromov--Wasserstein when
relationships are comparable.
\end{actuarialbox}

\subsection{Scores and predicted probabilities}\label{app:cost:scores-probabilities}

Let \(s,t\) denote two model scores, predicted conditional means, claim probabilities or risk rankings. The appropriate cost depends on the scale of the score.

\paragraph{Raw-score cost.}
\[
  c(s,t)=|s-t|^p.
\]
This is natural if the score is already on a meaningful monetary or risk scale. For a predicted conditional mean, \(|s-t|\) is a displacement on the scale of the target variable.

\paragraph{Probability-score cost.}
If \(s,t\in(0,1)\) are probabilities, raw differences may underweight changes
near 0 or 1. A logit-scale cost is
\[
  c_{\mathrm{logit}}(s,t)
  =
  \left|
  \log\frac{s}{1-s}
  -
  \log\frac{t}{1-t}
  \right|^p.
\]
This compares odds rather than probabilities.

\paragraph{Vector predicted probabilities.}
When a model outputs a full probability vector over categories, the state is
\(s,t\in\Delta_d\), not \(s,t\in(0,1)\). In that case, use the simplex costs of
Section~\ref{app:cost:simplex}: Euclidean cost for absolute probability changes,
Aitchison or log-ratio cost for relative category composition, Hellinger or
Jensen--Shannon cost for bounded comparisons, and Dirichlet transport cost for a
multiplicative simplex geometry.

\paragraph{Rank cost.}
Sometimes only ranking matters. If \(R_s\) and \(R_t\) denote empirical ranks,
use
\[
  c_{\mathrm{rank}}(s,t)=|R_s-R_t|^p.
\]
This cost is useful for audits focused on ordering rather than score calibration.

\paragraph{Calibration-aware cost.}
If the score \(s\) is interpreted as a conditional mean, it may be useful to
penalize changes in both score and calibration residual:
\[
  c((s,r),(t,u))
  =
  |s-t|^2 + \lambda |r-u|^2,
\]
where \(r=L-s\) and \(u=L'-t\) are residual-type quantities. This is useful
when score repair should not destroy loss-ratio diagnostics.

\iffalse
% To be restored in v2.
\begin{example}[Fairness post-processing]
Suppose \(Z\mid S=s\sim\mu_s\) is a score distribution and \(\bar\mu\) is a
barycentric target. The cost \(|z-z'|^2\) yields a minimal squared score repair.
A logit cost yields a repair that is smoother on the odds scale. A cost with
\(+\infty\) constraints can forbid changes that would reverse selected rank
orders or violate regulatory rules.
\end{example}
\fi

\subsection{Contracts and coverage terms}\label{app:cost:contracts}

Insurance contracts are not only points in \(\R^d\). A contract is a function
that maps a gross loss \(\ell\) to an indemnity payment. For example, a deductible
\(d\) and limit \(M\) define
\[
  I_{d,M}(\ell)=\min\{(\ell-d)_+,M\}.
\]
A natural distance between two contracts \(a\) and \(b\) is a distance between
their indemnity functions:
\[
  c_{\mathrm{contract}}(a,b)
  =
  \int_0^\infty
  \left|I_a(\ell)-I_b(\ell)\right|^p
  \omega(\dd \ell),
\]
where \(\omega\) is a reference loss measure. It may be a fitted severity law, an
exposure-weighted severity law, or a grid measure emphasizing relevant loss
levels.

\paragraph{Deductibles and limits.}
For two contracts \(a=(d,M)\) and \(b=(d',M')\), a simpler parameter cost is
\[
  c_{\mathrm{param}}(a,b)
  =
  \lambda_d |d-d'|^p + \lambda_M |M-M'|^p.
\]
The functional cost is often more actuarial because it compares actual payments
rather than contract labels.

\paragraph{Coverage compatibility.}
Some transformations should be forbidden. A liability contract should not be
matched to a property contract unless the application permits this. Write
\[
  c(a,b)=c_{\mathrm{contract}}(a,b)+\iota\{\mathrm{type}(a)=\mathrm{type}(b)\}.
\]

\begin{example}[Comparing product portfolios]
Let \(\mu\) be the distribution of contract terms in an old book and \(\nu\) the
distribution in a new book. Using the indemnity-function cost, OT compares
portfolios by the payments their contracts generate across a loss grid. This is
more meaningful than comparing deductibles and limits separately when contract
features interact.
\end{example}

\subsection{Reinsurance treaties}\label{app:cost:reinsurance-treaties}

A reinsurance treaty is also a transformation of losses. If \(L\) is the gross
loss and \(T\) is the ceded-loss function, then
\[
  L^{\mathrm{ceded}}=T(L),
  \qquad
  L^{\mathrm{net}}=L-T(L).
\]
A cost between treaties \(T_a\) and \(T_b\) may be defined by
\[
  c_{\mathrm{treaty}}(a,b)
  =
  \E_{L\sim\omega}
  \left[ |T_a(L)-T_b(L)|^p \right],
\]
where \(\omega\) is a reference gross-loss law.

For layer treaties, the cost should reflect attachment points, limits and tail
importance. For example,
\[
  c_{\mathrm{layer}}(a,b)
  =
  \int_0^\infty
  |T_a(\ell)-T_b(\ell)|^2 w(\ell)\,\omega(\dd \ell),
\]
with \(w(\ell)\) increasing in \(\ell\) if tail treaty differences are more important.

\begin{actuarialbox}
For contracts and reinsurance, costs defined on parameters can be misleading.
Two treaties with different parameters may have similar indemnity functions on
the relevant loss range. Conversely, small parameter changes near the tail can
have large capital effects.
\end{actuarialbox}

\subsection{Cash-flow vectors}\label{app:cost:cashflows}

Let
\[
  \bx=(x_1,\ldots,x_T)
  \text{ and }
  \by=(y_1,\ldots,y_T)
\]
be two future cash-flow vectors, for instance reserve payments by development
year or liability cash-flows by calendar year.

\paragraph{Discounted Euclidean cost.}
Let \(v_t=(1+i)^{-t}\) be a discount factor. A discounted cost is
\[
  c_{\mathrm{disc}}(\bx,\by)
  =
  \sum_{t=1}^T v_t |x_t-y_t|^p.
\]
This compares cash-flows in present-value units.

\paragraph{Cumulative cash-flow cost.}
Sometimes timing differences matter through cumulative payments. Define
\[
  C_t(\bx)=\sum_{s=1}^t x_s.
\]
A cumulative cost is
\[
  c_{\mathrm{cum}}(\bx,\by)
  =
  \sum_{t=1}^T w_t |C_t(\bx)-C_t(\by)|^p.
\]
This penalizes persistent differences in payment development.

\paragraph{Level-shape decomposition.}
Write total payment
\[
  S_{\bx}=\sum_{t=1}^T x_t
\]
and normalized shape
\[
  q_t(\bx)=\frac{x_t}{S_{\bx}},
\]
when \(S_{\bx}>0\). A cost may separate level and timing shape:
\[
  c_{\mathrm{shape}}(\bx,\by)
  =
  \lambda_S |S_{\bx}-S_{\by}|^p
  +
  \lambda_q \sum_{t=1}^T |q_t(\bx)-q_t(\by)|^p.
\]
This is useful when total reserve amount and payment pattern have different
actuarial meanings.

\paragraph{Duration-based cost.}
Define payment duration
\[
  D(\bx)=\frac{\sum_{t=1}^T t x_t}{\sum_{t=1}^T x_t}.
\]
A duration-aware cost is
\[
  c(\bx,\by)
  =
  \lambda_S |S_{\bx}-S_{\by}|^2
  +
  \lambda_D |D(\bx)-D(\by)|^2
  +
  \lambda_R \sum_{t=1}^T v_t|x_t-y_t|^2.
\]

\begin{example}[Reserving model comparison]
Let \(\mu\) and \(\nu\) be predictive distributions of future reserve cash-flow
vectors from two stochastic reserving models. A componentwise Euclidean cost
compares payment amounts year by year. A cumulative cost compares development
paths. A level-shape cost separates reserve level uncertainty from payment
timing uncertainty.
\end{example}

\subsection{Choosing weights}\label{app:cost:choosing-weights}

Composite costs require weights \(\lambda_k\). These weights are not innocuous.
They determine the exchange rate between actuarial dimensions. There are several
reasonable calibration principles.

\paragraph{Statistical scaling.}
Set \(\lambda_k=1/s_k^p\), where \(s_k\) is a standard deviation, median absolute
deviation or interquartile range. This makes all coordinates roughly comparable.

\paragraph{Materiality scaling.}
Choose \(s_k\) as a business materiality unit: 10,000 dollars of loss, one risk tier, five years of age, one development year or another documented threshold. Then
\[
  c_k(x_k,y_k)=\left|\frac{x_k-y_k}{s_k}\right|^p
\]
means ``distance measured in materiality units.''

\paragraph{Economic scaling.}
Choose \(\lambda_k\) so that each component has an approximate monetary impact.
For instance, if a one-unit change in variable \(x_k\) changes expected loss by
\(b_k\), set
\[
  \lambda_k\propto b_k^2.
\]
This turns covariate movement into expected-loss movement.

\paragraph{Governance scaling.}
Some weights reflect policy rather than data. For example, one may assign a large weight to changes in legally fixed or non-actionable variables, or set the cost to \(+\infty\) to forbid such changes entirely.

\begin{computebox}
A practical rule is to report results under at least three cost specifications:
a baseline cost, a tail-sensitive cost, and a governance-constrained cost. If
conclusions change dramatically, the cost is driving the result and should be
part of the actuarial discussion.
\end{computebox}

\subsection{Cost diagnostics}\label{app:cost:diagnostics}

A good OT analysis should not report only the optimal value. It should also
inspect how the cost shaped the coupling.

\paragraph{Distribution of matched costs.}
Given an optimal plan \(\gamma^\star\), examine the distribution of
\[
  c(\bX,\bY), \qquad (\bX,\bY)\sim\gamma^\star.
\]
Large quantiles identify expensive parts of the matching.

\paragraph{Component contributions.}
If
\[
  c(\bx,\by)=\sum_k \lambda_k c_k(x_k,y_k),
\]
report
\[
  C_k=\int \lambda_k c_k(x_k,y_k)\,\dd\gamma^\star(\bx,\by).
\]
This decomposes the transport cost by actuarial dimension.

\paragraph{Dual potentials.}
When available, Kantorovich potentials identify where the source and target
marginals exert pressure. In portfolio drift, high source potentials may
identify profiles that are expensive to remove; high target potentials may
identify profiles that are expensive to create.

\paragraph{Sensitivity to weights.}
Compute the optimal value and key actuarial quantities for a grid of weights
\(\lambda_k\). For example,
\[
  \lambda_{\mathrm{tail}}\in\{0,1,5,10\}
\]
shows how sensitive the plan is to tail emphasis.

\paragraph{Boundary and feasibility diagnostics.}
For log, logit and simplex costs, check how many observations are close to the
boundary. Report the pseudo-count, clipping rule or grouping rule used to make
the cost finite. If a cost contains infinite penalties, report the fraction of
forbidden pairs in the cost matrix and verify that the feasible transportation
polytope is non-empty.

\subsection{Mini-lab: building a mixed actuarial cost matrix in R}\label{app:cost:lab-mixed-cost}

The following code builds a cost matrix between two small portfolios with
continuous, categorical, geographical and model-output components.

\begin{lstlisting}[language=R]
set.seed(123)

n <- 200
m <- 180

A <- data.frame(
  age = rnorm(n, 45, 12),
  value = rlnorm(n, 10, .4),
  region = sample(c("urban", "suburban", "rural"), n, replace = TRUE),
  score = runif(n),
  lat = rnorm(n, 45.5, .3),
  lon = rnorm(n, -73.6, .3)
)

B <- data.frame(
  age = rnorm(m, 48, 13),
  value = rlnorm(m, 10.1, .45),
  region = sample(c("urban", "suburban", "rural"), m, replace = TRUE),
  score = pmin(pmax(rbeta(m, 2.2, 4), 0), 1),
  lat = rnorm(m, 45.6, .35),
  lon = rnorm(m, -73.5, .35)
)

# robust scales
s_age <- IQR(c(A$age, B$age))
s_logvalue <- IQR(log(c(A$value, B$value)))
s_score <- IQR(c(A$score, B$score))
s_geo <- 0.25  # roughly a chosen materiality unit in degrees

# pairwise squared differences helper
outer_diff2 <- function(x, y, scale = 1) {
  outer(x, y, function(a, b) ((a - b) / scale)^2)
}

C_age <- outer_diff2(A$age, B$age, s_age)
C_value <- outer_diff2(log(A$value), log(B$value), s_logvalue)
C_score <- outer_diff2(A$score, B$score, s_score)
C_geo <- outer_diff2(A$lat, B$lat, s_geo) +
         outer_diff2(A$lon, B$lon, s_geo)
C_region <- outer(A$region, B$region, function(a, b) as.numeric(a != b))

lambda <- list(age = .5, value = 1.0, score = 1.5,
               geo = .25, region = 2.0)

C <- lambda$age * C_age +
     lambda$value * C_value +
     lambda$score * C_score +
     lambda$geo * C_geo +
     lambda$region * C_region

summary(as.vector(C))
image(C, main = "Mixed actuarial ground cost")
\end{lstlisting}

This matrix can then be used in a discrete OT solver or an entropic Sinkhorn
solver. The key point is that the cost has been documented component by
component.

\subsection{Mini-lab: cash-flow cost between reserving scenarios}\label{app:cost:lab-cashflow-cost}

The following code compares two sets of cash-flow scenarios using discounted,
cumulative and level-shape components.

\begin{lstlisting}[language=R]
set.seed(1)

n <- 150
m <- 150
T <- 8

simulate_cf <- function(n, level_shift = 0, speed = 1) {
  base_pattern <- dgamma(1:T, shape = 2 * speed, rate = .6)
  base_pattern <- base_pattern / sum(base_pattern)
  total <- rlnorm(n, meanlog = 12 + level_shift, sdlog = .35)
  noise <- matrix(rgamma(n * T, shape = 20, rate = 20), n, T)
  cf <- total * sweep(noise, 2, base_pattern, "*")
  cf / rowSums(cf) * total
}

X <- simulate_cf(n, level_shift = 0, speed = 1.0)
Y <- simulate_cf(m, level_shift = .05, speed = 1.2)

cashflow_cost <- function(x, y, i = .03,
                          lambda_level = 1,
                          lambda_disc = 1,
                          lambda_cum = .5,
                          lambda_dur = .5) {
  T <- length(x)
  v <- (1 + i)^(-(1:T))

  Sx <- sum(x); Sy <- sum(y)
  qx <- x / Sx; qy <- y / Sy
  Dx <- sum((1:T) * x) / Sx
  Dy <- sum((1:T) * y) / Sy

  level <- (log(Sx) - log(Sy))^2
  disc <- sum(v * (qx - qy)^2)
  cum  <- sum((cumsum(qx) - cumsum(qy))^2)
  dur  <- ((Dx - Dy) / T)^2

  lambda_level * level +
    lambda_disc * disc +
    lambda_cum * cum +
    lambda_dur * dur
}

C_cf <- outer(1:n, 1:m, Vectorize(function(i, j) {
  cashflow_cost(X[i, ], Y[j, ])
}))

summary(as.vector(C_cf))
image(C_cf, main = "Cash-flow scenario ground cost")
\end{lstlisting}

The cost separates total reserve level, discounted payment pattern, cumulative
development and duration. These components answer different actuarial questions.

\subsection{Checklist for cost engineering}\label{app:cost:checklist}

Before using an OT cost in an actuarial report, it is useful to answer the
following questions.

\begin{enumerate}[label=(\arabic*)]
\item What is the state space? Losses, covariates, territories, model outputs, contracts, treaties, cash-flows or scenarios?

\item What does it mean for two states to be close? Monetary closeness,
statistical closeness, legal admissibility, policyholder impact, or physical
proximity?

\item Are all variables actionable? Should some coordinates be preserved with
infinite cost?

\item Are changes symmetric? Is an increase as acceptable as a decrease?

\item What are the units of the transport value? Dollars, squared dollars,
log-points, normalized units, kilometres or discounted cash-flow units?

\item How are component weights calibrated? Statistical scale, materiality,
economic impact, expert judgement or regulation?

\item Does the cost emphasize the body or the tail of the distribution?

\item Does the cost respect contractual constraints and product definitions?

\item If the state is a probability vector or composition, does the cost respect
the simplex geometry and the boundary convention?

\item If spaces do not align, are the internal distances for a Gromov--Wasserstein
comparison actuarially meaningful?

\item How sensitive are the conclusions to reasonable alternative costs?

\item Can the optimal coupling be explained to an actuarial, regulatory or
business audience?
\end{enumerate}

\begin{actuarialbox}
The cost function is part of the actuarial model. It should be versioned,
justified, stress-tested and documented. In many OT applications, the main
source of model risk is not the solver but the ground cost.
\end{actuarialbox}

\subsection{What to remember}\label{app:cost:remember}

The ground cost is not a technical nuisance to be chosen after the actuarial question has been stated. It is the way the actuarial question is stated. A reasonable cost should specify the state space, the units, the allowed moves, the relative weights, the treatment of tails, and the treatment of categorical or simplex-valued components. When these choices are unclear, the correct response is not to search for a universal cost, but to report sensitivity across several credible costs.

\subsection{Exercises}\label{app:cost:exercises}

\begin{exercise}[Loss scale]\label{ex:B:1}
Let \(\mu\) be the law of \(X\) and let \(\nu\) be the law of \((1+i)X\), with
\(X>0\). Compare the interpretation of \(W_1(\mu,\nu)\) using absolute loss
cost and the transport cost based on \(|\log x-\log y|\). Which cost is more
appropriate for inflation monitoring? Which one is more appropriate for capital
impact?
\end{exercise}

\iffalse
% To be restored in v2.
\begin{exercise}[Premium asymmetry]\label{ex:B:2}
Let
\[
  c(x,y)=\alpha(y-x)_+ + \beta(x-y)_+.
\]
Show that this cost is asymmetric when \(\alpha\ne\beta\). Give an actuarial
situation where \(\alpha>\beta\) is reasonable. What would \(\alpha<\beta\)
mean?
\end{exercise}
\fi

\begin{exercise}[Categorical substitution matrix]\label{ex:B:3}
Consider vehicle classes \(\{\text{sedan},\text{SUV},\text{truck}\}\). Propose
a substitution-cost matrix \(D(a,b)\). Explain why using only
\(\1\{a\ne b\}\) may be too crude.
\end{exercise}

\begin{exercise}[Simplex cost]\label{ex:B:simplex-cost}
Let
\(x=(0.80,0.10,0.10)\), \(y=(0.70,0.20,0.10)\), and
\(z=(0.40,0.30,0.30)\). Compare the Euclidean distance between \(x\) and \(y\)
with the Euclidean distance between \(y\) and \(z\). Then discuss whether the
same comparison is reasonable on a relative-composition scale. What changes if
you use clr or Aitchison distance?
\end{exercise}

\iffalse
% To be restored in v2.
\begin{exercise}[Protected variables]\label{ex:B:4}
Suppose \(x=(z,s)\), where \(z\) is a score and \(s\) is a protected group.
Define a cost that permits changing \(z\) but forbids changing \(s\). Interpret
the resulting OT problem in a fairness repair setting.
\end{exercise}
\fi

\begin{exercise}[Geographical cost]\label{ex:B:5}
Let territories be nodes of a graph. Define a graph-based cost and compare it
with a Euclidean coordinate cost. Give an example where graph distance is more
relevant than physical distance.
\end{exercise}

\begin{exercise}[Contract distance]\label{ex:B:6}
Let
\[
  I_{d,M}(\ell)=\min\{(\ell-d)_+,M\}.
\]
For two contracts \((d,M)\) and \((d',M')\), define
\[
  c((d,M),(d',M'))
  =
  \int_0^\infty |I_{d,M}(\ell)-I_{d',M'}(\ell)|\,\omega(\dd\ell).
\]
Compute this numerically on a finite loss grid. Compare with the simpler
parameter cost \(|d-d'|+|M-M'|\).
\end{exercise}

\begin{exercise}[Cash-flow shape]\label{ex:B:7}
Construct two cash-flow vectors with the same total payment but different
payment timing. Compare them under discounted Euclidean cost, cumulative cost
and duration-based cost.
\end{exercise}

\begin{exercise}[Weight sensitivity]\label{ex:B:8}
Take the mixed-cost R mini-lab and vary \(\lambda_{\mathrm{region}}\) from 0 to
10. How does the distribution of cost-matrix entries change? If you solve an
OT problem with this cost, how would you expect the optimal matching to change?
\end{exercise}

\begin{exercise}[Gromov--Wasserstein cost engineering]\label{ex:B:gw-cost}
Consider two insurers using different rating classes. Propose an internal
distance matrix for each insurer based on expected loss, lapse, claim frequency or exposure level.
Explain why a Gromov--Wasserstein comparison may be more appropriate than an
ordinary Wasserstein comparison. What modelling risk is moved into the choice of
the two internal distance matrices?
\end{exercise}

\begin{exercise}[Cost report]\label{ex:B:9}
Choose one actuarial application: portfolio drift, reserving model comparison, climate scenario matching, contract comparison or cash-flow comparison. Write a one-page
cost-engineering report specifying the state space, component costs, weights,
constraints, units and sensitivity checks.
\end{exercise}

\newpage
% Appendix / insertable section
% Deterministic Mapping versus Diffusion Processes
% Suggested placement: after the counterfactual chapter, after Dynamic OT,
% or after the appendix on Sliced and Projected Wasserstein distances.

\lecture{Deterministic Mapping versus Diffusion Processes}
\label{app:deterministic-diffusion}

\begin{readingbox}
This appendix compares two ways of generating actuarial scenarios from a
baseline distribution. The first uses a deterministic map or a deterministic
flow. The second uses a stochastic process, typically a diffusion or an
entropy-minimizing bridge. The distinction is central in recent generative AI:
normalizing flows and neural ODEs learn deterministic transformations, whereas
score-based diffusion models, Schr\"odinger bridges, stochastic interpolants and
flow-matching formulations work with stochastic paths, conditional probability
paths or velocity fields attached to distributional curves; see
\cite{rezende2015variational,dinh2017density,chen2018neuralode,
sohldickstein2015deep,ho2020denoising,song2021scorebased,
anderson1982reverse,hyvarinen2005scorematching,
lipman2023flowmatching,liu2023flowstraight,albergo2023stochastic,
debortoli2021diffusion,leonard2014survey}. The actuarial question is not which paradigm is more
fashionable, but which one gives the right form of uncertainty, traceability,
tail behaviour and governance for losses, reserves, mortality tables or climate scenarios.
\end{readingbox}

\subsection{Why this distinction matters}\label{app:diffusion:why-distinction}

Throughout this appendix, $t\in[0,1]$ denotes an artificial interpolation or
modelling time unless stated otherwise. It need not be calendar time. We write
$T$ for a static transport map, $\phi_t$ for an ODE flow, $K(x,\dd y)$ for a
Markov kernel, and $P$ for a law on paths. This keeps separate four objects that
are easily confused: endpoint map, marginal curve, conditional transition, and
complete path law.

Optimal transport is often introduced through a map. Given two laws
\(\mu_0\) and \(\mu_1\), one searches for a transformation
\[
  T:\mathcal X\to\mathcal X,
  \qquad T_\#\mu_0=\mu_1,
\]
which is optimal for a chosen cost. If \(\bX_0\sim\mu_0\), the generated target sample is
\[
  \bX_1=T(\bX_0).
\]
This is a deterministic construction. Conditional on the starting point
\(\bX_0=\bx\), the endpoint is fixed.

Generative AI has made a closely related distinction prominent. One may learn a
single invertible map from a simple latent law to the data law, as in normalizing
flows. Or one may learn a path of distributions, either through an ordinary
differential equation or through a stochastic differential equation. In the first
case, randomness enters only through the initial latent variable. In the second,
additional randomness can enter along the path.

For actuarial science, this is not a technical detail. It changes the meaning of
scenario generation.

\begin{itemize}
\item A deterministic transport map is appropriate when one wants a reproducible
rank-preserving stress, a policy-transformation audit trail or a counterfactual matching rule.

\item A stochastic diffusion is appropriate when the path itself contains genuine
uncertainty: claim arrival, cyber contagion, climate evolution, mortality
improvement, reserve runoff, lapse behaviour, or economic scenarios.

\item A deterministic flow can still create complex distributions, but it does not
represent process noise conditional on the current state.

\item A diffusion can represent uncertainty along the path, but it is less
immediately traceable at the individual observation level.
\end{itemize}

\begin{actuarialbox}
A deterministic map answers: given this baseline scenario, what is its matched
target scenario? A diffusion process answers: what random evolution, starting
from this baseline state, can lead to the target distribution while respecting a
chosen stochastic dynamics?
\end{actuarialbox}

\subsection{Three levels: map, ODE flow, diffusion}\label{app:diffusion:three-levels}

It is useful to separate three objects.

\begin{definition}[Deterministic generator and stochastic generator]\label{def:app-diffusion-generator}
Let $\mu_0$ and $\mu_1$ be probability measures on a state space $\mathcal X$.
A deterministic generator is a measurable map $T:\mathcal X\to\mathcal X$ such
that $T_\#\mu_0=\mu_1$. A stochastic generator is a Markov kernel
$K:\mathcal X\times\mathcal B(\mathcal X)\to[0,1]$ such that
\[
  \mu_1(B)=\int K(x,B)\,\mu_0(\dd x),\qquad B\in\mathcal B(\mathcal X).
\]
The deterministic case corresponds to the degenerate kernel
$K_T(x,\dd y)=\delta_{T(x)}(\dd y)$.
\end{definition}

\begin{proposition}[Maps as degenerate kernels]\label{prop:app-diffusion-map-kernel}
Let $\bX_0\sim\mu_0$. If $T_\#\mu_0=\mu_1$, then
$\gamma_T=(\Id,T)_\#\mu_0$ is a coupling in $\Coup(\mu_0,\mu_1)$ and can be
written as $\gamma_T(\dd x,\dd y)=\mu_0(\dd x)\delta_{T(x)}(\dd y)$. Conversely,
a kernel-induced coupling $\mu_0(\dd x)K(x,\dd y)$ is induced by a map if and
only if $K(x,\cdot)$ is a Dirac mass for $\mu_0$-almost every $x$.
\end{proposition}

\begin{proof}
The first marginal of $(\Id,T)_\#\mu_0$ is $\mu_0$ and its second marginal is
$T_\#\mu_0=\mu_1$. The disintegration with respect to the first coordinate is
$\delta_{T(x)}$. Conversely, if $K(x,\cdot)=\delta_{T(x)}$ almost surely for a
measurable selector $T$, then the joint law is $(\Id,T)_\#\mu_0$.
\end{proof}

\paragraph{Static deterministic map.}
A map \(T\) sends \(\mu_0\) directly to \(\mu_1\):
\[
  \bX_1=T(\bX_0), \qquad T_\#\mu_0=\mu_1.
\]
When \(T\) is the optimal transport map, it is selected by minimizing
\[
  \int c(x,T(x))\,\dd\mu_0(x).
\]
In dimension one, for continuous distributions,
\[
  T(x)=F_1^{-1}(F_0(x)).
\]
This is the usual quantile map.

\paragraph{Deterministic ODE flow.}
Instead of jumping directly from \(\bX_0\) to \(\bX_1\), one may define a trajectory
\[
  \frac{\dd}{\dd t}\phi_t(\bx)=\bv_t(\phi_t(\bx)),
  \qquad \phi_0(\bx)=\bx,
\]
and set
\[
  \bX_t=\phi_t(\bX_0),
  \qquad \mu_t=(\phi_t)_\#\mu_0.
\]
The marginal laws satisfy the continuity equation
\[
  \partial_t\mu_t+\nabla\cdot(\mu_t \bv_t)=0.
\]
This is the deterministic language behind Benamou--Brenier dynamic optimal
transport, continuous normalizing flows and much of flow matching.

\paragraph{Stochastic diffusion process.}
A diffusion adds pathwise noise:
\[
  \dd \bX_t=\bb_t(\bX_t)\,\dd t+\sigma_t(\bX_t)\,\dd \bB_t.
\]
If \(\rho_t\) is the density of \(\bX_t\), it satisfies the Fokker--Planck equation
\[
  \partial_t\rho_t
  =
  -\nabla\cdot(\bb_t\rho_t)
  +\frac12\sum_{i,j}\partial_{ij}\bigl(a_{ij,t}\rho_t\bigr),
  \qquad a_t=\sigma_t\sigma_t^\top.
\]
The diffusion term makes the path stochastic even after conditioning on the initial state. Score-based diffusion models and diffusion Schr\"odinger bridges
belong to this family.

\begin{proposition}[Weak Fokker--Planck equation]\label{prop:app-diffusion-fokker-planck}
Let $\bX_t$ solve $\dd \bX_t=\bb_t(\bX_t)\,\dd t+\sigma_t(\bX_t)\,\dd \bB_t$ and let $a_t=\sigma_t\sigma_t^\top$. For every smooth compactly supported $f$,
\[
  \frac{\dd}{\dd t}\Ebb[f(\bX_t)]
  =\Ebb\left[\bb_t(\bX_t)\cdot\nabla f(\bX_t)
  +\frac12\operatorname{tr}\{a_t(\bX_t)\nabla^2 f(\bX_t)\}\right].
\]
When $\bX_t$ has a smooth density $\rho_t$, this weak equation is equivalent to the
Fokker--Planck equation displayed above.
\end{proposition}

\begin{proof}
Apply It\^o's formula to $f(\bX_t)$ and take expectations. The martingale term has
mean zero. Integrating by parts yields the density form.
\end{proof}

\begin{computebox}
The same endpoint law \(\mu_1\) can be produced by many mechanisms: a quantile
map, an ODE flow, a stochastic diffusion, a jump process, a mixture model, or a
Markov chain. The endpoint distribution alone does not identify the generating
mechanism. This is why path assumptions matter.
\end{computebox}

\subsection{A dictionary from generative AI to actuarial OT}\label{app:diffusion:dictionary-ai-actuarial}

The following table gives a useful translation.

\begin{center}
\begin{tabular}{lll}
\toprule
Generative AI object & Mathematical object & Actuarial interpretation \\
\midrule
Latent law & \(\mu_0\) & baseline, simple prior, neutral scenario law \\
Data law & \(\mu_1\) & observed portfolio, target stress, empirical law \\
Generator map & \(T_\#\mu_0=\mu_1\) & scenario transformer or counterfactual map \\
Normalizing flow & invertible \(T_\theta\) & reversible stress or density model \\
Neural ODE & \(\dot X_t=v_t(X_t)\) & deterministic scenario trajectory \\
Flow matching & learned velocity field & learned distributional motion \\
Score model & \(\nabla\log \rho_t\) & local direction of denoising or concentration \\
Diffusion model & stochastic path law & noisy scenario generator \\
Schr\"odinger bridge & entropy-minimal path law & least-surprising stochastic stress path \\
Guidance / conditioning & conditional path law & scenario under constraints or observations \\
\bottomrule
\end{tabular}
\end{center}

The actuarial analogue is clear. A generative model does not merely generate
pictures or text. Abstractly, it generates samples from a probability law. In actuarial applications, that law may be the distribution of annual losses,
a vector of future reserve cash-flows, a portfolio of policyholder profiles, a mortality surface, a cyber incident path, or a climate loss scenario.

\subsection{Deterministic maps as actuarial audit trails}\label{app:diffusion:deterministic-audit-trails}

A deterministic map has a strong audit-trail interpretation. If \(X_0=x\), then
the target scenario is \(T(x)\). This is useful when one wants to explain how a
baseline object was transformed.

\begin{example}[Severity stress]
Let \(\mu_0\) be a baseline severity distribution and \(\mu_1\) a stressed
severity distribution. The monotone transport map
\[
  T(x)=F_1^{-1}(F_0(x))
\]
keeps the rank of every claim scenario. A claim at the 95th percentile remains a
95th percentile claim after stress, but its amount is replaced by the stressed
95th percentile. This is simple, transparent and easy to document.
\end{example}

\iffalse
% To be restored in v2.
\begin{example}[Premium revision]
Let \(P^{\mathrm{old}}\) and \(P^{\mathrm{new}}\) be old and new premiums. A map
from the old premium distribution to the new one can be used to summarize the
revision. Under a monotone map, old premium ranks are preserved. Under a
multivariate map, one may preserve a richer notion of rank involving premium,
exposure, risk score and affordability.
\end{example}
\fi

\iffalse
% To be restored in v2.
\begin{example}[Fairness repair]
For a score \(Z\) and sensitive group \(S=s\), one may transport the law
\(\Law(Z\mid S=s)\) to a common barycenter. The repaired score is
\[
  Z^{\mathrm{rep}}=T_s(Z).
\]
This is deterministic conditional on the group and original score. That is a
feature when the objective is a reproducible post-processing rule.
\end{example}
\fi

The main strength of a deterministic map is accountability. The main limitation is
that it gives a single endpoint for each starting point. If the intervention or
future evolution is genuinely random, the map may understate uncertainty.

\subsection{Normalizing flows: invertible deterministic generators}\label{app:diffusion:normalizing-flows}

Normalizing flows learn an invertible map
\[
  T_\theta:\R^d\to\R^d,
  \qquad \bZ\sim\mu_0,
  \qquad \bX=T_\theta(\bZ)\sim\mu_\theta.
\]
The invertibility allows likelihood computation through the change-of-variables
formula
\[
  \log p_\theta(\bx)
  =
  \log p_0(T_\theta^{-1}(\bx))
  +
  \log\left|\det D T_\theta^{-1}(\bx)\right|.
\]
This is attractive when density evaluation matters. In actuarial terms, one could
fit a flow to a vector of losses or cash-flows and then simulate new scenarios.

\begin{definition}[Normalizing flow]\label{def:app-diffusion-normalizing-flow}
A normalizing flow is an invertible differentiable map $T_\theta$ that transports
a tractable base law $\mu_0$ to a model law $\mu_\theta=(T_\theta)_\#\mu_0$ and
allows density evaluation through the Jacobian change-of-variables formula. A
continuous normalizing flow replaces $T_\theta$ by the time-one map of an ODE.
\end{definition}

However, invertibility is also a constraint. It can make it difficult to represent
distributions with very different geometry unless the architecture is sufficiently
expressive. For actuarial state spaces with atoms, zero masses, categorical
variables, constraints and extreme tails, a naive normalizing flow may be a poor
model unless the state representation is engineered carefully.

\begin{actuarialbox}
A normalizing flow is a deterministic generator with a likelihood. This is useful
for scenario simulation and density estimation, but actuarial data often contain
mixed variables, atoms, truncation, censoring, contractual constraints and heavy
tails. These features should be modelled explicitly rather than hidden inside a
black-box map.
\end{actuarialbox}

\subsection{ODE flows and flow matching}\label{app:diffusion:ode-flow-matching}

Continuous normalizing flows replace a single map by the time-one solution of an
ODE:
\[
  \dot{\bX}_t=\bv_\theta(t,\bX_t),
  \qquad \bX_0\sim\mu_0.
\]
The endpoint \(\bX_1\) is still deterministic conditional on \(\bX_0\), but the map is
constructed gradually. The marginal laws satisfy
\[
  \partial_t\mu_t+\nabla\cdot(\mu_t \bv_\theta)=0.
\]

Flow matching starts from a chosen family of probability paths
\(t\mapsto\mu_t\) and learns a velocity field that realizes or approximates that
path. The path can be induced by a coupling between \(\mu_0\) and \(\mu_1\), by a
conditional probability path, by a Markov kernel, or by a stochastic process. Once
the velocity field is learned, one samples by solving an ODE from latent noise to
data.

This is close in spirit to dynamic optimal transport. In Benamou--Brenier, the
path and velocity solve
\[
  \inf_{(\mu_t,\bv_t)}
  \int_0^1\int \|\bv_t(\bx)\|^2\,\dd\mu_t(\bx)\,\dd t,
  \qquad
  \partial_t\mu_t+\nabla\cdot(\mu_t \bv_t)=0.
\]
Flow matching does not necessarily minimize this exact kinetic energy, but it uses
the same geometric idea: distributions are moved through a velocity field. This
should be read together with Lecture~\ref{lec:dynamic-ot-actuarial-trajectories}:
there, $\bv_t$ is an actuarial velocity field; here, it is also a learnable generator.

\begin{example}[Cash-flow scenario generation]
Let \(\bX\in\R^T\) be a vector of future reserve cash-flows. A deterministic ODE
flow can transform a simple latent distribution, such as \(N(0,I_T)\), into a
model for realistic cash-flow vectors. The velocity field can be constrained or
post-processed so that cash-flows are nonnegative, discounted present values are
reasonable, and tail scenarios are not smoothed away.
\end{example}

\subsection{Diffusion models: generating by denoising}\label{app:diffusion:diffusion-denoising}

Score-based diffusion models use a different idea. The forward process gradually
corrupts data into noise. A simple version is
\[
  \dd \bX_t=f_t(\bX_t)\,\dd t+g_t\,\dd \bB_t,
\]
chosen so that \(\bX_1\) is close to a simple distribution, often Gaussian. The
reverse process then transports noise back to data. Under suitable conditions, the
reverse-time dynamics involve the score
\[
  \nabla_{\bx}\log \rho_t(\bx),
\]
where \(\rho_t\) is the density of the noised data at time \(t\). A neural network
is trained to approximate this score or an equivalent denoising vector field.

The important point is that the learned object is local in time. The model learns
how to remove a little noise at each step. This often scales better than learning a
single global map.

\begin{definition}[Score and score-based generator]\label{def:app-diffusion-score-generator}
If a distribution at noise level $t$ has density $\rho_t$, its score is the vector
field $\bs_t(\bx)=\nabla_{\bx}\log\rho_t(\bx)$. A score-based generator estimates $\bs_t$ and
uses it to construct a reverse-time dynamics from a simple noisy law to the target
law. In actuarial terms, the score is not a risk score; it is a density-gradient
field.
\end{definition}

\begin{actuarialbox}
For actuarial scenario generation, a diffusion model says: start from a simple
noisy scenario and iteratively denoise it until it resembles a realistic actuarial
scenario. This can be useful for high-dimensional objects such as spatial climate
loss fields, reserve triangles, mortality surfaces or cyber incident paths.
\end{actuarialbox}

\subsection{Probability flow ODE versus reverse SDE}\label{app:diffusion:probability-flow-vs-sde}

Score-based diffusion models often come with two related samplers.

The first is a reverse-time stochastic differential equation:
\[
  \dd \bX_t = \widetilde{\bb}_t(\bX_t)\,\dd t + \widetilde \sigma_t\,\dd \overline{\bB}_t,
\]
which generates random paths from noise to data.

The second is a probability flow ODE:
\[
  \dd \bX_t = \bu_t(\bX_t)\,\dd t,
\]
whose one-time marginals match those of the stochastic reverse process under the
ideal score. Thus the same distributional evolution can sometimes be represented
by either a stochastic sampler or a deterministic ODE sampler.

This creates an important distinction.

\begin{itemize}
\item The ODE sampler gives deterministic trajectories conditional on the latent
starting point. It is more convenient for inversion, interpolation and audit trails.

\item The SDE sampler keeps stochasticity along the path. It can explore more
variation and may be more robust when the learned score is imperfect.
\end{itemize}

In actuarial terms, one should not confuse equality of marginal laws with equality
of path laws. Two processes may produce the same distribution at each date while
having different conditional dynamics, different path volatility and different
scenario interpretation.

\begin{proposition}[Marginal curves do not identify path laws]\label{prop:app-diffusion-same-marginals-different-paths}
A curve of one-time marginals $(\mu_t)_{0\le t\le1}$ does not determine the law of
the full path $(X_t)_{0\le t\le1}$.
\end{proposition}

\begin{proof}
Let $Z$ be a Rademacher random variable and set $X_t=Z$ for all $t$. Let
$\varepsilon$ be an independent Rademacher random variable and define
$Y_t=Z$ for $t<1/2$ and $Y_t=\varepsilon Z$ for $t\ge1/2$. For every fixed $t$,
both $X_t$ and $Y_t$ are Rademacher, so the one-time marginals coincide. But
$X_0=X_1$ almost surely, whereas $Y_0=Y_1$ only with probability $1/2$. Hence the
path laws are different.
\end{proof}

\subsection{Schr\"odinger bridges as controlled diffusions}\label{app:diffusion:schrodinger-bridges}

Schr\"odinger bridges provide a particularly clean bridge between optimal
transport and diffusion models. Let \(R\) be a reference path law, for instance a
Brownian motion or a baseline actuarial scenario generator.

\begin{definition}[Schr\"odinger bridge]\label{def:app-diffusion-schrodinger-bridge}
Given a reference path law $R$ and endpoint distributions $\mu_0,\mu_1$, the
Schr\"odinger bridge is the path law, among all path laws with endpoints
$\mu_0$ and $\mu_1$, that minimizes relative entropy with respect to $R$.
\end{definition}

Given endpoint laws
\(\mu_0\) and \(\mu_1\), the Schr\"odinger bridge solves
\[
  \inf_P
  \left\{
    \KL(P\|R): P_0=\mu_0,\; P_1=\mu_1
  \right\}.
\]
It is the path law closest to the reference dynamics while matching the two
endpoint marginals.

\begin{proposition}[Entropy projection interpretation]\label{prop:app-diffusion-sb-projection}
If the feasible set $\{P:P_0=\mu_0,P_1=\mu_1\}$ is non-empty and the KL projection
exists, the Schr\"odinger bridge is the information projection of the reference
path law $R$ onto the affine set of path laws with prescribed endpoints. In
particular, changing $R$ changes the meaning of the bridge.
\end{proposition}

This proposition is often the most important modelling point: the bridge is not
``the'' stochastic path from $\mu_0$ to $\mu_1$; it is the least surprising one
relative to a reference dynamics.

For Brownian reference dynamics with noise level \(\varepsilon\), the bridge has a
controlled diffusion representation
\[
  \dd \bX_t=\bv_t(\bX_t)\,\dd t+\sqrt{\varepsilon}\,\dd \bB_t,
\]
and the density satisfies
\[
  \partial_t\rho_t+\nabla\cdot(\rho_t v_t)
  =
  \frac{\varepsilon}{2}\Delta \rho_t.
\]
Compare this with deterministic dynamic OT:
\[
  \partial_t\rho_t+\nabla\cdot(\rho_t v_t)=0.
\]
The difference is the diffusion term. When \(\varepsilon\) is small, the bridge is
close to deterministic OT. When \(\varepsilon\) is positive, the path keeps
stochastic dispersion.

\begin{actuarialbox}
A Schr\"odinger bridge is a natural model for a stochastic counterfactual stress:
match the baseline law and the target law, but otherwise stay as close as possible
to a documented reference dynamics.
\end{actuarialbox}

\begin{example}[Climate loss bridge]
Let \(\mu_0\) be the current annual loss distribution and \(\mu_1\) a target
climate-stressed distribution. A deterministic OT path gives the least-action
movement of losses. A Schr\"odinger bridge gives the least-informative stochastic
evolution from current to stressed losses relative to a baseline climate-loss
process.
\end{example}

\begin{example}[Longevity bridge]
Let \(\mu_0\) be the current distribution of future lifetimes and \(\mu_1\) a
longevity-improved target distribution. A deterministic quantile map shifts every
rank of lifetime. A diffusion bridge can represent uncertainty around the
transition path, which is more appropriate when mortality improvement itself is a
stochastic trend.
\end{example}

\subsection{Deterministic counterfactuals versus stochastic counterfactuals}\label{app:diffusion:deterministic-vs-stochastic-counterfactuals}

Counterfactual language makes the distinction especially clear.

A deterministic transport counterfactual has the form
\[
  \bX^{\mathrm{cf}}=T(\bX).
\]
It says that the observation \(\bX=\bx\) is matched with a unique counterfactual
state \(T(x)\). This is useful for recourse, audit and rank-preserving stress testing.

A stochastic counterfactual has the form
\[
  \bX^{\mathrm{cf}}\sim K(\bx,\dd \by),
\]
where \(K\) is a Markov kernel. Conditional on \(\bX=\bx\), several counterfactual
endpoints are possible. A diffusion path gives an even richer object:
\[
  (\bX_t^{\mathrm{cf}})_{0\le t\le 1}\sim P^{\bx}.
\]
It describes a distribution over complete counterfactual trajectories.

\begin{actuarialbox}
Deterministic counterfactuals are useful when the counterfactual rule is a
policy transformation. Stochastic counterfactuals are useful when the
counterfactual world contains residual uncertainty after the intervention is
specified.
\end{actuarialbox}

\subsection{A modelling table for actuarial use}\label{app:diffusion:modelling-table}

\begin{center}
\begin{tabular}{p{0.24\textwidth}p{0.33\textwidth}p{0.33\textwidth}}
\toprule
Question & Deterministic map / ODE & Diffusion / stochastic bridge \\
\midrule
\iffalse
% To be restored in v2.
Tariff revision & Excellent for audit trails and premium-change maps & Useful if future renewals contain stochastic behaviour \\
\fi
\iffalse
% To be restored in v2.
Fairness repair & Natural for reproducible post-processing & Useful for gradual randomized policies, but harder to govern \\
\fi
Reserve cash-flows & Useful for deterministic stress paths & Natural for runoff uncertainty and stochastic development \\
Climate scenarios & Useful for rank-preserving stress & Natural for stochastic climate transition and event noise \\
Mortality / longevity & Useful for quantile stress of lifetimes & Natural for stochastic mortality improvement \\
Cyber risk & Useful for scenario matching & Natural for contagion, response and recovery paths \\
Capital modelling & Useful for transparent worst-case maps & Useful for scenario generators and model uncertainty \\
\bottomrule
\end{tabular}
\end{center}

\subsection{What can go wrong?}\label{app:diffusion:failure-modes}

Both paradigms can fail actuarially.

\paragraph{Failure modes of deterministic maps.}
A deterministic map may be too rigid. It can preserve ranks when ranks should not
be preserved. It can create an artificial one-to-one matching between baseline and
target scenarios. It can understate uncertainty by replacing a range of plausible
outcomes with one matched outcome. In high dimension, the chosen cost may impose
an arbitrary geometry.

\paragraph{Failure modes of diffusions.}
A diffusion may generate plausible-looking but actuarially invalid scenarios. It may
violate accounting identities, cash-flow timing, positivity constraints, contractual
rules or regulatory restrictions. It may smooth extremes too much, or generate tails
that are visually rare but financially material. It may also be hard to explain why a
specific simulated scenario was produced.

\paragraph{Failure modes shared by both.}
Both maps and diffusions can learn the wrong distribution. Both can hide data bias,
model misspecification, insufficient tail information, changing exposure, or an
inappropriate cost. In actuarial applications, the learned generator must be audited
as a model, not merely used as a sampler.

\subsection{Tail behaviour and extreme scenarios}\label{app:diffusion:tail-behaviour}

Generative models are often judged by visual or average fit. Actuarial models must
also be judged in the tail. Suppose \(L\) is a generated loss and \(u\) a high
threshold. The tail law
\[
  \Law(L-u\mid L>u)
\]
may be more important for capital than the fit of the body of the distribution.

\begin{definition}[Tail validation law]\label{def:app-diffusion-tail-law}
For a generated loss $L$ and threshold $u$, the conditional excess law is
$\Law(L-u\mid L>u)$. A generator is tail-valid at level $u$ only if this excess
law, and not merely the unconditional law of $L$, is acceptably close to the
reference or target tail law.
\end{definition}

A deterministic quantile map can preserve tail ranks exactly in one dimension. This
is valuable for transparent stress testing. A diffusion model may generate more
diverse tail scenarios, but only if the training procedure gives sufficient attention
to extremes. Otherwise, repeated denoising can underrepresent rare but material
losses.

Practical diagnostics include:
\begin{itemize}
\item comparison of high quantiles and expected shortfall;
\item comparison of threshold exceedance counts;
\item comparison of tail-index estimates when relevant;
\item Wasserstein distances restricted to upper-tail conditional laws;
\item stress tests on generated scenarios rather than only likelihood or average fit.
\end{itemize}

\begin{actuarialbox}
For generative actuarial modelling, a model that looks good in the body but fails
in the tail may be worse than a simpler model whose limitations are transparent.
Tail validation is not optional.
\end{actuarialbox}

\subsection{Conditioning and guidance}\label{app:diffusion:conditioning-guidance}

Recent generative models often allow conditional generation. One specifies a
condition \(C=c\), and samples from an approximation to
\[
  \Law(X\mid C=c).
\]
In actuarial science, conditions may include:

\begin{itemize}
\item territory, line of business, policy form or coverage layer;
\item climate pathway or macroeconomic scenario;
\item target loss ratio or capital constraint;
\item mortality improvement scenario;
\item regulatory or accounting constraint;
\item observed partial cash-flow development.
\end{itemize}

In deterministic flow models, conditioning may be incorporated by learning a map
\[
  T_\theta(z,c).
\]
In diffusion models, conditioning can modify the reverse dynamics, either through a
conditional score or through guidance terms. In both cases, actuarial conditioning
must be checked for coherence. A model should not generate a reserve triangle that
violates cumulative development constraints simply because it satisfies a high-level
conditioning variable.

\subsection{Scenario generation versus causal interpretation}\label{app:diffusion:scenario-vs-causality}

Generative models produce samples. They do not automatically identify causal
effects. This warning is the path-level version of the warning in
Lecture~\ref{lec:counterfactual-worlds-transport}: a transport plan, map or bridge
is not automatically a structural causal model. A deterministic map from \(\mu_0\)
to \(\mu_1\) is a coupling, not a structural causal model. A diffusion bridge
from \(\mu_0\) to \(\mu_1\) is a path law, not proof that the path would occur
under an intervention.

To interpret generated paths causally, one needs additional assumptions:
\begin{itemize}
\item what intervention is being applied;
\item which variables are manipulable;
\item which variables are only descriptive;
\item which mechanisms remain invariant;
\item what information is available at each time;
\item how hidden confounding is ruled out or bounded.
\end{itemize}

This is particularly important for prevention and climate adaptation. A
transport or diffusion path can be a useful counterfactual scenario generator, but
causal claims require more than endpoint matching.

\subsection{Actuarial governance checklist}\label{app:diffusion:governance-checklist}

Before using a deterministic or diffusion generator in actuarial work, one should
answer the following questions.

\begin{enumerate}[label=(\roman*)]
\item \textbf{State space.} What exactly is generated: losses, contracts, cash-flows, mortality rates, triangles or paths?

\item \textbf{Units.} Are the coordinates monetary, temporal, categorical,
geographical, demographic or contractual?

\item \textbf{Constraints.} Are nonnegativity, monotonicity, accounting identities,
coverage limits and legal constraints enforced?

\item \textbf{Tail validation.} Are VaR, TVaR, threshold exceedances and extreme
scenarios correctly represented?

\item \textbf{Path meaning.} Does the path represent time, denoising time,
calibration time, renewal cycles, development years or an artificial interpolation
parameter?

\item \textbf{Noise meaning.} Does the diffusion noise represent process risk,
model uncertainty, posterior uncertainty or merely numerical smoothing?

\item \textbf{Audit trail.} Can one explain why a generated scenario was produced?

\item \textbf{Use case.} Is the generator used for pricing, capital, stress testing, reserving, prevention, mortality or communication? The validation standard depends on the use.
\end{enumerate}

\subsection{Mini-lab: deterministic quantile paths and noisy bridges}\label{app:diffusion:minilab-quantile-noisy}

This lab compares a deterministic quantile path with a simple noisy bridge. It is
not a full diffusion model. It is a pedagogical approximation showing how adding
path noise changes intermediate distributions while keeping endpoints fixed in each
paired scenario.

\begin{lstlisting}[language=R]
set.seed(123)

n <- 5000
u <- (seq_len(n) - 0.5) / n

# Baseline and stressed severity distributions
q0 <- qlnorm(u, meanlog = 8.0, sdlog = 0.55)
q1 <- qlnorm(u, meanlog = 8.2, sdlog = 0.85)

# Deterministic displacement interpolation in quantile space
path_det <- function(t) {
  (1 - t) * q0 + t * q1
}

# Noisy bridge around paired deterministic path.
# Noise vanishes at t=0 and t=1, so paired endpoints are preserved.
path_noisy <- function(t, eps = 0.20) {
  m <- (1 - t) * q0 + t * q1
  s <- sqrt(eps * t * (1 - t)) * sd(q1 - q0)
  pmax(m + s * rnorm(length(m)), 0)
}

emp_q <- function(x, alpha) {
  xs <- sort(x)
  xs[pmax(1, ceiling(alpha * length(xs)))]
}

VaR <- function(x, alpha = 0.99) {
  as.numeric(emp_q(x, alpha))
}

TVaR <- function(x, alpha = 0.99) {
  q <- VaR(x, alpha)
  mean(x[x >= q])
}

tail_mean_excess <- function(x, alpha = 0.99) {
  q <- VaR(x, alpha)
  mean(x[x >= q] - q)
}

grid <- seq(0, 1, by = 0.1)

out <- data.frame(
  t = grid,
  mean_det = NA_real_, VaR_det = NA_real_, TVaR_det = NA_real_,
  excess_det = NA_real_,
  mean_noisy = NA_real_, VaR_noisy = NA_real_, TVaR_noisy = NA_real_,
  excess_noisy = NA_real_
)

for (i in seq_along(grid)) {
  t <- grid[i]
  xd <- path_det(t)
  xn <- path_noisy(t, eps = 0.30)
  out$mean_det[i] <- mean(xd)
  out$VaR_det[i] <- VaR(xd)
  out$TVaR_det[i] <- TVaR(xd)
  out$excess_det[i] <- tail_mean_excess(xd)
  out$mean_noisy[i] <- mean(xn)
  out$VaR_noisy[i] <- VaR(xn)
  out$TVaR_noisy[i] <- TVaR(xn)
  out$excess_noisy[i] <- tail_mean_excess(xn)
}

print(out)

matplot(out$t,
        out[, c("VaR_det", "VaR_noisy", "TVaR_det", "TVaR_noisy")],
        type = "l", lty = 1, xlab = "t", ylab = "tail measure")
legend("topleft",
       legend = c("VaR deterministic", "VaR noisy",
                  "TVaR deterministic", "TVaR noisy"),
       lty = 1, bty = "n")
\end{lstlisting}

The deterministic path changes tail measures smoothly because every quantile is
moved along a straight line in quantile space. The noisy bridge may increase or
decrease intermediate tail measures depending on the noise scale. This illustrates
why a stochastic path is not merely a cosmetic version of a deterministic path: it
changes the interpretation of intermediate risk.

\subsection{Mini-lab: deterministic map versus stochastic Markov kernel}\label{app:diffusion:minilab-map-kernel}

The next toy example compares a deterministic repair map with a stochastic kernel.
Both target the same endpoint distribution, but their conditional interpretations are
different.

\begin{lstlisting}[language=R]
set.seed(321)

n <- 3000
x <- rnorm(n, mean = 0, sd = 1)

# Target distribution: shifted and more dispersed
u <- rank(x, ties.method = "random") / (n + 1)
y_det <- qnorm(u, mean = 1, sd = 1.5)

# Stochastic kernel: same broad target idea, but not deterministic conditional on x
# Here we use a noisy location-scale rule and then quantile-calibrate the marginal.
y_raw <- 1 + 1.1 * x + 0.9 * rnorm(n)
emp_q_vec <- function(x, u) {
  xs <- sort(x)
  xs[pmax(1, ceiling(u * length(xs)))]
}
u_raw <- rank(y_raw, ties.method = "random") / n
y_sto <- as.numeric(emp_q_vec(y_det, u_raw))

cor_det <- cor(x, y_det)
cor_sto <- cor(x, y_sto)

c(correlation_deterministic = cor_det,
  correlation_stochastic = cor_sto,
  mean_det = mean(y_det), mean_stochastic = mean(y_sto),
  sd_det = sd(y_det), sd_stochastic = sd(y_sto))

plot(x, y_det, pch = 16, cex = 0.35,
     xlab = "baseline x", ylab = "target y",
     main = "Deterministic map")
plot(x, y_sto, pch = 16, cex = 0.35,
     xlab = "baseline x", ylab = "target y",
     main = "Stochastic kernel with same marginal target")
\end{lstlisting}

The deterministic map creates a functional relation between \(x\) and \(y\). The
stochastic kernel gives the same type of target marginal but allows residual
variation conditional on \(x\). For auditability, the deterministic map is easier;
for process uncertainty, the stochastic kernel may be more realistic.

\subsection{Further reading and links to the notes}\label{app:diffusion:further-reading}

For deterministic maps and push-forwards, see Lectures~\ref{lec:couplings} and
\ref{lec:continuous-ot}. For velocity fields, action and distributional trajectories,
see Lecture~\ref{lec:dynamic-ot-actuarial-trajectories}. For stochastic counterfactuals and
Schr\"odinger bridges, see Lecture~\ref{lec:counterfactual-worlds-transport}. For cost design
on constrained actuarial spaces, see Appendix~\ref{app:cost}. For statistical
validation, noise floors and resampling, see Lecture~\ref{lec:statistical-ot}. The two mini-labs
in Sections~\ref{app:diffusion:minilab-quantile-noisy} and
\ref{app:diffusion:minilab-map-kernel} are deliberately simple; they are meant to accompany
the computational labs, where Wasserstein uncertainty and reserve cash-flow laws are studied more directly.

\subsection{Summary}\label{app:diffusion:summary}

Deterministic maps and diffusion processes are complementary.

\begin{itemize}
\item Deterministic OT maps give transparent scenario matching, rank-preserving
stress and reproducible counterfactual rules.

\item Deterministic ODE flows extend maps into smooth trajectories of
probability measures and connect naturally with flow matching and continuous
normalizing flows.

\item Diffusion models generate through stochastic paths and are natural when
intermediate uncertainty is meaningful.

\item Schr\"odinger bridges connect both worlds: they are entropy-regularized,
stochastic analogues of dynamic optimal transport.

\item In actuarial science, the choice should be driven by the use case: audit and
policy transformation often favour deterministic maps; scenario generation,
runoff uncertainty, mortality improvement, cyber contagion and climate paths often
favour stochastic processes.
\end{itemize}

\begin{actuarialbox}
Generative AI does not remove the need for actuarial modelling. It adds new ways
to represent distributions and paths. The actuarial task remains to decide what the
state space means, what constraints must hold, what tail behaviour is acceptable,
and whether the generated object should be a deterministic transformation or a
stochastic evolution.
\end{actuarialbox}

\subsection{Exercises}\label{app:diffusion:exercises}

\begin{exercise}[Map versus kernel]
\label{ex:DM:1}
Let \(X\sim\mu_0\) and let \(T_\#\mu_0=\mu_1\). Explain why
\(Y=T(X)\) defines a coupling between \(\mu_0\) and \(\mu_1\). Then give an
example of a Markov kernel \(K(x,\dd y)\) such that the joint law
\(\mu_0(\dd x)K(x,\dd y)\) has the same marginals but is not induced by a map.
\end{exercise}

\begin{exercise}[Continuity equation]
\label{ex:DM:2}
Let \(X_t=\phi_t(X_0)\), where \(\dot\phi_t(x)=v_t(\phi_t(x))\). Derive formally
that \(\mu_t=\Law(X_t)\) satisfies
\[
  \partial_t\mu_t+\nabla\cdot(\mu_t \bv_t)=0.
\]
Use a smooth test function argument.
\end{exercise}

\begin{exercise}[Fokker--Planck equation]
\label{ex:DM:3}
For the diffusion
\[
  \dd X_t=b(X_t)\,\dd t+\sigma\,\dd B_t,
\]
write the Fokker--Planck equation when \(\sigma\) is constant. Compare it with
the deterministic continuity equation.
\end{exercise}

\begin{exercise}[One-dimensional stress]
\label{ex:DM:4}
Let \(\mu_0\) and \(\mu_1\) be continuous loss distributions. Show that the
quantile map \(F_1^{-1}\circ F_0\) preserves ranks. Explain why this may be a
desirable property for a stress test, and why it may be undesirable for a stochastic
future scenario.
\end{exercise}

\begin{exercise}[Same marginals, different paths]
\label{ex:DM:5}
Construct two processes \((X_t)_{0\le t\le1}\) and \((Y_t)_{0\le t\le1}\) with the
same marginal law at \(t=0\) and \(t=1\), but different conditional dynamics. What
actuarial quantity could distinguish the two paths?
\end{exercise}

\begin{exercise}[Scenario validation]
\label{ex:DM:6}
Suppose a diffusion generator is trained to produce annual aggregate losses.
List five diagnostics that should be checked before using the generated scenarios
for capital estimation.
\end{exercise}

\begin{exercise}[Tail-aware generation]
\label{ex:DM:7}
Explain why a model with excellent average Wasserstein fit may still be unsuitable
for estimating \(\TVaR_{99.5\%}\). Propose a tail-specific validation criterion.
\end{exercise}

\begin{exercise}[Schr\"odinger bridge interpretation]
\label{ex:DM:8}
Explain why the Schr\"odinger bridge objective
\[
  \min\{\KL(P\|R):P_0=\mu_0,P_1=\mu_1\}
\]
can be interpreted as a least-informative stochastic update of a baseline scenario
generator.
\end{exercise}

\begin{exercise}[Governance choice]
\label{ex:DM:9}
For each of the following applications, say whether you would prefer a
deterministic map, an ODE flow, a diffusion process, or a Schr\"odinger bridge:
reserve runoff simulation, mortality improvement, cyber contagion, climate-loss transition, and rank-preserving severity stress. Justify your answers.
\end{exercise}

\begin{exercise}[Path time versus calendar time]\label{ex:DM:10}
In a diffusion generator for reserve cash-flow vectors, explain why the denoising time of the
model should not automatically be interpreted as accident year, development year or calendar
year. Give one example where confusing these meanings would lead to an invalid actuarial
interpretation.
\end{exercise}

\begin{exercise}[Reference dynamics in a bridge]\label{ex:DM:11}
Consider two Schr\"odinger bridges with the same endpoint laws $\mu_0$ and $\mu_1$, but two
different reference path laws $R$ and $\widetilde R$. Explain why the resulting bridges can have
different actuarial interpretations even if they match the same endpoints.
\end{exercise}

\begin{exercise}[Mini-lab extension]
\label{ex:DM:12}
Modify the first mini-lab by replacing the lognormal target with a Pareto target.
Compare the deterministic and noisy paths for \(\VaR_{99\%}\) and
\(\TVaR_{99\%}\). What changes in the tail?
\end{exercise}

\newpage
\stepcounter{moncompteur}
\section{Lab~\themoncompteur: Empirical Wasserstein Distances under Sampling Noise}
\label{sec:lab-empirical-wasserstein-noise}

Empirical Wasserstein distances are easy to compute, especially in one dimension. They are also easy to over-interpret. If two independent samples are drawn from the same continuous distribution, their empirical laws are almost surely different. Consequently, the empirical Wasserstein distance between the samples is positive, even when there is no business drift, no model change and no deterioration of the portfolio.

This lab studies that phenomenon. The aim is to distinguish a genuine distributional change from the distance that arises because two empirical samples are finite. The actuarial question is not only
\[
  \widehat W_p > 0,
\]
but rather whether the observed distance is large relative to its sampling noise.

\begin{actuarialbox}
A Wasserstein monitoring report should not display only the distance between two empirical loss distributions. It should also display a noise floor, a bootstrap uncertainty band, or a permutation reference distribution. Otherwise, small random fluctuations may be mistaken for portfolio drift.
\end{actuarialbox}

\subsection{Learning objectives}

After completing this lab, the reader should be able to:
\begin{itemize}
\item compute empirical \(W_1\) and \(W_2\) distances between one-dimensional loss samples by quantile matching;
\item simulate the null distribution of an empirical Wasserstein distance under no drift;
\item compare the noise floor to genuine shifts such as inflation, scale change or tail thickening;
\item implement a permutation test for equality of distributions;
\item implement bootstrap confidence intervals for a Wasserstein distance;
\item diagnose where two loss distributions differ using quantile-difference curves.
\end{itemize}

\subsection{The empirical object}

Let \(X_1,\ldots,X_n\) and \(Y_1,\ldots,Y_m\) be two one-dimensional samples of claim amounts, reserve outcomes or loss ratios. Their empirical laws are
\[
  \widehat\mu_n = \frac1n \sum_{i=1}^n \delta_{X_i},
  \qquad
  \widehat\nu_m = \frac1m \sum_{j=1}^m \delta_{Y_j}.
\]
The empirical Wasserstein distance is
\[
  W_p(\widehat\mu_n,\widehat\nu_m).
\]
In one dimension, it is computed from empirical quantiles:
\[
  W_p^p(\widehat\mu_n,\widehat\nu_m)
  \approx
  \frac1K\sum_{k=1}^K
  \left|\widehat Q_X(u_k)-\widehat Q_Y(u_k)\right|^p,
\]
where \(u_1,\ldots,u_K\) is a fine grid in \((0,1)\), and \(\widehat Q_X,\widehat Q_Y\) are empirical generalized inverse functions. In the code below, empirical quantiles are lower generalized inverses. This is the same convention as in the main text.

When \(n=m\) and the two empirical measures have equal weights, one can sort the samples:
\[
  W_p^p(\widehat\mu_n,\widehat\nu_n)
  =
  \frac1n\sum_{i=1}^n |X_{(i)}-Y_{(i)}|^p.
\]
The quantile-grid version below also works when the sample sizes differ.

\begin{computebox}
In this lab, everything is one-dimensional. No linear programming solver is needed. Sorting and empirical generalized inverses are enough.
\end{computebox}

\subsection{The noise floor}

Suppose first that both samples come from the same distribution:
\[
  X_i \sim \mu,
  \qquad
  Y_j \sim \mu.
\]
At the population level,
\[
  W_p(\mu,\mu)=0.
\]
But at the empirical level,
\[
  W_p(\widehat\mu_n,\widehat\nu_m)>0
\]
with high probability. This positive value is not evidence of drift. It is the sampling noise floor of the monitoring procedure.

A practical way to estimate this noise floor is repeated random splitting:
\begin{enumerate}
\item pool observations from a supposedly homogeneous portfolio;
\item split them into two pseudo-portfolios of comparable size;
\item compute the Wasserstein distance between the two empirical laws;
\item repeat many times.
\end{enumerate}
The resulting distribution is the reference scale for interpreting future observed distances.

\begin{actuarialbox}
A value such as \(W_1=120\) dollars is not meaningful without context. It may be large for a portfolio of small attritional claims and negligible for a portfolio of large liability claims. It may also be exactly what random splitting produces at the available sample size.
\end{actuarialbox}

\subsection{Three kinds of drift}

This lab compares three stylized alternatives.

\paragraph{Inflation drift.}
The second sample is multiplied by \(1+r\):
\[
  Y=(1+r)X.
\]
This shifts the severity distribution upward in a rank-preserving way.

\paragraph{Scale drift.}
The second distribution has higher dispersion, for example a larger lognormal standard deviation. This may leave the center of the distribution relatively stable while increasing both lower and upper tail dispersion.

\paragraph{Tail drift.}
Only the upper tail is modified. For example, a small proportion of observations is multiplied by a large factor. This creates a distributional change that may be modest for \(W_1\), larger for \(W_2\), and very visible in upper quantile differences.

\subsection{Permutation and bootstrap}

Two simple calibration tools are useful.

\paragraph{Permutation test.}
For the null hypothesis
\[
  H_0:\mu=\nu,
\]
the labels ``sample X'' and ``sample Y'' are exchangeable. A permutation test pools the data, randomly relabels them many times, and compares the observed statistic to the relabelled statistics.

\paragraph{Bootstrap uncertainty.}
For two independent samples, a nonparametric bootstrap resamples each sample with replacement and recomputes the Wasserstein distance. The resulting empirical distribution gives a standard error or percentile interval.

These two tools answer different questions. The permutation test asks whether two samples could plausibly have come from the same distribution. The bootstrap asks how variable the observed distance is as an estimator.

\subsection{R lab}
\label{lab:empirical-wasserstein-noise-r}

\subsubsection{Basic Wasserstein function}

\begin{lstlisting}[language=R]
# Lower empirical generalized inverse.
emp_q_left <- function(x, probs) {
  x <- sort(x[is.finite(x)])
  n <- length(x)
  if (n == 0) stop("x must contain at least one finite value")
  probs <- pmin(pmax(probs, 0), 1)
  idx <- ceiling(n * probs)
  idx[idx < 1] <- 1
  idx[idx > n] <- n
  x[idx]
}

# One-dimensional empirical Wasserstein distance.
# Works for unequal sample sizes by using a common quantile grid.
wass1d <- function(x, y, p = 1, grid_size = 1000, trim = 0) {
  stopifnot(p >= 1)
  x <- x[is.finite(x)]
  y <- y[is.finite(y)]
  if (length(x) == 0 || length(y) == 0) {
    stop("x and y must contain at least one finite value")
  }

  lo <- trim
  hi <- 1 - trim
  if (lo <= 0) lo <- 1 / (grid_size + 1)
  if (hi >= 1) hi <- grid_size / (grid_size + 1)
  if (lo >= hi) stop("trim is too large")

  u <- seq(lo, hi, length.out = grid_size)
  qx <- emp_q_left(x, u)
  qy <- emp_q_left(y, u)

  mean(abs(qx - qy)^p)^(1 / p)
}

# Quick check
x <- c(1, 2, 7, 10)
y <- c(0, 4, 6, 12)
wass1d(x, y, p = 1, grid_size = 4)
wass1d(x, y, p = 2, grid_size = 4)
\end{lstlisting}

The argument \texttt{trim} can be used to ignore extreme quantile levels. This is not a replacement for tail modelling, but it is useful for sensitivity checks.

\subsubsection{Simulated severity data}

\begin{lstlisting}[language=R]
set.seed(123)

# Baseline portfolio: attritional severities
n <- 1000
x <- rlnorm(n, meanlog = 8.0, sdlog = 0.75)

# Alternative 1: inflation drift
r <- 0.08
y_infl <- (1 + r) * rlnorm(n, meanlog = 8.0, sdlog = 0.75)

# Alternative 2: scale drift
y_scale <- rlnorm(n, meanlog = 8.0, sdlog = 0.95)

# Alternative 3: tail drift
base <- rlnorm(n, meanlog = 8.0, sdlog = 0.75)
tail_flag <- runif(n) < 0.06
y_tail <- base
y_tail[tail_flag] <- 2.5 * y_tail[tail_flag]

rbind(
  inflation_W1 = wass1d(x, y_infl, p = 1),
  inflation_W2 = wass1d(x, y_infl, p = 2),
  scale_W1     = wass1d(x, y_scale, p = 1),
  scale_W2     = wass1d(x, y_scale, p = 2),
  tail_W1      = wass1d(x, y_tail, p = 1),
  tail_W2      = wass1d(x, y_tail, p = 2)
)
\end{lstlisting}

\subsubsection{Noise floor by random splitting}

\begin{lstlisting}[language=R]
noise_floor_split <- function(z, p = 1, B = 500, split_frac = 0.5,
                              grid_size = 1000, seed = NULL) {
  if (!is.null(seed)) set.seed(seed)
  z <- z[is.finite(z)]
  N <- length(z)
  n1 <- floor(split_frac * N)
  if (n1 < 1 || n1 >= N) stop("split_frac gives an empty split")
  out <- numeric(B)

  for (b in seq_len(B)) {
    id <- sample.int(N, N, replace = FALSE)
    x <- z[id[seq_len(n1)]]
    y <- z[id[(n1 + 1):N]]
    out[b] <- wass1d(x, y, p = p, grid_size = grid_size)
  }
  out
}

# Estimate the null distribution from a homogeneous portfolio
z <- rlnorm(3000, meanlog = 8.0, sdlog = 0.75)
null_W1 <- noise_floor_split(z, p = 1, B = 500, seed = 1)
null_W2 <- noise_floor_split(z, p = 2, B = 500, seed = 1)

quantile(null_W1, c(0.50, 0.90, 0.95, 0.99))
quantile(null_W2, c(0.50, 0.90, 0.95, 0.99))
\end{lstlisting}

The 95\% or 99\% quantile of the null distribution can be used as a monitoring threshold. If an observed distance is below that threshold, it may be sampling variation. If it is far above the threshold, the distributional change is more credible.

\subsubsection{Permutation test}

\begin{lstlisting}[language=R]
perm_test_wass <- function(x, y, p = 1, B = 999, grid_size = 1000,
                           seed = NULL) {
  if (!is.null(seed)) set.seed(seed)
  x <- x[is.finite(x)]
  y <- y[is.finite(y)]
  n <- length(x)
  m <- length(y)
  if (n == 0 || m == 0) stop("x and y must be non-empty")

  z <- c(x, y)
  N <- n + m
  t_obs <- wass1d(x, y, p = p, grid_size = grid_size)
  t_perm <- numeric(B)

  for (b in seq_len(B)) {
    id <- sample.int(N, N, replace = FALSE)
    xb <- z[id[seq_len(n)]]
    yb <- z[id[(n + 1):N]]
    t_perm[b] <- wass1d(xb, yb, p = p, grid_size = grid_size)
  }

  p_value <- (1 + sum(t_perm >= t_obs)) / (B + 1)

  list(
    statistic = t_obs,
    p_value = p_value,
    permuted = t_perm
  )
}

ans_infl <- perm_test_wass(x, y_infl, p = 1, B = 499, seed = 12)
ans_scale <- perm_test_wass(x, y_scale, p = 1, B = 499, seed = 12)
ans_tail <- perm_test_wass(x, y_tail, p = 1, B = 499, seed = 12)

ans_infl$p_value
ans_scale$p_value
ans_tail$p_value
\end{lstlisting}

\subsubsection{Bootstrap confidence interval}

\begin{lstlisting}[language=R]
boot_wass_independent <- function(x, y, p = 1, B = 999, grid_size = 1000,
                                  seed = NULL) {
  if (!is.null(seed)) set.seed(seed)
  x <- x[is.finite(x)]
  y <- y[is.finite(y)]
  n <- length(x)
  m <- length(y)
  if (n == 0 || m == 0) stop("x and y must be non-empty")

  stat <- numeric(B)
  for (b in seq_len(B)) {
    xb <- sample(x, size = n, replace = TRUE)
    yb <- sample(y, size = m, replace = TRUE)
    stat[b] <- wass1d(xb, yb, p = p, grid_size = grid_size)
  }

  c(
    observed = wass1d(x, y, p = p, grid_size = grid_size),
    se = sd(stat),
    q025 = as.numeric(quantile(stat, 0.025)),
    q500 = as.numeric(quantile(stat, 0.500)),
    q975 = as.numeric(quantile(stat, 0.975))
  )
}

boot_wass_independent(x, y_infl, p = 1, B = 499, seed = 10)
boot_wass_independent(x, y_infl, p = 2, B = 499, seed = 10)
\end{lstlisting}

\iffalse
% To be restored in v2 with the material on premium drift and portfolio demutualization.
For paired tariff revisions, one should not resample old and new premiums independently. Resample policies, preserving the pair \((P_i^{old},P_i^{new})\).

\begin{lstlisting}[language=R]
boot_wass_paired <- function(old, new, p = 1, B = 999, grid_size = 1000,
                             seed = NULL) {
  if (!is.null(seed)) set.seed(seed)
  stopifnot(length(old) == length(new))
  n <- length(old)
  stat <- numeric(B)

  for (b in seq_len(B)) {
    id <- sample.int(n, n, replace = TRUE)
    stat[b] <- wass1d(old[id], new[id], p = p, grid_size = grid_size)
  }

  c(
    observed = wass1d(old, new, p = p, grid_size = grid_size),
    se = sd(stat),
    q025 = as.numeric(quantile(stat, 0.025)),
    q500 = as.numeric(quantile(stat, 0.500)),
    q975 = as.numeric(quantile(stat, 0.975))
  )
}
\end{lstlisting}
\fi

\subsubsection{Quantile-difference plots}

A single Wasserstein number does not tell us where the difference lies. The following function plots
\[
  u \longmapsto \widehat Q_Y(u)-\widehat Q_X(u).
\]

\begin{lstlisting}[language=R]
qdiff_curve <- function(x, y, grid_size = 200, trim = 0.005) {
  u <- seq(trim, 1 - trim, length.out = grid_size)
  qx <- emp_q_left(x, u)
  qy <- emp_q_left(y, u)
  data.frame(u = u, qx = qx, qy = qy, diff = qy - qx)
}

qd_infl <- qdiff_curve(x, y_infl)
qd_scale <- qdiff_curve(x, y_scale)
qd_tail <- qdiff_curve(x, y_tail)

plot(qd_infl$u, qd_infl$diff, type = "l", lwd = 2,
     xlab = "quantile level u", ylab = "Q_Y(u) - Q_X(u)",
     main = "Quantile difference curves")
lines(qd_scale$u, qd_scale$diff, lwd = 2, lty = 2)
lines(qd_tail$u, qd_tail$diff, lwd = 2, lty = 3)
abline(h = 0, lty = 3)
legend("topleft", lwd = 2, lty = 1:3,
       legend = c("inflation", "scale", "tail"), bty = "n")
\end{lstlisting}

The curves should have different shapes. Inflation tends to move the full curve upward. Scale drift often creates small changes near the center and larger changes away from the center. Tail drift appears mostly near high quantile levels.

\subsection{Suggested report template}

A short actuarial report based on this lab could contain the following table.

\begin{center}
\begin{tabular}{ll}
\toprule
Quantity & Interpretation \\
\midrule
\(W_1\) & Average monetary displacement between empirical quantile curves \\
\(W_2\) & Tail-sensitive squared displacement summary \\
Noise-floor median & Typical distance produced by random splitting \\
Noise-floor 95\% quantile & Monitoring threshold under no drift \\
Permutation p-value & Evidence against equality of the two distributions \\
Bootstrap interval & Sampling uncertainty around the observed distance \\
Quantile-difference curve & Location of the distributional change \\
\bottomrule
\end{tabular}
\end{center}

\begin{computebox}
For a production monitoring report, record the sample sizes, the scale of the variable, the chosen value of \(p\), the trimming convention, the observed distance, a null threshold, and a visual quantile-difference plot. The Wasserstein distance should be a diagnostic, not an isolated number.
\end{computebox}

\subsection{Exercises}

\begin{exercise}[Manual computation]
\label{ex:EW:1}
Let
\[
  \bx=(1,2,7,10),
  \qquad
  \by=(0,4,6,12).
\]
Compute \(W_1\) and \(W_2\) between the two empirical distributions with equal weights. Interpret the result as a comparison of two small claim samples.
\end{exercise}

\begin{exercise}[Noise floor]
\label{ex:EW:2}
Simulate \(3000\) observations from a lognormal severity distribution. Randomly split the sample into two groups of size \(1500\), and compute \(W_1\). Repeat this \(1000\) times. Plot the histogram of the resulting distances. What is the median? What is the 95\% quantile?
\end{exercise}

\begin{exercise}[Sample size]
\label{ex:EW:3}
Repeat Exercise~\ref{ex:EW:2} with total sample sizes \(N=500,1000,5000,10000\). How does the noise floor change with \(N\)? Does \(W_2\) decrease at the same speed as \(W_1\)?
\end{exercise}

\begin{exercise}[Inflation drift]
\label{ex:EW:4}
Let \(X\) be a lognormal severity sample and let \(Y=(1+r)X'\), where \(X'\) is an independent sample from the same distribution. Compute \(W_1(X,Y)\) for \(r=0,0.02,0.05,0.10,0.20\). Plot \(W_1\) as a function of \(r\). Compare it to the random-splitting noise floor.
\end{exercise}

\begin{exercise}[Tail drift]
\label{ex:EW:5}
Construct a sample \(Y\) from \(X\) by multiplying the largest 5\% of observations by a factor \(a>1\). Compare \(W_1\), \(W_2\), \(\VaR_{.95}\), \(\VaR_{.99}\) and \(\TVaR_{.99}\) before and after the perturbation. Which diagnostic is most sensitive?
\end{exercise}

\begin{exercise}[Permutation test]
\label{ex:EW:6}
Implement the permutation test from the lab. Apply it to two samples drawn from the same distribution, and then to two samples where the second distribution is inflated by 5\%. Repeat the experiment many times. How often does the test reject at the 5\% level?
\end{exercise}

\begin{exercise}[Bootstrap interval]
\label{ex:EW:7}
Implement a bootstrap confidence interval for \(W_1\) between two independent loss samples. How does the interval length change with sample size? How does it change when the loss distribution becomes heavier-tailed?
\end{exercise}

\iffalse
% To be restored in v2 with the material on premium drift and portfolio demutualization.
\begin{exercise}[Paired versus unpaired bootstrap]
\label{ex:EW:8}
Simulate old and new premiums for the same policies:
\[
  P_i^{new}=P_i^{old}(1+\epsilon_i),
\]
where \(\epsilon_i\) is a policy-level random change. Compare a paired bootstrap and an independent bootstrap for \(W_1(\widehat\mu_{old},\widehat\mu_{new})\). Why is the paired bootstrap more appropriate for tariff revisions?
\end{exercise}
\fi

\begin{exercise}[Where is the difference?]
\label{ex:EW:9}
For inflation drift, scale drift and tail drift, plot the quantile-difference curve
\[
  u\mapsto \widehat Q_Y(u)-\widehat Q_X(u).
\]
Explain how the shape of the curve helps identify the kind of drift.
\end{exercise}

\begin{exercise}[Actuarial governance]
\label{ex:EW:10}
Write a short paragraph explaining why an empirical Wasserstein distance should not be reported without a noise calibration. Include at least three items that a model-monitoring report should display alongside the raw distance.
\end{exercise}

\begin{exercise}[Trimming sensitivity]
\label{ex:EW:11}
Repeat the drift comparisons with \texttt{trim = 0}, \texttt{trim = 0.005} and \texttt{trim = 0.025}. Which conclusions are stable? Which conclusions depend on the most extreme quantiles?
\end{exercise}

\begin{exercise}[Choice of scale]
\label{ex:EW:12}
Compute Wasserstein distances on the original loss scale and on the log-loss scale. Explain how the interpretation changes. Which scale would you use for a portfolio with extremely heavy-tailed claims?
\end{exercise}

\newpage
\stepcounter{moncompteur}
\section{Lab~\themoncompteur: Sliced and Projected Wasserstein Distances}
\label{sec:sliced-projected-wasserstein}

Exact optimal transport is geometrically meaningful, but it can be expensive in moderate or high dimension. If the state variable is a vector of losses by line, a vector of policyholder characteristics, a cash-flow trajectory, or a vector of model outputs, solving the full multivariate transport problem may be numerically heavy and statistically unstable. Sliced and projected Wasserstein distances provide a useful alternative: they reduce a multivariate comparison to many one-dimensional comparisons.

The central idea is actuarially natural. Instead of comparing two multivariate portfolios only through the full cloud of points, we compare them through one-dimensional views: total loss, regional loss, first principal component, duration-weighted cash-flow, capital-weighted loss, or a random linear combination of features. Each projected comparison is one-dimensional, hence it can be computed by sorting and comparing empirical quantiles.

\begin{actuarialbox}
Sliced Wasserstein distances are useful when the actuarial object is multivariate but one-dimensional quantile comparisons remain desirable. They are fast, interpretable, and provide directional diagnostics: which projection of the portfolio, output vector or cash-flow vector has moved the most?
\end{actuarialbox}

\begin{actuarialbox}%[Quantile convention]
In this lab, empirical one-dimensional Wasserstein distances are computed with the lower generalized inverse convention. In R, the code below uses an explicit left-continuous empirical quantile function rather than the default interpolation rules of \texttt{quantile}. This makes the link with the convention used in the lectures transparent, especially when ties or atoms are present.
\end{actuarialbox}

\subsection{One-dimensional projections}

Let \(\mu,\nu\in\mathcal P_p(\R^d)\), with \(p\ge 1\). For a unit vector \(\btheta\in\mathbb S^{d-1}\), define the linear projection
\[
  P_{\btheta}(\bx)=\btheta^\top \bx, \text{ for } \bx\in\R^d.
\]
The projected laws are
\[
  \mu_{\btheta}=(P_{\btheta})_\#\mu, \text{ and } \nu_{\btheta}=(P_{\btheta})_\#\nu.
\]
They are probability measures on \(\R\). Hence their Wasserstein distance can be computed using the quantile formula
\[
  W_p^p(\mu_{\btheta},\nu_{\btheta})=\int_0^1 \left|F_{\mu_{\btheta}}^{-1}(u)-F_{\nu_{\btheta}}^{-1}(u)\right|^p\,\dd u.
\]

\begin{definition}[Projected Wasserstein distance in one direction]
For \(\btheta\in\mathbb S^{d-1}\), the projected \(p\)-Wasserstein distance in direction \(\btheta\) is
\[
  W_{p,\btheta}(\mu,\nu)=W_p\bigl((P_{\btheta})_\#\mu,(P_{\btheta})_\#\nu\bigr).
\]
\end{definition}

This directional distance answers a precise question: how far are the two laws when viewed through the scalar actuarial index \(\bx\mapsto\btheta^\top\bx\)?

\begin{example}[Line-of-business losses]
Let \(\bX=(X_1,\ldots,X_d)\) be a vector of losses by line of business. If \(\btheta=(1,\ldots,1)/\sqrt d\), then \(\btheta^\top\bX\) is proportional to total loss. If \(\btheta=\be_k\), then \(\btheta^\top\bX=X_k\) is the loss in line \(k\). If \(\btheta\) is a capital-allocation vector, then \(\btheta^\top\bX\) is a capital-weighted portfolio loss. Projected Wasserstein distances compare two multivariate models through these actuarially meaningful views.
\end{example}

\begin{example}[Cash-flow vectors]
Let \(\bX=(X_1,\ldots,X_T)\) be a future cash-flow vector. A projection \(\btheta^\top\bX\) may represent discounted present value, total nominal reserve, late-development exposure, or a contrast between short-tail and long-tail payments. A projected Wasserstein distance compares reserving distributions through the chosen cash-flow summary.
\end{example}

\subsection{Sliced Wasserstein distance}

Instead of selecting one direction, we may average over directions. Let \(\sigma\) denote the uniform probability measure on the unit sphere \(\mathbb S^{d-1}\).

\begin{definition}[Sliced Wasserstein distance]
For \(\mu,\nu\in\mathcal P_p(\R^d)\), the sliced Wasserstein distance is
\[
  SW_p(\mu,\nu)=\left(\int_{\mathbb S^{d-1}} W_p^p\bigl((P_{\btheta})_\#\mu,(P_{\btheta})_\#\nu\bigr)\,\dd\sigma(\btheta)\right)^{1/p}.
\]
\end{definition}

Thus \(SW_p^p\) is the average one-dimensional Wasserstein cost over all linear views of the distribution. In practice, the integral is approximated by random projections. If \(\btheta_1,\ldots,\btheta_M\) are independent uniform directions on the sphere, define
\[
  \widehat{SW}_{p,M}^p(\mu,\nu)=\frac1M\sum_{m=1}^M W_p^p\bigl((P_{\btheta_m})_\#\mu,(P_{\btheta_m})_\#\nu\bigr).
\]
For empirical distributions with equal sample size \(n\), each term is computed by sorting the projected samples:
\[
  W_p^p(\widehat\mu_{\btheta},\widehat\nu_{\btheta})=\frac1n\sum_{i=1}^n \left|X_{(i),\btheta}-Y_{(i),\btheta}\right|^p,
\]
where \(X_{(i),\btheta}\) and \(Y_{(i),\btheta}\) are the sorted projected values.

\begin{proposition}[Sliced Wasserstein is dominated by Wasserstein]
For every \(\btheta\in\mathbb S^{d-1}\),
\[
  W_p\bigl((P_{\btheta})_\#\mu,(P_{\btheta})_\#\nu\bigr)\le W_p(\mu,\nu).
\]
Consequently,
\[
  SW_p(\mu,\nu)\le W_p(\mu,\nu).
\]
\end{proposition}

\begin{proof}
The projection \(P_{\btheta}\) is 1-Lipschitz because \(|\btheta^\top\bx-\btheta^\top\by|\le \|\bx-\by\|\). Let \(\gamma\in\Pi(\mu,\nu)\). Then \((P_{\btheta},P_{\btheta})_\#\gamma\) is a coupling of \((P_{\btheta})_\#\mu\) and \((P_{\btheta})_\#\nu\). Hence
\[
  W_p^p\bigl((P_{\btheta})_\#\mu,(P_{\btheta})_\#\nu\bigr)\le \int |\btheta^\top\bx-\btheta^\top\by|^p\,\dd\gamma(\bx,\by)\le \int \|\bx-\by\|^p\,\dd\gamma(\bx,\by).
\]
Taking the infimum over \(\gamma\in\Pi(\mu,\nu)\) gives the first inequality. Integrating over \(\btheta\) gives the second.
\end{proof}

\begin{proposition}[Separation of distributions]
If \(SW_p(\mu,\nu)=0\), then \(\mu=\nu\).
\end{proposition}

\begin{proof}
If \(SW_p(\mu,\nu)=0\), then \(W_p((P_{\btheta})_\#\mu,(P_{\btheta})_\#\nu)=0\) for \(\sigma\)-almost every \(\btheta\). Hence the one-dimensional projected laws coincide for almost every \(\btheta\). By continuity in \(\btheta\), and by the Cram\'{e}r--Wold device, equality of all one-dimensional linear projections implies \(\mu=\nu\).
\end{proof}

\subsection{Max-sliced and selected-direction distances}

Averaging over directions is not always the actuarial goal. Sometimes one wants the most adverse direction, or a small set of interpretable directions.

\begin{definition}[Max-sliced Wasserstein distance]
The max-sliced Wasserstein distance is
\[
  MSW_p(\mu,\nu)=\sup_{\btheta\in\mathbb S^{d-1}} W_p\bigl((P_{\btheta})_\#\mu,(P_{\btheta})_\#\nu\bigr).
\]
\end{definition}

The sliced distance is an average discrepancy; the max-sliced distance is a worst-direction discrepancy. Both satisfy
\[
  SW_p(\mu,\nu)\le MSW_p(\mu,\nu)\le W_p(\mu,\nu).
\]

In actuarial applications, it is often useful to report three quantities:
\[
  \{W_{p,\be_k}(\mu,\nu):k=1,\ldots,d\}, \qquad SW_p(\mu,\nu), \qquad MSW_p(\mu,\nu).
\]
The coordinate projections identify marginal shifts, the sliced distance gives an average multivariate discrepancy, and the max-sliced distance reveals the largest directional change.

\begin{actuarialbox}
Sliced Wasserstein is a global average over views. Max-sliced Wasserstein is a stress diagnostic: it searches for the direction in which two distributions are most different. In portfolio monitoring, this direction may be more informative than the average discrepancy.
\end{actuarialbox}

\subsection{Choosing projections}

Random projections are convenient, but they are not the only choice. In actuarial problems, projections should be chosen with the application in mind.

\paragraph{Coordinate projections.}
The directions \(\be_1,\ldots,\be_d\) compare marginal distributions of each coordinate. They are easy to interpret, but they miss dependence changes.

\paragraph{Aggregate directions.}
The vector \((1,\ldots,1)/\sqrt d\) compares total loss or total exposure. More generally, a vector of present-value weights compares discounted cash-flows.

\paragraph{Capital directions.}
A vector \(\btheta\) may represent allocation weights, stress weights or a linearized capital sensitivity. Then \(\btheta^\top\bX\) is a capital-relevant projection.

\paragraph{Data-adaptive directions.}
One may use principal components, supervised directions, gradients of a prediction model, or directions found by maximizing projected Wasserstein distance. These are often informative, but they require governance: the projection is no longer purely neutral.

\paragraph{Random directions.}
Random directions provide a cheap approximation to the full spherical average. They are useful when one wants a stable summary rather than a named diagnostic. The number of projections \(M\) controls the Monte Carlo error.

\begin{computebox}
A practical report may include coordinate projected distances, an aggregate-loss projection, a present-value cash-flow projection, a random-projection sliced Wasserstein estimate, and the largest projected distance among a large random set of directions. This provides both interpretability and scalability.
\end{computebox}

\subsection{Mini-lab: sliced Wasserstein in R}
\label{lab:sliced-wasserstein-r}

The following code computes a Monte Carlo approximation of \(SW_p\) for two clouds of vectors. It uses only random projections and empirical lower quantiles.

\begin{lstlisting}[language=R]
# Empirical lower generalized inverse on a grid u in (0, 1)
emp_q_left <- function(x, u) {
  if (length(x) == 0L) stop("x must be non-empty")
  if (any(!is.finite(x))) stop("x must contain only finite values")
  if (any(u <= 0 | u >= 1)) stop("u must lie in (0, 1)")
  xs <- sort(as.numeric(x))
  n <- length(xs)
  xs[pmax(1L, ceiling(u * n))]
}

# Sliced Wasserstein distance between two point clouds
sliced_wasserstein <- function(X, Y, p = 2, M = 200, seed = NULL) {
  if (!is.null(seed)) set.seed(seed)
  X <- as.matrix(X)
  Y <- as.matrix(Y)
  if (ncol(X) != ncol(Y)) stop("X and Y must have the same number of columns")
  if (p < 1) stop("p must be at least 1")
  if (M < 1) stop("M must be positive")

  n <- nrow(X)
  m <- nrow(Y)
  d <- ncol(X)
  grid_size <- max(n, m)
  u <- seq(1 / grid_size, 1, length.out = grid_size)
  u <- pmin(u, 1 - .Machine$double.eps)
  vals <- numeric(M)

  for (k in seq_len(M)) {
    theta <- rnorm(d)
    theta <- theta / sqrt(sum(theta^2))
    xproj <- as.numeric(X %*% theta)
    yproj <- as.numeric(Y %*% theta)
    qx <- emp_q_left(xproj, u)
    qy <- emp_q_left(yproj, u)
    vals[k] <- mean(abs(qx - qy)^p)
  }

  mean(vals)^(1 / p)
}

set.seed(123)
n <- 1000

# Two simulated bivariate loss portfolios
Sigma1 <- matrix(c(1.0, 0.3, 0.3, 1.0), 2, 2)
Sigma2 <- matrix(c(1.2, 0.7, 0.7, 1.5), 2, 2)

X <- matrix(rnorm(2 * n), n, 2) %*% chol(Sigma1)
Y <- matrix(rnorm(2 * n), n, 2) %*% chol(Sigma2) + c(0.3, 0.1)

sliced_wasserstein(X, Y, p = 2, M = 500, seed = 1)
\end{lstlisting}

The same function can be used for cash-flow vectors, model-output vectors, simulated capital scenarios or policyholder profiles, after suitable scaling of the coordinates.

\subsection{Mini-lab: directional diagnostics}
\label{lab:projected-wasserstein-diagnostics}

The next code compares named directions and random directions.

\begin{lstlisting}[language=R]
projected_wasserstein <- function(X, Y, theta, p = 2, grid_size = 1000) {
  X <- as.matrix(X)
  Y <- as.matrix(Y)
  if (ncol(X) != ncol(Y)) stop("X and Y must have the same number of columns")
  if (length(theta) != ncol(X)) stop("theta must have one entry per column")
  if (sum(theta^2) == 0) stop("theta must be non-zero")
  if (p < 1) stop("p must be at least 1")

  theta <- theta / sqrt(sum(theta^2))
  xproj <- as.numeric(X %*% theta)
  yproj <- as.numeric(Y %*% theta)
  u <- seq(1 / grid_size, 1, length.out = grid_size)
  u <- pmin(u, 1 - .Machine$double.eps)
  qx <- emp_q_left(xproj, u)
  qy <- emp_q_left(yproj, u)

  mean(abs(qx - qy)^p)^(1 / p)
}

# Coordinate directions
projected_wasserstein(X, Y, c(1, 0), p = 2)
projected_wasserstein(X, Y, c(0, 1), p = 2)

# Aggregate-loss direction
projected_wasserstein(X, Y, c(1, 1), p = 2)

# Approximate max-sliced direction by random search
random_projected <- function(X, Y, p = 2, M = 500, seed = 123) {
  set.seed(seed)
  d <- ncol(X)
  out <- data.frame(value = numeric(M), theta1 = numeric(M), theta2 = numeric(M))

  for (k in seq_len(M)) {
    theta <- rnorm(d)
    theta <- theta / sqrt(sum(theta^2))
    out$value[k] <- projected_wasserstein(X, Y, theta, p = p)
    out$theta1[k] <- theta[1]
    out$theta2[k] <- theta[2]
  }
  out[order(-out$value), ][1:10, ]
}

random_projected(X, Y, p = 2, M = 1000)
\end{lstlisting}

\subsection{Limitations}

Sliced and projected distances are computationally attractive, but they should not be over-interpreted.

\begin{itemize}
\item A small number of projections may miss a localized multivariate change.
\item Coordinate projections can miss dependence shifts.
\item Random projections are less interpretable than actuarially named projections.
\item Max-sliced distances may be sensitive to sample noise if the maximizing direction is estimated from the same data.
\item Scaling of coordinates matters. Before projecting mixed actuarial features, one must engineer the ground geometry carefully.
\end{itemize}

The best practice is therefore not to replace full OT by sliced OT blindly, but to use sliced and projected distances as scalable diagnostics. They are especially valuable in exploratory analysis, model monitoring and high-dimensional scenario comparison.

\subsection{Exercises}

\begin{exercise}[Projection is non-expansive]
\label{ex:SW:1}
Let \(\btheta\in\mathbb S^{d-1}\). Prove that
\[
  W_p\bigl((P_{\btheta})_\#\mu,(P_{\btheta})_\#\nu\bigr)\le W_p(\mu,\nu).
\]
What changes if \(\btheta\) is not normalized?
\end{exercise}

\begin{exercise}[Coordinate projections]
\label{ex:SW:2}
Let \(\mu,\nu\in\mathcal P_p(\R^d)\). Show that
\[
  W_p\bigl((P_{\be_k})_\#\mu,(P_{\be_k})_\#\nu\bigr)
\]
only compares the \(k\)-th marginal distributions. Give an example where all coordinate projected distances are zero but \(\mu\neq\nu\).
\end{exercise}

\begin{exercise}[Sliced distance in dimension one]
\label{ex:SW:3}
Show that when \(d=1\), the sliced Wasserstein distance is equal to the ordinary Wasserstein distance, up to the convention used for the unit sphere \(\mathbb S^0=\{-1,1\}\).
\end{exercise}

\begin{exercise}[Average versus worst direction]
\label{ex:SW:4}
Prove that
\[
  SW_p(\mu,\nu)\le MSW_p(\mu,\nu)\le W_p(\mu,\nu).
\]
Give an actuarial example where the average projected discrepancy is small but one direction has a large discrepancy.
\end{exercise}

\begin{exercise}[Equal marginals, different dependence]
\label{ex:SW:5}
Construct two distributions on \(\R^2\) with the same two marginal distributions but different dependence. Compute or simulate coordinate projected distances. Then compare projected distances in directions \((1,1)/\sqrt2\) and \((1,-1)/\sqrt2\).
\end{exercise}

\begin{exercise}[Cash-flow projection]
\label{ex:SW:6}
Let \(\bX=(X_1,\ldots,X_T)\) and \(\bY=(Y_1,\ldots,Y_T)\) be two predictive cash-flow vectors. Let \(v_t=(1+r)^{-t}\) and \(\btheta=\bv/\|\bv\|\). Interpret
\[
  W_p\bigl((P_{\btheta})_\#\Law(\bX),(P_{\btheta})_\#\Law(\bY)\bigr)
\]
actuarially.
\end{exercise}

\begin{exercise}[Monte Carlo approximation]
\label{ex:SW:7}
Let \(\btheta_1,\ldots,\btheta_M\) be i.i.d. uniform on \(\mathbb S^{d-1}\). Show that
\[
  \frac1M\sum_{m=1}^M W_p^p\bigl((P_{\btheta_m})_\#\mu,(P_{\btheta_m})_\#\nu\bigr)
\]
converges almost surely to \(SW_p^p(\mu,\nu)\), provided the integrand is integrable.
\end{exercise}

\begin{exercise}[Random projection simulation]
\label{ex:SW:8}
Simulate two bivariate Gaussian samples with different covariance matrices. Compute coordinate projected distances, the aggregate projected distance in direction \((1,1)/\sqrt2\), and an approximation of \(SW_2\). Which diagnostic best detects the covariance change?
\end{exercise}

\begin{exercise}[Sensitivity to scaling]
\label{ex:SW:9}
Let one coordinate represent age and another represent claim amount. Explain why random projections are meaningless before scaling or cost engineering. Propose a normalization and discuss how it changes the sliced distance.
\end{exercise}

\begin{exercise}[Max-sliced search]
\label{ex:SW:10}
Implement a random search approximation of \(MSW_2\). Repeat the search with \(M=50,100,500,2000\) directions. Plot the maximum projected distance as a function of \(M\). Why is the curve nondecreasing?
\end{exercise}

\begin{exercise}[Two-dimensional output audit]
\label{ex:SW:11}
Suppose each policyholder has two model outputs, for instance a frequency output and a severity output. Compare two subportfolios using coordinate projected distances and random sliced Wasserstein distance. Explain what would be missed by auditing the two outputs separately.
\end{exercise}

\begin{exercise}[When full OT and sliced OT disagree]
\label{ex:SW:12}
Give a qualitative example where full \(W_2\) may be large but \(SW_2\) is moderate, or where a few projected directions reveal a discrepancy that an average over many directions dilutes. What should an actuarial report include in such a case?
\end{exercise}

\newpage
% Appendix / insertable lab
% Lab 3: Wasserstein fairness repair
% Suggested placement: after Lab 2: Sliced and Projected Wasserstein Distances,
% or after Section 12.16 Practical audit protocol.

\section{Lab 3: Wasserstein Fairness Repair}
\label{sec:lab-wasserstein-fairness-repair}

This lab studies one-dimensional Wasserstein post-processing for distributional fairness. The setting is deliberately simple: a real-valued actuarial score, predicted loss, technical premium or risk indicator is observed together with a group variable. The aim is to transform group-specific score distributions toward a common target distribution, while monitoring the actuarial cost of the transformation.

The lab illustrates three points. First, a fairness repair is a distributional operation, not a mean correction. Second, in one dimension Wasserstein barycenters provide an explicit common target through quantile averaging. Third, full distributional repair can conflict with prediction, calibration, profitability or transition constraints, so partial repair and diagnostic reporting are necessary.

\begin{actuarialbox}%[Distributional, not causal]
The maps used in this lab align empirical score distributions. They do not identify the causal effect of group membership on an individual outcome, and they do not decide which variables are legally or ethically admissible. Such questions require additional assumptions and institutional judgement.
\end{actuarialbox}

\begin{actuarialbox}%[Quantile convention]
Throughout this lab, empirical quantiles are computed using the lower generalized inverse convention. With atoms or ties, deterministic rank averaging is used in the code for reproducibility. Randomized ranks would be closer to the population quantile-coupling construction when exact treatment of atoms matters.
\end{actuarialbox}

\subsection{Learning objectives}

After completing this lab, the reader should be able to:
\begin{itemize}
\item compute one-dimensional Wasserstein disparities between group-specific score distributions;
\item construct a Wasserstein barycenter target by averaging quantile functions;
\item implement groupwise quantile maps sending each group distribution to the barycenter;
\item implement partial Wasserstein repair and trace a fairness and accuracy frontier;
\item distinguish distributional parity from calibration and predictive accuracy;
\item produce an actuarial report including disparity, movement, RMSE, loss-ratio and transition metrics.
\end{itemize}

\subsection{The statistical object}

Let \(S\in\mathcal S\) denote a group variable with \(\Pbb(S=s)>0\) for every group considered, and let \(Z\in\R\) be a score, predicted loss, technical premium or risk indicator. For each group \(s\), define
\[
  \mu_s=\Law(Z\mid S=s).
\]
Distributional demographic parity would require
\[
  \mu_s=\mu_{s'} \qquad\text{for all }s,s'\in\mathcal S.
\]
In practice, the groupwise laws are usually different. A Wasserstein disparity index can be based on pairwise distances
\[
  W_p(\mu_s,\mu_{s'}),
\]
or on distances to a common target distribution \(\bar\mu\):
\[
  \sum_{s\in\mathcal S}\alpha_s W_p^p(\mu_s,\bar\mu), \qquad \alpha_s=\Pbb(S=s).
\]

A natural Wasserstein target is the barycenter
\[
  \bar\mu\in\argmin_{\mu}\sum_{s\in\mathcal S}\alpha_s W_2^2(\mu_s,\mu).
\]
In one dimension, assuming finite second moments, its quantile function has the simple form
\[
  Q_{\bar\mu}(u)=\sum_{s\in\mathcal S}\alpha_s Q_s(u), \qquad u\in(0,1),
\]
where \(Q_s\) is the lower quantile function of \(\mu_s\).

\begin{computebox}
In one dimension, Wasserstein barycenters and repair maps can be computed by sorting and interpolating empirical quantiles. No multivariate OT solver is needed.
\end{computebox}

\subsection{Full and partial repair}

For each group \(s\), the monotone map from \(\mu_s\) to \(\bar\mu\) is
\[
  T_s(z)=Q_{\bar\mu}\bigl(F_s(z)\bigr),
\]
where \(F_s\) is the distribution function of \(Z\mid S=s\). With atoms, this formula should be read with the generalized-inverse convention and, if needed, randomized ranks inside atoms. The fully repaired score is
\[
  Z^{\mathrm{rep}}=T_S(Z).
\]
At the population level, under the usual one-dimensional quantile-coupling assumptions,
\[
  \Law(Z^{\mathrm{rep}}\mid S=s)=\bar\mu
\]
for every group \(s\). Thus the group-specific repaired score distributions are equal.

A partial repair interpolates between the original score and the fully repaired score:
\[
  Z_\lambda=(1-\lambda)Z+\lambda Z^{\mathrm{rep}}, \qquad \lambda\in[0,1].
\]
The parameter \(\lambda\) controls the strength of the repair. At \(\lambda=0\), nothing is changed. At \(\lambda=1\), the groupwise score laws are aligned with the barycenter. The quadratic decrease of the Wasserstein disparity, when used, is a property of this one-dimensional quantile interpolation, not of arbitrary post-processing rules.

\begin{remark}[What is preserved?]
The monotone repair preserves ranks within each group. If two individuals belong to the same group and one has a larger original score than the other, the same ordering holds after full repair. Ranks across groups may change, and the repaired score may no longer be calibrated for the outcome of interest.
\end{remark}

\subsection{Fairness is not calibration}

Let \(Y\) be the observed loss, claim indicator or response variable. If \(Z\) is a calibrated predicted loss, one may have approximately
\[
  \E[Y\mid Z=z]\approx z.
\]
After a groupwise transformation, calibration can fail:
\[
  \E[Y\mid Z^{\mathrm{rep}}=z]\not\approx z.
\]
This is not a numerical bug. It reflects a structural tension between making the score distribution independent of \(S\) and preserving the predictive meaning of the score. Therefore, a fairness repair should be evaluated not only through groupwise Wasserstein distances, but also through predictive and actuarial metrics.

For a technical premium \(P\), useful quantities include
\[
  \text{mean premium by group}, \qquad \text{loss ratio by group}, \qquad \E[|P_\lambda-P_0|], \qquad \Pbb(|P_\lambda-P_0|>\tau),
\]
and, when an outcome \(Y\) is available,
\[
  \mathrm{RMSE}(P_\lambda,Y), \qquad \mathrm{MAE}(P_\lambda,Y), \qquad \E[Y-P_\lambda\mid S=s].
\]

\begin{actuarialbox}
A distributional fairness repair may reduce group disparities while increasing mispricing, disrupting technical premiums or changing loss ratios. The relevant object is therefore a frontier, not a single corrected score.
\end{actuarialbox}

\subsection{R lab}
\label{lab:wasserstein-fairness-repair-r}

\subsubsection{Helper functions}

\begin{lstlisting}[language=R]
# Lower empirical quantile function.
emp_q_left <- function(x, probs) {
  x <- sort(x[is.finite(x)])
  if (length(x) == 0L) stop("x must contain at least one finite value")
  probs <- pmin(pmax(probs, 0), 1)
  idx <- ceiling(probs * length(x))
  idx <- pmax(1L, pmin(length(x), idx))
  x[idx]
}

# One-dimensional empirical Wasserstein distance on a common quantile grid.
wass1d <- function(x, y, p = 1, grid_size = 1000, trim = 0) {
  if (p < 1) stop("p must be at least 1")
  x <- x[is.finite(x)]
  y <- y[is.finite(y)]
  if (length(x) == 0L || length(y) == 0L) {
    stop("x and y must contain finite values")
  }
  if (grid_size < 2L) stop("grid_size must be at least 2")

  lo <- trim
  hi <- 1 - trim
  if (lo <= 0) lo <- 1 / (grid_size + 1)
  if (hi >= 1) hi <- grid_size / (grid_size + 1)
  if (lo >= hi) stop("trim is too large")

  u <- seq(lo, hi, length.out = grid_size)
  qx <- emp_q_left(x, u)
  qy <- emp_q_left(y, u)
  mean(abs(qx - qy)^p)^(1 / p)
}

# Empirical ranks in (0,1). Average ranks give a deterministic treatment of ties.
rank01 <- function(x) {
  if (length(x) == 0L) numeric(0)
  else (rank(x, ties.method = "average") - 0.5) / length(x)
}

# Weighted quantile barycenter on a grid.
barycenter_quantile <- function(z, group, weights = NULL, grid_size = 1000) {
  z <- as.numeric(z)
  group <- as.factor(group)
  lev <- levels(group)
  if (length(z) != length(group)) stop("z and group must have the same length")
  if (any(!is.finite(z))) stop("z must contain only finite values")
  if (any(table(group) == 0L)) stop("all groups must be nonempty")
  if (grid_size < 2L) stop("grid_size must be at least 2")

  u <- seq(1 / (grid_size + 1), grid_size / (grid_size + 1), length.out = grid_size)

  if (is.null(weights)) {
    weights <- as.numeric(prop.table(table(group)))
    names(weights) <- lev
  } else {
    weights <- as.numeric(weights)
    if (length(weights) != length(lev)) stop("weights must match the number of groups")
    names(weights) <- lev
    weights <- weights / sum(weights)
  }

  qmat <- sapply(lev, function(g) emp_q_left(z[group == g], u))
  qbar <- as.vector(qmat %*% weights[lev])

  list(u = u, qbar = qbar, qmat = qmat, weights = weights[lev])
}

# Linear interpolation of the barycenter quantile function.
qbar_eval <- function(u, bary) {
  approx(x = bary$u, y = bary$qbar, xout = u, rule = 2, ties = "ordered")$y
}

# Full Wasserstein repair to the barycenter target.
repair_to_barycenter <- function(z, group, bary) {
  z <- as.numeric(z)
  group <- as.factor(group)
  if (length(z) != length(group)) stop("z and group must have the same length")
  z_rep <- numeric(length(z))

  for (g in levels(group)) {
    idx <- which(group == g)
    u <- rank01(z[idx])
    z_rep[idx] <- qbar_eval(u, bary)
  }
  z_rep
}

# Partial repair.
partial_repair <- function(z, z_rep, lambda) {
  if (lambda < 0 || lambda > 1) stop("lambda must belong to [0,1]")
  (1 - lambda) * z + lambda * z_rep
}
\end{lstlisting}

\subsubsection{Simulating an actuarial score}

The following simulation creates two groups. The true loss depends on a risk covariate \(X\), a group effect and noise. The score \(Z\) is a noisy predicted loss. The example is a sandbox for distributional post-processing, not a causal or legal model.

\begin{lstlisting}[language=R]
set.seed(2026)

n <- 4000
S <- rbinom(n, size = 1, prob = 0.45)
S_fac <- factor(S, labels = c("A", "B"))

# Risk covariate with a different distribution by group.
X <- rnorm(n, mean = 0.4 * S, sd = 1)

# True expected loss and observed loss.
eta <- 7.7 + 0.55 * X + 0.25 * S
mu_true <- exp(eta)
Y <- rgamma(n, shape = 2.5, scale = mu_true / 2.5)

# A predicted premium or score, with model noise.
Z <- exp(7.7 + 0.55 * X + 0.25 * S + rnorm(n, sd = 0.12))

summary_by_group <- function(v, group) {
  tapply(v, group, function(x) {
    qs <- emp_q_left(x, c(.50, .90, .99))
    c(mean = mean(x), sd = sd(x), q50 = qs[1], q90 = qs[2], q99 = qs[3])
  })
}

summary_by_group(Z, S_fac)
wass1d(Z[S_fac == "A"], Z[S_fac == "B"], p = 1)
wass1d(Z[S_fac == "A"], Z[S_fac == "B"], p = 2)
\end{lstlisting}

\subsubsection{Computing the barycenter and full repair}

\begin{lstlisting}[language=R]
bary <- barycenter_quantile(Z, S_fac, grid_size = 1000)
Z_rep <- repair_to_barycenter(Z, S_fac, bary)

# Check groupwise distances before and after repair.
before_W1 <- wass1d(Z[S_fac == "A"], Z[S_fac == "B"], p = 1)
after_W1  <- wass1d(Z_rep[S_fac == "A"], Z_rep[S_fac == "B"], p = 1)

before_W2 <- wass1d(Z[S_fac == "A"], Z[S_fac == "B"], p = 2)
after_W2  <- wass1d(Z_rep[S_fac == "A"], Z_rep[S_fac == "B"], p = 2)

c(before_W1 = before_W1, after_W1 = after_W1,
  before_W2 = before_W2, after_W2 = after_W2)

summary_by_group(Z_rep, S_fac)
\end{lstlisting}

In finite samples, the after-repair distances are not exactly zero because of unequal group sizes, interpolation and empirical quantile approximations. They should nevertheless be much smaller than the before-repair distances.

\subsubsection{Visual diagnostics}

\begin{lstlisting}[language=R]
par(mfrow = c(1, 2))

plot(density(Z[S_fac == "A"]),
     main = "Original score distributions",
     xlab = "score", lwd = 2)
lines(density(Z[S_fac == "B"]), lwd = 2, lty = 2)
legend("topright", legend = c("A", "B"), lty = c(1, 2),
       lwd = 2, bty = "n")

plot(density(Z_rep[S_fac == "A"]),
     main = "After Wasserstein repair",
     xlab = "repaired score", lwd = 2)
lines(density(Z_rep[S_fac == "B"]), lwd = 2, lty = 2)
legend("topright", legend = c("A", "B"), lty = c(1, 2),
       lwd = 2, bty = "n")
\end{lstlisting}

\subsubsection{The fairness and accuracy frontier}

We now vary \(\lambda\) from 0 to 1 and compute a small set of diagnostics.

\begin{lstlisting}[language=R]
metrics_two_groups <- function(pred, original, y, group) {
  group <- as.factor(group)
  lev <- levels(group)
  if (length(lev) != 2L) stop("this helper expects exactly two groups")

  W1 <- wass1d(pred[group == lev[1]], pred[group == lev[2]], p = 1)
  W2 <- wass1d(pred[group == lev[1]], pred[group == lev[2]], p = 2)

  movement_abs <- mean(abs(pred - original))
  movement_rel <- mean(abs(pred - original) / pmax(original, 1))

  rmse <- sqrt(mean((y - pred)^2))
  mae <- mean(abs(y - pred))

  lr <- tapply(y, group, sum) / tapply(pred, group, sum)
  mean_pred <- tapply(pred, group, mean)
  mean_y <- tapply(y, group, mean)

  data.frame(W1 = W1, W2 = W2,
             mean_abs_movement = movement_abs,
             mean_rel_movement = movement_rel,
             RMSE = rmse, MAE = mae,
             LR_A = as.numeric(lr[1]), LR_B = as.numeric(lr[2]),
             mean_pred_A = as.numeric(mean_pred[1]),
             mean_pred_B = as.numeric(mean_pred[2]),
             mean_loss_A = as.numeric(mean_y[1]),
             mean_loss_B = as.numeric(mean_y[2]))
}

lambda_grid <- seq(0, 1, by = 0.05)
frontier <- do.call(rbind, lapply(lambda_grid, function(lambda) {
  Z_lam <- partial_repair(Z, Z_rep, lambda)
  cbind(lambda = lambda, metrics_two_groups(Z_lam, Z, Y, S_fac))
}))

frontier

par(mfrow = c(1, 2))
plot(frontier$lambda, frontier$W1, type = "b",
     xlab = expression(lambda), ylab = "W1 disparity",
     main = "Distributional disparity")
plot(frontier$W1, frontier$RMSE, type = "b",
     xlab = "W1 disparity", ylab = "RMSE",
     main = "Fairness and accuracy frontier")
\end{lstlisting}

The frontier shows how disparity decreases as the repair becomes stronger, while score movement and predictive error may increase. In practice, the choice of \(\lambda\) is a governance decision, not a numerical optimum alone.

\subsubsection{Transition constraints}

A repair may be unacceptable if too many policyholders experience a large movement in the score or technical premium. The following code computes simple transition metrics.

\begin{lstlisting}[language=R]
disruption <- function(pred, original, thresholds = c(.05, .10, .25)) {
  rel <- abs(pred - original) / pmax(original, 1)
  out <- sapply(thresholds, function(tau) mean(rel > tau))
  names(out) <- paste0("Pr_move_gt_", thresholds)
  c(mean_rel_move = mean(rel), out)
}

disruption_table <- do.call(rbind, lapply(lambda_grid, function(lambda) {
  Z_lam <- partial_repair(Z, Z_rep, lambda)
  c(lambda = lambda, disruption(Z_lam, Z))
}))

disruption_table
\end{lstlisting}

\begin{actuarialbox}
For insurance pricing, a fairness repair should be constrained by transition rules. A mathematically clean distributional repair may be operationally or legally unacceptable if it creates excessive individual movements.
\end{actuarialbox}

\begin{figure}[t]
\centering
\includegraphics[width=.82\textwidth]{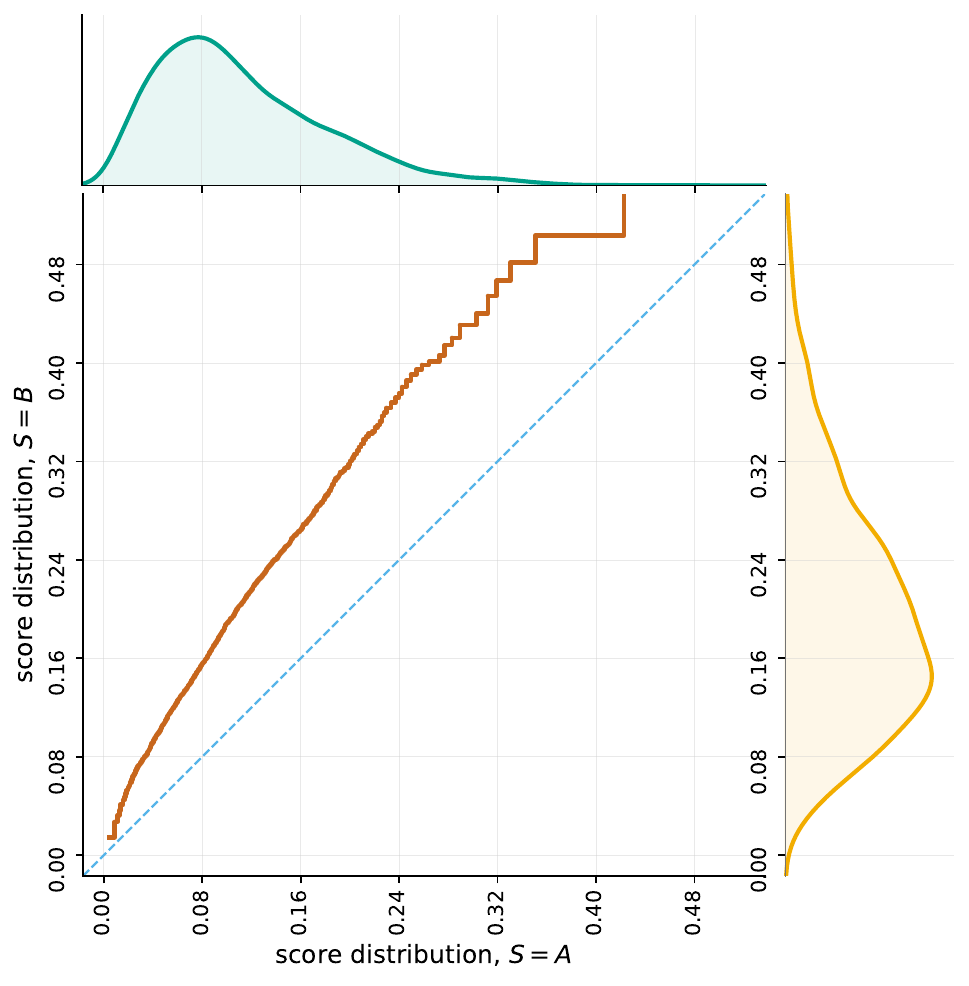}
\caption{Quantile transport between group-specific score distributions. The top and right panels show the marginal score densities for the two groups, while the red curve is the monotone transport map from group \(S=A\) to group \(S=B\). This is the basic one-dimensional object behind the Wasserstein repair used in this lab.}
\label{fig:transport-lab3-fairness}
\end{figure}

\subsection{Suggested report template}

A short lab report should include the following table.

\begin{center}
\begin{tabular}{lrrrrrr}
\toprule
Repair & \(\lambda\) & \(W_1\) disparity & RMSE & Mean move & LR A & LR B\\
\midrule
None & 0 & & & & & \\
Partial & 0.25 & & & & & \\
Partial & 0.50 & & & & & \\
Partial & 0.75 & & & & & \\
Full & 1 & & & & & \\
\bottomrule
\end{tabular}
\end{center}

The discussion should address three questions:
\begin{enumerate}
\item How much distributional disparity is removed?
\item What is the actuarial cost in prediction, calibration or loss ratio?
\item Are individual movements acceptable under the chosen transition constraints?
\end{enumerate}

\subsection{Extensions}

\paragraph{More than two groups.}
The code above can be extended to any finite number of groups. The barycenter quantile is the weighted average of all group quantile functions. For many small groups, empirical quantiles become noisy.

\iffalse
% To be restored in v2: discussion of intersectional fairness audits and multiple sensitive variables.
This small-cell issue is especially important for intersectional fairness audits, where the number of group cells can grow quickly and empirical quantiles may become unstable.
\fi

\paragraph{Protected attribute versus risk attribute.}
The lab treats \(S\) as a group used for auditing or post-processing. Whether it can be used, ignored, proxied or adjusted for depends on the legal and business context. The mathematical repair does not decide that question.

\paragraph{Outcome-aware repair.}
The barycenter repair aligns score distributions. It does not use the outcome \(Y\) except for evaluation. More advanced repairs may combine distributional fairness with calibration constraints or outcome loss functions.

\paragraph{Multivariate outputs.}
For multivariate outputs, coordinatewise repair may not align joint laws. One can use projected diagnostics or sliced Wasserstein distances; see Section~\ref{sec:sliced-projected-wasserstein}.

\subsection{What to remember}

Wasserstein fairness repair is a transparent way to align groupwise score distributions. In one dimension, it is computationally simple and has an explicit rank-preserving interpretation. Distributional parity is only one desideratum. An actuarial audit must also report calibration, predictive performance, loss ratios, individual movements and sensitivity to the repair strength.

\subsection{Exercises}

\begin{exercise}[Pairwise Wasserstein disparity]
Using the simulated data, compute \(W_1\) and \(W_2\) between the original score distributions of groups A and B. Repeat after full repair. Explain why the after-repair distances are small but not necessarily exactly zero in finite samples.
\label{ex:FR:1}
\end{exercise}

\begin{exercise}[Manual barycenter]
Let group A have empirical quantiles \((1,2,4)\), group B have empirical quantiles \((2,5,8)\), and weights \((0.6,0.4)\). Compute the barycenter quantiles by hand. Then map the three A-quantiles and the three B-quantiles to the barycenter.
\label{ex:FR:2}
\end{exercise}

\begin{exercise}[Rank preservation]
Show that the one-dimensional repair map \(T_s(z)=Q_{\bar\mu}(F_s(z))\) is nondecreasing. Explain what this implies for the ranking of individuals within the same group.
\label{ex:FR:3}
\end{exercise}

\begin{exercise}[Partial repair frontier]
Using \(\lambda\in\{0,.1,.2,\ldots,1\}\), compute the fairness and accuracy frontier. Plot \(W_1\) disparity against RMSE. Where does the largest marginal improvement in disparity occur?
\label{ex:FR:4}
\end{exercise}

\begin{exercise}[Loss-ratio diagnostics]
For each \(\lambda\), compute groupwise loss ratios
\[
  \frac{\sum_{i:S_i=s}Y_i}{\sum_{i:S_i=s}Z_{\lambda,i}}.
\]
Does the full repair improve or worsen groupwise loss-ratio balance in the simulation?
\label{ex:FR:5}
\end{exercise}

\begin{exercise}[Transition constraints]
For each \(\lambda\), compute
\[
  \Pbb\left(\frac{|Z_\lambda-Z|}{Z}>10\%\right).
\]
Find the largest \(\lambda\) such that this proportion remains below a threshold chosen by the instructor.
\label{ex:FR:6}
\end{exercise}

\begin{exercise}[Changing the barycenter weights]
The default barycenter weights are the empirical group proportions. Repeat the repair using equal group weights. Compare the target distribution, group mean movements and fairness and accuracy frontier.
\label{ex:FR:7}
\end{exercise}

\begin{exercise}[Tail disparity]
Compute the upper-tail Wasserstein disparity
\[
  \frac{1}{1-\alpha}\int_\alpha^1 |Q_A(u)-Q_B(u)|\,\dd u
\]
for \(\alpha=.90\) and \(\alpha=.95\), before and after repair. Compare with the global \(W_1\) disparity.
\label{ex:FR:8}
\end{exercise}

\begin{exercise}[Calibration curve]
Bin the repaired score \(Z_\lambda\) into deciles and compare average observed loss with average predicted loss in each bin. Repeat for \(\lambda=0\), \(\lambda=.5\) and \(\lambda=1\). What happens to calibration?
\label{ex:FR:9}
\end{exercise}

\begin{exercise}[More than two groups]
Modify the simulation to create three groups. Implement the barycenter repair and compute the maximum pairwise \(W_1\) disparity before and after repair.
\label{ex:FR:10}
\end{exercise}

\begin{exercise}[Group variable not used in the model]
Simulate a score \(Z\) that does not explicitly use \(S\), but where the covariate \(X\) has different distributions by group. Show that groupwise score distributions can still differ. Explain why unawareness of \(S\) is not, by itself, a distributional parity guarantee.
\label{ex:FR:11}
\end{exercise}

\begin{exercise}[Audit memo]
Write a one-page actuarial memo recommending a value of \(\lambda\). The memo should discuss distributional disparity, accuracy, calibration, loss ratios and transition constraints. It should also state which questions the Wasserstein repair cannot answer.
\label{ex:FR:12}
\end{exercise}

\newpage

\newpage
% Appendix / insertable lab
% Lab 5: Reserving cash-flow distributions
% Suggested placement: after Lab 4: Premium Drift and Tariff Disruption,
% or after Section 11.7 Transporting cash-flow scenarios / Section 11.16 Practical workflow.

\section{Lab 5: Reserving Cash-Flow Distributions}
\label{sec:lab-reserving-cash-flow-distributions}

This lab studies reserving models as generators of predictive cash-flow \emph{distributions}. A stochastic reserving method does not only produce a best estimate reserve. It produces a law on future payment vectors. Optimal transport then provides tools for comparing predictive laws, detecting timing changes, constructing barycentric model blends, and building distributional reserving reports.

The lab focuses on simulated cash-flow scenarios
\[
  \bC=(C_1,\ldots,C_T)\in\R_+^T,
\]
where \(C_t\) is the incremental payment at future development year \(t\). The same ideas apply to calendar-year cash flows, accident-year specific cash flows, claim-count/severity decompositions, or net cash flows including recoveries and expenses.

\begin{actuarialbox}
A reserve is often reported as one number. A stochastic reserving model gives a whole distribution of future cash-flow vectors. Comparing reserving models therefore means comparing laws on \(\R_+^T\), not only comparing best estimates.
\end{actuarialbox}

\subsection{Learning objectives}

After completing this lab, the reader should be able to:
\begin{itemize}
\item represent reserving outputs as empirical laws of cash-flow vectors;
\item compare present-value reserve distributions with one-dimensional Wasserstein distances;
\item compare full cash-flow distributions with selected projections and sliced Wasserstein diagnostics;
\item distinguish level changes, timing changes and tail changes;
\item use cumulative cash-flow transformations to detect payment acceleration or delay;
\item construct simple Wasserstein barycenters of present-value reserve distributions;
\item build a practical reserving report with best estimate, tail risk, timing and distributional distance metrics.
\end{itemize}

\subsection{The actuarial object}

Let
\[
  \bC=(C_1,\ldots,C_T)
\]
be the vector of future incremental payments generated by a stochastic reserving
model. Its law is denoted
\[
  \mu=\Law(\bC)\in\Prob(\R_+^T).
\]
A scalar reserve is usually obtained by applying a present-value map
\[
  g(\bC)=\sum_{t=1}^T v_t C_t,
\]
where \(v_t\) are discount factors. The present-value reserve distribution is the
push-forward
\[
  \Law(g(\bC)) = g_\#\mu.
\]
Thus two reserving models may be compared in at least two ways:
\[
  W_p\bigl(g_\#\mu_A,g_\#\mu_B\bigr),
\]
which compares scalar present-value reserves, and
\[
  W_p(\mu_A,\mu_B),
\]
which compares the full laws of cash-flow vectors.

\begin{remark}[Scalar reserves lose timing information]
Two cash-flow vectors can have the same present value but very different timing.
For example, an early payment vector and a late payment vector may have similar
best estimates under low discounting, while producing very different liquidity,
claims management and reporting consequences. A vector comparison is therefore
more informative than a scalar comparison when timing matters.
\end{remark}

\subsection{Cash-flow costs}

The simplest cost between two cash-flow vectors \(\bc,\bc'\in\R_+^T\) is a weighted
\(\ell_p\) cost:
\[
  d_{\bw}(\bc,\bc')^p
  =
  \sum_{t=1}^T w_t |c_t-c'_t|^p,
  \qquad w_t>0.
\]
The weights may be chosen as discount weights, reporting-period weights, or
business-priority weights. For example, setting \(w_t=v_t\) emphasizes present
value; setting larger weights for small \(t\) emphasizes near-term liquidity.

A timing-aware alternative compares cumulative cash flows. Define
\[
  A_t(\bc)=\sum_{s=1}^t c_s,
  \qquad t=1,\ldots,T.
\]
Then
\[
  d_{\mathrm{cum}}(\bc,\bc')^p
  =
  \sum_{t=1}^T w_t |A_t(\bc)-A_t(\bc')|^p
\]
penalizes differences in payment emergence. This cost is especially useful when
one model pays claims earlier or later than another model.

\begin{computebox}
For reserving, a cost between cash-flow vectors should be chosen according to the
diagnostic question:
\begin{itemize}
\item present-value cost for economic reserve comparisons;
\item near-term weighted cost for liquidity monitoring;
\item cumulative cost for payment-timing differences;
\item tail-projection costs for capital and risk adjustment diagnostics.
\end{itemize}
\end{computebox}

\subsection{Projection diagnostics}

Exact multivariate Wasserstein distances in \(\R^T\) may be expensive when the
number of scenarios is large. A practical alternative is to compare projections.
Important reserving projections include:
\[
  \btheta_{\mathrm{PV}}^\top \bC=\sum_{t=1}^T v_t C_t,
\]
\[
  \btheta_{\mathrm{tot}}^\top \bC=\sum_{t=1}^T C_t,
\]
\[
  \btheta_{\mathrm{early}}^\top \bC=\sum_{t=1}^{t_0} C_t,
  \qquad
  \btheta_{\mathrm{late}}^\top \bC=\sum_{t=t_0+1}^T C_t.
\]
Each projection gives a one-dimensional distribution that can be compared by
quantile formulas. A sliced Wasserstein diagnostic averages over many random
projections:
\[
  SW_p^p(\mu_A,\mu_B)
  =
  \int_{S^{T-1}}
  W_p^p\bigl((\btheta^\top)_\#\mu_A,(\btheta^\top)_\#\mu_B\bigr)
  \,\dd\btheta.
\]
In practice the integral is approximated by Monte Carlo directions.

\begin{actuarialbox}
Projected diagnostics are often easier to explain than a single high-dimensional
number. For reserves, useful projections are present value, undiscounted total,
early payments, late payments, calendar-year totals and cumulative paid-to-date
profiles.
\end{actuarialbox}

\subsection{A stability inequality for present values}

Let \(g(\bc)=\bv^\top \bc\) be the present-value map. If the cash-flow space is equipped
with the Euclidean norm, then
\[
  |g(\bc)-g(\bc')|
  \le
  \|\bv\|_2\,\|\bc-\bc'\|_2.
\]
Therefore present-value Wasserstein distances are controlled by vector
Wasserstein distances.

\begin{proposition}[Push-forward stability for reserves]
Let \(\mu_A,\mu_B\in\Probp{p}(\R^T)\), and let \(g(\bc)=\bv^\top \bc\). Then
\[
  W_p\bigl(g_\#\mu_A,g_\#\mu_B\bigr)
  \le
  \|\bv\|_q W_p(\mu_A,\mu_B),
\]
where \(q\) is the conjugate exponent of \(p\), with the usual convention \(q=\infty\) for \(p=1\). Here the vector-valued Wasserstein distance is computed with the \(\ell_p\) ground norm.
\end{proposition}

\begin{proof}
For any coupling \(\gamma\in\Coup(\mu_A,\mu_B)\), the pair
\((g(\bC),g(\bC'))\) is a coupling of \(g_\#\mu_A\) and \(g_\#\mu_B\). By H\"older's
inequality,
\[
  |g(\bC)-g(\bC')|^p
  \le
  \|\bv\|_q^p \|\bC-\bC'\|_p^p.
\]
Taking expectations under \(\gamma\), minimizing over \(\gamma\), and taking the
\(p\)-th root gives the result.
\end{proof}

\begin{actuarialbox}%[Quantile convention]
All one-dimensional Wasserstein computations in this lab use the lower empirical quantile function. This matches the generalized inverse convention used throughout the notes. In empirical reserving samples with ties or atoms, this convention should be stated explicitly rather than hidden in the software default.
\end{actuarialbox}

\subsection{R lab}
\label{lab:reserving-cash-flow-distributions-r}

\subsubsection{Helper functions}

\begin{lstlisting}[language=R]
emp_q_left <- function(x, probs) {
  x <- sort(x[is.finite(x)])
  if (length(x) == 0) stop("x has no finite observations")
  probs <- pmin(pmax(probs, 0), 1)
  idx <- ceiling(probs * length(x))
  idx[idx < 1] <- 1
  x[idx]
}

# One-dimensional empirical Wasserstein distance on a quantile grid.
wass1d <- function(x, y, p = 1, grid_size = 1000, trim = 0) {
  if (p < 1) stop("p must be at least 1")
  if (grid_size < 2) stop("grid_size must be at least 2")
  x <- x[is.finite(x)]
  y <- y[is.finite(y)]
  if (length(x) == 0 || length(y) == 0) stop("x and y must contain finite observations")

  lo <- trim
  hi <- 1 - trim
  if (lo <= 0) lo <- 1 / (grid_size + 1)
  if (hi >= 1) hi <- grid_size / (grid_size + 1)
  if (lo >= hi) stop("trim leaves an empty quantile range")

  u <- seq(lo, hi, length.out = grid_size)
  qx <- emp_q_left(x, u)
  qy <- emp_q_left(y, u)
  mean(abs(qx - qy)^p)^(1 / p)
}

wass1d_tail <- function(x, y, p = 1, alpha = 0.95, grid_size = 500) {
  if (alpha <= 0 || alpha >= 1) stop("alpha must be in (0, 1)")
  if (grid_size < 2) stop("grid_size must be at least 2")
  x <- x[is.finite(x)]
  y <- y[is.finite(y)]
  if (length(x) == 0 || length(y) == 0) stop("x and y must contain finite observations")
  upper <- grid_size / (grid_size + 1)
  if (alpha >= upper) stop("alpha is too high for this grid_size")
  u <- seq(alpha, upper, length.out = grid_size)
  qx <- emp_q_left(x, u)
  qy <- emp_q_left(y, u)
  mean(abs(qx - qy)^p)^(1 / p)
}

TVaR_emp <- function(x, alpha = 0.99) {
  if (alpha <= 0 || alpha >= 1) stop("alpha must be in (0, 1)")
  x <- x[is.finite(x)]
  if (length(x) == 0) stop("x has no finite observations")
  q <- emp_q_left(x, alpha)
  mean(x[x >= q])
}

pv_cashflows <- function(C, rate = 0.03) {
  if (!is.matrix(C)) stop("C must be a matrix")
  T <- ncol(C)
  v <- 1 / (1 + rate)^(seq_len(T))
  as.vector(C %*% v)
}

cum_cashflows <- function(C) {
  if (!is.matrix(C)) stop("C must be a matrix")
  t(apply(C, 1, cumsum))
}

mean_payment_time <- function(C) {
  if (!is.matrix(C)) stop("C must be a matrix")
  T <- ncol(C)
  tt <- seq_len(T)
  total <- rowSums(C)
  out <- as.vector(C %*% tt) / pmax(total, 1e-12)
  out[!is.finite(out)] <- NA_real_
  out
}

reserve_summary <- function(pv) {
  pv <- pv[is.finite(pv)]
  if (length(pv) == 0) stop("pv has no finite observations")
  c(mean = mean(pv), sd = sd(pv),
    q50 = emp_q_left(pv, .50),
    q75 = emp_q_left(pv, .75),
    q95 = emp_q_left(pv, .95),
    q99 = emp_q_left(pv, .99),
    TVaR99 = TVaR_emp(pv, .99))
}

projection_wass <- function(X, Y, theta, p = 1, grid_size = 1000) {
  if (!is.matrix(X) || !is.matrix(Y)) stop("X and Y must be matrices")
  if (ncol(X) != ncol(Y)) stop("X and Y must have the same number of columns")
  if (length(theta) != ncol(X)) stop("theta has incompatible dimension")
  norm_theta <- sqrt(sum(theta^2))
  if (!is.finite(norm_theta) || norm_theta == 0) stop("theta must be non-zero")
  theta <- theta / norm_theta
  x <- as.vector(X %*% theta)
  y <- as.vector(Y %*% theta)
  wass1d(x, y, p = p, grid_size = grid_size)
}

sliced_wass <- function(X, Y, L = 200, p = 2, grid_size = 500) {
  if (!is.matrix(X) || !is.matrix(Y)) stop("X and Y must be matrices")
  if (ncol(X) != ncol(Y)) stop("X and Y must have the same number of columns")
  if (L < 1) stop("L must be positive")
  d <- ncol(X)
  vals <- numeric(L)
  for (ell in seq_len(L)) {
    theta <- rnorm(d)
    theta <- theta / sqrt(sum(theta^2))
    vals[ell] <- projection_wass(X, Y, theta, p = p, grid_size = grid_size)^p
  }
  mean(vals)^(1 / p)
}

quantile_diff <- function(x, y, probs = seq(.01, .99, by = .01)) {
  qx <- emp_q_left(x, probs)
  qy <- emp_q_left(y, probs)
  data.frame(u = probs, q_A = qx, q_B = qy, diff = qy - qx)
}
\end{lstlisting}

\subsubsection{Simulating reserving cash-flow scenarios}

The following function simulates future incremental payment scenarios. The
baseline model has a moderate common shock. The late-payment model shifts mass
towards later years. The heavy-tail model keeps a similar mean pattern but adds a
stronger scenario-level shock.

\begin{lstlisting}[language=R]
set.seed(2026)

make_pattern <- function(T, type = c("baseline", "late", "early")) {
  type <- match.arg(type)
  x <- seq_len(T)
  if (type == "baseline") w <- dgamma(x, shape = 2.0, scale = 1.6)
  if (type == "late")     w <- dgamma(x, shape = 3.2, scale = 1.4)
  if (type == "early")    w <- dgamma(x, shape = 1.5, scale = 1.4)
  w / sum(w)
}

simulate_cashflows <- function(n = 5000, T = 8,
                               ultimate_mean = 1e6,
                               ultimate_cv = 0.35,
                               pattern_type = "baseline",
                               inflation = 0,
                               tail = c("moderate", "heavy")) {
  tail <- match.arg(tail)

  pat <- make_pattern(T, pattern_type)
  trend <- (1 + inflation)^(seq_len(T) - 1)
  pat <- pat * trend
  pat <- pat / sum(pat)

  # Scenario-level ultimate variability.
  sigma_u <- sqrt(log(1 + ultimate_cv^2))
  meanlog_u <- log(ultimate_mean) - 0.5 * sigma_u^2
  ultimate <- rlnorm(n, meanlog = meanlog_u, sdlog = sigma_u)

  # Common shock controls dependence between future development years.
  if (tail == "moderate") {
    shock <- rgamma(n, shape = 8, rate = 8)
  } else {
    shock <- rlnorm(n, meanlog = -0.5 * 0.55^2, sdlog = 0.55)
  }

  mu <- outer(ultimate * shock, pat)

  # Conditional gamma noise around the scenario mean.
  shape <- 12
  C <- matrix(rgamma(n * T, shape = shape,
                     scale = as.vector(mu) / shape),
              nrow = n, ncol = T)
  colnames(C) <- paste0("Y", seq_len(T))
  C
}

n <- 5000
T <- 8

C_A <- simulate_cashflows(n, T,
                          pattern_type = "baseline",
                          inflation = 0.02,
                          tail = "moderate")

C_B <- simulate_cashflows(n, T,
                          pattern_type = "late",
                          inflation = 0.04,
                          tail = "moderate")

C_C <- simulate_cashflows(n, T,
                          pattern_type = "baseline",
                          inflation = 0.02,
                          tail = "heavy")
\end{lstlisting}

\subsubsection{Present-value reserve distributions}

\begin{lstlisting}[language=R]
rate <- 0.03
PV_A <- pv_cashflows(C_A, rate = rate)
PV_B <- pv_cashflows(C_B, rate = rate)
PV_C <- pv_cashflows(C_C, rate = rate)

rbind(model_A = reserve_summary(PV_A),
      model_B = reserve_summary(PV_B),
      model_C = reserve_summary(PV_C))

# One-dimensional Wasserstein distances between present-value reserves.
pv_distances <- rbind(
  A_vs_B = c(W1 = wass1d(PV_A, PV_B, p = 1),
             W2 = wass1d(PV_A, PV_B, p = 2),
             W1_tail95 = wass1d_tail(PV_A, PV_B, p = 1, alpha = .95)),
  A_vs_C = c(W1 = wass1d(PV_A, PV_C, p = 1),
             W2 = wass1d(PV_A, PV_C, p = 2),
             W1_tail95 = wass1d_tail(PV_A, PV_C, p = 1, alpha = .95)),
  B_vs_C = c(W1 = wass1d(PV_B, PV_C, p = 1),
             W2 = wass1d(PV_B, PV_C, p = 2),
             W1_tail95 = wass1d_tail(PV_B, PV_C, p = 1, alpha = .95))
)

pv_distances
\end{lstlisting}

Here, model B should mainly differ through timing and inflation. Model C should
mainly differ through tail behaviour. The tail-restricted distance should be more
sensitive to model C than the global \(W_1\) distance.

\subsubsection{Timing diagnostics}

\begin{lstlisting}[language=R]
# Mean projected incremental payments by future year.
mean_A <- colMeans(C_A)
mean_B <- colMeans(C_B)
mean_C <- colMeans(C_C)

rbind(model_A = mean_A,
      model_B = mean_B,
      model_C = mean_C)

# Mean payment time distribution.
D_A <- mean_payment_time(C_A)
D_B <- mean_payment_time(C_B)
D_C <- mean_payment_time(C_C)

c(W1_duration_A_B = wass1d(D_A, D_B, p = 1),
  W1_duration_A_C = wass1d(D_A, D_C, p = 1),
  mean_duration_A = mean(D_A, na.rm = TRUE),
  mean_duration_B = mean(D_B, na.rm = TRUE),
  mean_duration_C = mean(D_C, na.rm = TRUE))

# Cumulative payment paths.
K_A <- cum_cashflows(C_A)
K_B <- cum_cashflows(C_B)
K_C <- cum_cashflows(C_C)

# Compare cumulative vectors through sliced Wasserstein.
c(SW_incremental_A_B = sliced_wass(C_A, C_B, L = 100, p = 2),
  SW_incremental_A_C = sliced_wass(C_A, C_C, L = 100, p = 2),
  SW_cumulative_A_B  = sliced_wass(K_A, K_B, L = 100, p = 2),
  SW_cumulative_A_C  = sliced_wass(K_A, K_C, L = 100, p = 2))
\end{lstlisting}

The cumulative comparison is often more sensitive to timing shifts than the raw
incremental comparison. A late-payment model may have a present value close to
the baseline but a visibly different cumulative emergence path.

\subsubsection{Selected projections}

\begin{lstlisting}[language=R]
T <- ncol(C_A)
v <- 1 / (1 + rate)^(seq_len(T))

theta_pv <- v
theta_tot <- rep(1, T)
theta_early <- c(rep(1, floor(T / 2)), rep(0, T - floor(T / 2)))
theta_late <- 1 - theta_early
theta_slope <- seq(-1, 1, length.out = T)

projection_table <- function(X, Y) {
  c(PV = projection_wass(X, Y, theta_pv, p = 1),
    Total = projection_wass(X, Y, theta_tot, p = 1),
    Early = projection_wass(X, Y, theta_early, p = 1),
    Late = projection_wass(X, Y, theta_late, p = 1),
    Slope = projection_wass(X, Y, theta_slope, p = 1))
}

rbind(A_vs_B = projection_table(C_A, C_B),
      A_vs_C = projection_table(C_A, C_C),
      B_vs_C = projection_table(C_B, C_C))
\end{lstlisting}

The selected projections should be interpreted. A large late projection indicates
differences in tail development years. A large slope projection indicates a timing
change, because the direction contrasts early and late payments.

\subsubsection{Barycentric present-value reserves}

In one dimension, Wasserstein barycenters are obtained by averaging quantile
functions. The following code constructs a barycentric present-value reserve
distribution from three model outputs.

\begin{lstlisting}[language=R]
barycenter_quantiles <- function(samples, weights = NULL,
                                 grid_size = 1000) {
  K <- length(samples)
  if (K == 0) stop("samples must be non-empty")
  if (is.null(weights)) weights <- rep(1 / K, K)
  if (length(weights) != K) stop("weights must have the same length as samples")
  if (any(weights < 0) || sum(weights) <= 0) stop("weights must be non-negative and non-zero")
  weights <- weights / sum(weights)

  u <- seq(1 / (grid_size + 1),
           grid_size / (grid_size + 1),
           length.out = grid_size)

  Q <- sapply(samples, function(z) {
    emp_q_left(z, u)
  })

  as.vector(Q %*% weights)
}

Q_bar <- barycenter_quantiles(list(PV_A, PV_B, PV_C),
                              weights = c(.5, .3, .2))

summary(Q_bar)

# Compare each model to the barycenter quantile sample.
c(A_to_bar = wass1d(PV_A, Q_bar, p = 2),
  B_to_bar = wass1d(PV_B, Q_bar, p = 2),
  C_to_bar = wass1d(PV_C, Q_bar, p = 2))
\end{lstlisting}

This barycenter is not a replacement for actuarial judgement. It is a transparent
way to summarize several predictive reserve distributions in the Wasserstein
geometry.

\subsubsection{Visual diagnostics}

\begin{lstlisting}[language=R]
par(mfrow = c(2, 2))

matplot(seq_len(T), cbind(mean_A, mean_B, mean_C), type = "l",
        lty = 1:3, lwd = 2,
        xlab = "future year", ylab = "mean incremental payment",
        main = "Mean cash-flow emergence")
legend("topright", legend = c("A", "B", "C"), lty = 1:3, lwd = 2,
       bty = "n")

plot(density(PV_A), lwd = 2, main = "PV reserve distributions",
     xlab = "present value")
lines(density(PV_B), lwd = 2, lty = 2)
lines(density(PV_C), lwd = 2, lty = 3)
legend("topright", legend = c("A", "B", "C"), lty = 1:3, lwd = 2,
       bty = "n")

qd_AB <- quantile_diff(PV_A, PV_B)
plot(qd_AB$u, qd_AB$diff, type = "l", lwd = 2,
     xlab = "quantile level", ylab = "Q_B(u) - Q_A(u)",
     main = "PV quantile difference: B - A")
abline(h = 0, lty = 2)

plot(density(na.omit(D_A)), lwd = 2,
     main = "Mean payment time", xlab = "payment year")
lines(density(na.omit(D_B)), lwd = 2, lty = 2)
lines(density(na.omit(D_C)), lwd = 2, lty = 3)
legend("topright", legend = c("A", "B", "C"), lty = 1:3, lwd = 2,
       bty = "n")
\end{lstlisting}

\subsection{Suggested report template}

A short reserving distributional report should contain at least the following
items.

\begin{center}
\begin{tabular}{lll}
\toprule
Diagnostic & Example quantity & Interpretation\\
\midrule
Best estimate & \(\E[\bv^\top \bC]\) & Mean present-value reserve\\
Tail reserve & \(\VaR_{.99}(\bv^\top \bC)\), \(\TVaR_{.99}(\bv^\top \bC)\) & Capital / risk adjustment view\\
PV distributional distance & \(W_1(g_\#\mu_A,g_\#\mu_B)\) & Scalar reserve movement\\
Timing distance & \(W_1(\Law(D_A),\Law(D_B))\) & Payment acceleration or delay\\
Selected projections & PV, total, early, late, slope & Interpretable vector diagnostics\\
Sliced distance & \(SW_2(\mu_A,\mu_B)\) & Multivariate cash-flow comparison\\
Cumulative distance & \(SW_2(A_\#\mu_A,A_\#\mu_B)\) & Emergence-pattern comparison\\
\bottomrule
\end{tabular}
\end{center}

\begin{actuarialbox}
A reserving model comparison should say whether differences reflect level, timing, dispersion or tail dependence. Wasserstein distances are useful precisely
because they can be computed after applying the relevant actuarial projection.
\end{actuarialbox}

\subsection{Extensions}

The lab can be extended in several directions.

\paragraph{Accident-year specific laws.}
Instead of one aggregate vector \(\bC\), let \(\bC^{(a)}\) be the future cash-flow vector
for accident year \(a\). One can compute Wasserstein distances by accident year,
then aggregate them with exposure weights.

\paragraph{Calendar-year liquidity.}
If the rows of a reserving scenario matrix correspond to accident years and the
columns to future calendar years, the projection onto calendar-year totals is a
liquidity diagnostic.

\paragraph{Net cash flows.}
Recoveries, reinsurance and expenses can be included by defining
\[
  C_t^{\mathrm{net}}
  =
  C_t^{\mathrm{gross}}-R_t^{\mathrm{recovery}}+E_t^{\mathrm{expense}}.
\]
The same transport diagnostics apply to the law of net cash-flow vectors.

\paragraph{IFRS 17 movement analysis.}
Distributional distances can be used as explanatory diagnostics when assumptions
change from one reporting date to the next. The Wasserstein distance is not an
accounting measurement rule; it is a tool for explaining how projected cash-flow
laws moved.

\subsection{Exercises}

\begin{exercise}[Present value as push-forward]
\label{ex:RCF:1}
Let \(\bC\in\R_+^T\) have law \(\mu\), and let \(g(\bc)=\sum_{t=1}^T v_t c_t\).
Show that the present-value reserve distribution is \(g_\#\mu\). Explain what
information is lost when replacing \(\mu\) by \(g_\#\mu\).
\end{exercise}

\begin{exercise}[Same total, different timing]
\label{ex:RCF:2}
Consider \(\bc=(100,0,0)\) and \(\bc'=(0,0,100)\). Compute their undiscounted totals
and their present values for discount rate \(i>0\). Then compute the cumulative
profiles \(A_t(\bc)\) and \(A_t(\bc')\). Explain why a cumulative cost detects timing
risk.
\end{exercise}

\begin{exercise}[Choice of weights]
\label{ex:RCF:3}
For the cost \(d_{\bw}(\bc,\bc')^2=\sum_t w_t(c_t-c'_t)^2\), compare the choices
\(w_t=1\), \(w_t=v_t\), and \(w_t=\1_{\{t\le 3\}}\). Give one reserving question
for which each choice is appropriate.
\end{exercise}

\begin{exercise}[PV stability]
\label{ex:RCF:4}
Prove the push-forward stability inequality for \(p=2\):
\[
  W_2(g_\#\mu_A,g_\#\mu_B)
  \le
  \|\bv\|_2 W_2(\mu_A,\mu_B).
\]
Interpret the result in terms of cash-flow model comparison.
\end{exercise}

\begin{exercise}[Selected projections]
\label{ex:RCF:5}
Using the R lab, compare models A and B along the PV, total, early, late and slope
projections. Which projection best detects the timing shift? Which projection best
detects inflation?
\end{exercise}

\begin{exercise}[Tail-heavy model]
\label{ex:RCF:6}
Compare models A and C. Report the mean, standard deviation, \(.99\) quantile and
\(\TVaR_{.99}\) of the present-value reserve. Does the global \(W_1\) distance or
the tail-restricted distance better reveal the change?
\end{exercise}

\begin{exercise}[Sliced Wasserstein sensitivity]
\label{ex:RCF:7}
Compute the sliced Wasserstein distance between models A and B for
\(L=20,50,100,500\) random projections. How stable is the estimate? Repeat for
models A and C.
\end{exercise}

\begin{exercise}[Cumulative versus incremental comparisons]
\label{ex:RCF:8}
Compute sliced distances using incremental cash-flow matrices and cumulative
cash-flow matrices. Explain why the two diagnostics may rank models differently.
\end{exercise}

\begin{exercise}[Barycentric reserve distribution]
\label{ex:RCF:9}
Construct a one-dimensional Wasserstein barycenter of the present-value reserve
distributions with weights \((.5,.3,.2)\). How does its mean and \(.99\) quantile
compare with the three input models?
\end{exercise}

\begin{exercise}[Bootstrap uncertainty]
\label{ex:RCF:10}
Bootstrap rows of the scenario matrices \(\bC_A\) and \(\bC_B\). Estimate a confidence
interval for \(W_1(g_\#\mu_A,g_\#\mu_B)\). Explain why simulation size matters for
reserving model comparisons.
\end{exercise}

\begin{exercise}[Calendar-year liquidity]
\label{ex:RCF:11}
Modify the simulation so that scenarios are generated separately for several
accident years. Compute future calendar-year totals. Propose a Wasserstein
diagnostic for liquidity stress.
\end{exercise}

\begin{exercise}[Governance report]
\label{ex:RCF:12}
Design a one-page report comparing two reserving models. Include at least one
best-estimate metric, one tail metric, one timing metric, one Wasserstein metric
and one graphical diagnostic.
\end{exercise}

\newpage
% Appendix / insertable lab
% Lab 6: Longevity transport
% Suggested placement: after Lab 5: Reserving Cash-Flow Distributions,
% or after Section 16 Life, Mortality and Longevity Transport.

\section{Lab 6: Longevity Transport}
\label{sec:lab-longevity-transport}

This lab studies mortality and longevity risk through probability laws on remaining lifetimes, ages at death and present values of life-contingent cash-flows. A mortality table, a mortality scenario or a longevity stress induces a distribution. Optimal transport then provides a way to compare mortality assumptions, construct longevity stresses, blend scenarios and quantify the financial effect of mortality improvement.

The lab is deliberately simple. It uses synthetic mortality tables generated from Gompertz--Makeham-type forces of mortality. The same workflow applies to population life tables, insured lives, annuitant tables, projected tables and stochastic mortality scenarios.

\begin{actuarialbox}
Life insurance and annuity models often start with a table of probabilities \(q_x\). Optimal transport asks a distributional question: how far does the law of the remaining lifetime move when mortality assumptions change? The same change can then be pushed forward through annuity, death-benefit or pension cash-flow maps.
\end{actuarialbox}

\begin{actuarialbox}
\textbf{Quantile convention.} Throughout this lab, quantiles are generalized lower inverses, \(Q(u)=\inf\{t:F(t)\ge u\}\). This matters because curtate lifetime distributions are discrete. In the R code, empirical quantiles are therefore computed with an explicit lower-inverse function rather than with an interpolating quantile convention.
\end{actuarialbox}

\begin{actuarialbox}
\textbf{Distributional, not causal.} A transport map between two lifetime laws is a distributional alignment. It is useful for stress testing and liability sensitivity, but it does not identify what a given individual lifetime would have been under another mortality table.
\end{actuarialbox}

\subsection{Learning objectives}

After completing this lab, the reader should be able to:
\begin{itemize}
\item convert a mortality table into a probability law of the curtate future lifetime;
\item compute one-dimensional Wasserstein distances between lifetime distributions;
\item build a rank-preserving longevity stress by quantile mapping;
\item compare mortality scenarios through their effect on life annuity and death-benefit present values;
\item distinguish a shift in lifetime distribution from a shift in financial liability distribution;
\item construct one-dimensional Wasserstein barycenters of longevity scenarios;
\item build a simple longevity report with life expectancy, high lifetime quantiles, annuity values, death-benefit values and Wasserstein diagnostics.
\end{itemize}

\subsection{The actuarial object}

Let \((x)\) be a life aged \(x\). Write
\[
  K_x\in\{0,1,2,\ldots\}
\]
for the curtate future lifetime, so that \(K_x=k\) means that the life survives \(k\) complete years and dies during the following policy year. If the mortality table provides one-year death probabilities
\[
  q_x,q_{x+1},q_{x+2},\ldots,
\]
then
\[
  \Pbb(K_x=k)={}_k p_x q_{x+k},\qquad {}_k p_x=\prod_{j=0}^{k-1}(1-q_{x+j}),
\]
with the convention \({}_0p_x=1\). The law of \(K_x\) is a probability measure on \(\N\):
\[
  \mu_x=\sum_{k\ge 0} {}_k p_x q_{x+k}\,\delta_k.
\]
The corresponding age-at-death law is the push-forward of \(\mu_x\) by \(k\mapsto x+k+1\) if death is recorded at the end of the year of death.

\begin{remark}[Lifetime law versus mortality table]
The table \((q_{x+k})_{k\ge0}\) is a sequence of conditional one-year death probabilities. The measure \(\mu_x\) is the unconditional distribution of the future lifetime. Transport distances in this lab compare the latter object.
\end{remark}

A change in mortality assumptions transforms \(\mu_x^0\) into another law \(\mu_x^1\). For example, \(\mu_x^0\) may be a baseline table and \(\mu_x^1\) a longevity-improved table. In one dimension, for \(p\ge1\) and finite \(p\)-th moments, the Wasserstein distance has the quantile representation
\[
  W_p^p(\mu_x^0,\mu_x^1)=\int_0^1 \left|Q_0(u)-Q_1(u)\right|^p\,\dd u,
\]
where \(Q_0\) and \(Q_1\) are the lower quantile functions of remaining lifetime.

\begin{actuarialbox}
For lifetime distributions, \(W_1\) is measured in years. It can be read as an average absolute displacement in remaining lifetime. \(W_2\) penalizes large movements more strongly and is more sensitive to changes in the old-age tail.
\end{actuarialbox}

\subsection{Rank-preserving longevity stress}

Let \(F_0\) and \(F_1\) be the distribution functions of two lifetime laws \(\mu_x^0\) and \(\mu_x^1\). If the source law is continuous, the monotone transport map can be written as
\[
  T(k)=Q_1(F_0(k)).
\]
For discrete lifetime laws, this pointwise formula must be read with care because \(F_0(K_x)\) is not uniform. The clean implementation uses a common probability rank:
\[
  K_0=Q_0(U),\qquad K_1=Q_1(U),\qquad U\sim \mathrm{Unif}(0,1).
\]
This construction preserves rank in distribution. A life at the 90th percentile of the baseline lifetime law is matched with the 90th percentile of the stressed lifetime law.

For longevity risk, the map often moves upper quantiles more than lower quantiles. This is important for annuity portfolios, pension liabilities and long-term care benefits, where late payments matter.

\subsection{Life-contingent present values as push-forwards}

Let \(v=(1+i)^{-1}\) be a discount factor. A whole-life annuity-due paying one unit per year while the life is alive has present value, as a function of the curtate future lifetime \(K\),
\[
  a(K)=\sum_{j=0}^{K} v^j=\frac{1-v^{K+1}}{1-v},
\]
with the limiting convention \(a(K)=K+1\) when \(i=0\). The distribution of the annuity present value is
\[
  \Law(a(K_x))=a_\#\mu_x.
\]
A death benefit of one unit paid at the end of the year of death has present value
\[
  A(K)=v^{K+1},
\]
and distribution
\[
  \Law(A(K_x))=A_\#\mu_x.
\]

Thus a mortality stress induces three related but different distributional changes:
\[
  \mu_x^0 \longrightarrow \mu_x^1,
\]
for lifetimes,
\[
  a_\#\mu_x^0 \longrightarrow a_\#\mu_x^1,
\]
for annuity present values, and
\[
  A_\#\mu_x^0 \longrightarrow A_\#\mu_x^1,
\]
for death-benefit present values.

\begin{actuarialbox}
Longevity improvement increases annuity values but may decrease death-benefit present values, because death payments are delayed. The same transport on lifetimes can therefore have opposite financial effects on different products.
\end{actuarialbox}

\subsection{Barycenters of mortality scenarios}

Suppose several mortality scenarios are available:
\[
  \mu_x^{(1)},\ldots,\mu_x^{(M)}.
\]
They may correspond to expert views, stochastic mortality projections, population versus insured tables, or different longevity-improvement assumptions. A one-dimensional Wasserstein barycenter with weights \(\lambda_m\ge0\), \(\sum_m\lambda_m=1\), has quantile function
\[
  Q_{\mathrm{bar}}(u)=\sum_{m=1}^M \lambda_m Q_m(u).
\]
This gives a distributional blend of mortality scenarios. It is not a mixture. A mixture first selects a scenario and then samples a lifetime. A Wasserstein barycenter averages the quantiles and therefore gives a central lifetime law in the transport geometry.

\begin{remark}[Barycenter versus mixture]
A mixture of longevity scenarios may be more dispersed because it keeps scenario uncertainty as a mixture component. A Wasserstein barycenter gives a consensus scenario. Both are useful, but they answer different actuarial questions.
\end{remark}

\subsection{R lab}
\label{lab:longevity-transport-r}

The following R code generates simple mortality tables, converts them into lifetime distributions, and compares baseline and longevity-improved scenarios. No external package is required.

\begin{lstlisting}[language=R]
# Helper functions

make_qx <- function(age0 = 65, omega = 120,
                    A = 0.0005, B = 0.00002, c = 1.105,
                    improvement = 0) {
  if (omega <= age0) stop("omega must be greater than age0")
  ages <- age0:omega

  # Gompertz-Makeham force. The improvement parameter reduces
  # mortality more strongly at future attained ages.
  force <- A + B * c^ages
  imp <- exp(-improvement * pmax(ages - age0, 0))
  qx <- 1 - exp(-force * imp)

  # Close the table at omega.
  qx[length(qx)] <- 1

  data.frame(age = ages, qx = pmin(pmax(qx, 0), 1))
}

curtate_pmf <- function(qx) {
  if (any(!is.finite(qx)) || any(qx < 0) || any(qx > 1)) {
    stop("qx must contain probabilities in [0, 1]")
  }
  n <- length(qx)
  prob <- numeric(n)
  surv <- 1
  for (k in seq_len(n)) {
    prob[k] <- surv * qx[k]
    surv <- surv * (1 - qx[k])
  }
  prob / sum(prob)
}

q_pmf <- function(prob, u) {
  if (any(!is.finite(prob)) || any(prob < 0) || sum(prob) <= 0) {
    stop("prob must be a nonnegative vector with positive mass")
  }
  prob <- prob / sum(prob)
  u <- pmin(pmax(u, 0), 1)
  cdf <- cumsum(prob)
  k <- seq_along(prob) - 1
  idx <- vapply(u, function(uu) which(cdf >= uu)[1], integer(1))
  k[idx]
}

emp_q_left <- function(x, u) {
  if (length(x) == 0 || any(!is.finite(x))) stop("x must be finite and non-empty")
  if (any(u < 0 | u > 1)) stop("u must lie in [0, 1]")
  xs <- sort(x)
  n <- length(xs)
  xs[pmax(1, ceiling(u * n))]
}

wass_pmf <- function(prob1, prob2, p = 1, grid_size = 5000) {
  if (p < 1) stop("p must be at least 1")
  u <- (seq_len(grid_size) - 0.5) / grid_size
  q1 <- q_pmf(prob1, u)
  q2 <- q_pmf(prob2, u)
  mean(abs(q1 - q2)^p)^(1/p)
}

annuity_pv <- function(k, i = 0.02) {
  if (abs(i) < 1e-12) return(k + 1)
  v <- 1 / (1 + i)
  (1 - v^(k + 1)) / (1 - v)
}

death_pv <- function(k, i = 0.02) {
  v <- 1 / (1 + i)
  v^(k + 1)
}

sample_pmf <- function(prob, n) {
  if (any(!is.finite(prob)) || any(prob < 0) || sum(prob) <= 0) {
    stop("prob must be a nonnegative vector with positive mass")
  }
  sample(seq_along(prob) - 1, size = n, replace = TRUE, prob = prob)
}

TVaR_emp <- function(x, alpha = 0.99) {
  q <- emp_q_left(x, alpha)
  mean(x[x >= q])
}

summ_lifetime <- function(prob) {
  k <- seq_along(prob) - 1
  c(
    mean = sum(k * prob),
    q50 = q_pmf(prob, 0.50),
    q90 = q_pmf(prob, 0.90),
    q95 = q_pmf(prob, 0.95),
    q99 = q_pmf(prob, 0.99)
  )
}

# Mortality scenarios

age0 <- 65
omega <- 120

base_table <- make_qx(age0 = age0, omega = omega, improvement = 0.000)
imp_table  <- make_qx(age0 = age0, omega = omega, improvement = 0.012)
sev_table  <- make_qx(age0 = age0, omega = omega, improvement = 0.020)

p_base <- curtate_pmf(base_table$qx)
p_imp  <- curtate_pmf(imp_table$qx)
p_sev  <- curtate_pmf(sev_table$qx)

rbind(
  baseline = summ_lifetime(p_base),
  improved = summ_lifetime(p_imp),
  severe_longevity = summ_lifetime(p_sev)
)

# Wasserstein distances between remaining lifetime laws
c(
  W1_base_improved = wass_pmf(p_base, p_imp, p = 1),
  W2_base_improved = wass_pmf(p_base, p_imp, p = 2),
  W1_base_severe = wass_pmf(p_base, p_sev, p = 1),
  W2_base_severe = wass_pmf(p_base, p_sev, p = 2)
)

# Push-forward to present values

set.seed(123)
n <- 20000
K_base <- sample_pmf(p_base, n)
K_imp  <- sample_pmf(p_imp, n)
K_sev  <- sample_pmf(p_sev, n)

i_rate <- 0.02

ann_base <- annuity_pv(K_base, i = i_rate)
ann_imp  <- annuity_pv(K_imp,  i = i_rate)
ann_sev  <- annuity_pv(K_sev,  i = i_rate)

death_base <- death_pv(K_base, i = i_rate)
death_imp  <- death_pv(K_imp,  i = i_rate)
death_sev  <- death_pv(K_sev,  i = i_rate)

W1_sample <- function(x, y) {
  if (length(x) != length(y)) stop("x and y must have the same length")
  mean(abs(sort(x) - sort(y)))
}

W2_sample <- function(x, y) {
  if (length(x) != length(y)) stop("x and y must have the same length")
  sqrt(mean((sort(x) - sort(y))^2))
}

rbind(
  annuity_baseline = c(mean = mean(ann_base), VaR99 = emp_q_left(ann_base, .99),
                       TVaR99 = TVaR_emp(ann_base, .99)),
  annuity_improved = c(mean = mean(ann_imp), VaR99 = emp_q_left(ann_imp, .99),
                       TVaR99 = TVaR_emp(ann_imp, .99)),
  annuity_severe = c(mean = mean(ann_sev), VaR99 = emp_q_left(ann_sev, .99),
                     TVaR99 = TVaR_emp(ann_sev, .99))
)

rbind(
  death_baseline = c(mean = mean(death_base), VaR99 = emp_q_left(death_base, .99),
                     TVaR99 = TVaR_emp(death_base, .99)),
  death_improved = c(mean = mean(death_imp), VaR99 = emp_q_left(death_imp, .99),
                     TVaR99 = TVaR_emp(death_imp, .99)),
  death_severe = c(mean = mean(death_sev), VaR99 = emp_q_left(death_sev, .99),
                   TVaR99 = TVaR_emp(death_sev, .99))
)

c(
  W1_annuity_base_imp = W1_sample(ann_base, ann_imp),
  W2_annuity_base_imp = W2_sample(ann_base, ann_imp),
  W1_death_base_imp = W1_sample(death_base, death_imp),
  W2_death_base_imp = W2_sample(death_base, death_imp)
)

# Quantile map and Wasserstein barycenter

u_grid <- (seq_len(2000) - 0.5) / 2000
q_base <- q_pmf(p_base, u_grid)
q_imp  <- q_pmf(p_imp,  u_grid)
q_sev  <- q_pmf(p_sev,  u_grid)

# A rank-preserving longevity stress map on the grid
stress_shift <- q_imp - q_base
summary(stress_shift)

# Wasserstein barycenter of three scenarios
lambda <- c(0.50, 0.30, 0.20)
q_bar <- lambda[1] * q_base + lambda[2] * q_imp + lambda[3] * q_sev

# Since q_bar may be non-integer, it is a continuous lifetime approximation.
# Push it forward through annuity and death-benefit maps.
ann_bar <- annuity_pv(q_bar, i = i_rate)
death_bar <- death_pv(q_bar, i = i_rate)

c(
  mean_lifetime_barycenter = mean(q_bar),
  mean_annuity_barycenter = mean(ann_bar),
  mean_death_barycenter = mean(death_bar)
)

# Simple plots

par(mfrow = c(2, 2))
plot(base_table$age, base_table$qx, type = "l", lwd = 2,
     xlab = "attained age", ylab = "q_x", main = "Mortality rates")
lines(imp_table$age, imp_table$qx, lwd = 2, lty = 2)
lines(sev_table$age, sev_table$qx, lwd = 2, lty = 3)
legend("topleft", legend = c("baseline", "improved", "severe"),
       lty = 1:3, lwd = 2, bty = "n")

plot(u_grid, q_base, type = "l", lwd = 2,
     xlab = "u", ylab = "Q_K(u)", main = "Lifetime quantiles")
lines(u_grid, q_imp, lwd = 2, lty = 2)
lines(u_grid, q_sev, lwd = 2, lty = 3)

plot(q_base, q_imp, type = "l", lwd = 2,
     xlab = "baseline lifetime quantile",
     ylab = "improved lifetime quantile",
     main = "Rank-preserving longevity map")
abline(0, 1, lty = 2)

plot(sort(ann_base), sort(ann_imp), type = "l", lwd = 2,
     xlab = "baseline annuity PV quantile",
     ylab = "improved annuity PV quantile",
     main = "Annuity PV transport")
abline(0, 1, lty = 2)
\end{lstlisting}

\subsection{Diagnostics to report}

A practical longevity transport report should not contain only a best-estimate annuity value. It should report lifetime and financial distributional effects. A minimal table is:

\begin{center}
\begin{tabular}{lrrrrrr}
\toprule
Scenario & \(\E[K_x]\) & \(Q_{.95}(K_x)\) & \(W_1(K)\) & \(\E[a(K)]\) & \(Q_{.99}(a(K))\) & \(\E[A(K)]\)\\
\midrule
Baseline & & & 0 & & & \\
Improved mortality & & & & & & \\
Severe longevity & & & & & & \\
Barycenter & & & & & & \\
\bottomrule
\end{tabular}
\end{center}

The table separates three layers:
\begin{itemize}
\item lifetime quantities, such as \(\E[K_x]\), \(Q_{.95}(K_x)\), and \(W_1(K)\);
\item annuity liability quantities, such as \(\E[a(K)]\) and \(Q_{.99}(a(K))\);
\item death-benefit quantities, such as \(\E[A(K)]\).
\end{itemize}

\begin{computebox}
In a longevity report, always state whether a Wasserstein distance is computed on remaining lifetimes, ages at death, annuity present values or death-benefit present values. These are different push-forwards of the mortality assumptions and they have different financial interpretations.
\end{computebox}

\subsection{Extensions}

The lab can be extended in several directions.

\paragraph{Different entry ages.} Repeat the analysis for \(x=55,65,75\). The same mortality improvement may have different financial effects depending on age.

\paragraph{Interest-rate sensitivity.} Repeat the annuity and death-benefit push-forwards for \(i=0\%,2\%,5\%\). Lower interest rates increase the value of late annuity payments and therefore amplify longevity risk.

\paragraph{Tail-aware longevity transport.} Focus on the upper tail of the lifetime distribution:
\[
  W_{1,\alpha}^{\mathrm{tail}}(\mu_x^0,\mu_x^1)=\frac{1}{1-\alpha}\int_\alpha^1 |Q_0(u)-Q_1(u)|\,\dd u.
\]
This is useful for annuity portfolios and pension plans that are sensitive to very old ages.

\paragraph{Multi-population mortality.} Compare male and female tables, insured and population tables, or two countries. The same transport workflow applies, but interpretation must account for selection, underwriting and population differences.

\paragraph{Health-state and long-term-care benefits.} Replace the lifetime law by a path law on health states. Then the state space is no longer one-dimensional, and dynamic or adapted transport becomes more relevant.

\subsection{Exercises}

\begin{exercise}[From \(q_x\) to a lifetime law]
\label{ex:LT:1}
Let \((q_{65},\ldots,q_{120})\) be a mortality table with \(q_{120}=1\). Derive the probabilities \(\Pbb(K_{65}=k)\) for \(k=0,\ldots,55\). Check that they sum to one.
\end{exercise}

\begin{exercise}[Life expectancy]
\label{ex:LT:2}
Using the probabilities from Exercise~\ref{ex:LT:1}, compute \(\E[K_{65}]\). Interpret the difference between \(\E[K_{65}]\) and complete life expectancy.
\end{exercise}

\begin{exercise}[Wasserstein in years]
\label{ex:LT:3}
Let \(\mu^0\) and \(\mu^1\) be two curtate lifetime laws. Explain why \(W_1(\mu^0,\mu^1)\) is measured in years. Compute it using a quantile grid.
\end{exercise}

\begin{exercise}[Rank-preserving stress]
\label{ex:LT:4}
Construct the quantile map from a baseline lifetime law to an improved mortality law. Plot \(Q_1(u)\) against \(Q_0(u)\). What does it mean if the curve lies above the diagonal?
\end{exercise}

\begin{exercise}[Annuity push-forward]
\label{ex:LT:5}
Let \(a(K)=(1-v^{K+1})/(1-v)\). Show that the distribution of the annuity present value is \(a_\#\mu_x\). Compute its mean and compare it under baseline and improved mortality.
\end{exercise}

\begin{exercise}[Death-benefit push-forward]
\label{ex:LT:6}
Let \(A(K)=v^{K+1}\). Explain why longevity improvement may decrease \(\E[A(K)]\) while increasing \(\E[a(K)]\).
\end{exercise}

\begin{exercise}[Interest-rate sensitivity]
\label{ex:LT:7}
Repeat the annuity comparison for interest rates \(0\%\), \(2\%\) and \(5\%\). How does the Wasserstein distance between annuity present-value distributions change?
\end{exercise}

\begin{exercise}[Tail lifetime distance]
\label{ex:LT:8}
Define and compute
\[
  W_{1,.90}^{\mathrm{tail}}(\mu^0,\mu^1).
\]
Compare it with global \(W_1(\mu^0,\mu^1)\). Which is more relevant for an annuity portfolio?
\end{exercise}

\begin{exercise}[Barycenter of scenarios]
\label{ex:LT:9}
Given three lifetime laws with quantile functions \(Q_1,Q_2,Q_3\), construct the Wasserstein barycenter with weights \((.5,.3,.2)\). Compare the barycenter with a mixture of the three scenarios.
\end{exercise}

\begin{exercise}[Selection effect]
\label{ex:LT:10}
Suppose annuitants have lower mortality than the general population. Describe how this should appear in the lifetime quantile functions and in \(W_1(\mu_{\mathrm{population}},\mu_{\mathrm{annuitant}})\).
\end{exercise}

\begin{exercise}[Product-specific interpretation]
\label{ex:LT:11}
Compute the transport distance between baseline and improved mortality on lifetimes, annuity present values and death-benefit present values. Explain why the three distances do not answer the same actuarial question.
\end{exercise}

\begin{exercise}[Governance report]
\label{ex:LT:12}
Design a one-page longevity transport report for an annuity portfolio. Include at least two lifetime metrics, two financial metrics, one Wasserstein diagnostic and one sensitivity analysis.
\end{exercise}

\newpage
% Appendix: hints and selected proof strategies for the exercises.
% Intended use: place this file after \appendix, e.g.
% \input{Structure/appendix_hints}
%
% Labelling convention assumed in the main text:
%   \label{ex:4:7} for Exercise 4.7,
%   \label{ex:15:10} for Exercise 15.10,
%   \label{ex:B:3} for Exercise B.3 in the cost-engineering appendix.

\section{Hints and selected proof strategies}\label{app:hints}

This appendix gives hints rather than full solutions. The goal is to indicate the main identity, standard theorem, or modelling argument that unlocks each exercise. A useful convention is to label exercises in the main text by chapter and number, for instance \verb|\label{ex:4:7}| for Exercise 4.7. The hints below use that convention.

\providecommand{\hyperref}[2][]{#2}

\newenvironment{exercisehint}[3]{%
  \par\smallskip
  \noindent\textbf{Hint for \hyperref[#1]{Exercise~\ref{#1}} (#3).}\quad
}{\par\smallskip}

\newenvironment{solutionhint}[3]{%
  \par\smallskip
  \noindent\textbf{Suggested solution path for \hyperref[#1]{Exercise~\ref{#1}} (#3).}\quad
}{\par\smallskip}

\begin{readingbox} Most proof-based exercises rely on a small number of recurring mechanisms: push-forward identities, construction of couplings, finite-dimensional linear-programming duality, the uncrossing argument on the real line, Kantorovich duality, Brenier's theorem, quantile representations, Jensen/Hölder/Minkowski inequalities, and basic convexity of risk measures. For the mathematical background, see \cite{villani2003topics,villani2008oldnew,figalli2021invitation,ambrosio2021lectures}. For computational OT and Sinkhorn algorithms, see \cite{peyre2019computational,solomon2018discrete,chewi2025statistical}. For actuarial dependence and risk measures, see \cite{denuit2005actuarial,charpentier2024insurance}. \end{readingbox}

\subsection*{General proof patterns}

\paragraph{Push-forward arguments.}
To prove that a map transports one law into another, show that for every measurable set
\(B\), \(\Prob(T(X)\in B)=\nu(B)\). Equivalently, test against bounded measurable functions:
\[
  \int \varphi(y)\,\dd(T_\#\mu)(y)=\int \varphi(T(x))\,\dd\mu(x).
\]
This is often the shortest path for deductible maps, tariff maps, treaty maps and scenario
transformations.

\paragraph{Coupling arguments.} To construct a coupling, build a joint random vector with the desired marginals. In finite spaces, write a matrix and check row and column sums. In one dimension, take a common \(U\sim\mathrm{Unif}(0,1)\) and set \(X=F_\mu^{-1}(U)\), \(Y=F_\nu^{-1}(U)\).

\paragraph{Duality arguments.}
For discrete OT, weak duality is usually one line:
\[
  \sum_i a_i u_i+\sum_j b_j v_j
  =\sum_{i,j}\gamma_{ij}(u_i+v_j)
  \le \sum_{i,j}\gamma_{ij}C_{ij}.
\]
For continuous OT, replace sums by integrals and \((u_i,v_j,C_{ij})\) by
\((\varphi(x),\psi(y),c(x,y))\).

\paragraph{One-dimensional transport.} When the cost is convex in \(x-y\), look for an uncrossing argument. If \(x_1<x_2\) and \(y_1<y_2\), the uncrossed pairing is cheaper than the crossed pairing. In the limit, this yields quantile transport.

\paragraph{Wasserstein inequalities.}
Most stability bounds use either the primal coupling inequality
\[
  |\E f(X)-\E f(Y)|\le \mathrm{Lip}(f)\E d(X,Y),
\]
or the quantile formula in dimension one, followed by Hölder's inequality.

\paragraph{Computational exercises.} When an exercise asks for a plot or a simulation, the important part is not the numerical value but the diagnostic curve: how does the Wasserstein distance, robust capital, residual diagnostic, tail index or transport cost change when a modelling parameter changes?

\subsection*{Lecture 2: Couplings, push-forwards and insurance dependence}

\begin{exercisehint}{ex:2:1}{2.1}{Push-forward of a deductible} Compute \(P((L-d)_+\le y)\) separately for \(y<0\), \(y=0\), and \(y>0\). The atom at zero is \(P(L\le d)=F(d)\), up to the convention at the deductible. \end{exercisehint}

\begin{exercisehint}{ex:2:2}{2.2}{Graph couplings} For the graph part, test the event \(\{(x,y):y=T(x)\}\) under \(\Law(X,T(X))\). For a non-graph coupling, take \(\mu=\delta_0\) and a non-Dirac \(\nu\): no deterministic map can split the single atom. \end{exercisehint}

\begin{exercisehint}{ex:2:3}{2.3}{Two-point Fréchet class} Use the parameter \(a=P(X=1,Y=1)\). Then the four probabilities are \(a\), \(p-a\), \(q-a\), and \(1-p-q+a\). Non-negativity gives the admissible interval. The law of \(X+Y\) follows by collecting masses at \(0,1,2\). \end{exercisehint}

\begin{exercisehint}{ex:2:4}{2.4}{Aggregation under three couplings} Use inverse-transform simulation. The same marginal quantile functions should be used in all three cases; only the dependence changes. Compare high empirical quantiles of \(X+Y\), not the marginal quantiles. \end{exercisehint}

\begin{exercisehint}{ex:2:5}{2.5}{Convexity of couplings} Interpret \((1-\theta)\gamma_0+\theta\gamma_1\) as: first flip a Bernoulli variable with probability \(\theta\), then sample from \(\gamma_1\) or \(\gamma_0\). It is a coupling because marginals are linear. It corresponds to a copula if the marginals are continuous, but not usually to a deterministic map. \end{exercisehint}

\begin{exercisehint}{ex:2:6}{2.6}{Worst-case expectation} With Bernoulli marginals, \(E[XY]=P(X=1,Y=1)=a\). Maximize and minimize \(a\) over the Fréchet interval. The endpoints are comonotonic and countermonotonic. \end{exercisehint}

\iffalse
% To be restored in v2 after the fairness material has been restored.
\begin{exercisehint}{ex:2:7}{2.7}{Coupling and fairness post-processing}
Emphasize that the coupling tells which raw-score quantiles or cells are sent to which repaired
score cells. Marginals only say before and after; the coupling is the audit trail.
\end{exercisehint}
\fi

\subsection*{Lecture 3: Discrete optimal transport}

\begin{exercisehint}{ex:3:1}{3.1}{The \(2\times2\) polytope} Use the upper-left entry \(t\). The remaining entries are forced by row and column sums. The objective is affine in \(t\), so the optimum is one endpoint unless the coefficient of \(t\) is zero. \end{exercisehint}

\begin{exercisehint}{ex:3:2}{3.2}{Duality by hand} The primal optimum is the diagonal plan. For dual potentials, try \(u=(0,0)\), \(v=(0,0)\), or shifted variants, and then check the inequalities \(u_i+v_j\le C_{ij}\). Complementary slackness only needs equality on positive-mass entries. \end{exercisehint}

\begin{exercisehint}{ex:3:3}{3.3}{Independence is usually not optimal} Take \(\bx=\by=(0,1)\) and uniform weights. Independence spreads mass over all four entries, while the diagonal plan has zero quadratic cost. \end{exercisehint}

\begin{exercisehint}{ex:3:4}{3.4}{No-crossing argument} For \(p=1\), draw the four points on the line and use the triangle inequality. For \(p>1\), use that \(|x-y|^p\) has the Monge property on ordered points, or differentiate the function obtained by sliding the two intervals. \end{exercisehint}

\begin{exercisehint}{ex:3:5}{3.5}{Worst-case stop-loss} Again parameterize the \(2\times2\) coupling by \(a=P(X=10,Y=10)\). Then compute \(S=X+Y\) and \((S-12)_+\). The worst case puts as much mass as possible on the pair \((10,10)\). \end{exercisehint}

\iffalse
% To be restored in v2 with the tariff-distribution material.
\begin{exercisehint}{ex:3:6}{3.6}{Tariff distribution versus policy disruption}
Use four policies and swap premiums pairwise. For example, old premiums \((1,2,3,4)\) and new
premiums \((2,1,4,3)\) have the same empirical distribution but every policy changes.
\end{exercisehint}
\fi

\begin{exercisehint}{ex:3:7}{3.7}{Empirical computation} For one-dimensional empirical laws with equal sample sizes, sort both samples and compute \(n^{-1}\sum_i |x_{(i)}-y_{(i)}|^p\). Multiplying one sample by \(1+r\) should produce an increasing curve in \(|r|\), with stronger tail effects for \(W_2\) than \(W_1\). \end{exercisehint}

\begin{exercisehint}{ex:3:8}{3.8}{Dual potentials as sensitivity} The dual potentials are subgradients of the optimal value with respect to marginal weights. Because total mass must remain one, moving a small mass from source cell \(k\) to source cell \(i\) changes the value approximately by \(u_i-u_k\). \end{exercisehint}

\subsection*{Lecture 4: Continuous optimal transport and Kantorovich duality}

\begin{exercisehint}{ex:4:1}{4.1}{Monge plans are Kantorovich plans} Use the definition of \((\Id,T)_\#\mu\). The first marginal is automatic; the second marginal is \(T_\#\mu\). Then apply the change-of-variables formula with \(h(x,y)=c(x,y)\). \end{exercisehint}

\begin{exercisehint}{ex:4:2}{4.2}{No deterministic splitting} A deterministic image of \(\delta_0\) is always a Dirac mass. For the Kantorovich coupling, take \(p\delta_{(0,0)}+(1-p)\delta_{(0,1)}\). \end{exercisehint}

\begin{exercisehint}{ex:4:3}{4.3}{Weak duality} Integrate \(\varphi(x)+\psi(y)\le c(x,y)\) with respect to \(\gamma\). Use the marginal constraints to replace the two integrals by \(\int\varphi\,\dd\mu\) and \(\int\psi\,\dd\nu\). For maximization of a payoff \(h\), apply the same argument to \(-h\) or reverse the bound accordingly. \end{exercisehint}

\begin{exercisehint}{ex:4:4}{4.4}{The \(c\)-transform} The inequality \(\varphi(x)+\varphi^c(y)\le c(x,y)\) is immediate from the definition of the infimum. If \((\varphi,\psi)\) is feasible, then \(\psi(y)\le c(x,y)-\varphi(x)\) for every \(x\), hence \(\psi(y)\le \inf_x\{c(x,y)-\varphi(x)\}\). \end{exercisehint}

\begin{exercisehint}{ex:4:5}{4.5}{Quantile transport} Show \(P(T(X)\le y)=P(F_\mu(X)\le F_\nu(y))\). For exponentials, \(F_\lambda(x)=1-e^{-\lambda x}\), so the map is \(T(x)=\lambda x/\theta\). \end{exercisehint}

\begin{exercisehint}{ex:4:6}{4.6}{One-dimensional \(W_1\)} Use the quantile formula with \(p=1\). For uniform laws, write the quantile functions as affine functions and integrate the absolute value of an affine function on \([0,1]\). If the intervals have the same length, this collapses to the distance between left endpoints. \end{exercisehint}

\begin{exercisehint}{ex:4:7}{4.7}{Gaussian transport} Write \(Q_\mu(u)=m_0+\sigma_0\Phi^{-1}(u)\) and \(Q_\nu(u)=m_1+\sigma_1\Phi^{-1}(u)\). Therefore \(T(x)=m_1+(\sigma_1/\sigma_0)(x-m_0)\). For \(W_2^2\), integrate the square of \((m_0-m_1)+(\sigma_0-\sigma_1)\Phi^{-1}(u)\); the cross term vanishes. \end{exercisehint}

\begin{exercisehint}{ex:4:8}{4.8}{Actuarial severity stress} The rank is \(F_\mu(x)\). The map preserves this rank by construction. The advantage is interpretability and monotone tail reshaping; the limitation is that it assumes rank preservation, which is not a causal fact. \end{exercisehint}

\subsection*{Lecture 5: Wasserstein distances, quantiles and barycenters}

\begin{exercisehint}{ex:5:1}{5.1}{Dirac and two-point laws} The only mass that must move is the mass \(1-p\) from \(0\) to \(a\). For \(W_p^p\), the cost is \((1-p)a^p\). \end{exercisehint}

\begin{exercisehint}{ex:5:2}{5.2}{Uniform laws} Use \(Q_\mu(u)=u\) and \(Q_\nu(u)=au\). Then \(W_p^p=\int_0^1 |1-a|^p u^p\,\dd u=|1-a|^p/(p+1)\). \end{exercisehint}

\begin{exercisehint}{ex:5:3}{5.3}{TVaR stability} Write TVaR as an average of quantiles on \([\alpha,1]\). Apply Hölder's inequality to \(Q_\mu-Q_\nu\) on that interval, then bound by the full \(L^p(0,1)\) norm. \end{exercisehint}

\begin{exercisehint}{ex:5:4}{5.4}{Gaussian laws} Use the same calculation as \hyperref[ex:4:7]{Exercise~4.7}. The one-dimensional Gaussian formula is a useful check on the multivariate trace formula. \end{exercisehint}

\begin{exercisehint}{ex:5:5}{5.5}{Barycenter of two distributions} In quantile coordinates, the problem is pointwise least squares: \(\min_q (1-\lambda)|q-Q_0(u)|^2+\lambda|q-Q_1(u)|^2\). The minimizer is the weighted average of quantiles. \end{exercisehint}

\begin{exercisehint}{ex:5:6}{5.6}{Mixture versus barycenter} The mixture is \(\frac12\delta_0+\frac12\delta_2\). The barycenter has quantile \(\frac12 Q_1(u)+\frac12 Q_2(u)=1\), hence is \(\delta_1\). This illustrates heterogeneity versus centrality. \end{exercisehint}

\begin{exercisehint}{ex:5:7}{5.7}{Empirical implementation} Use a common probability grid \(u_k\). Compute empirical quantiles for each sample, then take means of \(|Q_x(u_k)-Q_y(u_k)|^p\). Check sensitivity to the grid and to tail trimming. \end{exercisehint}

\begin{exercisehint}{ex:5:8}{5.8}{Vector-rank stress} Use the definition of push-forward twice. If \((R_\mu)_{\push}\mu=\mu_0\) and \((Q_\nu)_{\push}\mu_0=\nu\), then \((Q_\nu\circ R_\mu)_{\push}\mu=\nu\). The reference law \(\mu_0\) fixes what rank means in dimension \(d>1\), so it is a modelling choice rather than a canonical object. \end{exercisehint}

\subsection*{Lecture 6: Entropic regularization and Sinkhorn algorithms}

\begin{exercisehint}{ex:6:1}{6.1}{Scaling form} Write first-order conditions for the Lagrangian with row and column multipliers. The derivative of \(\gamma\log\gamma-\gamma\) is \(\log\gamma\). Exponentiating gives \(\gamma_{ij}=u_i K_{ij}v_j\), with \(K_{ij}=e^{-C_{ij}/\varepsilon}\). \end{exercisehint}

\begin{exercisehint}{ex:6:2}{6.2}{Two-by-two Sinkhorn} Do the row scaling and column scaling explicitly. This exercise is mainly about seeing that Sinkhorn alternates between satisfying one marginal and disturbing the other, with convergence through repeated rescaling. \end{exercisehint}

\begin{exercisehint}{ex:6:3}{6.3}{Independent coupling limit} As \(\varepsilon\to\infty\), \(K_{ij}=e^{-C_{ij}/\varepsilon}\to1\). Scaling a matrix of ones to marginals \((\ba,\bb)\) yields \(\ba\bb^\top\), the independent coupling. \end{exercisehint}

\begin{exercisehint}{ex:6:4}{6.4}{Bias and self-cost} Compute entropic OT between a measure and itself. Unless \(\varepsilon\) is very small, the regularized plan is diffuse and not concentrated on the diagonal, so the regularized self-cost is positive. The Sinkhorn divergence subtracts this bias. \end{exercisehint}

\begin{exercisehint}{ex:6:5}{6.5}{Implementation} Start with the transparent update \(\bu=\ba/(\bK\bv)\), \(\bv=\bb/(\bK^\top\bu)\), understood componentwise. Then compare row and column residuals. Use log-domain updates if costs are large or \(\varepsilon\) is small. \end{exercisehint}

\iffalse
% To be restored in v2 after submission of the portfolio mutualization and demutualization paper.
\begin{exercisehint}{ex:6:6}{6.6}{Premium drift}
Use premiums as one-dimensional support points or empirical samples. Compare exact sorting
\(W_p\) to entropic OT on a grid. The regularized value should be smoother and biased upward
unless debiased.
\end{exercisehint}
\fi

\begin{exercisehint}{ex:6:7}{6.7}{Unbalanced prevention} Let the prevention program reduce total expected loss or exposure. Balanced OT forces all mass to match; unbalanced OT permits part of the mass to disappear. Interpret disappearance as prevented exposure or avoided loss mass. \end{exercisehint}

\iffalse
% To be restored in v2 after submission of the multi-sensitive fairness paper.
\begin{exercisehint}{ex:6:8}{6.8}{Fairness correction}
Use Sinkhorn to transport each group distribution to a target barycenter. The smaller
\(\varepsilon\), the closer to exact OT; the larger \(\varepsilon\), the smoother and more randomized
the correction.
\end{exercisehint}
\fi

\subsection*{Lecture 7: Statistical optimal transport}

\begin{exercisehint}{ex:7:1}{7.1}{Empirical consistency} Use the triangle inequality: \(W_p(\mu_n,\nu_n)\le W_p(\mu_n,\mu)+W_p(\mu,\nu)+W_p(\nu,\nu_n)\), and similarly in the other direction. Conclude by convergence of the two empirical laws. \end{exercisehint}

\begin{exercisehint}{ex:7:2}{7.2}{One-dimensional computation} Sort the samples or use empirical quantiles on a common grid. The key is that in dimension one OT does not require solving a linear program. \end{exercisehint}

\begin{exercisehint}{ex:7:3}{7.3}{Noise floor under the null} Repeated random splits estimate the sampling distribution of \(W_1\) when the two samples come from the same population. This is the empirical noise floor: observed distances below this level are not strong evidence of drift. \end{exercisehint}

\iffalse
% To be restored in v2 with the premium-distribution material.
\begin{exercisehint}{ex:7:4}{7.4}{Permutation test for premium drift}
Pool the two premium samples, randomly permute date labels, recompute the distance, and
compare the observed distance to the permutation distribution. The null hypothesis is equality
of distributions.
\end{exercisehint}
\fi

\iffalse
% To be restored in v2 with the premium-distribution material.
\begin{exercisehint}{ex:7:5}{7.5}{Bootstrap for a tariff revision}
Resample policies, keeping old and new premiums paired. For each bootstrap sample, recompute
the paired disruption and the distributional Wasserstein distance. Pairing matters because both
tariffs are evaluated on the same policies.
\end{exercisehint}
\fi

\iffalse
% To be restored in v2 with the fairness material.
\begin{exercisehint}{ex:7:6}{7.6}{Small group fairness}
A small group has high sampling variability. Use bootstrap confidence bands or permutation
tests before interpreting a large empirical Wasserstein distance as structural disparity.
\end{exercisehint}
\fi

\begin{exercisehint}{ex:7:7}{7.7}{Sliced Wasserstein for reserve cash flows} Project each ten-year vector onto random directions \(\btheta\), compute one-dimensional Wasserstein distances, and average over directions. Increasing the number of projections reduces Monte Carlo variability. \end{exercisehint}

\begin{exercisehint}{ex:7:8}{7.8}{Entropic smoothing} Compare exact empirical OT with entropic OT for several \(\varepsilon\). Small \(\varepsilon\) has low bias but high numerical instability; large \(\varepsilon\) is smoother but more biased. \end{exercisehint}

\subsection*{Lecture 8: Risk measures, quantiles and Wasserstein robustness}

\begin{exercisehint}{ex:8:1}{8.1}{Robust mean under \(W_1\)} Use the Kantorovich--Rubinstein dual or simply note that \(h(x)=x\) is 1-Lipschitz on \(\R\). The upper bound is mean plus radius, and it is approached by shifting mass upward. \end{exercisehint}

\begin{exercisehint}{ex:8:2}{8.2}{Sensitivity of robust TVaR} Apply the stated Lipschitz inequality and take the supremum over \(W_1(\nu,\mu)\le r\). The factor \((1-\alpha)^{-1}\) explains the explosion as \(\alpha\uparrow1\). \end{exercisehint}

\begin{exercisehint}{ex:8:3}{8.3}{Radius by bootstrap} Resample the losses with replacement, compute \(W_1(\widehat\mu_n^*,\widehat\mu_n)\), and take a high quantile. Repeat for several \(n\) to see the radius decrease with sample size. \end{exercisehint}

\begin{exercisehint}{ex:8:4}{8.4}{Pricing versus capital ambiguity} The pricing ambiguity set usually lives on covariates \(\bX\); the capital ambiguity set lives on losses \(L\) or aggregate losses. Give an example where \(\Law(\bX)\) changes but the conditional model \(\Law(L\mid \bX)\) is unchanged. \end{exercisehint}

\begin{exercisehint}{ex:8:5}{8.5}{Ground cost} Write two costs, e.g. \(c((A,V),(A',V'))=\lambda_A|A-A'|+\lambda_V|\log V-\log V'|\). Changing the weights changes which perturbations are considered cheap. \end{exercisehint}

\begin{exercisehint}{ex:8:6}{8.6}{Expectation and \(W_1\)} For any coupling \((X,Y)\), use \(|\E X-\E Y|\le \E|X-Y|\). Then minimize over couplings. This is the primal proof of the 1-Lipschitz stability of the mean. \end{exercisehint}

\begin{exercisehint}{ex:8:7}{8.7}{TVaR Lipschitz constant} Use the quantile representation of \(\TVaR_\alpha\), then Hölder on \([\alpha,1]\). The constant is \((1-\alpha)^{-1/p}\) for \(W_p\). \end{exercisehint}

\begin{exercisehint}{ex:8:8}{8.8}{Robust TVaR} Write the adversarial quantile perturbation as a nonnegative function on \([\alpha,1]\) subject to an \(L^p\) budget. Hölder identifies the worst direction: concentrate according to the dual norm. \end{exercisehint}

\begin{exercisehint}{ex:8:9}{8.9}{VaR sensitivity} Choose \(\varepsilon\) just above or below \(1-\alpha\). Then a tiny amount of mass at a very large loss can leave weak convergence almost unchanged while making \(\VaR_\alpha\) jump. \end{exercisehint}

\begin{exercisehint}{ex:8:10}{8.10}{Stop-loss stability} The map \(x\mapsto(x-r)_+\) is 1-Lipschitz. Apply the primal coupling inequality or the Kantorovich--Rubinstein dual. \end{exercisehint}

\begin{exercisehint}{ex:8:11}{8.11}{Dependence-robust stop-loss} Represent the expectation as an integral over a coupling of the marginal quantiles. The objective is linear in the coupling, so the problem is a transport problem with reward \((Q_1(u)+Q_2(v)-d)_+\). \end{exercisehint}

\begin{exercisehint}{ex:8:12}{8.12}{Computational exercise} Use empirical quantiles for \(\VaR\) and tail averages for \(\TVaR\). For robustness, plot the bounds as functions of the radius rather than reporting only one number. \end{exercisehint}

\subsection*{Lecture 9: Pricing, segmentation and mutualization}

\begin{exercisehint}{ex:9:1}{9.1}{Variance and Wasserstein distance} The optimal coupling between \(\Law(Z)\) and \(\delta_{\bar z}\) is forced: pair \(Z\) with the constant \(\bar z\). Therefore \(W_2^2(\Law(Z),\delta_{\bar z})=\E[(Z-\bar z)^2]\). \end{exercisehint}

\begin{exercisehint}{ex:9:2}{9.2}{Nested rating information} Use the Hilbert-space projection property of conditional expectation. Since \(\mathcal G_1\subseteq \mathcal G_2\), the \(L^2\) projection on the larger space has smaller residual variance. \end{exercisehint}

\begin{exercisehint}{ex:9:3}{9.3}{Shrinkage and distance to pooling} The random variable \(\Pi_\lambda-\bar\pi=\lambda(\Pi-\bar\pi)\). Use homogeneity of \(W_p\) under scaling around the same center, or compute the quantile functions directly. \end{exercisehint}

\begin{exercisehint}{ex:9:4}{9.4}{Policy-level versus distributional movement} The paired coupling \((\Pi_0,\Pi_1)\) is feasible for the two premium laws. Therefore the Wasserstein distance is bounded above by the paired \(L^p\) displacement. Equality need not hold. \end{exercisehint}

\begin{exercisehint}{ex:9:5}{9.5}{Two-class tariff} Write the tariff distribution as a two-point law. The distance to the pooled premium is forced by coupling each class premium to the common mean. \end{exercisehint}

\begin{exercisehint}{ex:9:6}{9.6}{Affordability distribution} Work with \(A=\Pi/Y\). A tariff change induces a push-forward of the joint law of \((\Pi,Y)\). Compare affordability distributions, not only premium distributions. \end{exercisehint}

\begin{exercisehint}{ex:9:7}{9.7}{Premium caps} The map \(x\mapsto \min(x,M)\) is 1-Lipschitz and equals the identity below \(M\). Use the coupling \((\Pi,\Pi_M)\) to bound the distance by \(\E[(\Pi-M)_+^p]^{1/p}\). \end{exercisehint}

\iffalse
% To be restored in v2 after submission of the multi-sensitive fairness paper.
\begin{exercisehint}{ex:9:8}{9.8}{Computational exercise}
Track three curves: predictive accuracy, Wasserstein distance to the pooled benchmark, and
protected-group distributional disparity. The exercise is about the trade-off, not the specific
simulated coefficients.
\end{exercisehint}
\fi

\subsection*{Lecture 10: Portfolio drift, calibration and domain adaptation}

\begin{exercisehint}{ex:10:1}{10.1}{Covariate shift and expected loss} Use the law of total expectation. Under covariate shift, \(m(\bx)=\E[L\mid \bX=\bx]\) is unchanged, so only the marginal law of \(X\) changes. \end{exercisehint}

\begin{exercisehint}{ex:10:2}{10.2}{Push-forward drift} If \(f\) is \(K\)-Lipschitz, then \(|f(x)-f(y)|\le Kd(x,y)\). Apply this inequality inside any coupling and minimize over couplings. \end{exercisehint}

\iffalse
% To be restored in v2 with the premium-distribution and portfolio demutualization material.
\begin{exercisehint}{ex:10:3}{10.3}{Premium drift decomposition}
Add and subtract a common term, such as the old tariff applied to the new portfolio. This
separates model/tariff change from portfolio-composition change.
\end{exercisehint}
\fi

\begin{exercisehint}{ex:10:4}{10.4}{Transport-transfer predictor} Use barycentric projection: for each target point, average the source responses according to the conditional transport weights. Check whether the plan is source-to-target or target-to-source before writing the weights. \end{exercisehint}

\begin{exercisehint}{ex:10:5}{10.5}{Residual drift} Compute residual distributions by time period. Drift in fitted means and drift in residuals answer different questions: the first concerns portfolio mix, the second concerns conditional model validity. \end{exercisehint}

\begin{exercisehint}{ex:10:6}{10.6}{Choice of metric} Standardize continuous variables, use a substitution cost for categorical variables, and choose weights according to actuarial materiality. Explain how geography or vehicle value can dominate the metric if left unscaled. \end{exercisehint}

\begin{exercisehint}{ex:10:7}{10.7}{Scenario transport} Since \(L=\exp(H)\), the map from hazard to loss is not globally Lipschitz. Work either on a bounded hazard range or use local Lipschitz constants; then compare hazard-space and loss-space displacements. \end{exercisehint}

\subsection*{Lecture 11: Reserving, cash-flow distributions and climate prevention}

\begin{exercisehint}{ex:11:1}{11.1}{Best estimate versus predictive law} Construct two distributions with the same mean but different variance or tail, e.g. a Dirac mass at the mean and a two-point distribution around it. The best estimate is the same; capital is not. \end{exercisehint}

\begin{exercisehint}{ex:11:2}{11.2}{Aggregation contraction} For \(p=2\), use that the sum map \(a(\br)=\sum_t r_t\) is Lipschitz with constant \(\sqrt T\) under the Euclidean norm. Then apply push-forward contraction of Wasserstein distances under Lipschitz maps. \end{exercisehint}

\begin{exercisehint}{ex:11:3}{11.3}{Reserve barycenter} In dimension one, average quantile functions. For three reserve laws, the barycenter quantile is \(\sum_k \lambda_k Q_k(u)\). \end{exercisehint}

\begin{exercisehint}{ex:11:4}{11.4}{Inflation stress} The map is diagonal: \(T_{\ba}(\br)=(a_1r_1,a_2r_2,a_3r_3)\). Its Lipschitz constant is \(\max_t a_t\) for the Euclidean norm. Use this to bound the Wasserstein movement of cash-flow vectors. \end{exercisehint}

\begin{exercisehint}{ex:11:5}{11.5}{Prevention profile} Compute the transformed tail by splitting \(x\le M\) and \(x>M\). The map \(T\) caps or reduces large losses, so compare its effect on the mean, VaR and TVaR. \end{exercisehint}

\begin{exercisehint}{ex:11:6}{11.6}{Balanced versus unbalanced comparison} Use an example where the normalized cash-flow shape is identical but total payment volume changes. Balanced OT sees no shape change; unbalanced OT sees volume change. \end{exercisehint}

\begin{exercisehint}{ex:11:7}{11.7}{Mini-project} Keep the project simple: simulate two sets of future cash-flow vectors, compute distances for total reserves and for full vectors, then compare what is lost by aggregation. \end{exercisehint}

\subsection*{Lecture 12: Fairness, discrimination and Wasserstein projections}

\begin{exercisehint}{ex:12:1}{12.1}{Two-group barycenter} Use the one-dimensional barycenter formula: \(\bar Q=(1-p)Q_0+pQ_1\). The group repair maps are \(T_s=\bar Q\circ F_s\). \end{exercisehint}

\begin{exercisehint}{ex:12:2}{12.2}{Gaussian score repair} For Gaussian groups, the one-dimensional OT map is affine. Send each group mean and standard deviation to the barycenter mean and standard deviation. \end{exercisehint}

\begin{exercisehint}{ex:12:3}{12.3}{Calibration tension} Use the fact that distributional parity of scores and group calibration impose restrictions on the joint law of \((Y,Z,S)\). Unless base rates coincide or the score is uninformative/perfect, both conditions cannot generally hold. \end{exercisehint}

\begin{exercisehint}{ex:12:4}{12.4}{Partial repair path} The map \(Z^{(\lambda)}=(1-\lambda)Z+\lambda T_S(Z)\) is a displacement interpolation in one dimension when \(T_S\) is monotone. Distances to the target scale by \((1-\lambda)\). \end{exercisehint}

\begin{exercisehint}{ex:12:5}{12.5}{Empirical implementation} Sort scores within each group, compute a barycenter quantile grid, then map each observation through its empirical rank. Check ties carefully. \end{exercisehint}

\begin{exercisehint}{ex:12:6}{12.6}{Residual audit} A score repair changes \(Z\), hence changes residuals \(Y-Z\). Fair score distributions do not imply fair residual or loss-ratio distributions. Compare both before and after repair. \end{exercisehint}

\begin{exercisehint}{ex:12:7}{12.7}{Conditional disparity} Condition on each admissible risk class \(A=a\). Compute Wasserstein disparities within cells, then aggregate with cell weights. This separates distributional disparity from legitimate risk-class mix differences. \end{exercisehint}

\subsection*{Lecture 13: Reinsurance, capital allocation and multi-marginal transport}

\begin{exercisehint}{ex:13:1}{13.1}{Push-forward of a quota-share treaty} A quota-share treaty is the map \(T_q(L)=qL\) or \((1-q)L\), depending on whether you track ceded or retained loss. The law is the push-forward by a scaling map; moments scale accordingly. \end{exercisehint}

\begin{exercisehint}{ex:13:2}{13.2}{Layer distribution} For \(C=\min\{(L-a)_+,\ell\}\), identify three regions: \(L\le a\), \(a<L<a+\ell\), and \(L\ge a+\ell\). Atoms occur at \(0\) and \(\ell\). Use the stop-loss identity to compute the mean. \end{exercisehint}

\begin{exercisehint}{ex:13:3}{13.3}{Worst-case expected layer loss} For discrete marginals, write a transportation matrix and maximize a linear objective with reward \(\min\{(x_i+y_j-a)_+,\ell\}\). This is a discrete OT problem with negative cost. \end{exercisehint}

\begin{exercisehint}{ex:13:4}{13.4}{Comonotonic aggregation} With \(X_k=Q_k(U)\), the aggregate quantile is the sum of marginal quantiles when the quantile functions are continuous at the level considered. This is the defining rank-aligned dependence. \end{exercisehint}

\begin{exercisehint}{ex:13:5}{13.5}{Expected shortfall and multi-marginal transport} For fixed \(t\), the objective \(E[(S-t)_+]\) is an expectation over a multi-marginal coupling. Solve or approximate that problem for several \(t\), then minimize over \(t\). Tail numerical difficulty comes from rare but high-impact configurations. \end{exercisehint}

\begin{exercisehint}{ex:13:6}{13.6}{Euler allocation under two dependence models} Estimate conditional expectations given tail events. Under comonotonicity, large lines are large together, so allocations tend to reflect aligned marginal tail heaviness. \end{exercisehint}

\begin{exercisehint}{ex:13:7}{13.7}{Treaty sensitivity} The map \(L\mapsto\min(L,a)\) changes the whole distribution smoothly in \(a\), whereas \(\VaR_\alpha\) reacts to whether the cap crosses the relevant quantile. Thus the two curves need not have the same shape. \end{exercisehint}

\begin{exercisehint}{ex:13:8}{13.8}{Scenario reweighting} Let \(w_i\ge0\), \(\sum_i w_i=1\). Maximize \(\sum_i w_i \ell(\bx^{(i)})\) subject to moment constraints. Add \(-\eta\sum_i w_i\log w_i\) to penalize concentration, or equivalently constrain KL divergence from uniform weights. \end{exercisehint}

\subsection*{Lecture 14: Counterfactual worlds and transport}

\begin{exercisehint}{ex:14:1}{14.1}{Two-point counterfactual worlds} Use the Bernoulli Fréchet interval again. The counterfactual coupling is determined by \(a=P(X=1,Y=1)\); endpoints give maximal persistence or maximal switching. \end{exercisehint}

\begin{exercisehint}{ex:14:2}{14.2}{Non-identification} Take the identity coupling \(Y=X\) and the flip coupling \(Y=1-X\) when both marginals are Bernoulli\((1/2)\). They have the same marginals but opposite individual counterfactual stories. \end{exercisehint}

\begin{exercisehint}{ex:14:3}{14.3}{Premium recourse under a linear tariff} Project the point \(\bx\) onto the half-space \(\{\bz:m(\bz)\le p^\star\}\). The Euclidean projection formula is along the normal vector \(\bbeta\). \end{exercisehint}

\begin{exercisehint}{ex:14:4}{14.4}{Quantile transport between normal laws} Use the same affine map as in \hyperref[ex:4:7]{Exercise~4.7}. Interpret it as rank-preserving: standardized normal scores are kept fixed. \end{exercisehint}

\iffalse
% To be restored in v2 with the fairness counterfactual material.
\begin{exercisehint}{ex:14:5}{14.5}{Transport counterfactual fairness}
Compute the barycenter of the two Gaussian score distributions, then write the group-specific
affine maps to that barycenter. Check whether means and variances align after the maps.
\end{exercisehint}
\fi

\begin{exercisehint}{ex:14:6}{14.6}{Counterfactual catastrophe catalogue} Compute expected counterfactual loss for each event and for the catalogue. For tail quantities, combine all event-loss atoms with their probabilities and sort them. \end{exercisehint}

\begin{exercisehint}{ex:14:7}{14.7}{Prevention and coupling} The marginals alone do not identify which original \(100\)-loss states become \(80\)-loss states. Construct at least two couplings with the same before/after laws and compare their individual stories. \end{exercisehint}

\begin{exercisehint}{ex:14:8}{14.8}{Hard actionability constraints} Only optimize over \(\bx_A\); keep \(\bx_N\) fixed. In a linear model, this is the projection of the actionable subvector onto a half-space whose normal is the actionable coefficient vector. \end{exercisehint}

\begin{exercisehint}{ex:14:9}{14.9}{Sequential counterfactuals} Apply the first transport map to \(\bX_1\), then update \(\bX_2\) conditionally or through the graph mechanism. The main point is that changing parent variables and then children differs from transporting all marginals independently. \end{exercisehint}

\subsection*{Lecture 15: Dynamic optimal transport and actuarial trajectories}

\begin{exercisehint}{ex:15:1}{15.1}{Inflation velocity} Differentiate \(L_t=a_tL_0\). The velocity in state space is \(\dot a_t/a_t\) times the current position. The action involves \(\E[L_t^2]\). \end{exercisehint}

\begin{exercisehint}{ex:15:2}{15.2}{Quantile geodesic} In one dimension, set \(Q_t=(1-t)Q_0+tQ_1\). Then \(\|Q_t-Q_s\|_{L^2}=|t-s|\|Q_1-Q_0\|_{L^2}\), which is the constant-speed geodesic property. \end{exercisehint}

\begin{exercisehint}{ex:15:3}{15.3}{Uniform-to-uniform trajectory} The optimal map from \([0,1]\) to \([0,2]\) is \(T(x)=2x\). Hence \(T_t(x)=(1+t)x\), so \(\mu_t=U[0,1+t]\). \end{exercisehint}

\begin{exercisehint}{ex:15:4}{15.4}{Dirac-to-two-point target} No deterministic map can split \(\delta_0\). The Kantorovich plan can split mass from \(0\) to \(-1\) and \(1\). Dynamic paths may require stochastic branching. \end{exercisehint}

\begin{exercisehint}{ex:15:5}{15.5}{TVaR along a geodesic} TVaR is linear in the quantile function. Substitute \(Q_t=(1-t)Q_0+tQ_1\) into the integral over \([\alpha,1]\). \end{exercisehint}

\begin{exercisehint}{ex:15:6}{15.6}{Gaussian covariance path} In one dimension, Gaussian quantiles are \(\sigma_i\Phi^{-1}(u)\). The geodesic quantile is \(((1-t)\sigma_0+t\sigma_1)\Phi^{-1}(u)\), so standard deviations interpolate linearly, not variances. \end{exercisehint}

\begin{exercisehint}{ex:15:7}{15.7}{Path length and churn} Take \(\mu_0=\mu_2=\delta_0\) and \(\mu_1=\delta_1\). Endpoint distance is zero, but the path length through \(\mu_1\) is positive. Interpret as movement out and back. \end{exercisehint}

\iffalse
% To be restored in v2 with the premium-distribution and demutualization material.
\begin{exercisehint}{ex:15:8}{15.8}{Tariff phase-in}
If \(P_1=T(P_0)\) is induced by monotone OT, then \(P_t=(1-t)P_0+tP_1\) is the displacement
interpolation. Randomly switching a fraction of policies gives a mixture path instead.
\end{exercisehint}
\fi

\begin{exercisehint}{ex:15:9}{15.9}{Discrete dynamic action} Use the triangle inequality \(W_2(\mu_0,\mu_T)\le\sum_t W_2(\mu_t,\mu_{t+1})\), then Cauchy--Schwarz: \((\sum_t a_t)^2\le T\sum_t a_t^2\). \end{exercisehint}

\iffalse
% To be restored in v2 with the fairness material.
\begin{exercisehint}{ex:15:10}{15.10}{Fairness repair path}
Compute the target distance in quantile coordinates:
\(\bar Q-Q_{s,t}=(1-t)(\bar Q-Q_s)\). Square and integrate.
\end{exercisehint}
\fi

\begin{exercisehint}{ex:15:11}{15.11}{Exposure growth} Balanced OT between normalized laws reports no change if composition is unchanged. It loses total exposure. Use finite measures \(\alpha_t=E_t\mu_t\) and compare with unbalanced OT or report exposure-scaled mass separately. \end{exercisehint}

\begin{exercisehint}{ex:15:12}{15.12}{Mini-lab extension} The Pareto target should mostly affect high quantiles and TVaR. Plot the curves over \(t\); the mean may move moderately while \(\VaR_{0.995}\) and \(\TVaR_{0.99}\) move sharply. \end{exercisehint}

\subsection*{Lecture 16: Unbalanced transport, exposure and portfolio growth}

\begin{exercisehint}{ex:16:1}{16.1}{Finite-measure KL} Use the finite-measure entropy formula \(\KL(\rho\mid\alpha)=q\log(q/m)-q+m\), with the usual conventions at \(q=0\). \end{exercisehint}

\begin{exercisehint}{ex:16:2}{16.2}{One-point unbalanced OT} The only transport variable is the transported mass \(\gamma\ge0\). Write the objective as \(c_0\gamma+\tau_1\KL(\gamma\mid m)+\tau_2\KL(\gamma\mid n)\), then differentiate in \(\gamma\). \end{exercisehint}

\begin{exercisehint}{ex:16:3}{16.3}{Balanced limit} As the marginal penalties go to infinity, deviations of the transported marginals from \(\alpha\) and \(\beta\) become infinitely expensive. Any finite-limit minimizer must satisfy the balanced marginal constraints. \end{exercisehint}

\begin{exercisehint}{ex:16:4}{16.4}{Normalization can hide growth} The normalized laws are both \(\delta_0\), so balanced OT reports zero. The finite measures have different total masses, so unbalanced OT or a mass-difference diagnostic detects growth. \end{exercisehint}

\begin{exercisehint}{ex:16:5}{16.5}{Two-cell exposure} Write the transport matrix and compare row/column sums to the finite masses. Report which cell is transported, which source mass is unused, and which target mass is newly created. \end{exercisehint}

\begin{exercisehint}{ex:16:6}{16.6}{Generalized Sinkhorn} Start from the scaling form \(\Gamma=\operatorname{diag}(\bu)\bK\operatorname{diag}(\bv)\). The KL marginal penalties yield damped updates with exponents \(\theta_k=\tau_k/(\tau_k+\varepsilon)\). \end{exercisehint}

\begin{exercisehint}{ex:16:7}{16.7}{Penalty sensitivity} Plot transported mass, source usage and target explanation. As \(\tau\) increases, the solution should behave more like balanced transport; as \(\tau\) decreases, creation/destruction becomes cheaper. \end{exercisehint}

\begin{exercisehint}{ex:16:8}{16.8}{Asymmetric penalties} If \(\tau_1\) is large and \(\tau_2\) small, losing source mass is expensive but creating target mass is cheap. Reverse the interpretation when \(\tau_1\) is small and \(\tau_2\) large. \end{exercisehint}

\begin{exercisehint}{ex:16:9}{16.9}{Claim-intensity measures} The normalized severities are identical, but the finite intensity measures differ by total mass. Balanced OT sees zero severity-shape change; unbalanced comparison sees frequency growth. \end{exercisehint}

\begin{exercisehint}{ex:16:10}{16.10}{Climate exposure} Use three zones with both spatial movement and total insured-value growth. Let one coastal zone receive new target mass; unbalanced OT should interpret part of it as created exposure rather than migrated old exposure. \end{exercisehint}

\iffalse
% To be restored in v2 with the fairness material.
\begin{exercisehint}{ex:16:11}{16.11}{Fairness with group masses}
Separate two questions: distributional parity conditional on group membership and disparity in
group mass or exposure. Balanced repair handles conditional distributions; unbalanced repair can
also account for unequal masses.
\end{exercisehint}
\fi

\begin{exercisehint}{ex:16:12}{16.12}{Dynamic source term} Integrate the equation over space. The divergence term vanishes under no-flux or decay conditions, leaving \(\frac{d}{dt}|\rho_t|=\int \zeta_t\,\dd\rho_t\). Interpret this as growth or lapse of total exposure. \end{exercisehint}

\subsection*{Lecture 17: Life, mortality and longevity transport}

\begin{exercisehint}{ex:17:1}{17.1}{Life table as a measure} Use survival to the beginning of each curtate year. The mass at \(k\) is \(p_x p_{x+1}\cdots p_{x+k-1}q_{x+k}\), with the empty product equal to one. If the table is closed at \(\omega\), add the remaining survivor mass to the final age. \end{exercisehint}

\begin{exercisehint}{ex:17:2}{17.2}{Residual lifetime Wasserstein distance} Use the lower generalized inverse of the two discrete distributions. On \((0,1)\), identify the intervals on which each quantile is \(0\), \(1\), or \(2\), then integrate the absolute difference of the two step functions. \end{exercisehint}

\begin{exercisehint}{ex:17:3}{17.3}{Quantile longevity stress} For an exponential law, \(F_\lambda(t)=1-e^{-\lambda t}\) and \(Q_\lambda(u)=-\log(1-u)/\lambda\). Hence \(T(t)=Q_{\lambda_1}(F_{\lambda_0}(t))=(\lambda_0/\lambda_1)t\). If \(\lambda_1<\lambda_0\), the map stretches lifetimes. \end{exercisehint}

\begin{exercisehint}{ex:17:4}{17.4}{Annuity push-forward} If \(p_k=\P(K_x=k)\), then the law of \(g(K_x)\) is \(\sum_k p_k\delta_{g(k)}\). If several \(k\)'s give the same present value, collect the corresponding masses. This is exactly the definition of a push-forward by \(g\). \end{exercisehint}

\begin{exercisehint}{ex:17:5}{17.5}{Opposite effects} The annuity present value \(g(k)=\sum_{j=0}^k v^j\) is increasing in \(k\). A death benefit paid at the end of the year of death has present value \(v^{k+1}\), which is decreasing in \(k\) when \(0<v<1\). Apply first-order stochastic dominance to increasing and decreasing functions. \end{exercisehint}

\begin{exercisehint}{ex:17:6}{17.6}{Wasserstein barycenter} In dimension one, solve the barycenter problem pointwise in quantile space. The barycenter quantile is \(Q_\lambda=(1-\lambda)Q_0+\lambda Q_1\). Interpret this as a mortality scenario whose lifetime percentiles are blended rather than mixed at the model level. \end{exercisehint}

\begin{exercisehint}{ex:17:7}{17.7}{Liability push-forward} For lifetime vector \(\bK=(K_1,\ldots,K_n)\), write the portfolio present value as \(G(\bK)=\sum_i b_i\sum_{j=0}^{K_i}v^j\), or the appropriate contract-specific version. If \(\mu_{\bK}\) is the joint law of lifetimes, the liability law is \(G_{\push}\mu_{\bK}\). \end{exercisehint}

\begin{exercisehint}{ex:17:8}{17.8}{Joint-life dependence} The function \(\max(T_x,T_y)\) depends on the joint law, not only the two marginals. Under comonotonicity, high lifetime ranks occur together; under independence, one life can be long while the other is not. Last-survivor annuities are sensitive to this dependence. \end{exercisehint}

\begin{exercisehint}{ex:17:9}{17.9}{Tail-aware longevity} Define the tail distance by integrating quantile differences over \([0.95,1]\), with the normalizing factor used in the lecture. This focuses on the longest-lived percentiles, which drive old-age annuity risk more directly than a global average distance. \end{exercisehint}

\begin{exercisehint}{ex:17:10}{17.10}{Multi-state transport} Build a cost matrix with zero diagonal and larger costs for clinically or contractually more distant moves. For example, healthy to disabled may be cheaper than healthy to long-term care, while moves involving death may be absorbing or very costly depending on the modelling question. \end{exercisehint}

\begin{exercisehint}{ex:17:11}{17.11}{Prudent annuity price} Write \(\sup_{\nu:\Wone(\nu,\widehat\mu)\le r}\int g\,\dd\nu\). If \(g\) is \(K\)-Lipschitz, Kantorovich--Rubinstein gives the upper bound \(\int g\,\dd\widehat\mu+Kr\). The bound is conservative but transparent. \end{exercisehint}

\begin{exercisehint}{ex:17:12}{17.12}{Federated quantile sketches} Approximate \(\Wone\) by a Riemann sum of absolute differences between reported quantiles on the common grid. The sketch preserves marginal lifetime information at reported probabilities, but loses dependence between lives, health-state paths and within-bin tail shape. \end{exercisehint}

\subsection*{Appendix B: Cost engineering for actuarial OT}

\begin{exercisehint}{ex:B:1}{B.1}{Loss scale} For \(\nu=\Law((1+i)X)\), use the monotone coupling \((X,(1+i)X)\). The \(p\)-cost is \(i^p\E[X^p]\) for the raw loss metric. Compare this with a log-scale cost, where multiplicative changes become additive. \end{exercisehint}

\iffalse
% To be restored in v2 with the premium-cost material.
\begin{exercisehint}{ex:B:2}{B.2}{Premium asymmetry}
A symmetric cost treats increases and decreases equally. To model regulatory or commercial
asymmetry, use different slopes for \((y-x)_+\) and \((x-y)_+\), or use a piecewise-linear cost with
a steeper upward-premium penalty.
\end{exercisehint}
\fi

\begin{exercisehint}{ex:B:3}{B.3}{Categorical substitution matrix} Construct a symmetric matrix with zero diagonal. The triangle inequality is optional unless you need a true metric. Interpret low sedan--SUV cost and higher sedan--truck cost, for instance. \end{exercisehint}

\begin{exercisehint}{ex:B:simplex-cost}{B.4}{Simplex cost} Work inside the simplex. For probability vectors, compare Euclidean, Hellinger, log-ratio or Fisher-Rao type costs. State which parts of the simplex are allowed: log-ratio costs require positive components, while Euclidean costs can handle boundary points more directly. \end{exercisehint}

\iffalse
% To be restored in v2 with the protected-variable fairness material.
\begin{exercisehint}{ex:B:4}{B.4}{Protected variables}
Setting infinite cost for changing \(s\) forbids cross-group matching. Setting zero cost for changing
\(s\) may hide protected-group displacement. The choice is normative and should match the audit
question.
\end{exercisehint}
\fi

\begin{exercisehint}{ex:B:5}{B.5}{Geographical cost} Use shortest-path distance on the territory graph. This captures adjacency and travel across territories better than arbitrary numeric territory codes. \end{exercisehint}

\begin{exercisehint}{ex:B:6}{B.6}{Contract distance} Decompose contract terms into deductible, limit, coinsurance and endorsements. Use scaled absolute differences for numerical terms and substitution costs for categorical endorsements. \end{exercisehint}

\begin{exercisehint}{ex:B:7}{B.7}{Cash-flow shape} Take two vectors with the same sum but different timing, e.g. \((100,0,0)\) and \((0,0,100)\). A total-reserve metric sees no difference; a discounted or time-weighted vector metric does. \end{exercisehint}

\begin{exercisehint}{ex:B:8}{B.8}{Weight sensitivity} Plot the transport value and plan diagnostics as the regional weight changes. Large changes in matched pairs indicate that conclusions are cost-engineering sensitive. \end{exercisehint}

 \begin{exercisehint}{ex:B:gw-cost}{B.8}{Gromov--Wasserstein cost engineering} Do not try to define a direct cost between points in two unrelated spaces. Define within-space dissimilarities first, for example distances between classes inside each portfolio, then compare how well a coupling preserves these internal distances. The conclusion depends on those internal dissimilarities, so they must be documented. \end{exercisehint}

\begin{exercisehint}{ex:B:9}{B.9}{Cost report} A good report states the state space, units, scaling, categorical substitution choices, fixed or non-actionable variables, sensitivity analyses, and what the cost is not allowed to mean. Include at least one alternative cost specification. \end{exercisehint}

\subsection*{Appendix C: Deterministic mapping versus diffusion processes}

\begin{exercisehint}{ex:DM:1}{C.1}{Map versus kernel} For the map case, use the graph coupling \((\Id,T)_{\push}\mu_0\). For a non-map kernel, take a source atom and split it between two target states, or take any kernel \(K(\bx,\dd\by)\) that is not almost surely a Dirac mass. \end{exercisehint}

\begin{exercisehint}{ex:DM:2}{C.2}{Continuity equation} Test the law \(\mu_t\) against a smooth compactly supported function \(\varphi\). Differentiate \(\E[\varphi(\bX_t)]\), use \(\dot\bX_t=\bv_t(\bX_t)\), then integrate by parts to identify \(\partial_t\mu_t+\nabla\cdot(\mu_t\bv_t)=0\). \end{exercisehint}

\begin{exercisehint}{ex:DM:3}{C.3}{Fokker--Planck equation} Apply It\^o's formula to a smooth test function. For constant \(\sigma\), the weak form gives \(\partial_t\rho_t=-\nabla\cdot(b\rho_t)+(\sigma^2/2)\Delta\rho_t\). The deterministic continuity equation is recovered when \(\sigma=0\). \end{exercisehint}

\begin{exercisehint}{ex:DM:4}{C.4}{One-dimensional stress} Use the probability integral transform: \(F_0(X)\) is the rank of the baseline loss, and \(F_1^{-1}(F_0(X))\) has target law. Rank preservation is useful for ordered stress tests, but it rules out random branching and should not be read as a causal mechanism. \end{exercisehint}

\begin{exercisehint}{ex:DM:5}{C.5}{Same marginals, different paths} Construct one process by linear interpolation with small noise and another by moving out and back while keeping the same endpoints. Path-dependent quantities such as running maxima, hitting probabilities, cumulative losses or discounted cash-flow timing can distinguish them. \end{exercisehint}

\begin{exercisehint}{ex:DM:6}{C.6}{Scenario validation} Check at least marginal fit, tail fit, dependence, temporal dynamics, calibration to historical or expert constraints, stability under resampling and sensitivity to conditioning variables. For capital use, tail diagnostics matter more than average visual fit. \end{exercisehint}

\begin{exercisehint}{ex:DM:7}{C.7}{Tail-aware generation} A small average Wasserstein distance can coexist with poor extreme-tail behaviour. Compare empirical and generated \(\VaR\), \(\TVaR\), exceedance probabilities, tail quantile slopes or distances restricted to high quantile levels. \end{exercisehint}

\begin{exercisehint}{ex:DM:8}{C.8}{Schr\"odinger bridge interpretation} The bridge minimizes relative entropy from a reference path law subject to endpoint constraints. It changes the baseline generator only as much as needed, in KL terms, to match the prescribed initial and terminal laws. \end{exercisehint}

\begin{exercisehint}{ex:DM:9}{C.9}{Governance choice} Choose the modelling object first. Deterministic maps suit auditable stress transformations; ODE flows suit smooth deterministic evolution; diffusions suit random temporal mechanisms; Schr\"odinger bridges suit stochastic paths constrained by endpoint laws and a chosen reference dynamics. \end{exercisehint}

\begin{exercisehint}{ex:DM:10}{C.10}{Path time versus calendar time} Denoising time is an algorithmic time index. It need not correspond to accident year, development year or calendar year unless the model is trained and constrained that way. Confusing the two can create artificial claims-development or reserve-runoff interpretations. \end{exercisehint}

\begin{exercisehint}{ex:DM:11}{C.11}{Reference dynamics in a bridge} The endpoint constraints fix only the marginals at times \(0\) and \(1\). The reference law controls what intermediate paths are considered natural. Changing it can alter volatility, persistence, jumps or dependence even when endpoints are unchanged. \end{exercisehint}

\begin{exercisehint}{ex:DM:12}{C.12}{Mini-lab extension} A Pareto target changes the upper quantiles much more than the centre of the distribution. Compare deterministic and noisy paths using tail curves, \(\VaR_{99\%}\), \(\TVaR_{99\%}\) and the stability of simulated extremes across seeds. \end{exercisehint}

\label{LastPage}
\end{document}